\documentclass[12pt,a4paper,twoside]{book}

\usepackage{amssymb}
\usepackage{feynarts}
\usepackage[dvips]{graphicx,color}
\usepackage{cancel}
\usepackage{cite}
\usepackage[greek,english]{babel}
\usepackage[utf8x]{inputenc}
\usepackage{xspace}   
\usepackage{snapshot}

\def\URLtilde{\lower0.2em\hbox{$\tilde{\phantom{a}}$}}

\def\beq{\begin{equation}}
\def\eeq{\end{equation}}

\def\beqn{\begin{eqnarray}}
\def\eeqn{\end{eqnarray}}

\def\u#1{\tt#1\endgroup}
\def\red{
}
\def\black{
}
\def\mycomm#1{\hfill\break\strut\kern-3em{\red\tt ====> #1\black}\hfill\break}

\def\slashchar#1{\setbox0=\hbox{$#1$}           
   \dimen0=\wd0                                 
   \setbox1=\hbox{/} \dimen1=\wd1               
   \ifdim\dimen0>\dimen1                        
      \rlap{\hbox to \dimen0{\hfil/\hfil}}#1
   \else                                        
      \rlap{\hbox to \dimen1{\hfil$#1$\hfil}}/
       \fi}

\newcommand{\pmiss}{\ensuremath{\cancel{p}}}

\definecolor{red}{rgb}{1,0,0}
\definecolor{blue}{rgb}{0,0,1}
\definecolor{purple}{rgb}{.5,0,.5}
\definecolor{newpurple}{rgb}{.5,0,.3}

\def\mytitle{{Phenomenological aspects of new physics at high energy hadron colliders}}
\def\myname{{Andreas Papaefstathiou}}
\def\mycollege{{Robinson College}}

\usepackage{afterpage}
\usepackage[hang,small,bf]{caption}
\usepackage{fancyhdr}
\usepackage{rotating}
\usepackage{setspace}

\usepackage{calc}
\newcommand{\Herwigpp}{H\protect\scalebox{0.8}{ERWIG++}\xspace}
\newcommand{\bea}{\begin{eqnarray}}
\newcommand{\eea}{\end{eqnarray}}
\newcommand{\met}{\not \!\! E_T}
\newcommand{\etam}{\eta_{\rm max}}
\newcommand{\thc}{\theta_c}
\newcommand{\alps}{\alpha _s}
\newcommand{\as}{\alpha _s}
\def\vecex#1{\mbox{\boldmath\scriptsize $#1$}}
\def\beeq{\begin{eqnarray}}
\def\res{{\rm res.}}
\def\nn{\nonumber}
\def\eeeq{\end{eqnarray}}
\def\ms{${\overline {\rm MS}}$}
\def\msbar{${\overline {\rm MS}}$\xspace}
\def\tC{\widetilde{C}}
\def\mrd{\mathrm{d}}
\def\etamax{\eta_{\mbox{\scriptsize max}}}
\def\rs{\sqrt{s}}

\renewcommand{\chaptermark}[1]{\markboth{\chaptername\ \thechapter.\ #1}{}}

\fancypagestyle{plain}{%
  \fancyhead{}  
  \fancyfoot[C]{\bfseries \thepage}  
  \renewcommand{\headrulewidth}{0pt}  
}

\newcommand{\Pythia}{P\protect\scalebox{0.8}{YTHIA}\xspace}
\def\ptmx{p_{\rm miss}^x\,}
\def\ptmy{p_{\rm miss}^y\,}
\def\pmiss{{\bf p}_{\rm miss}}

\def\gev{~{\rm GeV}}
\def\tev{~{\rm TeV}}

\def\gev{~{\rm GeV}}

\def\lsim{\mathrel{\raise.3ex\hbox{$<$\kern-.75em\lower1ex\hbox{$\sim$}}}}
\def\gsim{\mathrel{\raise.3ex\hbox{$>$\kern-.75em\lower1ex\hbox{$\sim$}}}}
\def\ifmath#1{\relax\ifmmode #1\else $#1$\fi}

\def\gtap{\lower .7ex\hbox{$\;\stackrel{\textstyle >}{\sim}\;$}}
\def\gmu{\gamma_{\mu}}
\def\gnu{\gamma_{\nu}}

\definecolor{orange}{cmyk}{0,0.7,1,0}
\definecolor{darkyellow}{cmyk}{0,0.4,1,0}
\definecolor{darkgreen}{cmyk}{1,0.4,1,0}
\definecolor{plum}{cmyk}{0.3,1,0,0}
\definecolor{pink}{cmyk}{0,1,0.5,0}
\definecolor{brown}{cmyk}{0.2,0.75,1,0}

\usepackage[todo,colour,check,front]{optional}
\usepackage{slashed}
\usepackage{multirow}
\usepackage{url}

\begin{document}
\frontmatter
\opt{front}{











\setcounter{tocdepth}{2} 
\setcounter{secnumdepth}{3} 


\begin{titlepage}
\begin{center}
  ${}^{}$ \\
  \vspace{20mm}
  {\Huge \bf \mytitle} \\
  \vspace{35mm} \large
  \myname\\
  ${}^{}$\\
  \mycollege\\
  \vspace{1.5cm}
  \includegraphics[height=3.0cm]{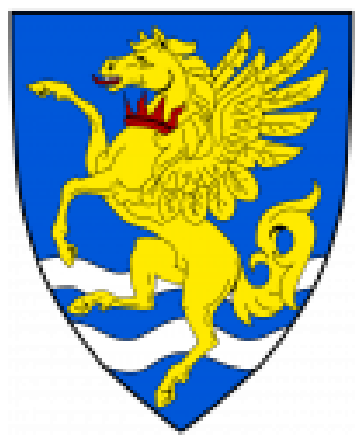}\\
  \vspace{4.0cm} \normalsize
  A dissertation submitted to the University of Cambridge \\
  for the degree of Doctor of Philosophy \\
  April 2011\\

  \vspace{2cm}
\end{center}
\end{titlepage}

\clearpage{\pagestyle{empty}\cleardoublepage}
\renewcommand{\baselinestretch}{1.3}\tiny\normalsize
\thispagestyle{plain}
\begin{center}
{\Large\bf \mytitle} \\
\vspace{0.2cm}
{\large \myname}\\
\vspace{0.2cm}
{\Large\bf Abstract} \\
\vspace{-0.1cm}
\end{center}
\normalsize 




This thesis contains studies of phenomenological aspects of new
physics at hadron colliders, such as the Large Hadron Collider
(LHC). After a general introduction in chapter~\ref{sec:introduction}, in
chapter~\ref{cha:smbsm} we outline the main features of the Standard Model
(SM) of particle physics and the theoretical motivations for going
beyond it. We subsequently provide brief
descriptions of a few popular models that aim to solve the issues that
arise within the SM.

In chapter~\ref{cha:mc} we describe the general Monte Carlo method for
evaluating multidimensional integrals and show how it can be used to
construct a class of computational tools called Monte Carlo
event generators. We describe the main generic features of event
generators and how these are implemented in the \Herwigpp event generator. 

By applying resummation techniques, we provide, in
chapter~\ref{cha:qcdrad}, analytical calculations of two types of
hadron collider observables. The first, global inclusive variables, are
observables that make use of all measured particle momenta and can provide useful
information on the scale of new physics. The second observable is the
transverse energy of the QCD initial state radiation ($E_T$),
associated with the either Drell-Yan gauge boson production or Higgs boson
production. In both cases we provide comparisons to results obtained from Monte Carlo
event generators.

In chapter~\ref{cha:NEWPHYS} we examine two well-motivated models for
new physics: one of new heavy charged vector bosons
($W$ prime),
similar to the SM $W$ gauge bosons, and a model motivated by strong dynamics
electroweak symmetry breaking that contains new
resonances, leptoquarks, that couple primarily to quarks and leptons
of the third generation. In the prior model, we improve the current treatment of the $W'$ by
considering interference effects with the SM $W$ and construct an
event generator accurate to next-to-leading order which we use to
conduct a phenomenological analysis. For the leptoquark model, starting from an effective
Lagrangian for production and decay, we provide an implementation in the \Herwigpp event
generator and use it to form a strategy for mass reconstruction. The thesis ends with some conclusions and suggestions for extensions
of the work presented. Further details and
useful formul\ae~are given in the appendices.




\clearpage{\pagestyle{empty}\cleardoublepage}

\renewcommand{\baselinestretch}{1}\tiny\normalsize
\thispagestyle{plain}

\vspace*{50mm}
\centerline{\Huge {\bf Declaration}}
\vspace{10mm}
This dissertation is the result of my own work, except where
explicit reference is made to the work of others, and has not been
submitted for another qualification to this or any other
university.
\\

The original work in chapter~\ref{cha:qcdrad} was done in collaboration with Bryan
Webber and Jennifer Smillie and appears in~\cite{Papaefstathiou:2009hp,
  Papaefstathiou:2010ru, Papaefstathiou:2010bw}. The original work in chapter~\ref{cha:NEWPHYS} was done in collaboration with
Oluseyi Latunde-Dada, published in~\cite{Papaefstathiou:2009sr}, and with Bryan
Webber, Kazuki Sakurai and Ben Gripaios, published
in~\cite{Gripaios:2010hv}.
\\

This thesis does not exceed the 60,000 word limit prescribed by the Degree Committee for
Physics and Chemistry.
\\
\\

\vspace{3cm} \hspace{8cm} \myname

\clearpage{\pagestyle{empty}\cleardoublepage}
\thispagestyle{plain}
{\ }
\begin{center}
{\onehalfspacing 

\vspace{4.0cm}
\Large{\textgreek{``t`a >'onta >i`enai te p'anta ka`i m'enein o>ud'en.''}}\\
-- \textgreek{<Hr'akleitos <o >Ef'esios}\\
\vspace{0.5cm}
`All things move and nothing remains still.'\\
-- Heracletus of Ephesus\\
\vspace{2.0cm}
\centerline{********}
\vspace{2.0cm}
Dedicated to my family, for their love and support.
}


\end{center}

\clearpage{\pagestyle{empty}\cleardoublepage}
\renewcommand{\baselinestretch}{1}\tiny\normalsize
\thispagestyle{plain}
{\ }
\begin{center}
{\Large\bf Acknowledgements}
\end{center}
\vspace{5mm}
{\onehalfspacing 

I will begin by thanking my supervisor, Professor Bryan Webber, for
his guidance, encouragement and support throughout the past four
years. He has not only been a brilliant teacher, but also an
inspiring collaborator. Moreover I am grateful for his help, patience
and time spent on preparing this thesis. I have been fortunate to have collaborated with very energetic and
inspiring people who have taught me a lot during our work together. These include Seyi Latunde-Dada, Kazuki Sakurai, Ben
Gripaios and Jennie Smillie during my time in Cambridge, as well as Jeff Forshaw and Andy
Pilkington for provoking my interest in particle physics while at
Manchester. 

In addition, I've learned a lot through discussions with members of
the Cambridge Supersymmetry Group, such as Ben Allanach, Andy Parker,
Chris Lester and Are Raklev. A great amount of knowledge on computational issues has also come through my participation in
the \Herwigpp phone meetings, and I would like
to particularly thank Peter Richardson and David Grellscheid for
providing me with assistance whenever required. It has been an honour and a privilege to have been part of the Cavendish HEP
group and interacted with lively, interesting and hard-working
people. Moreover, as a phenomenologist, the interaction with the
experimentalists has given me invaluable insight to the more
`practical' aspects of our field.

My research has been funded by the Science and Technology Facilities
Council. I have also received funding from the European Training
Network MCnet and the Cavendish HEP group which have allowed me to
attend conferences and visit collaborators. 

On a more personal note, I would like to thank my (near-)officemates,
Seyi Latunde-Dada, Marco Sampaio, Jo Gaunt, Lucian Harland-Lang, Eleni
Vryonidou and Edwin Stansfield, as well as Professor James Stirling and Steve Kom, for
creating such an inspiring working environment. I would also like to
thank the group's former members, Deirdre Black and Are Raklev, who
have both provided me with guidance in terms of physics, as well as career advice.
 
A great amount of love, support and patience throughout the years has come from
my family, and I would like to thank them all for
believing in me always. I would never have made it to this point without them.

Finally, I will end by thanking Bryan, Lucian, Eleni and Ed for proofreading the thesis. Any
remaining errors are of course my sole responsibility.
}


\clearpage{\pagestyle{empty}\cleardoublepage}
\renewcommand{\baselinestretch}{1}\tiny\normalsize






}

\tableofcontents


\mainmatter
\onehalfspacing
\chapter{Introduction}
\label{sec:introduction}
The Human kind, by good fortune, has developed the ability to ponder and investigate natural phenomena. We have been intrinsically
acting as scientists for thousands of years. However, the form of what
we call `science', a term originating from the Latin \textit{sciencia},
meaning knowledge, has evolved dramatically through the ages. The
Greek philosopher Aristotle, who lived in the 4$^\mathrm{th}$ century
BC, profoundly affected those who followed him with his views on
natural phenomena. His persistent beliefs included that substances that make up
the Earth (`earth', `air', `water', `fire') were different than those
that made up the heavens (`aether') and that
objects moved as long as they were being pushed. His writings were largely
qualitative and, although he had basic ideas regarding a few of the fundamental concepts
of nature, such as speed and temperature, he was lacking the proper instruments to make quantitative
statements about them. It was not until the 16$^\mathrm{th}$ century AD,
and the ideas of Galileo Galilei, an Italian natural
scientist, that things began to change. Galileo, with the aid of the
newly-invented telescope which he had improved, made detailed
astronomical observations that made it more plausible that the heavens
and the Earth were made from the same materials. He also proposed the
`law of inertia', whereby objects tended to maintain their state of
motion instead of preferring to be at rest. Aristotle's misconceptions
originated from his lack of understanding of frictional forces. Isaac
Newton, unarguably one of the most important scientists of recent centuries,
effectively weaved his theories of motion and gravitation based on the
groundwork laid by scientists like Galileo. 

Newton's Universe was like clockwork: mechanical and perfectly deterministic. Space was absolute: the scene in which the heavenly bodies and the
Earth executed their eternal motion, a rigid grid of three
dimensions. Time, according to Newton, was flowing always at the same rate (`equably'), and
`without regard to anything external'. These concepts persisted until
the advent of further scientific revolutions that occurred in the early
20$^\mathrm{th}$ century: Albert Einstein's theory of relativity, doing
away with absolute space and time, and quantum mechanics, a theory of
subatomic particles, whose results for
the evolution of a physical system were of probabilistic nature. The
revolutions were either instigated by experimental facts
(e.g. Einstein's explanation of the photoelectric effect) or guided by
the philosophy that natural laws should be `beautiful' (e.g. the Dirac
equation). It is important to emphasise, however, that Newton's
theories were not discarded completely; rather, they were shown to be
specific limits of the theories that encompass them.  In the words of the mathematician David Hilbert, in a
lecture delivered before the International Congress of Mathematicians
at Paris in 1900, 
\begin{quotation}
`History teaches the continuity of the development of science. We know
that every age has its own problems, which the following age either
solves or casts aside as profitless and replaces by new ones.'
\end{quotation}
Old problems are viewed from different perspectives by scientists of the
following generations, in a different framework of thought, possessing
more powerful analytical and experimental tools. Some are solved, some discarded, and new questions are posed.

Nowadays, the study of the fundamental nature of matter is called
`particle physics', or `high energy physics'. Particle physicists are currently
faced with a multitude of unsolved puzzles. The quest to address them
may lead to a revolution of
our view of the fundamental principles of equal magnitude as the ones that
have occurred before. We present an overview of the
current understanding of the subatomic world in chapter~\ref{cha:smbsm}. This
framework is called the `Standard Model' of particle physics. We will
also examine the issues that are thought to plague this framework and
outline some suggestions that have been put forward to address some of
them. 

Science is based on careful observations, known as experiments. Through
experiments, we put our predictions to a test in a controlled
environment, in a reproducible way. In particle physics, the most
common form of experiment is rudimentary: we `throw' particles onto
one another and study the scattering process. Though basic, the idea
is powerful: very detailed quantitative predictions can be made and
theories can be put to a rigorous test. The Large Hadron Collider (LHC), at
the European Organisation for Nuclear Research (CERN), near Geneva,
Switzerland, involves such experiments. It will primarily\footnote{The
  LHC is also a heavy-ion collider.} collide
protons to protons at energies we have never examined before, about 14000 times the rest mass energy of the proton
($\sim 1\gev$). There, we expect to \textit{at least} observe a hypothetical
particle that is required for the consistency of the Standard Model,
the Higgs boson.\footnote{Or, if it is absent, we expect to observe a
  mechanism that explains that absence.} 
\begin{figure}[!htb]
  \centering
  \vspace{0.5cm}
  \includegraphics[scale=0.45, angle=270]{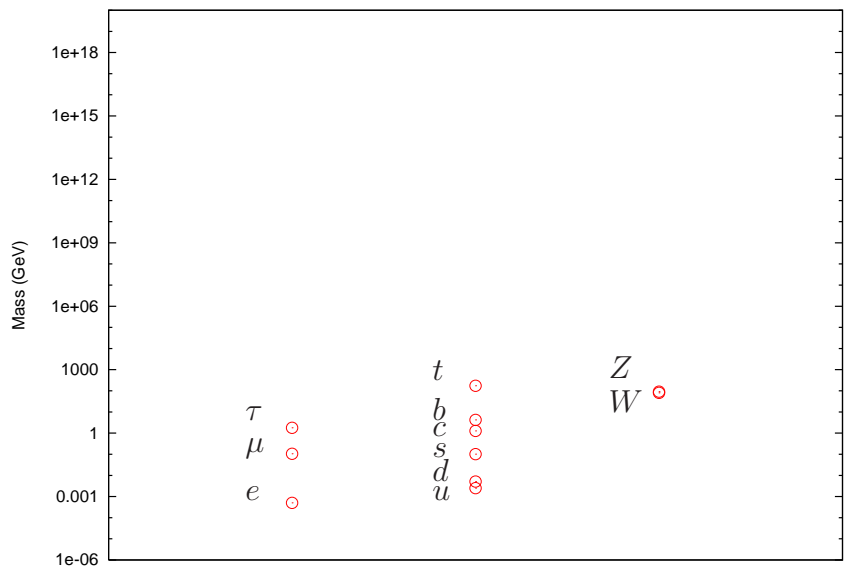}
  \put(-212,-155){$\tau$}  
  \put(-212,-167){$\mu$}
  \put(-212,-185){$e$}
 \put(-142,-140){$t$}  
  \put(-142,-155){$b$}
  \put(-142,-161){$c$}
\put(-142,-169){$s$}  
  \put(-142,-178){$d$}
  \put(-142,-185){$u$}
  \put(-76,-139){$Z$}
  \put(-76,-152){$W$}
  \vspace{0.75cm}
  \caption{The Standard Model spectrum of known massive particles. The mass is shown on
    the vertical axis in logarithmic scale, in units of GeV, from
    $10^{-6}\gev$ to $10^{19} \gev$. The photon, gluon and the
    neutrinos are not shown.}
  \label{fig:smspectrum}
\end{figure}

Forgetting for the moment the more concrete theoretical
reasons (which we will review in the following chapter) to expect the
observation of new phenomena at the energy scales of the
LHC, we can make a plausibility argument for their existence by a
simple observation. Figure~\ref{fig:smspectrum}
demonstrates the spectrum of known fundamental particles present in
the SM that possess mass.\footnote{The neutrinos also possess mass,
  albeit a small one. However, only mass differences are known and we
  do not show them here.} The vertical axis shows the mass in
logarithmic scale, extending from $10^{-6}
\gev$ to the scale $10^{19}\gev$, the fundamental scale of gravity,
known as the Planck scale.\footnote{The Planck scale is the scale at
  which the strength of the gravitational interaction between particles
  becomes of order one. An estimate is given by $M_{Pl} = G_N^{-1/2}\sim 10^{19}\gev $, where $G_N$ is Newton's gravitational constant.} The Planck
scale is the highest scale at which our current
understanding of physics makes sense. By examining
Fig.~\ref{fig:smspectrum}, one may observe something bizarre and slightly
suspicious: many particles have been discovered with masses ranging from
fractions of a GeV to fractions of a TeV, and one more (the Higgs
boson) is predicted below
the TeV scale, but no known particles exist above that scale! The
question that arises is: why should there be such a `desert' of energy
(or mass)
scales between the TeV scale and the Planck scale? If we do not accept
the existence of such a puzzling absence of particles, then we can only
conclude one of the following:
\begin{itemize}
\item{We do not currently possess the correct understanding of the
    fundamental scale of gravity, the Planck scale. Either there
    exists some specific mechanism explaining why such a large hierarchy arises between the
    TeV scale and the Planck scale, or the Planck scale is much lower
    than $10^{19} \gev$.}
\item{There exists a multitude of new particles and new interactions,
    waiting for potential discovery by future experiments.}
\end{itemize}
Both of these prospects are very exciting. The LHC will be
instrumental in exploring the TeV scale, perhaps revealing a whole new
set of particles or phenomena that will provide an explanation for the
above observation.

This thesis focuses on hadron colliders, of which the LHC is the most
`powerful' ever built. It is an extraordinary experiment, based on a collaboration
which transcends borders, involving tens of thousands of researchers
from more than a hundred nations. The task that particle
physicists are facing at the LHC is non-trivial. Certain ingenuity is
required if we wish to maximise the amount of physics results obtained
from the experiments. A solid bridge of communication between the theorists and the
experimentalists needs to be established. This is where phenomenology
comes into play: it provides an interface between theory and experiment, for example through
powerful computational tools, Monte Carlo programs, which we will be
discussing in detail in chapter~\ref{cha:mc}. These tools are
indispensable to both experimentalists and theorists. In
chapter~\ref{cha:qcdrad}, we make analytical phenomenological predictions of the effects of the
theory of the strong force, quantum chromodynamics (QCD), to observables
that experimenters will be using, either in their searches for new
physics, or for better understanding of known SM processes. In the final
chapter,~\ref{cha:NEWPHYS}, we take two well-motivated new physics scenarios and
provide a phenomenological analysis to act as a guide for
experimentalists in their search. 

In another part of his 1900 talk in Paris, Hilbert asks the following:
\begin{quote}
`Who of us would not be glad to lift the veil behind which the future lies hidden; to cast a glance at the next advances of our science and at the secrets of its development during future centuries?'
\end{quote}
We hope that soon we will have reached such a point. 
{\ }

\begin{boldmath}
\chapter{The Standard Model and beyond} 
\label{cha:smbsm}
\end{boldmath}
\section{Introduction}\label{sm:intro}
In his popular science book `QED: The Strange Theory of Light and
Matter'~\cite{book:qedfeynman}, Richard Feynman writes:
\begin{quote}
Therefore the possibility exists that the three $W$'s and the photon are
all different aspects of the same thing. Stephen Weinberg and Abdus
Salam tried to combine quantum electrodynamics with what's called the
`weak interactions' into one quantum theory, and they did it. But if
you just look at the results they get you can see the glue, so to
speak. It's very clear that the photon and the three $W$'s are
interconnected somehow, but at the present level of understanding, the
connection is difficult to see clearly -- you can still see the
`seams' in the theories; they have not yet been smoothed out so that
the connection becomes more beautiful and, therefore, probably more
correct.
\end{quote}
Despite Feynman's observation, the so-called `Standard Model' of particle
physics (abbreviated as SM) has been tremendously successful at describing experimental
data. In fact, the SM can be arguably considered as the quantitatively most
successful theory our species has ever constructed, with astounding
agreement between theory and experiment: the anomalous magnetic
moments of the electron and of the muon are amongst the most precise
measurements in the whole of physics.

The `glue'-ing that Feynman refers to is the fact that the SM is a
gauge theory of a \textit{product} of gauge symmetry groups:
\begin{equation}
SU(3)_c \times SU(2)_L \times U(1)_Y\;,
\end{equation}
where the $SU(3)_c$ describes quantum chromodynamics (QCD) and the
$SU(2)_L \times U(1)_Y$ the electroweak (EW) theory. The
$SU(3)_c$ symmetry, known as colour, is explicitly manifest in Nature,
whereas the $SU(2)_L \times U(1)_Y$ symmetry is broken down to
$U(1)_{em}$ via the Higgs Mechanism. 

The Higgs Mechanism, introduced to explain the masses of fermions and bosons in the theory by
breaking the electroweak symmetry, is currently the least understood part of the
SM and is thus a source of inspiration to many
extensions to the SM. Many of these extensions
attempt to explain the so-called `hierarchy problem', which can be briefly
described as the lack of explanation for the stability of the Higgs
boson mass against radiative corrections. Due to this issue and other
open theoretical questions, the SM is often believed to be an
incomplete description of particle physics. Its amazing success at
describing experiments, however, ensures that it will be a subset of a
`true' theory, understood in the framework of effective field
theories as capturing the low-energy limit of some more fundamental,
microscopic physics~\cite{book:smprimer}.

In section~\ref{sec:sm} we give a general introduction to the SM at its
current state, starting from the general principles for writing down a
relativistic quantum field theory. We will be focusing on the
phenomenological aspects of the EW theory and QCD and mathematical details will be kept to a minimum. In section~\ref{sec:bsm} we
examine some important open theoretical and experimental questions that suggest the
need for extensions to the SM and present a brief overview of BSM
theories. 

The reader is referred to Ref.~\cite{book:peskin} for further details on the
fundamentals of quantum field theory. For further details on the basic
principles of QCD and particularly on
QCD phenomenology,~\cite{Seymour:2010dn, Ellis:1996qj} are recommended and for a general introduction
to gauge theories and the Standard Model~\cite{book:smprimer,
  book:dynamicsofsm, book:gaugefieldtheories}.

\section{The Standard Model}
\label{sec:sm}
\subsection{Gauge theories}
\label{sec:sm:gauge}
Symmetries provide good candidates for underlying first principles in
Physics. The gauge principle is an economical guide for using \textit{local}
symmetry to construct renormalisable quantum field theories. The fact
that local invariance is required is motivated by the relativistic
viewpoint: each observer, at every space-time point, possesses some
freedom of convention. This of course may be considered by some as an
aesthetical argument; however, the gauge principle has been proven to produce
phenomenologically successful field theories and accounts for quantum
electrodynamics (QED) as well as EW theory and QCD. 

The gauge principle is most simply illustrated in the case of
invariance under the unitary group $U(1)$, which yields an abelian
gauge theory that describes QED. Consider the classical Lagrangian
density\footnote{In what follows, and the rest of this thesis, we will
  be referring to the `Lagrangian density' simply as the
  `Lagrangian'.} describing the interaction of a spin 1/2, charged fermion field $\psi$ with the gauge
field $A_\mu$:
\begin{equation}
\mathcal{L}_{\mathrm{em}} = -\frac{1}{4} F_{\mu\nu} F^{\mu\nu} +
\bar{\psi} ( i \slashed{D} - m ) \psi \;,
\end{equation}
where $F^{\mu \nu} = \partial^\mu A^\nu - \partial^\nu A^\mu$ is the
field strength tensor and the covariant derivative is defined by
$\slashed{D} \psi \equiv \gamma^\mu D_\mu \psi \equiv \gamma^\mu ( \partial_\mu
+ i e A_\mu ) \psi$, where $m$ and $e$ are respectively the mass and
electric charge of $\psi$ and we have also defined the Feynman slash
convention. The Lagrangian is invariant under the simultaneous gauge
transformations:
\begin{eqnarray}
A ^{\mu} \rightarrow A'^{\mu}  \equiv A^\mu - \partial^\mu \chi\;, \label{eq:sm:gauge:amu}\\
\psi \rightarrow \psi' \equiv e^{ie\chi} \psi \;,  \label{eq:sm:gauge:psi}
\end{eqnarray}
where $\chi = \chi (x,t)$ is an arbitrary function of space-time. Had we not included the field $A^\mu$, the Lagrangian would
not have remained invariant under the local transformation of the
field $\psi$ alone, owing to the derivatives present in the fermion
kinetic terms. This can be seen explicitly if we examine the
transformation of the term $\bar{\psi} \gamma^\mu \partial_\mu  \psi
$ under Eq.~(\ref{eq:sm:gauge:psi}):
\begin{eqnarray}
\bar{\psi} \gamma^\mu \partial_\mu  \psi \rightarrow \bar{\psi}
e^{-ie\chi} \gamma^\mu \partial_\mu (
e^{ie\chi} \psi) = \bar{\psi}( \gamma^\mu \partial_\mu +i e
\gamma^\mu (\partial_\mu \chi) )\psi \;,
\end{eqnarray}
which is evidently not invariant unless we use the covariant
derivative, which introduces $A_\mu$, to
cancel out the extra term:
\begin{eqnarray}
\bar{\psi} \gamma^\mu D_\mu  \psi \rightarrow \bar{\psi}
e^{-ie\chi} \gamma^\mu D_\mu (
e^{ie\chi} \psi) &=& \bar{\psi}( \gamma^\mu \partial_\mu +i e\gamma^\mu
A_\mu -i e\gamma^\mu (\partial_\mu \chi) \psi +i e
\gamma^\mu (\partial_\mu \chi) )\psi \; \nonumber \\
&=& \bar{\psi} \gamma^\mu D_\mu \psi~\mathrm{(invariant)}\;.
\end{eqnarray}
Thus, the requirement that the Lagrangian (and
hence the equations of motion) is invariant under local $U(1)$
transformations requires the existence of a \textit{gauge} field
$A^\mu$. This field corresponds to the electromagnetic field and hence
to the photon.  
\subsection{Electroweak theory} 
\label{sec:sm:ew}
As we alluded in the introduction, the electroweak sector of the SM can be described by a
non-abelian gauge theory based on the group $SU(2)_L \times U(1)_Y$,
where the $L$ subscript indicates the left-handed chiral nature of the coupling of
the gauge fields and $Y$ is the hypercharge, to be distinguished from
the electromagnetic charge, $Q$. The $SU(2)_L$ quantum number is referred
to as the weak isospin. We can separate the Lagrangian into three
parts as
\begin{equation}
\mathcal{L}_{\mathrm{EW}} = \mathcal{L}_{\mathrm{bosons}} +
\mathcal{L}_{\mathrm{Higgs}} + \mathcal{L}_{\mathrm{fermions}} \;,
\end{equation}
where $\mathcal{L}_{\mathrm{bosons}}$, $\mathcal{L}_{\mathrm{Higgs}}$
and $\mathcal{L}_{\mathrm{fermions}}$ correspond to the gauge bosons,
the Higgs field and the fermions respectively. 
\subsubsection{Boson masses}
The gauge boson Lagrangian now contains two gauge fields:
\begin{equation}
\mathcal{L}_{\mathrm{bosons}} = - \frac{1}{4} F^A_{\mu\nu} F^{A
  \mu\nu}  - \frac{1}{4} B_{\mu\nu} B^{
  \mu\nu} \;,
\end{equation}
where the $F^{A\mu\nu}$ and $B^{\mu\nu}$ are the field tensors
corresponding to the $SU(2)_L$ and $U(1)_Y$ symmetries and the index
$A$ labels the $SU(2)_L$ weak isospin quantum numbers:
\begin{eqnarray}
F^{A\mu\nu} &=& \partial^\mu W^{A\nu} - \partial^\nu W^{A\mu} - g
f^{ABC} W^{B\mu} W^{C\nu} \;, \nonumber \\
B^{\mu\nu} &=& \partial^\mu B^{\nu} - \partial^\nu B^{\mu}\;,
\end{eqnarray}
where the $f^{ABC}$ are the group structure constants (the alternating tensor in
the $SU(2)$ case) and $g$ is the $SU(2)_L$ charge. It is important to note that the non-abelian field tensor $F^{A\mu\nu}$ now
contains a self-interaction term $\propto W^{B\mu} W^{C\nu}$, a
feature that is even more significant in $SU(3)$ non-abelian gauge
theory, as we shall see.

\begin{figure}[!t]
  \centering 
    \includegraphics[scale=0.60, angle=270]{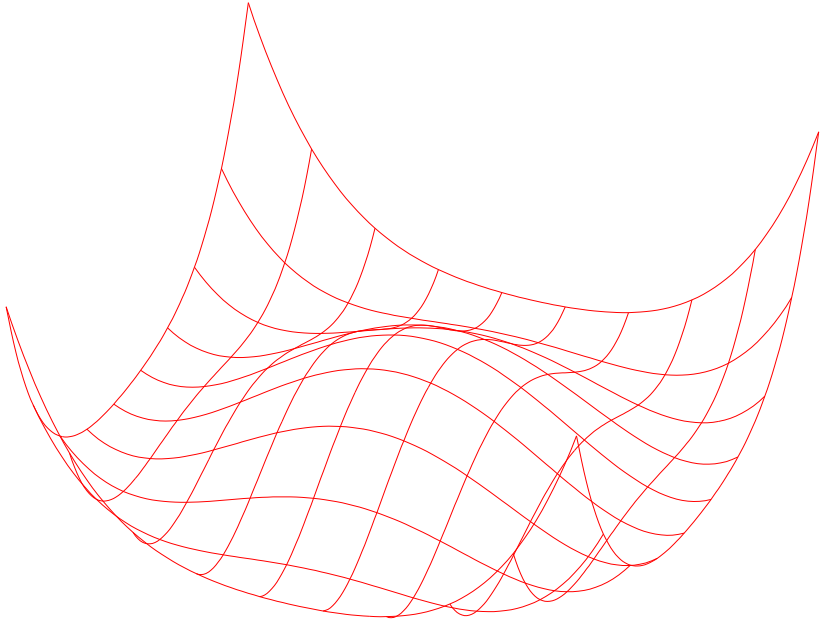}
\caption[]{The `mexican hat' potential given by
  Eq.~(\ref{sm:ew:mexpot}), with $\mu^2, \lambda >
0$.} 
\label{fig:sm:ew:mexhat}
\end{figure}

The $SU(2)_L \times U(1)_Y$ symmetry is not manifested in nature. It is in
fact, a \textit{spontaneously} broken symmetry: the $W$ and $Z$ gauge
bosons are massive. The minimal way to
break it within the SM and give masses to the gauge bosons, while preserving the gauge-invariant nature of the theory, is
to introduce the complex scalar Higgs field that is an $SU(2)_L$
doublet and possesses hypercharge $Y=1/2$:\footnote{The hypercharge
  for the Higgs field could
  have been chosen $Y=1$. The choice affects
  the hypercharges for the rest of the matter content in the
  theory. The relation between electric charge, hypercharge and the
  third component of weak isospin will contain factors of 2 accordingly.}
\begin{eqnarray}
\phi &=& \left(
\begin{array}{c}
\phi^+ \\ \phi^0 \end{array}
\right) \;, \nonumber \\
\phi^\dagger&=& \left(
\begin{array}{cc}
\bar{\phi^0} &\phi^-\end{array}
\right) \;, \nonumber \\
\end{eqnarray}
where the meaning of the labels on the components will become apparent
subsequently. The corresponding Lagrangian,
$\mathcal{L}_{\mathrm{Higgs}}$ is given by
\begin{equation}
\mathcal{L}_{\mathrm{Higgs}} = (D_\mu\phi)^\dagger (D^\mu \phi) -
V(\phi^\dagger \phi)\;,
\label{sm:ew:higgsL}
\end{equation}
where the covariant derivative, which introduces the interaction between
the Higgs field and the gauge fields, is defined by
\begin{equation}
D^\mu \equiv \partial^\mu + i g ( T \cdot W ^\mu) + i Y g' B^\mu\;.
\end{equation}
In the above, we have suppressed weak isospin indices, the $T$ are matrix
representations of the $SU(2)_L$ generators and $g$ and $g'$ are the
$SU(2)_L$ and $U(1)_Y$ gauge charges respectively. The potential term is given a
special form containing quadratic and quartic terms, commonly referred to as the `mexican hat' potential:
\begin{equation}
V(\phi^\dagger \phi) = - \mu^2 (\phi^\dagger \phi) + \lambda
(\phi^\dagger \phi)^2\;,
\label{sm:ew:mexpot}
\end{equation}
and the constants are chosen such that $\mu^2, \lambda > 0$. This potential possesses a minimum at $(\phi^\dagger \phi)_\mathrm{min} =
\mu^2 / 2 \lambda \equiv v^2$, an unstable maximum at the origin and
goes off to positive infinity as $(\phi^\dagger \phi) \rightarrow
\infty$ (hence the resemblance to the mexican
hat). Figure~\ref{fig:sm:ew:mexhat} illustrates the shape of the
potential on the complex $\phi$ plane.
To break the symmetry, a particular direction (referred to as a `vacuum expectation
value' or VEV) in the $SU(2)$ space for
the minimum of $\phi$ is chosen:
\begin{eqnarray}
\left<\phi\right> &=& \frac{1}{\sqrt{2}} \left(
\begin{array}{c}
0 \\ v \end{array}
\right) \;.
\label{eq:sm:ew:vev}
\end{eqnarray}
Notice that, at the minimum, the
theory still possesses a residual $U(1)$ symmetry. To determine the
properties of the theory at this minimum, we need to expand the fields
about it, using an $SU(2)$ transformation of the field:
\begin{eqnarray}
U(\xi) &=& e^{-i T\cdot \xi /v }\;,\\
\phi &=& \left(
\begin{array}{c}
0 \\ (H+v)/\sqrt{2}\end{array}
\right) U(\xi)\;,
\end{eqnarray}
where we have three $\xi$ components and one scalar degree of
freedom, $H$. At first order this is just an expansion of the field
about the minimum. Since the Lagrangian of Eq.~(\ref{sm:ew:higgsL}) is
gauge-invariant, we should simultaneously perform an $SU(2)$ gauge
transformation:
\begin{eqnarray}
T\cdot W^\mu &\rightarrow& U T \cdot W^\mu U^{-1} + \frac{i}{g}
(\partial^\mu U) U^{-1}\;.
\end{eqnarray}
We obtain the following form for the Higgs boson Lagrangian:
\begin{equation}
\mathcal{L} = \frac{1}{2} \partial_\mu H\partial^\mu H - V \left(
  \frac{ (v+H)^2 } { 2} \right) + \frac{(v+H)^2}{8} \chi^\dagger ( 2 g
T\cdot W_\mu + g' B_\mu ) ( 2 g T \cdot W^\mu + g' B^\mu) \chi \;,
\end{equation}
where we have used the vector $\chi = (0,1)$, a unit vector along the
direction of the Higgs boson VEV. Evidently the three degrees of
freedom from the $\xi$ field do not appear explicitly in the
Lagrangian. These degrees of freedom have been absorbed by the gauge bosons and provide
the longitudinal degree of freedom: a massive vector boson has three
polarisation states whereas a massless one has two. 

We now consider the terms quadratic in the vector boson fields:
\begin{eqnarray}
\mathcal{L}_{\mathrm{M}} = \frac{v^2}{8} \left[ (g W^3_\mu - g' B_\mu)
  ( g W^{3\mu} - g' B^\mu ) + 2 g^2 W^-_\mu W^{+\mu} \right]\;.
\end{eqnarray}
We need to rewrite this in `diagonal form', i.e. in terms of mass eigenstates. We define two electrically
neutral fields $Z_\mu$ and $A_\mu$:
\begin{eqnarray}
Z_\mu = \cos \theta_w W^3_\mu - \sin \theta_w B_\mu \;,\nonumber\\\
A_\mu = \sin \theta_w W^3_\mu + \cos \theta_w B_\mu \;,
\label{eq:sm:ew:zafields}
\end{eqnarray}
where we have also defined the electroweak mixing angle, $\sin^2 \theta_w =
g'^2 / (g^2 + g'^2) \simeq 0.23$.\footnote{The current Particle Data Group value is
  $\sin ^2 \hat{\theta} (M_Z) ({\overline {\rm MS}}) =
  0.23116(13)$~\cite{Nakamura:2010zzi}.} Using the definitions of
Eq.~(\ref{eq:sm:ew:zafields}) we have
\begin{equation}
\mathcal{L}_{\mathrm{M}} = \frac{g^2 v^2}{4} W^+_\mu W^{-\mu} + \frac{
  (g^2 + g'^2) v^2 } { 8 } Z_\mu Z^\mu \;,
\end{equation} 
from which we can now deduce the vector boson masses:
\begin{eqnarray}
M_A = 0\;,\;\; M_W = \frac{1}{2} v g\;,\;\; M_Z = \frac{1}{2} v\sqrt{g^2 +
  g'^2}   \;.
\end{eqnarray}
Thus, with the particular choice for the Higgs boson representation, masses
have been generated for the weak vector bosons, $Z$ and $W^\pm$, while
one field, $A$, has remained massless. This corresponds to a $U(1)$-symmetric force which we identify with
QED and a boson which we identify with the photon, $\gamma$. We can
thus write the following symmetry breaking relation:
\begin{equation}
SU(2)_L \times U(1)_Y \rightarrow U(1)_{\mathrm{em}}\;,
\end{equation}
and we can associate the different charges of fields by
\begin{equation}
Q = T^3 + Y\;,
\end{equation}
where $Q$ is the electromagnetic charge, $T^3$ is the third component
of weak isospin and $Y$ is the hypercharge. $Q$ is essentially the only remaining unbroken
generator. We can now see that the upper component of the Higgs doublet has $Q = 1$, whereas the lower one has charge $Q = 0$, which
explains why the VEV was chosen as in Eq.~(\ref{eq:sm:ew:vev}).
\subsubsection{Fermion masses and couplings}\label{sec:sm:fermions}
Introducing fermion masses explicitly into the fermion Lagrangian
would break gauge invariance. The Higgs mechanism in its
simplest form is the conventional method to give masses to the SM fermions without
any adverse effects. The choices of fermion representations in the SM
and their charges are given in Table~\ref{tb:ew:fermionrep}, for the
first generation. The SM has been observed to contain three
`copies' of this structure: three generations. 
\begin{table}[!t]
\begin{center}
\begin{tabular}{|c|c|c|c|c|} \hline
Representation & $T^3$ & $Y$ & $Q$ \\\hline \hline
\rule{0cm}{0.50cm} $e^-_R$ & 0  &  -1 & -1 \rule[-0.25cm]{0cm}{0.50cm}\\ \hline 
\rule{0cm}{0.75cm}\multirow{3}{*}{$\left(\begin{array}{c}
\nu_{eL} \\\\ e^-_L\end{array} \right)$}
& $+1/2$&  & 0\\ 
& & - 1/2 &   \\ 
&$-1/2$  &  & -1\rule[-0.375cm]{0cm}{0.75cm} \\\hline
\rule{0cm}{0.5cm}  $\nu_{eR}$ & 0 & 0& 0\rule[-0.25cm]{0cm}{0.50cm}\\ \hline\hline
\rule{0cm}{0.5cm}  $d'_R$ & 0  &  $-1/3$  & $-1/3$ \rule[-0.25cm]{0cm}{0.50cm}\\ \hline 
\rule{0cm}{0.75cm} \multirow{3}{*}{$\left(\begin{array}{c}
u_L \\\\ d'_L\end{array}\right)$}
 & $+1/2$&  & $+2/3$\\ 
 & & $+1/6$ &  \\
 &$1/2$  &  & $-1/3$\rule[-0.375cm]{0cm}{0.75cm} \\\hline
\rule{0cm}{0.5cm} $u_{R}$ & 0 & $+2/3$& +2/3 \rule[-0.25cm]{0cm}{0.50cm}\\ \hline
\end{tabular}
\end{center}
\caption{The first-generation fermion representations in the electroweak $SU(2)_L
  \times U(1)_Y$ theory. The third component of weak isospin, $T^3$,
  the hypercharge, $Y$ and the resulting electromagnetic charge $Q$
  are given. The primes on the quark sector indicate that they are not mass eigenstates. The
  right-handed neutrino is hypothetical, does not couple to the SM
  particles and is shown for completeness.}
\label{tb:ew:fermionrep}
\end{table}
Notice that the difference between the hyperchages of the singlet
(right-handed) and doublet (left-handed) is $\pm 1/2$, which allows us
to use the Higgs doublet (with $Y=1/2$) to form gauge-invariant terms:
\begin{eqnarray}
\mathcal{L} &=& g_{ee} ( \bar{\nu}_{eL} , ~\bar{e}_L ) \left(\begin{array}{c}\phi^+
    \\ \phi^0 \end{array}\right) e_R + g_{dd} ( \bar{u}_L, ~\bar{d}'_L )
\left(\begin{array}{c}\phi^+ 
    \\ \phi^0 \end{array}\right) d'_R  \nonumber \\
&+& g_{\nu\nu} ( \bar{\nu}_{eL} ,~\bar{e}_L ) \left(\begin{array}{c}\bar{\phi^0}
    \\\phi^-\end{array}\right) \bar{\nu}_{eR}+ g_{uu} ( \bar{u}_L ,~\bar{d}'_L )
\left(\begin{array}{c}\bar{\phi^0}
    \\ \phi^-\end{array}\right) u_R \nonumber \\
 &+& (\mathrm{other~flavours}) + \mathrm{h.c.}\;,
\end{eqnarray}
Thus, when the Higgs field obtains a VEV and the symmetry is broken,
we obtain terms quadratic in the fermion fields: mass terms
proportional to the VEV, $m_{ff} = g_{ff} v / \sqrt{2}$, and mixing
terms between the fermions. To take care of this mixing we
conventionally define
\begin{eqnarray}
\left(\begin{array}{c} d'
    \\ s' \\ b' \end{array}\right) = V \left(\begin{array}{c} d
    \\ s \\ b \end{array}\right) \;,
\end{eqnarray}
where $V$ is known as the Cabbibo-Kobayashi-Maskawa (CKM)
matrix.\footnote{Kobayashi and Maskawa were awarded the Nobel Prize in 2008,
  on work related to the CKM matrix, and specifically `for the
  discovery of the origin of the broken symmetry which predicts the
  existence of at least three families of quarks in
  nature'~\cite{nobelwebsite}.} $V$ is a $3\times3$
unitary matrix, given by
\begin{eqnarray}
V &=& \left(\begin{array}{ccc} 
    V_{ud} & V_{us} & V_{ub} \\
    V_{cd} & V_{cs} & V_{cb} \\
    V_{td} & V_{ts} & V_{tb} \\ 
\end{array}\right)\;,
\label{eq:sm:ckm}
\end{eqnarray}
and can be parametrized by three mixing angles and a charge-parity (CP)
violating phase. In the SM, this phase provides a
source of CP violation, observed experimentally via mixing in
the neutral kaon system ($K^0$-$\bar{K}^0$
mixing)~\cite{book:dynamicsofsm}. See appendix~\ref{sec:ckm} for
parametrizations of the CKM matrix and measured values of its matrix elements.

In practical calculations, the CKM matrix elements $V_{ij}$ can be
inserted into the amplitude of diagrams where a $W$ boson couples to
quarks $i$ and $j$. If we define the right- and left-handed projection
operator for fermions, $P_{R,L} = \frac{1}{2} ( 1 \pm \gamma^5 )$, we
may write down the interaction term between the SM fermions $f$ and
$f'$ and the $W$ boson:
\begin{equation}\label{eq:sm:wff}
\mathcal{L}_{Wff} = \frac{g}{\sqrt{2}} V_{ff'} \bar{f} \gamma_\mu
P_L f' W^\mu + \mathrm{h.c.}\;,
\end{equation}
where $V_{ff'} = 1$ for leptons ($\ell\nu_\ell$) and $V_{ff'} = V_{ij}$ for quarks $i$
and $j$. 
\subsection{Quantum chromodynamics}
\label{sec:sm:qcd}
Quantum chromodynamics (QCD) is also formulated in terms of a gauge
theory, based on the non-abelian group $SU(3)$. It possesses several
distinct features: it is unbroken in Nature, contains self-interacting
degrees of freedom, the gluons, and exhibits \textit{asymptotic
  freedom}, which reveals that only in the short-distance limit we can
use perturbative methods legitimately. 

We will briefly review the construction of the theory and subsequently present perturbative tools that
will be employed in calculations that will follow in this thesis. 
\subsubsection{SU(3) gauge theory}
\label{sec:sm:qcd:su3}
The fermions that carry $SU(3)$ charge, or colour charge, are the quarks. The
full quantum Lagrangian is given by
\begin{equation}
\mathcal{L}_{\mathrm{QCD}} = \mathcal{L}_{\mathrm{classical}} +
\mathcal{L}_{\mathrm{gauge-fixing}} + \mathcal{L}_{\mathrm{ghost}}\;.
\end{equation}
The expression for the classical Lagrangian is similar to what we have
written down for the QED and EW theories:
\begin{equation}
\mathcal{L}_{\mathrm{classical}} = -\frac{1}{4} F^A_{\mu\nu} F^{A\mu \nu} + \sum_{\mathrm{flavours}} \bar{q}_a ( i \slashed{D} - m_q
)_{ab} q_b \;,
\label{eq:sm:qcd:lclass}
\end{equation}
where the sum over the index $A$ is over the eight colour degrees of freedom of
the gluon field $G^A_\mu$, the sum over flavours is over the $n_f$
quark flavours and the field strength tensor $F^{A\mu\nu}$ is
defined as
\begin{equation}
F^{A\mu\nu} \equiv \partial^\mu G^{A\nu} - \partial^\nu G^{A\mu} - g_s
f^{ABC} G^{B\mu} G^{C\nu}\;,
\end{equation}
where $g_s$ is the strong charge and the $f^{ABC}$ are the
structure constants of $SU(3)$. We may also define the strong coupling
constant $\alpha_s \equiv g_s^2 / 4\pi$. The covariant derivative $D^\mu$ is defined according to whether it
acts on triplet or octet fields:
\begin{equation}
(D^\mu)_{ab} = \partial^\mu \delta_{ab} + i g (t \cdot G^\mu)_{ab} \;,\;\; (D^\mu)_{AB} = \partial^\mu \delta_{AB} + i g (T \cdot G^\mu)_{AB}\;,
\end{equation}
where $t$ and $T$ are generators (matrices) in the fundamental and adjoint
 representations of $SU(3)$ respectively. They satisfy the following relations:
\begin{equation}
[t^A, t^B] = i f^{ABC} t^C\;,\;\; [T^A, T^B] = i f^{ABC} T^C\;,\;\;
(T^A)_{BC} = -i f^{ABC}\;.
\end{equation}
The following identities are true for $SU(N)$ gauge theories:
\begin{eqnarray}
\mathrm{Tr}(t^At^B) = \frac{1}{2} \delta^{AB} \equiv T_R \delta^{AB}
\;, \nonumber\\
\sum_A t^{A}_{ab} t^A_{bc} = \frac{N^2 - 1} {2N} \delta_{ac} \equiv
C_F \delta_{ac} \nonumber \;,\\
\mathrm{Tr}( T^C T^D ) = \sum_{A,B} f^{ABC} f^{ABD} = N \delta^{CD}
\equiv C_A \delta^{CD}\;,
\label{eq:sm:qcd:identities}
\end{eqnarray}
which imply that for QCD, for which $N=3$: $T_R = 1/2$, $C_F = 4/3$ and $C_A = 3$. In
practical calculations an explicit representation for the $t^A$ is not
necessary, and the identities of Eq.~(\ref{eq:sm:qcd:identities}) are used.

The classical Lagrangian, Eq.~(\ref{eq:sm:qcd:lclass}), is invariant
under the simultaneous $SU(3)$ transformations:
\begin{eqnarray}
q_a &\rightarrow& q'_a = \left(e^{i t \cdot \theta} \right)_{ab} q_b \equiv
U_{ab} q_b\;,\nonumber \\
t.G_\mu &\rightarrow& t.G'_\mu =  U t\cdot G_\mu U^{-1} + \frac{i}{g_s}
(\partial_\mu U) U^{-1} \;,
\end{eqnarray}
where $\theta^A = \theta^A (x,t)$ are eight arbitrary real functions of space-time. 

The Lagrangian of Eq.~(\ref{eq:sm:qcd:lclass}) cannot be used
immediately to calculate Feynman rules for QCD: in this form, a
propagator for the gluon field cannot be defined.\footnote{The same
  issue arises when defining the photon propagator in QED.} We can
exploit gauge invariance to add a gauge-fixing term to the QCD
Lagrangian which amends this issue:
\begin{equation}
\mathcal{L}_\mathrm{gauge-fixing} = -\frac{1}{\lambda} (\partial^\mu G^A_\mu
)^2 \;,
\end{equation} 
where $\lambda$ is an arbitrary parameter. Provided we work in the
\textit{covariant} gauge, that is, a gauge in which we choose $\partial^\mu
G^A_\mu = 0$,  we have not made any changes in the physics and we can
now define a propagator for the gluon fields.

Finally, it is necessary to add an extra term to the Lagrangian which
is related to the need for ghost particles, $\mathcal{L}_\mathrm{ghost}$, whose purpose is to cancel
unphysical degrees of freedom that may arise when renormalising a
non-abelian gauge theory. For further details on non-abelian gauge
theory renormalisation and the need for ghost fields,
see~\cite{book:peskin}.
\subsubsection{Renormalisation and the running of $\alpha_s$}\label{sec:qcd:renorm}
Besides the masses of the quarks, the only other parameter which appears in the QCD
Lagrangian is the strong charge,\footnote{In practical calculations one
usually employs $\alpha_s$ rather than $g_s$.} $g_s$. One should be cautious
however: parameters in a Lagrangian are not necessarily physical
quantities. Physical observables can be calculated as functions of
these parameters, in this case of $g_s$. What we would like to do is
reformulate the theory so we can write a physical observable as a
function of another. This process is called \textit{renormalisation}. 

\begin{figure}[!t]
  \centering 
    \includegraphics[scale=0.40, angle=0]{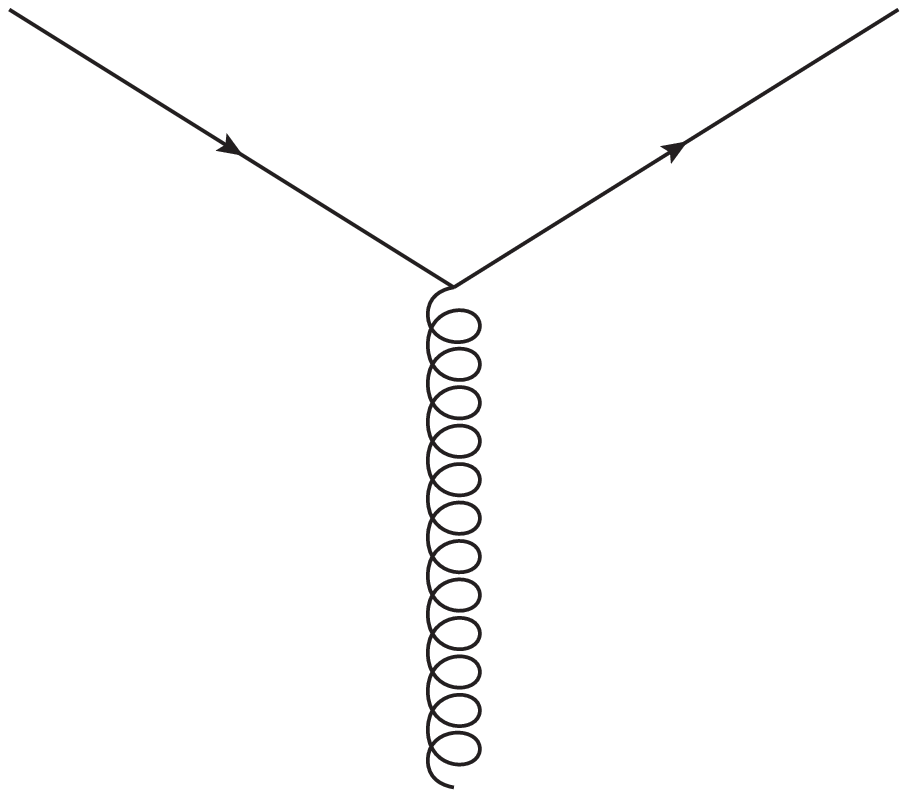}
    \hspace{0.2cm}
    \includegraphics[scale=0.40,angle=0]{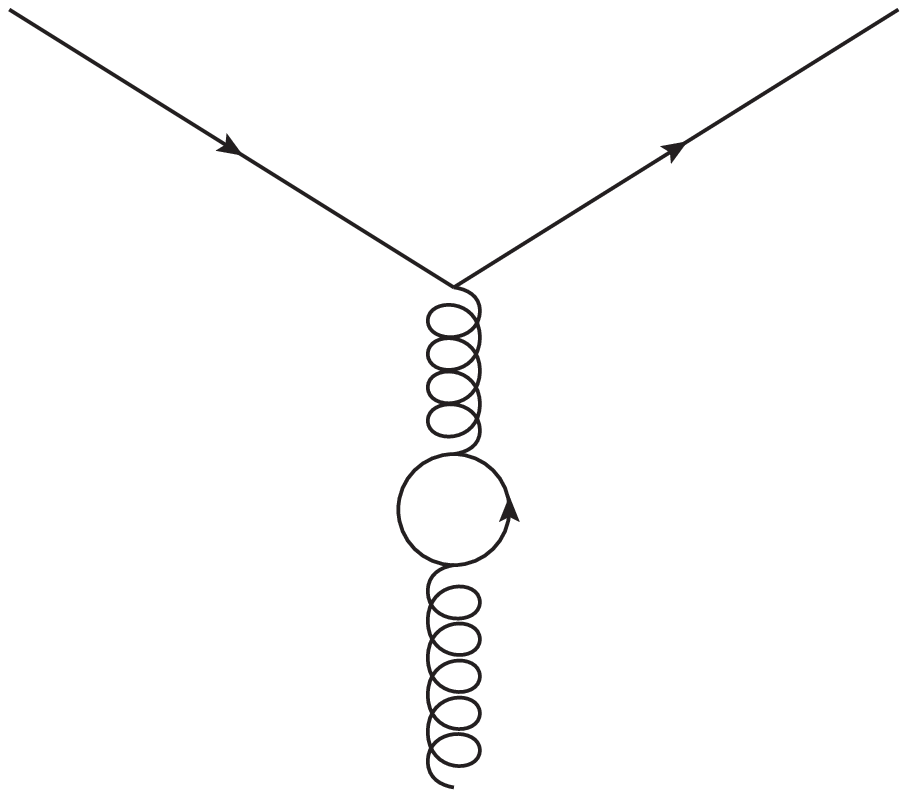}
    \hspace{0.2cm}
    \includegraphics[scale=0.40, angle=0]{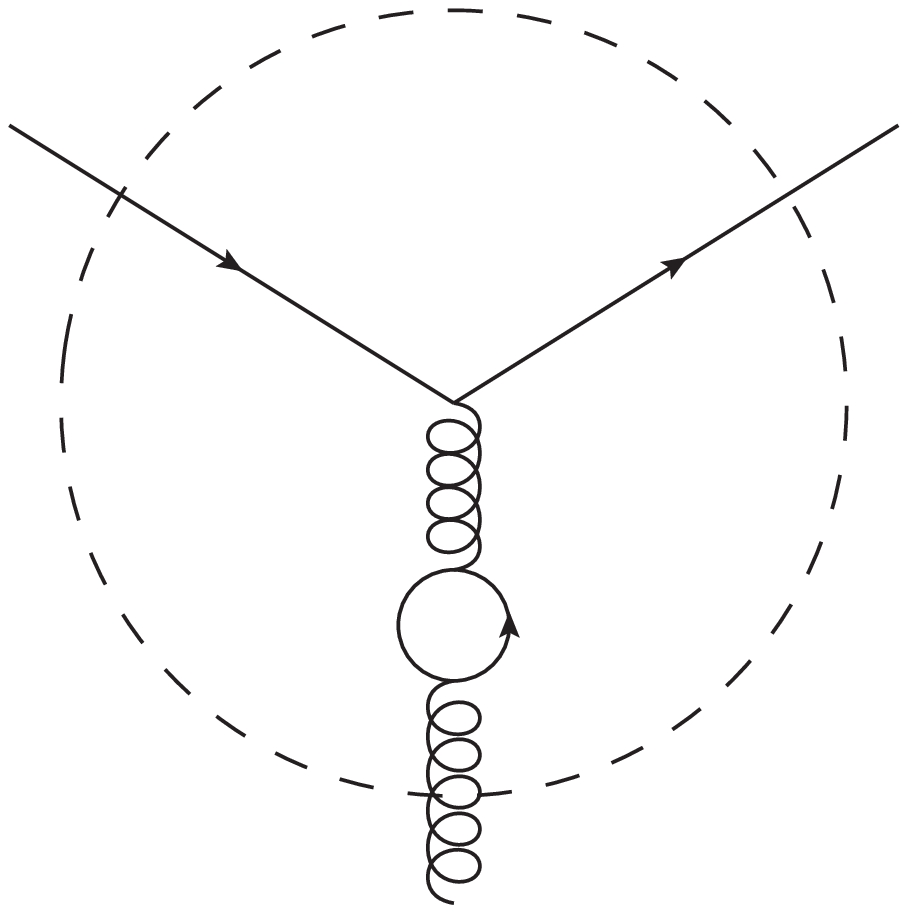}
    \hspace{0.2cm}
    \includegraphics[scale=0.40,angle=0]{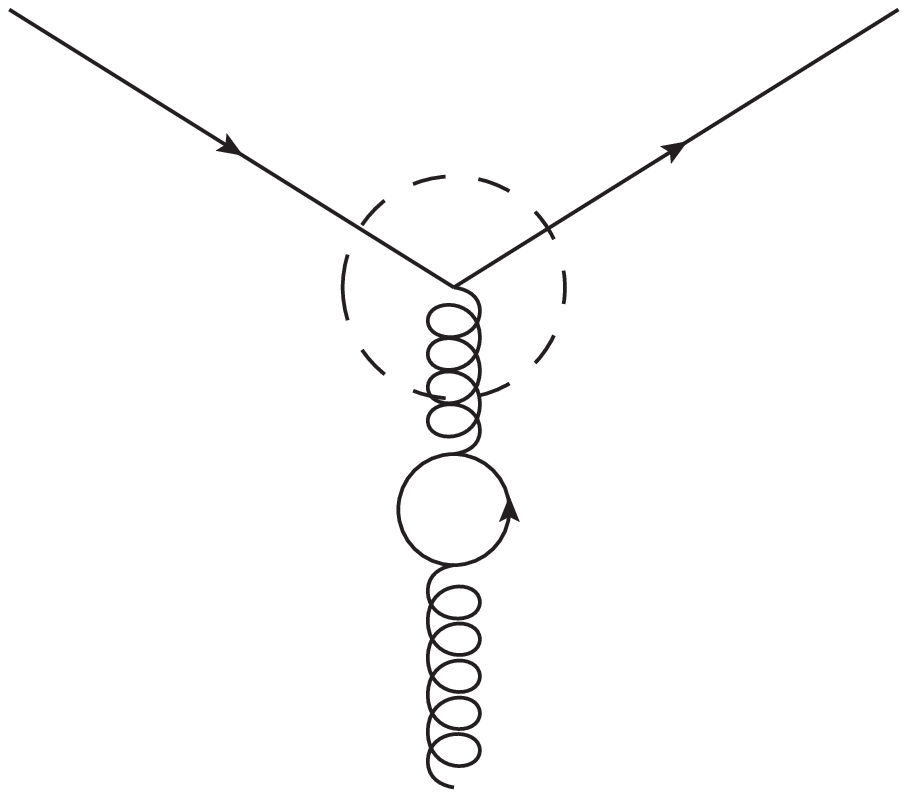}
    \put(-408,-25){(a)}
    \put(-290,-25){(b)}
    \put(-173,-25){(c)}
    \put(-59,-25){(d)}
\caption[]{The quark-gluon vertex, (a), and part of the one-loop correction to
  it, (b), are shown. We need to decide whether we assign the loop
  correction as a correction to the vertex, as in (c), or to the rest of the
  diagram, as in (d). This is done at some scale $\mu_R$, above which the
  loop is part of the vertex and below which it is part of the rest of
the diagram.} 
\label{fig:sm:qcd:quarkgluon}
\end{figure}

As an illustrative example, as given in Ref.~\cite{Seymour:2010dn}, consider the
quark-gluon vertex shown in Fig.~\ref{fig:sm:qcd:quarkgluon}.
Figure~\ref{fig:sm:qcd:quarkgluon}a shows the lowest order of the vertex in
perturbation theory, for which we can define $g_s$ to be the
strength. If we now start considering higher orders, and specifically
part of the one-loop correction shown in Fig.~\ref{fig:sm:qcd:quarkgluon}b, we must decide whether we will
include this as part of the vertex, as in
Fig.~\ref{fig:sm:qcd:quarkgluon}c, or as part of the rest of the
diagram, as in Fig.~\ref{fig:sm:qcd:quarkgluon}d. To do this we choose
some scale $\mu_R$, called the renormalisation scale, above which the
  loop is part of the vertex, as in Fig.~\ref{fig:sm:qcd:quarkgluon}c,
   and below which it is part of the rest of
the diagram, as in Fig.~\ref{fig:sm:qcd:quarkgluon}d. This scale should
not have any physical significance: it is simply a device we
introduce to avoid double-counting. Due to this, $\mu_R$ should not
affect any physical prediction. However, the strength $g_s$ now becomes a
function of the scale $\mu_R$. 

We can see the effect of this procedure on the theory
by considering a dimensionless physical observable $R$ that is a function of a
physical scale $Q^2$. After applying the above renormalisation
procedure, the observable $R$ can only be a function of $Q^2$,
$\mu_R^2$ and $\alpha_s(\mu^2_R)$. Dimensional analysis restricts $R$ to
depend on $Q^2$ and $\mu_R^2$ only through their ratio, giving
\begin{equation}
R = R(Q^2/\mu_R^2, \alpha_s(\mu^2_R))\;.
\end{equation}
We can now employ the fact that a physical observable cannot depend on
the value of the renormalisation scale, $\mu_R$ and write, using the
chain rule:
\begin{eqnarray}
\mu^2_R \frac{\mathrm{d}} {\mathrm{d} \mu_R^2 } R(Q^2/\mu_R^2,
\alpha_s) &=& 0\;, \nonumber\\
\Rightarrow \left[ \mu^2_R \frac{\partial} {\partial \mu_R^2 } +
  \beta(\alpha_s) \frac{\partial}
  {\partial \alpha_s } \right] R &=& 0 \;,
\end{eqnarray}
where we have defined
\begin{equation}
\beta(\alpha_s) \equiv \mu_R^2  \frac{\partial
  \alpha_s} {\partial \mu_R^2 } \;.
\label{eq:sm:qcd:beta}
\end{equation}
An important observation is that the
$Q$-dependence of the quantity $R$ would not have come about in a classical
theory: it is a feature of the renormalised theory which arises due to
the introduction of the scale $\mu_R$.

Even though the $\beta$-function of QCD is currently known to four-loop accuracy,
\begin{equation}
\beta(\alpha_s) = - \alpha_s^2 ( \beta_0 + \beta_1 \alpha_s + \beta_2
\alpha_s^2 + \beta_3 \alpha_s^3 + ... ) \;,
\end{equation}
only the one-loop result is necessary for qualitative understanding of
QCD:
\begin{equation}
\beta_0 = \frac{ 11 C_A - 4 T_R n_f }{ 12 \pi }\;.
\label{eq:sm:qcd:betazero}
\end{equation}
For most phenomenology the number of active quark flavours can be
taken to be $n_f = 5$ and hence the $\beta$-function is
negative when $\alpha_s$ is small. This implies that the running
coupling $\alpha_s$ decreases to zero as an inverse power of $\ln
Q^2$. To one-loop order, we have
\begin{equation}
\alpha_s(Q^2) = \frac{ \alpha_s(\mu_R^2) } { 1 + \alpha_s(\mu_R^2)
  \beta_0 \ln (Q^2/\mu_R^2) }\;.
\label{eq:sm:qcd:alphasrun}
\end{equation}
Hence QCD interactions become weak at high energy,
a feature called \textit{asymptotic freedom}, and strong at low
energy. This is in contrast to QED, where the observed charge of the
electron is smaller at large distances. This can be thought of as
being due to the self-interactions of the gluons: emission of virtual gluons, which are
themselves charged, causes the colour charge of the source that emitted
them to `leak out' to the vacuum. In effect, this reduces the colour
force at short distances since the interaction between distributed
charges that overlap is weaker than that between point charges.
 
\subsubsection{Parton evolution}
\label{sec:sm:qcd:partevolution}
Free quarks or gluons have not been observed in Nature. This has led to the
confinement hypothesis: the only energy eigenstates of QCD that can
exist in Nature have to be colour-neutral (or colour-singlets). There is strong
circumstantial evidence in favour of the hadrons as bound states of
quarks and anti-quarks. For example, there exists quantitative understanding of
high energy inelastic scattering of hadrons once they are taken to
be composites of quarks and gluons. In fact, hadron-hadron and hadron-lepton scattering can be
described in terms of perturbation theory and the parton model, a
direct consequence of the property of asymptotic freedom. The
basic assumption of the parton model is that interactions of hadrons
are due to interactions of more elementary entities, called partons, which turn out to be the quarks
and gluons of QCD. The number and momenta of partons are most conveniently described in
terms of parton density functions (PDFs). Theoretical and experimental
details on the parton model can be found in~\cite{Ellis:1996qj}. 

The PDFs are fundamentally non-perturbative and at present cannot be
predicted from first principles. However, the evolution
equations for their scale-dependence can be derived (see, for
example,~\cite{Ellis:1996qj}). We denote the fraction of
momentum of the proton that a parton $i$ possesses at scale $Q^2$ as
$x$. The momentum fraction distribution is then denoted by
$f_i(x,Q^2)$. The equation describing the evolution of $f_i(x,Q^2)$,
known as the DGLAP (Dokshitzer-Gribov-Lipatov-Altarelli-Parisi)
equation, takes the following form:
\begin{equation}
Q^2 \frac{ \partial } { \partial Q^2 } f_i(x,Q^2) = \sum_j \int_x^1
\frac{\mathrm{d} z} {z } \frac{\alpha_s} {2\pi} P_{ij} (z)
f_j (x/z,Q^2) \;,
\label{eq:DGLAP}
\end{equation}
where $P_{ij}(z)$ are the so called (regularised) splitting functions,
related to the probability of finding a parton $i$ in a parton $j$ and
the integral is taken over all possible momentum fractions for the
splittings, $z$. The regularised splitting functions at leading order
are given by
\begin{eqnarray} 
P_{qq}(z) &=& C_F \left[ \frac{1+z^2} { (1 - z)_+ } + \frac{3}{2}
  \delta (1-z) \right]  \;,\nonumber\\
P_{qg}(z) &=& T_R [ z^2 + (1 - z)^2 ] \;,\nonumber \\
P_{gg}(z) &=& 2 C_A \left[ \frac{z} { (1-z)_+ } + \frac{1-z} { z} + z ( 1
  - z) \right] \nonumber \\
&+& \frac{1}{6} ( 11 C_A - 4 n_f T_R ) \delta ( 1 -z )
\; ,
\end{eqnarray}
where we have used the so-called `plus' prescription, 
\begin{equation}
\int_0^1 \mathrm{d} x \frac{ f(x) } { (1-x)_+ } = \int_0^1 \mathrm{d}
x \frac{ f(x)
  - f(1) } { 1 - x } \;,
\end{equation}
and the values of the constants $C_A$, $C_F$ and $T_R$ for QCD have
been given in section~\ref{sec:sm:qcd:su3}.
\subsubsection{Parton branching}
\label{sec:sm:qcd:partonbranch}
\begin{figure}[!t]
  \centering 
  \includegraphics[scale=0.60, angle=0]{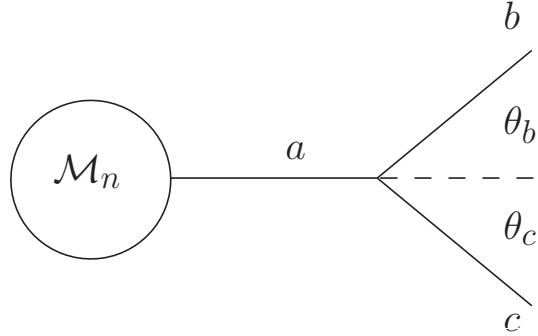}
  \caption[]{The kinematics of time-like branching: an outgoing
    parton, part of an $n$-body matrix element $\mathcal{M}_n$, branches
    into partons $c$ and $b$ at angles $\theta_b$ and $\theta_c$.} 
\label{fig:sm:qcd:branchtime}
\end{figure}

\begin{figure}[!t]
  \centering 
  \includegraphics[scale=0.60, angle=0]{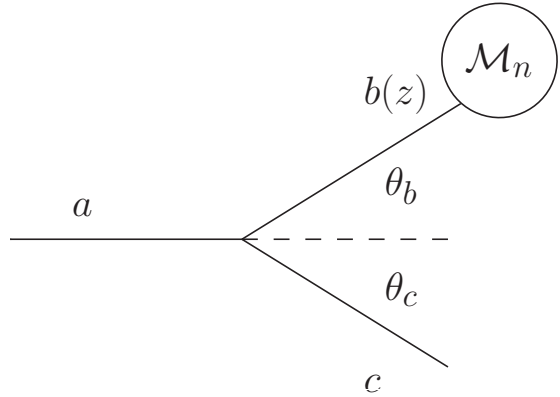}
  \caption[]{The kinematics of space-like branching: an incoming
    parton $a$ branches into partons $b$ and $c$. Parton $b$ carries a
  momentum fraction $z$ of parton $a$. Parton $b$ then takes part
 in an interaction in some $n$-body matrix element $\mathcal{M}_n$.} 
\label{fig:sm:qcd:branchspace}
\end{figure}

Perturbative calculations in QCD are hard beyond leading
order: the work involved increases roughly factorially with the
order. However, there are cases when we cannot truncate the series to
a fixed order since there are higher order terms that are
enhanced in certain  regions of phase space. Such a region is collinear parton
emission from a parton involved in a scattering process. This can be
either an incoming parton or an outgoing parton. Branchings of
outgoing partons are called `time-like' (Fig.~\ref{fig:sm:qcd:branchtime}) and branchings of incoming
partons are called `space-like' (Fig.~\ref{fig:sm:qcd:branchspace}). 

We consider the kinematics for the time-like branching shown in
Fig.~\ref{fig:sm:qcd:branchtime} first. Our aim is to calculate the
modification to the $n$-body cross section when we have a single
branching in the collinear approximation. We define the opening angle
between the outgoing partons $b$ and $c$ as $\theta = \theta_b +
\theta_c$. The collinear approximation implies that $\theta \rightarrow
0$. If we neglect the parton masses then $p_b^2 \approx 0$ and $p_c^2
\approx 0$. If we then define $t \equiv p_a^2$, using $p_a = p_b + p_c$, we may write
\begin{equation}
t = 2 p_b \cdot p_c = 2 E_b E_c (1 - \cos\theta)\;,
\end{equation}
where $E_b$ and $E_c$ are the corresponding energies of the partons
$b$ and $c$. Defining the energy fraction of
the splitting, $z$, by
\begin{equation}
z \equiv \frac{E_b}{E_a} = 1 - \frac{E_c}{E_a}\;,
\end{equation}
$t$ can then be approximated for small $\theta$ as
\begin{equation}
t \approx z (1-z )E_a^2 \theta^2\;.
\label{eq:sm:qcd:tproptothetasq}
\end{equation}
Due to the propagator factor, the $(n+1)$-body matrix element squared, $\left|\mathcal{M}_{n+1}
\right|^2$, is proportional to $1/t$. In fact, it can be shown that
it can be written in terms of the $n$-body matrix element squared as
\begin{equation}
\left|\mathcal{M}_{n+1} \right|^2 = \frac{8\pi \alpha_s } { t }
\hat{P}_{ba}(z) \left|\mathcal{M}_{n}\right|^2\;.
\label{eq:sm:qcd:mnplusone}
\end{equation}
 The splitting function in this case, $\hat{P}_{ba}$, is
`unregularised', i.e. it may diverge as $z\rightarrow 0$ or
$z\rightarrow 1$. It will be shown later how to obtain a regularised
splitting function when including the virtual corrections to the
differential cross section, for the case of space-like branching (section~\ref{sec:qcdeff:globalinc}). The
origin of the collinear enhancement is now obvious: it is due to the
$1/t$ factor which diverges as $t \rightarrow 0$, i.e. when the internal
parton line is on-shell. 

We can use Eq.~(\ref{eq:sm:qcd:mnplusone}) to compute the cross
section for one branching in terms of that for no branchings, using the
following relations:
\begin{eqnarray}
\mathrm{d} \sigma_{n+1} &=& \frac{ 8 \pi \alpha_s } {t} \hat{P}_{ba}(z)
\mathcal{F} \left|\mathcal{M}_{n}\right|^2 \mathrm{d} \Phi_{n+1}\;,
\nonumber \\
\mathrm{d} \sigma_{n} &=& \mathcal{F} \left|\mathcal{M}_{n}\right|^2
\mathrm{d} \Phi_{n} \;,
\end{eqnarray}
where $\mathcal{F}$ is the flux factor for the initial state and
$\Phi_{m}$ is the relevant $m$-body phase space. It can be shown that
\begin{equation}
\mathrm{d} \Phi_{n+1} = \mathrm{d} \Phi_{n} \frac{1}{ 4 (2\pi)^3 }
\mathrm{d}t \mathrm{d} z \mathrm{d} \phi\;,
\end{equation}
where $\phi$ is the azimuthal angle related to the branching. 
Hence, in the collinear approximation, when all angles are small, we have
derived the relation between the $n$-body differential cross section and
the differential cross section for one time-like emission,
\begin{equation}
\mathrm{d} \sigma_{n+1} = \mathrm{d} \sigma _n \frac{ \mathrm{d} t} {
  t } \mathrm{d} z \frac{ \alpha_s } { 2 \pi } \hat{P}_{ba}(z) \;,
\label{eq:sm:qcd:singlebranch}
\end{equation}
where an average/sum over initial/final spins has been taken. We can repeat the same procedure for space-like
branchings (Fig.~\ref{fig:sm:qcd:branchspace})  for which we define
$\left|p_b^2\right| \equiv t$. We then have $t = E_a E_b \theta^2_c$
and the phase space factor becomes
\begin{equation}
\mathrm{d} \Phi_{n+1} = \mathrm{d} \Phi_{n} \frac{1}{ 4 (2\pi)^3 }
\mathrm{d}t \frac{\mathrm{d} z}{z} \mathrm{d} \phi\;,
\end{equation}
where $z = E_b / E_a = 1 - E_c / E_a$. But in this case the
initial-state flux factor, $\mathcal{F}$, changes by a factor of $z$
because of the change of energy of the incoming parton from $E_b$ to
$E_a$. Thus, it turns out that the
expression for space-like branching is identical to the one for
time-like branching, Eq.~(\ref{eq:sm:qcd:singlebranch}).
\section{Beyond the Standard Model}
\label{sec:bsm}

\subsection{The need for BSM physics}
\label{sec:bsmmotivation}
The standard model is phenomenally successful at describing strong,
weak and electromagnetic interactions, with precision results up to
$\mathcal{O}(100~\mathrm{GeV})$. Given its success, why don't we just look for the `missing'
Higgs particle and declare the end of particle physics once we
discover it? The reason is
that there are conceptual and phenomenological hints that the SM is
incomplete. The conceptual issues include the multitude of unexplained parameters,
family replication and flavour hierarchies, the inability of the SM to
incorporate gravity and the hierarchy problem. Phenomenological
hints are neutrino masses, Dark Matter, the cosmological vacuum energy
(also known as Dark Energy) and the quest for Grand Unification and
coupling constant merging. We discuss a few of these issues here.
\subsubsection{The hierarchy problem and new physics}
\label{sec:bsmhierarchy}
\begin{figure}[!t]
  \centering 
  \includegraphics[scale=0.75, angle=0]{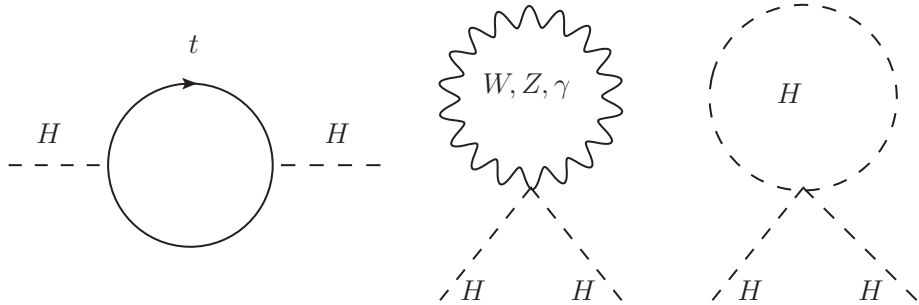}
  \caption[]{The three most `dangerous' quadratically divergent one-loop corrections to the
    Higgs boson mass in the Standard Model. From left to right: the top
    contribution, the gauge boson contribution and the Higgs boson contribution.} 
\label{fig:bsm:motiv:higgscorr}
\end{figure}

An important conceptual issue, known as the \textit{hierarchy problem}, concerns the
Higgs field. Essentially the problem arises since, in its usual
Standard Model incarnation, the Higgs boson is considered to be a fundamental scalar. Fundamental
scalars suffer from radiative instability in their masses due to
radiative corrections. The three most `dangerous' contributions from
radiative corrections to the Higgs boson
mass in the SM come from one-loop diagrams with top quarks, gauge
bosons and the Higgs boson itself, as shown in
Fig~\ref{fig:bsm:motiv:higgscorr}. 
The contributions to the Higgs boson mass squared are
proportional to a cut-off scale squared, $\Lambda^2$, with the top contribution
being negative and the gauge and Higgs boson contributions being
positive. If $\Lambda$ is set to be the Planck scale, $M_{Pl} \sim
10^{19}~\mathrm{GeV}$, then the Higgs boson mass ought naturally to be of that
order, barring any unnatural cancellations between the positive and
negative contributions. On the other hand, experimental evidence shows
that the Higgs boson should
be light: its mass should be $\lesssim 245~\mathrm{GeV}$, assuming no
new physics~\cite{Nakamura:2010zzi}. To achieve this,
the cut-off scale $\Lambda$ should be much lower than
the Planck scale and new
physics, possibly in the form of new resonances, is expected to appear at the electroweak scale $\Lambda_{EW} \sim
\mathcal{O}(1~\mathrm{TeV})$. 

\subsubsection{The little hierarchy problem}\label{sec:bsm:motiv:littlehierarchy}
If new particles are indeed responsible for cancelling the quadratic
divergences to the square of the Higgs boson mass, their masses should be $\mathcal{O}(1~\mathrm{TeV})$ by
naturalness. However, current experimental data already set some
constraints on possible new physics at the TeV scale. For example, absence of
nucleon decays and strong bounds on flavour-changing neutral currents
indicate that these effects cannot receive any significant
contributions from TeV-scale physics. Precision electroweak
measurements put constraints on operators arising by exchanging new
heavy particles and scales which suppress them are required to be
larger than $2-7~\mathrm{TeV}$~\cite{Barbieri:1999tm}. Therefore,
there exists an issue of compatibility between the experimental data and
the expectation of the masses of the new particles required in order to satisfy the
naturalness of the low Higgs boson mass. This is sometimes referred to as the
`little hierarchy problem'. 

A possible `solution' to this problem relies on the fact that the quadratic
sensitivity to the high energy physics of the Higgs boson mass is a result
of loop contributions: to cancel the quadratic divergences, the new
TeV-scale particles only need to appear in interaction vertices in
pairs or more. Therefore, we can suppress the tree-level contributions,
while maintaining the cancellation of the loop contributions, by introducing a new symmetry acting on the new particles,
under which all the SM fields are neutral. The simplest, and most
common, choice is a $Z_2$ symmetry, or parity. 
\subsubsection{Dark Matter}
The nature of Dark Matter (DM) has been an open question in astrophysics
since the initial observations by Fritz Zwicky in 1933 which required the
existence of non-luminous, weakly interacting matter to explain the
orbital velocities of galaxies in clusters.  Following the initial discovery, observations have indicated that galaxy rotation curves do not fall
off with radial distance from the galactic centre and evidence from
the cosmic microwave background (CMB) indicates that DM makes
up about 25\% of the energy density of the
Universe~\cite{Spergel:2006hy}. 

There are two basic types of candidates for DM. The first is Massive Compact Halo
Objects (MACHOs). MACHOs are objects of baryonic origin such as black holes,
brown dwarf stars and giant planets. It has been shown, however that
MACHOS cannot account for more than 20\% of DM~\cite{Alcock:2000ph}. The second type of candidates are weakly interacting particles. The only candidates within the SM, the neutrinos, do not possess the necessary density to
compose the Dark Matter. In addition, the existence of hot (i.e. relativistic) DM is not
consistent with observations and hence any candidates should be
relatively heavy. Thus DM is thought to consist of cold
(i.e. non-relativistic), stable, or at least meta-stable~\cite{Raklev:2009mg,Ellis:2006vu}, massive particles which appear
in theories beyond the SM. These are usually called Weakly Interacting
Massive Particles, or WIMPs. There are good arguments that these
particles might appear at the TeV scale~\cite{Dimopoulos:1990gf}.
\subsubsection{Dark Energy}
Strong evidence for the existence of Dark Energy comes
from distant supernovae observations, indicating that our Universe is
currently undergoing an 
accelerated expansion. The question of the nature of Dark Energy is considered
by theorists to be even more severe than the
Dark Matter problem, not only because Dark Energy is thought to
contribute about 70\% of the energy density of the Universe, but also
because we currently have no strong theoretical explanations for it: it is
totally mysterious. It could possibly be the vacuum energy, in the form of a
cosmological constant. The issue is much worse than the
previously mentioned hierarchy problem: calculations of the energy
density of vacuum in quantum field theory, with a cut-off at
$10^{14}~\mathrm{GeV}$,\footnote{This scale could be the Grand
  Unification scale, for example.} give a
result for a cosmological constant of the order of $10^{54}~\mathrm{GeV}^4$,
whereas the measured value of the dark-energy density is
$10^{-47}~\mathrm{GeV}^4$~\cite{book:dine}. It is likely that
solving the problem requires a complete quantum theory of
gravity~\cite{Bustamante:2009us}.
\subsubsection{GUTs and coupling constant merging}
A troubling feature of the Standard Model is the fact that it contains
almost 20 parameters. It is thought that a fundamental theory of nature
should be able to explain the origin of their values through some
underlying principle. We would also need to
address the issue Feynman wasn't comfortable with at the conception
of the Standard Model: the fact that the different components `have
not yet been smoothed out'. It basically boils down to the fact that
we have three gauge couplings of different magnitudes. In mathematical terms, the question is why
is the gauge group of the Standard-Model semi-simple and not simple? 

A resolution to this issue is to assume that the underlying gauge
symmetry of nature is in fact a simple group, but it is broken at some
high energy scale down to the gauge group of the SM. For example, the
underlying gauge group could be the $SU(5)$ group, which can be shown
to contain an $SU(3)\times SU(2) \times U(1)$ subgroup. The $SU(5)$
group can then be easily broken down to the SM group via the introduction of a set
of Higgs fields~\cite{book:dine}. Although $SU(5)$ can successfully
incorporate the SM, it predicts proton decay with a lifetime that has
already been excluded~\cite{book:mohapatra}. Nevertheless, it is a
simple example of a Grand Unified Theory (GUT). 

\begin{figure}[!t]
  \centering 
  \includegraphics[scale=0.55, angle=0]{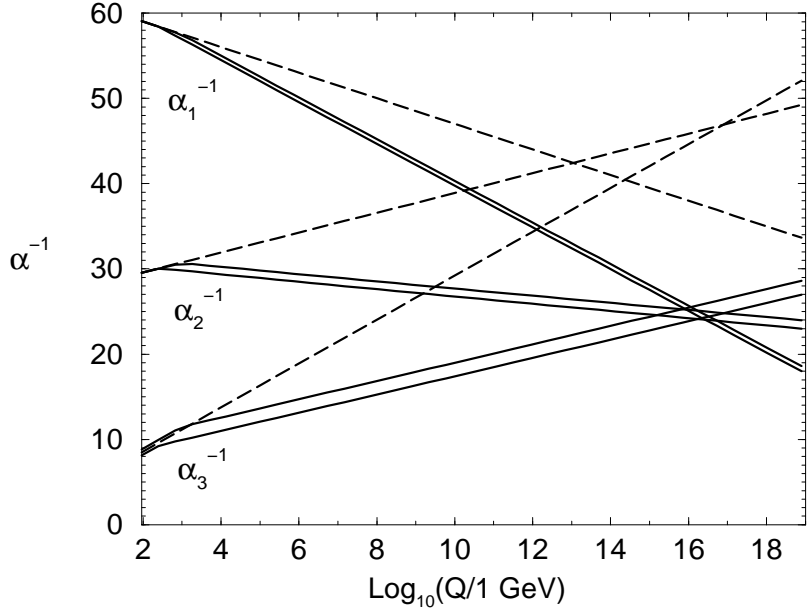}
  \caption[]{Renormalisation group evolution of the inverse gauge
    couplings for $SU(3)$, $SU(2)$ and $U(1)$ ($\alpha^{-1}_{3,2,1}(Q)$
    respectively) in the Standard Model (dashed lines) and the Minimal
  Supersymmetric Standard Model (solid lines). The super-particle mass
thresholds are varied between $250~\mathrm{GeV}$ and $1~\mathrm{TeV}$,
$\alpha_3(m_Z)$ between $0.113$ and $0.123$ and two-loop effects are included~\cite{Martin:1997ns}.}
\label{fig:bsm:motiv:gaugeunification}
\end{figure}

A GUT can possibly solve the problem of seemingly separate gauge
couplings. As we have already demonstrated in the case of the strong
coupling constant, $\alpha_s$, the gauge couplings are scale-dependent
quantities. If the hypothesis of unification is to hold, they must all
equal each other at some scale, the grand unification scale. Below
this scale, the simple group (e.g. $SU(5)$) is broken down to the SM
group and the separate couplings have different behaviour. In fact,
below the grand unification scale, the couplings for each group
$SU(3)$, $SU(2)$ and $U(1)$ evolve according to the respective
$\beta$-functions with no memory of the simple group they originated
from. If the exercise is performed, evolving from low to high
scale, however, it is found that the unification of the couplings in
the Standard Model is a `near-miss', as
can be seen by the dashed lines in Fig.~\ref{fig:bsm:motiv:gaugeunification}. 

\subsection{Survey of BSM theories}
\label{sec:bsmsurvey}

\subsubsection{Supersymmetry}
Supersymmetry is one of the most popular extensions to the Standard
Model, and not without good reasons. Conceptually it is very
appealing: it is the only possible extension to the spatial symmetries
of the theory in flat, four-dimensional space. It appears to be able to accommodate a
solution to the hierarchy problem, contains natural candidates for Dark
Matter and solves the `near-miss' of the gauge-coupling unification in
the SM. It is also a powerful tool for understanding quantum field
theories, especially in the non-perturbative
regime~\cite{Krippendorf:2010ui}.

Supersymmetry introduces fermionic operators $Q_\alpha$ and
$\bar{Q}_{\dot{\alpha}}$ to the Poincar\'e generators $P^\mu$ (corresponding to translations) and
$M^{\mu\nu}$ (corresponding to rotations and Lorentz
boosts).\footnote{Hence, the Poincar\'e
  group corresponds to the basic symmetries of special
  relativity.} The new operators satisfy the following anti-commutation relation:
\begin{equation}
\{Q_\alpha, \bar{Q}_{\dot{\beta}}\} = 2 (\sigma^\mu)_{\alpha
  \dot{\beta}} P_\mu\;, 
\end{equation}
where $\sigma^\mu$ is the 4-vector of Pauli matrices. The
anti-commutation relation implies that two symmetry transformations $Q_\alpha
\bar{Q}_{\dot{\beta}}$ have the effect of a translation. This is to be
expected since the operators themselves carry spin angular momentum,
so it is clear that supersymmetry must be a space-time symmetry.

Let $\left| B \right>$ be a bosonic state and $\left| F \right>$ a fermionic
state. Then we have, schematically,
\begin{eqnarray}
Q_\alpha \left| F \right> = \left| B \right>\nonumber \;,\\
\bar{Q}_{\dot{\beta}} \left| B \right>  =  \left| F \right> \;,
\label{eq:bsm:susy:bosonfermion}
\end{eqnarray}
and, as a result,
\begin{equation}
Q\bar{Q} : \left| B \right>  \rightarrow \left| B (\mathrm{translated}) \right>\;.
\end{equation}
Supersymmetry is a symmetry between bosonic particles and fermionic particles. The matter fields (in the
fundamental representations) of the Standard Model, fit into chiral
multiplets, while the gauge fields fit into vector multiplets. In
supersymmetry, a chiral multiplet contains a fermion field and a
scalar field, related by a supersymmetry transformation such as the one described by
Eqs.~(\ref{eq:bsm:susy:bosonfermion}) and a vector multiplet contains a
vector field and a fermion field also related by the same
transformation. If supersymmetry is exact (i.e. unbroken) the
particles related to fields in the same multiplet should possess the same mass.

The minimal additional content of a supersymmetric theory to the matter fields
of the SM consists of a (super-)scalar for each chirality of the SM
fermions and  a (super-)fermion for each SM
vector boson. The Higgs boson sector for electroweak symmetry breaking becomes more complicated, requiring the introduction of a second Higgs
doublet, otherwise the electroweak gauge symmetry would suffer a gauge
anomaly. The conditions for cancellation of gauge anomalies are
already miraculously satisfied in the SM by the known quarks and
leptons, but in a supersymmetric theory, the multiplet which contains
the SM Higgs boson will now contain a fermionic partner. This would spoil
the anomaly cancellation, but can be avoided if a second doublet which
has $Y=-1/2$ is introduced, so that the contributions from the two
fermionic members of the Higgs multiplets are cancelled out. Furthermore, the $Y=-1/2$ doublet is
required so that masses can be given to the $-1/3$ down-type quarks and charged
leptons~\cite{Martin:1997ns}.

\begin{figure}[!t]
  \centering 
  \includegraphics[scale=0.75, angle=0]{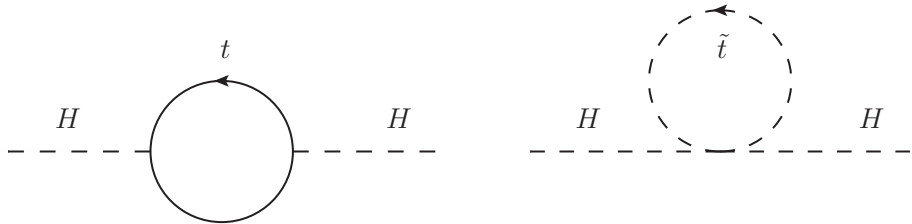}
  \caption[]{Cancellation between the quadratically divergent contributions
  to the Higgs boson mass squared from top and stop loops in the MSSM
  render its mass finite.}
\label{fig:bsm:susy:divergentcancel}
\end{figure}
The Minimal Supersymmetric Standard Model (MSSM) provides
an elegant solution to the hierarchy problem of the SM: the
supersymmetric partner corrections cancel out the quadratically
divergent SM particle corrections to the square of the Higgs boson mass, as
illustrated in Fig.~\ref{fig:bsm:susy:divergentcancel}. Supersymmetry can also accommodate candidates for Dark Matter. As we have
mentioned in section~\ref{sec:bsm:motiv:littlehierarchy}, to solve the
little hierarchy problem we need to introduce a new discrete $Z_2$
symmetry, or parity. In the MSSM, and in other supersymmetric models,
$R$-Parity is introduced, under which the super-partners are odd and
the SM particles are even. If $R$-parity is exact, the lightest
supersymmetric particle (LSP)
cannot decay to lighter SM particles and thus may be stable and contribute
to Dark Matter.

Exact supersymmetry is of course not manifested in nature, otherwise
we would have already observed the super-multiplets. A method to break
supersymmetry is thus required; in fact this is where the elegance
of the theory ends and the complications begin: as a result of the breaking, many free parameters are
added to the model. To maintain the successful cure for the hierarchy
problem, resulting in the observed $W$ and $Z$ masses, the masses of
super-partners should be around the TeV scale, with the lightest ones
at most about $1~\mathrm{TeV}$~\cite{Martin:1997ns}. 
\subsubsection{Extra Dimensions}\label{sec:bsm:extradim}
Another possibility of extending the space-time symmetries of Nature
is to introduce extra spatial dimensions. This has been a
long-discussed idea and has re-surfaced in different
contexts. There are many flavours of models with extra dimensions, each
attempting to address different issues that plague the SM: e.g. Universal Extra Dimensions~\cite{Appelquist:2000nn} or Randall-Sundrum type
scenarios~\cite{Randall:1999ee}. 

The Randall-Sundrum scenario is particularly interesting. In its
simplest form, it is a 5-dimensional theory. The extra dimension is an
interval, with the surfaces (or `branes') at the end of each interval being
(3+1)-dimensional. One surface is at $y=0$ and the other at $y
= \pi R$, where $R$ is a length related to the extra dimensions. The
metric changes from $y=0$ to $y=\pi R$ as $\eta_{\mu\nu} \rightarrow
e^{-k\pi R} \eta_{\mu\nu}$ where $k$ is a constant. This implies that
all the length and energy scales change with $y$. If the fundamental
scale is the Planck scale, $M_{Pl}$, the $y=0$ brane carries physics
at $M_{Pl}$, but all energy scales are `red-shifted' by the
exponential factor until the other brane is reached, where we would
have an exponentially smaller scale. In particular, this scale can be
the electroweak scale: $M_{EW} \approx M_{Pl} e^{-k\pi R} \sim
1~\mathrm{TeV}$. In fact this method `shifts' the hierarchy problem to
a problem of finding the proper mechanism to fix the size of the
extra dimensions. The Randall-Sundrum scenario can also potentially
incorporate a mechanism that explains the flavour structure and mass
hierarchy of the SM~\cite{Fitzpatrick:2007sa}.

Very often, extra-dimensional models suffer from a `little hierarchy'
problem and a discrete parity needs to be introduced to restrict the
production of heavier resonances to pairs. This is sometimes called
`Kaluza-Klein' parity (KK parity). The lightest KK resonance is
sometimes also considered to be a viable Dark Matter candidate. 

\subsubsection{Strong dynamics}
Strong coupling dynamics can provide an elegant and natural solution to the hierarchy
problem. It is natural in a literal sense since we have already
observed an example in Nature in which a large hierarchy arises: between the Planck
scale and the proton mass scales ($1~\mathrm{GeV}$). This is a result of the running of
the QCD coupling constant and the strong coupling regime in the
infrared. 

In technicolour models, as some strong dynamics models are usually
called, mass is given to the electroweak gauge bosons via some
\textit{new} strong dynamics. Extensions to these models allow mass to
be given to the SM fermions as well. In typical types of extensions we obtain
large flavour-changing neutral currents which are highly constrained
by experiment. Technicolour models also have
trouble facing the electroweak precision tests. Minimal Walking
Technicolour (MWTC)~\cite{Foadi:2007ue} is a model which has the smallest deviation from
precision data, with the most economical particle
content. In MWTC theories the coupling remains large and nearly
constant over a wide range of energy scales. Technicolour theories have also been combined with
supersymmetry to alleviate some of the issues present~\cite{Antola:2010zz}.

Strong dynamics theories may contain composite fermionic or bosonic
resonances or composite Higgs particles. We will examine a model which
contains scalar leptoquark resonances in section~\ref{sec:newphys:leptoquarks}.

\subsubsection{Dimensional deconstruction and Little Higgs models}
Dimensional deconstruction involves building extra dimensions instead
of starting with them~\cite{ArkaniHamed:2001ca}. The idea can be used
to construct renormalisable four-dimensional gauge theories that
dynamically generate extra dimensions. In this framework, extra
dimensions can be used purely as an inspiration and may be discarded
in the end, together with all the issues they introduce (for example,
without the need of justification of the size of the extra dimension). As
a result, realistic theories of electroweak symmetry breaking in
four-dimensions can be constructed, with the new feature that they are
perturbative (as supersymmetry is) and insensitive to high energy
details up to a cut-off scale much larger than
$\mathcal{O}(\mathrm{TeV})$. 

`Little Higgs' models are such models, inspired by the method of
dimensional deconstruction~\cite{ArkaniHamed:2001nc}, although most do
not have a simple five-dimensional or `theory space'
interpretation. In Little Higgs models a special `collective' pattern
is created in which the gauge and Yukawa couplings break some global
symmetries. As a consequence of this special pattern, the one-loop
contribution to the Higgs boson mass is not generated. This can be viewed as
the cancellation between divergences caused by the SM particles
(Fig.~\ref{fig:bsm:motiv:higgscorr}) and new resonances, as in
supersymmetry. Unlike supersymmetry, the new resonances have the same statistics as
their SM counterparts. The remaining
corrections to the Higgs boson mass parameter are smaller and no fine tuning is required to keep the
Higgs boson light. At energies of $\mathcal{O}(10~\mathrm{TeV})$, the Little
Higgs description becomes strongly-coupled and the model needs to be
completed in the ultraviolet regime, for example by a QCD-like gauge theory with a
confinement scale around
$10~\mathrm{TeV}$~\cite{Perelstein:2005ka}. Note that in
phenomenologically viable Little Higgs models, a discrete parity, called $T$-parity, needs to be
introduced to solve strains that arise from the
electroweak precision tests~\cite{Cheng:2003ju,Perelstein:2006uw}. The phenomenological consequences are
similar to those in supersymmetry: new heavy
resonances will be pair-produced and the theory may contain Dark
Matter candidates.

\subsubsection{String theory and all that}
String theory~\cite{book:dine} was `discovered' by accident in the late 1960s, first
proposed as a theory of strong interactions. It was later dismissed as
a valid theory of hadronic physics, but some theorists considered the
fact that it contained spin-2 resonances as an indication that it could
possibly lead to a theory of gravity. A lot of work has been done in
the `framework' of string theory and it has since become a popular candidate
for a quantum theory of gravity. However, organising
principles for the `theory' do not exist and at present it seems that the number of
possible solutions is practically infinite.

Another candidate for a theory of quantum gravity is `loop quantum
gravity'. It is a non-perturbative approach to a quantum theory of
gravity, in which no classical background metric is used. One of the
consequences is that quantities such as area and volume are quantised
in units of the Planck length. It has modest aims, not attempting unification; rather, its goal is to quantise Einstein's general
theory of relativity in four dimensions~\cite{Sahlmann:2010zf}. 

Whatever the theory of quantum gravity looks like, what is
certain at this point is that we need a major revision of our
understanding of the nature of space-time to discover a path towards it. 

\begin{boldmath}
\chapter{Monte Carlo methods and event generators}
\label{cha:mc}
\end{boldmath}
\section{Introduction}
\label{sec:mc:intro}
Theories of physical systems are formulated in terms of
`equations of motion'. These are usually differential equations which describe
the rate of change of variables with respect to system parameters,
such as time. For concreteness, let us assume that the rate of change
of a particle's position in one dimension, $X$, is given by the
following differential equation:
\begin{eqnarray}
\frac{ \mathrm{d} X } { \mathrm{d} t } = f(t) \;,
\end{eqnarray}
where $f(t)$ is a given function of time. To calculate the
displacement of the system from $t_1$ to $t_2$, we need to integrate
the differential equation:
\begin{equation}
X = \int_{t_1}^{t_2} f(t) \mathrm{d}t \;.
\end{equation}
In fact, a vast majority of problems in Physics can be reduced to
solving particular integrals. Most of these integrals cannot be solved
analytically, forcing us to resort to numerical techniques to evaluate them.

In one dimension, calculating integrals numerically is just a matter of applying
certain well-known techniques, such as the trapezium rule, Simpson's
rule, gaussian quadrature and so on. These take the values of $f$ on $N$ points
$\{t_1,...,t_N\}$, at certain fixed intervals, and yield an estimate
of the integral. These `quadrature' methods are based on approximating
the function $f(t)$ with some polynomial. An alternative technique is
based on the fact that the value of the integral can be recast as the
average of the integrand:
\begin{equation}
X = (t_2 - t_1) \left< f(t) \right>\;.
\end{equation}
We can approximate the average of the integrand by taking $N$
values of $t$, uniformly distributed on $(t_1,t_2)$, and hence obtain a
reasonable estimate of $X$:
\begin{equation}
X \approx  (t_2 - t_1) \frac{1}{N} \sum_{i=1}^N f(t_i) \;.
\label{eq:mc:Xaverage}
\end{equation}
The order in the sum in Eq.~(\ref{eq:mc:Xaverage}) is not of any
significance so it is possible to draw the $N$ values of $t_i$
randomly. We can then write
\begin{eqnarray} 
X &\approx& (t_1 - t_2) \left< f(t_i) \right> \nonumber \;, \\
t_i &=& (t_2 - t_1 ) \rho_i + t_1 \;,
\label{eq:mc:Xmc}
\end{eqnarray}
where $\rho_i$ is a random number\footnote{A random number is one
  whose value is unpredictable from any initial information. In
  practice \textit{pseudo}-random numbers are used in Monte Carlo event
  generators. These are sequences of numbers which are fully
  deterministic, but are supposed to be indistinguishable from random
  numbers. We will assume that a pseudo-random number generator has
been provided and we will not make the distinction in this thesis.} on
the interval $(0,1)$. This method of using random numbers to aid the
calculation of integrals is known as `Monte Carlo' integration.\footnote{The name is, of course, inspired by the `Casino de
  Monte Carlo' which is a `legendary
  casino, a jewel of the arts `Belle Epoque', the absolute
  reference for all players. Its wide range of table games is the most
  prestigious and the most complete in Europe'~\cite{casino}. The only association
  of the author of the thesis to the famous casino appears at
  \textit{http://www.hep.phy.cam.ac.uk/theory/andreas/mc.jpg}.} If we
assume $N \gg 1$, we can estimate the accuracy of the method using the
Central Limit Theorem. The distribution of $\left<f(t_i)\right>$ will tend to a Gaussian with
standard deviation $\sigma = \sigma_i / \sqrt{N}$, where $\sigma_i$ is
the standard deviation of the values of $f(x_i)$. This implies that
the inaccuracy of our estimate simply decreases as $1 / \sqrt{N}$. 

In particle physics the dimensionality of the integrals is usually very
large and variable: for an $n$-particle final state, there are $3n - 4$ dimensions, coming from the
three components of momentum and taking into account the total
4-momentum conservation, plus flavour and spin labels. Table~\ref{tb:mc:accuracy} shows the
rate of convergence of the various integral approximation techniques
in $d$-dimensions: the `quadrature' rules all suffer from the problem
that they converge in accordance to $N^{1/d}$, the number of points
along each axis.\footnote{However this is only true for the quadrature
  methods if the derivatives
  exist and are bounded.} The convergence of the Monte Carlo technique overtakes each of
the methods mentioned at $d = 4,8$ and $d = 4m - 2$ respectively and
hence it is well-suited for the high-dimensional integrals that appear
in particle physics.
\begin{table}[!t]
\begin{center}
\begin{tabular}{|c|c|} \hline
Technique & Convergence \\\hline \hline
trapezium & $1/N^{2/d}$ \\ \hline
Simpson's & $1/N^{4/d}$ \\ \hline 
$m$th-order gaussian quadrature & $1/N^{(2m-1)/d}$ \\ \hline
Monte Carlo & $1 / \sqrt{N}$ \\ \hline
\end{tabular}
\end{center}
\caption{The rate of convergence with the number of points $N$ used for
  each method in $d$-dimensions.}
\label{tb:mc:accuracy}
\end{table}
The Monte Carlo integration method in fact possesses many advantages
over numerical quadrature methods: it converges fast in many
dimensions, it can handle complex integration regions, it has a small
`feasibility limit' (the number of function evaluations which must be
made for the method to work) and it is easy to estimate the accuracy
of the result.  Moreover, it is useful in the study of fundamentally
random processes since there can be a direct correspondence between
the parameter space points and actual events being modelled. This is one
of the main reasons why the Monte Carlo method has become an important
tool for  collider experiments, through its use in constructing event generators and detector simulators.
\section{Monte Carlo event generators}
\label{sec:mc:generators}
An event generator can be defined as any program which aims to
simulate individual events, rather than the bulk properties of a
physical process. Using the Monte Carlo method, an event generator is capable of simulating a wide range of interesting processes
that are expected at hadron colliders such as the LHC. 

As we have already noted at the end of the previous section, in particle physics the Monte Carlo method is closely related to the physical process under study in a way
that allows us to make a direct connection between phase space points
and events. The Monte Carlo method can be used to generate
`unweighted' events which can be manipulated like those obtained by experiment. 

\subsection{Unweighted event generation}
\label{sec:mc:eventgen}
The unweighting of events (or phase space points) is performed by a
method called `hit-or-miss'. The method can be briefly described in
the following steps:
\begin{itemize}
\item{We find the maximum of the distribution $f(t)$ in the interval $(t_1,t_2)$, $f_{\mathrm{max}}$, during an initial
  sampling. This is taken to be the maximum weight in the integration region.}
\item{In a subsequent sampling process, we choose whether we keep
    (accept) or throw away (reject) a randomly chosen event with
    properties given by $t$, with probability
    $f(t)/f_{\mathrm{max}}$. We thus end up with a sample of events
    which have been `accepted'.}
\item{During the initial sampling, the value of the integral, $X$, can also be
    calculated using Monte Carlo integration as described by Eq.~(\ref{eq:mc:Xmc})}.
\end{itemize}
It should be understood that $t$ is now a multidimensional
phase space point. The above steps can be translated to particle physics
`language' readily: the point $t$ would be a set of particle
momenta and other quantum numbers, $f(t)$ would be the matrix element
squared for the configuration $t$ and $X$ would be the total cross
section within the cuts set by $t_1$ and $t_2$.

Both the convergence of the integral and the event
generation efficiency can be improved if importance sampling is
employed. The basic idea is to perform a Jacobian transformation so
that the integrand is flatter in the new integration variable, thus
reducing the standard deviation of the weights. Using importance
sampling can improve the efficiency by many orders of magnitude.

\subsection{General features of event generators}
\label{sec:mc:genfeatures}
We outline the components of an event generator. For further
details the recent review~\cite{Buckley:2011ms} is useful. We also
discuss jets: although not explicitly part of the event generation, they are important
for making the connection between the parton picture, arising from
the theoretical calculations, and the hadrons observed in experiments. 
\subsubsection{Hard subprocesses}
A particle physics event generator has at its core the simulation of
`hard subprocesses': particle scattering at large invariant momentum
transfer. These processes are calculable in the framework of QCD, as it becomes asymptotically free at high
energy and hence perturbative.

Unweighted hard scattering events are produced using the `hit-or-miss' method as
described in section~\ref{sec:mc:eventgen}. In the case of hadron
colliders, with incoming hadrons $h_1$ and $h_2$, partons $i$ and $j$
are `extracted' from each respectively, with corresponding momentum fractions $x_i$
and $x_j$. The probability density functions for the momentum fractions are then given by the
relevant Parton Density Functions (PDFs): $f_{i/h_1}(x_i,\mu^2)$ and
$f_{j/h_2}(x_j,\mu^2)$, as defined in section~\ref{sec:sm:qcd:partevolution}. The cross section for the hard
subprocess $h_1 h_2 \rightarrow n + X$, can then be
calculated by assuming that the non-perturbative hadron physics and
the short-distance hard physics can be factorized~\cite{Buckley:2011ms}:
\begin{eqnarray}
\sigma &=& \sum_{i,j} \int_0^1 \mathrm{d} x_i \mathrm{d} x_j \int
f_{i/h_1} (x_i, \mu_F^2) f_{j/h_2} (x_j, \mu_F^2)  \mathrm{d}
\hat{\sigma}_{ij\rightarrow n} (\mu_F, \mu_R) \;,
\end{eqnarray}
where $\mu_F$ and $\mu_R$ are the relevant factorisation~\cite{Ellis:1996qj} and
renormalisation scales and the parton-level differential cross section
may be written as
\begin{eqnarray}
\mathrm{d} \hat{\sigma}_{ij\rightarrow n }(\mu_F, \mu_R) =
  \frac{1}{2\hat{s}} \left| \mathcal{M}_{ij \rightarrow n} \right| ^2
  (\Phi; \mu_F, \mu_R) \mathrm{d} \Phi_n \;,
\end{eqnarray}
where $\Phi$ is the final-state phase space, $\hat{s} = x_i x_j s$ is
the partonic centre-of-mass energy squared in terms of the hadronic
centre-of-mass energy squared, $s$, and $\left| \mathcal{M}_{ij
    \rightarrow n} \right| ^2$ is the matrix element squared for the process, averaged over initial
spin and colour. The phase space $\mathrm{d} \Phi_n$ over the $n$
final-state particles is given by
\begin{equation}
\mathrm{d} \Phi_n = \prod_{k=1}^n \frac{ \mathrm{d} ^3 p_k } { (2\pi)^3
2E_i } (2\pi)^4 \delta ( p_i + p_j - \sum_{k = 1}^n p_k )\;,
\end{equation}
where $p_i$ and $p_j$ are the initial-state momenta. 
\subsubsection{Parton showers}
The particles which participate in the hard subprocesses at hadron
colliders are likely to carry QCD charge. Of course, this is always true for the incoming
partons. Inevitably, colour-charged particles will radiate via the QCD
interaction: quarks and gluons can radiate gluons, gluons can branch out to
quark-anti-quark pairs. This is in accordance with the parton branching picture we
presented in section~\ref{sec:sm:qcd:partonbranch}, where we wrote
down the expression for the differential cross section for a single
branching,
Eq.~(\ref{eq:sm:qcd:singlebranch}). This equation
contains a $1/t$ factor, and hence the $t \rightarrow 0$ phase space
region is enhanced. This is in fact a  collinear-enhanced region,
since, as we have already shown (Eq.~\ref{eq:sm:qcd:tproptothetasq}), $t
\propto \theta^2$, where $\theta$ is the opening angle for the
branching. 

The Monte Carlo method is well-suited for describing parton
branching. This is done in terms of the Sudakov form factor, which
forms the basis of the `parton shower'. The Sudakov form factor,
\begin{equation}
\Delta_{ba} (t_0, t) = \exp \left[ - \int_{t_0}^t \frac{\mathrm{d} t' } {t'} \int
  \mathrm{d} z \frac{ \alpha_s } { 2 \pi } \hat P_{ba} (z) \right]\;,
\label{eq:mc:sudakov}
\end{equation}
is simply the probability of a parton evolving from scale (squared) $t_0$ to $t$
without branching, for a certain type of evolution $ba$ (a parton $a$
evolving to a parton $b$). The integral over $z$ would diverge at $z=0$
and $z=1$ and so needs to be cut-off at appropriate values, functions
of the cut-off scale $t_0$. The Monte Carlo branching algorithm, details
of which can be found in~\cite{Ellis:1996qj}, evolves the parton from
squared scale and momentum fraction $(t_1, x_1)$ to $(t_2,x_2)$
in steps in $(t,x)$ space. The algorithm takes into account collinear
enhancements to all orders in perturbation theory, a procedure called
`resummation' to which we will return in chapter~\ref{cha:qcdrad}.

In addition to the collinear enhancements, there are enhancements due
to the emission of soft gluons. These are treated in detail
in~\cite{Ellis:1996qj}. It turns out that these follow a remarkable
property called angular ordering: an example of a coherence effect
common to all gauge theories. The upshot of this effect is that parton
emissions occur at successively smaller angles as one evolves from the
high scale of the hard subprocess to lower scales. In the case of
final-state showers, this means as one evolves `forwards' from the
hard subprocess to the hadronization scale at which the outgoing
hadrons are formed. But for initial-state showers, it means as one
evolves `backwards' from the hard subprocess to the constituent
partons of the incoming hadrons. It is convenient to generate the
initial-state showers by backwards evolution because the hard
subprocess kinematics must be specified first and the shower must
evolve to satisfy them.

\subsubsection{Hadronization}
\label{sec:mc:hadronization}
The hard subprocess in conjunction with the parton shower will produce a set of partons
(quarks/gluons) from the scattering of the incoming hadrons, possibly
associated with some heavier resonances (heavy quarks, gauge bosons or
new heavy particles) which would then subsequently be decayed. Yet quarks and gluons are never visible in their own
right.

After the parton shower has terminated, we are left with a set of
partons with virtualities (square of 4-momenta) of the order of the cut-off scale, which
lies in the low momentum transfer, long-distance regime. This regime is dominated by
non-perturbative effects, such as the conversion of the partons into
hadrons, or \textit{hadronization}. As a non-perturbative effect, hadronization
can currently only be described by phenomenological models that are inspired by
QCD. The general approach assumes the hypothesis of local
parton-hadron duality, which associates the flow of momentum and
quantum numbers at parton level with those at hadron level. This
hypothesis is important as it allows us to associate the theoretical
parton-level calculations with what is observed at collider
experiments, the hadrons. Here we briefly discuss the main features of
two popular hadronization models: the \textit{string model} and
the \textit{cluster model}. Further details on these can be found
in~\cite{Ellis:1996qj, Buckley:2011ms}.

The \textit{string model} is based on the
assumption of linear confinement at large distances. For example, for
the production of a $q\bar{q}$ pair, the model assumes the
physical picture of a `colour flux tube' being stretched between the
$q$ and the $\bar{q}$. For a uniform tube, this leads to a confinement
picture with a linearly rising potential $V(r) = \kappa r$, where
$\kappa$ is the string constant, phenomenologically taken to have the
value $\kappa \approx 0.2~\mathrm{GeV}^2$. As the $q$ and $\bar{q}$ move apart
the string may break, creating a new $q' \bar{q}'$ pair, and the
system breaks into two colour-singlet systems $q\bar{q}'$ and
$q'\bar{q}$. The string model offers a very predictive framework and
applies to complicated multiparton distributions but contains many
parameters related to flavour properties which need to be determined
by data.

The \textit{cluster model} is based on the so-called preconfinement property of
parton showers: the colour structure of the shower at any evolution
scale is such that the colour-singlet combinations of partons (i.e.
clusters) can be formed with an asymptotically universal (i.e. dependent
only on the evolution scale and the QCD scale) invariant mass distribution. In this model, cluster hadronization starts with
non-perturbative splitting of gluons into $q\bar{q}$ pairs (or
diquark-anti-diquark) and the formation of clusters from
colour-connected pairs. Clusters above a certain (flavour-dependent) maximum mass are first fragmented in a string-like fashion into
lighter clusters. Most clusters then undergo sequential two-body
phase space decays. In the cluster model, heavy flavour, strangeness
and baryon production are suppressed and transverse momenta are
limited by the cluster mass distribution. The model, in combination with the angular-ordered
shower, gives a fairly good overall description of high-energy
collider data, less good than the string model but with fewer
parameters. 

\subsubsection{Jets}
As we have seen, almost immediately after they are produced, quarks and gluons
fragment and hadronize, leading to sprays of energetic hadrons, which
we may call `jets'. However, the definition of a jet is not a simple one,
even though we may easily identify the structure on an event
display. Defining a jet is complicated for several reasons:
\begin{itemize}
\item{Partons have divergent branching probabilities in perturbative
    QCD.}
\item{A jet may
originate from the hadronic decay of a heavy particle or it may be
radiative, arising through the emission of a parton via parton branching.}
\item{It is
also never possible to identify a jet with a single parton: jets are
colour-singlets whereas partons are not.}
\end{itemize}
To address these issues and create a consistent definition of what a jet
is we need to define a `jet algorithm'. These algorithms provide a set
of rules for grouping particles into jets. They involve a 
parameter that defines a distance measure between the particles, used
as a criterion to judge whether they belong to the same jet or not. A
`recombination scheme' is also required, which indicates what the
momentum of the combination of two particles is when they are grouped
together. A jet algorithm together with a recombination scheme form a `jet
definition'. A jet definition should possess the following
properties, the `Snowmass accord'~\cite{Huth:1990mi}, set in 1990 by a group of
influential theorists and experimentalists: 
\begin{itemize}
\item{It has to be simple to implement in an experimental analysis and
    a theoretical calculation.}
\item{It has to be defined at any order of perturbation theory and
    yield a finite cross section at any order of perturbation theory.}
\item{It has to yield a cross section that is relatively insensitive
    to hadronization.}
\end{itemize}
Another important and desirable property not explicitly mentioned but
implied by the above list is `infrared and
collinear safety' (IRC): if one modifies an event by either adding a
collinear or a soft emission, the hard jets that are found
in the event should remain unchanged. Some modern jet definitions do satisfy the Snowmass criteria and are
IRC safe. Examples are the
$k_t$ and anti-$k_t$ algorithms, the Cambridge/Aachen algorithm and
seedless cone algorithms (e.g. SISCone). For a further, more detailed
discussion on carefully defining and using IRC-safe and Snowmass-accord jet algorithms
see Ref.~\cite{Salam:2009jx}.

\subsubsection{The underlying event}\label{sec:mc:ue}
Strong evidence for the existence of the `underlying event' has
existed since the CERN ISR experiment, through measurements of
momentum imbalance~\cite{Akesson:1986iv}. Experimentally, what is called the `underlying event' (UE) includes all
activity in a hadronic collision that is not related to the `signal'
particles from the hard subprocess (for example, the leptons in the
Drell-Yan process). This
definition will obviously include initial- and final-state radiation
described by the parton shower, but for the sake of modelling, these
extra emissions can be attributed to the hard
subprocess. It is then thought that the UE originates from additional
soft scatters that occur during a hadron-hadron collision. It is a
significant effect to consider when it comes to measuring jet
properties: jet algorithms will inevitably gather together any energy
deposits from the soft particles in the vicinity when constructing a jet. This will alter
the energy and internal structure of the jets formed by the algorithm.

Currently, description and understanding of the UE comes mainly through the use of
phenomenological models. The simplest model, called the UA5 model~\cite{Alner:1986is}, treats the UE as additional
soft hadronic activity generated by additional clusters which have
been formed flat in rapidity, with an exponentially falling transverse
momentum distribution. 

A more recent, and phenomenologically more successful, model treats the UE as a sequence
of more or less independent parton interactions which include full
parton showers~\cite{Butterworth:1996zw, Bahr:2008dy}. This is capable of
describing the jet-like structure of the UE. The additional scatters
are always modelled as simple $2\rightarrow2$ QCD scatterings as long
as the scattering contains a hard jet of at least a few GeV. 

We will be examining the effect of the UE described by a model of the
latter philosophy of multiple parton interactions on certain hadron collider
variables in chapter~\ref{cha:qcdrad}. 

\subsection{The \Herwigpp event generator}\label{sec:mc:hpp}
There is a healthy choice of general-purpose Monte Carlo event generators on the market. The
latest incarnations of the more popular generators are \texttt{Pythia
  8}~\cite{Sjostrand:2007gs} and \Herwigpp~\cite{Bahr:2008pv}, and a `new-comer', \texttt{SHERPA}~\cite{Gleisberg:2008ta}. These
differ mainly in physics, philosophy and implementation of the various
components described in section~\ref{sec:mc:genfeatures}, a variety
which allows cross-checking of models and implementations. 

The author of the present thesis has been a heavy user of the
\Herwigpp event generator as well as member of
the \Herwigpp collaboration and hence this thesis will inevitably be heavily
biased towards that event generator. In this section we briefly review
the main features of the \Herwigpp event generator, in reference to
the general features we have presented in
section~\ref{sec:mc:genfeatures}. For further details, one can consult
the manual~\cite{Bahr:2008pv} and public webpages~\cite{herwigweb} which are updated to keep up with the latest developments.

\Herwigpp is based on the \texttt{FORTRAN} event generator \texttt{HERWIG} (which stands for Hadron Emission Reactions
With Interfering Gluons), first published in
1986~\cite{Marchesini:1987cf}. \Herwigpp is not just a rewrite of the
earlier \texttt{FORTRAN} version in the \texttt{C++} language, but also introduces
physics improvements whenever necessary and feasible.
\subsubsection{\texttt{ThePEG}}
\Herwigpp is distributed as a comprehensive collection of plugin
modules to \texttt{ThePEG}, the `Toolkit for High Energy Physics Event
Generation'~\cite{Lonnblad:2006pt}, which provides all the infrastructure that is necessary
to construct an event generator. It can handle random number
generation, the event record and a mechanism for creating physics
implementations for all steps of event generation. It also provides a
reader for the  Les Houches Accord event format~\cite{Alwall:2006yp},
a feature we make use
of in section~\ref{sec:newphys:wprime} to perform a phenomenological
analysis of a heavy charged vector boson model.
\subsubsection{Hard process generation, parton shower and decays}
Three main mechanisms for simulation of hard processes are available
in \Herwigpp:
\begin{enumerate}
\item{A hand-coded set of matrix
    elements for common processes in hadron, lepton and deep inelastic
    scattering collisions. These are written using a reimplementation
    of the \texttt{HELAS} helicity amplitude formalism, which allows
    the spin correlations to be carried forward to the remaining
    event simulation consistently.}
\item{A generic matrix element calculator for $2\rightarrow 2$
    processes, mainly used for Beyond-the-Standard Model physics,
    which automatically determines the permitted diagrams for a set of
    given external legs from a list of active vertices.}
\item{As already mentioned, it is possible to read Les Houches Accord
    event format files at parton level, with any number of legs to be
    read from external sources.}
\end{enumerate}

The parton shower associated with the hard processes is based on a new
evolution variable $\tilde{q}$, motivated from the branching of gluons
off heavy quarks~\cite{Gieseke:2003rz}. The evolution in this variable
ensures the angular ordering of emissions, to take into account the
colour coherence effects. Prior to the parton shower, heavy unstable
particles (e.g. Higgs bosons, $W$,
$Z$, top quarks and other BSM particles) are decayed. All intermediate coloured lines are also showered.  

\subsubsection{Hadronization and hadron decays} 
\Herwigpp uses the cluster hadronization model, described in
section~\ref{sec:mc:hadronization}. The unstable hadrons that are formed are decayed via the same framework as fundamental
unstable particles: using either a general matrix element based on the
spin structure, or a specific matrix element for the important decay
modes. 

\subsubsection{The underlying event}
The implementation of the multiparton scattering in \Herwigpp is
connected to the parton shower and hadronization models. Event
generation starts with sampling the hard process according its matrix
element and PDFs. The parton shower evolves the final-state partons
from the scale of the hard interaction down to the cut-off scale for
the hadronization. The incoming partons are extracted out of the
hadrons and the chosen number of secondary interactions is sampled
according to the QCD $2\rightarrow 2$ matrix elements and the same
PDFs as for the hard process. The partons involved in the additional
hard scatters also undergo parton showers. Secondary interactions are
showered in an identical way to the
hard process. However, backward
evolution has to be modified: for example, an additional scattering
may lead to partons with more energy than the remaining energy of the
hadron remnants, and has to be vetoed. A further modification is that
any incoming partons are always evolved back to an initial gluon, with
a gluon distribution identical to the one in the initial hadron. Further details of the implementation can be found in~\cite{Bahr:2008dy}.

\subsubsection{Beyond the Standard Model}
Each new physics model in \Herwigpp is implemented in a model class which includes
the model parameters and vertex structure. A model input text file
allows for changes in the parameters and all possible production and
decay matrix elements with up to four external legs are
constructed. The BSM models currently available in \Herwigpp (version $2.5.0$) are~\cite{Gieseke:2011na}:
\begin{itemize}
\item{Supersymmetric models: MSSM and NMSSM implemented with
    flexibility in the parameters via the Susy Les Houches Accord file
    reader~\cite{Allanach:2008qq}.}
\item{A model for universal extra dimensions and an implementation of
    the Randall-Sundrum model and ADD-type gravitons.}
\item{A model for transplanckian scattering.}
\item{A model involving non-derivatively and derivatively-coupled
    leptoquark resonances. }
\end{itemize}
We will study the implementation and phenomenology of the leptoquark model in detail in section~\ref{sec:newphys:leptoquarks}.
\section{Next-to-leading order matching} 
\label{sec:mc:nlomatching}
Fixed-order matrix elements are excellent for simulating well
separated, hard partons, but are unable to describe collinear and
soft partons, which have logarithmically divergent probabilities. Parton
showers are good in the opposite region: hard, wide-angled emissions
are handled poorly while the enhanced soft and collinear emissions are well-described, even for multiple emissions. Generic Monte Carlo event generators start the parton shower from a
leading order (LO) distribution of partons to produce a high
multiplicity hadronic state with relatively low transverse
momenta. The fact that the shower starts from the LO distribution
implies that the total cross section is also accurate to that order. 

However, for many processes there exist next-to-leading order (NLO)
perturbative calculations. These may provide significant corrections to
both the total cross section and the shape of distributions of
observables. They are also essential in providing control over the
scale dependence of our calculations, absent from a LO calculation. We would thus like to combine the NLO matrix elements with the
parton shower. The task is non-trivial for several reasons, which we
shall discuss. We
present a brief overview of two popular NLO `matching' methods, the
`Monte Carlo at Next-to-leading Order'
(\texttt{MC@NLO})~\cite{Frixione:2002ik, Frixione:2010wd} method and
the `Positive Weight Hardest Emission Generation'
(\texttt{POWHEG})~\cite{Nason:2004rx, Frixione:2007vw} method. The
discussion in this section has been adapted from~\cite{Buckley:2011ms}
and~\cite{seyi}. We use these two methods to simulate the
production of heavy charged vector bosons at NLO in section~\ref{sec:newphys:wprime}. 

\subsection{MC@NLO}
We begin by describing \texttt{MC@NLO} method for combining the NLO matrix
element with the parton shower. The parton shower and the NLO result
contain terms of the same order and hence when defining any matching
method we have to take care to avoid double-counting. The
\texttt{MC@NLO} method, as we shall see, exhibits
the complication that a small fraction of the generated events possess
negative weights. These are few enough so that the number of events
required for constructing smooth distributions is comparable to that
for an ordinary LO process. 

We demonstrate the procedure by applying it to a toy model, by
assuming the emission of only photons for simplicity. Consider a system that radiates particles with energy $x$, such that $0
\leq x \leq 1$. In perturbation
theory, the cross section for a process at NLO (one photon emission),
after dimensional regularisation in $d = 4 - 2\epsilon$ dimensions is
given by\footnote{Dimensional regularisation
  is a technique used when renormalising a theory: the idea is to compute a
  Feynman diagram as an analytic function of the dimensionality of
  space $d$. Any loop-momentum integral will converge for
  sufficiently large $d$ (in the case of infrared divergences). Hence
  in this case $\epsilon < 0$ is required.}  
\begin{equation}
\sigma_{\mathrm{NLO}} = \lim_{\epsilon \rightarrow 0} \int_0^1
\mathrm{d} x x^{-2\epsilon} \left[ \left(\frac{\mathrm{d} \sigma} {
    \mathrm{d} x}\right)_B + \left(\frac{\mathrm{d} \sigma} {
    \mathrm{d} x}\right)_V + \left(\frac{\mathrm{d} \sigma} {
    \mathrm{d} x}\right)_R \right] \;,    
\label{eq:mc:mcnlo:sigmanlo1}
\end{equation}
where the factor $x^{-2\epsilon}$ has been retained from the phase
space factor and
\begin{eqnarray}
\left(\frac{\mathrm{d} \sigma} {
    \mathrm{d} x}\right)_B &=& B \delta (x)\;, \nonumber \\
\left(\frac{\mathrm{d} \sigma} {
    \mathrm{d} x}\right)_V &=& a \left( \frac{B}{2 \epsilon} + V \right)
\delta (x) \;, \nonumber \\ 
\left(\frac{\mathrm{d} \sigma} {
    \mathrm{d} x}\right)_R &=& a \frac{R(x)} { x }  \;,
\label{eq:mc:mcnlo:sigmanloterms} 
\end{eqnarray}
where $a$ is the coupling constant (analogous to $\alpha_s$ in QCD),
$B$ and $V$ are constant with respect to $x$ and represent the Born
(leading order) and virtual contributions, and $R(x) \rightarrow B$ as
$x\rightarrow 0$, where $R$ is the real contribution.  

Now, at leading order, an infrared-safe observable $O(x)$ has an expectation value
given by
\begin{eqnarray}
\left< O \right>_{\mathrm{LO}} &=& BO(0) \nonumber \\
 &=& \lim_{\epsilon \rightarrow 0} \int
\mathrm{d} xx^{-2 \epsilon} B \delta(x) O(x) \;,
\label{eq:mc:mcnlo:loobs}
\end{eqnarray}
where we have written the second line in a way convenient to be used below.
We can then write down the equivalent NLO prediction for $O$ using
Eqs.~(\ref{eq:mc:mcnlo:sigmanlo1}) and~(\ref{eq:mc:mcnlo:sigmanloterms}):
\begin{eqnarray}
\left< O \right>_{\mathrm{NLO}} &=& \lim_{\epsilon \rightarrow 0} \int
\mathrm{d} xx^{-2 \epsilon} O(x) \left[ B \delta(x) + a \left(
    \frac{B}{2\epsilon} + V\right) \delta (x) + a \frac{ R(x) } { x }
\right] \nonumber \\
&=& (B + aV) O(0) +  \lim _{\epsilon \rightarrow 0} \left[ a
  \frac{B}{2 \epsilon} O(0) + \int_0^1 \mathrm{d}x x^{-2 \epsilon} O(x)
  a \frac{R(x)} {x} \right]\;.
\label{eq:mc:mcnlo:nlo2}
\end{eqnarray}
For an observable $O(x)$, the $\epsilon$ parameter gives rise to poles
of opposite sign in the virtual and real contributions which cancel to
give a finite integral. To show this explicitly we start by using the
fact that
\begin{equation}
a B O(0) \int_0^1 \mathrm{d}x \frac{x^{-2 \epsilon}}{x} = -a \frac{B} {
  2 \epsilon }O(0)\;,   
\label{eq:mc:mcnlo:singular}
\end{equation}
to write the expectation value of $O(x)$ as
\begin{equation}
\left< O \right>_{\mathrm{NLO}} = (B + aV) O(0) + \lim_{\epsilon
  \rightarrow 0} \int_0^1  \mathrm{d} x x^{-2 \epsilon} \left[ a \frac{
    O(x) R(x) - B O(0) }{x} \right] \;, 
\end{equation}
where Eq.~(\ref{eq:mc:mcnlo:singular}) has been added and subtracted
from Eq.~(\ref{eq:mc:mcnlo:nlo2}). The integrand in the second term
does not contain any singularities and hence we can set $\epsilon = 0
$ to get:
\begin{equation}
\left< O \right>_{\mathrm{NLO}} = \int_0^1 \mathrm{d} x \left[ O(0)
  \left( B + a V - \frac{aB}{x} \right) + O(x) \frac{ R(x) } { x }
\right] \;. 
\label{eq:mc:mcnlo:explicitsub}
\end{equation}
This method of making the finiteness of the NLO expression
explicit is called `subtraction'. It yields an NLO-accurate expression
for an observable. The \texttt{MC@NLO} formalism aims to match the shower
Monte Carlo to the NLO calculation, reproducing the expression of
Eq.~(\ref{eq:mc:mcnlo:explicitsub}). 

An observable at leading order, i.e. with \textit{no} radiation from the system before
being interfaced to the shower Monte Carlo, is given by the first line
in Eq.~(\ref{eq:mc:mcnlo:loobs}). The energy of the system in this
case is $x_M = 1$. A shower MC sums the enhanced higher order terms to
all orders to give the following for the distribution of the
observable $O$:
\begin{equation}
\frac{ \mathrm{d} \sigma } { \mathrm{d} O_{\mathrm{LO} + \mathrm{MC}
  }} = B I_{\mathrm{MC}}( O, x_M ) = B I_{\mathrm{MC}} (O,1)\;,
\label{eq:mc:mcnlo:showerlo}
\end{equation} 
where $I_{\mathrm{MC}}(O, x_M)$ is the distribution of the observable
after an MC shower starting with energy $x_M$. The above
expression implies that the total rate at LO is given by $B$, which is
due to the unitary nature of the parton shower. We can make an attempt
to extend Eq.~(\ref{eq:mc:mcnlo:showerlo}) to NLO by simply replacing $O(0)$
and $O(x)$ in Eq.~(\ref{eq:mc:mcnlo:explicitsub}) with the MC shower observables $I_{\mathrm{MC}} ( O,1 ) $
and $I_{\mathrm{MC}} (O,1 - x)$ respectively:
\begin{equation}
\left< O \right>_{\mathrm{naive}} = \int_0^1 \mathrm{d} x \left[
  I_{\mathrm{MC}} (O,1) \left( B + a V - \frac{ aB } { x } \right) +
    I_{\mathrm{MC}} (O, 1 - x) \frac{ R(x) } { x } \right]\;.
\label{eq:mc:mcnlo:naive}
\end{equation}
We have assumed that the energy of the system is $x_M = 1 - x$ after
the radiation of a photon of energy $x$. The na\"ive expression of
Eq.~(\ref{eq:mc:mcnlo:naive}) implies that we need to generate two
events for a randomly chosen $x$: 
\begin{itemize}
\item{A `no emission + shower MC'-type event with $x_M = 1$ and weight
    given by the integrand in the first term, $B + aV - aB /x$, and}
\item{a `one emission + shower MC'-type event with $x_M = 1-x$ and
    weight given by the integrand in the second term, $a R(x) / x$. }
\end{itemize}
We have of course called Eq.~(\ref{eq:mc:mcnlo:naive}) the `na\"ive'
result since it is plagued by two issues. The first is that as
$x\rightarrow 0$, the weights for the two types of events diverge even
though the integral is finite. The second issue is that this procedure
introduces double-counting: terms that appear in the NLO emission also
appear in the shower MC. For an explicit demonstration of how double
counting occurs see Ref.~\cite{seyi}. 

Both identified problems of the na\"ive subtraction can be solved by introducing a
`modified subtraction':
\begin{equation}
\left< O \right>_{\mathrm{mod}} = \int_0^1 \mathrm{d} x \left[
  I_{\mathrm{MC}} (O,1) \left( B + a V - \frac{ aB( 1 - Q(x) ) } { x } \right) +
    I_{\mathrm{MC}} (O, 1 - x) \frac{ R(x) - aBQ(x) } { x }
  \right]\;,
\label{eq:mc:mcnlo:mod}
\end{equation}
where we have added to the first term the following:
\begin{equation}
I_{\mathrm{MC}} (O, 1)\frac{ a B Q (x ) } { x } \;,
\end{equation}
and subtracted from the second term:
\begin{equation}
I_{\mathrm{MC}} (O, 1-x) \frac{ a B Q (x ) } { x } \;,
\end{equation}
where $a Q(x) / x$ is analogous to a splitting function $ \alpha_s
P(z)$ in QCD and $Q(x)$ is a monotonic function that satisfies 
\begin{equation}
0 \leq Q(x) \leq 1\;,\;\; \lim_{x \rightarrow 0} Q(x) = 1\;,\; \; \lim_{x
  \rightarrow 1 } Q(x)= 0\;.
\end{equation}
The difference between the added and subtracted terms does not
contribute to the observable at $\mathcal{O}(a)$ since
$I_{\mathrm{MC}} (O, x_M)$ is independent of $x_M$ at $\mathcal{O}(a^0)$. The function $Q(x)$ is dependent on the shower MC used, which
effectively requires the \texttt{MC@NLO} method to be customised for
each MC. The modified procedure can be shown to coincide with the NLO result at
$\mathcal{O}(a)$ and the integrand for the two `types' of events can
be shown to be finite as $x\rightarrow0$~\cite{seyi}, hence solving
both issues. The extension to QCD from the toy model does not require
any significant changes to the procedure we have described. A
necessary modification involves an extra term
related to the initial-state collinear divergences. 

Notice that the weights in Eq.~(\ref{eq:mc:mcnlo:mod}) can be either positive or
negative. These lead to `unphysical' negative weight events, which can
be handled during the unweighting process by assigning to them a weight
-1 instead of +1. This can be easily taken into account when producing
histograms of distributions, by removing an event from the
histogram bin it corresponds to rather than adding it there. The
events with negative weights, however, can be manipulated in the
same way as the
positively-weighted ones: e.g. cuts can be applied to them and they
can be processed by detector simulation software.
\subsection{POWHEG}
The `Positive Weight Hardest Emission Generation' (\texttt{POWHEG}) method was proposed
to overcome the problem of negatively-weighted events generated by
\texttt{MC@NLO}. The goal is to generate the hardest emission first
using the exact NLO matrix element and yield only positively-weighted
events. Another advantage of the \texttt{POWHEG} method is that it
does not depend upon the subsequent shower MC. 

In Monte Carlo event generators with angular-ordered showers, such as
\Herwigpp, the first emission is not necessarily the hardest
one. Hence, implementing the \texttt{POWHEG} method requires the use
of a transverse momentum veto to ensure that any emissions that follow
the first one are softer. Also, a `truncated shower', extra soft
radiation, must be generated to recover the double-logarithm accuracy
of the shower. 

We first write down the inclusive differential cross section as given
by the first emission in a parton shower,
\begin{equation}
\mathrm{d} \sigma_{\mathrm{PS}} = \mathrm{d} \Phi_0 B \left[
  \Delta(Q^2, Q_0^2) + \int_{Q^2_0} \frac{ \mathrm{d} q^2 } { q^2
  }  \int \mathrm{d} z \frac{ \alpha_s } { 2 \pi } P(z) \Delta(Q^2 ,
  q^2 ) \right]\;,
\label{eq:mc:powheg:ps}
\end{equation}
where $\Delta(q_1^2, q_2^2)$ is the Sudakov form factor (see
Eq.~(\ref{eq:mc:sudakov})), $Q^2$ and $Q^2_0$ are the starting scale
and cut-off scales for the shower respectively and $\mathrm{d} \Phi_0$ is the Born phase
space. Performing the integral would give a
total cross section equal to the Born cross section, as previously
mentioned. 

Note that the real emission matrix element, $R$, can be split into a
singular part and a non-singular part: $R = R^s + R^{ns}$. We can
replace the splitting function in Eq.~(\ref{eq:mc:powheg:ps}) with the
singular part of the real emission matrix element,
\begin{equation}
\frac{ \mathrm{d} q^2 } { q^2
  }  \mathrm{d} z \frac{ \alpha_s } { 2 \pi } P(z) \rightarrow
  \mathrm{d} \Phi_r \frac{R^s}{B} \;, 
\end{equation}
where $\mathrm{d} \Phi_r$ is the phase space of the radiation
variables. The replacement can be carried over to the Sudakov form
factor by defining~\cite{Buckley:2011ms}
\begin{equation}
\bar{\Delta}(Q^2,q^2) = \exp \left[ - \int \mathrm{d} \Phi_r \alpha_s
  \frac{ R^s } { B } \right] \;.
\end{equation}
For angular-ordered parton showers, a hard matrix element correction, related to the non-singular term $R^{ns}$,
is necessary to cover the whole phase space and we get the following
result:
\begin{equation}
\mathrm{d} \sigma_{\mathrm{PScorr}} = \mathrm{d} \Phi_0 B \left[
  \bar{\Delta} (Q^2 , Q^2_0) + \int \mathrm{d} \Phi_r \alpha_s \frac{
    R^s } { B } \bar{\Delta} ( Q^2 , q^2 ) \right]+ \mathrm{d} \Phi_{0,r}
  \alpha_s R^{ns}\;.
\label{eq:mc:powheg:pscorr}
\end{equation}
In the \texttt{POWHEG} method, we define $R^s = R$ and hence $R^{ns} =
0$. We then write
\begin{equation}
\mathrm{d} \sigma_{\mathrm{POWHEG}} = \mathrm{d} \Phi_0 \bar{B} \left[
  \bar{\Delta} (Q^2 , Q^2_0) + \int \mathrm{d} \Phi_r \alpha_s \frac{
    R^s } { B } \bar{\Delta} ( Q^2 , q^2 ) \right] \;,
\label{eq:mc:powheg:powheg}
\end{equation}
where we have now used the NLO-weighted Born matrix element, $\bar{B}$, defined by
\begin{equation}
\bar{B} \equiv B + \alpha_s V + \int \left( R - C \right) \mathrm{d} \Phi _r\;,
\end{equation}
where $C$ are counter-terms chosen to approximate $R$ with the same
singularities so that the integral over four dimensions is
finite. Parton showering using Eq.~(\ref{eq:mc:powheg:powheg}) will
give similar emissions as the first term in
Eq.~(\ref{eq:mc:powheg:pscorr}), but with a global NLO reweighing
$\bar{B}/B$, which can be considered to be a local
$k$-factor.\footnote{A $k$-factor is the ratio between a cross section
  at LO and one at higher order: $k = \sigma_{\mathrm{h.o.}} /
  \sigma_{\mathrm{LO}}$.} The replacement $B \rightarrow \bar{B}$ will
result in the integrated cross section being correct to NLO.

We note that the term $ \alpha_s V + \int \left( R - C \right) $,
which can be negative, is formally of order $\alpha_s$ and therefore
it is always overcome by the positive-definite term $B$ (otherwise the
perturbation expansion would not be valid). This results
in positive weights being assigned to each event, since the
NLO-weighted $\bar{B}$ is positive-definite. 

\begin{boldmath}
\chapter{Effects of QCD radiation on hadron collider observables}
\label{cha:qcdrad}
\end{boldmath}
The original work in this chapter was done in collaboration with Bryan
Webber and Jennifer Smillie and appears in~\cite{Papaefstathiou:2009hp,
  Papaefstathiou:2010ru, Papaefstathiou:2010bw}.
\section{Introduction}
\label{sec:qcdeff:intro}
We have already laid down a few convincing arguments for why we should
expect new particles or phenomena at the TeV scale in
section~\ref{sec:bsmmotivation}. These include the naturalness of the expected Higgs mass and the electroweak
    symmetry breaking scale, and arguments that dark matter should naturally occur at the TeV scale.
The LHC has been designed to investigate the electroweak scale and
hence it is hopeful that it will shed light to some of the puzzles of
the Standard Model within the next few years,  by uncovering new phenomena.

\begin{figure}[!t]
  \centering 
  \includegraphics[scale=0.75, angle=0]{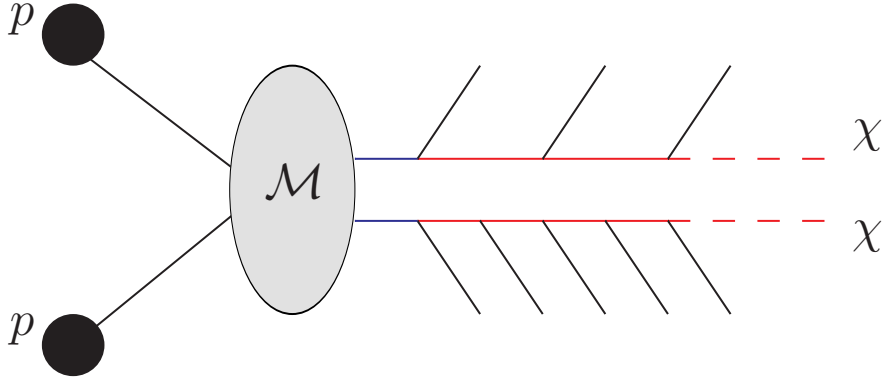}
  \caption[]{New physics processes are expected to provide complex
    signals: multiple jets and/or leptons resulting in combinatorial
    issues plus missing energy. The diagram shows the production of
    two heavy resonances (blue) via some matrix element $\mathcal{M}$
    and their subsequent decay chains into SM particles (black), new
    resonances (red) and finally into some weakly interacting
    particles that escape detection (red dashes). This could be, for
    example, the production of a gluino and a squark with subsequent
    decay into a $\chi$ in supersymmetry. } 
\label{fig:qcdrad:challenges1}
\end{figure}

But searching for new physics at a hadron collider such as the LHC is a
non-trivial task. There are two main challenges we need to
face. First of all, possible new physics
signals can be quite complex (see Fig.~\ref{fig:qcdrad:challenges1}
for an illustration). New
heavy resonances can decay into multiple jets and/or leptons. If the
new dynamics possesses some discrete parity, such as $R$-parity in
supersymmetry, then the new particles will be pair-produced, resulting in
two identical (or similar) decay chains and introducing a
combinatorial problem. Moreover, the end-point of likely long decay
chains can very possibly be a particle which interacts weakly with ordinary matter,
especially since we expect it to be a dark matter candidate (a
WIMP). This would result in missing energy in our detector which would complicate
the issue further,
especially in combination with the SM neutrinos which would also escape the
detector. Moreover, signals are expected to have low production rates,
especially in comparison against backgrounds usually occurring orders of magnitude more often. 

\begin{figure}[!t]
  \centering 
  \includegraphics[scale=0.75, angle=0]{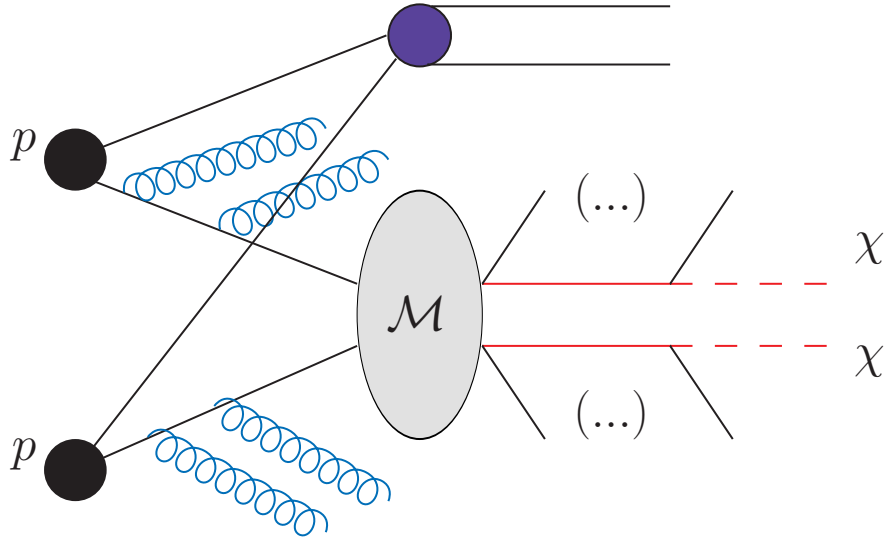}
  \caption[]{Physics searches at hadron colliders are complex: high energy processes are always associated with copious
  initial-state radiation (blue gluon lines), and secondary
  interactions are very likely (secondary partons linking to small
  blue interaction). We do not show secondary proton-proton
  interactions (pile-up).} 
\label{fig:qcdrad:challenges2}
\end{figure}

The second set of challenges we need to face is intrinsic to a high energy hadron
collider (see Fig~\ref{fig:qcdrad:challenges2}). These are mainly due to the fact that the incoming hadrons are
`bags' of quarks and gluons and the high energy interacting partons
radiate gluons before they interact, according to the parton branching picture we presented
in section~\ref{sec:sm:qcd:partonbranch}. Thus, processes are always associated with copious
collinear initial-state radiation (ISR) which affects the longitudinal
momentum components of particles. Moreover, it is not possible to measure
the initial longitudinal (i.e. along the $z$-axis, defined parallel to the
detector's beam pipe) components of the interacting partons in hadron
colliders, since what is left of the colliding hadrons (the `remnants')
travels down the beam pipe and is not detected.\footnote{Note,
  however, that central exclusive production, in which the protons
  remain intact and forward detectors can be used to measure the
  $z$-component of the central system, has been investigated. See, for example,~\cite{Forshaw:2007ra}.}  But the issues are not
only limited to ISR and our inability to measure the $z$-components
of incoming partons: secondary partonic interactions between the interacting hadrons also play an
important role, contaminating the signal with soft particles. It is
also worthwhile to note that since the LHC is destined to become a
high-luminosity machine,\footnote{Here we refer to the instantaneous
  luminosity of a collider with bunched beams. If two bunches containing $n_1$ and $n_2$
  particles collide with frequency $f$ and the beam profiles are
  characterised by $\sigma_x$ and $\sigma_y$, the transverse beam
  profiles in the horizontal and vertical directions respectively, then the instantaneous luminosity is given by
$\mathcal{L} = (f n_1 n_2) / (4 \pi \sigma_x \sigma_y)$. The rate of
events for a certain process of cross section $\sigma$ is then given by $\mathrm{d}N/\mathrm{d}t = \mathcal{L} \times \sigma$~\cite{Nakamura:2010zzi}.} secondary proton-proton interactions (or
`pile-up') are also expected to add soft particles to the events. Of
course, one cannot forget the various experimental challenges due to
beam halo, noise and so on.\footnote{See, for
  example, Ref.~\cite{BouhovaThacker:2010zza} for details of the expected performance
  of the ATLAS experiment at the LHC.}

Several search variables of varying complexity and assumptions have been constructed to tackle the
aforementioned issues. This is done, for example, either through cleverly defining
variables that have discriminatory power against backgrounds, after
cuts have been imposed to the event sample, or variables that are
insensitive to QCD effects in certain ways. 

Kinematic variables can be constructed in order
to constrain the momenta of particles and hence search for new
physics. They represent the first step in understanding the
observations at hadron colliders, since they make very few assumptions
about the details of the underlying physical model, such as the gauge
groups, spins or couplings, providing model-independent, robust
information~\cite{Barr:2010zj}. The very simplest class of the kinematic variables are
\textit{global inclusive} variables, i.e. those that make use of all
observed momenta, without hypothesising any particular structure of
the final state. Since the longitudinal momentum of the hard process
is unknown, transverse variables of this class have been generally
investigated so far. Examples are the observed transverse energy $E_T$,
the missing transverse energy, $\slashed{E}_T$, and their sum, $H_T =
E_T + \slashed{E}_T$. The distributions of these quantities can
provide information on the energy scales of new processes such as
supersymmetric particle production~\cite{Hinchliffe:1996iu,
  Paige:1997xb, Tovey:2000wk}. 

Although longitudinal components of final-state momenta are
strongly influenced by ISR, they do contain information about the
underlying hard process. Indeed, the amount of ISR emitted is
determined by the energy scale of the subprocess. This has motivated
several studies of global inclusive variables that contain
longitudinal components, such as the $\hat{s}^{1/2}_{\mathrm{min}}$
and the total visible mass $M$
  variables, which will be investigated in
  section~\ref{sec:qcdeff:globalinc}. 

Initial-state radiation also modifies the distributions of the
products of the hard process. This effect has been studied in great
detail for the processes of electroweak boson production, with the
result that the transverse momentum and rapidity distributions of $W$,
$Z$ and Higgs bosons at the Tevatron and LHC are predicted with good
precision~\cite{Bozzi:2007pn, Bozzi:2008bb, Mantry:2009qz}. In
contrast, the equivalent $E_T$ distributions have received little
attention. In section~\ref{sec:qcdeff:etres} we will be investigating
the $E_T$ distribution in vector boson production and Higgs boson production.

\section{Effects of QCD radiation on global inclusive variables}
\label{sec:qcdeff:globalinc}
\subsection{Global inclusive variables}
\label{sec:qcdeff:globaldef}
\begin{figure}[!t]
  \centering 
  \includegraphics[scale=0.75, angle=0]{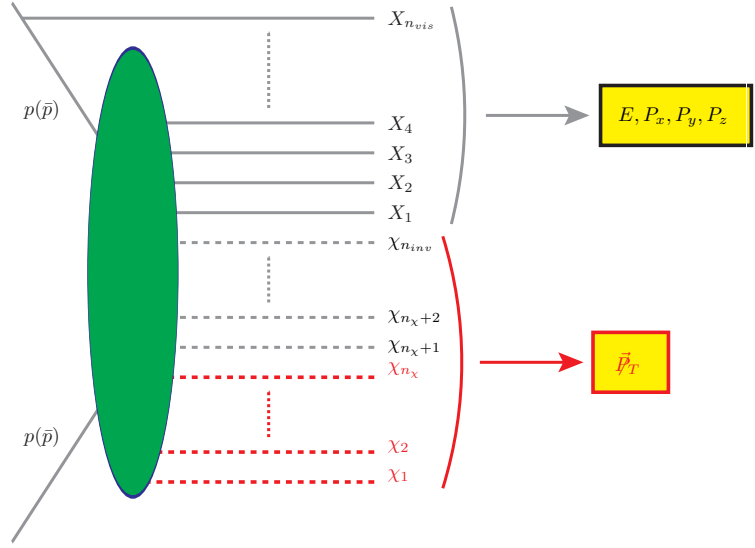}
  \caption[]{Global inclusive variables can be defined with reference
    to the above figure, taken from~\cite{Konar:2008ei}. Black (red) lines correspond to SM (BSM) particles.
 The solid lines denote SM particles $X_i$, $i=1,2,\ldots, n_{vis}$,
which are visible in the detector, e.g.~jets, electrons, muons and photons.
The SM particles may originate either from initial-state radiation, or from the hard scattering and subsequent 
cascade decays (indicated with the green-shaded ellipse).
The dashed lines denote neutral stable invisible particles 
$\chi_i$, $i=1,2,\ldots, n_{inv}$ which consist of some number 
$n_{\chi}$ of BSM particles (indicated with the red dashed lines),
as well as some number $n_{\nu}=n_{inv}-n_{\chi}$ of SM neutrinos 
(denoted with the black dashed lines). The global event variables describing the visible
particles are: the total energy $E$, the transverse components 
$P_x$ and $P_y$ and the longitudinal component $P_z$ 
of the total visible momentum $\mathbf{P}$. The only experimentally available
information regarding the invisible particles is the missing transverse
2-momentum $\slashed{P}_T$.} 
\label{fig:qcdrad:konarmatchevfig}
\end{figure}

Global inclusive variables are easily defined with reference to
Fig.~\ref{fig:qcdrad:konarmatchevfig}, as found in~\cite{Konar:2008ei}. We focus on a
specific subprocess, formed by the interaction of two incoming partons
from the hadrons (protons or anti-protons). The resulting final-state
particles can be either visible, $X_i$, or invisible, $\chi_i$. The
visible particles $X_i$ can originate either from the hard process
itself or from ISR and can be jets, electrons, muons and photons. The
invisible particles consist of SM neutrinos, which we take to be
massless, and some new massive BSM particles, not necessarily of the
same type. We can define, using the sum of the 4-momenta of all
the visible particles, the total 3-momentum $\mathbf{P}$ and the total
energy $E$. The only experimental information we possess about the invisibles is the
total missing transverse momentum 2-vector,
$\mathbf{\slashed{P}}_T$. Any global inclusive variable can be
defined by using $E$, components of $\mathbf{P}$ and
$\mathbf{\slashed{P}}_T$. 

\subsection{The variable $\hat{s}_{\mathrm{min}}$}
In Ref.~\cite{Konar:2008ei} various global variables were investigated, including those that make
use of longitudinal as well as transverse momentum components.  The quantities studied included
the total energy $E$ visible in the detector (as defined in
section~\ref{sec:qcdeff:globaldef}) and the visible invariant mass
$M$,
\beq\label{eq:Mdef}
M = \sqrt{E^2-P_z^2-\slashed{P}_T^2}\;,
\eeq
where $P_z$ is the visible longitudinal momentum.  In addition a new variable was
introduced, defined as
\beq
\hat{s}^{1/2}_{\rm min}(M_{\rm inv}) \equiv \sqrt{M^2+\slashed{P}_T^2}+\sqrt{M_{\rm inv}^2+\slashed{P}_T^2}\ ,
\label{eq:smin_def}
\eeq
where the parameter $M_{\rm inv}$ is a variable estimating the sum of masses of all
invisible particles in the event:
\beq
M_{\rm inv} \equiv \sum_{i=1}^{n_{\rm inv}} m_i\ .
\label{eq:minv}
\eeq
The variable $\hat{s}^{1/2}_{\rm min}$ is simply the \textit{minimum} value of
the parton-level Mandelstam variable $\hat{s}$ (the square of the
partonic centre-of-mass energy) which is consistent with the observed
values of the total energy $E$, $P_z$ and $\slashed{P}_T$ in a given
event. It was argued that the peak in the distribution of
$\hat{s}^{1/2}_{\rm min}$ is a good indicator of the mass scale of new
physics processes involving heavy particle production in the absence
of ISR and multiple parton interactions. In Ref.~\cite{Konar:2008ei} specific
examples were given in $t\bar{t}$, gluino pair-production and
gluino-LSP associated production and the dependence on
$M_{\rm inv}$ was also investigated. 

It was also recognised that the
effects of the ISR and the multiple parton interactions make this
measurement impossible, since $\hat{s}^{1/2}_{\rm min}$ would measure
the total energy of the full system, whereas the interest is on the
energy of the hard scattering. In Ref.~\cite{Konar:2010ma} an alternative approach
was proposed, preserving the definition, but instead of using
information from the calorimeters to construct the quantities $E$,
$\mathbf{P}$ and $\slashed{P}_T$, reconstructed objects were
used. That means, objects that have been recognised via some algorithm
as jets, muons, electrons or photons were used. This seems to recover
some of the attributes of $\hat{s}^{1/2}_{\rm min}$, but the
distributions are not calculable analytically and hence we do not
pursue this method here. 

\subsection{ISR effects without invisible particle emission}
\label{sec:qcdrad:isrnoinv}
In the present section we examine the effects of ISR on global inclusive variables, first in an
approximate fixed-order treatment, taking into account collinear-enhanced terms, and then
in an all-orders resummation of such terms.  We quantify the way the distributions of
quantities that involve longitudinal momenta depend on the scale of the underlying hard
subprocess and on the properties of the detector, in particular the maximum visible
pseudorapidity $\etam$. Initially, we ignore the effect of
invisible final-state particles: we assume that all the final-state
particles from the hard subprocess are detected. We will present the
treatment of invisibles in section~\ref{sec:qcdrad:isrinvis}.

The Monte Carlo results presented in Ref.~\cite{Konar:2008ei} show that the second term on
the right-hand side of Eq.~(\ref{eq:smin_def}) is not strongly affected by ISR. The first term is
intended to add extra longitudinal information about the hard subprocess, allowing a more reliable
determination of its mass scale. The extra longitudinal information
enters through the visible mass $M$, and we therefore concentrate on this quantity.
\subsubsection{Born approximation}\label{sec:qcdrad:born}
In the Born (or leading-order) approximation, assuming that no beam remnants are detected, $M$ yields a perfect
estimate of the centre-of-mass energy of the hard subprocess with no invisibles. For incoming partons with
momentum fractions $x_{1,2}$:
\beq\label{eq:EPBorn}
E=\frac 12 \sqrt{S} (x_1+x_2)\;,\;\;
P_z=\frac 12 \sqrt{S} (x_1-x_2)\;,
\eeq
where $\sqrt S$ is the hadron-hadron centre-of-mass energy, so that
\beq\label{eq:QYBorn}
M^2 = x_1x_2S\;,\;\;
Y \equiv \frac 12 \ln\left(\frac{E+P_z}{E-P_z}\right)=\frac 12 \ln\frac{x_1}{x_2}\ \;,
\eeq
where we have also defined the rapidity of the visible system, $Y$.
The differential cross section for parton flavours $a,b$ is thus:
\beq\label{eq:sigMY}
\frac{\mrd \sigma_{ab}}{\mrd M^2 \mrd Y} = \int \mrd x_1\,\mrd x_2\,f_a(x_1)f_b(x_2)\delta(M^2-x_1x_2S)
\delta\left(Y-\frac 12
  \ln\frac{x_1}{x_2}\right)\hat\sigma_{ab}(x_1x_2S) \;,
\eeq
where $f_{a,b}$ are the relevant parton distribution functions for the incoming hadrons and
$\hat\sigma_{ab}$ is the hard subprocess cross section. Hence, at Born level we find
\beq
S\frac{\mrd \sigma_{ab}}{\mrd M^2 \mrd Y} =
f_a\left(\frac{M}{\sqrt S}e^{Y}\right)f_b\left(\frac{M}{\sqrt S}e^{-Y}\right)\hat\sigma_{ab}(M^2)\ .
\eeq

The parton distributions are sometimes given as $F_i(x)=xf_i(x)$, in terms of which we have
\beq\label{eq:dsigabBorn}
M^2\frac{\mrd \sigma_{ab}}{\mrd M^2 \mrd Y} =
F_a\left(\frac{M}{\sqrt S}e^{Y}\right)F_b\left(\frac{M}{\sqrt S}e^{-Y}\right)\hat\sigma_{ab}(M^2)\ .
\eeq
If the partonic cross section $\hat\sigma_{ab}$ has a threshold or peak, indicating that
the $ab$ subprocess has a characteristic scale $Q$, then this is also manifest
in the Born cross section (\ref{eq:dsigabBorn}) at $M\sim Q$, provided the relevant
parton distributions are large enough for that subprocess to contribute significantly. 
\subsubsection{Quasi-collinear NLO correction}\label{sec:qcdrad:nlo}
To examine the sensitivity of the above results to ISR, let us first compute the NLO contribution due to
quasi-collinear gluon emission and the associated virtual corrections.  Consider first the emission of
a gluon from parton $a$.  If the emission angle $\theta$, defined with
respect to the beam direction in the lab frame, is large enough, say $\theta>\thc$, the
gluon enters the detector and contributes to $M$. In the small-angle approximation we then have
\beq\label{eq:EPnlo}
E=\frac 12 \sqrt{S} (x_1/z+x_2)\;,\;\;
P_z=\frac 12 \sqrt{S} (x_1/z-x_2)\;,
\eeq
where $x_1/z$ is the momentum fraction of parton $a$ before the emission, so that
\beq\label{eq:MYnlo}
M^2= x_1x_2S/z\;,\;\;
Y=\frac 12 \ln\frac{x_1}{zx_2}\ .
\eeq
The correction associated with a detected emission from parton $a$ is then:
\beq
\frac{\alps}\pi\int_{\thc}\frac{\mrd\theta}\theta \frac{\mrd z}z dx_1\,dx_2
\hat P_a(z)f_a(x_1/z)f_b(x_2)\delta(M^2-x_1x_2S/z)
\delta\left(Y-\frac 12 \ln\frac{x_1}{zx_2}\right)\hat\sigma_{ab}(x_1x_2S)
\eeq
 where $\hat P_a(z)$ is the unregularised $a\to ag$ splitting function
 and we have neglected the running of $\alpha_s$ for the moment. 

On the other hand if the gluon misses the detector ($\theta<\thc$),
$E$ and $P_z$ are still given by Eq.~(\ref{eq:EPBorn}), so the contribution is
\beq
\frac{\alps}\pi\int_0^{\thc}\frac{\mrd \theta}\theta \frac{\mrd z}z
\mrd x_1\,\mrd x_2
\hat P_a(z)f_a(x_1/z)f_b(x_2)\delta(M^2-x_1x_2S)
\delta\left(Y-\frac 12 \ln\frac{x_1}{x_2}\right)\hat\sigma_{ab}(x_1x_2S)\,.
\eeq
Finally the associated virtual correction is the term that regularises the splitting function,
which in this approximation is
\beq
-\frac{\alps}\pi\int\frac{\mrd\theta}\theta\,\mrd z\,\mrd x_1\,\mrd x_2
\hat P_a(z)f_a(x_1)f_b(x_2)\delta(M^2-x_1x_2S)
\delta\left(Y-\frac 12 \ln\frac{x_1}{x_2}\right)\hat\sigma_{ab}(x_1x_2S)\,.
\eeq
Adding everything together gives a correction
\bea
\delta\left(\frac{\mrd\sigma_{ab}}{\mrd M^2 \mrd Y}\right) &=&\frac{\alps}\pi
\int\frac{\mrd\theta}\theta\,\mrd z\,\mrd x_1\,\mrd x_2\,\hat P_a(z)f_b(x_2)
\hat\sigma_{ab}(x_1x_2S)\nonumber\\
&\times&\Bigl[\frac 1z f_a(x_1/z)\delta\left(Y-\frac 12 \ln\frac{x_1}{zx_2}\right)
\delta(M^2-x_1x_2S/z)\Theta(\theta-\thc)\nonumber\\
&&+\Bigl\{\frac 1z f_a(x_1/z)\Theta(\thc-\theta)
-f_a(x_1)\Bigr\}\delta\left(Y-\frac 12 \ln\frac{x_1}{x_2}\right)
\delta(M^2-x_1x_2S)\Bigr]\ .\nonumber\\
\eea
Setting aside for the moment the possibility of splittings other than $a\to ag$,
the DGLAP evolution equation for $f_a(x_1)$ is
\beq\label{eq:evol}
q\frac{\partial}{\partial q}f_a(x_1) =\frac{\alps}\pi\int \mrd z\,\hat P_a(z)
\left[\frac 1z f_a(x_1/z) - f_a(x_1)\right]\;,
\eeq
where $q$ represents the scale at which the parton distribution is measured.
Hence the correction may be written as
\bea
\delta\left(\frac{\mrd\sigma_{ab}}{\mrd M^2\mrd Y}\right) &=&
\int\frac{\mrd\theta}\theta\,\mrd x_1\,\mrd x_2\,f_b(x_2)\hat\sigma_{ab}(x_1x_2S)
\Biggl[q\frac{\partial f_a}{\partial q}\delta\left(Y-\frac 12 \ln\frac{x_1}{x_2}
\right)\delta(M^2-x_1x_2S)\nonumber\\
&+&\frac{\alps}\pi\int\frac{\mrd z}z \hat P_a(z)\,f_a(x_1/z)\Biggl\{\delta\left(Y-\frac 12 \ln\frac{x_1}{zx_2}\right)\delta(M^2-x_1x_2S/z)\nonumber\\
&-&\delta\left(Y-\frac 12 \ln\frac{x_1}{x_2}\right)\delta(M^2-x_1x_2S)
\Biggr\}\Theta(\theta-\thc)\Biggr]\ .
\eea
Since $\mrd \theta/\theta = \mrd q/q$, the first term represents a change of scale in the Born term.
It replaces the reference scale in $f_a$ by the scale $Q$ of the hard subprocess. 
 The remaining terms give a correction
\bea
\delta\left(\frac{\mrd \sigma_{ab}}{\mrd M^2 \mrd Y}\right) &=&
\frac{\alps}{\pi S}\int_{\thc}\frac{\mrd \theta}\theta\int\frac{\mrd z}z \hat P_a(z)
f_b\left(\frac{M}{\sqrt S}e^{-Y}\right)\nonumber\\
&\times&\left[f_a\left(\frac{M}{\sqrt S}e^{Y}\right)z\hat\sigma_{ab}(zM^2)
-f_a\left(\frac{M}{z\sqrt S}e^{Y}\right)\hat\sigma_{ab}(M^2)\right]\ .
\eea
In leading-log approximation the $\theta$ integration just gives a factor of $-\ln\thc$.
In the same approximation, we may set $-\ln\thc =
\etam$,\footnote{Note that this is an approximation to the conventional definition
  of the pseudorapidity $\eta = - \ln \tan (\frac{\theta}{2})$,
  consistent with the leading-long approximation.}  the maximum
pseudorapidity seen by the detector.  Note that this is a different quantity from $Y$,
the true rapidity of the visible system.
The correction associated with parton $b$ gives the same expression with
$a\leftrightarrow b$ and $Y\to -Y$.  Thus, defining
\beq\label{eq:qcdrad:xbar}
\bar x_1 = \frac{M}{\sqrt S}e^{Y}\;,\;\;
\bar x_2 = \frac{M}{\sqrt S}e^{-Y}\;,
\eeq
we have
\bea
S\frac{\mrd\sigma_{ab}}{\mrd M^2 \mrd Y} &=&
f_a(\bar x_1,Q)f_b(\bar x_2,Q)\hat\sigma_{ab}(M^2)\nonumber\\
&+&\etam\frac{\alps}\pi\int\frac{\mrd z}z
\Bigl[z\{\hat P_a(z)+\hat P_b(z)\}f_a(\bar x_1,Q)f_b(\bar x_2,Q)
\hat\sigma_{ab}(zM^2)\nonumber\\
&-&\{\hat P_a(z)f_a(\bar x_1/z,Q)f_b(\bar x_2,Q)+
\hat P_b(z)f_a(\bar x_1,Q)f_b(\bar
x_2/z,Q)\}\hat\sigma_{ab}(M^2)\Bigr]. \nonumber \\
\eea
Expressing this in terms of $F_i(x)=xf_i(x)$, as in Eq.~(\ref{eq:dsigabBorn}),
\bea\label{eq:M2sig}
M^2\frac{\mrd\sigma_{ab}}{\mrd M^2 \mrd Y} &=&
F_a(\bar x_1,Q)F_b(\bar x_2,Q)\hat\sigma_{ab}(M^2)\nonumber\\
&+&\etam\frac{\alps}\pi\int \mrd z
\Bigl[\{\hat P_a(z)+\hat P_b(z)\}F_a(\bar x_1,Q)F_b(\bar x_2,Q)\hat\sigma_{ab}(zM^2)\\
&-&\{\hat P_a(z)F_a(\bar x_1/z,Q)F_b(\bar x_2,Q)+\hat P_b(z)
F_a(\bar x_1,Q)F_b(\bar x_2/z,Q)\}\hat\sigma_{ab}(M^2)\Bigr]\ .\nonumber
\eea

\begin{figure}[t]
  \centering 
  \vspace{6.0cm}
  \includegraphics[scale=0.75, angle=0]{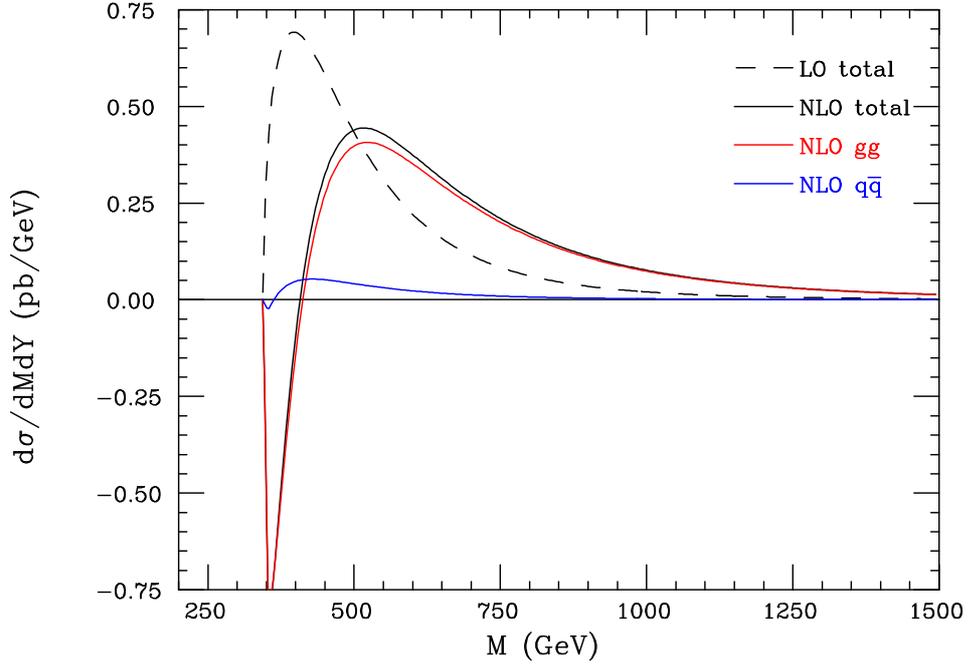}
\caption[]{Distribution of visible mass $M$ in $t\bar t$ production at LHC for $\etam=5$ and
$Y=0$: leading and approximate next-to-leading order.}\label{fig:ttbar}
\end{figure}

Results for $t\bar t$ production at the LHC ($pp$ at $\sqrt S=14$ TeV) with $\etam=5$ and $Y=0$
are shown in Fig.~\ref{fig:ttbar}.  Leading-order MSTW parton distributions~\cite{Martin:2009iq}
were used.  For simplicity we have taken $Q=M$.
Recall that the simplifying assumption made here is that all $t\bar t$ decay products are detected,
so the $M$ distribution vanishes below $t\bar t$ threshold.  We see that there is a large negative
NLO correction near threshold, followed by a broad positive peak.
To understand these qualitative features, consider the case $a=b$, as in $gg\to t\bar t$,
and $Y=0$, so that $\bar x_{1,2} = M/\sqrt S\equiv \bar x$. Then the NLO correction
becomes simply:
\beq
\delta\left(M^2\frac{\mrd \sigma}{\mrd M^2\mrd Y}\right) =
2\etam\frac{\alps}\pi F(\bar x) \int \mrd z\hat P_a(z)\left[F(\bar x)\hat\sigma(zM^2)-F(\bar x/z)\hat\sigma(M^2)\right]\;.
\eeq
The first term is positive-definite, contributes only above threshold,
and diverges at threshold.  It produces the broad positive peak.  The
second term is negative-definite, contributes around threshold, and
has a divergent coefficient. It provides the sharp negative peak.
\subsubsection{Resummation}\label{sec:qcdrad:resumm}
By adding and subtracting the expression
\beq
\{\tilde P_a(z)+\tilde P_b(z)\}F_a(\bar x_1,Q)F_b(\bar x_2,Q)\hat\sigma_{ab}(M^2)\;,
\eeq
in the integrand of Eq.~(\ref{eq:M2sig}) and comparing with Eq.~(\ref{eq:evol}), we see that the last line of 
that equation corresponds to a change of scale $Q\to Q_c \sim \thc Q$ in the parton distributions, leading to
\beq\label{eq:M2sigFFS}
M^2\frac{\mrd\sigma_{ab}}{\mrd M^2 \mrd Y} = F_a(\bar x_1,Q_c)F_b(\bar x_2,Q_c) \Sigma_{ab}(M^2)\;.
\eeq
The above definition of $\Sigma_{ab}$ includes the approximation
$Q_c \approx \theta_c M$ in the evaluation of the PDFs, since
they do not vary substantially with scale. To first order we can then
write
\bea\label{eq:SigOa}
\Sigma_{ab}(M^2) &=& \hat\sigma_{ab}(M^2)\nonumber\\
&+&\etam\frac{\alps}\pi\int \mrd z\{\hat P_a(z)+\hat P_b(z)\}\{\hat\sigma_{ab}(zM^2)-\hat\sigma_{ab}(M^2)\}\;.
\eea
The interpretation of this result is simple:  undetected ISR at angles less than $\thc$, corresponding to scales less
than $\sim \theta_c Q$, is absorbed into the structure of the incoming
hadrons.  

To resum the effects of gluons at angles
greater than $\thc$, consider first the real emission of $n$ such gluons from parton $a$.  In the quasi-collinear
approximation these form an angular-ordered sequence, giving rise to a
contribution to $\Sigma_{ab}$ of
\bea\label{eq:nreal}
\delta^{\mathcal{R},n} (\Sigma_{ab})&=&
\left(\frac{\alps}\pi\right)^n\int_{\thc}\frac{\mrd
  \theta_1}{\theta_1}\int_{\theta_1}\frac{\mrd\theta_2}{\theta_2}
\ldots \int_{\theta_{n-1}}\frac{\mrd \theta_n}{\theta_n}\int_0^1 \mrd
z_1\ldots \mrd z_n \hat P_a(z_1)\ldots \hat P_a(z_n)\hat\sigma_{ab}(z_1\ldots z_n M^2)\nonumber\\
&=&\frac 1{n!}\left(\etam\frac{\alps}\pi\right)^n
\int_0^1 \mrd z_1\ldots \mrd z_n \hat P_a(z_1)\ldots \hat P_a(z_n)\hat\sigma_{ab}(z_1\ldots z_n M^2)\;,
\eea
where again we have made the identification $-\ln\thc = \etam$ and we
are still neglecting the running of $\alpha_s$. The multiple
convolution of the momentum fractions $z_i$ can be transformed into a
product by taking \textit{Mellin} moments.
Defining
\beq\label{eq:momdef}
\int_0^\infty \mrd M^2 \left(M^2\right)^{-N} \hat\sigma_{ab}(M^2) \equiv \hat\sigma^{ab}_N\;,
\eeq
we have
\beq
\left(\frac{\alps}\pi\right)^n
\int_0^\infty \mrd M^2 \left(M^2\right)^{-N} \int\mrd z_1\ldots \mrd z_n \hat P_a(z_1)\ldots  \hat P_a(z_n)
\hat\sigma(z_1\ldots z_n M^2) = \left(\hat\gamma^a_N\right)^n\hat\sigma^{ab}_N\;,
\eeq
where
\beq
\hat{\gamma}^a_N = \frac{\alps}\pi\int_0^1 \mrd z\,z^{N-1}\hat P_a(z)\ .
\eeq 
Therefore defining correspondingly,
\beq\label{eq:SigNdef}
\int_0^\infty \mrd M^2 \left(M^2\right)^{-N} \Sigma_{ab}(M^2) \equiv \Sigma^{ab}_N\;,
\eeq
the contribution (\ref{eq:nreal}) to this quantity will be
\beq
\delta^{\mathcal{R},n} ( \Sigma^{ab}_N ) =\frac 1{n!}\left(\etam\hat{\gamma}^a_N\right)^n \hat\sigma^{ab}_N\;,
\eeq
which summed over $n$ gives
\beq
\sum_n \delta^{\mathcal{R},n} ( \Sigma^{ab}_N ) = \exp\left(\etam\hat{\gamma}^a_N\right)\hat\sigma^{ab}_N\ .
\eeq
The corresponding virtual contributions give a Sudakov-like form factor,
\beq\label{eq:qcdrad:virtsudakov}
\exp\left(-\frac{\alps}\pi\int_{\thc}\frac{\mrd \theta}\theta \int_0^1
  \mrd z\, \hat{P}_a(z)\right)\;.
\eeq
and therefore the total contribution from parton $a$ is
\beq
\exp\left(\etam\gamma^a_N\right)\hat\sigma^{ab}_N \;,
\eeq
where $\gamma^a_N$ is the anomalous dimension,
\beq
\gamma^a_N = \frac{\alps}\pi\int_0^1 \mrd z\left(z^{N-1}-1\right)\hat P_a(z)
=\frac{\alps}\pi\int_0^1 \mrd z\,z^{N-1} P_a(z)\;,
\eeq
$P_a(z)$ being the regularised $a\to ag$ splitting function. Note that
it is the virtual correction, introduced via the form factor in
Eq.~(\ref{eq:qcdrad:virtsudakov}), that has regularised the splitting
function. Parton $b$ gives a similar factor with $\gamma^b_N$ in place of $\gamma^a_N$, so the
result for the quantity  (\ref{eq:SigNdef}) is simply
\beq\label{eq:SigN}
\Sigma^{ab}_N  = e^{\etam(\gamma^a_N+\gamma^b_N)}\hat\sigma^{ab}_N\;.
\eeq
We can see as follows that this result is qualitatively correct.  The anomalous dimensions are positive
for small $N$ and negative for large $N$.  Thus, for $\thc\ll 1$, $\Sigma^{ab}_N$ is enhanced relative
to $\hat\sigma^{ab}_N$ at small $N$ and suppressed at large $N$.   Now from the moment definition
(\ref{eq:momdef}) small $N$ corresponds to large $M$ and vice versa.  Hence the distribution of $M$
is suppressed at small $M$ and enhanced at large $M$ relative to the Born term, as observed in the
Monte Carlo~\cite{Konar:2008ei} and NLO results.

The emission of partons other than gluons is included by introducing the anomalous dimension
matrix $\Gamma_N$ with elements given by
\beq
(\Gamma_N)_{ba} = \frac{\alps}\pi\int_0^1 \mrd z\,z^{N-1}P_{ba}(z) \;,
\eeq
where $P_{ba}(z)$ is the regularised $a\to b$ splitting function.  Then
\beq\label{eq:SigabN}
\Sigma^{ab}_N  =  \hat\sigma^{a'b'}_N\left(e^{\etam\Gamma_N}\right)_{a'a}
\left(e^{\etam\Gamma_N}\right)_{b'b}\ .
\eeq
The corresponding generalisation of the evolution equation (\ref{eq:evol}) is
\beq\label{eq:evolba}
q\frac{\partial}{\partial q}f_b(x) =\frac{\alps}\pi\int\frac{\mrd z}z P_{ba}(z) f_a(x/z)\;.
\eeq
Defining the moments of the parton distribution functions
\beq
f^a_N = \int_0^1 \mrd x\,x^{N-1} f_a(x)\;,
\eeq
we see that
\beq\label{eq:evolN}
q\frac{\partial}{\partial q}f^b_N= (\Gamma_N)_{ba} f^a_N\;,
\eeq
with solution
\beq
f^b_N(q) = \left([q/q_0]^{\Gamma_N}\right)_{ba} f^a_N(q_0)\ .
\eeq
Hence
\beq\label{eq:fbN}
f^b_N(Q) = \left(e^{\etam\Gamma_N}\right)_{ba} f^a_N(Q_c)\;,
\eeq
where
\beq\label{eq:Qc}
Q_c=\thc Q= Qe^{-\etam}\;,
\eeq
showing that the evolution of the visible mass distribution
 is related to that of the parton distributions over the same range of
 scales. 

Taking into account the running of the strong coupling $\alps(q)$ in the
evolution equation (\ref{eq:evolba}), Eq.~(\ref{eq:fbN}) becomes
\beq\label{eq:fbNK}
f^b_N(Q) = K^{ba}_N f^a_N(Q_c) \;,
\eeq
where
\beq\label{eq:Kba}
 K^{ba}_N =
 \left(\left[\frac{\alps(Q_c)}{\alps(Q)}\right]^{p\Delta_N}\right)_{ba} \;,
\eeq
with the factor $p = 6/(11C_A-2n_f)$, coming from the QCD one-loop
$\beta$-function, Eq.~(\ref{eq:sm:qcd:betazero}), given in
section~\ref{sec:qcd:renorm}, and
\beq
(\Delta_N)_{ba} = \frac\pi{\alps}(\Gamma_N)_{ba} = \int_0^1 \mrd z\,z^{N-1}P_{ba}(z) \ .
\eeq
The running of $\alps$ will affect Eq.~(\ref{eq:SigabN}) similarly, giving
\beq\label{eq:SigabNK}
\Sigma^{ab}_N  =  \hat\sigma^{a'b'}_N K^{a'a}_N K^{b'b}_N\ .
\eeq
To invert the above we can write the following double convolution
\beq\label{eq:SigabKK}
\Sigma_{ab}(M^2) =  \int_0^1
\mrd z_1\,\mrd z_2\,\hat\sigma_{a'b'}(z_1z_2M^2)\, K_{a'a}(z_1)\, K_{b'b}(z_2) \;,
\eeq
where
\beq\label{eq:qcrad:bromwich}
K_{b'b}(z) = \frac 1{2\pi i}\int_C \mrd N\,z^{-N} K^{b'b}_N\;.
\eeq
In the above inversion, performed on the complex plane, the contour
$C$ is formally defined to be to the right of all singularities of the
integrand and runs parallel to the imaginary axis from $-i\infty$ to
$+i\infty$. We will be discussing the details of the inversion in section~\ref{sec:qcdrad:mellininversion}.
It then follows from Eq.~(\ref{eq:Kba}) that $K_{b'b}(z)$ obeys an evolution equation like that of the parton distributions,
\beq\label{eq:evolK}
Q\frac{\partial}{\partial Q}K_{b'b}(z)
=\frac{\alps(Q)}\pi\int\frac{\mrd z'}{z'} P_{b'a}(z') K_{ab}(z/z')\;.
\eeq

Putting everything together, the visible mass distribution is related to the hard subprocess
cross section (in the absence of invisible final-state particles) as follows:
\beq\label{eq:M2sigz1z2}
M^2\frac{\mrd\sigma_{ab}}{\mrd M^2 \mrd Y} = \int \mrd z_1\,dz_2\,\hat\sigma_{a'b'}(z_1 z_2 M^2)\,K_{a'a}(z_1)
F_a(\bar x_1,Q_c)\,K_{b'b}(z_2)F_b(\bar x_2,Q_c) \;,
\eeq
where the kernel functions $K_{a'a}(z)$ and $K_{b'b}(z)$ can be obtained by solving the evolution
equation (\ref{eq:evolK}) with the initial condition that
$K_{ab}(z)=\delta_{ab}\delta(1-z)$ at $Q=Q_c$ or by directly using
Eq.~(\ref{eq:qcrad:bromwich}) to invert the Mellin transform. The
results shown in this section use the former method whereas in section~\ref{sec:qcdrad:isrinvis} results given by the latter
method will be shown. Note also that the assumption $Q_c \approx
\theta_c M$ that we had made in Eq.~(\ref{eq:M2sigFFS}) has been alleviated in Eq.~(\ref{eq:M2sigz1z2}).

To verify that the integrated cross section is not affected by resummation, define
$x_{1,2}=z_{1,2}\bar x_{1,2}$ and write Eq.~(\ref{eq:M2sigz1z2}) as
\beq\label{eq:M2sigx1x2}
M^2\frac{\mrd \sigma_{ab}}{\mrd M^2 \mrd Y} = \int \mrd x_1\,\mrd x_2\,\hat\sigma_{a'b'}(x_1x_2S)\,K_{a'a}(x_1/\bar x_1)
f_a(\bar x_1,Q_c)\,K_{b'b}(x_2/\bar x_2)f_b(\bar x_2,Q_c)\ .
\eeq
Now
\beq
\frac{\mrd M^2}{M^2} \mrd Y = \frac{\mrd\bar x_1}{\bar x_1}\frac{\mrd\bar x_2}{\bar x_2}\;,
\eeq
and
\beq\label{eq:qcdrad:evolkernel}
\sum_a\int\frac{\mrd \bar x_1}{\bar x_1}K_{a'a}(x_1/\bar x_1)f_a(\bar x_1,Q_c) = f_{a'}(x_1,Q)\ .
\eeq
Hence
\beq\label{eq:intMY}
\sum_{ab}\int \mrd M^2\mrd Y\frac{\mrd\sigma_{ab}}{\mrd M^2 \mrd Y}
= \sum_{a'b'}\int \mrd x_1\,\mrd x_2\,\hat\sigma_{a'b'}(x_1x_2S) f_{a'}(x_1,Q)f_{b'}(x_2,Q)\;,
\eeq
in agreement with Eq.~(\ref{eq:sigMY}).

\begin{figure}[!htb]
\begin{center}
  \vspace{2cm}
  \includegraphics[scale=0.55, angle=90]{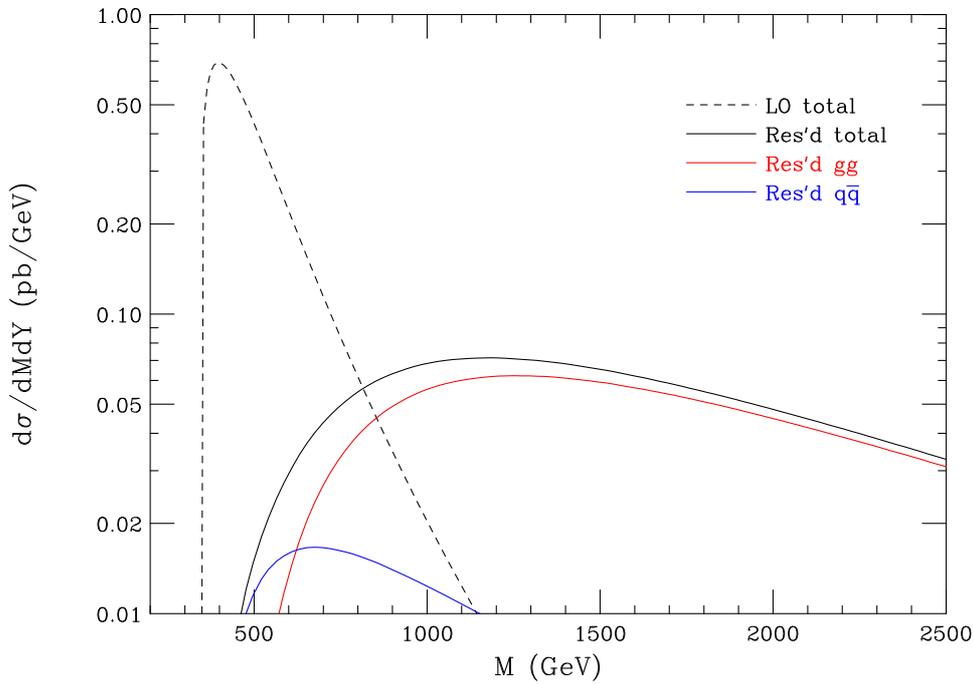}
\end{center}
\caption{Resummed distribution of visible mass $M$ in $t\bar t$ production at LHC for $\etam=5$
and $Y=0$.\label{fig:tresum}}
\end{figure}

\begin{figure}[!htb]
\begin{center}
  \vspace{4.0cm}
  \includegraphics[scale=0.45, angle=0]{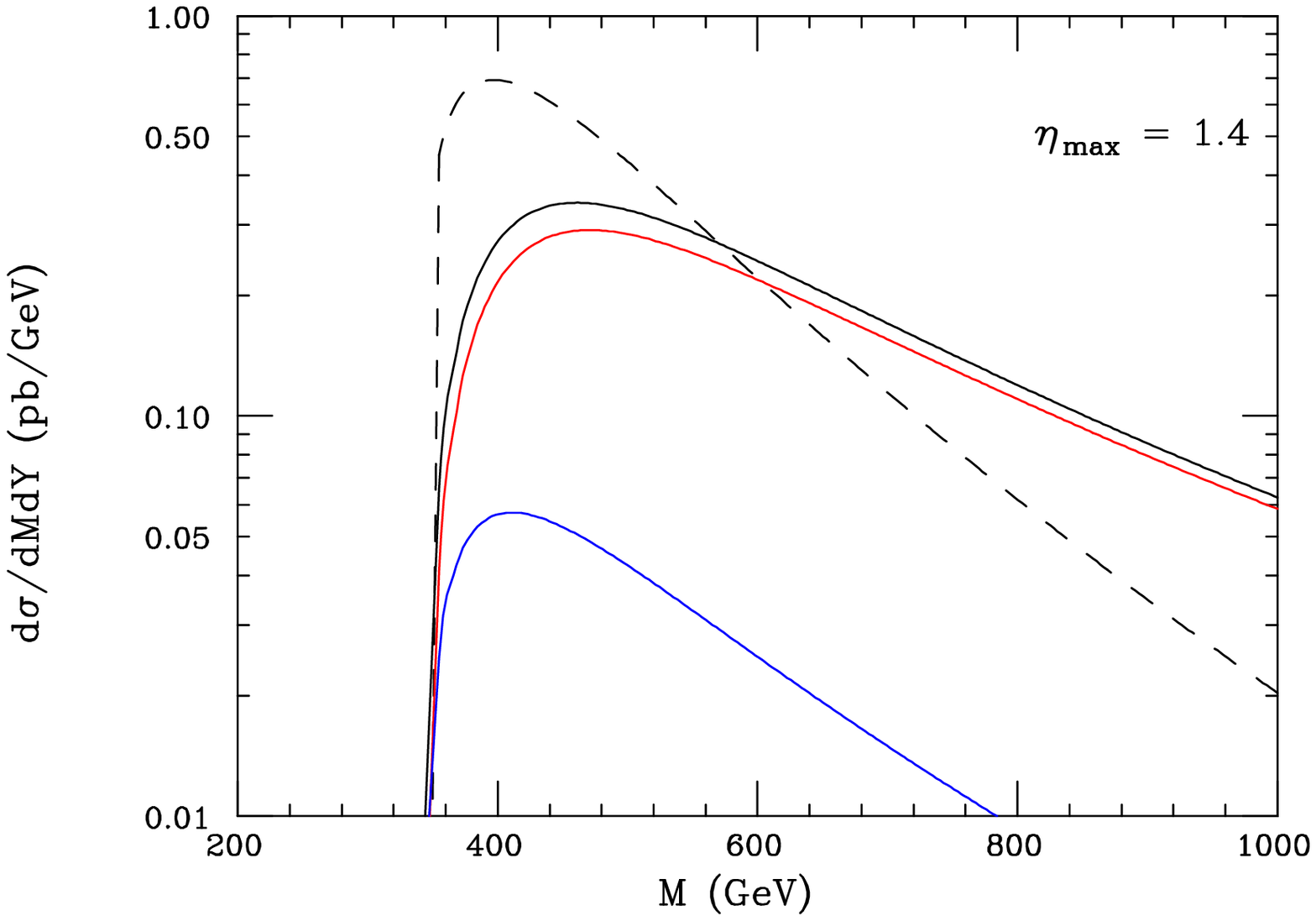}
  \hspace{2cm}
  \includegraphics[scale=0.45, angle=0]{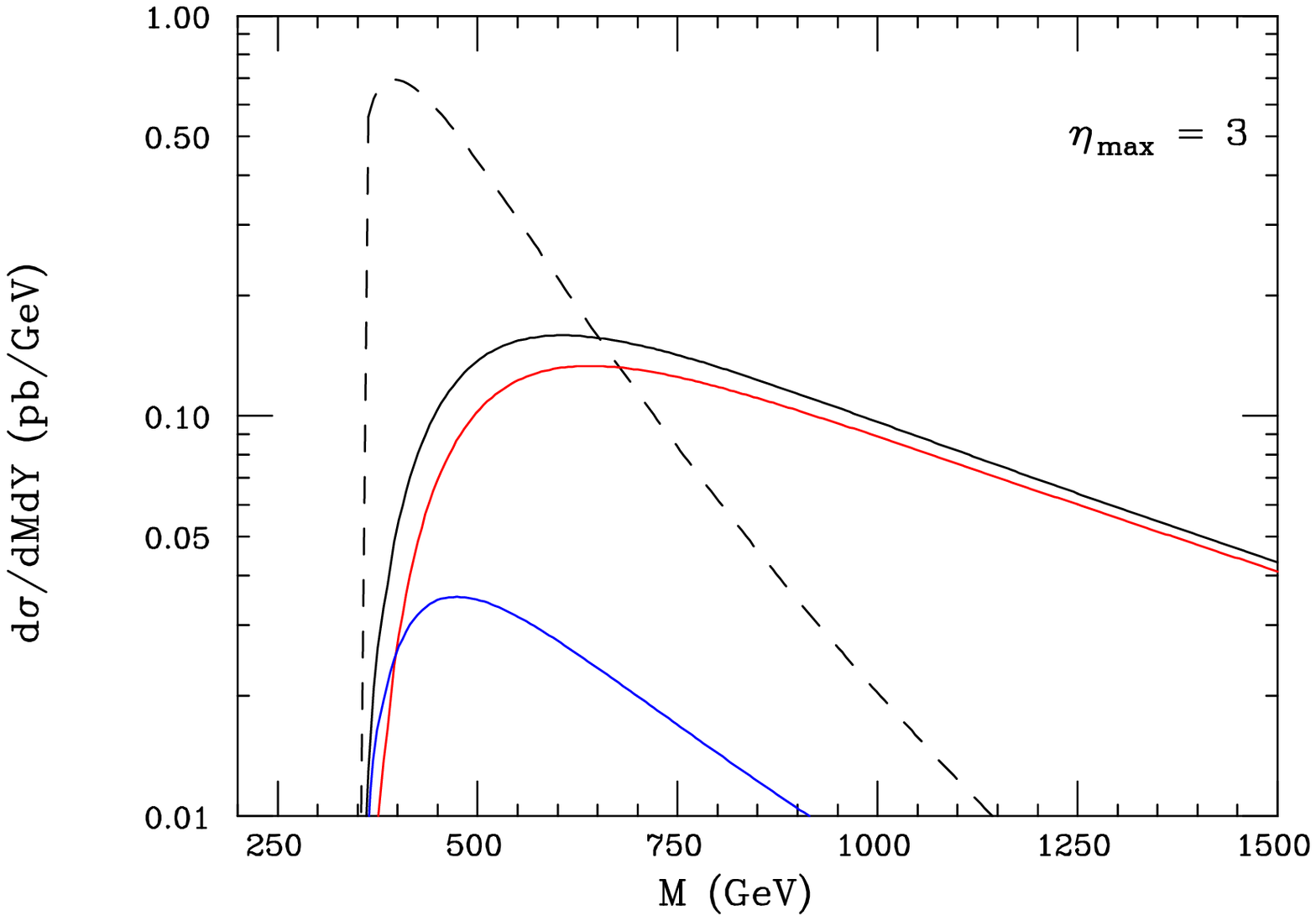}
\end{center}
\caption{Resummed distribution of visible mass $M$ in $t\bar t$ production at LHC for $Y=0$
and lower values of $\etam$: colour scheme as in
Fig.~\ref{fig:tresum}.}
\label{fig:tresumEta}
\end{figure}

\begin{figure}[!t]
\begin{center} 
  \vspace{4.0cm}
  \includegraphics[scale=0.45, angle=0]{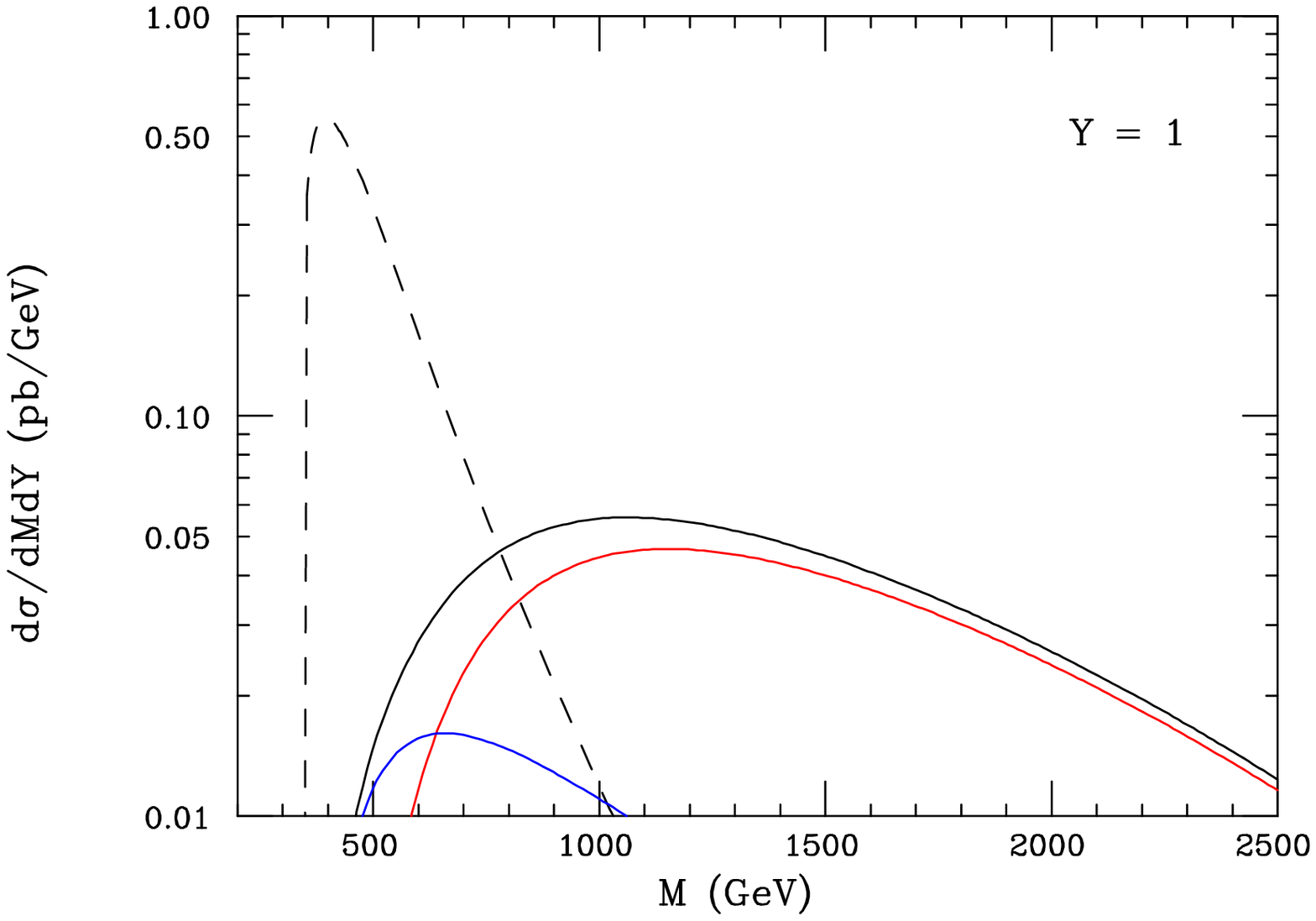}
  \hspace{2cm}
  \includegraphics[scale=0.45, angle=0]{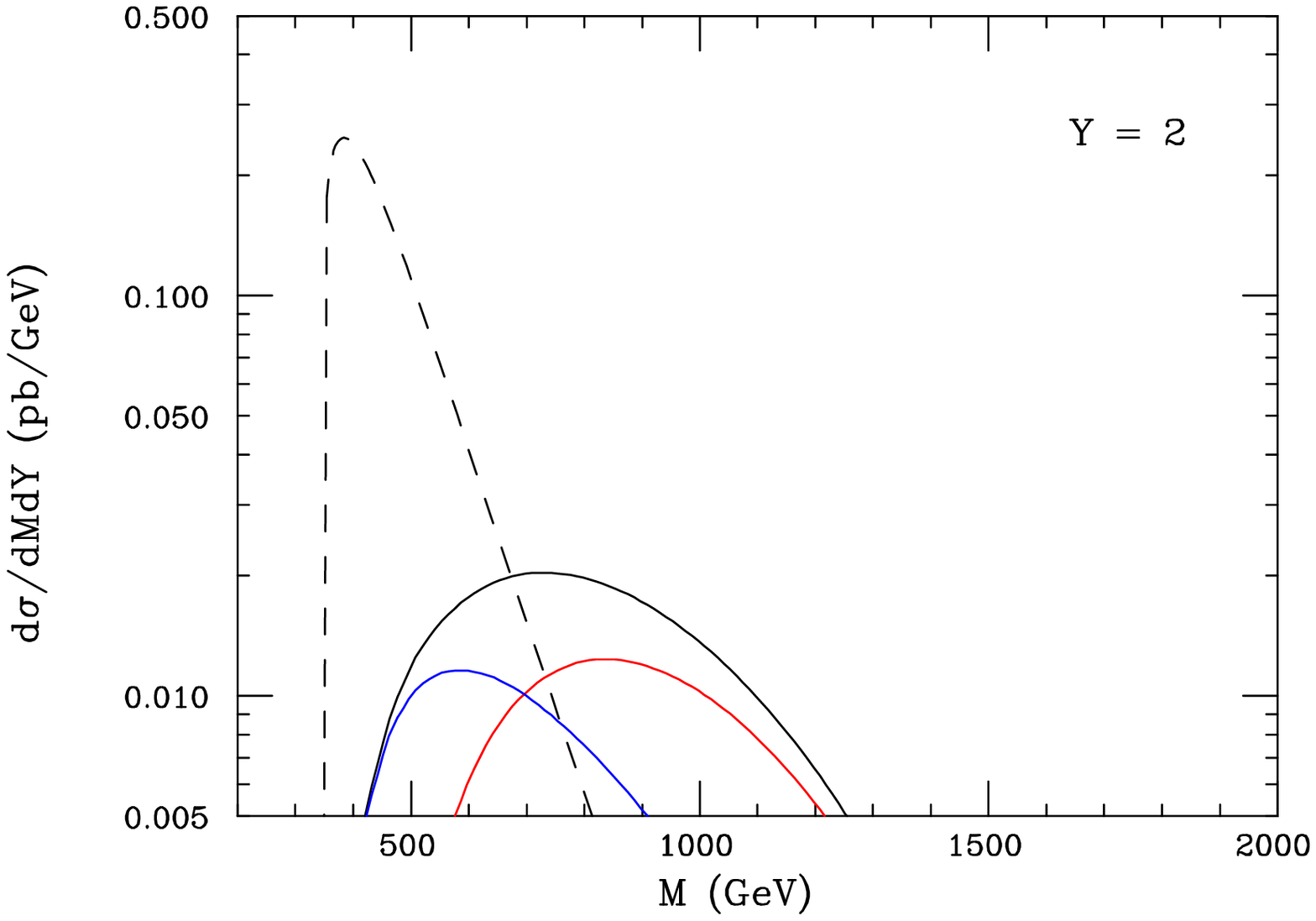}
\end{center}
\caption{Resummed distribution of visible mass $M$ in $t\bar t$ production at LHC for $\etam=5$:
results at non-zero visible rapidity $Y$.\label{fig:tresumY}}
\end{figure}

Resummed results corresponding to Fig.~\ref{fig:ttbar} are shown in Fig.~\ref{fig:tresum}.  We see
that the peak of the distribution has moved to much higher mass, beyond 1 TeV.  This is due to
multiple emission of ISR partons in the evolution of the initial state from the detection scale $Q_c$
to the hard subprocess scale $Q$.  As the value of $\etam$ is reduced, the range of evolution becomes
smaller, less ISR is emitted, and the peak moves closer to the hard subprocess scale, as illustrated in
Fig.~\ref{fig:tresumEta}. Results for higher values of the visible rapidity $Y$ are shown in Fig.~\ref{fig:tresumY}.  The peak moves to
lower mass as $Y$ increases, as a consequence of the suppression of high masses by the rapid fall-off of
the parton distributions at high $x$. All the results shown in the
present section have been obtained by solving the evolution equation
(\ref{eq:evolK}) directly.

\subsection{Mellin transform inversion}\label{sec:qcdrad:mellininversion}
In the previous section we presented results using the direct
integration of Eq.~(\ref{eq:evolK}). In the following sections we will
be using the inverse Mellin transform to calculate the kernel
functions. We outline the procedure in this subsection. 

\begin{figure}[!htb]
  \centering 
  \vspace{4.0cm}
  \begin{picture}(300,120)
    \put(0,0){\includegraphics[scale=0.50,
      angle=90]{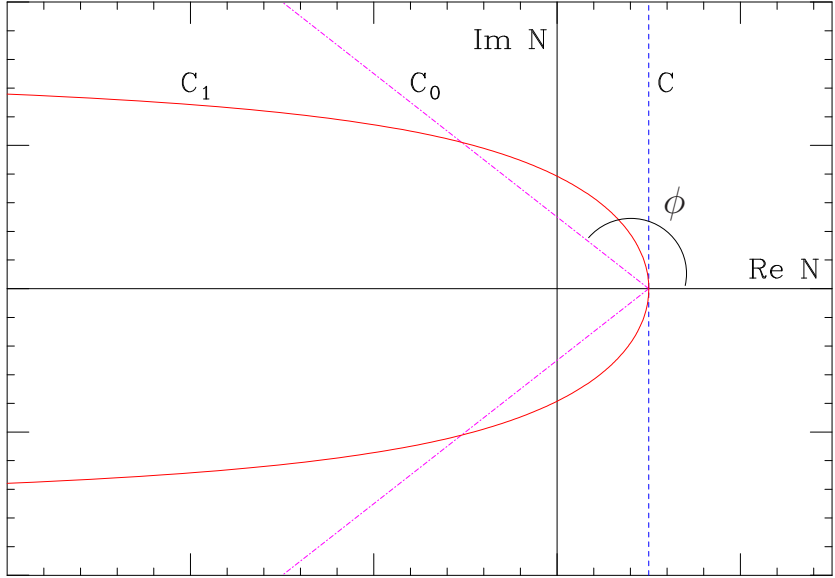}}
    \put(246,137){\large{$\phi$}}
  \end{picture}
  \caption[]{Integration contours, $C$, $C_0$ and
    $C_1$, for the inverse Mellin transform as given by the
    Bromwich integral, Eq.~(\ref{eq:qcrad:bromwich}). The angle $\phi$
  is used in the parametrization of $C_0$.}  
\label{fig:qcdrad:mellincontours}
\end{figure}

Equation~(\ref{eq:qcrad:bromwich}), sometimes called the Bromwich integral, defines the inverse of a Mellin
transform. Formally, the contour $C$ on the complex $N$-plane is to the right of all
singularities in the integrand and runs parallel to the imaginary axis
from $-i\infty$ to +$i\infty$. The formal contour $C$ is shown in
Fig.~\ref{fig:qcdrad:mellincontours} in blue dashes.
However, we often need to perform the inversion on the complex
plane numerically. The formal contour $C$ turns out to be inappropriate for a
numerical calculation. To see this, consider the parametrization of
the contour $C$ passing through the point $a$ on the real axis: $N = a + i t$, where
$t$ is a parameter which runs from $-\infty$ to $+\infty$. Then the
Bromwich integral becomes
\beq\label{eq:qcrad:bromwichC}
K_{b'b}(z) = \frac {z^{-a}} {2\pi}\int_{-\infty}^{+\infty}
\mrd t\,e^{-it\ln z} K^{b'b}_{N(t)}\;.
\eeq
The above expression will not converge numerically because of the
oscillatory nature of the factor $e^{-it \ln z}$. In PDF evolution this is
ameliorated by deforming the contour $C$ to contour $C_0$, shown in
Fig.~\ref{fig:qcdrad:mellincontours} in purple dot-dashes. If there
are no singularities in the integrand in the region $C_0 - C$, we
expect the result of the integration not to
change by the deformation. This is indeed the case for all the functions we will be
considering. The contour $C_0$ can be parametrized as $N = a + t e^{i
  \phi}$ where $\phi$ is a constant related to the slope of the
straight lines as shown in Fig.~\ref{fig:qcdrad:mellincontours}. It is
easy to see that if $\phi > \pi/2$, then a damping term of the form
$e^{t \cos \phi \ln (1/z) }$ is introduced in the
integrand and hence numerical convergence can be achieved. The contour $C_0$ is
usually used to evolve parton density
functions~\cite{Vogt:2004ns}, according to
Eq.~(\ref{eq:fbNK}). 

\begin{figure}[!t]
  \centering
  \vspace{4.0cm}
 \begin{picture}(300,120)
    \put(-40,0){\includegraphics[scale=0.50,
      angle=0]{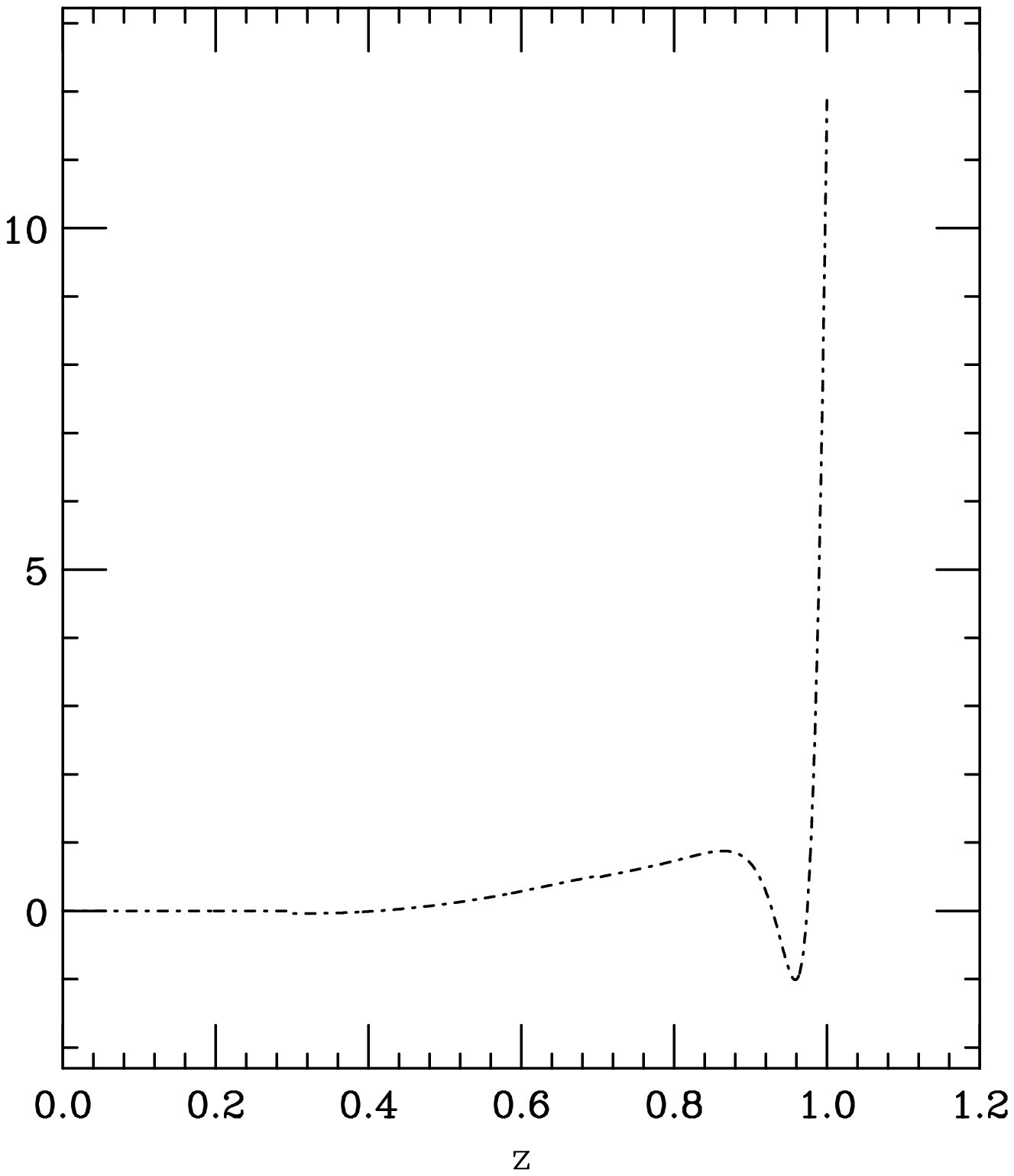}}
    \put(160,0){\includegraphics[scale=0.50,
      angle=0]{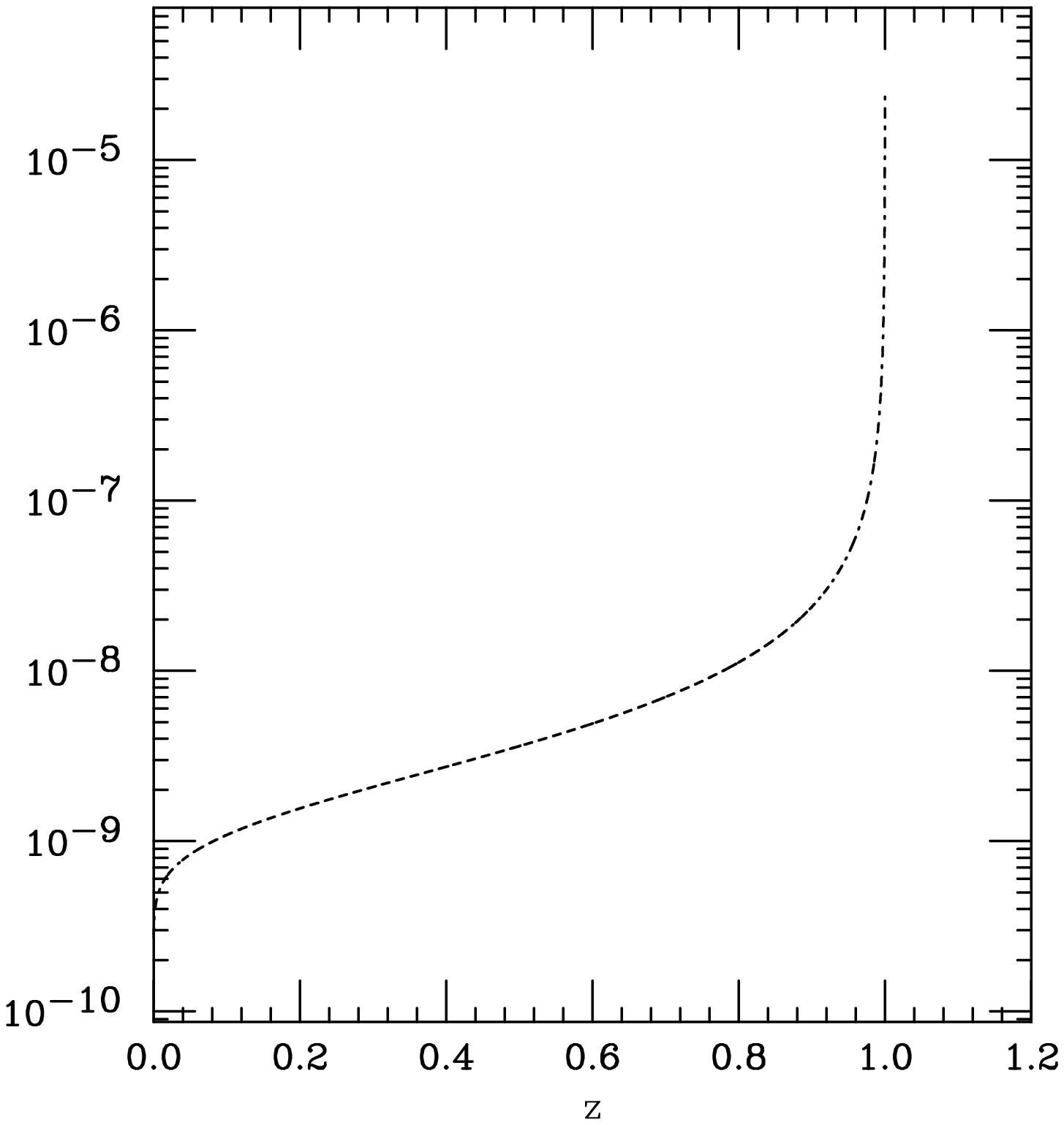}}
    \put(40,127){\large{$\delta_{C_0}(z)$}}
    \put(255,127){\large{$\delta_{C_1}(z)$}}
  \end{picture}
\caption[]{A comparison
of the resulting `delta' functions obtained by using $C_0$ (left) and
$C_1$ (right). It is obvious that the
function obtained by using $C_1$ behaves much more like a delta
function: going from $z=0.8$ to $z\approx1$, $\delta_{C_0}$ goes from
a value of $\sim 1$ to a value of $\sim 10$, whereas $\delta_{C_1}$
goes from $\sim10^{-8}$ to $\sim10^{-5}$. Note the logarithmic scale on the
vertical axis in the right-hand figure.}\label{fig:qcdrad:deltafunc}
\end{figure}

However, in the case of the evolution kernels $K_{a'a}(z)$, the
linear contour $C_0$ does not  provide sufficient accuracy to reproduce the
function from its transform. This is due to its inability to
accurately invert a constant function $f_N = c$ to the correct analytic result, a
delta function. This implies that the inversion does not
reproduce the necessary initial condition,
$K_{a'a}(z, Q=Q_c) \varpropto \delta(1-z)$. A numerically more
accurate contour is available in the literature, used in the so-called
`Fixed-Talbot algorithm'. This contour, $C_1$ is shown in solid red in
Fig.~\ref{fig:qcdrad:mellincontours}. It has the form $\mathrm{Re}(N) =
\mathrm{Im} N \cot ( \mathrm{Im} N / r )$, where $r$ is a parameter
which we will set to $r = 0.4 m/\log(1/z)$ during the computation, $m$
being the required precision in number of decimal digits, a value
derived from numerical experiments. The contour is related to the
`steepest descent' path for a certain class of functions. For further
details on its origin and accuracy see Ref.~\cite{Abate:2004}. A comparison
of the resulting `delta' functions obtained by using $C_0$ and $C_1$
can be seen in Fig.~\ref{fig:qcdrad:deltafunc}. It is obvious that the
function obtained by using $C_1$ behaves much more like a delta
function.
We first
rewrite the evolution kernel in a more convenient basis:
\begin{equation}\label{eq:matrixexponential}
K^{a'a}_N = \left(\mathcal{O}^{-1} \left[  \frac{\alpha_S(Q_c)}{\alpha_S(Q)} \right]^{p\,\mathrm{diag}(\lambda_{N,i})} \mathcal{O} \right)_{a'a}\;,
\end{equation}
where $\mathcal{O}$ is the matrix of eigenvectors of $\Delta_N$ and
$\mathrm{diag}(\lambda_N,i)$ is the diagonal matrix of its
eigenvalues. This is equivalent to using, implicitly, the singlet and
non-singlet basis~\cite{Ellis:1996qj}. 

\begin{figure}[!htb]
  \vspace{3.0cm}
  \centering
  \begin{picture}(300,120)
    \put(-90,200){\includegraphics[scale=0.37,
      angle=270]{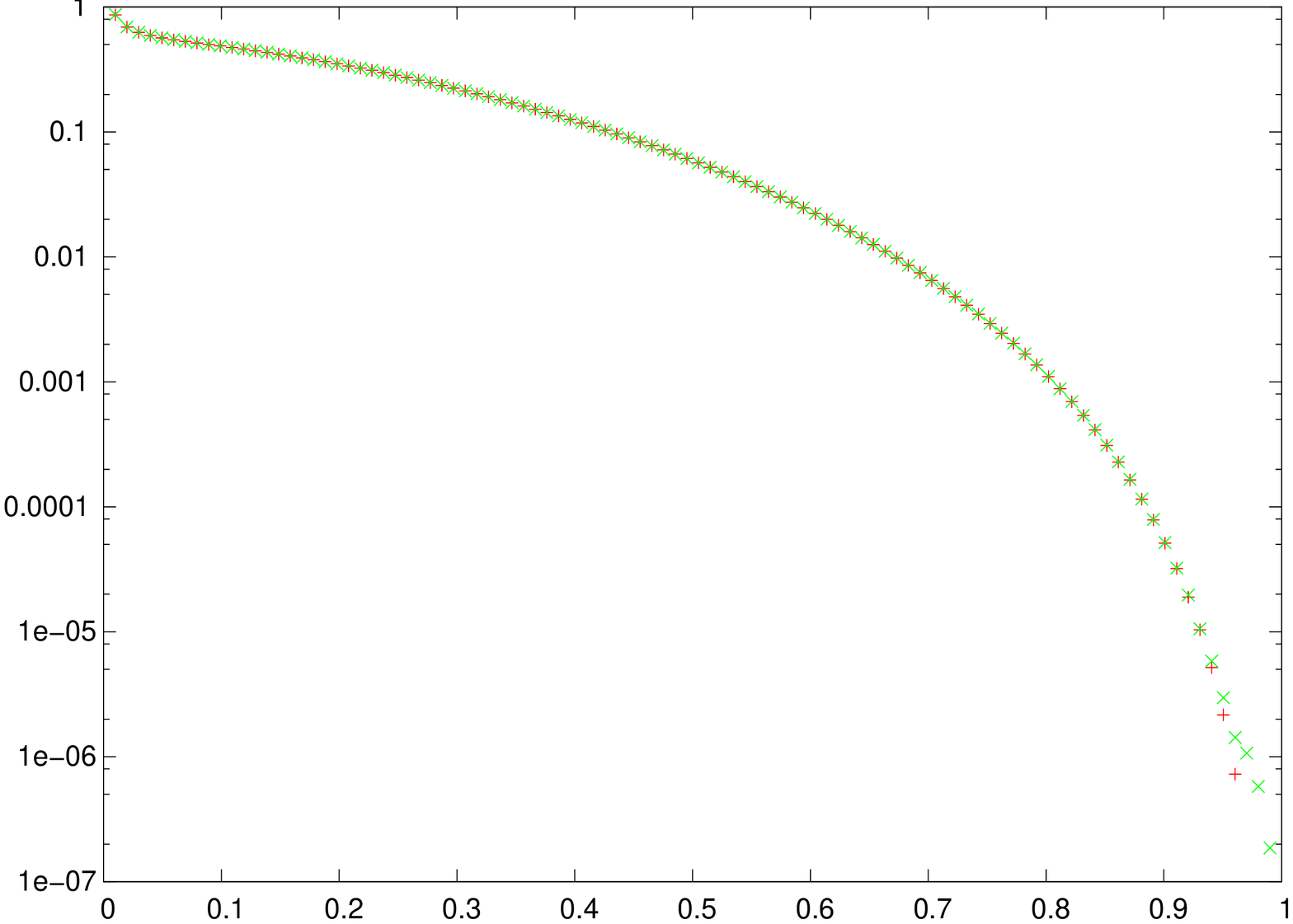}}
    \put(175,200){\includegraphics[scale=0.37,
      angle=270]{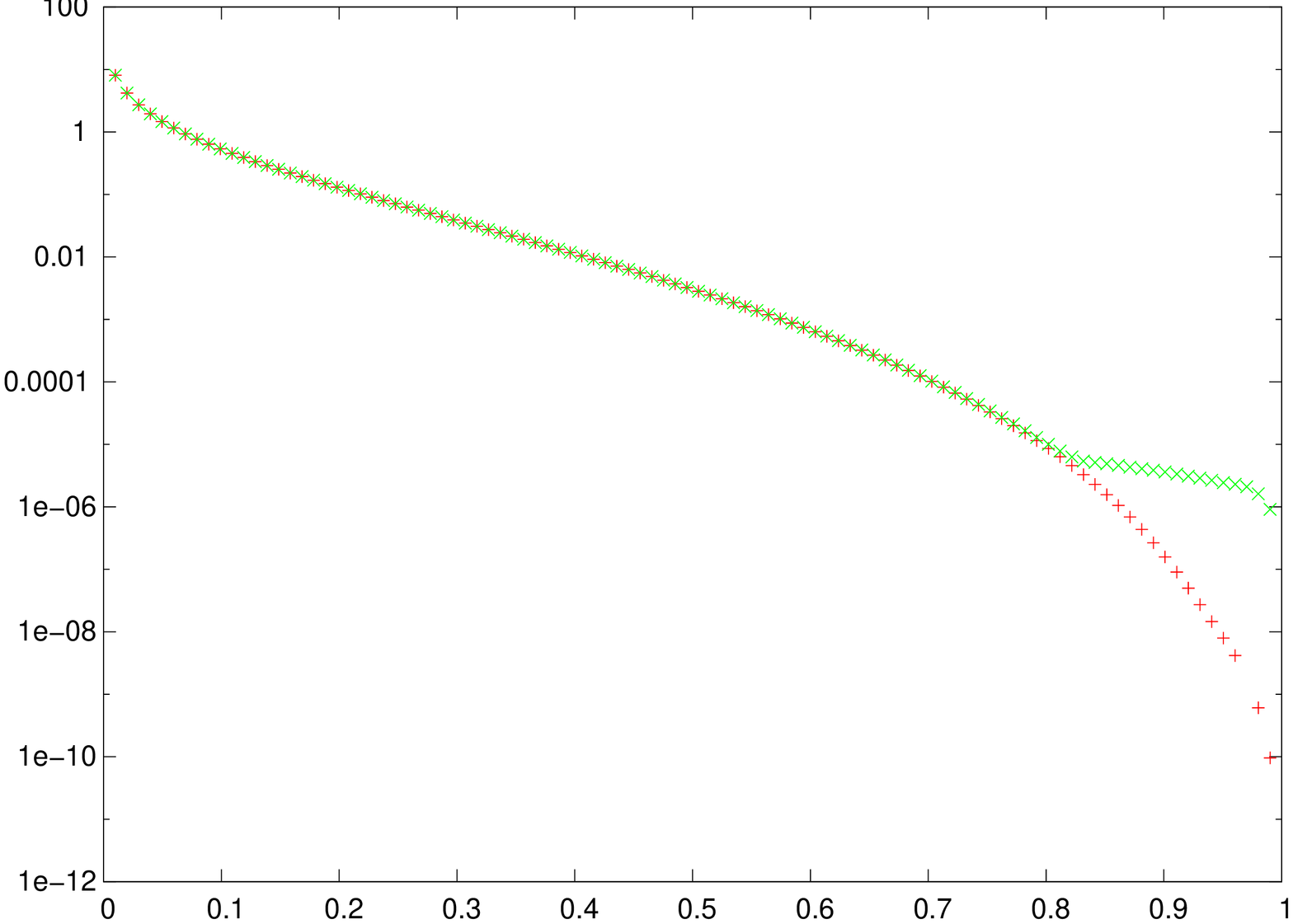}}
    \put(38,117){\large{$f_u(x)$}}
    \put(305,117){\large{$f_g(x)$}}
    \put(38,10){\large{$x$}}
    \put(305,10){\large{$x$}}
  \end{picture}
  \caption[]{The MSTW PDFs for the $u$-quark PDF (left) and the gluon PDF
  (right) evolved from their given form at $Q_c =
  10~\mathrm{GeV}$ to $Q = 10^4~\mathrm{GeV}$ using Mellin inversion
  on the complex plane via the $C_1$ contour (red points). The actual values from the MSTW
  PDFs are given for comparison (green points).}\label{fig:qcdrad:pdfevol}
\end{figure}

As a test of the evolution using Mellin inversion, we used the contour
$C_1$ to evolve $u$-quark and gluon PDFs from a starting scale of $Q_c
= 10~\mathrm{GeV}$ to
$Q = 10^4~\mathrm{GeV}$. The form of $f_a(x,Q_c)$ at $Q_c = 10~\mathrm{GeV}$ was extracted
directly from the the leading-order MSTW parton
distributions~\cite{Martin:2009iq} and the evolved results at $Q = 10^4~\mathrm{GeV}$ were
compared to the actual values given by the MSTW PDFs. The results are
shown in Fig.~\ref{fig:qcdrad:pdfevol}, exhibiting good agreement for
most of the range of values
of $x$. The discrepancy at high $x$ is due to the difference in
treatment of $f_a(x)$ as $x\rightarrow1$. 

\subsection{ISR effects including invisible particle emission}\label{sec:qcdrad:isrinvis}
Suppose now that an invisible 4-momentum $p^{\mu}_{inv}$ is emitted from
the hard subprocess. If we define the total lab frame 4-momentum of the incoming
partons $a$ and $b$ as $P^{\mu} = (E,\vec{P})$, then
\begin{equation}\label{eq:total4mom}
P^{\mu} = \frac{1}{2} \sqrt{S} [ (\bar{x}_1 + \bar{x}_2), 0, 0, (\bar{x}_1 - \bar{x}_2) ] \;,
\end{equation}
and the visible 4-momentum will be given by $P^{\mu} - p^{\mu}_{inv}$.
By definition, the visible mass is then
\begin{equation}\label{eq:vismassdef}
M^2 = (P -  p_{inv})^2 = P^{\mu} P_{\mu} + p^{\mu}_{inv} p_{inv,\mu} - 2 p^{\mu}_{inv} P_{\mu}\;.
\end{equation}
Equation~(\ref{eq:vismassdef}) demonstrates the interplay between two effects: on one hand ISR increases the `true' scale of the hard process $Q$, to the `apparent' scale $M$ by contaminating the detector with extra particles, and on the other hand the invisible particle emission decreases $M$ by the loss of particles. In the case of gluino pair-production both effects are equally important, as we will show. 

Substituting from Eq.~(\ref{eq:total4mom}) in Eq.~(\ref{eq:vismassdef}) and defining
$p^{\pm}_{inv} \equiv p^0_{inv} \pm p^3_{inv}$, we obtain
\begin{equation}\label{eq:vismass3}
M^2 = \bar{x}_1 \bar{x}_2 S + m_{inv}^2 - \sqrt{S} [ \bar{x}_1 p^-_{inv} + \bar{x}_2 p^+_{inv} ]\;,
\end{equation}
where $m_{inv}$ represents the total invariant mass of the invisibles,
$m_{inv}^2 = p^{\mu}_{inv} p_{inv,\mu}$.

The momenta $p^\mu_{inv}$ are defined in the lab frame, relative to which
the centre-of-mass frame of the hard subprocess is boosted by an
amount defined by the momentum fractions $x_1$ and $x_2$ of the
partons entering the subprocess. This implies that the $p^{\pm}_{inv}$ transform as
\begin{eqnarray}\label{eq:pplusminus}
p^+_{inv} = \sqrt{\frac{x_1}{x_2}} q^+_{inv} \;,\;\; p^-_{inv} = \sqrt{\frac{x_2}{x_1}} q^-_{inv}\;,
\end{eqnarray}
where $q^{\pm}_{inv} \equiv q^0_{inv} \pm q^3_{inv}$, defined in terms of the invisible momentum, $q^{\mu}_{inv}$, in the centre-of-mass frame of the hard subprocess. Substituting the expressions of Eq.~(\ref{eq:pplusminus}) in Eq.~(\ref{eq:vismass3}), we find an expression for the visible invariant mass:
\begin{eqnarray}
M^2 = m_{inv}^2 + \bar{x}_1 \bar{x}_2S \left[  1 - z_1 f^+_{inv} - z_2 f^-_{inv} \right]\;,
\label{eq:invmsq}
\end{eqnarray}
where we have defined $f^{\pm}_{inv} = q^{\pm}_{inv}/Q$ and
used $Q^2 = \bar{x}_1 \bar{x}_2 z_1 z_2 S$. We may now solve
Eq.~(\ref{eq:invmsq}) for $Q^2$ to obtain $Q^2$ in terms of $M^2$:
\begin{eqnarray}\label{eq:s}
Q^2 = \frac{ z_1 z_2 (M^2 - m_{inv}^2)}{1 - z_1 f^+_{inv} - z_2 f^-_{inv}}\;.
\end{eqnarray}
The above expression for the hard subprocess scale now becomes the argument
of the parton-level cross section, $\hat{\sigma}_{a'b'}$ in Eq.~(\ref{eq:M2sigz1z2}):
\beq\label{eq:SigmaDefinv}
S\frac{\mrd \sigma_{ab}}{\mrd M^2 \mrd Y} = \int \mrd z_1\,\mrd z_2\,K_{a'a}(z_1)f_a(\bar x_1,Q_c)\,K_{b'b}(z_2)f_b(\bar x_2,Q_c) 
\,\hat\sigma_{a'b'}\left( \frac{ z_1 z_2 (M^2 - m_{inv}^2)}{1 - z_1 f^+_{inv} - z_2 f^-_{inv}} \right)\,.
\eeq
The functions $f^{\pm}_{inv}$, which are related to the invisible
particle 4-momenta, remain to be determined. The visible system
rapidity, $Y$, is also modified by the presence of invisible particles
as
\begin{equation}\label{eq:Yinv}
Y = \frac{1}{2} \log \left( \frac{\bar{x}_1 ( 1 - z_1 f^+_{inv} )} { \bar{x}_2 ( 1 - z_2 f^-_{inv} ) } \right)\;,
\end{equation}
and therefore Eqs. (\ref{eq:qcdrad:xbar}) for $\bar{x}_{1,2}$ become:
\bea\label{eq:xbarinv}
\bar x_1 &=& \sqrt{\frac{(M^2 - m_{inv}^2)(1 - z_2 f^-_{inv})}{S(1 -
   z_1 f^+_{inv} - z_2 f^-_{inv})(1- z_1 f^+_{inv})}} e^{Y}\;,\nonumber\\
\bar x_2 &=& \sqrt{\frac{(M^2 - m_{inv}^2)(1 - z_1 f^+_{inv})}{S(1 -
   z_1 f^+_{inv} - z_2 f^-_{inv})(1- z_2 f^-_{inv})}} e^{-Y}\;.
\eea
The kinematic constraints restrict $Q^2$ to be greater than the
threshold energy squared for the process and the true invariant mass,
$M_{\mathrm{true}}^2 \equiv \bar{x}_1 \bar{x}_2 S = Q^2 / (z_1 z_2)$,
to be greater than the visible invariant mass, $M^2$. These result in
the following two constraints for $Q^2$:
\begin{equation}
Q^2 > Q_{\mathrm{threshold}} ^2 \;,\;\; Q^2 > z_1 z_2 M^2 \;.
\end{equation}

\subsubsection{Single-invisible decays}\label{sec:1inv}
The benchmark scenario for a single invisible decay originating from the hard process is $t\bar{t}$ production in which one of the two tops decays into $bqq'$ (hadronic) and the other into $b\ell\nu$ (semi-leptonic), the neutrino comprising the missing 4-momentum. Excluding the proton remnants, we assume that all other particles within the pseudorapidity coverage are detected. We will refer to the neutrino as the invisible particle and the $W$ as the intermediate particle in the $t\bar{t}$ case, but the treatment is readily applicable to the gluino case where the invisible particle is the $\chi_1^0$ and the intermediate particle is a squark (treated in section~\ref{sec:2inv}). 

\begin{figure}
 \begin{centering}
   \includegraphics[scale=0.70]{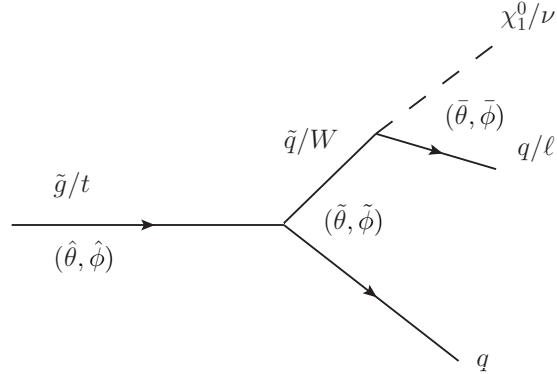}
   \caption{The sequential two-body decay chain under consideration in
     the invisible particle treatment. The relevant production angles
     in the parent centre-of-mass frame are also shown in
     parentheses.}
   \label{fig:decchain}
\end{centering}
\end{figure}

To calculate the functions $f^{\pm}_{inv}$ and obtain $Q^2$, we need to calculate the neutrino 4-momentum in the hard process frame. This is done by choosing the neutrino 4-momentum in the frame of its parent $W$ and then applying two subsequent Lorentz boosts: one going from the $W$ frame to the top frame, and one from the top frame to the hard process frame. The decay chain is shown in Fig.~\ref{fig:decchain}. Each of these boosts involves two angular variables which originate from the `decay' of the parent particle. Hence the 4-momentum $q^{\mu}_{inv}$ of the neutrino may be written as
\begin{equation}\label{eq:qinv}
q^{\mu}_{inv} = \Lambda^{\mu}_\kappa\left(Q, \hat{\theta}, \hat{\phi} \right) \Lambda^{\kappa}_\lambda \left( \tilde{\theta} , \tilde{\phi} \right) \bar{p}^\lambda_\nu(\bar{\theta}, \bar{\phi})\;,
\end{equation}
where the $\Lambda$'s are Lorentz boost matrices and where quantities
with a hat refer to the hard process frame, quantities with a tilde
refer to the top frame and quantities with a bar refer to the $W$
frame. The angles $\theta$ and $\phi$ represent the usual polar
angles, defined with respect to the direction of the `sister'
particle (see Fig.~\ref{fig:decchain}).
For example, in the case $W^+ \rightarrow \ell ^+ \nu_\ell$,
where the $W^+$ was produced from the top decay along with a bottom
quark, the angles $(\bar{\theta},\bar{\phi})$ are defined with respect
to the direction of motion of the $b$ quark in the $W^+$ frame. The two
boost vectors have magnitudes given by $| \vec{\beta}_i| = |\vec{p}_i|
/ E_i$ ($i=t,W$), the ratio of the parent 3-momentum magnitude and its
energy. The boosts, as well as the magnitude of the invisible particle
4-momentum, can be obtained by considering the kinematics in each frame as:
\begin{eqnarray}
&\bar{p}^\lambda_\nu(\bar{\theta}, \bar{\phi}) = \frac{m_W}{2}  ( 1, \vec{\bar{r}}) \;,\\\nonumber
&\vec{\beta}_W = \frac{m_t^2-m_W^2}{m_t^2 + m_W^2}  \vec{\tilde{r}} \;,\\\nonumber 
&\vec{\beta}_t = \sqrt{ 1 - \frac{4 m_t^2}{Q^2}}  \vec{\hat{r}} \;,
\end{eqnarray}
where $\vec{r} = (\sin\theta \cos \phi, \sin\theta \sin \phi, \cos
\phi)$ is the unit vector in spherical polar coordinates in the
appropriate frame and $m_{W}$, $m_t$ are the $W$ and top quark masses respectively. The 4-vector $f^\mu_{inv}$, and hence the functions $f^{\pm}_{inv}$, are calculated by $f^{\pm}_{inv} = q^{\pm}_{inv}/Q$. Evidently, the functions $f^{\pm}_{inv}$ are functions of $Q^2$, giving an implicit equation for $Q^2$. To make this more explicit, we rewrite Eq.~(\ref{eq:s}):
\begin{equation}\label{eq:implicits}
Q^2 = \frac{ z_1 z_2 \left[M^2 - m_{inv}^2(Q^2, \Omega)\right]}{1 - z_1 f^+_{inv}(Q^2, \Omega) - z_2 f^-_{inv} (Q^2, \Omega)} \;,
\end{equation}
and analogously for Eq.~(\ref{eq:xbarinv}),
where $\Omega$ represents the set of all angular
variables. In the present case $m_{inv}(Q^2, \Omega)=m_\nu\simeq 0$
but for multiple invisible particles it will also be a function as
indicated.

Equation (\ref{eq:implicits}) needs to be solved numerically for
each set ($z_1, z_2, \Omega$) in the region $(4m_{t / \tilde{g}}^2,
z_1 z_2 S)$, where $S$ is the square of the proton centre-of-mass
energy, along with the restriction that the visible invariant mass
should be lower than the `true' invariant mass, $M \leq M_{\mathrm{true}}$. The
numerical solution was found using the Van Wijngaarden-Dekker-Brent
method~\cite{brent, gslmanual}, a bracketing method for finding roots
of one-dimensional equations. Since $Q$ is not uniquely determined for
each $M$, different values of the `true' centre-of-mass energy $Q$
contribute to the cross section. Note that not all possible
configurations ($z_1, z_2, \Omega$) are kinematically allowed to
contribute to the cross section at $M$ and hence some configurations
do not yield roots of Eq.~(\ref{eq:implicits}). 
Once $Q^2$ is obtained, the parton-level cross section for the hard process partons, $\hat{\sigma}_{a'b'}(Q^2)$, is calculated. This result is then multiplied with the parton density functions for the incoming partons, $f_{a,b}(\bar{x}_{1,2}, Q_c)$, and the kernels for evolution from incoming partons $a$ and $b$ to hard process partons $a'$ and $b'$ ($K_{a'a}(z_1)$ and $K_{b'b}(z_2)$). We then integrate over all possible values of $z_1$ and $z_2$, according to Eq.~(\ref{eq:SigmaDefinv}). Finally, to obtain the full resummed result we have to integrate over the distribution of the angular variables $\Omega$.
Notice that the visible invariant mass distribution becomes non-zero below the
threshold for production, $M < 2 m_ {t / \tilde{g}}$, owing to the
loss of invisible particles.
\subsubsection{Double-invisible decays}\label{sec:2inv}
We now turn to the case where both particles produced in the hard
process decay invisibly. For illustration we refer to sequential decays of the gluino: $\tilde{g} \rightarrow \tilde{q} q \rightarrow \chi^0_1 qq$. Although this decay mode is generally not the dominant one, it is useful for illustration of the procedure. We extend the treatment given in the semi-leptonic/hadronic top case by writing out functions related to the two invisible particle 4-momenta in the decay chain (which we call $\chi$ and $\chi '$):
\begin{eqnarray}\label{eq:qnuqnubar}
q^{\mu}_{\chi} = \Lambda^{\mu}_\kappa\left(Q, \hat{\theta}, \hat{\phi} \right) \Lambda^{\kappa}_\lambda \left( \tilde{\theta} , \tilde{\phi} \right) \bar{p}^\lambda_\chi(\bar{\theta}, \bar{\phi})\;,\\
q^{\mu}_{\chi '} = \Lambda^{\mu}_\kappa\left(Q, \hat{\theta'}, \hat{\phi'} \right) \Lambda^{\kappa}_\lambda \left( \tilde{\theta'} , \tilde{\phi'} \right) \bar{p}^\lambda_{\chi '}(\bar{\theta'}, \bar{\phi'})\;,
\end{eqnarray}
where the primed quantities now distinguish between the two
invisibles. Since both of these 4-vectors are defined in the
hard subprocess frame, we simply have
\begin{equation}\label{eq:finv2}
f^{\pm}_{inv} = \frac{1}{Q}\left( q^{\pm}_{\chi} + q^{\pm}_{\chi '} \right)\;.
\end{equation}
The rest of the treatment is identical to the case for one invisible particle: an
implicit equation has to be solved to obtain $Q^2$ for each ($z_1$,
$z_2$, $\Omega$) set and then an 
integral over $\Omega$ is taken to obtain the resummed result.

\subsubsection{Angular distributions}\label{sec:angular}
\begin{figure}[htb]
  \centering 
  \vspace{1.0cm}
  \hspace{2.9cm}
  \begin{picture}(300,120)
  \put(0,0){\includegraphics[scale=0.34, angle=90]{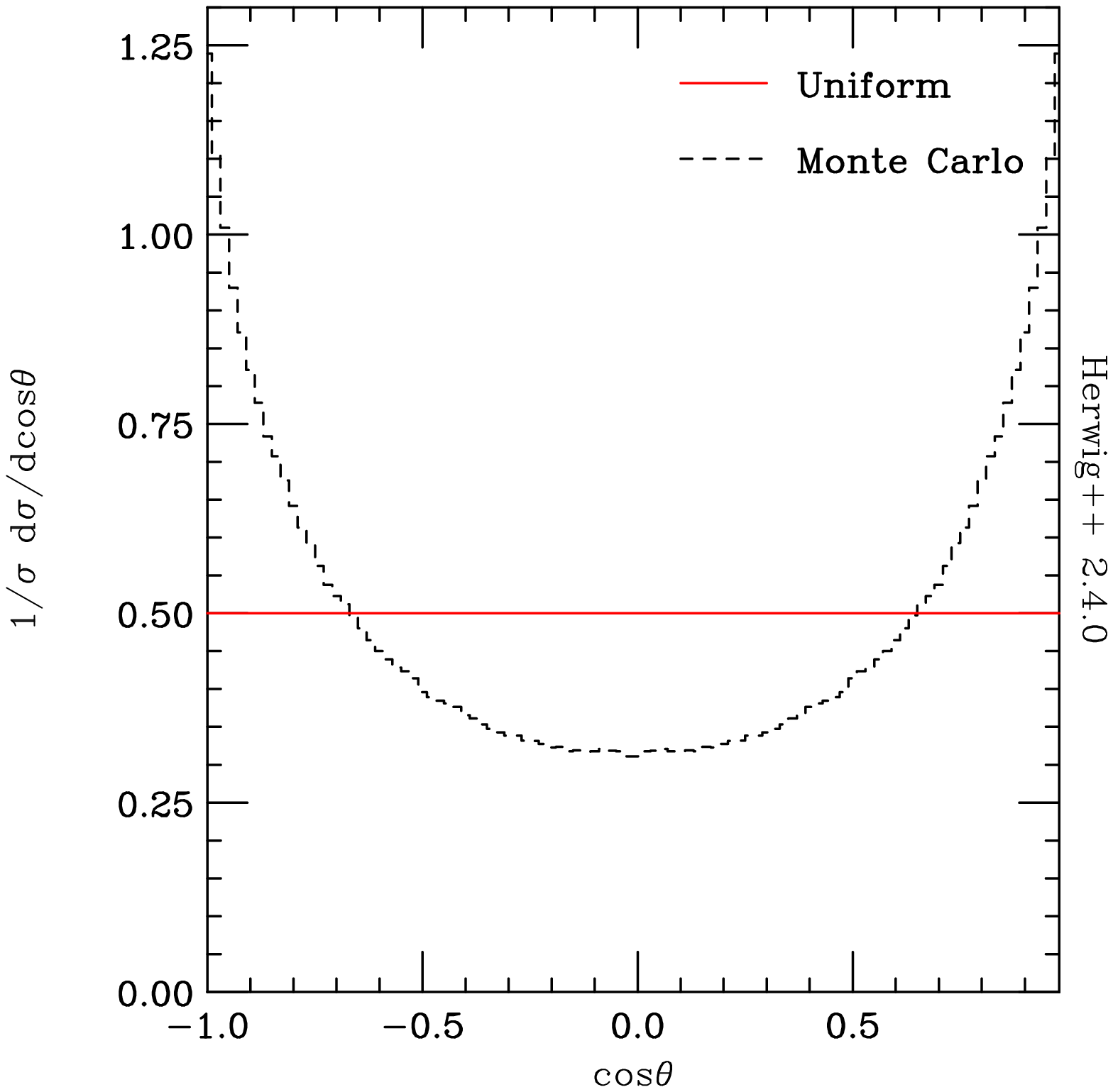}}
  \put(150,0){\includegraphics[scale=0.34, angle=90]{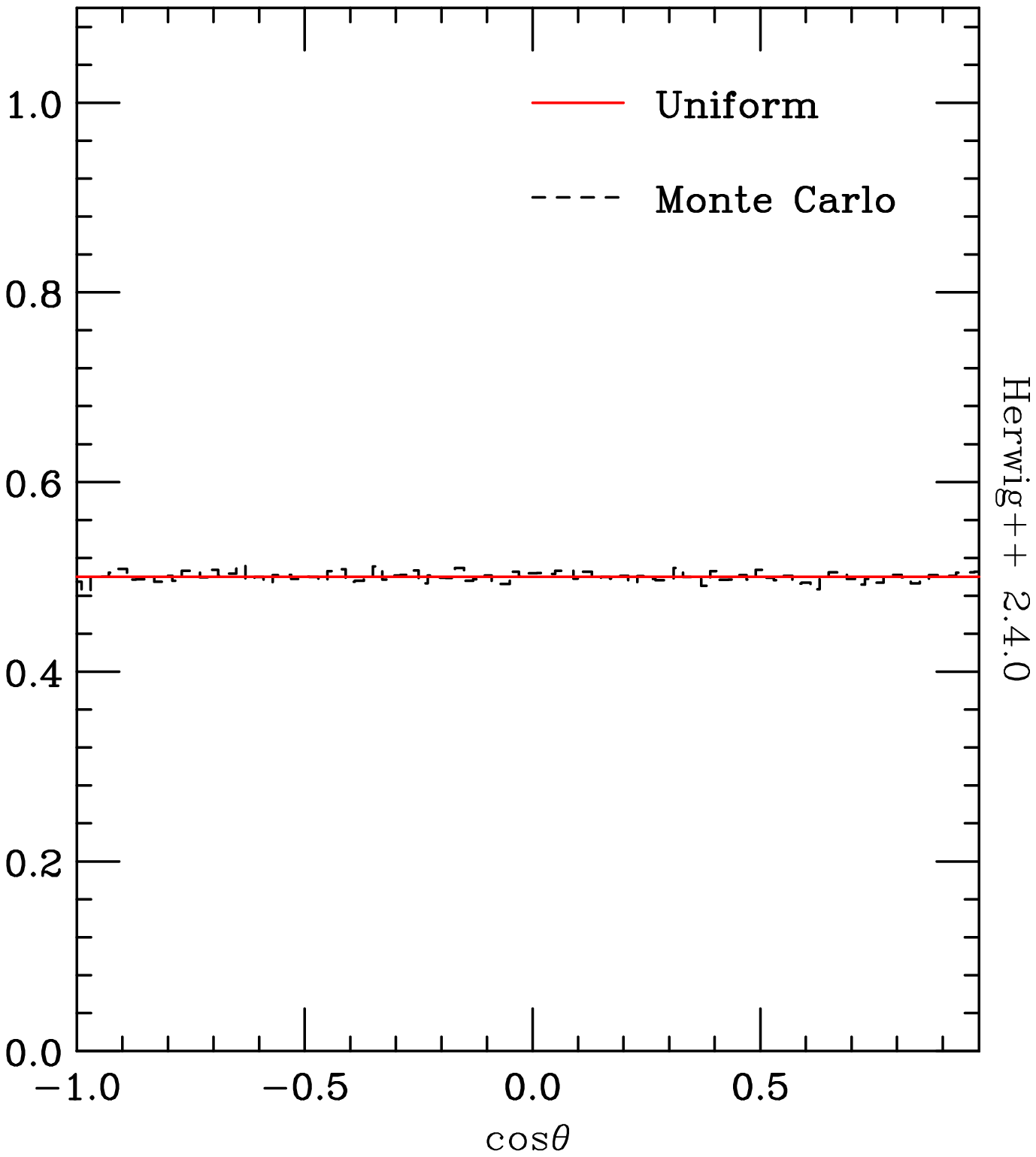}}
  \put(300,0){\includegraphics[scale=0.34, angle=90]{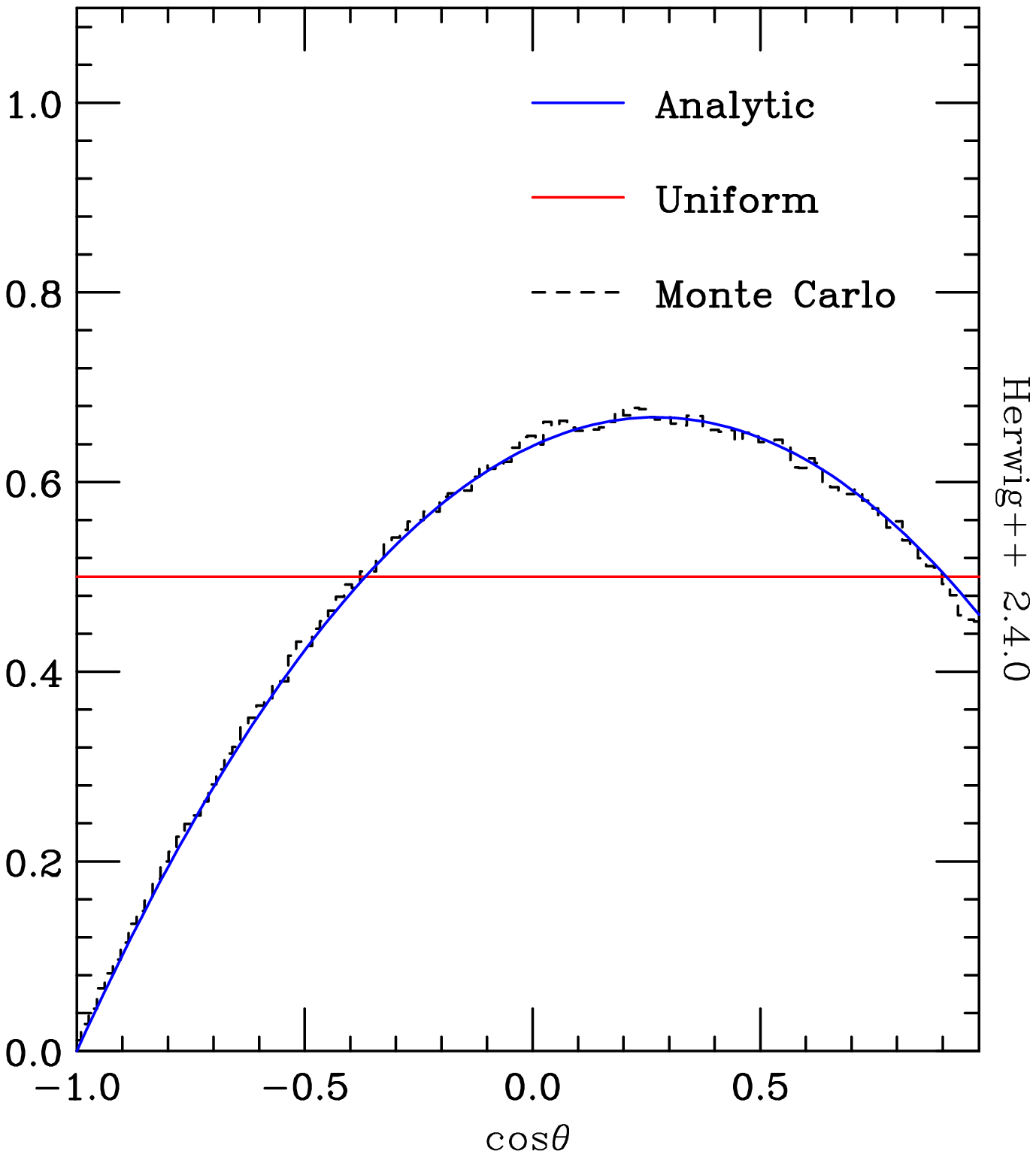}}
  \put(-25.5,-8.5){\small{$\hat{}$}}
  \put(124.5,-8.5){\small{$\tilde{}$}}
  \put(274.5,-8.5){\small{$\bar{}$}}
  \end{picture}
\caption{Monte Carlo results for the top pair-production decay chain angles. From left to right: the production angle of the top in the hard process frame, the angle of the outgoing $W$ boson in the top frame and the angle of the outgoing neutrino in the $W$ frame. The uniform distributions are shown for comparison. The neutrino angle in the $W$ frame is also compared to the analytic calculation. }
\label{fig:ttangles}
\end{figure}

The distributions of the angular variables $\Omega =
(\hat{\theta},\hat{\phi}, \tilde{\theta}, \tilde{\phi}, \bar{\theta},
\bar{\phi})$, appearing in the treatment of invisibles given in the
previous sections, are process-dependent. They represent the angles at
which the daughter particle is emitted in the frame of the parent
particle. We investigated the angular distributions using \Herwigpp
version 2.4.0 and subsequently used the results in calculating the $f^{\pm}_{inv}$ functions. The results for SPS1a gluino pair-production are shown in Fig.~\ref{fig:ggangles}, where the uniform distributions are shown for comparison (red horizontal line). Figure~\ref{fig:ttangles} shows the distributions as obtained for top pair-production. The neutrino angle in the $W$ frame is also compared to the analytic calculation. As expected, all the $\phi$ angles, in both cases, were found to be uniform (not shown). The form of all the distributions can be justified using general spin considerations:

\begin{itemize}
\item[$\hat{\theta}_i$:] The angular distribution of the angle $\hat{\theta} _i$ at which the fermions are produced in the hard process frame is expected to have the form $\sim 1 + \beta \cos ^2 \hat{\theta}_i$, where $\beta$ is a process-dependent constant.

\item[$\tilde{\theta}_i$:] The angle $\tilde{\theta} _i$, is defined between the direction of the daughter boson ($W$ or $\tilde{q}$) with respect to the direction of polarisation of the parent ($t$ or $\tilde{g}$). The angular distribution for a spin-up fermion parent is then given by~\cite{Hubaut:2005er}
\begin{equation}
\frac{1}{N_\uparrow} \frac{\mathrm{d}N_\uparrow}{\mathrm{d} \cos \tilde{\theta} _i} = \frac{1}{2} ( 1 + P \alpha _i \cos \tilde{\theta} _i ) \;,
\end{equation}
where $\alpha _i$ is a constant and $P$ is the modulus of the
polarization of the parent. Since the production processes for both
$t\bar{t}$ and $\tilde{g} \tilde{g}$ are parity-conserving, there is also an equal spin-down ($N_\downarrow$) contribution to the total distribution with the sign of $\alpha _i$ reversed. This results in a uniform distribution for $\cos \tilde{\theta}_i$. 
\begin{figure}[htb]
  \centering 
  \vspace{1.0cm}
  \hspace{2.9cm}
  \begin{picture}(300,120)
  \put(0,0){\includegraphics[scale=0.34, angle=90]{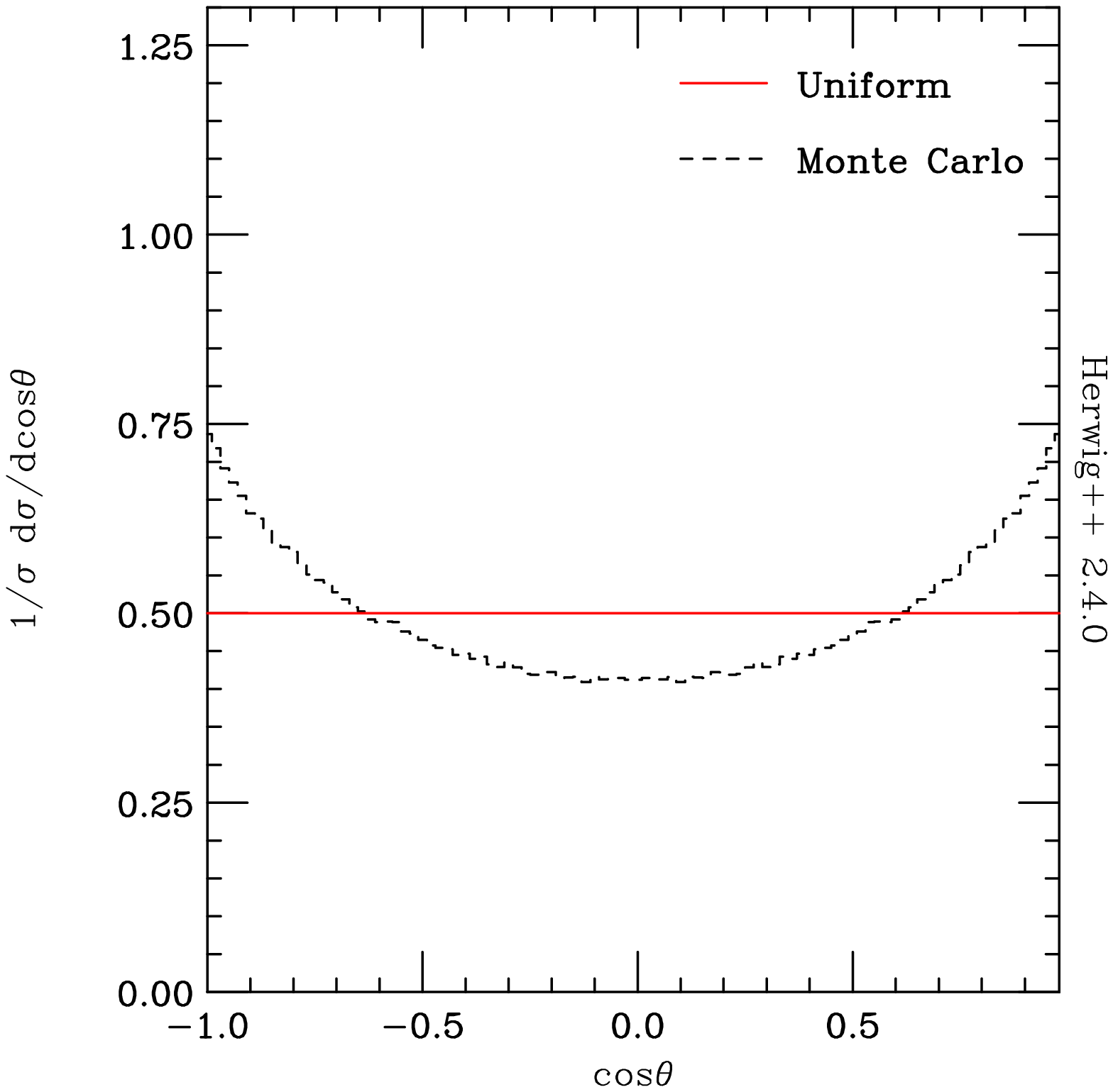}}
  \put(150,0){\includegraphics[scale=0.34, angle=90]{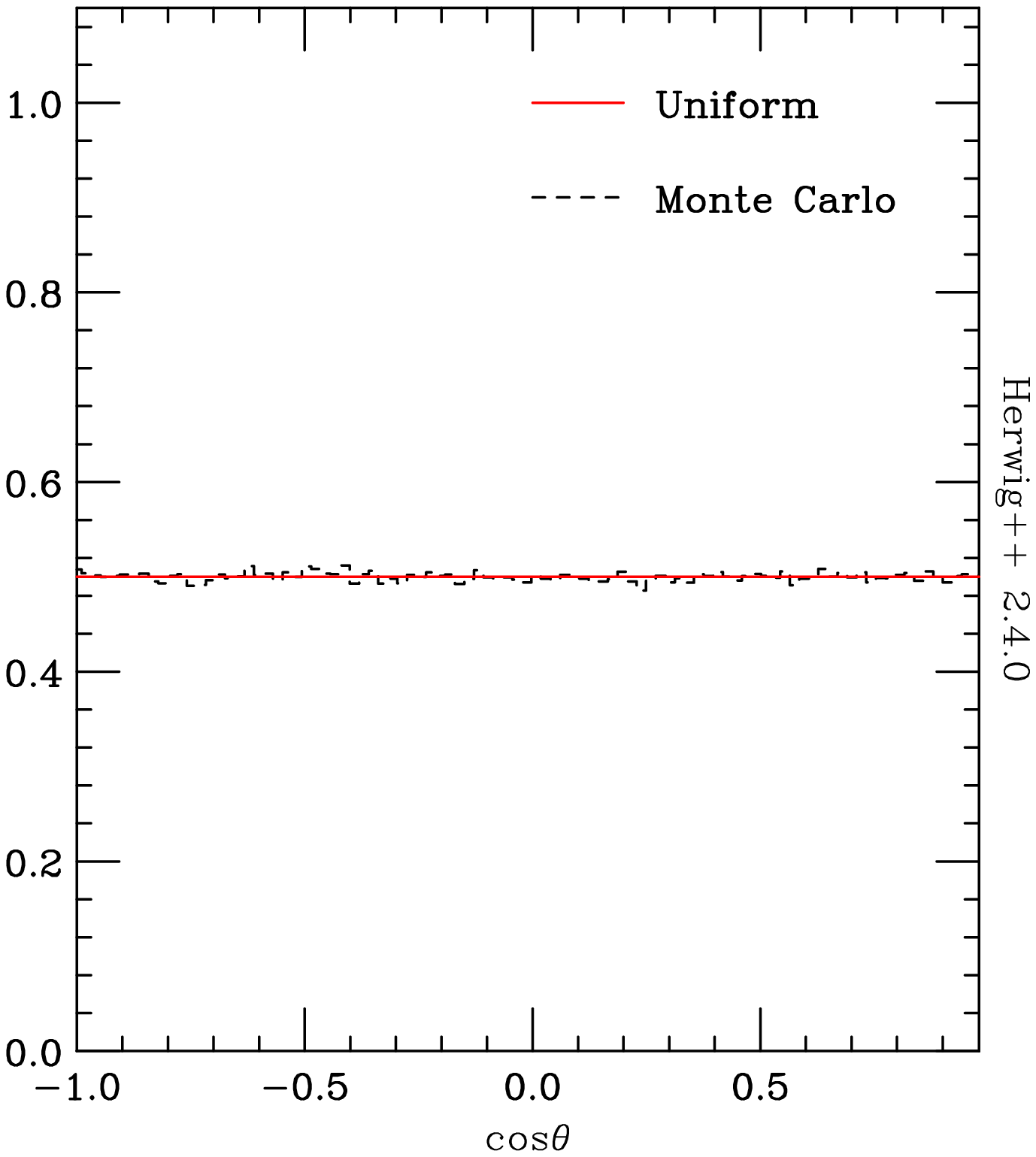}}
  \put(300,0){\includegraphics[scale=0.34, angle=90]{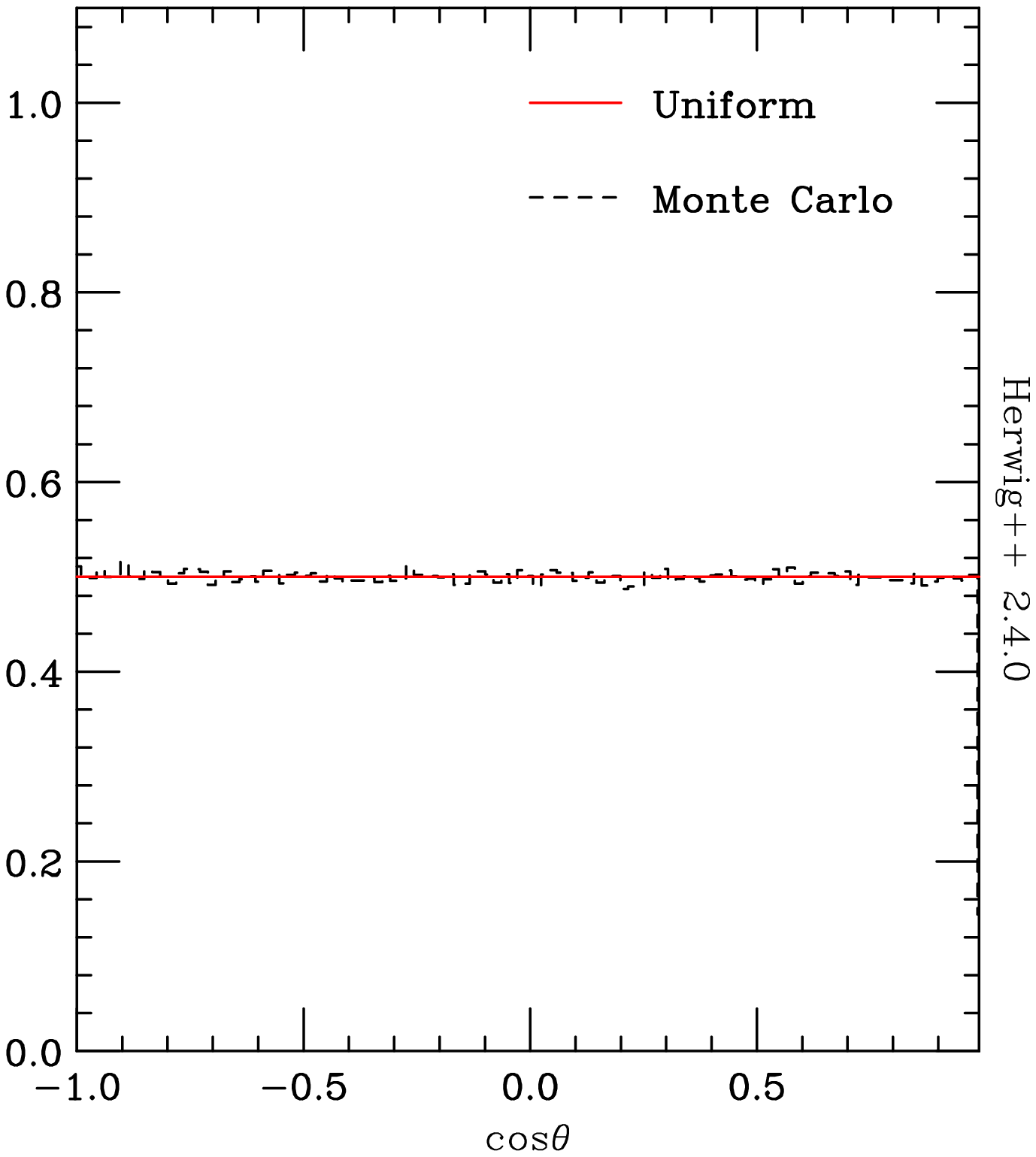}}
  \put(-25.5,-8.5){\small{$\hat{}$}}
  \put(124.5,-8.5){\small{$\tilde{}$}}
  \put(274.5,-8.5){\small{$\bar{}$}}
  \end{picture}
\caption{Monte Carlo results for the gluino pair-production decay chain angles. From left to right: the production angle of the gluino in the hard process frame, the angle of the outgoing squark in the gluino frame and the angle of the outgoing neutralino in the squark frame. The uniform distributions are shown for comparison.}
\label{fig:ggangles}
\end{figure}
\item[$\bar{\theta}_i$:] In gluino pair-production, the decay products
  of the squark, $\tilde{q}$, which is a scalar, are uniformly distributed in $\cos
  \bar{\theta}$.  In top pair-production, on the other hand, the decay
  $W\rightarrow \ell \nu _{\ell}$ is parity-violating and the
  distribution of $\cos \bar{\theta}$ is forward-backward asymmetric
  in the $W$ frame~\cite{Ellis:1996qj}.  The angle $\bar{\theta}$
  (sometimes called $\Psi$, see e.g.~\cite{Aad:2009wy}) can be used
  experimentally to infer helicity information on the $W$. The
  distribution may be written as
\begin{eqnarray}
\frac{1}{N} \frac{\mathrm{d} N} { \mathrm{d} \cos \bar{\theta} } = \frac{3}{2} \left[ F_0 \left( \frac{ \sin \bar{\theta} } { \sqrt{2} } \right)^2 + F_L \left( \frac{ 1 - \cos \bar{ \theta} } { 2} \right) ^2 +  F_R \left( \frac{ 1 + \cos \bar{ \theta} } { 2} \right) ^2  \right]\;,\nonumber \\
\end{eqnarray}
where $F_L$, $F_R$ and $F_0$ are the probabilities for left-handed,
right-handed and longitudinal helicities of the $W$ in top quark decay
respectively. The SM predictions, $(F_L, F_R, F_0) = (0.304, 0.001,
0.695)$, yield the blue solid curve shown on the right in Fig.~\ref{fig:ttangles}.
\end{itemize}

The spins of the two produced fermions (tops or gluinos) are correlated and
this may cause a degree of correlation between the distributions of
particles in the decay chains. We investigated whether these
correlations play an important role in the calculation of the
invisible particle effects on the visible mass. By comparing the
invariant mass distributions with and without the spin correlations in
the Monte Carlo we concluded that the effect is small in both top and
gluino pair-production and can be safely neglected.

\subsection{Results}\label{sec:invresults}
We present the resummed distributions obtained for $t\bar{t}$ and
$\tilde{g}\tilde{g}$ production according to
Eq.~(\ref{eq:SigmaDefinv}). All results are for the LHC at design
energy, i.e.\ $pp$ collisions at $\sqrt s = 14$ TeV.  We have
integrated over the visible system rapidity, $Y$, in the range
$|Y|<5$. We first compare our results to those obtained using the
\Herwigpp event generator at parton level (i.e. no
hadronization or underlying event) and excluding the proton
remnants.\footnote{We  verified, using the event generator, that the
  contribution of the proton remnants to the total invariant mass in
  the considered rapidity range is negligible.}  In
sections~\ref{sec:hadronization} and \ref{sec:MPI} we examine the
effects of hadronization and the underlying event. Parton-level top
and gluino pair-production cross section formulae are given in
appendix~\ref{app:cross-sections}. The PDF set used both in the
resummation calculation and \Herwigpp is the MRST LO** (MRSTMCal)
set~\cite{Sherstnev:2007nd, Sherstnev:2008dm}.

\subsubsection{Top quark pair-production}\label{sec:topresults}
\begin{figure}
  \centering 
  \vspace{1.5cm}
    \includegraphics[scale=0.50, angle=90]{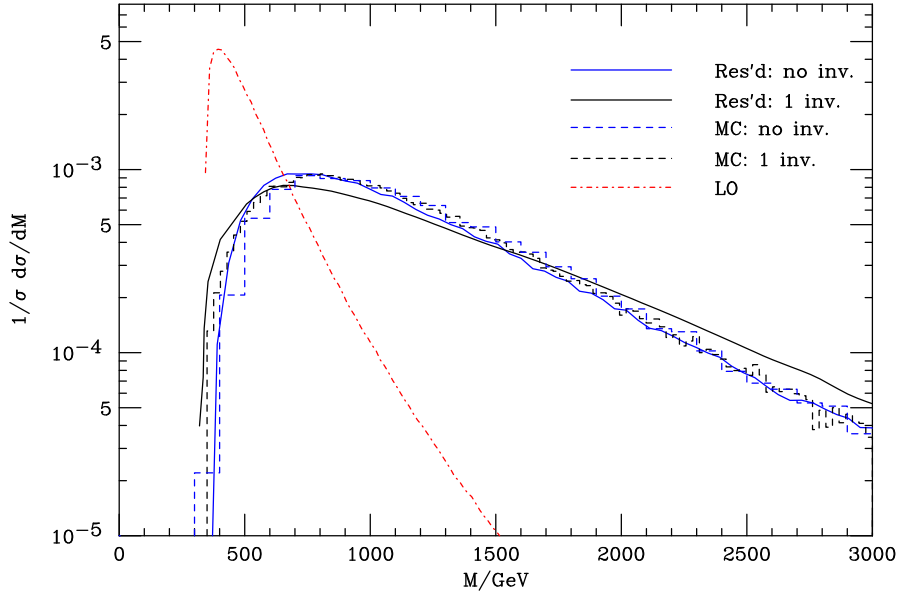}
\caption{The $t\bar{t}$ visible mass distributions for a
  pseudorapidity cut $\etam=5$, comparing
hadronic (no invisibles) and semi-leptonic (one invisible) decays. The leading-order $t\bar{t}$ invariant mass distribution is shown (red dot-dashes) for comparison.}
\label{fig:tte5a}
\end{figure}

\begin{figure}
  \centering 
  \vspace{1.5cm}
    \includegraphics[scale=0.50, angle=90]{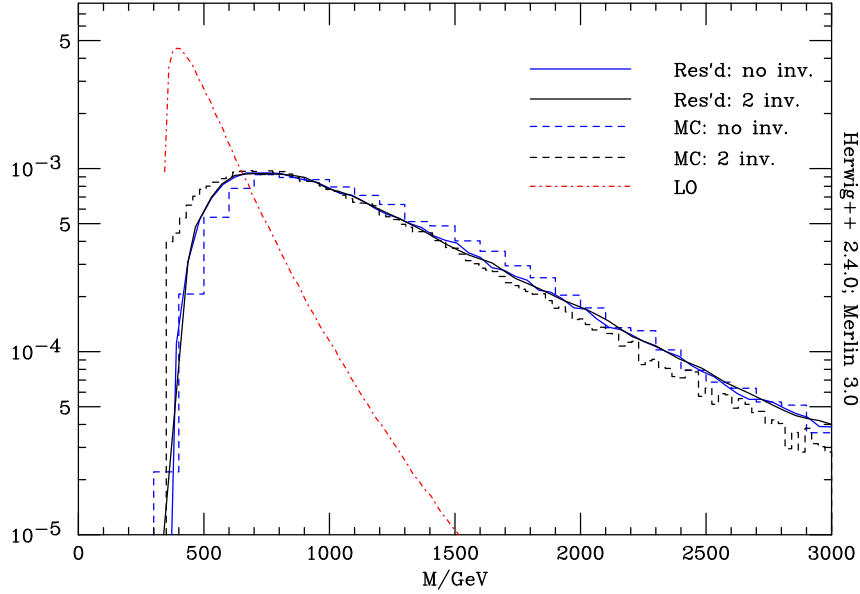}
\caption{The $t\bar{t}$ visible mass distributions for a
  pseudorapidity cut $\etam=5$, comparing
hadronic (no invisibles) and fully leptonic (two invisibles) decays.
The leading-order $t\bar{t}$ invariant mass distribution is shown (red dot-dashes) for comparison.}
\label{fig:tte5b}
\end{figure}

We present resummed results in comparison to Monte Carlo for Standard
Model $t\bar{t}$ production, where we include particles with maximum
pseudorapidity $\etam = 5$. In Figs.~\ref{fig:tte5a} and~\ref{fig:tte5b} we show
separate results for combinations of hadronic and semi-leptonic decays
of the top, leading to zero, one or two invisible neutrinos from the
hard process. The effect of the invisibles in both the fully
semi-leptonic case and the hadronic/semi-leptonic case are small
compared to the effects of hadronization, to be discussed in
section~\ref{sec:hadronization}. The differences between the Monte
Carlo and resummed curves in Figs.~\ref{fig:tte5a} and~\ref{fig:tte5b} may be attributed to sensitivity to the
behaviour of the PDFs and parton showering at low
scales, since
$Q_c$ can be as low as $ 2 m_t \times e^{-5} \sim 2~\mathrm{GeV}$ in the case
of $t\bar{t}$ production, and the precise definition of $Q_c$ in terms of $\etam$.
\subsubsection{Gluino pair-production}\label{sec:gluinoresults}
\begin{figure}[!htb]
  \centering 
  \vspace{1.2cm}
    \includegraphics[scale=0.34, angle=90]{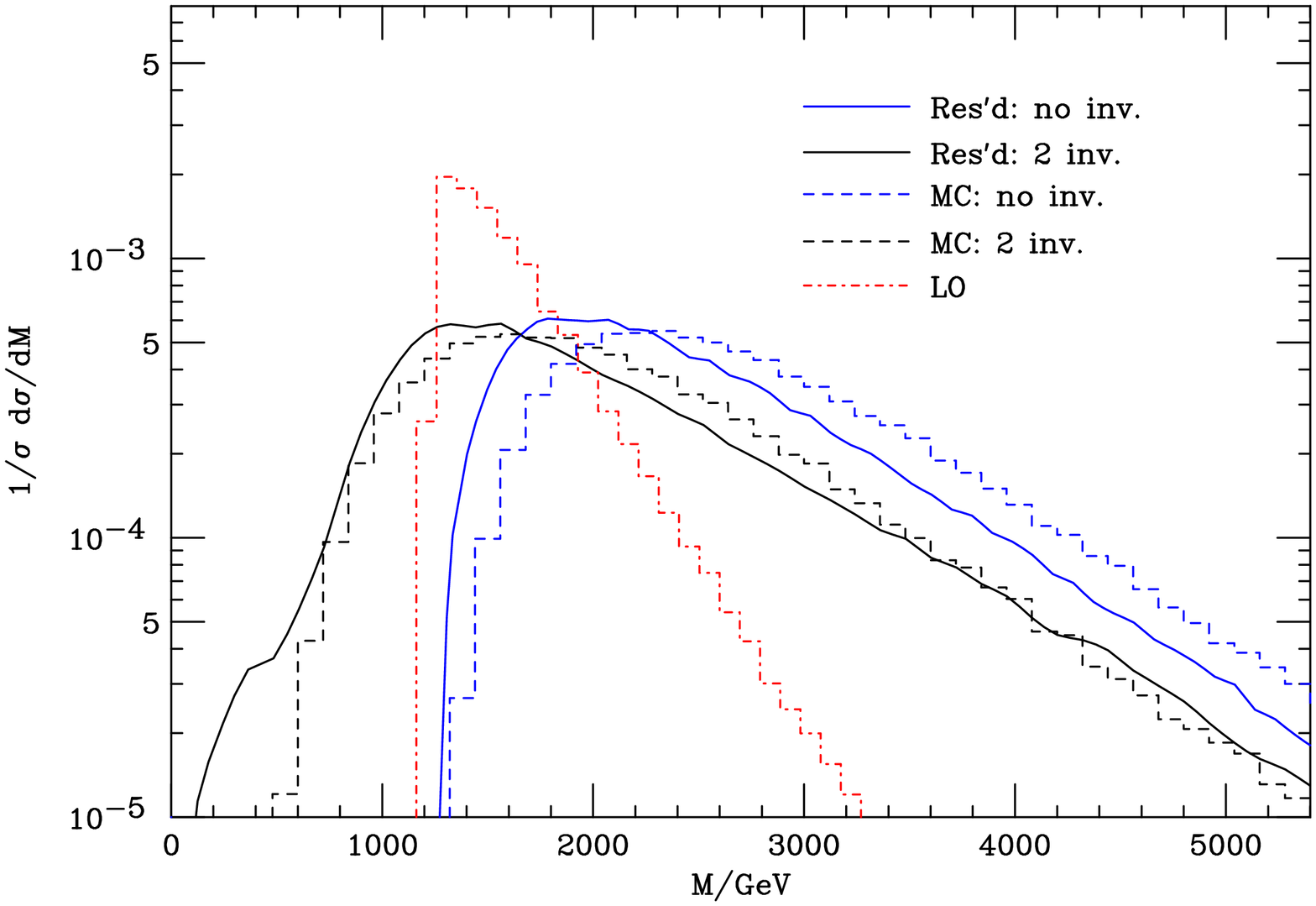}
    \hspace{0.5cm}
    \includegraphics[scale=0.34, angle=90]{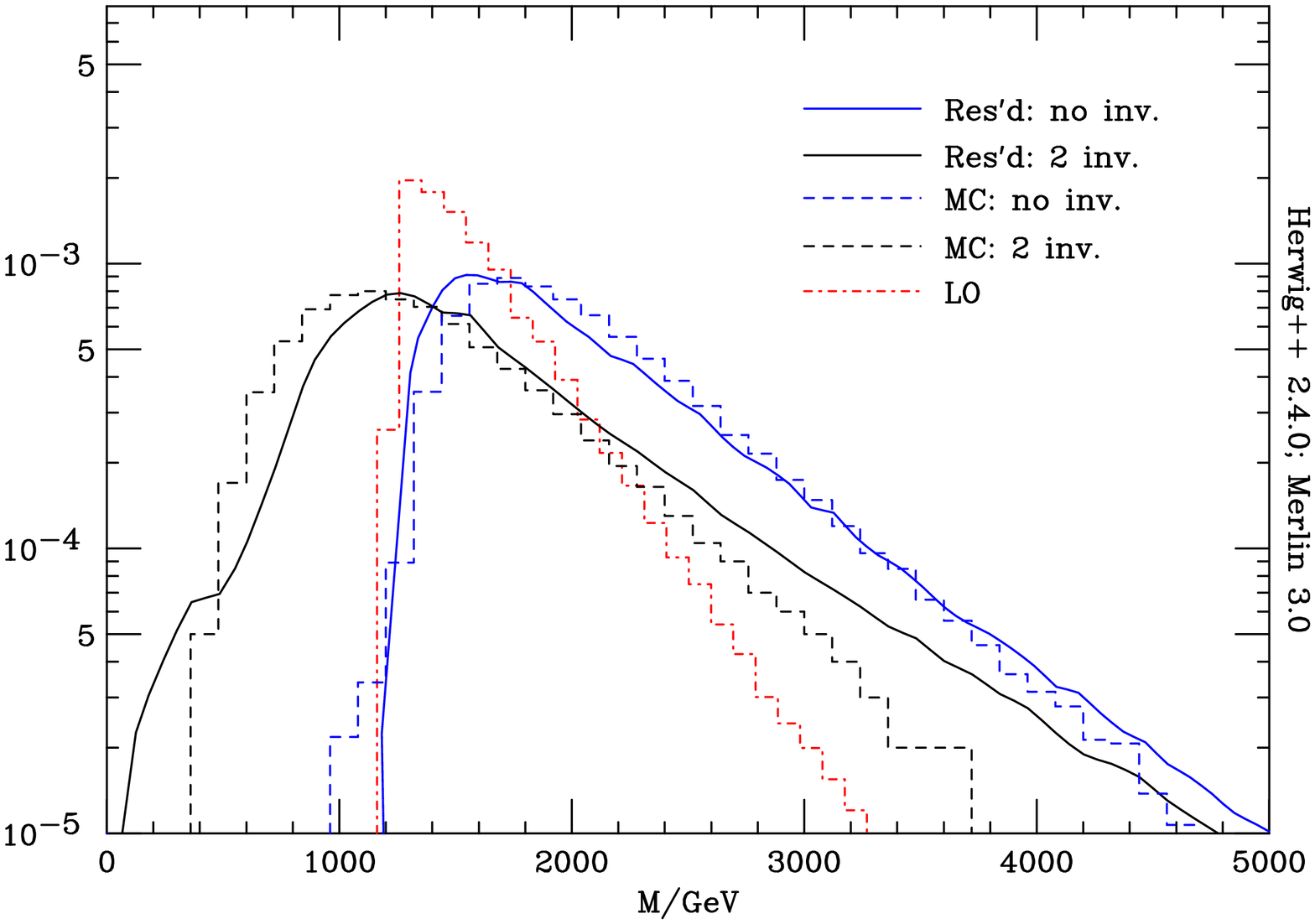}
\caption{The SPS1a gluino pair-production visible mass distributions for pseudorapidity cuts $\etam=5$ (left) and $\etam=3$ (right). The leading-order distribution is shown (red dot-dashes) for comparison.}
\label{fig:gge53}
\end{figure}

\begin{figure}[!htb]
  \centering 
  \vspace{1.2cm}
  \includegraphics[scale=0.34, angle=90]{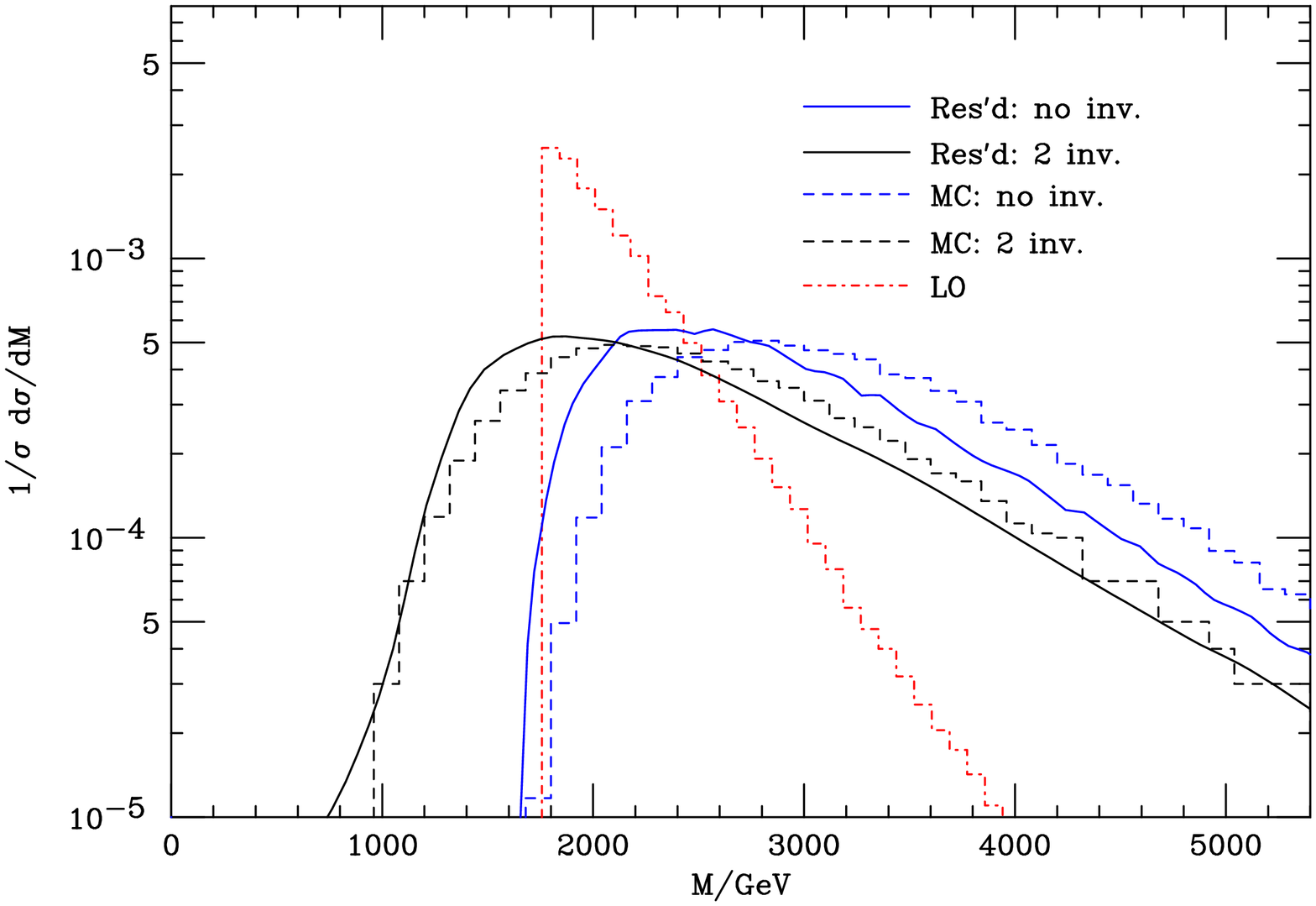}
    \hspace{0.6cm}
    \includegraphics[scale=0.34, angle=90]{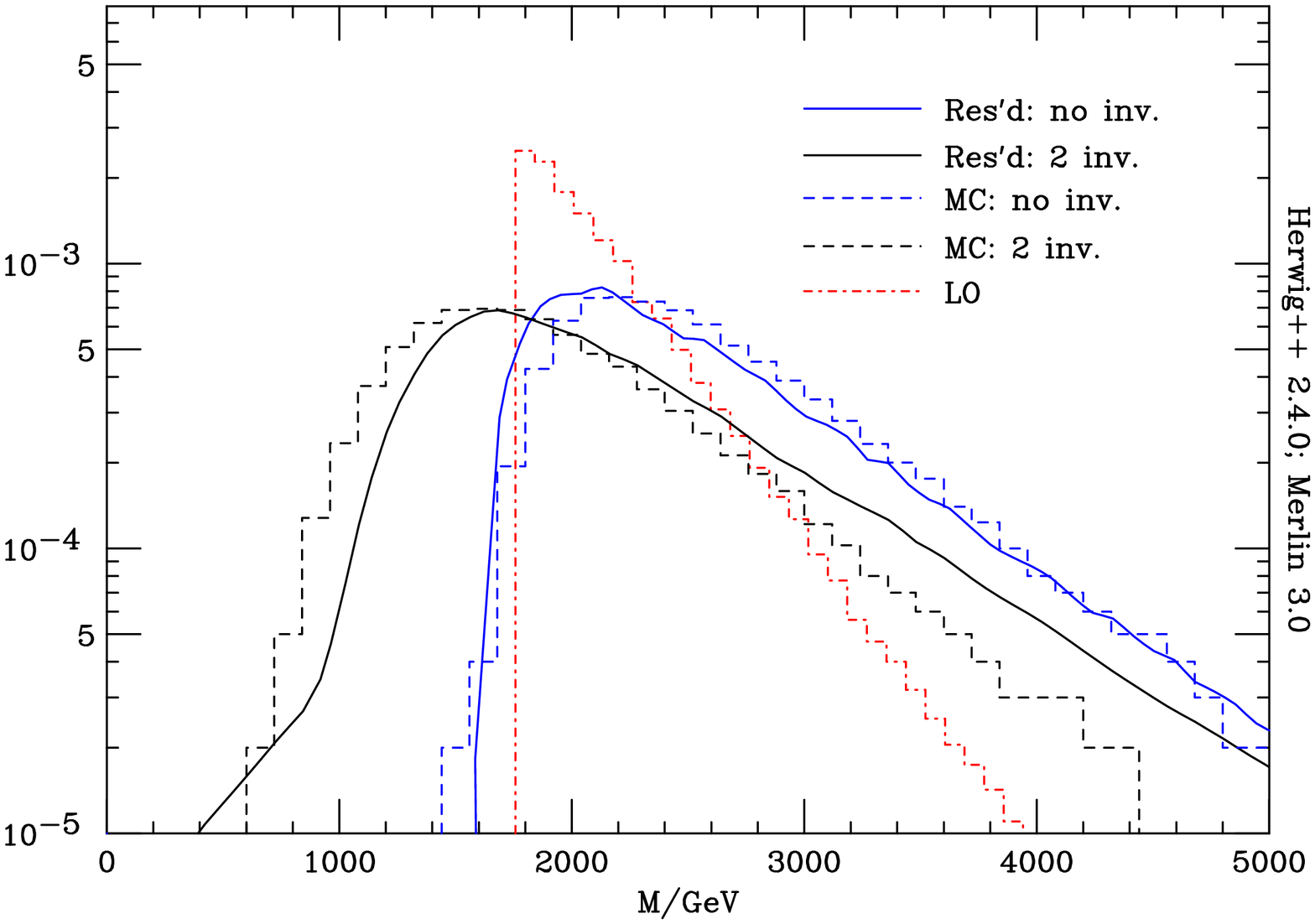}
\caption{The modified SPS1a gluino pair-production (with $m_{\tilde{g}} = 800 ~\mathrm{GeV}$) results for pseudorapidity cuts $\etam=5$ (left) and $\etam=3$ (right). The leading-order distribution is shown (red) for comparison.}
\label{fig:gg800e53}
\end{figure}

We focus on the SPS1a point~\cite{Allanach:2002nj}, which
has gluino and lightest neutralino masses $m_ {\tilde{g}} = 604.5~\mathrm{GeV}$ and $m _{\chi _1 ^0 } = 97.0~\mathrm{GeV}$ respectively (and see
table~\ref{tb:masses} for the squark masses). For simplicity we set
the squark mass in the invisible particle treatment to $550~\mathrm{GeV}$. We
also present results for a modified SPS1a point, with $m_ {\tilde{g}}
= 800 ~\mathrm{GeV}$.  In this process only the two-invisibles case is
realistic, but for comparison we also show results for no invisibles,
i.e. imagining that the two lightest neutralinos are also detected.
\begin{table}[htb]
\begin{center}
\begin{tabular}
{|c|c|c|c|} \hline
Particle& Mass (GeV)& Particle& Mass (GeV) \\ \hline
$\tilde{g}$&604.5&$\tilde{s}_L$&570.7 \\ \hline
$\chi _1 ^0$& 97.0&$\tilde{s}_R$&547.9 \\ \hline
$\tilde{u}_L$&562.3&$\tilde{b}_1$&515.3 \\ \hline
$\tilde{u}_R$&548.2&$\tilde{b}_2$&547.7 \\ \hline
$\tilde{d}_L$&570.7&$\tilde{t}_1$&400.7 \\ \hline
$\tilde{d}_R$&547.9&$\tilde{t}_2$&586.3 \\ \hline
\end{tabular}
\end{center}
\caption{The relevant particle masses in the supersymmetric model used
  in the invisible study, SPS1a. The modified SPS1a point differs in that it has $m_{\tilde{g}} = 800 ~\mathrm{GeV}$.}
\label{tb:masses}
\end{table}
When $\etam = 5,3$, there is fairly good agreement between the Monte Carlo and resummation
predictions in both the two-invisibles and no-invisibles cases, and
for both gluino masses, as shown in Figs.~\ref{fig:gge53}
and~\ref{fig:gg800e53}, where one should compare the dashed histograms
(Monte Carlo) to the solid curves of the same colour (resummation).

The shift in the peak of the visible mass distribution in going from
no  to two invisibles is much larger than that in top pair-production,
amounting to 600-700 GeV, roughly independent of $\etam$ and the
gluino mass.  This results mainly from the higher masses of the
intermediate particles in the decays ($m_{\tilde q}\simeq 550$ GeV
vs. $m_W=80$ GeV), which implies a higher energy release, rather than
the masses of the invisible particles themselves ($m_{\chi_1^0} = 97$
GeV vs. $m_\nu=0$).

\begin{figure}[!htb]
  \centering 
  \vspace{1.5cm}
    \includegraphics[scale=0.35, angle=90]{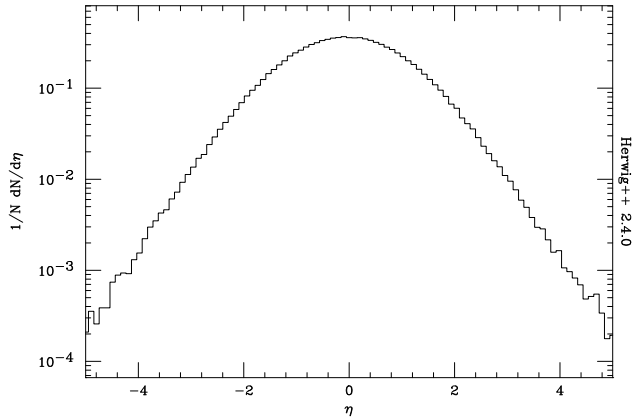}
\caption{The SPS1a gluino pair-production pseudorapidity distribution
  of gluino decay products, for the case $m_{\tilde{g}} = 604.5 ~\mathrm{GeV}$.}
\label{fig:gprodrap}
\end{figure}

\begin{figure}[!htb]
  \centering 
  \vspace{1.2cm}
    \includegraphics[scale=0.34, angle=90]{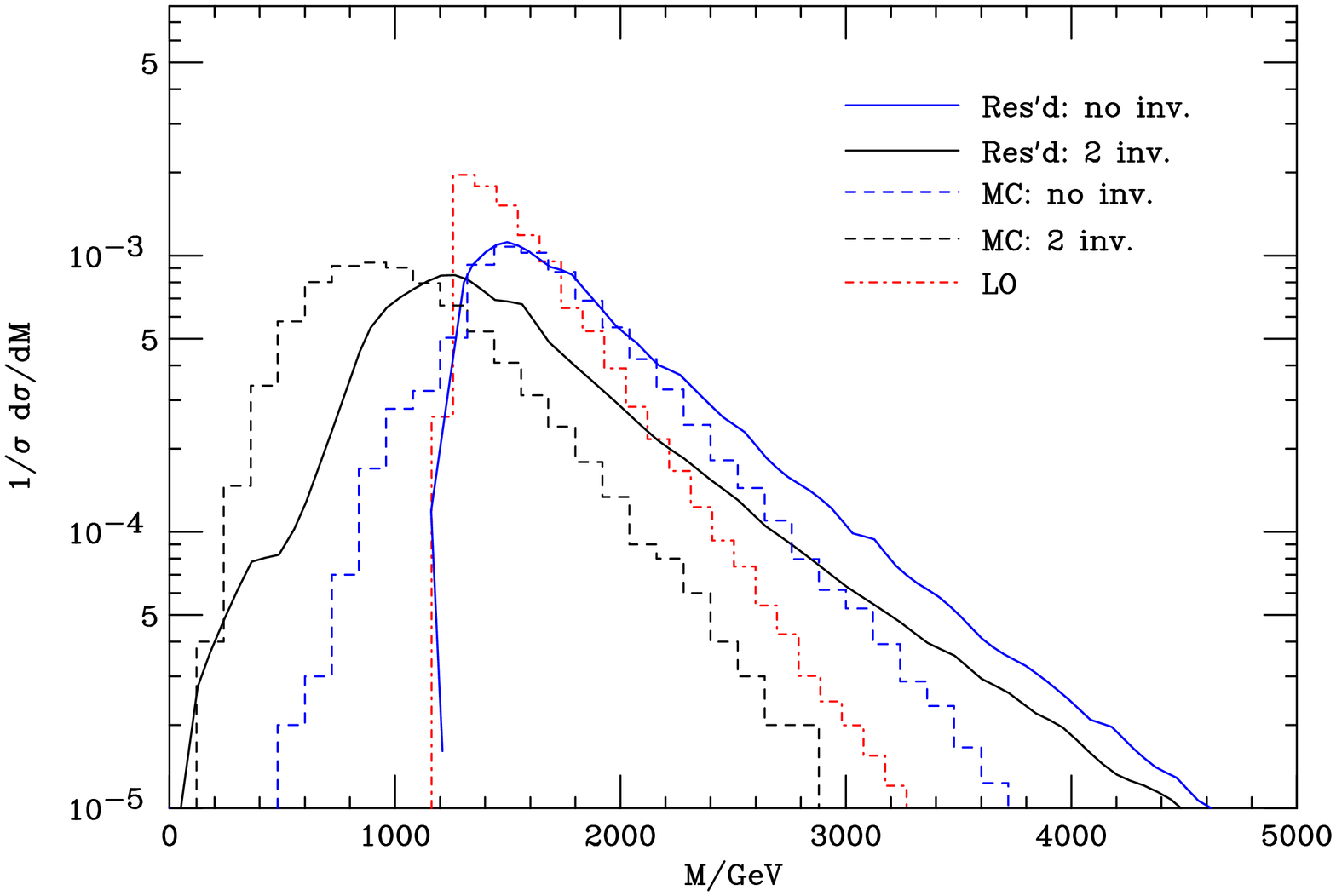}
    \hspace{0.5cm}
    \includegraphics[scale=0.34, angle=90]{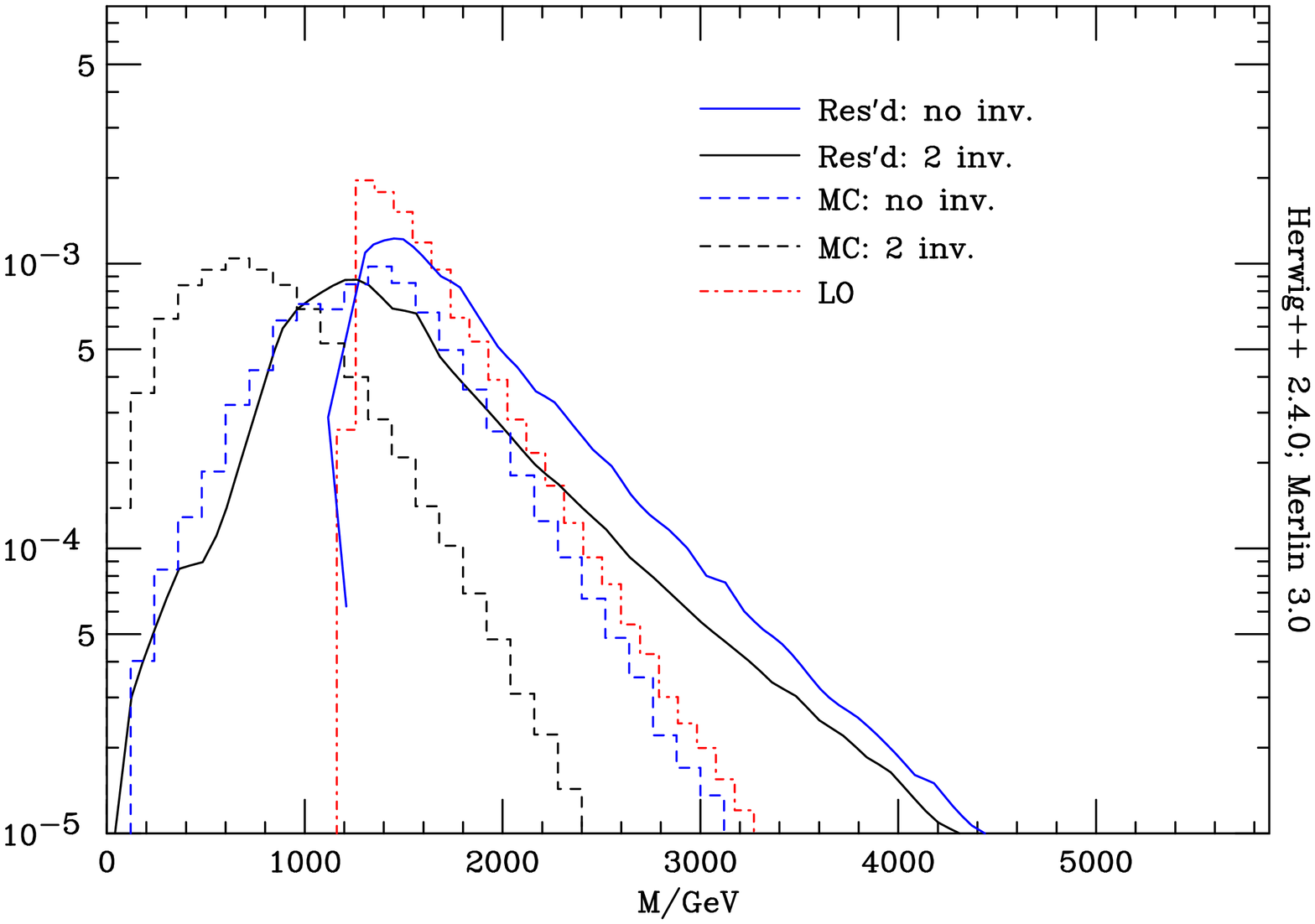}
\caption{The SPS1a gluino pair-production results for pseudorapidity cuts $\etam=2$ (left) and $\etam=1.4$ (right). The leading-order distribution is shown (red dot-dashes) for comparison.}
\label{fig:gge214}
\end{figure}

\begin{figure}[!htb]
  \centering 
  \vspace{1.2cm}
    \includegraphics[scale=0.34, angle=90]{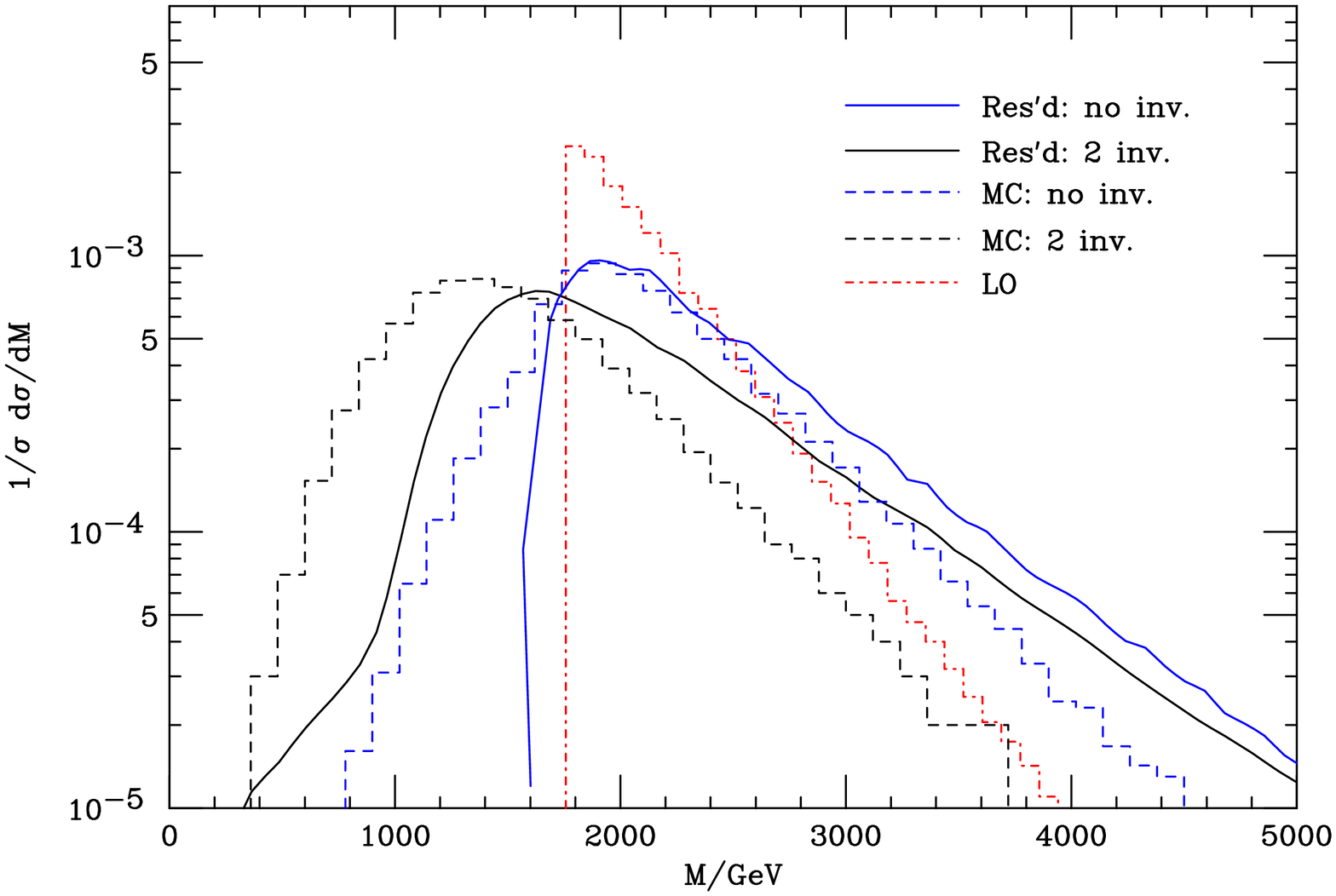}
    \hspace{0.5cm}
    \includegraphics[scale=0.34, angle=90]{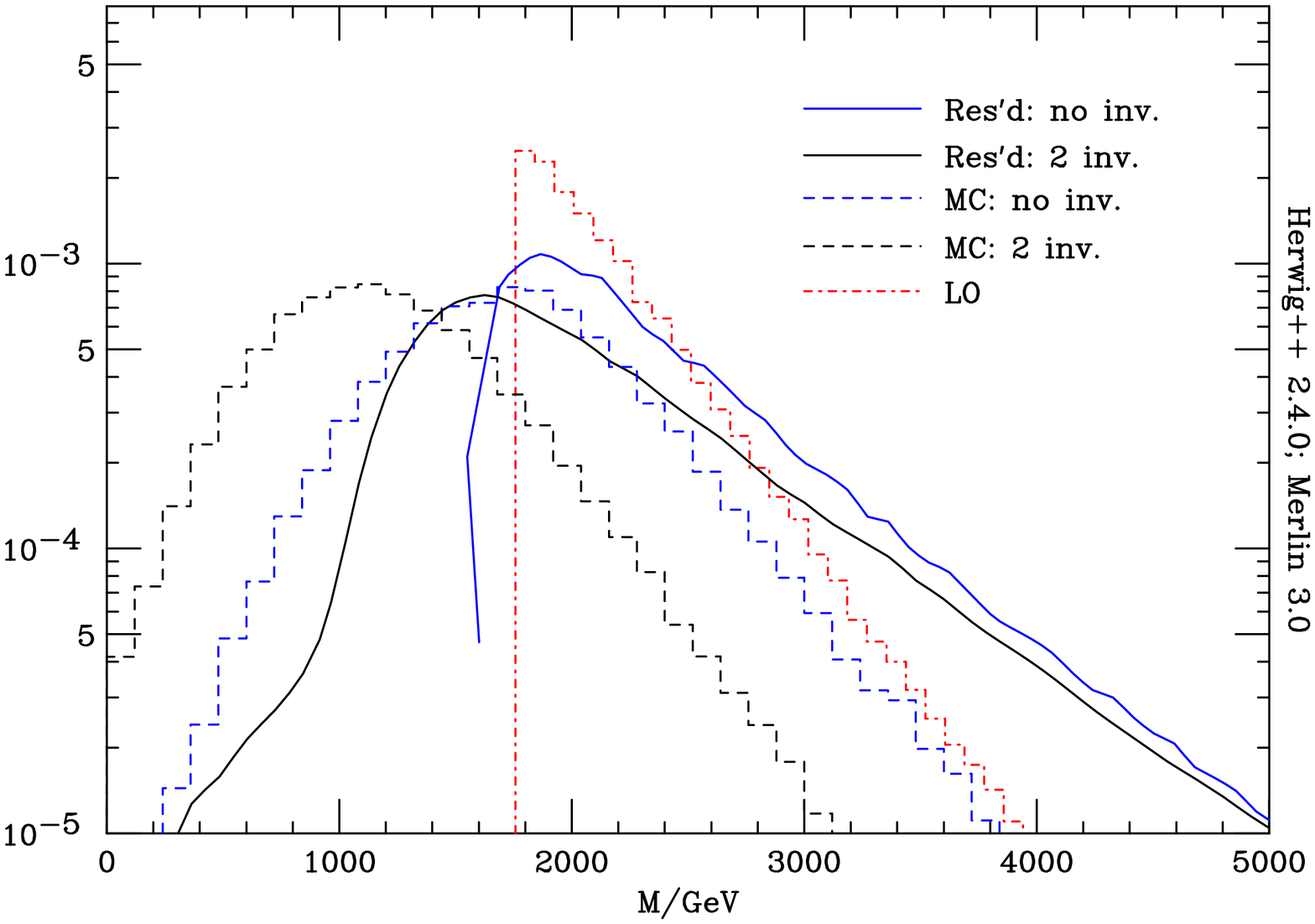}
\caption{The modified SPS1a gluino pair-production (with $m_{\tilde{g}} = 800 ~\mathrm{GeV}$) results for pseudorapidity cuts $\etam=2$ (left) and $\etam=1.4$ (right). The leading-order distribution is shown (red) for comparison.}
\label{fig:gg800e2}
\end{figure}

One of the assumptions of the resummation is that all the visible hard
process decay products are detected, which is not true when the
maximum pseudorapidity $\etam$ is restricted to lower
values. When $\etam \sim 2$ in the Monte Carlo analysis, a significant
number of hard process particles begin to be excluded and hence the
curves shift to lower values compared to the resummed
predictions. Figure~\ref{fig:gprodrap} shows the rapidity distribution
of the decay products of the gluino at parton level for $m_
{\tilde{g}} = 604.5 ~\mathrm{GeV}$. For the case shown, cuts of $\etam = 5,3,2$
and $1.4$ correspond to exclusion of, respectively, $\sim$0.002\%,
1.1\%, 7.5\% and 20.0\% of the gluino decay products from the
detector. The effect of this appears in Figs.~\ref{fig:gge214}
and~\ref{fig:gg800e2}, where the Monte Carlo distributions are
narrower and peak at lower masses than the resummed predictions.  The
variation between the resummed $\etam = 2$ and 1.4 curves is
smaller than that between $\etam = 5$ and 3,
since they correspond to smaller differences in $Q_c$.

The heavy and light gluino scenarios exhibit similar behaviour when
varying the pseudorapidity coverage and the number of invisibles,
showing the lack of dependence of the resummation on the mass of the
pair-produced particle. The sensitivity to low-scale PDF behaviour and
showering is reduced compared to the $t\bar{t}$ case since we are
considering higher centre-of-mass energies, with the lowest possible
$Q_c$ now being of the the order $2 m_{\tilde{g}} \times e^{-5} \sim 8
~\mathrm{GeV}$. The position of the curves is again also sensitive to the precise
definition of $Q_c$ in terms of $\etam$.

\begin{table}[!htb]
\begin{center}
\begin{tabular}
{|c|c|c|c|} \hline
$m_{\tilde{g}}$ (GeV).& $\eta_{\mathrm{max}}$& MC (GeV) (0 inv./2 inv.) & Resum. (GeV) (0 inv./2 inv.) \\ \hline
604.5&5& 2280/1560 & 1785/1620 \\ \hline
604.5&3& 1680/1080& 1593/1204 \\ \hline
604.5&2& 1440/840& 1497/1204\\ \hline
604.5&1.4& 1380/660 & 1497/1204 \\ \hline
800.0&5& 2820/2100& 2569/1870 \\ \hline
800.0&3& 2220/1620& 2128/1684 \\ \hline
800.0&2& 1920/1380& 1865/1683 \\ \hline
800.0&1.4& 1740/1140& 1865/1683 \\ \hline
\end{tabular}
\end{center}
\caption{Summary of the positions of the
  peaks of the gluino pair-production visible mass distributions as
  given by the Monte Carlo and the resummation, for different values
  of the maximum pseudorapidity and for no and two invisibles.}
\label{tb:peaks}
\end{table}

Table~\ref{tb:peaks} shows a summary of the peak
positions for all cases and different pseudorapidity cuts.  For the
higher values of $\etam$, the agreement between the Monte Carlo and
resummation is satisfactory.  There is a large difference in the peak
positions for no invisibles and $\etam=5$, but this is
mainly due to the broad shape of the peak in this case, while the
overall distributions agree better.  For $\etam\leq 2$ there is a
growing discrepancy, especially for the realistic case of two
invisibles, due to the loss of particles coming from the hard process.

\subsubsection{Hadronization effects}\label{sec:hadronization}
We have assumed that ISR partons emitted at pseudorapidities above
$\etam$ do not contribute to the visible invariant mass.  This would
be true if the hadronization process were perfectly local in
angle. However, as a result of  hadronization  high rapidity ISR
partons can produce lower rapidity hadrons and thus `contaminate' the
detector and shift the visible mass to higher values.

\begin{figure}[!htb]
  \centering 
  \vspace{1.0cm}
  \hspace{0.5cm}
    \includegraphics[scale=0.33, angle=90]{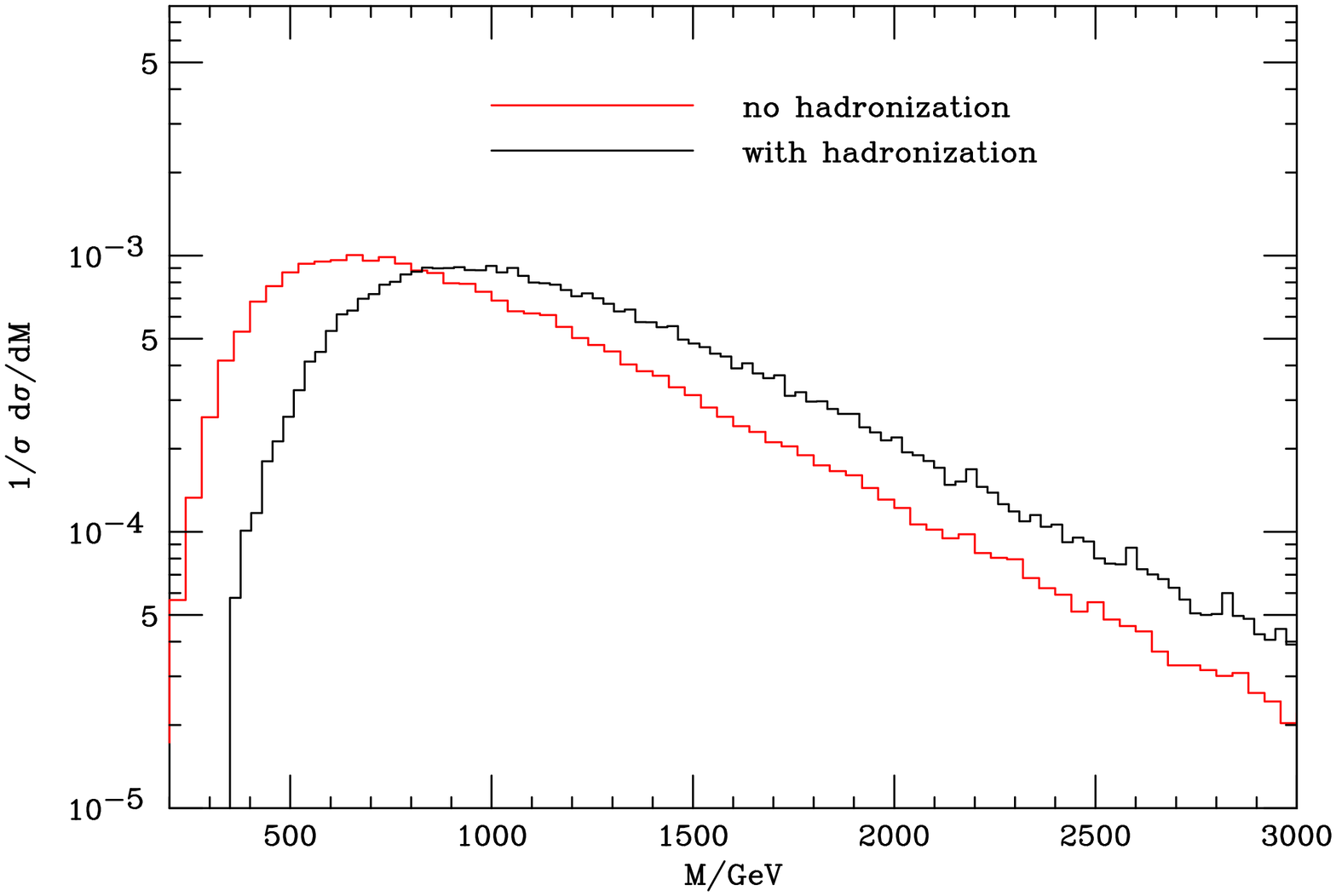}
   \hspace{1.4cm}
    \includegraphics[scale=0.33, angle=90]{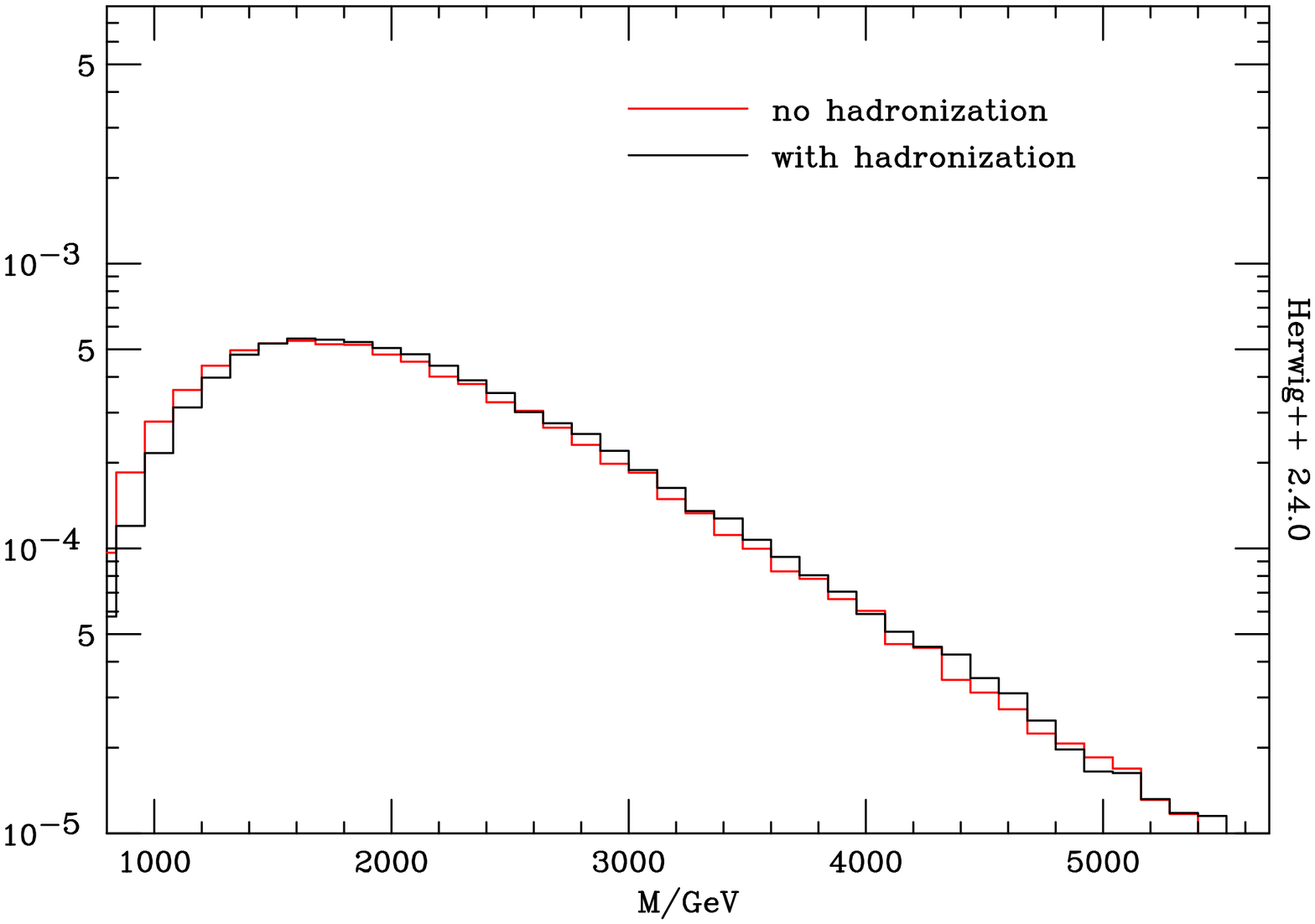}
    \caption{The $t\bar{t}$ fully semi-leptonic (left) and SPS1a gluino pair-production (right, with $m_{\tilde{g}} = 604.5~\mathrm{GeV}$) visible mass distributions for a pseudorapidity cut $\etam=5$ with and without hadronization (black and red respectively).}
\label{fig:hadroniz}
\end{figure}

As we have already discussed in section~\ref{sec:mc:hpp}, the hadronization model employed in the \Herwigpp Monte Carlo
is a refinement of the cluster model. The model involves clustering of partons
into colour-singlet objects that decay into hadrons, resulting in a smearing of the
pseudorapidity distribution which causes the increase in the visible
mass described above.  The effect is shown in Fig.~\ref{fig:hadroniz}
for gluino and top pair-production (excluding the invisible particles
from the hard process). The effect was found to be larger for
$t\bar{t}$ production where the mass distribution is shifted
significantly, whereas in gluino pair-production the shift is
negligible.\footnote{This was found to dependent solely on the mass of
the pair-produced particle, with a similar effect to the gluino case
appearing if the top mass is increased to $\sim 605~\mathrm{GeV}$.}

\subsubsection{Underlying event}\label{sec:MPI}
\begin{figure}[!htb]
  \centering 
  \vspace{1.0cm}
  \hspace{0.5cm}
    \includegraphics[scale=0.33, angle=90]{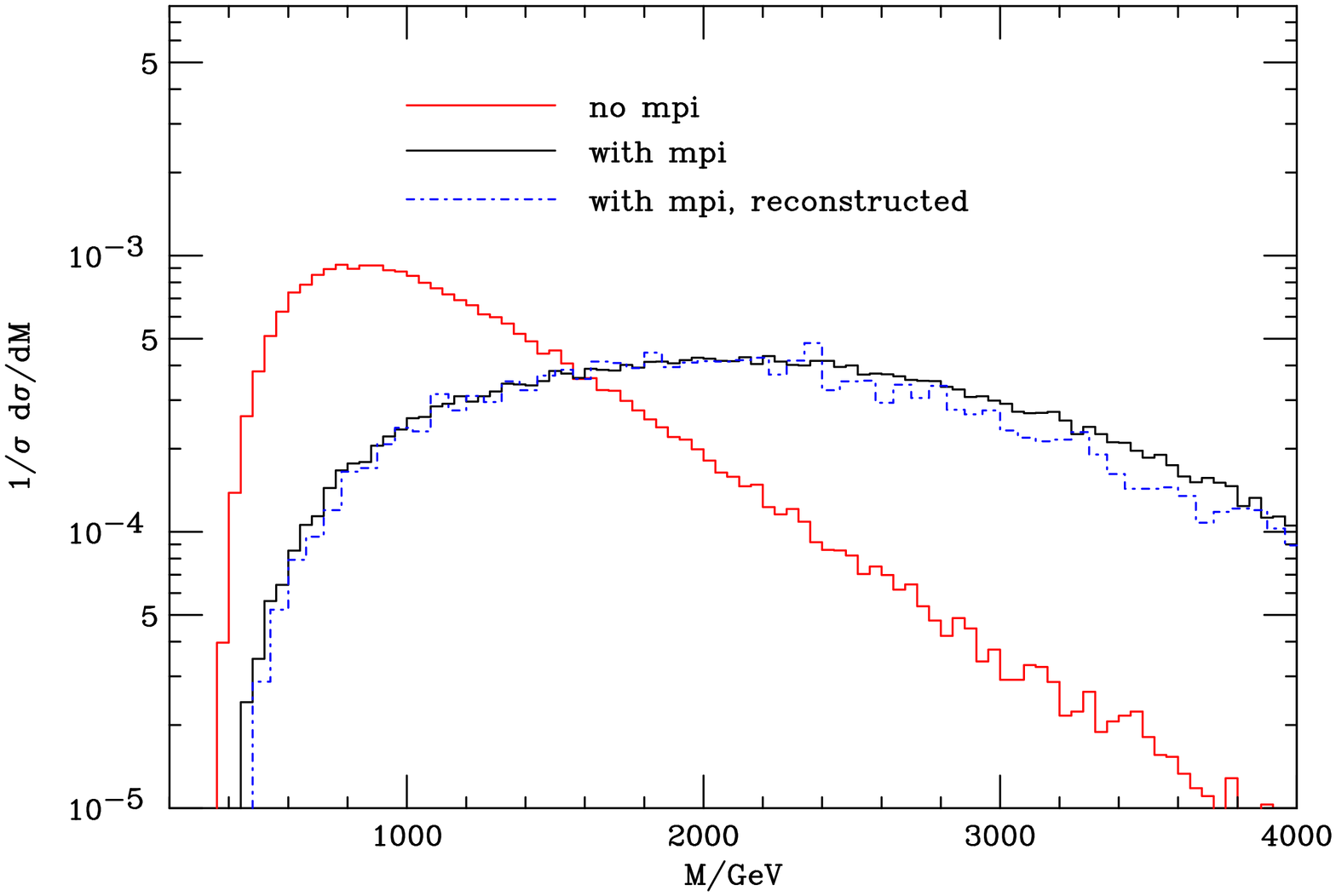}
    \hspace{1.4cm}
    \includegraphics[scale=0.33, angle=90]{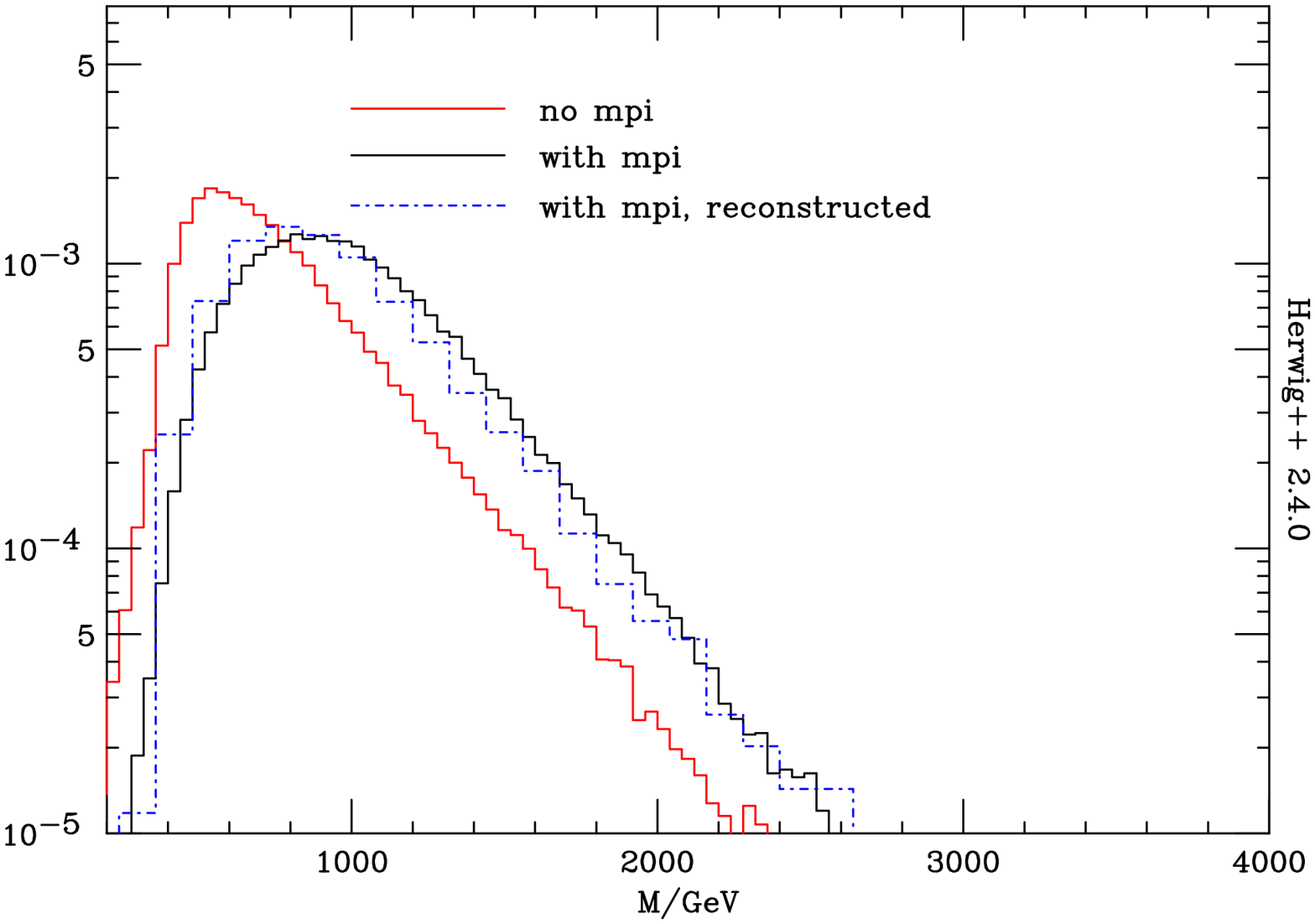}
    \caption{The $t\bar{t}$ fully hadronic visible mass
      distributions for pseudorapidity cuts $\eta_{max}=5$ (left) and
      $\eta_{max}=3$ (right),  with and without multiple parton
      interactions (black and red respectively) and the reconstructed
      curves (blue dot-dashes). The $\etam = 5$ curve was
      reconstructed using the resummed results for the visible mass
      and rapidity, whereas the $\etam = 3$ curve was reconstructed
      using the Monte Carlo visible mass and rapidity.}
\label{fig:mpi_tt}

\end{figure}
\begin{figure}[!htb]
  \centering 
  \vspace{1.0cm}
  \hspace{0.5cm}
    \includegraphics[scale=0.33, angle=90]{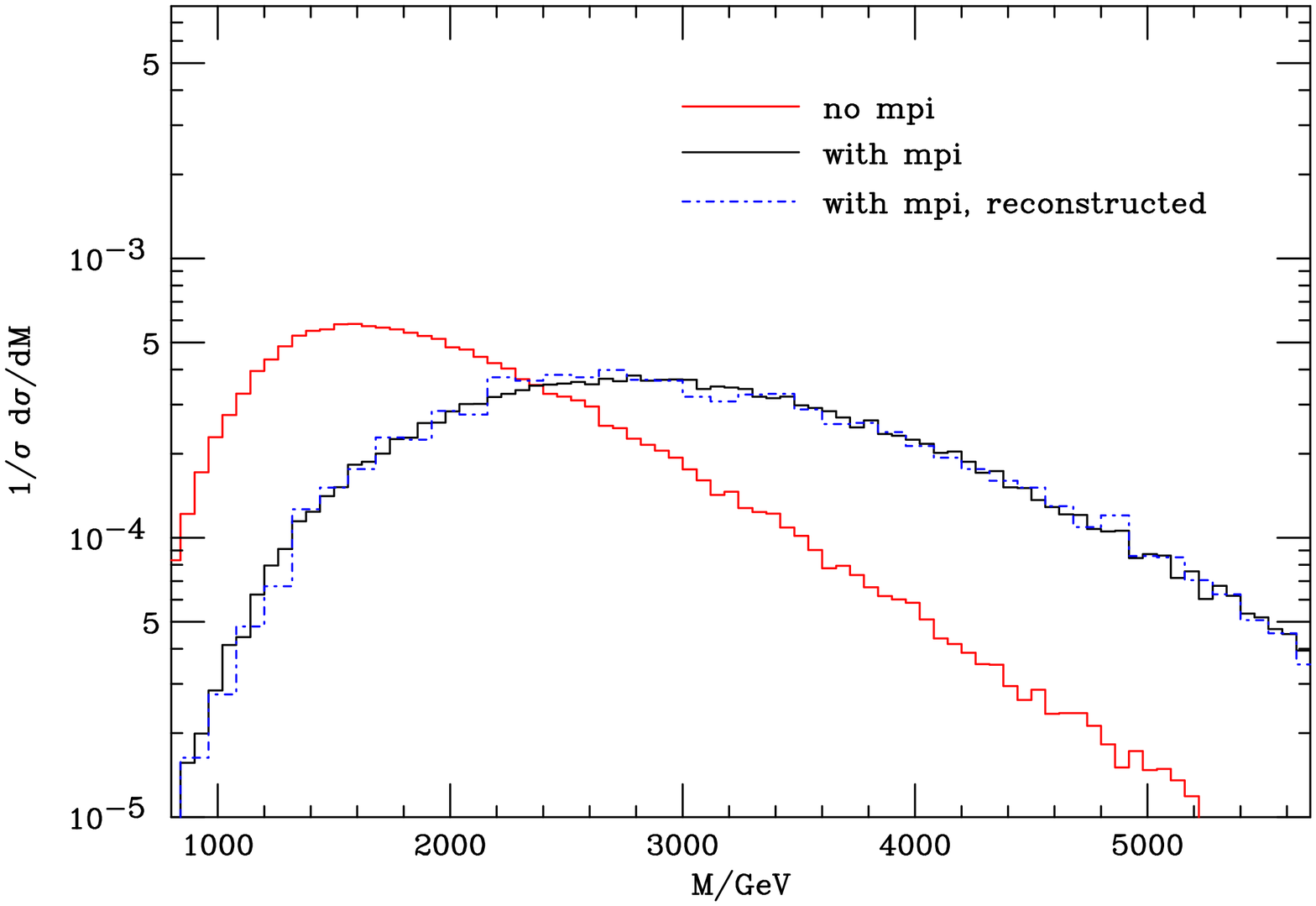}
    \hspace{1.4cm}
    \includegraphics[scale=0.33, angle=90]{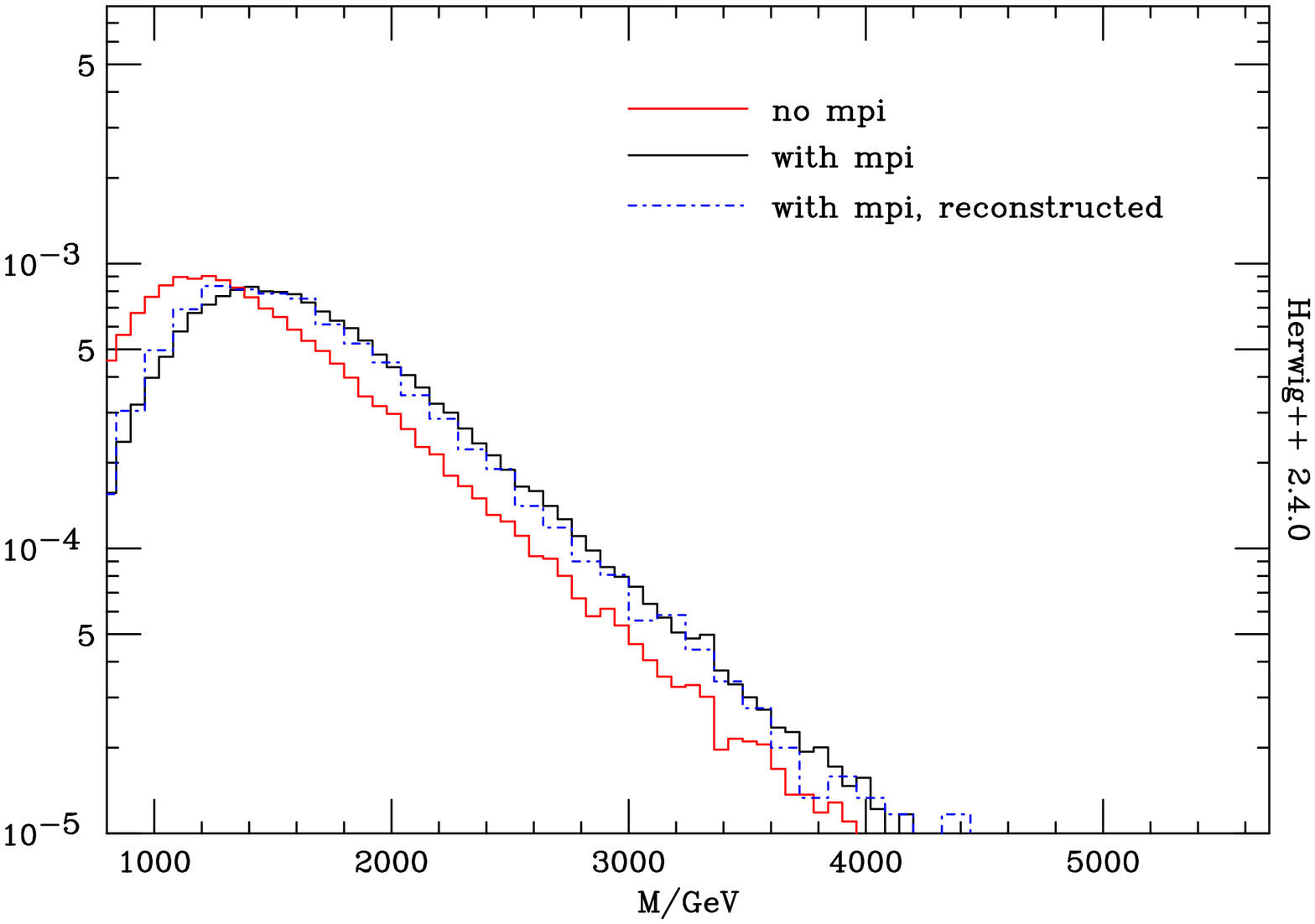}
    \caption{The SPS1a gluino pair-production (with $m_{\tilde{g}} =
      604.5 ~\mathrm{GeV}$) visible mass distributions for pseudorapidity cuts
      $\eta_{max}=5$ (left) and $\eta_{max}=3$ (right),  with and
      without multiple parton interactions (black and red
      respectively) and the reconstructed curves from the Monte Carlo visible masses and rapidities (blue dot-dashes).}
\label{fig:mpi_gg}
\end{figure}

The underlying event, which we have discussed in section~\ref{sec:mc:ue}, is a further source of
non-perturbative contributions to the visible mass.  If $P_H^\mu$
represents the `hard' visible 4-momentum studied in earlier sections
and $P_U^\mu$ represents that due to the underlying event, the total
visible mass is given by
\bea\label{eq:Mtot}
M^2 = (P_H+P_U)^2 &=& M_H^2 + M_U^2 +2 (E_HE_U-P_{zH}
P_{zU})\nonumber\\
&=& M_H^2 + M_U^2 +2 M_U\sqrt{M_H^2+\met^2} \cosh(Y_H-Y_U)\;.
\eea
where we neglect transverse momentum associated with the underlying
event. Thus, even if the visible invariant mass due to the underlying event
is small, its effect on the overall visible mass may be enhanced through
the last term on the right-hand side.

As we have already mentioned, the underlying event is simulated in \Herwigpp by a multiple
parton interaction model. In this model, for the rapidity
ranges considered here, the underlying event is approximately
process-independent and exhibits little correlation
with the rest of the event.  Therefore, to a good approximation, the
distributions of the variables related to the underlying event, $Y_U$
and $M_U$, can be determined once and for all at each collider energy. The
process-dependence comes primarily through the dependence on $Y_H$ and
$M_H$, which can be calculated using the resummation formula given in
Eq.~(\ref{eq:SigmaDefinv}). The overall visible mass distribution can
then be obtained by convolution using Eq.~(\ref{eq:Mtot}).

\begin{figure}[!htb]
  \centering 
  \vspace{1.0cm}
  \hspace{0.5cm}
    \includegraphics[scale=0.33, angle=90]{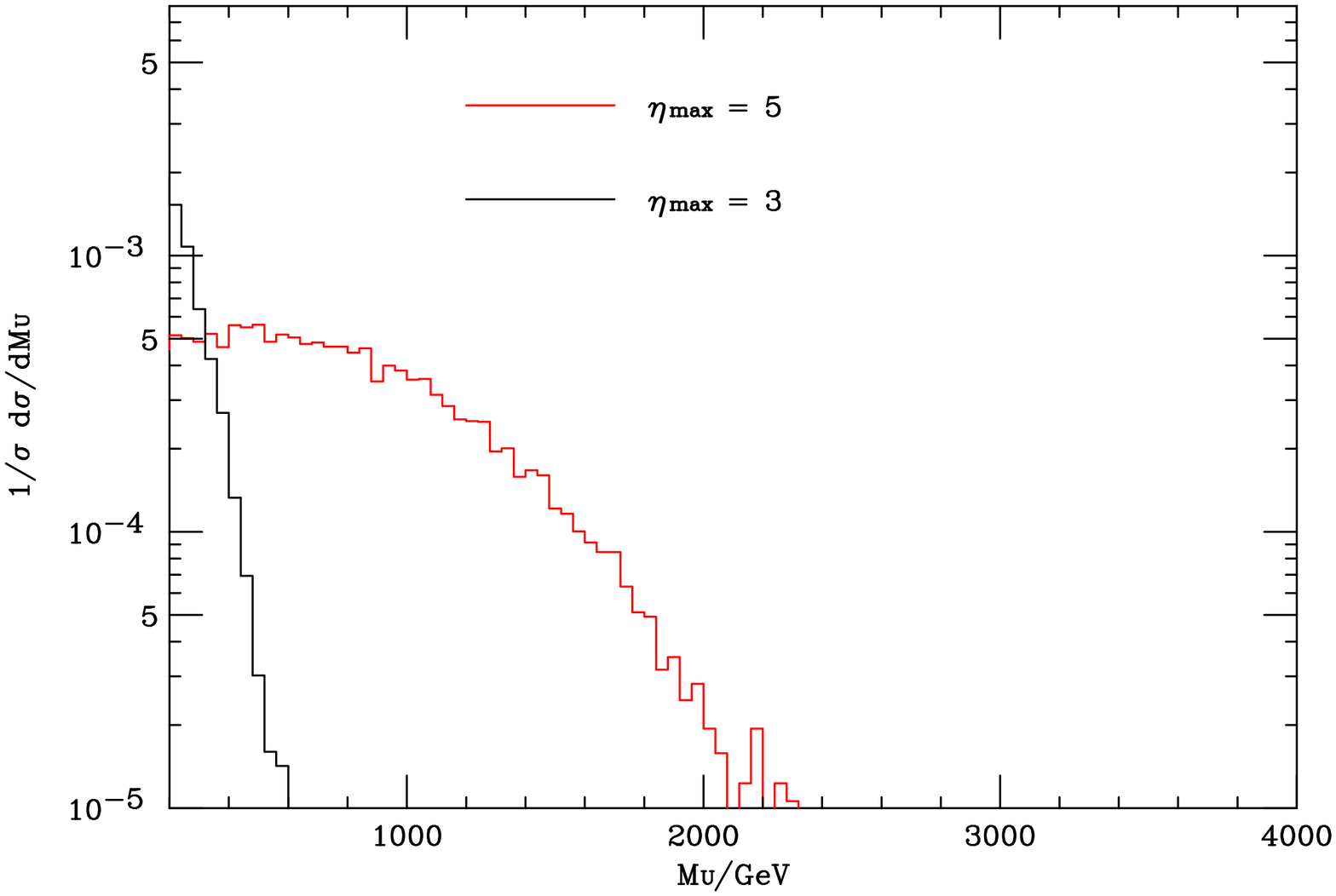}
    \hspace{1.4cm}
    \includegraphics[scale=0.33, angle=90]{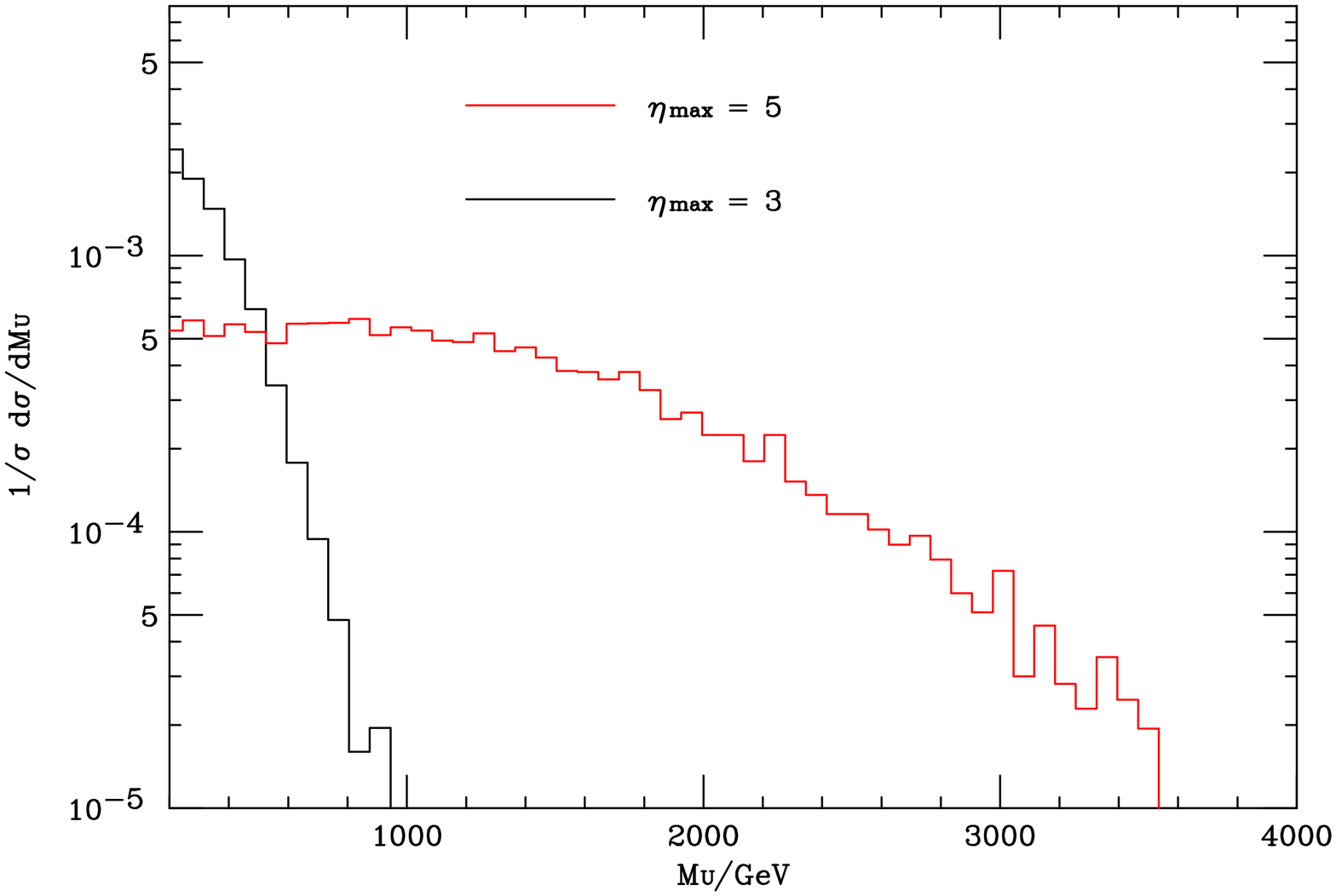}
    \caption{The $t\bar{t}$ (left) and SPS1a gluino pair-production
      (right, with $m_{\tilde{g}} =
      604.5 ~\mathrm{GeV}$) underlying event $M_U$ distributions for pseudorapidity cuts
      $\eta_{max}=5$ (red) and $\eta_{max}=3$ (black).}
\label{fig:mpi_mu}
\end{figure}

The effects of including the underlying event in the visible mass
distribution are shown in Figs.~\ref{fig:mpi_tt} and \ref{fig:mpi_gg}
for $t\bar t$ and gluino pair-production, respectively. The multiple
parton interactions push the peak value to substantially higher
masses.  The shift amounts to about 250 GeV at $\eta_{max}=3$ and 1.2
TeV at $\eta_{max}=5$, and is roughly process-independent.  However,
since the underlying event is approximately uncorrelated with the hard
process, the visible mass distributions can be reconstructed well by
the convolution procedure outlined above, as shown by the blue
dot-dashed curves in Figs.~\ref{fig:mpi_tt} and \ref{fig:mpi_gg}. The 
distributions for the underlying event, $M_U$, used to obtain
Figs.~\ref{fig:mpi_tt} and \ref{fig:mpi_gg}, are shown in Fig.~\ref{fig:mpi_mu}.
These features of the underlying event will need to be validated by
LHC data on a variety of processes.  Accurate modelling of the
underlying event is important for practically all aspects of hadron
collider physics.

\subsection{Conclusions}\label{sec:qcdrad:conc}
We have presented detailed predictions on the total
invariant mass $M$ of the final-state particles registered in a detector,
as a function of its pseudorapidity coverage $\etam$ by considering
the effects of QCD initial-state radiation, first in the
quasi-collinear NLO approximation and then in an all-orders
resummation of the collinear-enhanced terms. This quantity
provides the dominant contribution to many global inclusive observables
such as the new variable $\hat{s}^{1/2}_{\rm min}$ (Eq.~(\ref{eq:smin_def})), which can provide
information on the energy scales of hard processes.
We have extended the resummation method presented to include the effects of invisible
particle emission from the hard process.
We have considered the case of one or two invisible
particles and presented results for Standard Model top quark pair-production and SPS1a gluino pair-production, obtained using a
numerical Mellin moment inversion method.

In the case of $t\bar{t}$ production the invisible particles are
neutrinos from $W$ boson decays and their effect on the visible
invariant mass distribution is small, even when both decays are
leptonic.  This is mainly a consequence of the small $W$ boson mass
compared to the overall invariant mass, rather than the negligible
neutrino mass. For gluino pair-production the invisibles are a pair of
massive LSPs  from squark decays.  The LSP mass is again small
compared to the overall invariant mass, but the squark masses are not,
leading to a substantial downward shift in the visible mass
distribution, of the order of the squark mass. In both cases the
resummed predictions are in fair agreement with Monte Carlo estimates
of the position of the peak in the distribution, provided the
pseudorapidity range covered by the detector is large enough
($\etam\gtap 3$). For $\etam\sim 3$, the difference between the Monte
Carlo prediction and resummed predictions is of the order of 100 GeV
for both the heavy and light gluino SPS1a points. The agreement
becomes worse when the pseudorapidity range is restricted, due to
particle loss from the hard process. Table~\ref{tb:peaks} shows the
positions of the peaks of the distributions for the Monte Carlo
results from \Herwigpp and the resummation.
 
These comparisons were made with
Monte Carlo visible mass distributions at parton
level. We found that non-perturbative effects,
especially the underlying event, tend to shift the invariant mass
distributions to significantly higher values than expected from a
purely perturbative calculation.
 According to the underlying event model used in \Herwigpp,
 the shift amounts to about 250 GeV at $\eta_{\mathrm{max}}=3$ and 1.2
 TeV at $\eta_\mathrm{{max}}=5$.
This effect is also expected in other observables sensitive to longitudinal momentum
components, such as $\hat{s}^{1/2}_{\rm min}$.  However, in the model
found in \Herwigpp version 2.4.x, the underlying event is only weakly correlated with the rest of the
event and hence its effects can be determined once and for all at each
collider energy. The modelling of the underlying event is an
important feature of the Monte Carlo programs that needs to be validated by
comparison with experiment.  Once this has been done, a wide range of
global inclusive observables, including the visible invariant mass,
will be reliably predicted and useful for establishing the scales of
contributing hard subprocesses.

It is important to note that recent UE results from the LHC experiments have shown that
the model present in \Herwigpp 2.4.x does not describe the data
adequately~\cite{Aad:2011qe}. A more recent version of the event
generator, 2.5.0~\cite{Gieseke:2011na}, which includes a
model for colour reconnection, an extension of the cluster model of
hadronization, achieves major improvements in the description of the UE
LHC data.
\section{Resummation of $E_T$ in vector boson and Higgs boson
  production at hadron colliders}
\label{sec:qcdeff:etres}
QCD radiation from incoming partons changes
the distributions of the products of the hard process. This effect has been studied in great detail for the processes of
electroweak boson production, with the result that the transverse
momentum and rapidity distributions of $W$, $Z$ and Higgs bosons at the
Tevatron and LHC are predicted with good
precision.\footnote{See~\cite{Bozzi:2007pn,Bozzi:2008bb,Mantry:2009qz}
  and references therein.} The predictions
for the transverse momentum ($q_T$) distributions in particular include
resummation of terms enhanced at small $q_T$ to all orders in $\as$,
matched with fixed-order calculations at higher $q_T$ values.
The transverse momentum of the boson arises (neglecting
the small intrinsic transverse momenta of the partons in the
colliding hadrons) from its recoil against the transverse momenta of
the radiated partons: $q_T=|\vec q_T|$, where:
\beq\label{eq:qt}
\vec q_T = -\sum_i\vec p_{Ti}\;.
\eeq
The resummation of enhanced terms therefore requires a sum over
emissions $i$ subject to the constraint (\ref{eq:qt}), which is most
conveniently carried out in the transverse space of the impact
parameter $\vec b$ Fourier conjugate to $\vec q_T$:
\beq\label{eq:deltaqt}
\delta(\vec q_T +\sum\vec p_{Ti}) = \frac 1{(2\pi)^2}\int \mrd ^2\vec b\,{\rm e}^{i\vecex
  q_T\cdot\vecex b}\prod_i {\rm e}^{i\vecex p_{Ti}\cdot\vecex b}\;.
\eeq
One then finds that the cumulative distribution in $b=|\vec b|$
contains terms of the form $\as^n\ln^p(Qb)$, where $Q$ is the
scale of the hard process, set in this case by the mass of the
electroweak boson, and $p\leq 2n$.  These terms, which spoil the convergence
of the perturbation series at large $b$, corresponding to small $q_T$,
are found to
exponentiate~\cite{Dokshitzer:1978hw,Parisi:1979se,Curci:1979bg,Bassetto:1979nt,Kodaira:1981nh,Kodaira:1982az,Collins:1984kg}: that is, they can be assembled into an
exponential function of terms that are limited to $p\leq n+1$.
This resummation procedure improves the convergence of the perturbation
series at large values of $b$ and hence allows one to extend predictions of the
$q_T$ distribution to smaller values. 

Together with its vector transverse momentum $\vec p_{Ti}$, every
emission generates a contribution to the total hadronic transverse energy
of the final state, $E_T$, which, neglecting parton masses, is given by
\beq\label{eq:Et}
E_T = \sum_i|\vec p_{Ti}|\;.
\eeq
To first order in $\as$ (0 or 1 emissions) this quantity coincides
with $q_T$, but they differ in higher orders.  In particular, at small
$q_T$ there is the possibility of vectorial cancellation between the
contributions of different emissions, whereas this cannot happen for
the scalar $E_T$.  Thus the distribution of $E_T$ vanishes faster at
the origin, and its peak is pushed to higher values.  To resum these
contributions at small $E_T$, one should perform a one-dimensional
Fourier transformation and work in terms of a `transverse time'
variable $\tau$ conjugate to $E_T$:
\beq\label{eq:deltaEt}
\delta(E_T -\sum|\vec p_{Ti}|) = \frac 1{2\pi}\int \mrd \tau\,{\rm e}^{-iE_T\tau}\prod_i {\rm e}^{i|\vecex p_{Ti}|\tau}\;.
\eeq
Since the matrix elements involved are the same, one finds a similar
pattern of enhanced terms at large $\tau$ as was the case for large
$b$: terms of the form $\as^n\ln^p(Q\tau)$ with $p\leq 2n$, which
arise from an exponential function of terms with $p\leq n+1$.
Evaluation of the exponent to a certain level of precision
(leading-logarithmic, LL, for $p=n+1$, next-to-leading, NLL, for
$p=n$, etc.) resums a corresponding class of enhanced terms and
extends the validity of predictions to lower values of $E_T$.

The resummation of $E_T$ in this way has received little attention
since the first papers on this topic, over 20 years
ago~\cite{Halzen:1982cb,Davies:1983di,Altarelli:1986zk}.  This is surprising, as
most of the effects of QCD radiation from incoming partons mentioned
above depend on this variable rather than $q_T$.  A possible reason is
that, unlike $q_T$, $E_T$ also receives an important contribution from
the underlying event, which is thought to arise from
secondary interactions between spectator partons, as discussed in section~\ref{sec:mc:ue}.  At present this can
only be estimated from Monte Carlo simulations that include multiple
parton interactions (MPI).  Nevertheless it is worthwhile to predict
as accurately as possible the component coming from the primary
interaction, which carries important information about the hard
process.  For example, we expect the $E_T$ distributions in Higgs and
vector boson production to be different, as they involve
primarily gluon-gluon and quark-anti-quark annihilation, respectively.
Accurate estimates of the primary $E_T$ distribution are also
important for improving the modelling of the underlying event.

In the present section we extend the resummation of $E_T$ in vector
boson production to next-to-leading order (NLO) in the resummed exponent,
parton distributions and coefficient functions, and present for the
first time the corresponding predictions for Higgs boson production.  
In section~\ref{sec:etres:method} the resummation procedure is reviewed and extended to
NLO; results on the resummed component are presented in section~\ref{sec:etres:resum}.
This component alone is not expected to describe the region of larger $E_T$
values, of the order of the boson mass; in section~\ref{sec:etres:match} we
describe and apply a simple procedure for including the
unresummed component at order $\as$. Section~\ref{sec:etres:MC} presents
$E_T$ distributions generated using  the parton shower Monte Carlo
programs {\tt HERWIG} version 6.510~\cite{Corcella:2000bw} and \Herwigpp version 2.4.0~\cite{Bahr:2008pv}, which are compared with the
analytical results and used to estimate of the effects of hadronization and
the underlying event.  Our conclusions are summarised in
section~\ref{sec:etres:conc}.  Appendix~\ref{app:etres1} gives mathematical details of a
comparison between the resummation of the transverse energy $E_T$
and transverse momentum $q_T$ and appendix~\ref{app:etres2} shows results for the
LHC at lower centre-of-mass energy ($7\tev$).

\subsection{Resummation method}\label{sec:etres:method}
\subsubsection{General procedure}
Here we generalise the results of Ref.~\cite{Davies:1983di} to NLO resummation.
The resummed component of the transverse energy distribution in the
process $h_1h_2\to FX$ at scale $Q$ takes the form:
\beeq
\label{resgen}
\left[ \frac{\mrd \sigma_{F}}{\mrd Q^2\; \mrd E_T} \right]_{\res} &=& \frac 1{2\pi}\sum_{a,b}
\int_0^1 \mrd x_1 \int_0^1 \mrd x_2 \int_{-\infty}^{+\infty} \mrd\tau \; {\rm e}^{-i\tau E_T} 
\;f_{a/h_1}(x_1,\mu) \; f_{b/h_2}(x_2,\mu) \nn \\
&\times& W_{ab}^{F}(x_1 x_2 s; Q, \tau,\mu)\;,
\eeeq
where $f_{a/h}(x,\mu)$ is the parton distribution function (PDF) of
parton $a$ in hadron $h$ at factorisation scale $\mu$, taken to be
the same as the renormalisation scale here.  In what follows we use
the \ms\ renormalisation scheme.   As mentioned earlier,  to take into
account the constraint that the transverse energies of emitted partons
should sum to $E_T$,  the resummation procedure is carried out in the
domain that is Fourier conjugate to $E_T$, using
Eq.~(\ref{eq:deltaEt}).  The transverse energy distribution
(\ref{resgen}) is thus obtained by performing the inverse Fourier
transformation with respect to the transverse time, $\tau$.  The
factor $W_{ab}^{F}$ is the perturbative and  process-dependent
partonic cross section that embodies the all-order  resummation of the
large logarithms $\ln (Q\tau)$.  Since $\tau$ is  conjugate to $E_T$,
the limit $E_T\ll Q$ corresponds to $Q\tau \gg 1$.

As in the case of transverse momentum resummation~\cite{Catani:2000vq},
the resummed partonic cross section can be written in the
following universal form:
\beeq
\label{eq:Wab}
W_{ab}^{F}(s; Q, \tau,\mu) &=& \sum_c \int_0^1 dz_1 \int_0^1 \mrd z_2 
\; C_{ca}(\as(\mu), z_1;\tau,\mu) \; C_{{\bar c}b}(\as(\mu), z_2;\tau,\mu)
\; \delta(Q^2 - z_1 z_2 s) \nn\\
&\times& \sigma_{c{\bar c}}^F(Q,\as(Q)) \;S_c(Q,\tau) \;.
\eeeq
Here $\sigma_{c{\bar c}}^F$ is the cross section
for the partonic subprocess
$c + {\bar c} \to F$, where $c,{\bar c}=q,{\bar q}$ (the quark $q_f$ and
the anti-quark ${\bar q}_{f'}$ can possibly have different flavours $f,f'$)
or $c,{\bar c}=g,g$. The term $S_c(Q,\tau)$ is the quark $(c=q)$ or
gluon $(c=g)$ Sudakov form factor.  In the case of $E_T$ resummation,
this takes the form~\cite{Davies:1983di,Altarelli:1986zk}:
\beq
\label{formfact}
S_c(Q,\tau) = \exp \left\{-2\int_0^Q \frac{\mrd q}q 
\left[ 2A_c(\as(q)) \;\ln \frac{Q}{q} + B_c(\as(q)) \right] 
\left(1-{\rm e}^{iq\tau}\right)\right\} \;, 
\eeq
with $c=q$ or $g$. The functions $A_c(\as), B_c(\as)$, as well as the 
coefficient functions $C_{ab}$ in Eq.~(\ref{eq:Wab}), contain no
$\ln (Q\tau)$ terms and are perturbatively computable as power
expansions with constant coefficients:
\beeq
\label{aexp}
A_c(\as) &=& \sum_{n=1}^\infty \left( \frac{\as}{\pi} \right)^n A_c^{(n)} 
\;, \\
\label{bexp}
B_c(\as) &= &\sum_{n=1}^\infty \left( \frac{\as}{\pi} \right)^n B_c^{(n)}
\;, \\
\label{cexp}
C_{ab}(\as,z) &=& \delta_{ab} \,\delta(1-z) + 
\sum_{n=1}^\infty \left( \frac{\as}{\pi} \right)^n C_{ab}^{(n)}(z) \;.
\eeeq
Thus a calculation to NLO in $\as$ involves the coefficients
$A_c^{(1)}$,  $A_c^{(2)}$,  $B_c^{(1)}$,  $B_c^{(2)}$ and  $C_{ab}^{(1)}$.
All these quantities are known for both the quark and gluon form
factors and associated coefficient functions. Knowledge of the
coefficients $A^{(1)}$ leads to the resummation of
the leading logarithmic (LL) contributions at small $E_T$,
which in the differential distribution are of the form
$\as^n\ln^p(Q/E_T)/E_T$ where $p=2n-1$.
The coefficients $B^{(1)}$ give the next-to-leading logarithmic (NLL)
terms with $p=2n-2$, $A^{(2)}$ and $C^{(1)}$ give the 
next-to-next-to-leading logarithmic (N$^2$LL) terms with $p=2n-3$,
and $B^{(2)}$ gives the N$^3$LL terms with $p=2n-4$. 
With knowledge of all these terms, the first term neglected
in the resummed part of the distribution is of order
$\as^3\ln(Q/E_T)/E_T$.

In general the coefficient functions in Eq.~(\ref{eq:Wab}) contain logarithms of
$\mu\tau$, which are eliminated by a suitable choice of factorisation
scale. To find the optimal factorisation scale, we note that, to NLL accuracy,
\beq\label{eq:nllint}
\int_0^Q\frac{\mrd q}q \ln^p q \left(1-{\rm e}^{iq\tau}\right)
\simeq \int_{i\tau_0/\tau}^Q\frac{\mrd q}q \ln^p q \;,
\eeq
where $\tau_0 = \exp(-\gamma_E)=0.56146\ldots$, $\gamma_E$ being the
Euler-Mascheroni constant. See appendix~\ref{app:etres1} for a derivation. The effective lower limit of the
soft resummation becomes $i\tau_0/\tau$, and the parton distributions and
coefficient functions should be evaluated at this scale.  However,
evaluation of parton distribution functions at an imaginary scale
using the standard parametrizations is not feasible.  We avoid this by
noting that
\beq
 f_{a/h}(x,q') = \sum_b\int_x^1 \frac{ \mrd z}z K_{ab}(z;q',q) f_{b/h}(x/z,q)\;,
\eeq
where $K_{ab}$ is the DGLAP evolution kernel, also used in
section~\ref{sec:qcdrad:resumm} (see, e.g. Eq.~(\ref{eq:qcdrad:evolkernel})).  Therefore:
\beq\label{eq:fabi}
 f_{a/h}(x,i\mu) = \int_x^1 \frac{\mathrm{d}z}z K_{ab}(z;i\mu,\mu) f_{b/h}(x/z,\mu)\;,
\eeq
where the evolution kernel $K_{ab}(z;i\mu,\mu)$ is given to NLO by
\beq\label{eq:Uabi}
K_{ab}(z;i\mu,\mu) = \delta_{ab} + \frac i2 \as(\mu)\,P_{ab}(z)\;,
\eeq
where $P_{ab}(z)$ is the leading-order DGLAP splitting function.
Similarly, in the coefficient functions we can write $\as(i\mu)$ in terms of  $\as(\mu)$
using the definition of the running coupling, given in section~\ref{sec:qcd:renorm}:
\beq
\int_{\mu}^{i\mu}\frac{\mrd \as}{\beta(\as)} =
2\int_{\mu}^{i\mu}\frac{\mrd q}q = i\pi\;,
\eeq
where $\beta(\as)=-b\as^2 +{\cal O}(\as^3)$, so that
\beq
\as(i\mu) = \as(\mu) -i\pi b[\as(\mu)]^2 +{\cal O}(\as^3)\;.
\eeq
Furthermore, as the expressions (\ref{resgen}) and (\ref{eq:Wab}) are
convolutions, we can transfer the extra terms from (\ref{eq:fabi})
into the coefficient functions to obtain
\beeq
\label{eq:Wfin}
W_{ab}^{F}(s; Q, \tau) &=& \sum_c \int_0^1 \mrd z_1 \int_0^1 \mrd z_2 
\; \tC_{ca}(\as(\tau_0/\tau), z_1) \; \tC_{{\bar c}b}(\as(\tau_0/\tau), z_2)
\; \delta(Q^2 - z_1 z_2 s) \nn\\
&\times& \sigma_{c{\bar c}}^F(Q,\as(Q)) \;S_c(Q,\tau)\;,
\eeeq
where
\beq\label{eq:tCca}
\tC_{ca}(\as(\mu),z) = \sum_d \int_z^1 \frac{\mrd z'}{z'}C_{cd}(\as(i\mu),z/z')\,K_{da}(z';i\mu,\mu)\;.
\eeq
Now the lowest-order coefficient function is of the form:
\beq\label{eq:tC0}
\tC_{ca}^{(0)}(z) = C_{ca}^{(0)}(z) = \delta_{ca}\delta(1-z)\;,
\eeq
and therefore
\beq\label{eq:tC1}
\tC_{ca}^{(1)}(z) = C_{ca}^{(1)}(z) +i\frac{\pi}2 P_{ca}(z)\;.
\eeq

Putting everything together, we have
\beq\label{eq:resSR}
\left[ \frac{\mrd\sigma_{F}}{\mrd Q^2\;\mrd E_T} \right]_{\res}  = \frac 1{2\pi s}\sum_{c}
\int_{-\infty}^{+\infty} \mrd\tau \; {\rm e}^{-i\tau E_T} S_c(Q,\tau)\;R_c(s;Q,\tau)
\;\sigma_{c{\bar c}}^F(Q,\as(Q))\;,
\eeq
where, taking all PDFs and coefficient functions to be evaluated at scale $\mu=\tau_0/\tau$,
\beq\label{eq:Rcdef}
R_c(s;Q,\tau) =\sum_{a,b}
\int_0^1 \frac{\mrd x_1}{x_1} \frac{\mrd x_2}{x_2} \frac{\mrd z_1}{z_1} 
f_{a/h_1}(x_1)\,f_{b/h_2}(x_2)\,\tC_{ca}(z_1)\,\tC_{\bar cb}\left(\frac{Q^2}{z_1x_1x_2 s}\right)\;.
\eeq
To write (\ref{eq:resSR}) as an integral over $\tau>0$ only, we note from (\ref{eq:fabi}) and (\ref{eq:Uabi})
that when $\tau\to -\tau$, to NLO the real parts of $f_{a/h_1}$ and $f_{b/h_2}$ are unchanged but
the imaginary parts change sign.  All other changes in (\ref{eq:Rcdef}) are beyond NLO.  Thus, writing
\beq
R_c= R_c^{(R)}  + iR_c^{(I)}\;,
\eeq
$R_c^{(R)}$ is symmetric with respect to $\tau$ and $R_c^{(I)}$ is antisymmetric.
Defining
\beeq\label{eq:Fcs}
F_c^{(R)}(Q,\tau) &=& 2\int_0^Q \frac{\mrd q}q 
\left[ 2A_c(\as(q)) \;\ln \frac{Q}{q} + B_c(\as(q)) \right] 
\left(1-\cos q\tau\right)\;,\nn\\
F_c^{(I)}(Q,\tau) &=& 2\int_0^Q \frac{\mrd q}q 
\left[ 2A_c(\as(q)) \;\ln \frac{Q}{q} + B_c(\as(q)) \right]
\sin q\tau\;,
\eeeq
we therefore obtain
\beeq\label{eq:resF}
\left[ \frac{d\sigma_F}{\mrd Q^2\;dE_T} \right]_{\res} &=&
\frac 1{\pi s}\sum_{c} \int_{0}^{\infty}
 \mrd \tau \; {\rm e}^{-F_c^{(R)}(Q,\tau)} \Bigl[
 R_c^{(R)}(s;Q,\tau)\cos\{F_c^{(I)}(Q,\tau)-\tau E_T\}\nn\\
&&-R_c^{(I)}(s;Q,\tau)\sin\{F_c^{(I)}(Q,\tau)-\tau E_T\}\Bigr]
\;\sigma_{c{\bar c}}^F(Q,\as(Q))\;,
\eeeq
where, inserting (\ref{eq:tC0}) and (\ref{eq:tC1}) in (\ref{eq:Rcdef})
and defining $\xi=Q^2/s$, we have to NLO,
\beeq\label{eq:RcNLO}
&&R_c^{(R)}(s;Q,\tau) = R_c^{(R)}(\xi=Q^2/s,\tau) \nn\\
&&=\int \frac{\mrd x_1}{x_1} \frac{\mrd x_2}{x_2}
\Bigl\{f_{c/h_1}(x_1)f_{\bar c/h_2}(x_2)+\frac{\as}{\pi}\sum_a\Bigl[
f_{a/h_1}(x_1)f_{\bar c/h_2}(x_2)C_{ca}^{(1)}\left(\frac{\xi}{x_1x_2}\right)\nn\\
&&\quad +f_{c/h_1}(x_1)f_{a/h_2}(x_2)C_{\bar ca}^{(1)}\left(\frac{\xi}{x_1x_2}\right)
\Bigr]\Bigr\}\;,\nn\\
&&R_c^{(I)}(s;Q,\tau) = R_c^{(I)}(\xi=Q^2/s,\tau) \nn\\
&&=\frac{\as}2\sum_a\int \frac{\mrd x_1}{x_1} \frac{\mrd x_2}{x_2}
\Bigl[f_{a/h_1}(x_1)f_{\bar c/h_2}(x_2)P_{ca}\left(\frac{\xi}{x_1x_2}\right)\nn\\
&&\quad +f_{c/h_1}(x_1)f_{a/h_2}(x_2)P_{\bar ca}\left(\frac{\xi}{x_1x_2}\right)
\Bigr]\;.
\eeeq
It will be more useful to write, for example,
\beeq\label{eq:ffPz}
&&\quad \int \frac{\mrd x_1}{x_1} \frac{\mrd x_2}{x_2}
f_{a/h_1}(x_1)f_{\bar c/h_2}(x_2)P_{ca}\left(\frac{\xi}{x_1x_2}\right)\nn\\
&&=\int \frac{\mrd x_1}{x_1} \frac{\mrd x_2}{x_2}\mrd z\,\delta\left(z-\frac{\xi}{x_1x_2}\right)
f_{a/h_1}(x_1)f_{\bar c/h_2}(x_2)P_{ca}(z)\nn\\
&&=\int \frac{\mrd x_1}{x_1} \frac{\mrd z}{z} f_{a/h_1}(x_1)f_{\bar c/h_2}\left(\frac{\xi}{zx_1}\right)P_{ca}(z)\;.
\eeeq
This makes it more straightforward to interpret the plus prescription, which appears in some splitting functions, as
\beeq\label{eq:ffPplus}
&& \int \frac{\mrd x_1}{x_1} \frac{\mrd z}{z} f_{a/h_1}(x_1)f_{\bar c/h_2}\left(\frac{\xi}{zx_1}\right)P(z)_+\nn\\
&=&\int \frac{\mrd x_1}{x_1} f_{a/h_1}(x_1)\int_0^1 \mrd z\left[\frac 1z f_{\bar c/h_2}\left(\frac{\xi}
{zx_1}\right) - f_{\bar c/h_2}\left(\frac{\xi}{x_1}\right) \right] P(z)\nn\\
&=&\int_{\xi}^1 \frac{\mrd x_1}{x_1} f_{a/h_1}(x_1)\int_{\xi/x_1}^1
\mrd z\,\left[\frac 1z
f_{\bar c/h_2}\left(\frac{\xi}{zx_1}\right) - f_{\bar c/h_2}\left(\frac{\xi}{x_1}\right) \right] P(z)\nn\\
&&\quad -\int_{\xi}^1 \frac{\mrd x_1}{x_1} f_{a/h_1}(x_1)f_{\bar
  c/h_2}\left(\frac{\xi}{x_1}\right) \int_0^{\xi/x_1} \mrd z\,P(z)\;.
\eeeq

We show in appendix~\ref{app:etres1} that the results of resummation of the scalar
transverse energy are identical  to those of the more familiar
resummation of vector transverse momentum at order $\as$, as they
should be since at most one parton is emitted at this order.

The transverse energy computed here is the resummed component of
hadronic initial-state radiation integrated over the full range of
pseudorapidities $\eta$.  In Ref.~\cite{Davies:1983di} the $E_T$ distribution
of radiation emitted in a restricted rapidity range $|\eta|< \etamax$ was
also estimated.  This was done by replacing the lower limit of
integration in Eqs.~(\ref{eq:Fcs}) by $Q_c=Q\exp(-\etamax)$, i.e.\
assuming that radiation at $q<Q_c$ does not enter the detected region.
This is justified at the leading-logarithmic level, where
$q/Q\sim\theta\sim\exp(-\eta)$ and the
scale dependence of the parton distributions and
coefficient functions in Eq.~(\ref{eq:Rcdef}) can be neglected.
Then when $\etamax=0$ the form factor $S_c$ is replaced by unity
and Eq.~(\ref{eq:resSR}) correctly predicts a delta function at
$E_T=0$ times the Born cross section.  However, this simple
prescription cannot be correct at the NLO level, where the $\tau$-dependence of the scale must be taken into account.  Therefore we do
not consider the $E_T$ distribution in a restricted rapidity range here.

It is worth noting at this point, an existing related resummation of a variable
called the `beam thrust', essentially defined as the $E_T$ weighted by
$\exp( -\eta )$~\cite{Stewart:2010pd}.

\subsubsection{Vector boson production}
One of the best studied examples of resummation is in vector boson
production through the partonic subprocess $q + \bar q'\to V$
($V=W$ or $Z$):
\beq\label{eq:DYLO}
\sigma_{c{\bar c}}^F(Q,\as(Q)) = \delta_{cq}\delta_{\bar c\bar q'}\delta(Q^2-M_V^2)\sigma_{qq'}^V\;,
\eeq
where, at lowest order,
\beeq
\sigma_{qq'}^W &=& \frac \pi 3\sqrt 2 G_F M_W^2 |V_{qq'}|^2\;,\nn\\
\sigma_{qq'}^Z &=& \frac \pi 3\sqrt 2 G_F M_Z^2 (V_q^2+A_q^2)\delta_{qq'}\;,
\eeeq
with $V_{qq'}$ the appropriate CKM matrix element and $V_q,A_q$ the
vector and axial couplings to the Z$^0$.
The coefficients in the quark form factor $S_q(Q,\tau)$ are~\cite{Kodaira:1981nh,Davies:1984hs}:
\beeq\label{eq:Aqetc}
&&A_q^{(1)} = C_F \;, \quad A_q^{(2)} = \frac{1}{2} C_F K \;, \quad B_q^{(1)} = - \frac{3}{2} C_F \;, \nn\\
&&B_q^{(2)} = 
C_F^2\left(\frac{\pi^2}{4}-\frac{3}{16}-3\zeta_3\right)
+C_F\,C_A\left(\frac{11}{36}\pi^2-\frac{193}{48}+\frac{3}{2}\zeta_3\right)
+ C_F\,n_f\left(\frac{17}{24}-\frac{\pi^2}{18}\right)\;,\nn\\
\eeeq
where $\zeta_n$ is the Riemann $\zeta$-function $(\zeta_3=1.202\dots)$,
$C_F=4/3$, $C_A=3$, $n_f$ is the number of light flavours, and:
\beq
K = \left( \frac{67}{18} - \frac{\pi^2}{6} \right) C_A - \frac{5}{9} n_f \;.
\eeq

The above expression for $B_q^{(2)}$ is in a scheme where the subprocess
cross section is given by the leading-order expression (\ref{eq:DYLO}).  In the same
scheme the NLO coefficient functions are~\cite{Davies:1984hs,Balazs:1995nz}:
\beeq\label{eq:CDY}
C_{qq}(\as,z)&=&\left\{1+\frac{\as}{4\pi}C_F(\pi^2-8)\right\}\delta(1-z)
+\frac{\as}{2\pi}C_F(1-z)\nn\\
&\equiv& \left(1+\frac{\as}{\pi}c^{(1)}_q\right)\delta(1-z)
+\frac{\as}{2\pi}C_F(1-z)\;,\nn\\
C_{qg}(\as,z)&=&\frac{\as}{2\pi}z(1-z)\;,
\eeeq
where the second line defines $c_q^{(1)}$.  The corresponding splitting functions are:
\beeq\label{eq:PDY}
P_{qq}(z)&=&C_F\left[\frac{1+z^2}{(1-z)_+}+\frac 32 \delta(1-z)\right] \;,\nn\\
P_{qg}(z)&=&\frac 12\left[z^2+(1-z)^2\right]\;.
\eeeq
Equations (\ref{eq:RcNLO})--(\ref{eq:ffPplus}) therefore give
\beeq\label{eq:RqNLO}
R_q^{(R)}(\xi,\tau)&=&
\int_{\xi}^1 \frac{\mrd x_1}{x_1}\Biggl\{f_{q/h_1}(x_1)f_{\bar q/h_2}\left(
\frac{\xi}{x_1}\right)\left(1+\frac{\as}{\pi}2c^{(1)}_q\right)\nn\\
&+&  \frac{\as}{\pi}\int_{\xi/x_1}^1\frac{\mrd z}z
\Biggl[f_{q/h_1}(x_1)f_{\bar q/h_2}\left(\frac{\xi}{zx_1}\right)C_F(1-z)\nn\\
&+&\left\{f_{g/h_1}(x_1)f_{\bar q/h_2}\left(\frac{\xi}{zx_1}\right)+ f_{q/h_1}(x_1)
f_{g/h_2}\left(\frac{\xi}{zx_1}\right)\right\}\frac 12 z(1-z)\Biggr]\Biggr\}\;,\nn\\
R_q^{(I)}(\xi,\tau)&=&
\frac{\as}2\int_{\xi}^1 \frac{\mrd x_1}{x_1} \int_0^1\frac{\mrd z}z
\Biggl\{2f_{q/h_1}(x_1)f_{\bar q/h_2}\left(\frac{\xi}{zx_1}\right)P_{qq}(z)\\
&+&\left[f_{g/h_1}(x_1)f_{\bar q/h_2}\left(\frac{\xi}{zx_1}\right)
+f_{q/h_1}(x_1)f_{g/h_2}\left(\frac{\xi}{zx_1}\right)\right]P_{qg}(z)\Biggr\}\nn\\
&=&\frac{\as}2\int_{\xi}^1 \frac{\mrd x_1}{x_1}\Biggl\{
2C_F f_{q/h_1}(x_1)f_{\bar q/h_2}\left(\frac{\xi}{x_1}\right)
\left[2\ln\left(1-\frac{\xi}{x_1}\right)+\frac 32\right]\nn\\
&+&\int_{\xi/x_1}^1\frac{\mrd lz}z\Biggl[2C_F f_{q/h_1}(x_1)\left\{f_{\bar q/h_2}\left(\frac{\xi}{zx_1}\right)
\frac{1+z^2}{1-z}
- f_{\bar q/h_2}\left(\frac{\xi}{x_1}\right)\frac{2z}{1-z}\right\} \nn\\
&+&\left\{f_{g/h_1}(x_1)f_{\bar q/h_2}\left(\frac{\xi}{zx_1}\right)
+f_{q/h_1}(x_1)f_{g/h_2}\left(\frac{\xi}{zx_1}\right)\right\}
\frac 12\left\{z^2+(1-z)^2\right\}\Biggr]\Biggr\}\;. \nn
\eeeq

\subsubsection{Higgs boson production}

In the case of Higgs boson production the 
corresponding LO partonic subprocess is gluon
fusion, $g + g \to H$, through a massive-quark loop:
\beq\label{eq:higgsLO}
\sigma_{c{\bar c}}^F(Q,\as(Q)) = \delta_{cg}\delta_{\bar cg}\delta(Q^2-m_H^2)\sigma_0^H\;,
\eeq
where in the limit of infinite quark mass:
\beq
\sigma_0^H = \frac{\as^2(m_H)G_Fm_H^2}{288\pi\sqrt{2}}\;.
\eeq
The coefficients in the gluon form factor $S_g(Q,\tau)$ are~\cite{Catani:1988vd,deFlorian:2000pr,deFlorian:2001zd}:
\beeq\label{eq:Agetc}
&&A_g^{(1)} = C_A \;,\quad A_g^{(2)} = \frac{1}{2} C_A K \;,\quad
B_g^{(1)} = - \frac{1}{6} (11 C_A - 2 n_f) \;,\nn\\
&&B_g^{(2) \,H}=C_A^2\left(\frac{23}{24}+
\frac{11}{18}\pi^2-\frac{3}{2}\zeta_3\right)
+\frac{1}{2} C_F\,n_f-C_A\,n_f\left(\frac{1}{12}+\frac{\pi^2}{9} \right)
-\frac{11}{8} C_F C_A\, .\nn\\
\eeeq

Here again, the above expression for $B_g^{(2)}$ is in a scheme where
the Higgs boson subprocess
cross section is given by the leading-order expression (\ref{eq:higgsLO}).
In the same scheme the NLO coefficient functions are~\cite{Kauffman:1991cx}:
\beeq\label{eq:Chiggs}
C_{gg}(\as,z)&=&\left\{1+\frac{\as}{4\pi}
\left[C_A\left(2-\frac{\pi^2}3\right)+5+4\pi^2\right]\right\}\delta(1-z)\nn\\
&\equiv& \left(1+\frac{\as}{\pi}c^{(1)}_g\right)\delta(1-z)\;,\nn\\
C_{gq}(\as,z)&=&C_{g\bar q}(\as,z)=\frac{\as}{2\pi}C_F z\;.
\eeeq
The corresponding splitting functions are:
\beeq\label{eq:Phiggs}
P_{gg}(z)&=&2C_A\left[\frac z{(1-z)_+}+\frac{1-z}z+z(1-z)\right]
+\frac 16 (11C_A-2n_f)\delta(1-z)\;,\nn\\
P_{gq}(z)&=&P_{g\bar q}(z)= C_F\frac{1+(1-z)^2}z\;.
\eeeq
Equations (\ref{eq:RcNLO})--(\ref{eq:ffPplus}) therefore give
\beeq\label{eq:RgNLO}
&R_g^{(R)}(\xi,\tau)&=
\int_{\xi}^1 \frac{\mrd x_1}{x_1}\Biggl\{f_{g/h_1}(x_1)f_{g/h_2}\left(
\frac{\xi}{x_1}\right)\left(1+\frac{\as}{\pi}2c^{(1)}_g\right)\nn\\
&& + \frac{\as}{\pi}\int_{\xi/x_1}^1\frac{dz}z
\left[f_{g/h_1}(x_1)f_{s/h_2}\left(\frac{\xi}{zx_1}\right)+ f_{s/h_1}(x_1)
f_{g/h_2}\left(\frac{\xi}{zx_1}\right)\right]\frac 12 C_F z\Biggr\}\;,\nn\\
&R_g^{(I)}(\xi,\tau)&=
\frac{\as}2\int_{\xi}^1 \frac{\mrd x_1}{x_1} \int_0^1\frac{\mrd z}z
\Biggl\{2f_{g/h_1}(x_1)f_{g/h_2}\left(\frac{\xi}{zx_1}\right)P_{gg}(z)\nn\\
&&+\left[f_{g/h_1}(x_1)f_{s/h_2}\left(\frac{\xi}{zx_1}\right)
+f_{s/h_1}(x_1)f_{g/h_2}\left(\frac{\xi}{zx_1}\right)\right]P_{gq}(z)\Biggr\}\nn\\
&&=\frac{\as}2\int_{\xi}^1 \frac{\mrd x_1}{x_1}\Biggl\{
2f_{g/h_1}(x_1)f_{g/h_2}\left(\frac{\xi}{x_1}\right)
\left[2C_A\ln\left(1-\frac{\xi}{x_1}\right)+\frac 16 (11C_A-2n_f)\right]\nn\\
&&+\int_{\xi/x_1}^1\hspace{-0.2cm}\frac{\mrd z}z\Biggl[4C_Af_{g/h_1}(x_1)\left\{f_{g/h_2}\left(\frac{\xi}{zx_1}\right)
\left[\frac{z}{1-z}+\frac{1-z}z+z(1-z)\right]
- f_{g/h_2}\left(\frac{\xi}{x_1}\right)\frac{z}{1-z}\right\} \nn\\
&&+\left\{f_{g/h_1}(x_1)f_{s/h_2}\left(\frac{\xi}{zx_1}\right)
+f_{s/h_1}(x_1)f_{g/h_2}\left(\frac{\xi}{zx_1}\right)\right\}C_F\frac{1+(1-z)^2}z\Biggr]\Biggr\} \;,
\eeeq
where $f_s = \sum_q(f_q+f_{\bar q})$.

\subsection{Resummed distributions}\label{sec:etres:resum}
\subsubsection{Vector boson production}

\begin{figure}[!htb]
\begin{center}
  \vspace{1.2cm}
  \includegraphics[scale=0.35, angle=90]{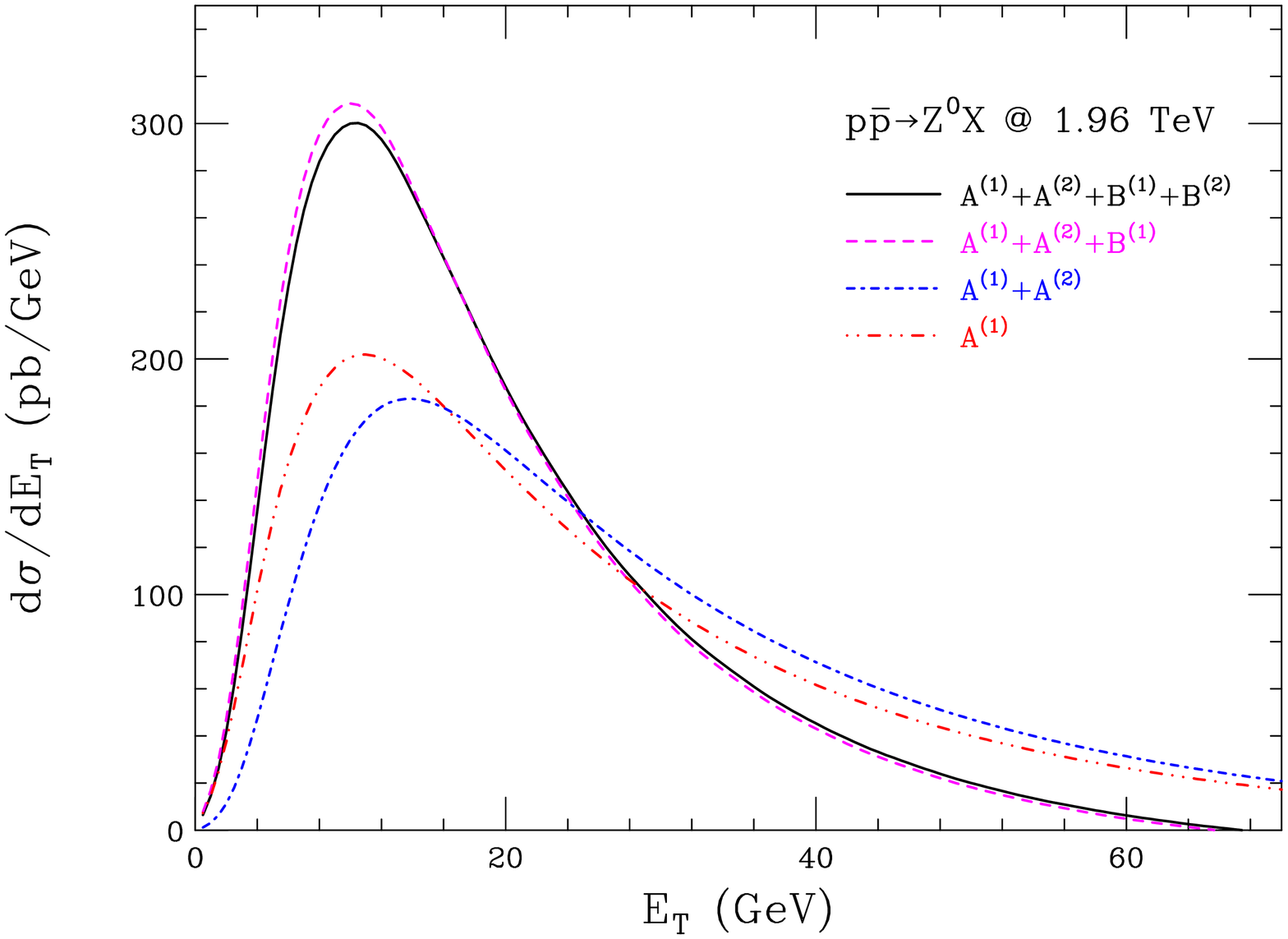}
  \hspace{1.2cm}
  \includegraphics[scale=0.35, angle=90]{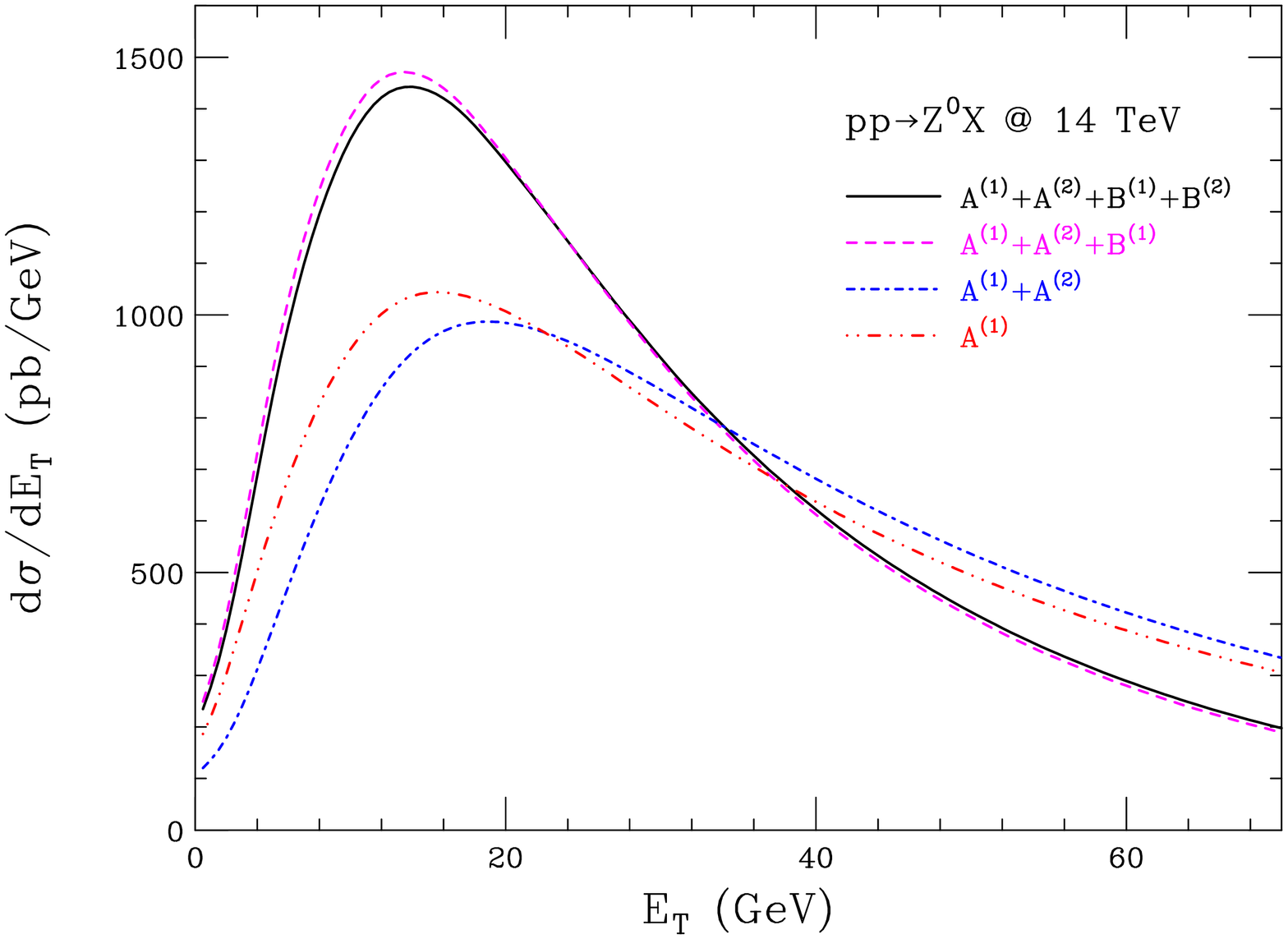}
\end{center}
\caption{Resummed component of the transverse energy distribution in $Z^0$ boson production at the Tevatron
and LHC.  The curves show the effects of the coefficients in the quark form factor:
black, all coefficients; magenta omitting $B_q^{(2)}$; blue $A_q^{(1)}$ and $A_q^{(2)}$ only; red $A_q^{(1)}$ only.
\label{fig:ETdrelyTev} }
\end{figure}

Figure~\ref{fig:ETdrelyTev} shows the resummed component of the
transverse energy distribution in $Z^0$ boson production at the
Tevatron  ($p\bar p$ at $\rs =1.96$ TeV) and LHC ($pp$ at $\rs
=14$ TeV).\footnote{Results for $pp$ at $\rs =7$ TeV are given in
  appendix~\ref{app:etres2}.}  For all calculations, we use the MSTW 2008 NLO parton
distributions~\cite{Martin:2009iq}.  The different curves show the
effects of the subleading coefficients (\ref{eq:Aqetc}) in the quark
form factor.  We see that while $B_q^{(1)}$ has a large effect (the
difference between the blue and magenta curves), the effects of the
other subleading coefficients are quite small.

The peak of the resummed distribution lies at around
$E_T\sim 10$ GeV at the Tevatron, rising to $\sim 14$ GeV at the LHC.
This is comfortably below $M_Z$, justifying the resummation of
logarithms of $E_T/M_Z$ in the peak region.  However, at LHC energy
the predicted distribution has a substantial tail at larger values of $E_T$,
indicating that the higher-order terms generated by the resummation
formula remain significant even when the logarithms are not large.
In addition, the LHC prediction does not go to zero as it should at
small $E_T$.  However, this region is sensitive to the treatment of
non-perturbative effects such as the behaviour of the strong coupling
at low scales (we freeze its value below 1 GeV) and the upper limit in
the integral over transverse time (we set $\tau_{\mbox{\scriptsize
  max}}=1/\Lambda$ where $\Lambda$ is the two-loop
QCD scale parameter, set to 200 MeV here).

The resummed component for $W^\pm$ boson production looks very
similar, apart of course from the overall normalisation, and therefore
we do not show it here.  Predictions with matching to fixed order will
be presented in section~\ref{sec:etres:match}.

\subsubsection{Higgs boson production}

\begin{figure}[!htb]
\begin{center}
\vspace{1.2cm}
  \includegraphics[scale=0.35, angle=90]{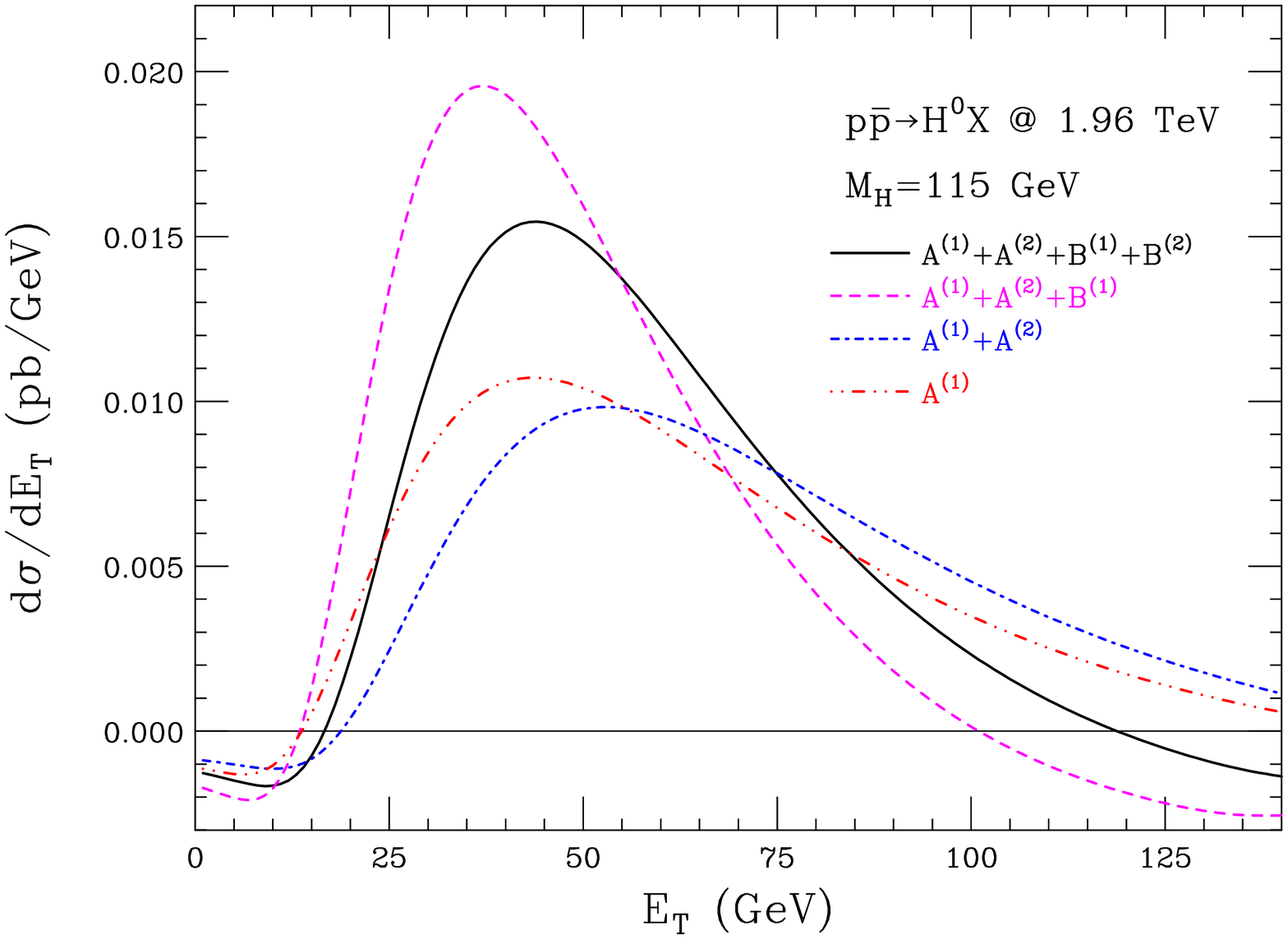}
  \hspace{1.2cm}
  \includegraphics[scale=0.35, angle=90]{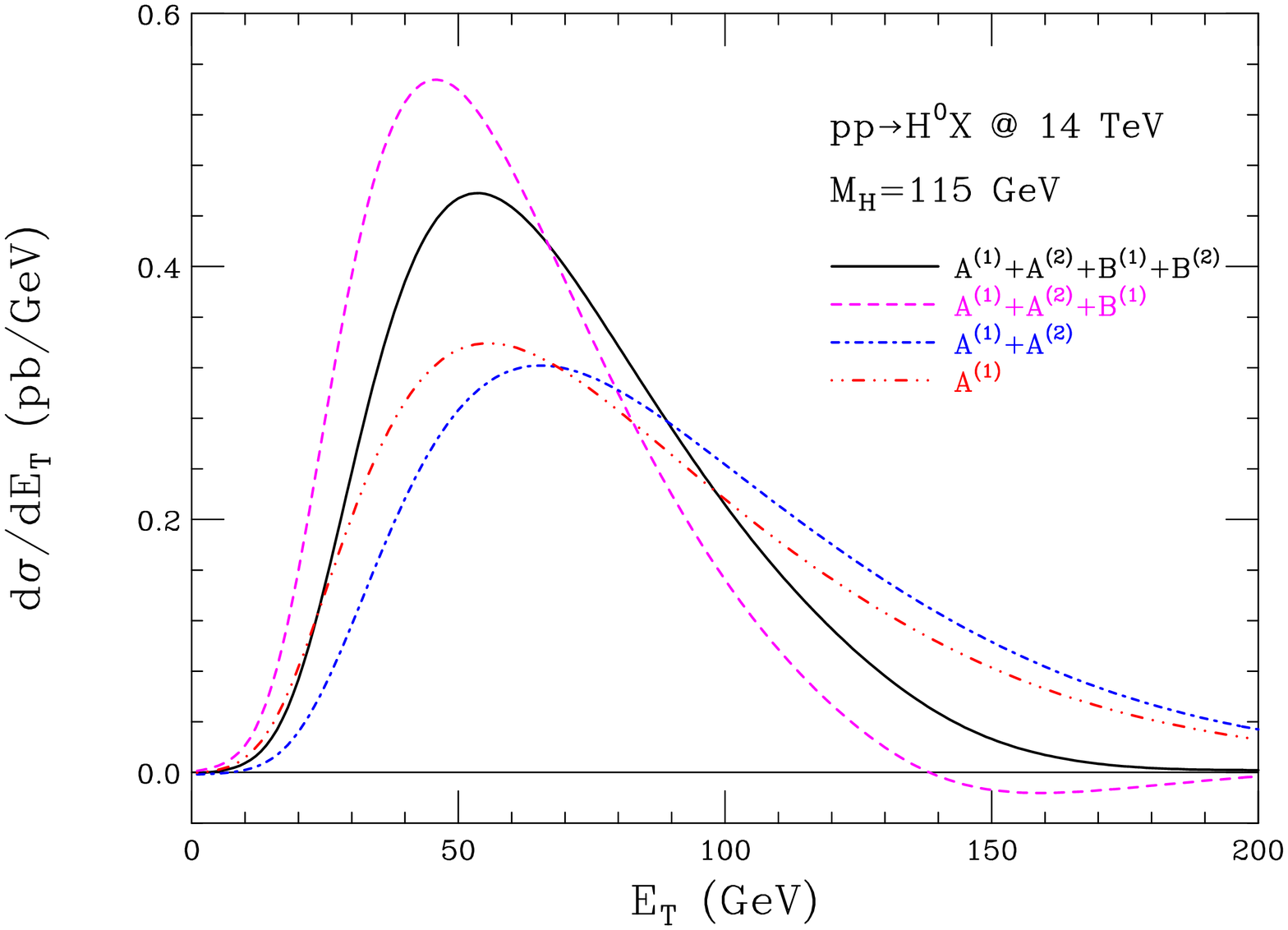}
\end{center}
\caption{Resummed component of the transverse energy distribution in
  Higgs boson production at the Tevatron and LHC. The curves shows
the effects of the coefficients in the gluon form factor: black, all
coefficients; magenta omitting $B_g^{(2)}$; blue $A_g^{(1)}$
and $A_g^{(2)}$ only; red $A_g^{(1)}$ only.
\label{fig:ETh115Tev} }
\end{figure}

Figure~\ref{fig:ETh115Tev} shows the resummed component of the
transverse energy distribution in Higgs boson production at the
Tevatron and LHC, for a Higgs boson mass of 115 GeV.  The effects
of subleading terms in the gluon form factor (\ref{eq:Agetc}) are more
marked than those of the quark form factor discussed above.  The
distribution peaks at large values of $E_T$, around 40 GeV at
the Tevatron, rising to $\sim 50$ GeV at the LHC.  This is due to the
larger colour charge of the gluon.  However, together with the large
effects of subleading terms, it does make the reliability of the
resummed predictions more questionable.  Also in contrast to
the vector boson case, the suppression at low and high $E_T$ is if
anything too great, resulting in negative values below 16 GeV and
above 120 GeV at Tevatron energy. We verified, by cutting-off the
gluon PDFs at zero, that the negative values are due to the
resummation and not due to the gluon PDF becoming negative at low $x$ values.

\subsection{Matching to fixed order}\label{sec:etres:match}

The resummed distributions presented above include only terms that are
logarithmically enhanced at small $E_T$.  To extend the predictions to
larger $E_T$ we must match the resummation to fixed-order
calculations.  To avoid double counting of the resummed terms, the
corresponding contribution must be subtracted from the fixed-order
result.

We consider here only matching to first order in $\as$.  To this order
the $E_T$ distribution for $E_T>0$ has the form:
\beq\label{eq:Oas}
\frac{\mrd \sigma}{\mrd E_T} = \frac 1{E_T}(A\ln E_T +B) + C(E_T)\;,
\eeq
where $A$ and $B$ are constants (for a given process and collision
energy) and the function $C(E_T)$ is regular at $E_T=0$.  The terms
involving $A$ and $B$ are already included in the resummed prediction,
and therefore we have only to add the regular function $C$ to it to obtain a
prediction that is matched to the ${\cal O}(\as)$ result.  This
function is determined by fitting the ${\cal O}(\as)$ prediction
for $E_T\,\mrd\sigma/\mrd E_T$ to a linear function of $\ln E_T$ at small
$E_T$, extracting the coefficients $A$ and $B$, and then subtracting the
enhanced terms in Eq.~(\ref{eq:Oas}). 

\subsubsection{Vector boson production}

\begin{figure}[!htb]
\begin{center}
\vspace{1.2cm}
  \includegraphics[scale=0.35, angle=90]{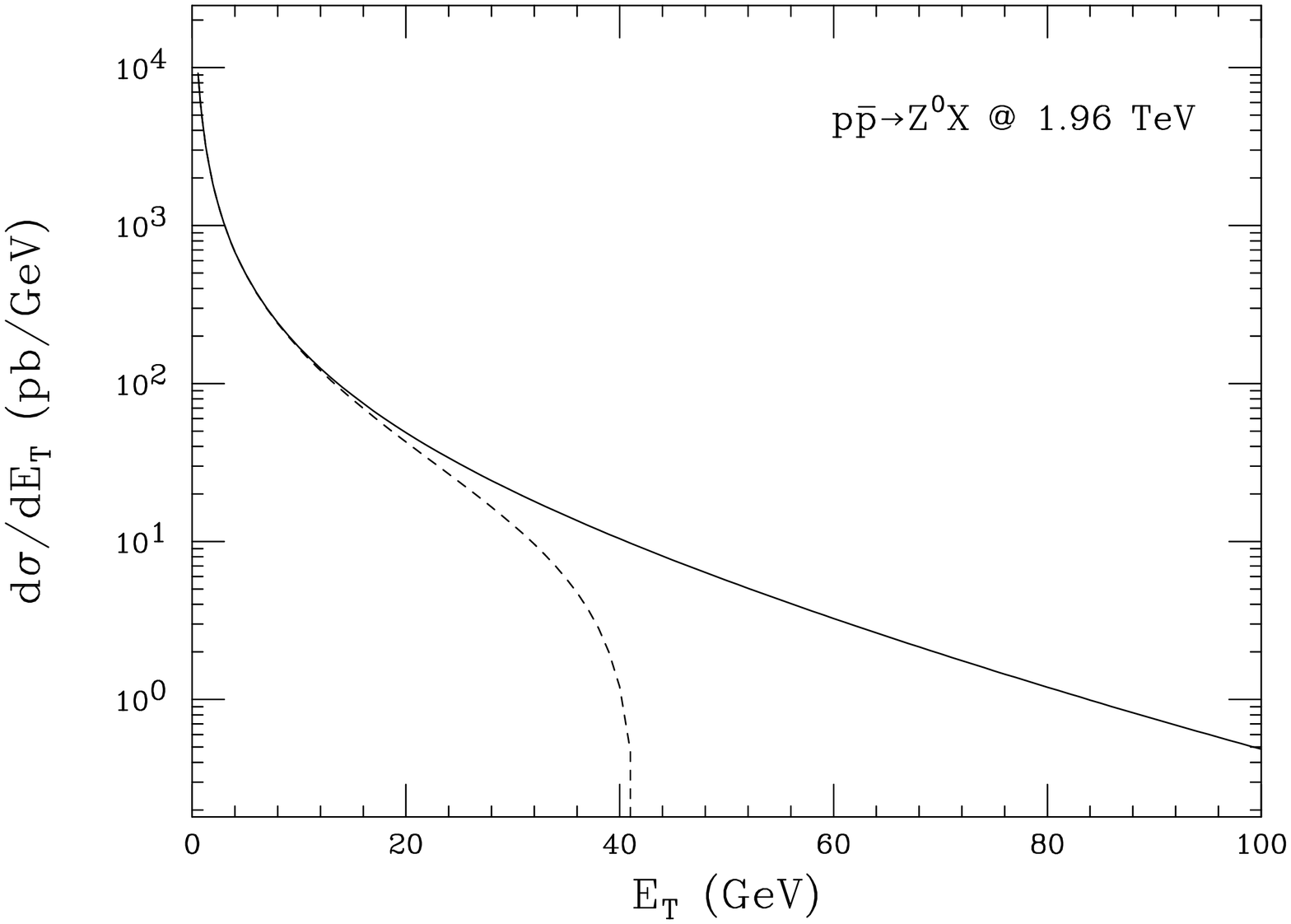}
  \hspace{1.2cm}
  \includegraphics[scale=0.35, angle=90]{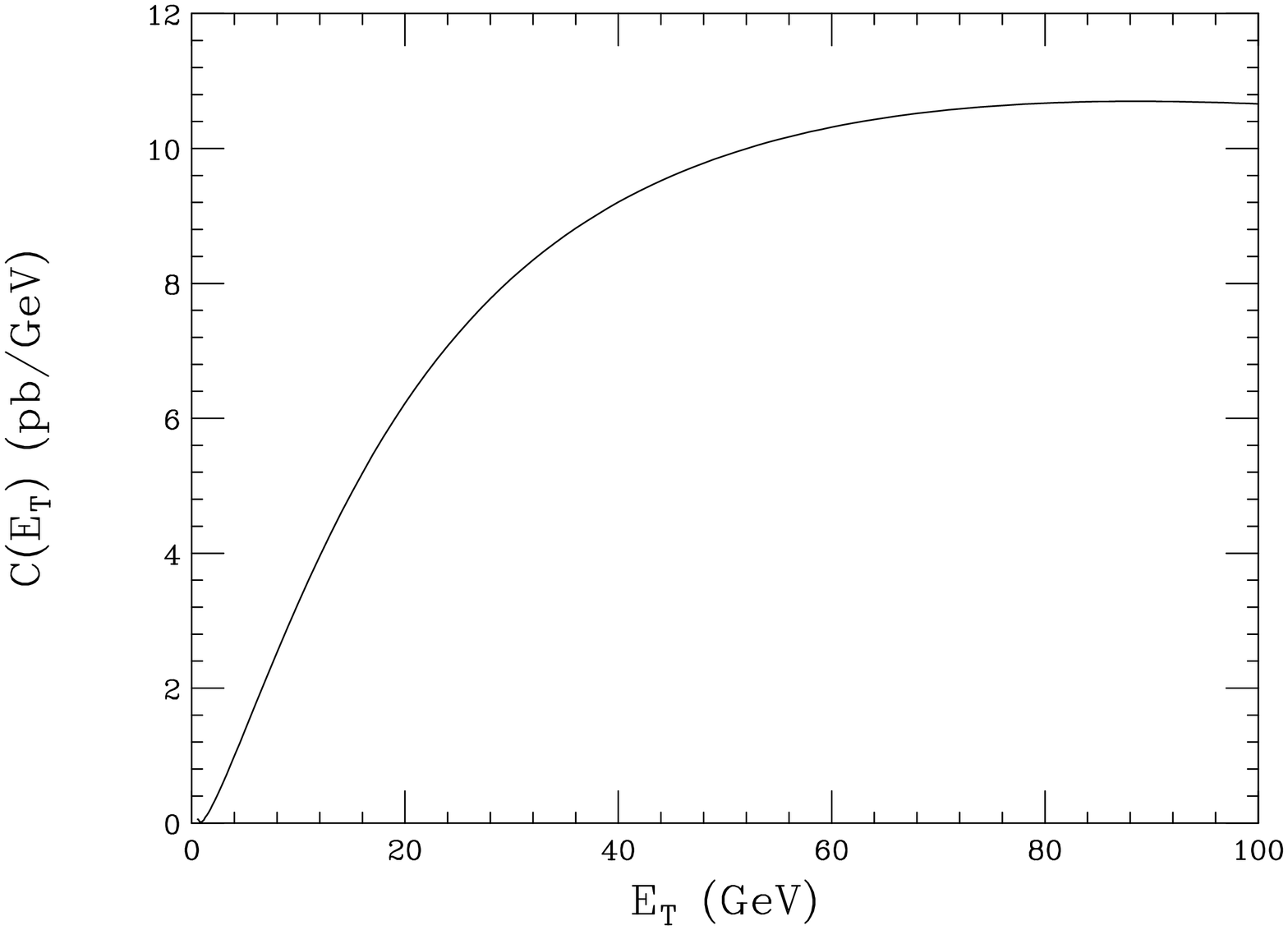}
\end{center}
\caption{Left: $\mathcal{O}(\as)$ $E_T$ distribution in $Z^0$ production at the Tevatron;
solid, full prediction; dashed, fit to enhanced terms.
Right: difference between full prediction and fit to enhanced terms.
\label{fig:ZfitTev1} }
\end{figure}

\begin{figure}[!htb]
\begin{center}
\vspace{1.2cm}
  \includegraphics[scale=0.35, angle=90]{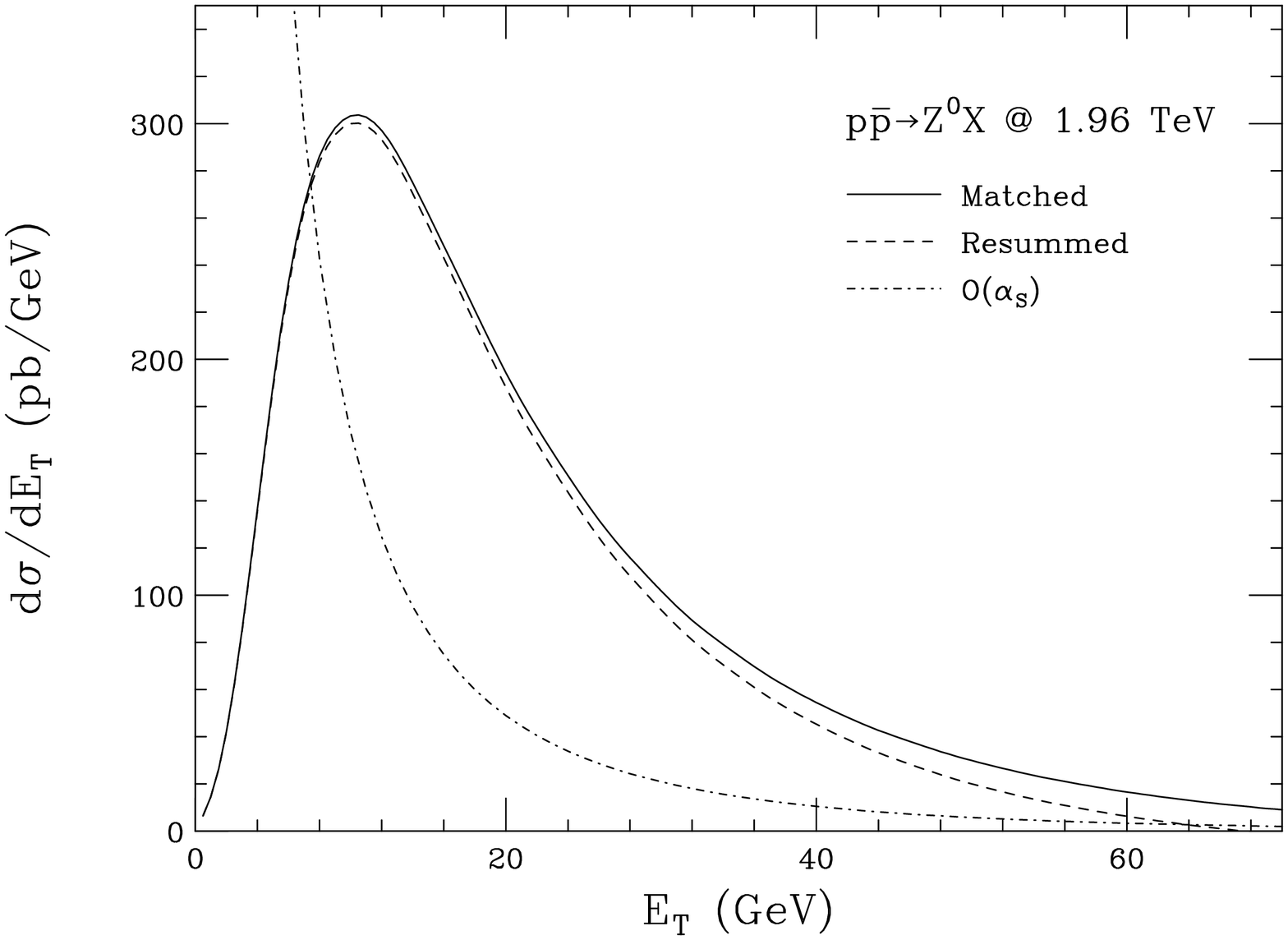}
  \hspace{1.2cm}
  \includegraphics[scale=0.35, angle=90]{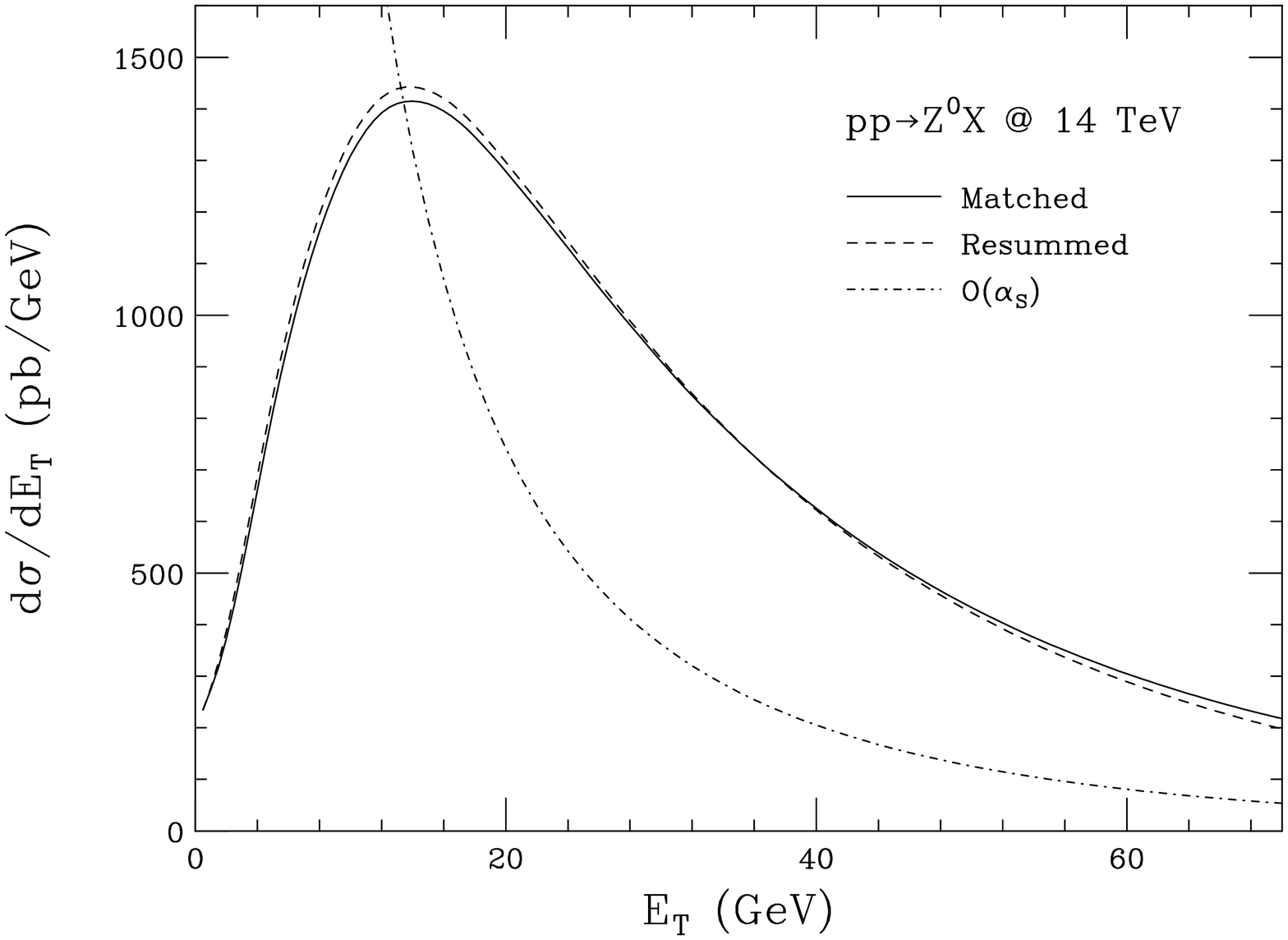}
\end{center}
\caption{Predicted $E_T$ distribution in $Z^0$ production  at the
  Tevatron and LHC.  Solid: resummed prediction matched to ${\cal
    O}(\as)$.  Dashed: resummed only. Dot-dashed: ${\cal O}(\as)$ only.
\label{fig:ZfitTev3} }
\end{figure}

The above matching procedure is illustrated for  $Z^0$ production at
the Tevatron in Fig.~\ref{fig:ZfitTev1}.  The fit to the
logarithmically enhanced terms gives excellent agreement with the
order-$\as$ result out to around 20 GeV, confirming the dominance of
such terms throughout the region of the peak in
Fig.~\ref{fig:ETdrelyTev}.  The remainder function $C(E_T)$ vanishes
at small $E_T$ and rises to around 10 pb/GeV, falling off slowly at
large $E_T$. Consequently the matching correction to the resummed
prediction is small and roughly constant throughout the region 40--100
GeV, as shown in Fig.~\ref{fig:ZfitTev3}. 

As shown on the right in Fig.~\ref{fig:ZfitTev3}, the situation is
similar at LHC energy: the matching correction is small, although in
this case it is negative below about 40 GeV.  The large tail at high
$E_T$ and the bad behaviour at low $E_T$, due to uncompensated
higher-order terms generated by resummation, are not much affected by
matching to this order.

\begin{figure}[!htb]
\begin{center}
\vspace{1.2cm}
  \includegraphics[scale=0.35, angle=90]{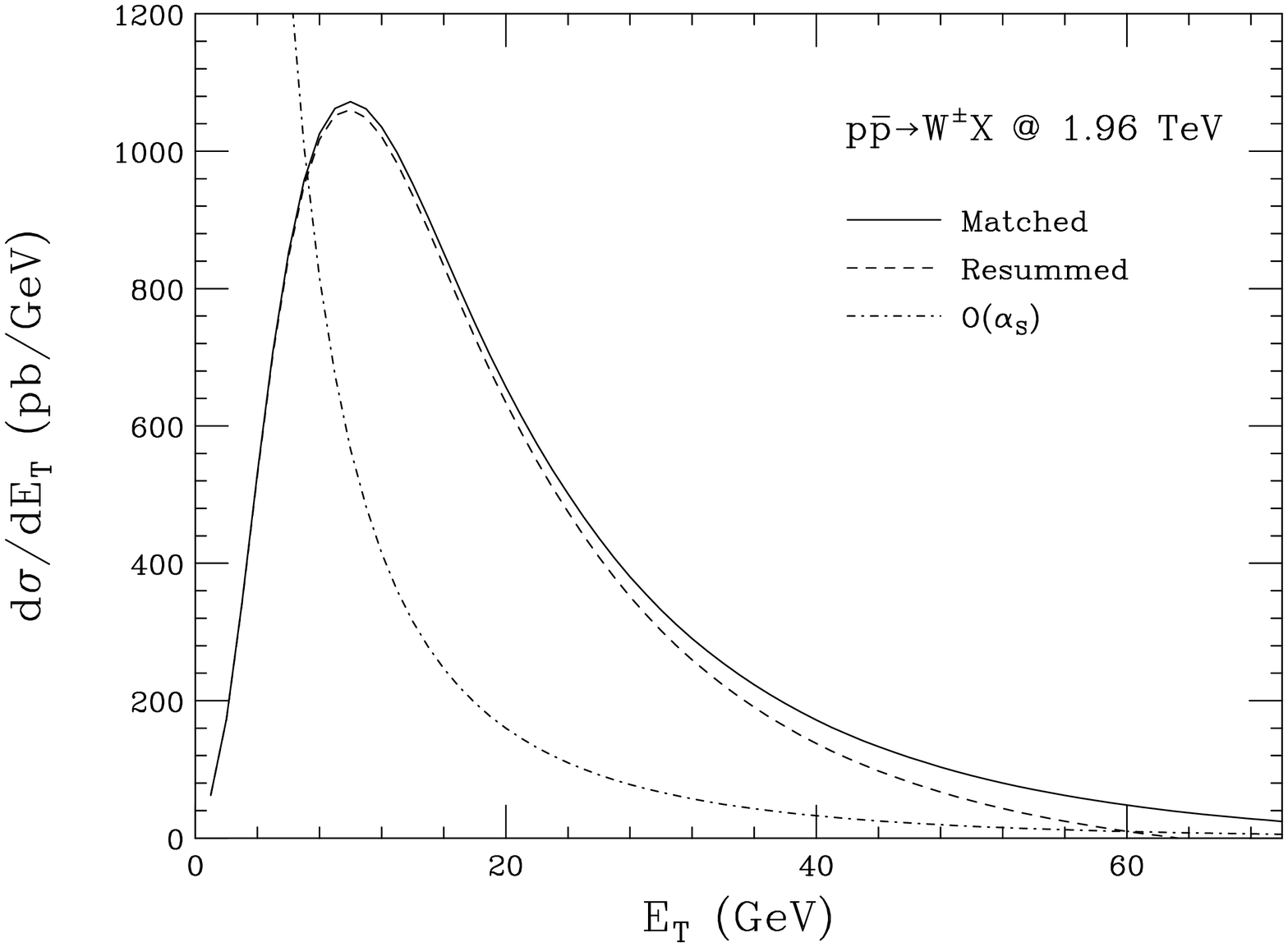}
  \hspace{1.2cm}
  \includegraphics[scale=0.35, angle=90]{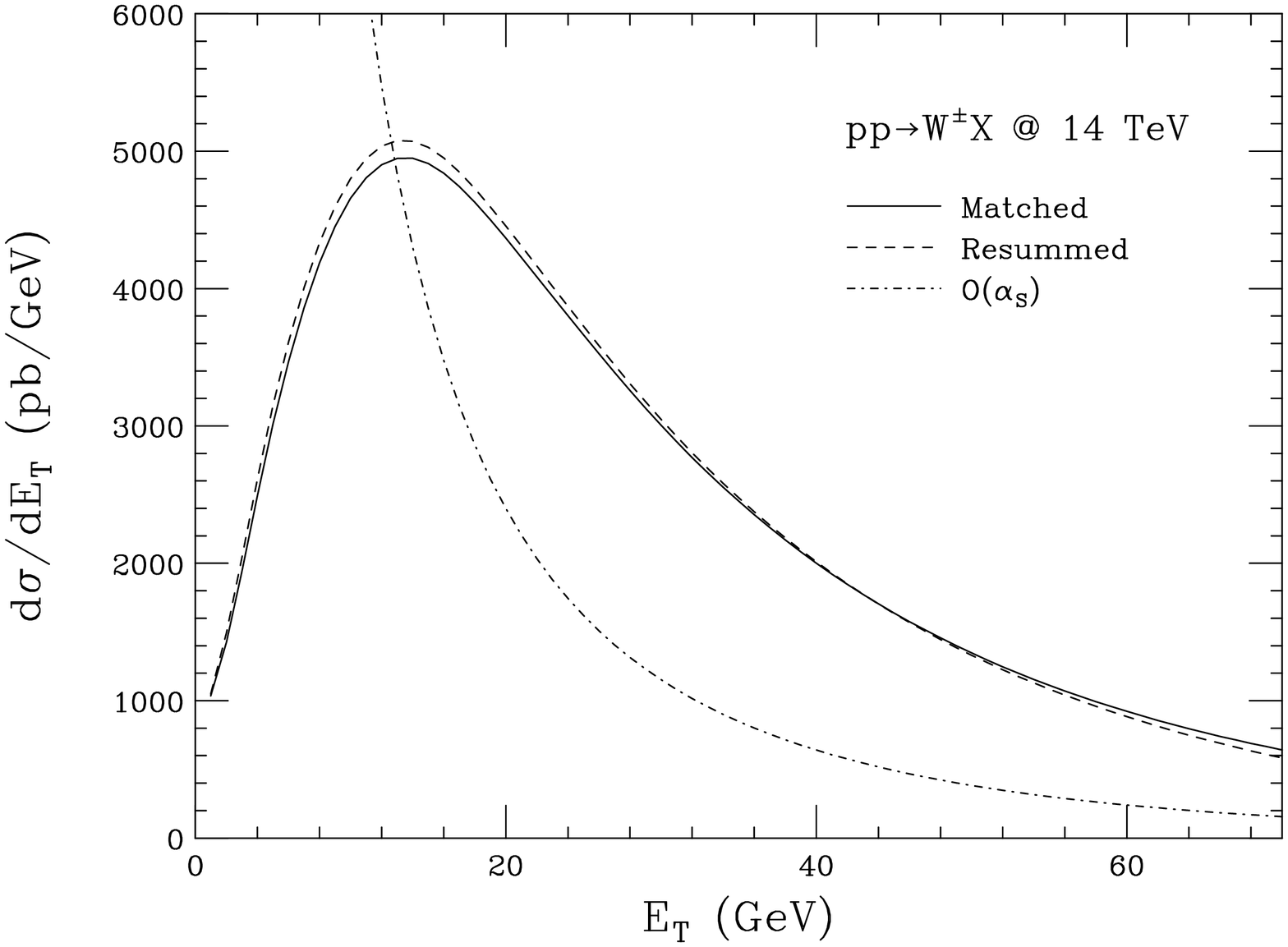}
\end{center}
\caption{Predicted $E_T$ distribution in $W^+$+$W^-$ production  at the
  Tevatron and LHC.  Solid: resummed prediction matched to ${\cal
    O}(\as)$.  Dashed: resummed only. Dot-dashed: ${\cal O}(\as)$ only.
\label{fig:WfitLHC3} }
\end{figure}

The corresponding matched predictions for $W^\pm$ boson production are shown
in Fig.~\ref{fig:WfitLHC3}.  As remarked earlier, the form of the
resummed distribution is very similar to that for $Z^0$ boson production,
and again the matching correction is small.

Note that at high $E_T$ the
$\mathcal{O}(\as)$ distributions should approximate the matched
distributions, although this is not apparent in the figures.

\subsubsection{Higgs boson production}

\begin{figure}[!htb]
\begin{center}
\vspace{1.2cm}
  \includegraphics[scale=0.38, angle=90]{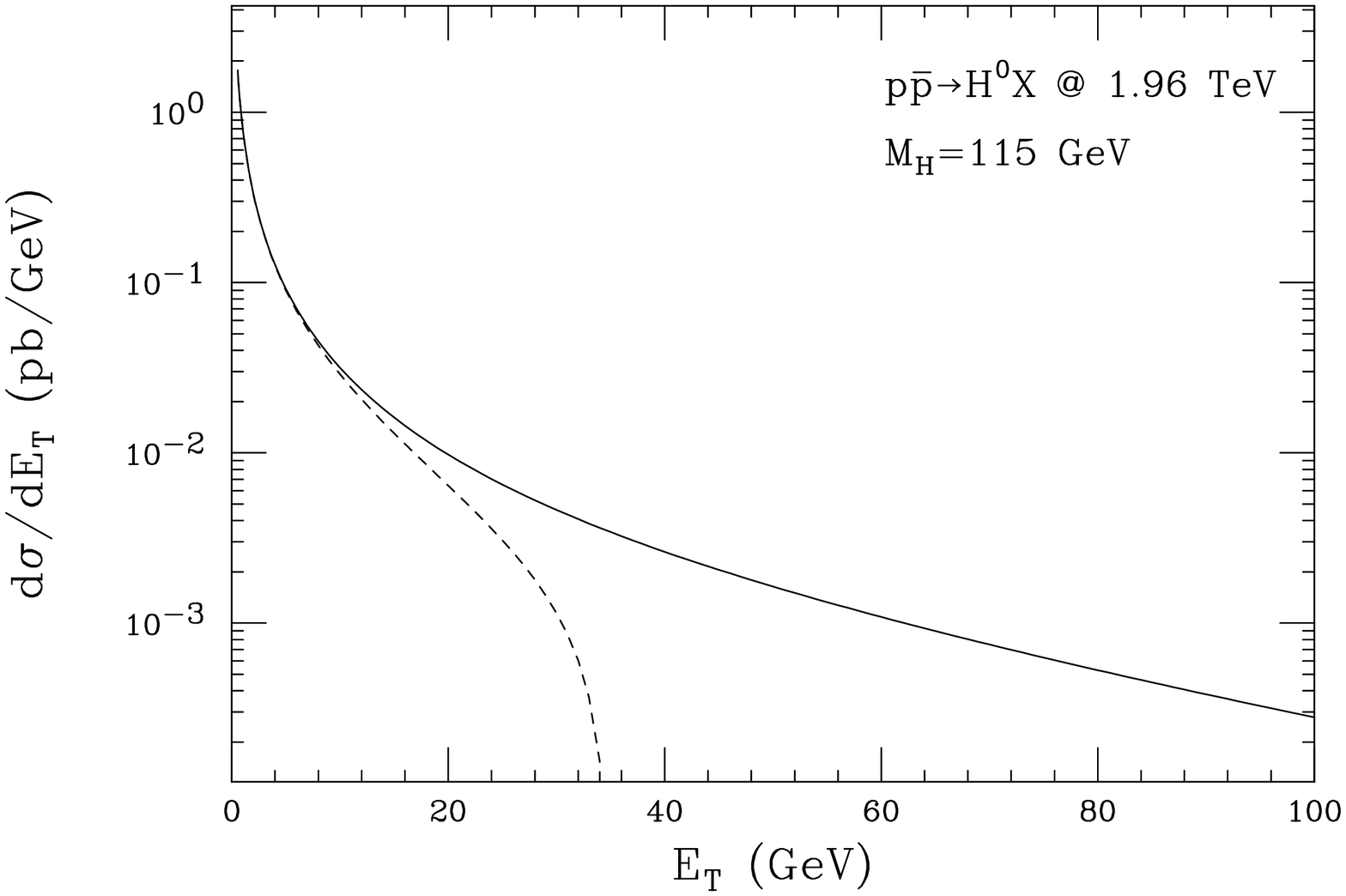}
  \hspace{1.2cm}
  \includegraphics[scale=0.34, angle=90]{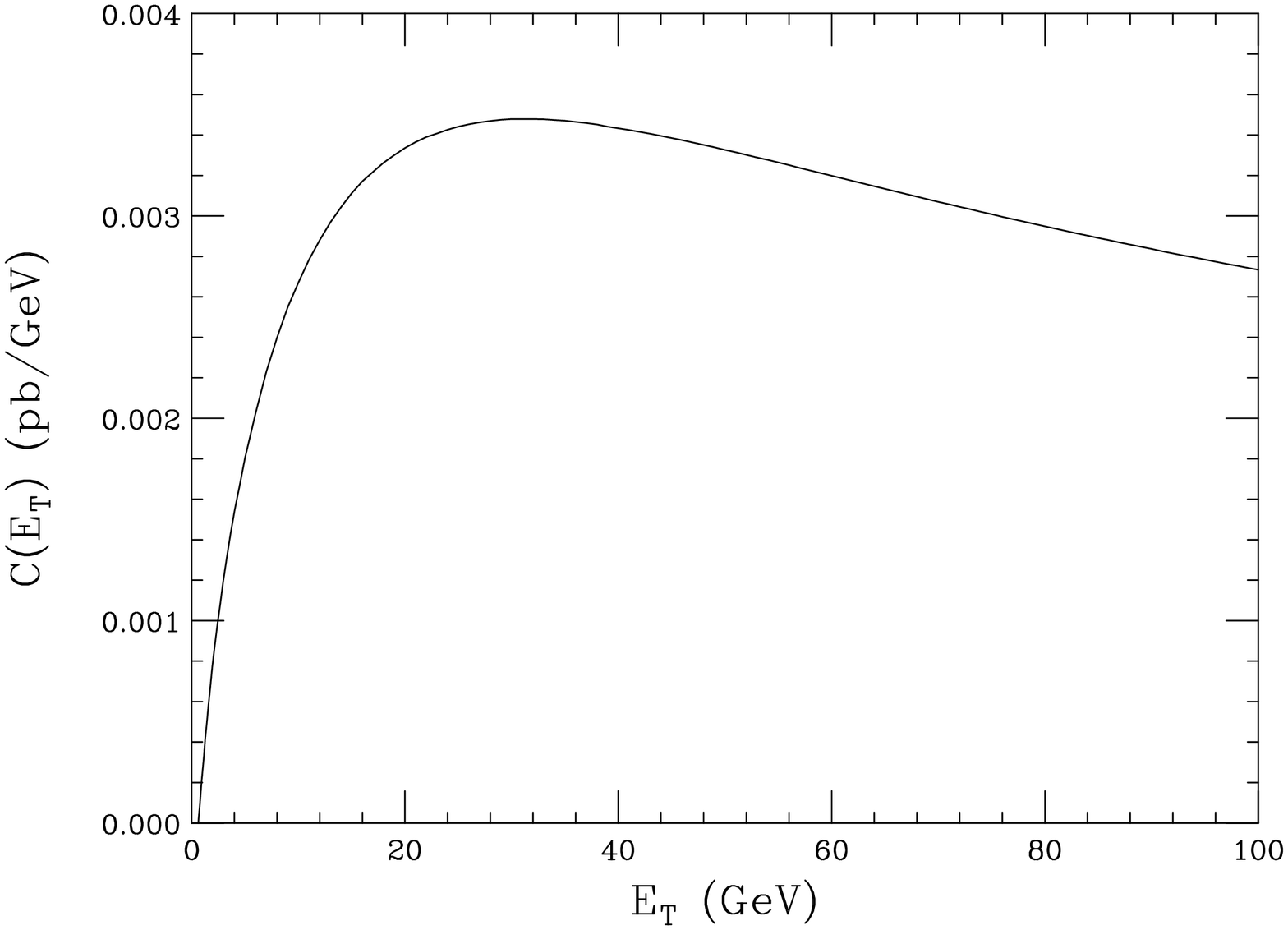}
\end{center}
\caption{ Left: $\mathcal{O}(\as)$ $E_T$ distribution in Higgs boson
  production at the Tevatron;  solid, full prediction; dashed, fit to enhanced terms.
Right: difference between full prediction and fit to enhanced terms.
\label{fig:HfitTev1} }
\end{figure}

\begin{figure}[!htb]
\begin{center}
\vspace{0.6cm} 
 \includegraphics[scale=0.38, angle=90]{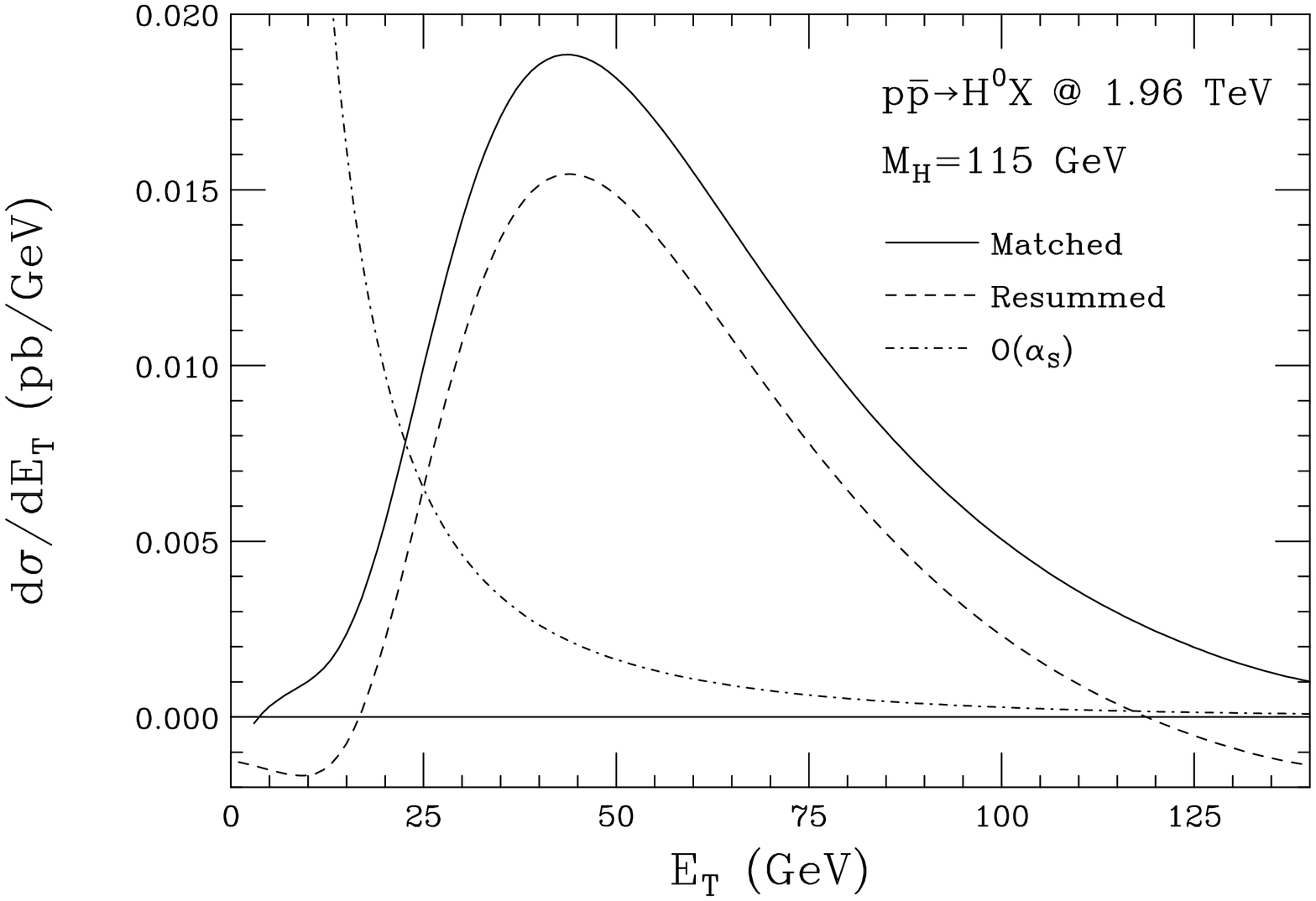}
  \hspace{1.2cm}
  \includegraphics[scale=0.34, angle=90]{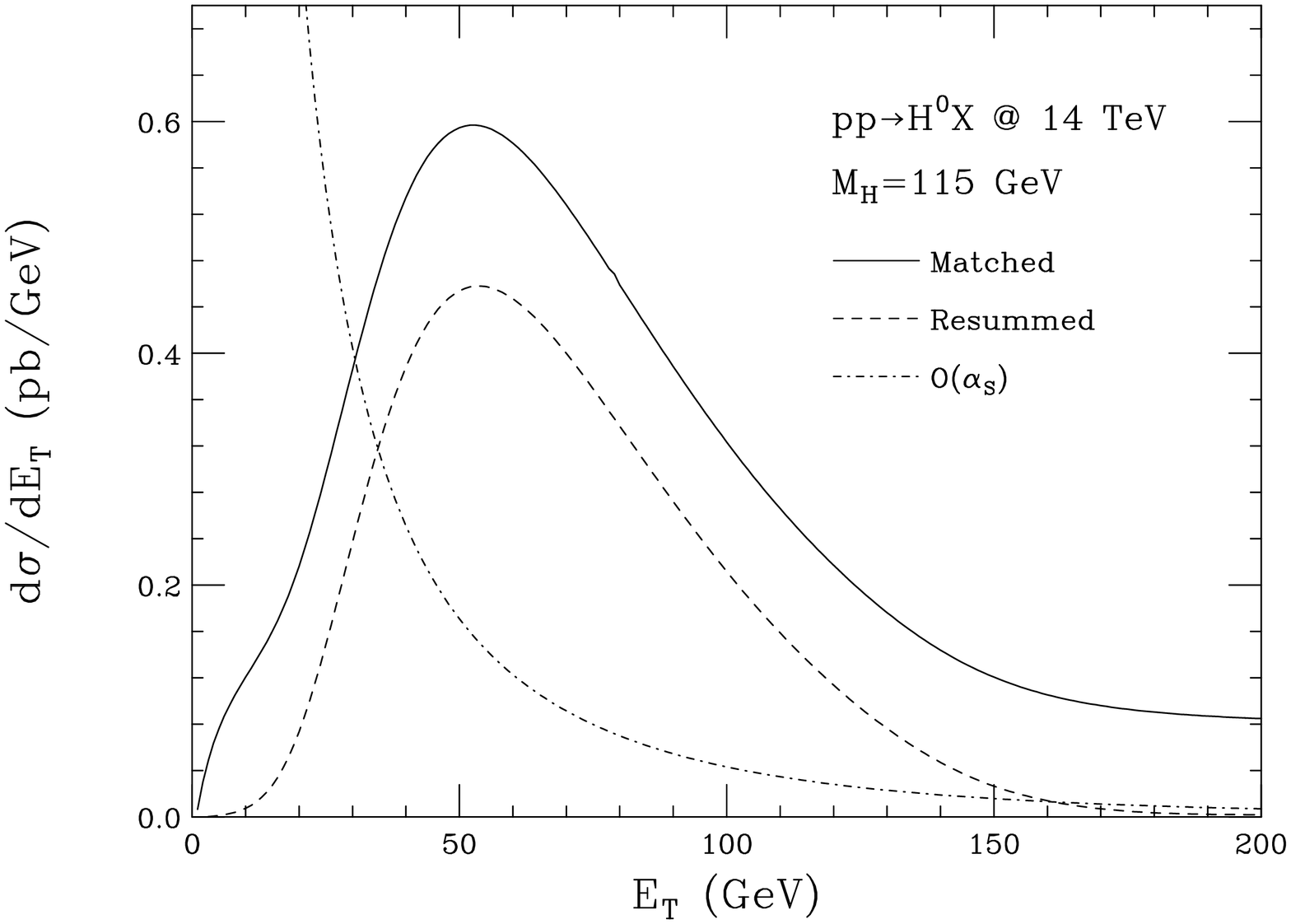}
\end{center}
\caption{Predicted $E_T$ distribution in Higgs boson production at the
  Tevatron and LHC.  Solid: resummed prediction matched to ${\cal
    O}(\as)$.  Dashed: resummed only. Dot-dashed: ${\cal O}(\as)$ only.
\label{fig:HfitTev3} }
\end{figure}

Adopting the same matching procedure for Higgs boson production, we
find the results shown in Figs.~\ref{fig:HfitTev1} and
\ref{fig:HfitTev3}.  The form of the matching correction is similar to
that for vector bosons, but its effect is rather different. The
roughly constant, then slowly decreasing, correction in the region 20--100 GeV is
not small compared to the resummed result and therefore it raises the whole
distribution by a significant amount throughout this region.
This has the beneficial effect of compensating the negative values
at low and high $E_T$ at Tevatron energy.  However, it further
enhances the high $E_T$ tail of the distribution at LHC energy.  This,
together with the relatively large correction in the peak region,
casts further doubt on the reliability of the predictions in the case
of Higgs boson production.

\subsection{Monte Carlo comparisons}\label{sec:etres:MC}

In this section we compare the resummed and matched distributions
obtained above with the predictions of the parton shower Monte Carlo
programs {\tt HERWIG}~\cite{Corcella:2000bw} and \Herwigpp~\cite{Bahr:2008pv}.

Comparisons are performed first at the parton level, that is, after QCD showering
from the incoming and outgoing partons of the hard subprocess.
We say `incoming and outgoing' because both programs apply hard matrix
element corrections: in addition to the Born process, $\mathcal{O}(\as)$ real
emission hard subprocesses are included in phase space regions not covered
by showering from the Born process.

After showering, the Monte Carlo programs apply a hadronization model
to convert the partonic final state to a hadronic one.  We show the
effects of hadronization in the case of  {\tt HERWIG} only; those in
\Herwigpp are broadly similar since both programs use basically
the same cluster hadronization model we described in section~\ref{sec:mc:hadronization}.  The programs also model the
underlying event, which arises from the interactions of spectator
partons (see section~\ref{sec:mc:ue}) and makes a significant contribution to the hadronic
transverse energy.  In this case we show only the underlying event
prediction of  \Herwigpp, since the default model used in {\tt HERWIG}
has been found to give an unsatisfactory description of Tevatron data.
For an improved simulation of the underlying event, {\tt HERWIG} can
be interfaced to the multiple interaction package
{\tt JIMMY}~\cite{Butterworth:1996zw}, which
is similar to the model built into \Herwigpp.

\subsubsection{Vector boson production}

\begin{figure}[!htb]
\begin{center}
\vspace{0.6cm}
  \includegraphics[scale=0.35, angle=90]{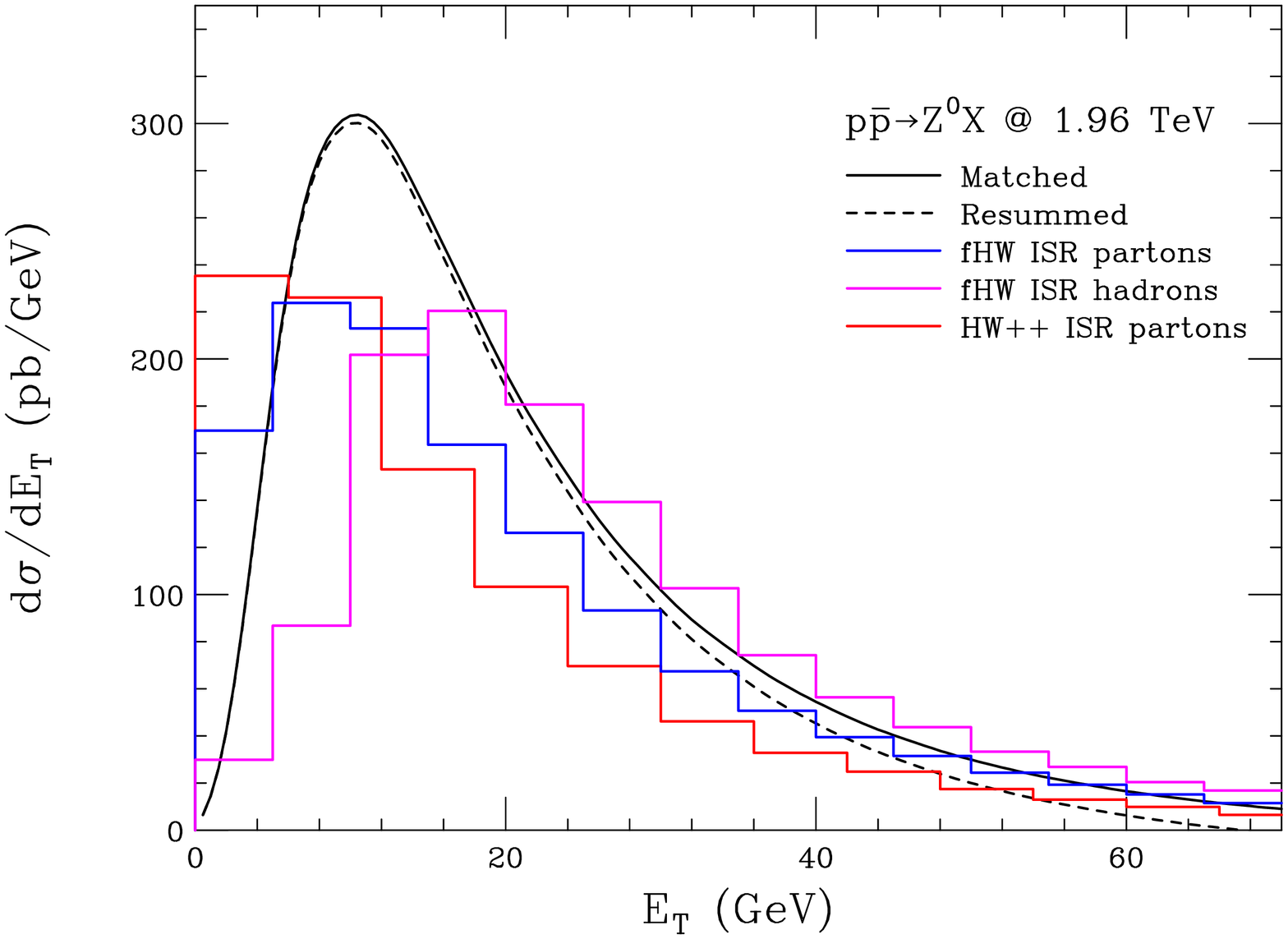}
  \hspace{1.2cm}
  \includegraphics[scale=0.35, angle=90]{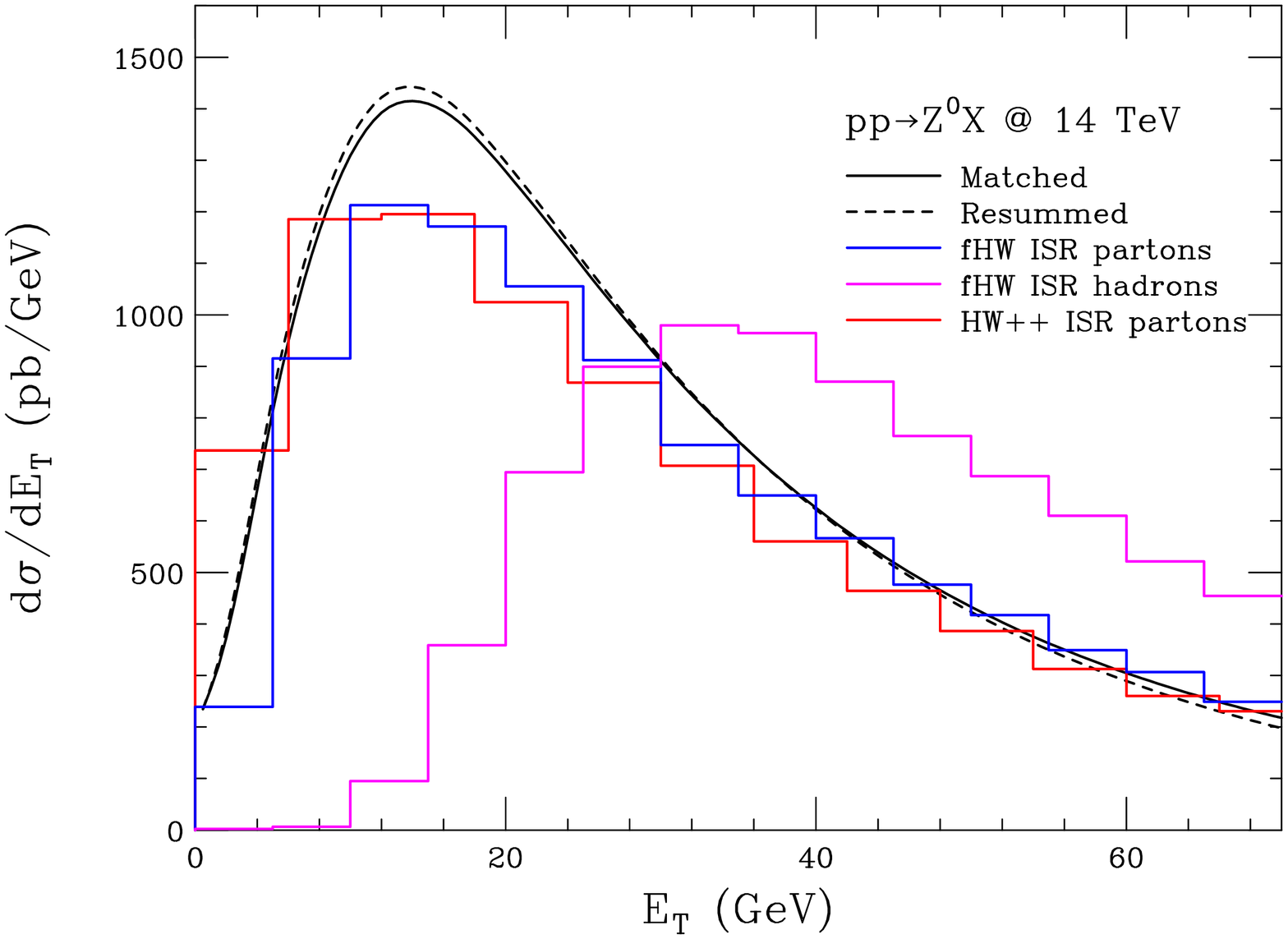}
\end{center}
\caption{Predicted $E_T$ distribution in $Z^0$ boson production  at the
 Tevatron and LHC. Comparison of resummed and Monte Carlo results.
\label{fig:ZfHWall} }
\end{figure}

\begin{figure}[!htb]
\begin{center}
\vspace{1.2cm}
  \includegraphics[scale=0.35, angle=90]{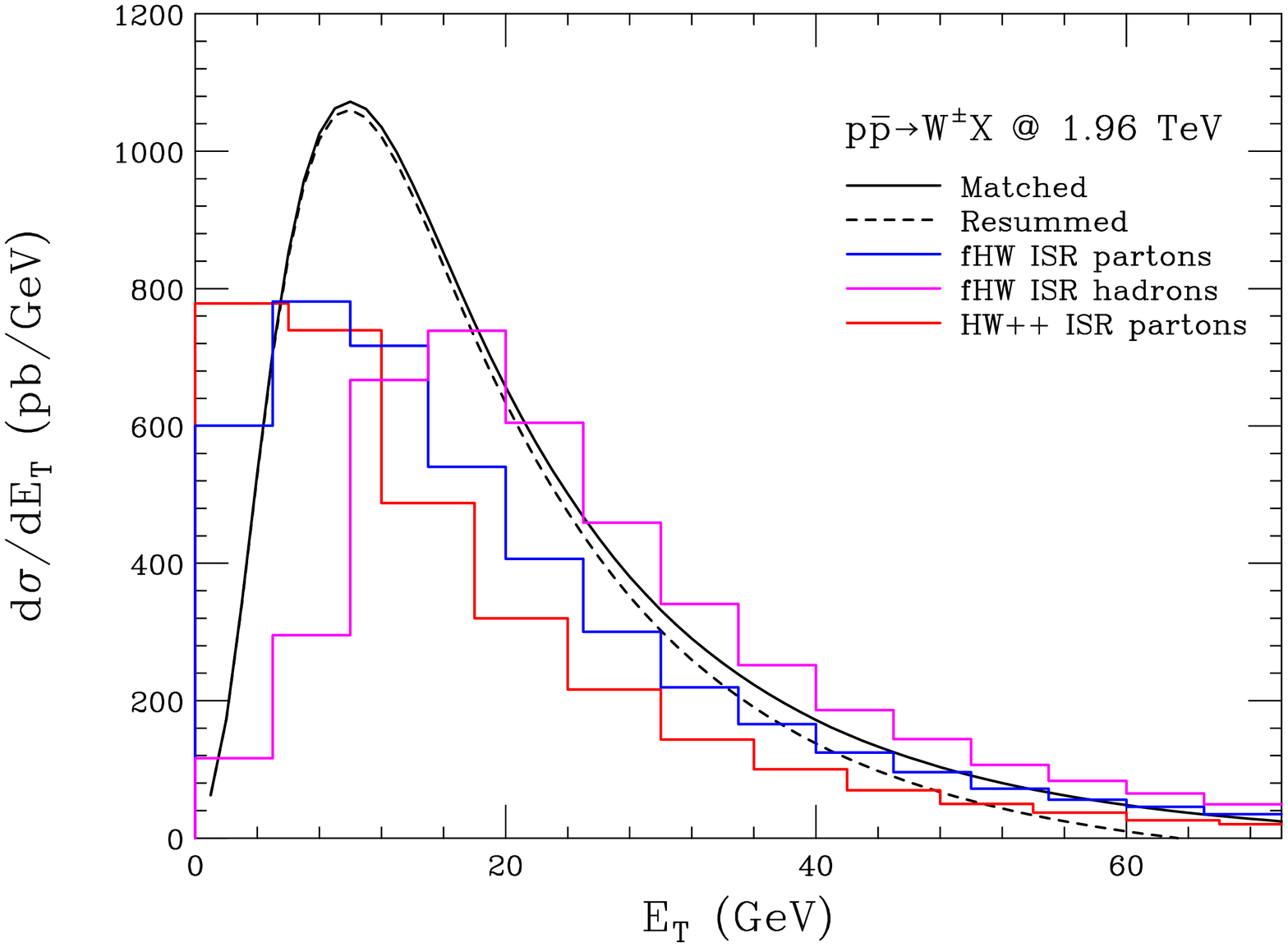}
  \hspace{1.2cm}
  \includegraphics[scale=0.35, angle=90]{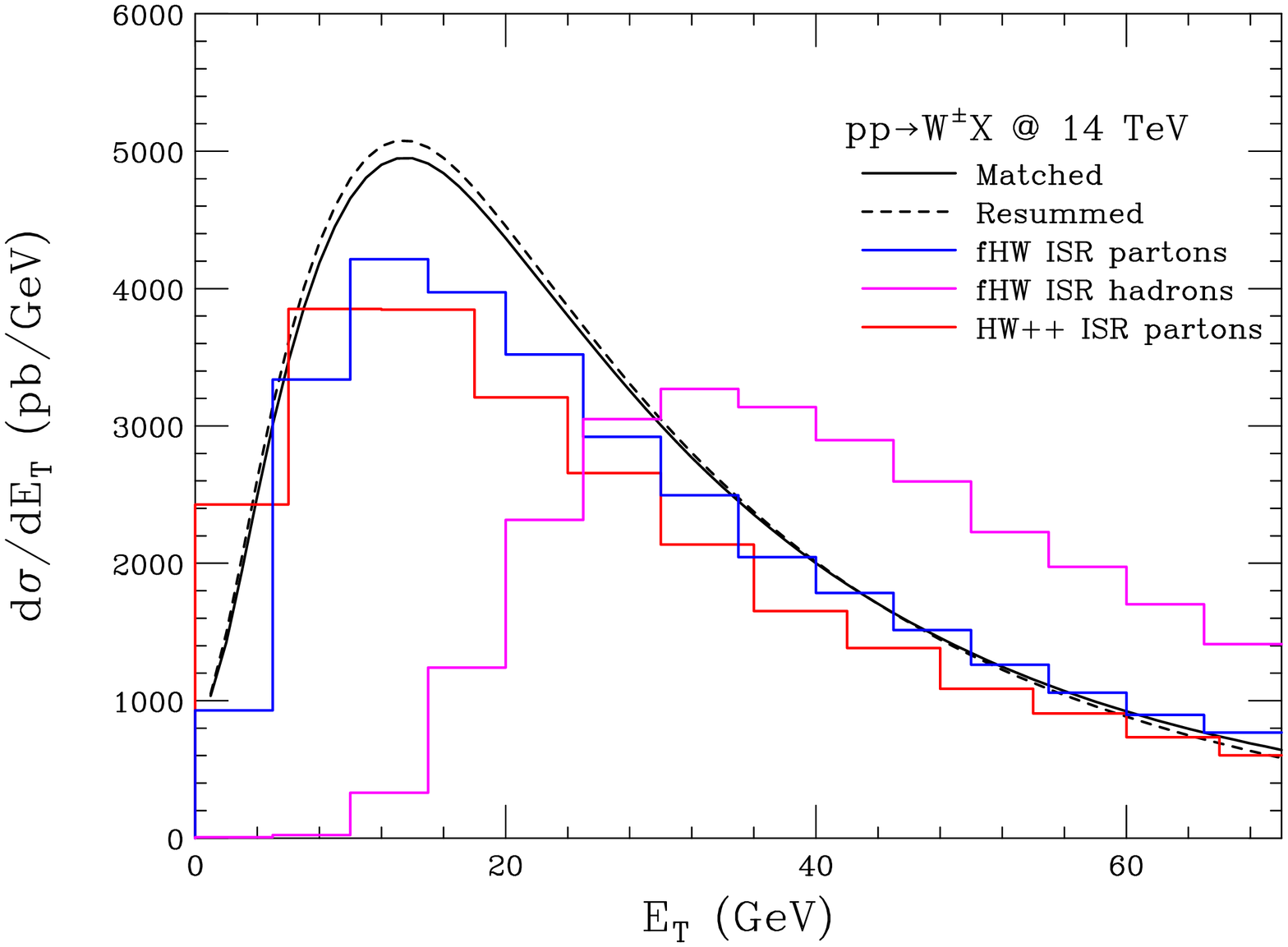}
\end{center}
\caption{Predicted $E_T$ distribution in $W^+$+ $W^-$ boson production  at the
 Tevatron and LHC. Comparison of resummed and Monte Carlo results.
\label{fig:WfHWall} }
\end{figure}

Figure~\ref{fig:ZfHWall} shows the comparisons for $Z^0$ production at
the Tevatron and LHC.  The {\tt HERWIG} predictions are renormalised by
a factor of 1.3 to account for the increase in the cross section from
LO to NLO.  The  \Herwigpp results were not renormalised, because they
were obtained using LO** parton distributions \cite{Sherstnev:2007nd}, 
which aim to reproduce the NLO cross section.  We see that the parton-level 
Monte Carlo predictions of both programs agree fairly well with the matched
resummed results above about 15 GeV, but  \Herwigpp generates a
substantially higher number of events with low values of $E_T$.  A
similar pattern is evident in the results on $W^\pm$ boson production,
shown in Fig.~\ref{fig:WfHWall}.  The effects of hadronization, shown by the
difference between the blue and magenta histograms, are also similar for
both vector bosons.  They generate a significant shift in the
distribution, of around 10 GeV at Tevatron energy and 20 GeV at LHC.

\subsubsection{Higgs boson production}

\begin{figure}[!htb]
\begin{center}
\vspace{1.2cm}
  \includegraphics[scale=0.35, angle=90]{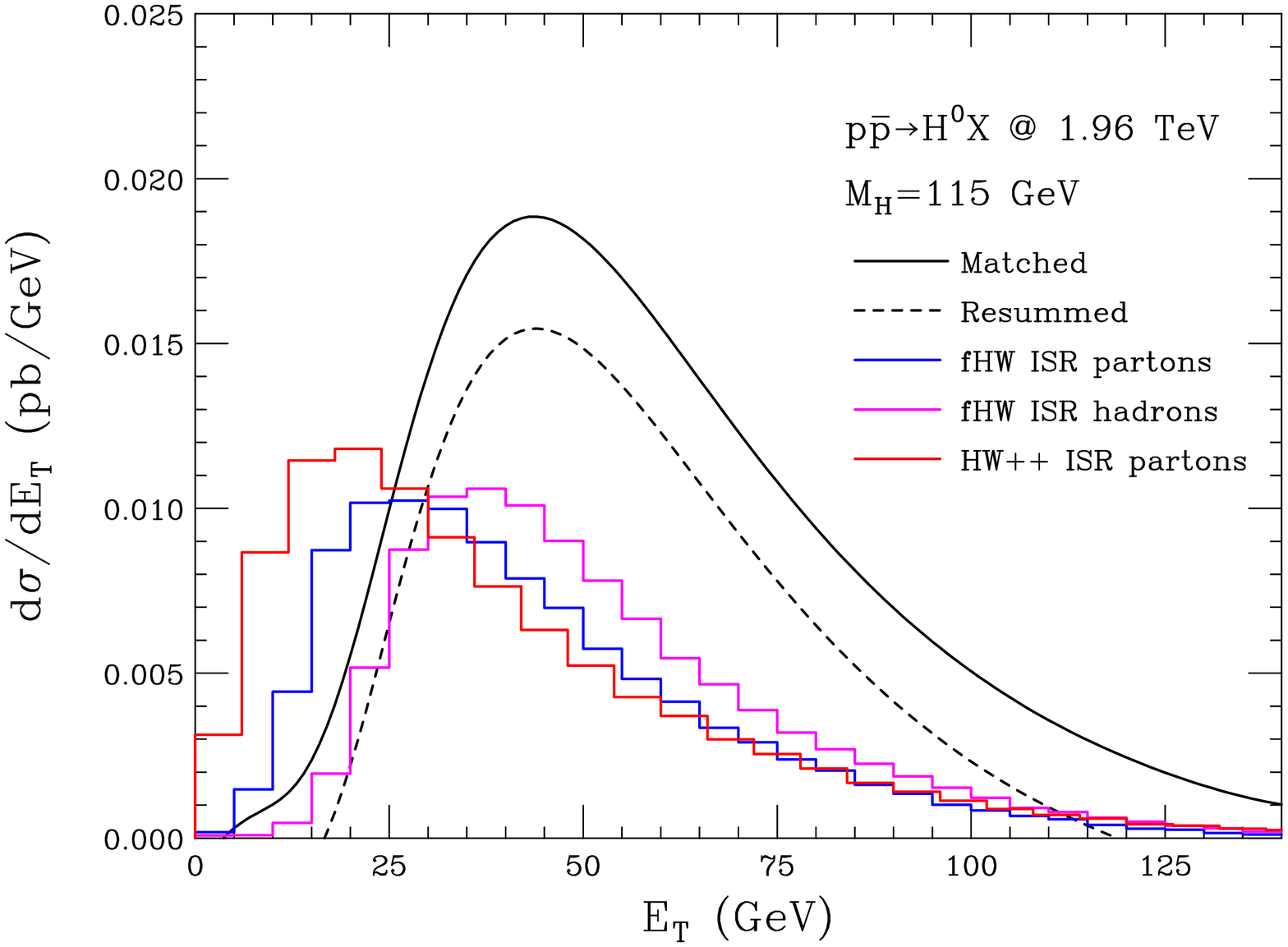}
  \hspace{1.2cm}
  \includegraphics[scale=0.35, angle=90]{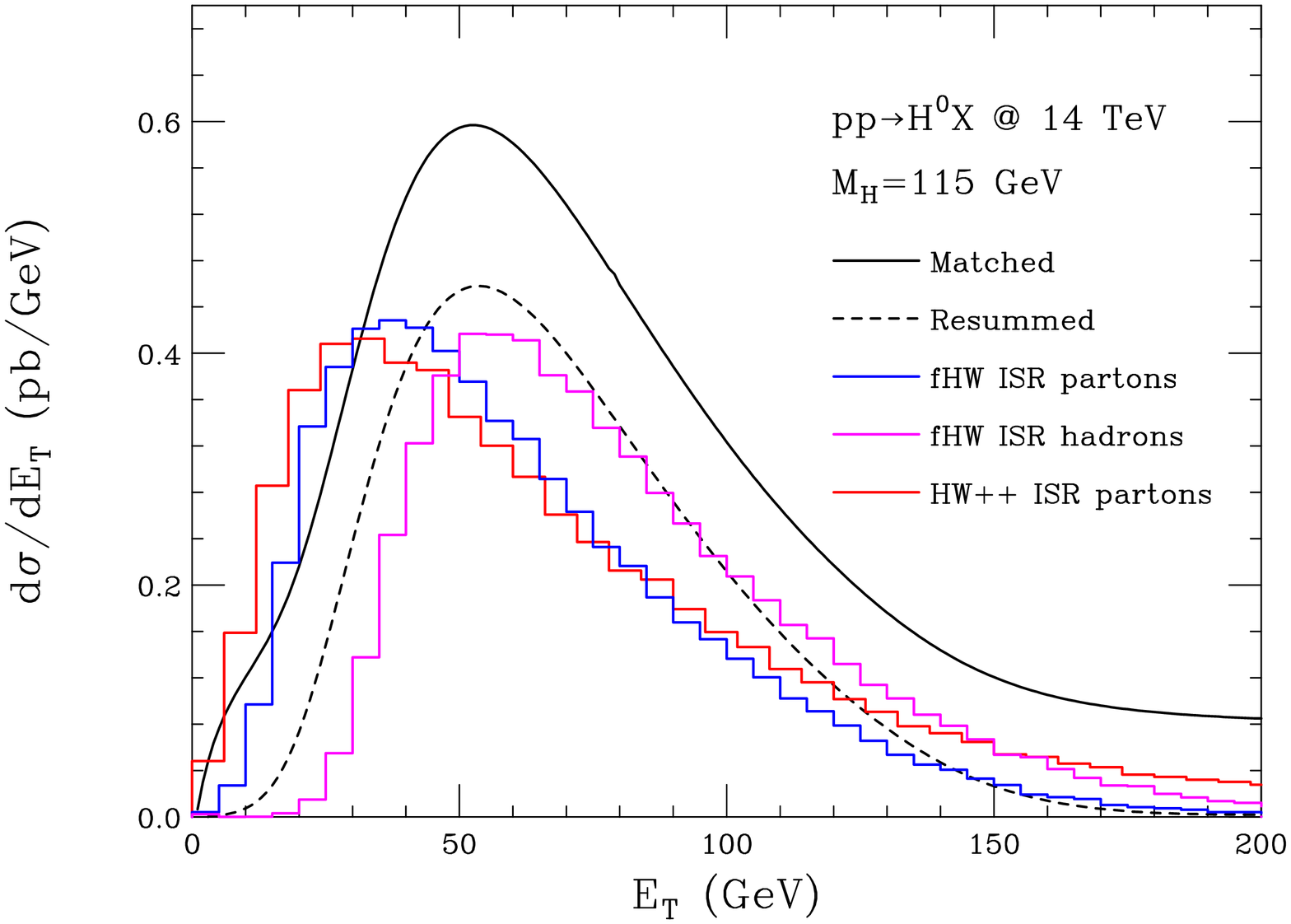}
\end{center}
\caption{Predicted $E_T$ distribution in Higgs boson production  at the
 Tevatron and LHC. Comparison of resummed and Monte Carlo results.
\label{fig:HfHWall} }
\end{figure}

As may be seen from Fig.~\ref{fig:HfHWall}, the agreement between the
resummed and parton-level Monte Carlo results is less good in the case
of Higgs boson production than it was for vector bosons.  Here we have
renormalised the  {\tt HERWIG} predictions by a factor of 2 to allow for
the larger NLO correction to the cross section.  Then the Monte Carlo $E_T$
distributions agree quite well with each other but fall well below the
matched resummed predictions.  Fair agreement above about 40 GeV can
be achieved by adjusting the normalisation, but then the Monte Carlo
programs predict more events at lower $E_T$. The effect of
 hadronization is similar to that in vector boson production, viz.\ a
 shift of about 10 GeV at the Tevatron rising to 20 GeV at the LHC,
 which actually brings the {\tt HERWIG} distribution into
 somewhat better agreement with the resummed result.

\subsubsection{Modelling the underlying event}

\begin{figure}[!htb]
\begin{center}
\vspace{1.0cm}
  \includegraphics[scale=0.35, angle=90]{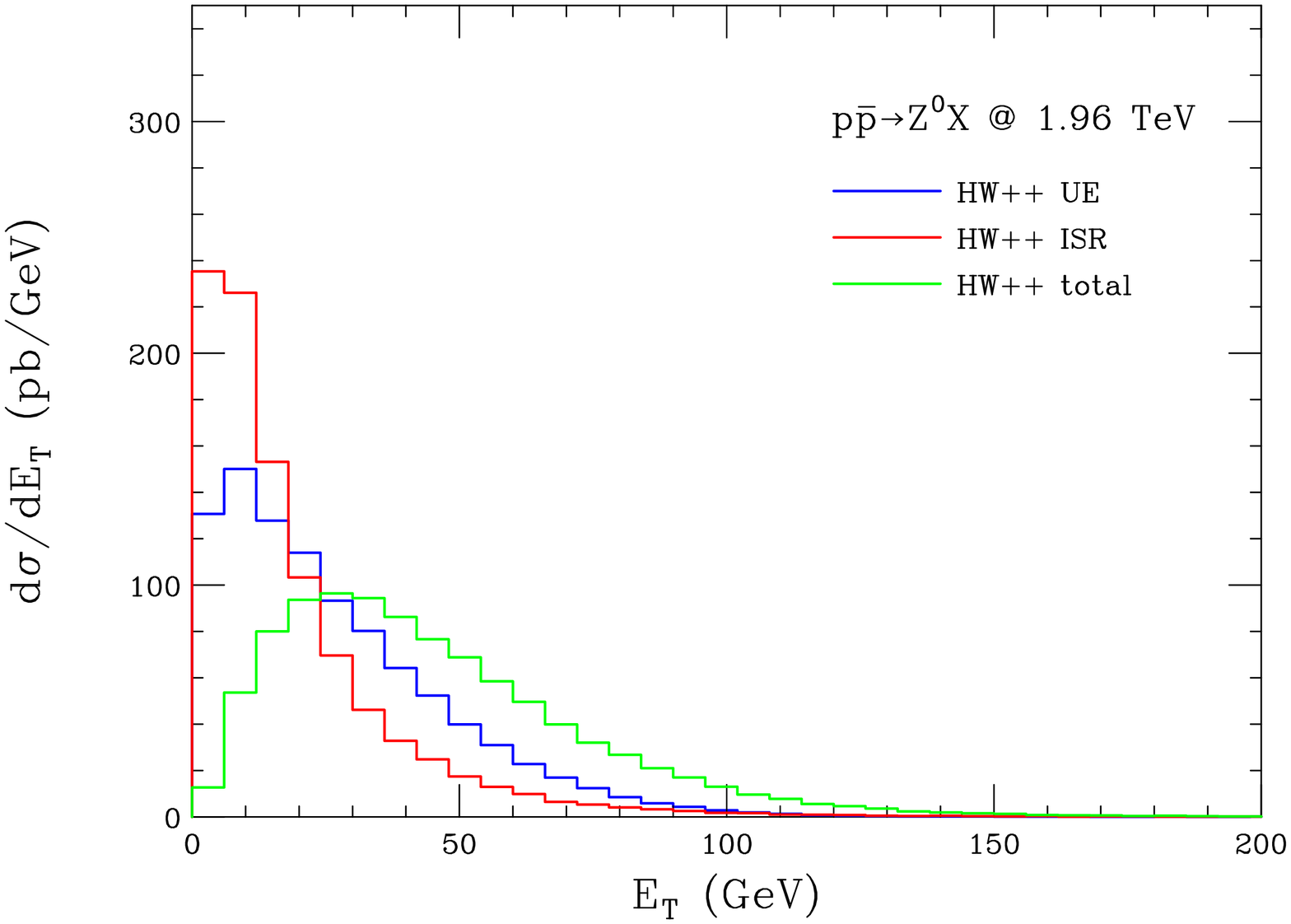}
  \hspace{1.2cm}
  \includegraphics[scale=0.35, angle=90]{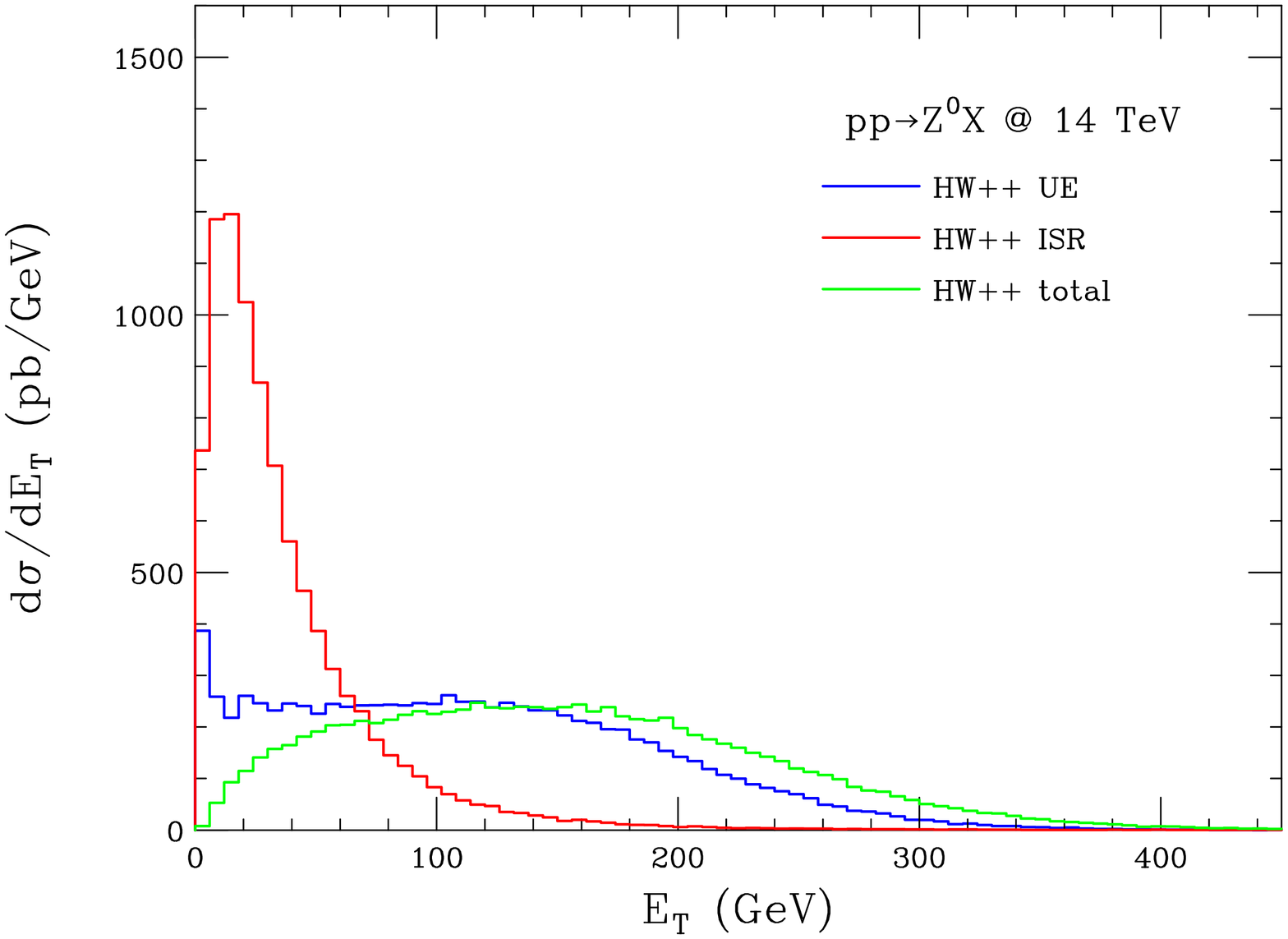}
\end{center}
\caption{Predicted $E_T$ distribution in $Z^0$ boson production  at the
 Tevatron and LHC. Monte Carlo results including underlying event.
\label{fig:ZTevall} }
\end{figure}

\begin{figure}[!htb]
\begin{center}
\vspace{0.7cm}
  \includegraphics[scale=0.35, angle=90]{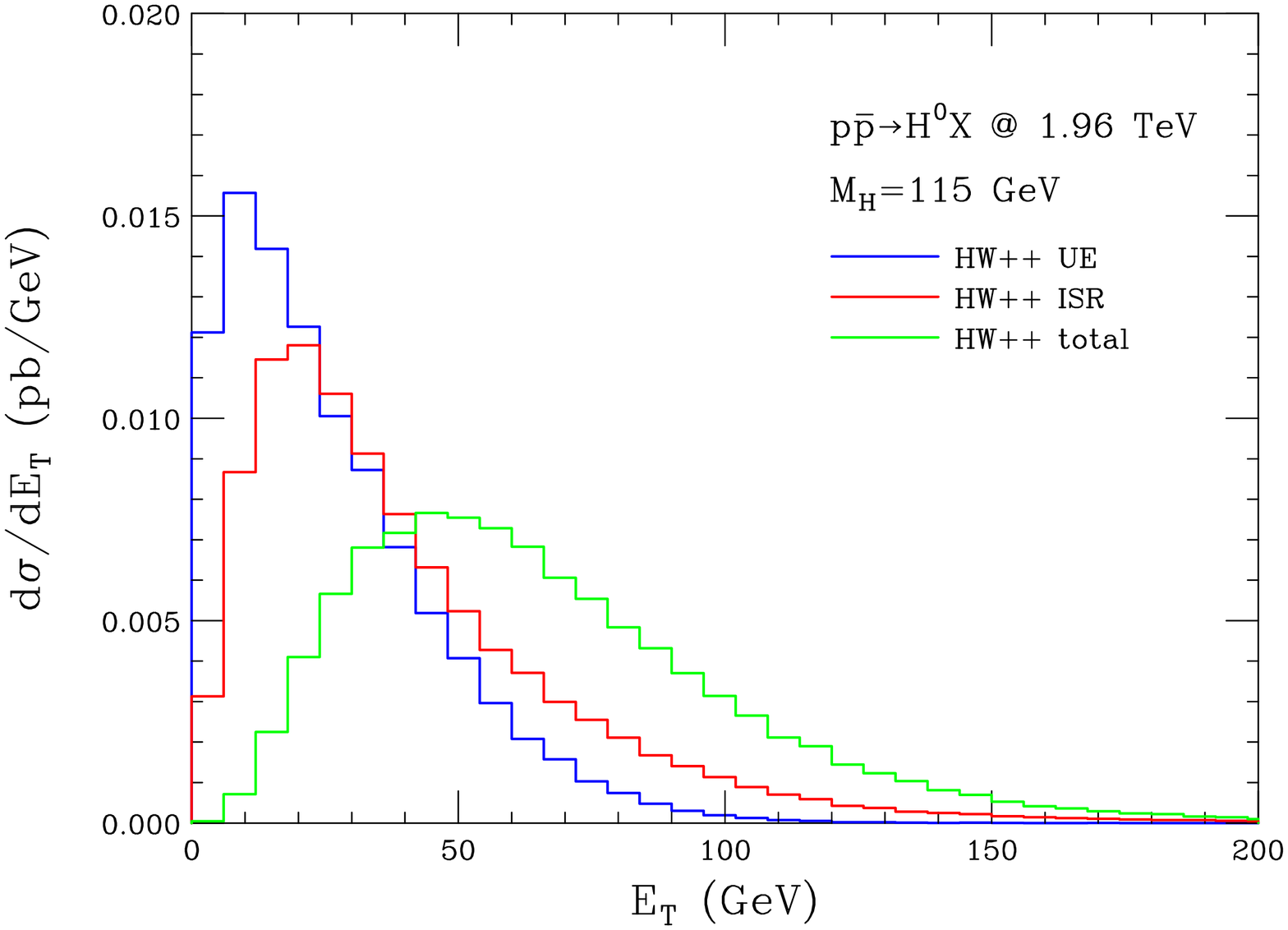}
  \hspace{1.2cm}
  \includegraphics[scale=0.35, angle=90]{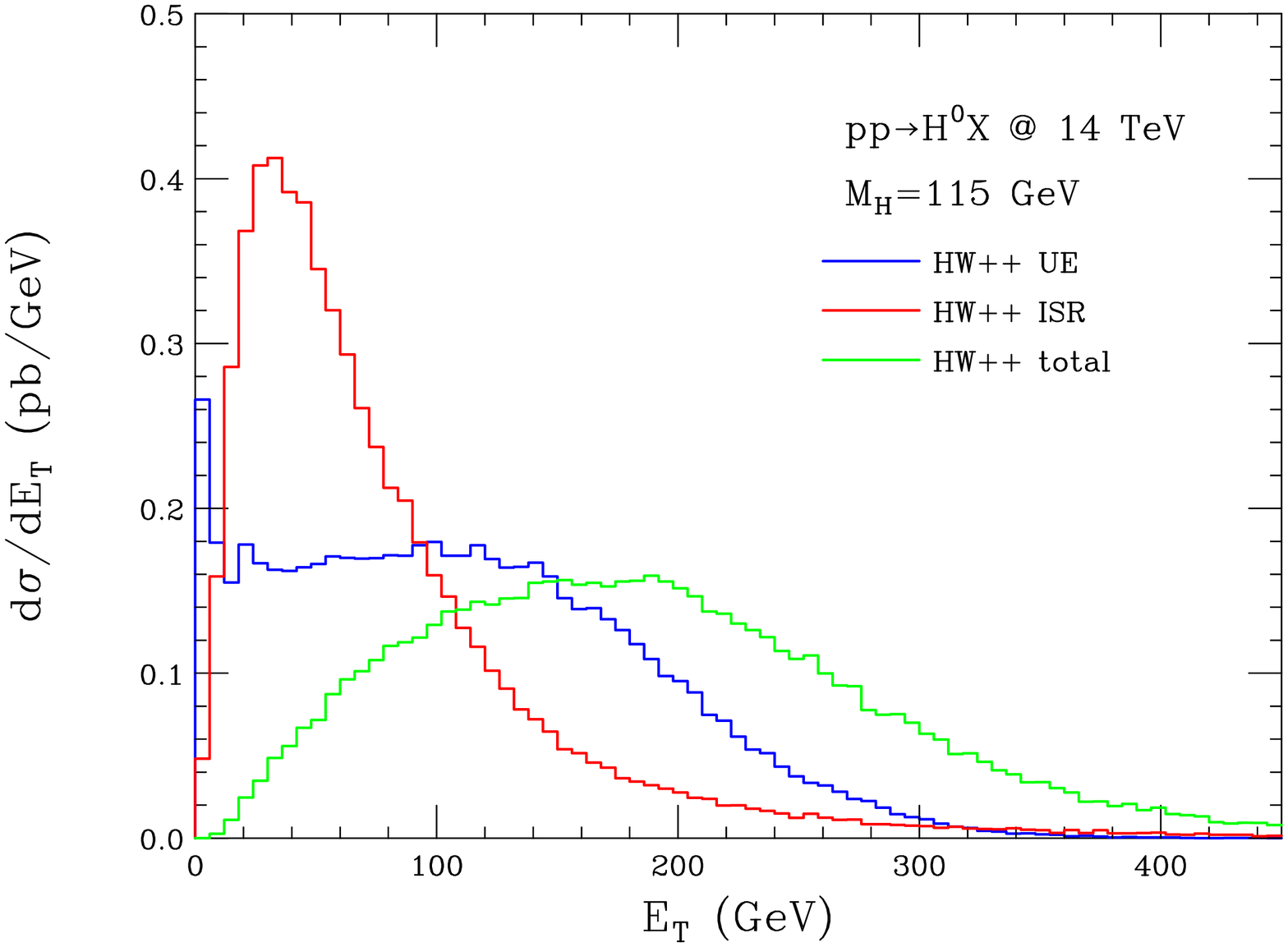}
\end{center}
\caption{Predicted $E_T$ distribution in Higgs boson production  at the
 Tevatron and LHC. Monte Carlo results including underlying event.
\label{fig:HTevall} }
\end{figure}

\begin{figure}[!htb]
\begin{center}
\vspace{2.0cm}
\hspace{5.0cm}
  \includegraphics[scale=0.62, angle=90]{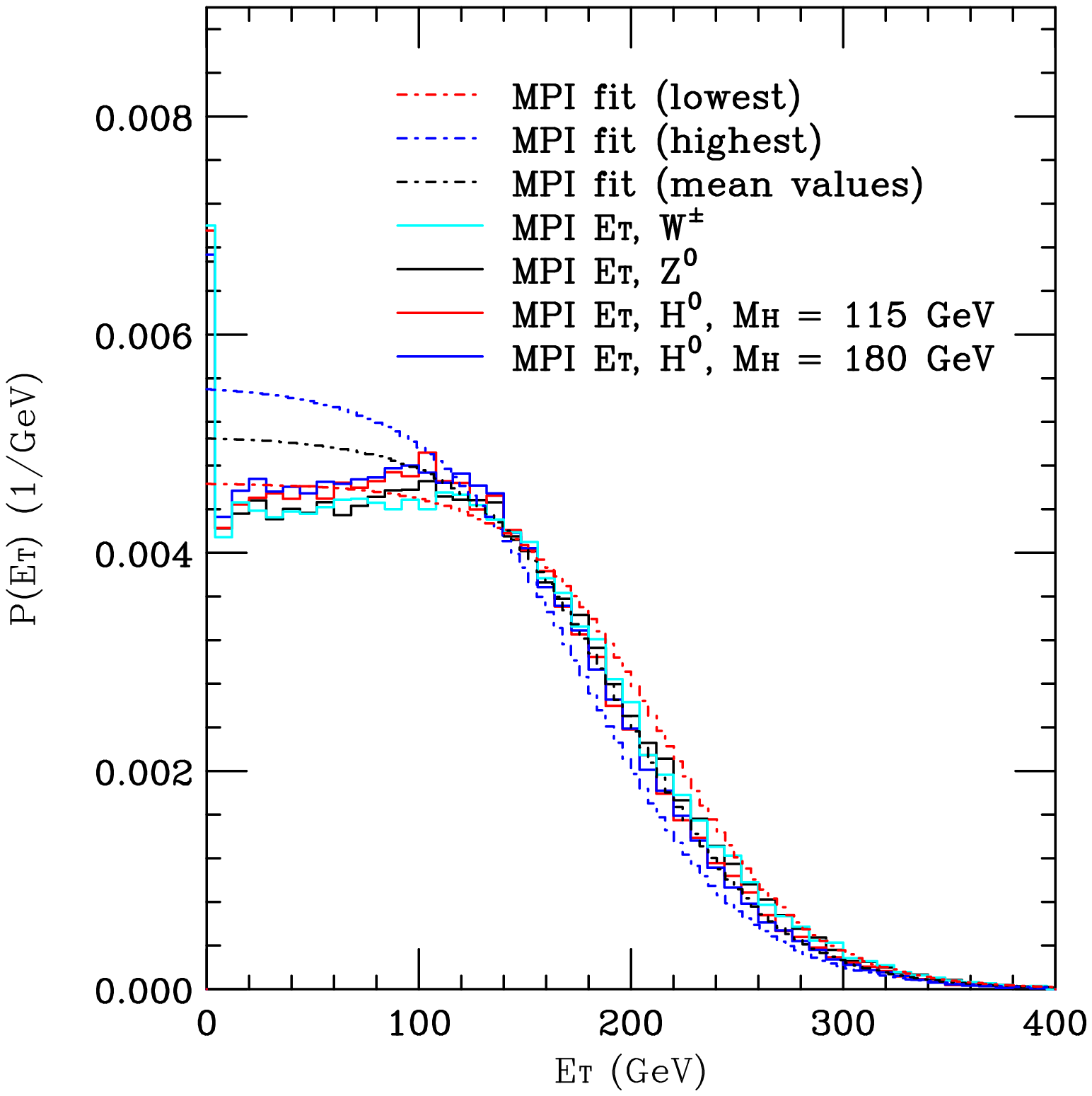}
\end{center}
\caption{Comparison of $E_T$ distributions of the \Herwigpp underlying event
in different subprocesses at the LHC. Fits obtained using the Fermi
distribution of Eq.~\ref{eq:Fermi} for the mean values of the
parameters given below Eq.~\ref{eq:tmu} are shown, as well as two
example fits, obtained by varying the parameters by one standard deviation in
different directions (lowest: $A = 21$, $B = 0.026$, $q = 34$, $r = 0.31$, highest: $A = 19$, $B = 0.034$, $q = 38$, $r = 0.25$).}
\label{fig:MPIcomp}
\end{figure}

Figures~\ref{fig:ZTevall} and \ref{fig:HTevall}  show the parton-level  \Herwigpp
predictions for the $E_T$ distribution in $Z^0$ and Higgs boson
production, respectively, with the contributions from initial-state
radiation (in red, already shown in Figs.~\ref{fig:ZfHWall}
and~\ref{fig:HfHWall}), the underlying event (blue) and the
combination of the two (green).  As we have already seen in section~\ref{sec:mc:ue}, the underlying event is modelled
using multiple parton interactions.
Clearly it has a very significant effect on the $E_T$ distribution.
However, this effect is substantially independent of the hard
subprocess, as we have already found in section~\ref{sec:MPI} when
examining the total invariant mass, $M$. This can also be seen from the
comparison of the $E_T$ of the UE associated to different
subprocesses in Fig.~\ref{fig:MPIcomp}.

\begin{figure}[!htb]
\begin{center}
\vspace{1.2cm}
  \includegraphics[scale=0.35, angle=90]{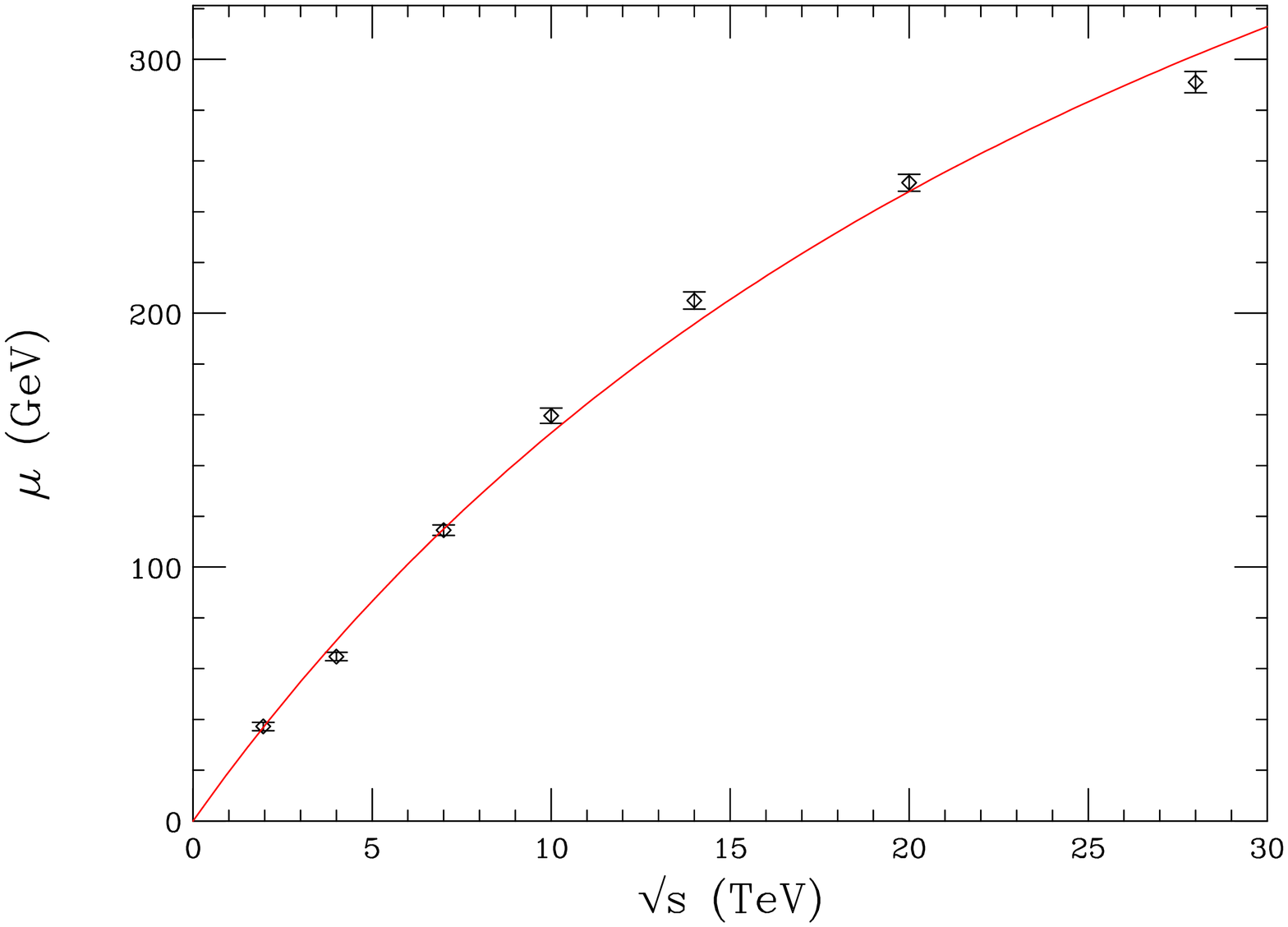}
  \hspace{1.2cm}
  \includegraphics[scale=0.35, angle=90]{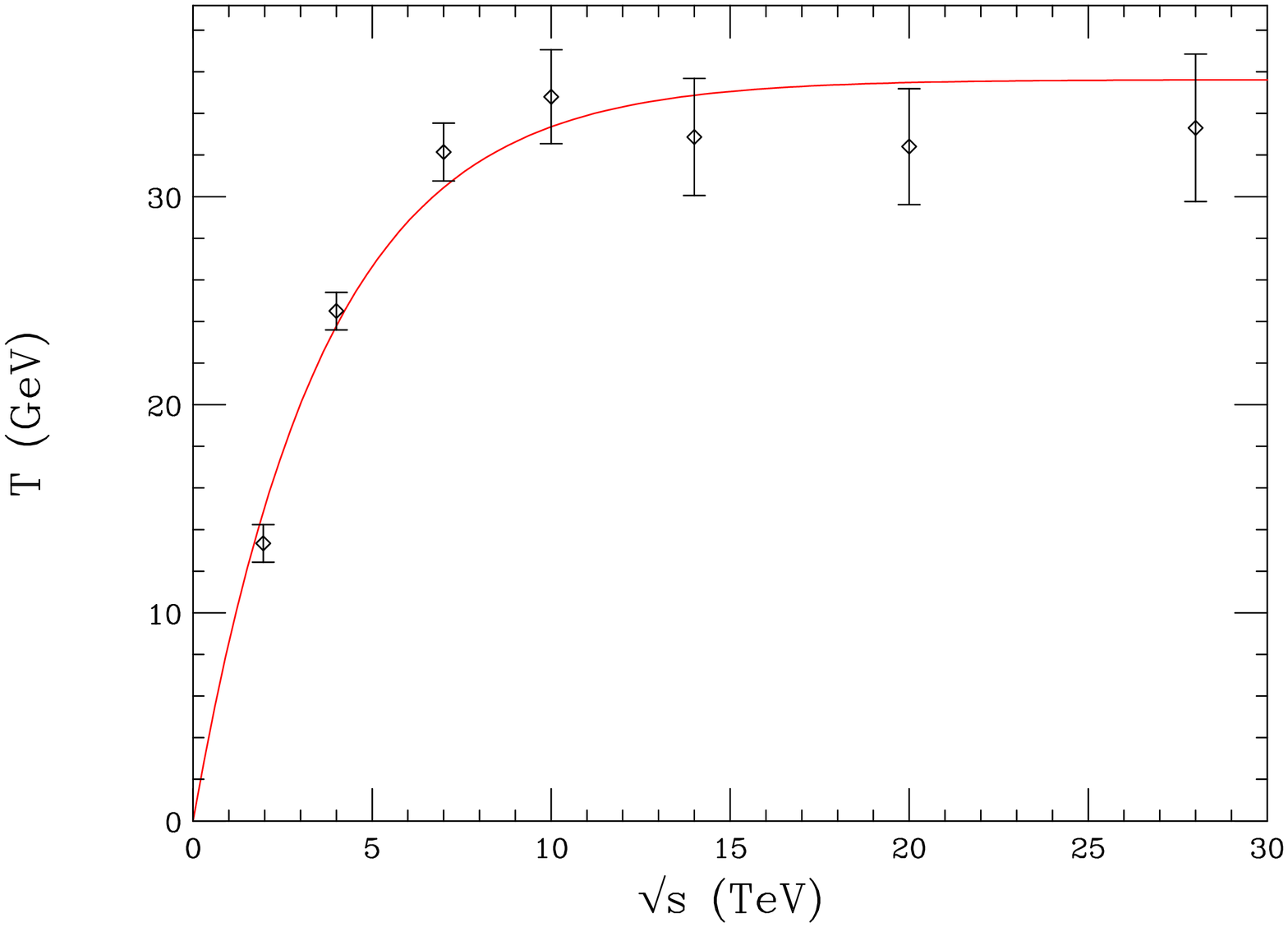}
\end{center}
\caption{Fitted values of the parameters of the \Herwigpp underlying event in
  Higgs production in $pp$ collisions at various energies.
\label{fig:MPIfits} }
\end{figure}

We find that the probability distribution of the $E_T$ contribution of
the underlying event in the \Herwigpp Monte Carlo can be
represented quite well by a Fermi distribution:
\beq\label{eq:Fermi}
P(E_T) = \frac 1{\cal N}\frac 1{\exp\left(\frac{E_T-\mu}T\right)+1}\;,
\eeq
for which the normalisation, $\mathcal{N}$, is given by
\beq
{\cal N} = T\ln\left[\exp\left(\frac{\mu}T\right)+1\right]\;.
\eeq
The dependence of the `chemical potential' $\mu$ and `temperature'
$T$ on the hadronic collision energy is shown in
Fig.~\ref{fig:MPIfits}.  The red curves show fits to the energy
dependence of the form:
\beq\label{eq:tmu}
\mu =\frac{A\rs}{1+B\rs}\,,\quad
T = q\left(1-{\rm e}^{-r\rs}\right)\;,
\eeq
where the coefficients in the fits are $A = 20(1)$, $B = 0.030(4)$,
$q = 36(2)$, $r = 0.28(3)$. Example fits for the LHC case, $\rs = 14
\tev $, are shown in Fig.~\ref{fig:MPIcomp}.
\subsection{Conclusions}\label{sec:etres:conc}
We have extended the resummation of the hadronic transverse energy $E_T$ in vector
boson production to next-to-leading order (NLO) in the resummed exponent,
parton distributions and coefficient functions, and also presented for the
first time the corresponding predictions for Higgs boson production.
We have matched the resummed results to the corresponding ${\cal O}(\as)$
predictions, by adding the contributions in that order which are not included
in the resummation.  In addition we have compared our results to parton shower Monte Carlo
predictions and illustrated the effects of hadronization and the
underlying event.

In the case of vector boson production, the resummation procedure
appears stable and the parton-level results should be quite reliable.
The leading-order mechanism of quark-anti-quark annihilation typically
generates a moderate amount of transverse energy in initial-state QCD
radiation.  Consequently  the effects of subleading resummed terms and
fixed-order matching are small and the peak of the $E_T$ distribution
lies well below the boson mass scale, where resummation makes good
sense.  The comparisons with Monte Carlo programs reveal some
discrepancies but these are at the level of disagreements between
different programs; in this case the resummed predictions should be
more reliable (at parton level) than existing Monte Carlo programs.  The
programs suggest that the non-perturbative effects of hadronization
and the underlying event are substantial.  These effects can however
be modelled in a process-independent way.  We have suggested a
simple parametrization of the contribution of the underlying event
through the model given in the \Herwigpp event generator. We stress once
again, however, that recent UE results from the LHC experiments have shown that
this model does not describe the data adequately~\cite{Aad:2011qe},
and as of version 2.5.0, \Herwigpp includes an improved
description~\cite{Gieseke:2011na} via a colour reconnection model. The
effect of the improved model on the $E_T$ distributions remains to be
investigated in future work.

The situation in Higgs boson production is not so good.  The dominant
mechanism of gluon fusion generates copious ISR and the effects of
subleading terms and matching are large.  The resummed $E_T$
distribution peaks at a value that is not parametrically smaller than
the Higgs boson mass and the behaviour at low and high $E_T$ is unphysical
before matching. The discrepancies between the matched resummed and
Monte Carlo predictions are substantially greater than those between
different programs, even allowing for uncertainties in the overall
cross section. All this suggests that there are significant higher-order
corrections that are not taken into account, either further subleading
logarithms or unenhanced terms beyond NLO.  It would be
interesting (but very challenging) to attempt to extract such terms
from the available NNLO calculations of Higgs boson production.

\begin{boldmath}
\chapter{New physics searches at hadron colliders}
\label{cha:NEWPHYS}
\end{boldmath}
The original work in this chapter was done in collaboration with
Oluseyi Latunde-Dada~\cite{Papaefstathiou:2009sr}, Bryan
Webber, Kazuki Sakurai and Ben Gripaios~\cite{Gripaios:2010hv}.
\section{Introduction}
\label{sec:newphys:intro}
The challenges present
at hadron colliders have been discussed in
section~\ref{sec:qcdeff:intro}, where we emphasised that new physics signals
can be difficult to observe and interpret. The difficulties arise due to multiple jets and/or
leptons, the presence of invisible particles and the huge backgrounds
that may potentially imitate the topology and kinematics of a signal
of relatively low rate. It is thus extremely important to investigate
the phenomenology of concrete models of new physics. At the same time,
we need to ensure that we do not
weaken our reach due to biases introduced by focusing
on specific models. The strategies that we develop must allow us to explore the possibilities for generic new physics signatures,
motivated by theoretically and experimentally plausible
models. Furthermore, it is important for the observables that we
construct to be well-defined and calculable so they can provide
unambiguous information for the discovery of new physics
and determination of its properties.

In this chapter we investigate two different new physics
scenarios. In section~\ref{sec:newphys:wprime} we present an investigation
on a generic model that involves the production of a new heavy charged
vector boson, called $W'$ ($W$ prime), essentially a heavy version of
the standard model $W$ gauge boson. Such heavy bosons may arise from
the breaking of a large symmetry group to the SM symmetry group, as
excitations of the SM $W$ in models with extra dimensions, or as a
composite particle in a strong dynamics theory. We study the
interference effects of a $W'$ with the SM $W$ and extend the treatment
to next-to-leading order by using the \texttt{MC@NLO} and
\texttt{POWHEG} methods described in
section~\ref{sec:mc:nlomatching}. We incorporate these features in a
publicly available event generator and use this to examine the detection reach at the Tevatron and the LHC. In
section~\ref{sec:newphys:leptoquarks}  we examine a model that
contains composite leptoquarks, particles that couple to leptons and
quarks (as the name suggests), which may arise in
strong dynamics theories. These couple primarily to the third
generation of fermions, a feature motivated by a model of fermion mass
generation which aims to solve the problem of flavour-changing neutral
currents in theories where electroweak symmetry is broken via strong
dynamics. We propose a general search
strategy for discovery and mass reconstruction of leptoquarks.\footnote{It is interesting to mention that these two scenarios of new physics are not
mutually exclusive. Both leptoquarks and heavy vector bosons may arise
in the same model. This could be, for example, a strong dynamics theory
that contains composite scalar particles, acting as leptoquarks, and composite vector
particles, acting as $W'$'s. In these models the $W'$ bosons would
also potentially couple preferentially to the fermions of the third generation.} 

\section{NLO production of heavy charged vector bosons}
\label{sec:newphys:wprime}
\subsection{Introduction}\label{sec:wprime:intro}
There exists a proliferation of theories which contain new heavy,
electrically neutral or charged, gauge bosons referred to as $Z'$ and $W'$
respectively. Both the $Z'$ and $W'$  have been studied extensively and
reviews can be found in~\cite{Nakamura:2010zzi}
and~\cite{Langacker:2008yv} respectively. The study of this section focuses on $W'$
bosons. 

The new charged vector bosons may or may not have similar
properties to the SM bosons, depending on the underlying theory. In particular they may have right-handed instead of left-handed
couplings, may couple to new fermions, or may even be
fermiophobic. Models which predict new charged vector bosons
may be based on extensions of the electroweak gauge group, $SU(2)\times
U(1)$, for example to the gauge group $SU(2)_1 \times SU(2)_2\times
U(1)$~\cite{Mohapatra:1974hk}, or groups that contain the electroweak
symmetry, such as $SU(3) \times U(1)$ or $SU(4) \times U(1)$~\cite{Pisano:1991ee}. Several models with extra dimensions contain
$W'$ bosons as Kaluza-Klein excitations in the bulk. Examples of these
models include the Randall-Sundrum model (section~\ref{sec:bsm:extradim}) with bulk gauge fields~\cite{Randall:1999ee} and Universal Extra Dimensions~\cite{Appelquist:2000nn, Cheng:2002ab}. Theories which break the
electroweak sector via strong dynamics may also contain the $W'$ as a
composite spin-1 particle~\cite{Hill:2002ap, Chivukula:2003wj}.  

Current Monte Carlo simulations of Drell-Yan-type $W'$ production at
hadron colliders rely mainly on leading-order matrix elements and
parton showers.  There exists no treatment of next-to-leading (NLO)
QCD effects which simultaneously includes the interference
effects for the $W'$. Here, we present the results of the event
generator package \texttt{Wpnlo}~\cite{wpnlo} which improves the
treatment of Drell-Yan production of heavy charged gauge bosons. We
consider the interference effects with the Standard Model $W$, which
have been shown to provide valuable information~\cite{Rizzo:2007xs},
but have not been considered in experimental searches. We use the
\texttt{MC@NLO} and \texttt{POWHEG} methods, discussed in
section~\ref{sec:mc:nlomatching} to match the NLO QCD calculation to
the parton shower, producing fully exclusive events using the \Herwigpp
event generator. Note that a similar implementation for the $Z'$ exists for the NLO \texttt{MC@NLO} event generator, which matches the complete NLO matrix elements with the parton shower and cluster hadronization model of the Fortran \texttt{HERWIG} event generator~\cite{Fuks:2007gk}.
\subsection[$W'$ at leading order]{\boldmath $W'$ at leading order}\label{sec:wprime:refmodel}

\begin{figure}[t]
  \centering 
  \vspace{0.5cm}
  \includegraphics[scale=0.70, angle=0]{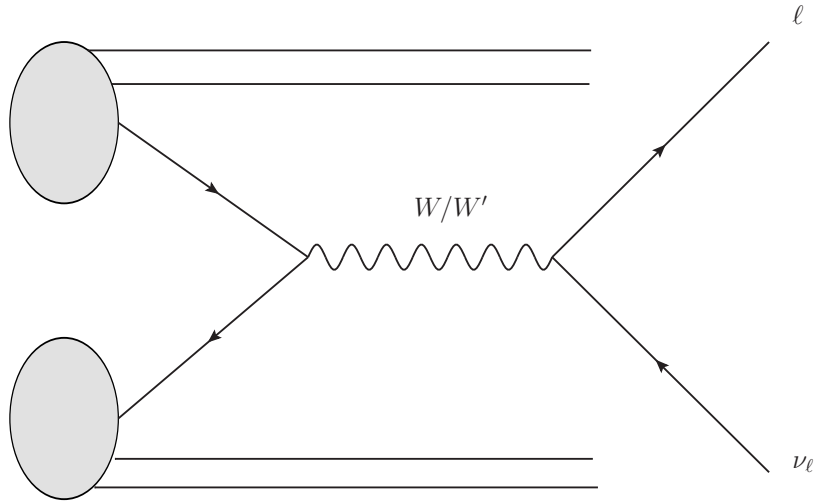}
\caption{Schematic diagram for the Drell-Yan process $pp \rightarrow W/W' \rightarrow \ell \nu X$.}
\label{fig:ppWlnu}
\end{figure}
The $W'$ reference model is based on the one which originally appeared
in Ref.~\cite{Altarelli:1989ff}. In the model described therein,
sometimes referred to as the Sequential Standard Model, the magnitudes
of the $W'$ couplings to fermions are directly transcribed from the SM
$W$, i.e. it is a heavy copy of the SM $W$. In the present treatment
we allow both right- and left-handed couplings, $\propto (1\pm \gamma_5)$ respectively. In the case of right-handed couplings, we assume that the right-handed neutrinos are light compared to the $W'$, but not light enough for the $Z$ boson to decay into them. The $W'$ and $W$ couplings to fermions are given by
\begin{equation}\label{eq:coupling}
\mathcal{L}_{W_iff'} = \left(\frac{G_FM_W^2}{\sqrt{2}}\right)^{1/2}V_{ff'}C_i^{\ell,q}\bar{f}\gamma_{\mu}(k_i
- h_i \gamma_5)f'W^{\mu}_i + \mathrm{h.c.}\;,\;\;\;i = \{W, W'\},
\end{equation}
where $G_F$ is the Fermi coupling constant,\footnote{The constant
  $G_F$ is related to the weak coupling constant $g$, which appears in
 Eq.~(\ref{eq:sm:wff}), by $G_F/ \sqrt{2} = g^2 / (8 M_W^2)$.} $M_W$ is the SM $W$ mass, $C^{\ell,q}_i$ are the coupling
  strengths of boson $i$ to leptons and quarks respectively, $W^{\mu}$
  is the massive boson polarisation vector, $f$ and $f'$ are the Dirac
  spinors for the fermions and $V_{ff'}$ is the unit matrix when $ff'$
  are leptons and the CKM matrix, given in appendix~\ref{sec:ckm},
  when $ff'$ are quarks. The $k_i$ and $h_i$ represent the structure
  of the vector-axial vector (V-A) coupling of the bosons, where for
  the case $i=W$ we have $k_W=h_W = 1$, i.e. purely left-handed
  coupling. Using the above coupling to fermions, we show in 
  appendix~\ref{app:wprime:crosssection} that the hadronic
  differential cross section for the process $pp \rightarrow
  W^+/W'^+ \rightarrow \ell \nu X$ (Fig.~\ref{fig:ppWlnu})  is given
  by
\begin{equation}\label{eq:xsection}
\frac{\mathrm{d} \sigma}{ \mathrm{d} \tau \mathrm{d} y \mathrm{d} z} = \frac{G_F^2M_W^4}{192 \pi} \sum_{qq'}|V_{qq'}|^2 \left[S G^{+}_{qq'}(1+z^2) + 2AG^{-}_{qq'}z\right]\;,
\end{equation}
where $z=\cos \theta$ is defined as the scattering angle between the
$u$-type quark and the outgoing neutrino (both being fermions) in the
centre-of-mass frame, $y$ is the rapidity of the intermediate
boson, $\tau = \hat{s} / s$ is the ratio of the squares of the quark
centre-of-mass energy to the proton centre-of-mass energy. $S=S(\hat{s})$ and $A=A(\hat{s})$
are functions of the quark centre-of-mass energy:
\begin{equation}\label{eq:Sterm}
S = \sum_{i,j} S_{i,j} = \sum_{i,j}P_{ij} (C_iC_j)^\ell (C_iC_j)^q (k_ik_j + h_ih_j)^2\;,
\end{equation}
\begin{equation}\label{eq:Aterm}
A = \sum_{i,j}A_{i,j} = \sum_{i,j}P_{ij} (C_iC_j)^\ell (C_iC_j)^q (k_jh_i + h_jk_i)^2\;,
\end{equation}
with
\begin{equation}\label{eq:pterm}
P_{ij} = \hat{s} \frac{ (\hat{s} - M_i^2) (\hat{s} - M_j^2) + \Gamma_i \Gamma_j M_i M_j } { [(\hat{s} - M_i^2)^2 + \Gamma_i^2 M_i^2][i\rightarrow j]}\;,
\end{equation}
where $i,j$ can be either $W$ or $W'$ and $M_i$, $\Gamma_i$ are the
mass and width of boson $i$ respectively. The functions
$G^{\pm}_{qq'}$ that appear in the differential cross section are even or odd products of parton density functions for the relevant hadrons, given by
\begin{equation}\label{eq:Gpm}
G^{\pm}_{qq'} = \left[f_{q/A}(x_a, \hat{s})f_{q'/B}(x_b, \hat{s}) \pm f_{q/B}(x_b, \hat{s})f_{q'/A}(x_a, \hat{s})\right]\;,
\end{equation} 
where $f_{q/h}(x,\hat{s})$ is the parton density function for a quark
$q$ in a hadron $h$ carrying hadron momentum fraction $x$, in a
collision in which the quark pair centre-of-mass energy squared is
$\hat{s}$. The $A,B$ indices represent the type of the `left'
(travelling in the positive $z$-direction) or `right' (travelling in
the negative $z$-direction) hadrons respectively. This definition
allows for easy modification of the $pp \rightarrow W/W' \rightarrow
\ell \nu X$ cross section to the $p\bar{p} \rightarrow W/W'
\rightarrow \ell \nu X$, by changing the PDFs accordingly. Analogous
expressions can also be obtained for the case of ($W^{-},~W'^{-}$), by appropriately modifying the functions $G^{\pm}_{qq'}$ and taking $z\rightarrow-z$.
The width can be taken to be a free parameter in the reference model:
the couplings of the $W'$ to other gauge bosons or the Higgs boson are
model-dependent.\footnote{An exception is the photon, for which the
  coupling is fixed by gauge invariance.} Here we shall assume for
illustration that the fermionic decay width\footnote{The fermionic
  decay width is thus also taken to be the lower bound on the total width.} scales with the mass as
$\Gamma_{W'\rightarrow ff'} = (4\Gamma_{W}M_{W'}/ 3M_{W})$ (provided
that $M_{W'} \gg M_{t}$, the mass of the top quark)\footnote{The factor of 4/3 comes from
  the extra decay channel that opens up when $M_{W'} > M_{\mathrm top}$, e.g. $W'^+ \rightarrow t\bar{b}$.} and that the
tri-boson $W'WZ$ vertex is suppressed by a small mixing angle and can
be neglected in the analysis.

\subsection[$W-W'$ interference]{\boldmath $W-W'$ interference}
The narrow width approximation (NWA) is often used when discussing the
production of new vector bosons. This approximation is usually claimed
to be valid up to $\mathcal{O}(\Gamma_{W'}/M_{W'})$ corrections. However,
$W$-$W'$ interference effects can become important in certain regions
even as the width $\Gamma_{W'}\rightarrow 0$, see for
example~\cite{Rizzo:2007xs}, and as we also show below. Use of the NWA
may thus lead to invalid conclusions, as pointed out
in~\cite{Berdine:2007uv}.

\begin{figure}[!htb]
  \begin{center}
  \includegraphics[scale=0.75,angle=0]{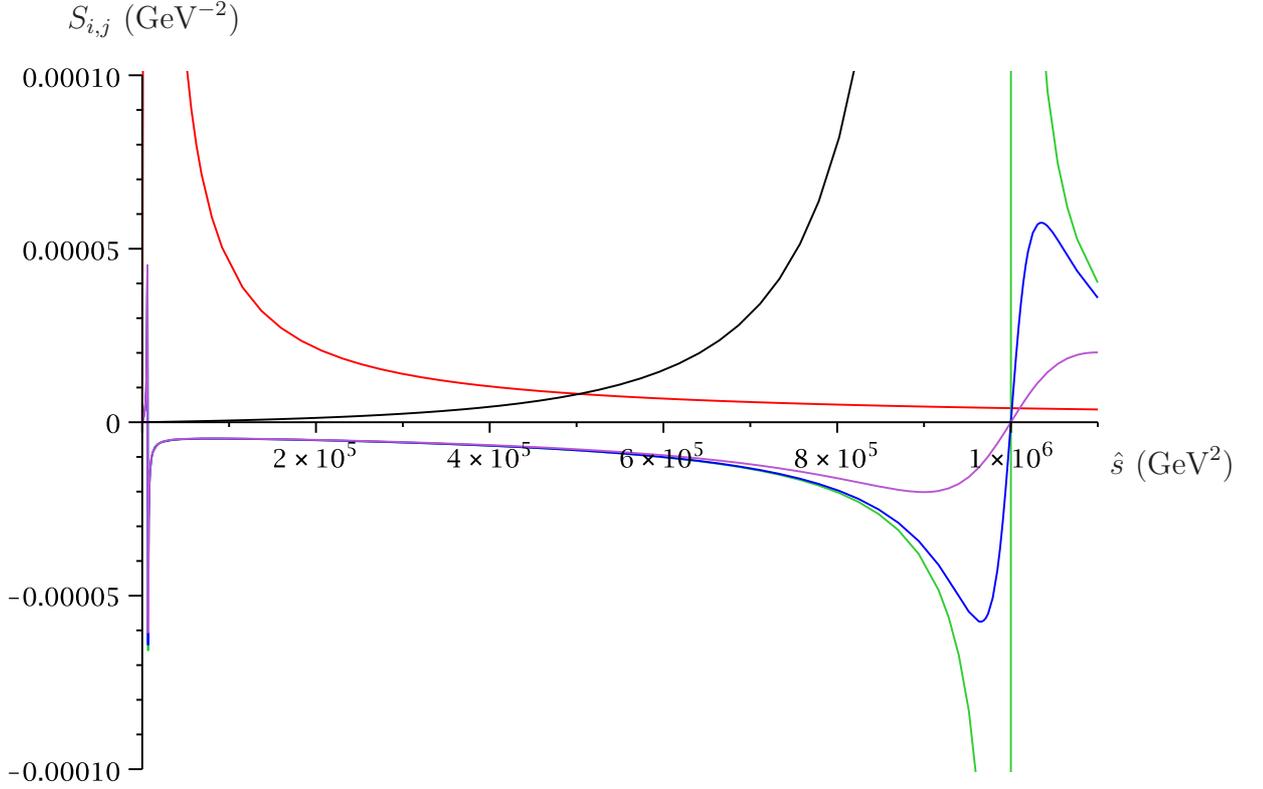}
  \put(0,122){$\hat{s}~(\mathrm{GeV}^2)$}  
  \put(-390,290){$S_{i,j}~(\mathrm{GeV}^{-2})$}

  \end{center}
\caption{The interference term $S_{int}$ for $h_{W'} = 1$, $M_{W'} = 1~\mathrm{TeV}$, plotted against $\hat{s}$, for different widths: $\Gamma_{W'} = 1, 35, 100~\mathrm{GeV}$ (green, blue, purple respectively). The terms $S_{W,W}$ (red) and $S_{W',W'}$ (black) are shown for comparison. It is evident that $S_{int}$ is negative in the intermediate region $(M_W^2,M_{W'}^2)=
(\sim6400~\mathrm{GeV}^2,10^6~\mathrm{GeV}^2)$. It is also clear that
the magnitude of the interference term is comparable to $S_{W,W}$ and
$S_{W',W'}$. As the width decreases the negative peak becomes
narrower, but there always exists a portion of the curve in the
intermediate region which is independent of the width.}
\label{fig:sint}
\end{figure}

Interference effects arise because the Drell-Yan process $pp
\rightarrow W/W' \rightarrow \ell \nu X$ can proceed either via an intermediate SM $W$ or a $W'$ in the reference model. The matrix element squared for the process may be decomposed in the following way:
\begin{equation}\label{eq:mesquared}
\left|\mathcal{M}\right|^2= \left|\mathcal{M}_{W}\right|^2 + \left|\mathcal{M}_{W'}\right|^2 + 2 \mathrm{Re}(\mathcal{M}_W^* \mathcal{M}_{W'})\;.
\end{equation}
The last term, which contains the interference, depends on the
functions $S(\hat{s})$ and $A(\hat{s})$ (Eqs.~(\ref{eq:Sterm}) and~(\ref{eq:Aterm})). Here we discuss the function $S(\hat{s})$ when
studying interference effects, although the arguments for $A(\hat{s})$
are equivalent. The function $S(\hat{s})$ can be decomposed into
pieces that arise individually due to the $W$ or the $W'$, and an interference piece:
\begin{equation}\label{eq:stermdecomp}
S = S_{W,W} + S_{W',W'} + S_{W,W'} + S_{W',W} = S_{W,W} + S_{W',W'} + 2 S_{int}\;,
\end{equation} 
where we have defined the interference term $S_{int} \equiv S_{W,W'} = S_{W',W}$. Explicitly, this term may be written as
\begin{equation}\label{eq:sint}
S_{int} = \left[\hat{s} \frac{ (\hat{s} - M_W^2) (\hat{s} - M_{W'}^2) + \Gamma_{W} \Gamma_{W'} M_{W} M_{W'} } { [(\hat{s} - M_W^2)^2 + \Gamma_W^2 M_W^2][W\rightarrow W']}\right] (1 + h_W h_{W'})^2\;,
\end{equation}
where we have set all the couplings $C^{\ell,q}_{W/W'} = 1$ and
$k_W=k_{W'}=1$. It is evident that since $h_W = 1$ for the SM $W$,
when we set $h_{W'} = -1$ then the interference term vanishes: $S_{int} = 0$. This implies that
there is \textit{no} interference for the case of the SM $W$ and a
purely right-handed $W'$, and the square of the total matrix element for the process can be written as the sum of the squares of the individual matrix elements for the $W$ and $W'$:
\begin{equation}\label{eq:merhwp}
\left|\mathcal{M}(h_{W'} = -1)\right|^2=
\left|\mathcal{M}_{W}\right|^2 + \left|\mathcal{M}_{W'}\right|^2 \;.
\end{equation}
It is simple to see why this is so: the $W'$ and SM $W$
decay to final-state particles of different helicities, which are
distinguishable, and hence the two processes cannot interfere. However, when $h_W = 1$
and $h_{W'} = 1$, i.e. both bosons possessing left-handed couplings,
we have $S_{int} \ne 0$. In fact, by examining of the expression for
$S_{i,j}$ (Eq.~(\ref{eq:Sterm})), we observe that $S_{int}$ should be
of the same order of magnitude as $S_{W',W'}$ and
$S_{W,W}$. Figure~\ref{fig:sint} shows the variation of the
interference term, as well as the $W$ and $W'$ terms $S_{W,W}$ and
$S_{W',W'}$, for the case $M_{W'} = 1~\mathrm{TeV}$. We observe that $S_{int}$
is negative (green, blue, purple) in the intermediate mass squared
region $\hat{s} \in (M_W^2,M_{W'}^2) =
(\sim6400~\mathrm{GeV}^2,10^6~\mathrm{GeV}^2)$. The total
cross section in this region is \textit{less} than the sum of the
individual $W$ and $W'$ cross sections. We note that the possibility of a reduced
cross section is seldom considered in experimental searches.
It is important to realise that
the interference term is non-vanishing and comparable in magnitude to
the other terms in $S(\hat{s})$ \textit{even} as
$\Gamma_{W'}\rightarrow 0$, a clear indication that the narrow width
approximation is not justified in the intermediate region. 
\subsection{Extension to NLO}\label{sec:wprime:nlo}
Next, we extend the simulation to NLO using the {\tt MC@NLO} and the {\tt POWHEG}
methods. We briefly discuss their application to $W'$ boson production. Full
details of the application of the {\tt MC@NLO} method to vector boson
production can be found in section~$6$ of Ref.~\cite{LatundeDada:2007jg}. Details of the application of the {\tt POWHEG} method can be
found in chapter~$4$ of Ref.~\cite{seyi}, where vector boson production is discussed in
detail. 
\subsubsection{The {\tt MC@NLO} method}
The NLO cross section for the production of $W'$ bosons
can be written as a sum of two contributions:
\begin{equation}
\label{sigg}
\sigma_{\rm NLO} = \sigma_{q\bar{q}'} + \sigma_{(q\bar{q}')g} \;,
\end{equation}
where $\sigma_{q\bar{q}'}$ is the contribution from $q\bar{q}'$ annihilation and
$\sigma_{(q\bar{q}')g}$ is the contribution from the Compton subprocesses. In the modified minimal subtraction (\msbar) factorisation
scheme, these are:
\begin{eqnarray}
\label{qbg}
\sigma_{q\bar{q}'} &=& \sigma_0 \sum_q \int \mrd x_1 \mrd x_2 \frac{x[D_q(x_1)D_{\bar{q}'}(x_2)+q\leftrightarrow \bar{q'}]}{D_q(x_q)D_{\bar{q}'}(x_{\bar{q}'})} \left[\delta(1-x)
    +\frac{\alpha_S}{2\pi} C_F \left \{
      -2\frac{1+x^2}{1-x}\ln x \right. \right. \nonumber \\
&+& \left. \left. 4(1+x^2)\left(\frac{\ln(1-x)}{1-x}\right)_++\left(-8+\frac{2\
}{3}\pi^2\right)\delta(1-x)\right\}\right] \;,\nonumber \\
\sigma_{(q\bar{q}')g} &=& \sigma_0 \sum_{q,\bar{q}'} \int \mrd x_1 \mrd x_2\frac{x[D_{(q,\bar{q}')}(x_1)D_g(x_2)+ (q,\bar{q}')\leftrightarrow
g]}{D_q(x_q)D_{\bar{q}'}(x_{\bar{q}'})}\frac{\alpha_S}{2\pi}T_R\left [
\frac{1}{2}+3x-\frac{7}{2}x^2 \right.
 \nonumber \\
&+& \left.(x^2+(1+x^2))\ln\frac{(1-x)^2}{x}\right] \,,
\end{eqnarray}
where $\sigma_0$ is the Born differential cross section
$\frac{\mrd ^2\sigma_0}{\mrd Q^2 \mrd Y}$ with $Q$ the invariant mass and $Y$ the rapidity of the vector boson. The $x_1,x_2$ are the NLO momentum fractions and
$x_q,x_{\bar{q}'}$ are the Born momentum fractions with $Q^2 = x_qx_{\bar{q}'}S$, if $S$ denotes the hadronic centre-of-mass energy. Also, $x = \frac{x_qx_{\bar{q}'}}{x_1x_2}$ and $D_q(x_1) = x_1f_q(x_1)$ etc., with $f_q(x_1)$ being the parton distribution
function of parton $q$. 

Focusing on the $q\bar{q}'$ annihilation process for the moment, if we introduce the variable
\begin{equation}
y = \cos \theta \;,
\end{equation}
where $\theta$ is the scattering angle of the emitted parton in the
partonic centre-of-mass frame, we can
rewrite $\frac{\sigma_{q\bar{q}'}}{\sigma_0}$ as an integral over $x$ and $y$:
\begin{eqnarray}
\label{sub}
\frac{\sigma_{q\bar{q}'}}{\sigma_0} &=& \sum_q \int \mrd x \mrd y \left[\left\{
\frac{x[D_q(x_1)D_{\bar{q}'}(x_2)+q\leftrightarrow \bar{q'}]}{D_q(x_q)D_{\bar{q}'}(x_{\bar{q}'})}\frac{1}{2} \left(\delta(1-x)
    +\frac{\alpha_S}{2\pi} C_F \left (-2
      \frac{1+x^2}{1-x}\ln x \right. \right.\right. \right. \nonumber \\
&+&
\left. \left. \left.
  \left. 4(1+x^2)\left(\frac{\ln(1-x)}{1-x}\right)_++\left(-8+\frac{2}{3}\pi^2\right)\delta(1-x)\right)\right) -M_{q\bar{q}'}(x,y) \right\}+ M_{q\bar{q}'}(x,y) \right] \;,\nonumber \\
\end{eqnarray}
where $M_{q\bar{q}'}(x,y)$ is the real emission matrix element. Since we have subtracted this
contribution from the total cross section, in the curly brackets we are left with the sum of the Born, virtual
and QCD PDF correction contributions. Now we can define an infrared-safe observable $O$
whose NLO expectation value is given by
\begin {eqnarray}
\langle O^{q\bar{q}'} \rangle&=&\sum_{q}\int
\mrd x \mrd y\left[O_{W'}\left\{\frac{x[D_q(x_1)D_{\bar{q}'}(x_2)+q\leftrightarrow
\bar{q}]}{D_q(x_q)D_{\bar{q}'}(x_{\bar{q}'})}\frac{1}{2}\left (\delta(1-x)
  \right. \right.\right.\nonumber \\
&+&\left.\left.\frac{\alpha_S}{2\pi}C_F\left(-2\frac{1+x^2}{1-x}\ln x+4(1+x^2)\left(\frac{\ln(1-x)}{1-x}\right)_+
      \right.\right.\right. \nonumber \\
&+& \left. \left. \left.\left.\left(-8+\frac{2}{3}\pi^2\right)\delta(1-x)\right)\
\right)-M_{q\bar{q}'}(x,y)\right\}+O_{W'g}M_{q\bar{q}'}(x,y)\right] \,,
\end{eqnarray}
where $O_{W'}$ and $O_{W'g}$ are observables arising from hadronic final states generated from  $q+\bar{q} \rightarrow W'$ and
$q+\bar{q} \rightarrow W'+g$ starting configurations respectively. This however is not
entirely correct because of double counting in the final states represented by $O_{W'}$
which are already included in the states arising from $O_{W'g}$. The
solution to this is to subtract the parton shower contributions, which
we denote $M_{C_{q\bar{q}'}}(x,y)$, from the regions in which the parton shower contributes (the jet region $J$) and integrate the full matrix
element in the hard emission region $D$, left untouched by the shower. This gives for
$\langle O^{q\bar{q}'} \rangle$:
\begin {eqnarray}
\langle O^{q\bar{q}'} \rangle&=&\sum_{q}\int_J\left[O_{W'}\left\{\frac{x[D_q(x_1)D_{\bar{q}'}(x_2)+q\leftrightarrow
\bar{q}']}{D_q(x_q)D_{\bar{q}'}(x_{\bar{q}'})}\frac{1}{2}\left (\delta(1-x)+\frac{\alpha_S}{2\pi}C_F\left(-2\frac{1+x^2}{1-x}\ln x
  \right. \right. \right.\right.\nonumber \\
&+&\left.\left.\left.4(1+x^2)\left(\frac{\ln(1-x)}{1-x}\right)_+
+\left(-8+\frac{2}{3}\pi^2\right)\delta(1-x)\right)\right)-M_{q\bar{q}'}+M_{{C}_{q\bar{q}'}}\right\}\nonumber \\
&+&\left.O_{W'g}\left \{M_{q\bar{q}'}-M_{{C}_{q\bar{q}'}}\right\}\right] \nonumber\\
&+&\sum_{q}\int_D \left[O_{W'}\left\{\frac{x[D_q(x_1)D_{\bar{q}'}(x_2)+\
q\leftrightarrow
\bar{q}']}{D_q(x_q)D_{\bar{q}'}(x_{\bar{q}'})}\frac{1}{2}\left (\delta(1-x)+\frac{\
\alpha_S}{2\pi}C_F\left(-2\frac{1+x^2}{1-x}\ln x
  \right. \right. \right.\right.\nonumber \\
&+&\left.\left.\left.\left.4(1+x^2)\left(\frac{\ln(1-x)}{1-x}\right)_++\left(-8+\frac{2}{3}\pi^2\right)\delta(1-x)\right)\right)-M_{q\bar{q}'}\right\}
+ O_{W'g}M_{q\bar{q}'}\right] \;.\nonumber \\
\end{eqnarray}
A similar functional $\langle O^{(q\bar{q}')g} \rangle$ can be generated for the Compton
subprocesses. Events can then be generated in the different regions of phase space according
to their contributions to the above integrals. These events are then interfaced with
\Herwigpp and showered. Full details of the algorithm for event generation can be found in
\cite{LatundeDada:2007jg}.
\subsubsection{The {\tt POWHEG} method}
The {\tt POWHEG} method, as described in section~\ref{sec:mc:nlomatching}, involves the generation of the hardest radiation from
the parton shower according to the real emission matrix element and independently of the
shower Monte Carlo generator used. If we introduce
\begin{equation}
\centering
\label{eq:M}
R_{v,r}=M_{q\bar{q}'}+M_{(q\bar{q}')g} \,,
\end{equation}
where $M_{q\bar{q}}$ and $M_{(q\bar{q}')g}$ are real emission matrix elements for $q\bar{q}'$
annihilation and the Compton subprocesses respectively, we can write
the cross section for the hardest gluon emission event as
\begin{equation}
\centering
\label{eq:sig2}
\mrd \sigma=\sum_q \bar {B}^q_{v} \mrd \Phi_{v}\left[\Delta^q(0)+\Delta^q(p_{\rm T})R_{v,r}d
  \Phi_{r}\right] \;.
\end{equation}
The index $q$ runs over all quarks and anti-quarks. The subscript $v$ represents the Born
variables, which in this case are the invariant mass $Q$ and
the rapidity $Y$ of the boson, $r$ represents the
radiation variables $x, y$ and $\mrd \Phi_{v}$, $\mrd\Phi_{r}$ are the Born and real emission phase spaces respectively.

$\Delta^q(p_{\rm T})$ is the modified Sudakov form factor for the hardest emission with
transverse momentum $p_{\rm T}$, as indicated by the Heaviside function in the exponent of
Eq.~(\ref{eq:dnloo}):
\begin{equation}
\centering
\label{eq:dnloo}
\Delta^q(p_{\rm T})=\exp \left[-\int \mrd \Phi_{r} R_{v,r}\Theta(k_{\rm T}(v,r)-p_{\rm T})\right]\;,
\end{equation}
where $k_{\rm T}$ is the transverse momentum of
the hardest emission relative to the splitting axis and in this case is given by
\begin{equation}
\centering
\label{eq:kT2}
k_{\rm T}(x,y)=\sqrt{\frac{Q^2}{4x}(1-x)^2(1-y^2)} \;.
\end{equation}
Furthermore:
\begin{equation}
\centering
\label{eq:Bbarr}
\bar{B}^q_{v}=B^q_{v}+V^q_{v}+\int (R_{v,r}-C_{v,r})d \Phi_{r}\;.
\end{equation}
$\bar{B}^q_{v}$ is the sum of the Born ($B^q_{v}$), virtual
($V^q_{v}$) and real ($R_{v,r}$) terms (with some
counter-terms, $C_{v,r}$). The Born variables are generated with distribution
$\bar{B}^q_{v}$, with the radiation variables of the first emission generated according to $[\Delta^q(0)+\Delta^q(p_{\rm T})R_{v,r}d \Phi_{r}]$.

In the \msbar factorisation scheme, the contribution to the order-$\alpha_S$
cross section for $W'$ production is given by Eqs.~(\ref{sigg}) and~(\ref{qbg}).
The function $\bar{B}^q$ in Eq.~(\ref{eq:Bbarr}) can then be written as a
sum of finite terms using the subtraction method. Here we borrow the \texttt{MC@NLO}
subtraction formula introduced in 
Eq.~(\ref{sub}) and write the function $\tilde{B}^q(Q^2,Y)$ as
\begin{eqnarray}
\label{Bbar2}
\tilde{B}^q(Q^2,Y)&=&\sum_{q} \int \mrd x \mrd y \mrd Q^2 \mrd Y \,
\frac{\mrd^2\sigma_0}{\mrd Q^2 \mrd Y}\left[ \
\frac{x[D_q(x_1)D_{\bar{q}'}(x_2)+q\leftrightarrow
\bar{q}]}{D_q(x_q)D_{\bar{q}'}(x_{\bar{q}})}\frac{1}{2}\left[\delta(1-x)
  \right. \right.\nonumber \\
&+&\frac{\alpha_S}{2\pi}C_F\left\{-2\frac{1+x^2}{1-x}\ln x\left.+4(1+x^2)\left(\frac{\ln(1-x)}{1-x}\right)_+
  \right. \right.
\nonumber \\
&+& \left. \left.\left(-8+\frac{2}{3}\pi^2\right)\delta(1-x)\right\}\right]-M_{q
  \bar{q}} +M_{C_{q \bar{q}'}}+\left\{M_{q\bar{q}'} -M_{C_{q \bar{q}'}}\right\}\nonumber \\
&+& \frac{x[D_{(q, \bar{q}')}(x_1)D_g(x_2)+ (q,\bar{q}')  \leftrightarrow
g]}{D_q(x_q)D_{\bar{q}}(x_{\bar{q}})}\frac{\alpha_S}{2\pi}T_F\frac{1}{2}\left [\
\frac{1}{2}+3x - \frac{7}{2}x^2 \right.
 \nonumber \\
&+&\left.\left. (x^2+(1+x^2))\ln\frac{(1-x)^2}{x}\right] 
-M_{(q,\bar{q}')g} +M_{C_{(q,\bar{q}')g}} +\left\{M_{(q,\bar{q}')g}-M_{C_{(q,\bar{q}') g}}\right\}\right] \,, \nonumber \\
\end{eqnarray}
where we have written the virtual and PDF corrections in terms
of the real emission matrix elements and $M_C$ are the subtracted parton shower approximation terms in
the \Herwigpp jet regions. Note that the above prescription does not
imply that the {\tt POWHEG} method depends on the shower MC used. We
have simply used the shower approximation terms to define a
subtraction scheme for the definition of the NLO cross section.

The flavour of the event, the Born variables $Q^2$ and $Y$, as well as
the the radiation variables
$x$ and $y$ are then generated
according to the integrand in Eq.~(\ref{Bbar2}). The radiation variables are ignored, which
amounts to integrating away these variables, leaving the Born variables distributed
according to $\bar{B}^q(Q^2,Y)$. The radiation variables $x,y$ are generated according to
\begin{equation}
\Delta^q(p_{\rm T})R(x,y)\mrd x \mrd y \;.
\end{equation}
Details of the algorithm used can be found in Ref.~\cite{seyi}.
\subsection[Experimental bounds]{\boldmath Experimental bounds}\label{sec:wprime:bounds}
We provide a brief overview of direct and indirect searches for $W'$
before presenting our results.

At the Fermilab Tevatron\footnote{The Tevatron is a proton-antiproton
  collider located at Fermilab, USA, with a hadronic centre-of-mass energy of $1.96\tev$.}  both the D0 and CDF collaborations have studied the $e \nu$~\cite{Abulencia:2006kh,Abazov:2007bs} and $t\bar{b}$~\cite{Abazov:2006aj,Abazov:2008vj,Aaltonen:2009qu} channels. The $W'$ was
assumed to have narrow width and SM-like couplings to fermions. In the
$e \nu$ channel the signal consists of a high-energy electron and
missing transverse energy, with an edge in the transverse mass
distribution at $M_{W'}$. In the $t\bar{b}$ channel the signal
consists of a $W$ boson decaying leptonically and two
$b$-jets~\cite{Nakamura:2010zzi}. The limits obtained from the $e\nu$
searches, corresponding to 1 fb$^{-1}$ of data for D0 and 205
pb$^{-1}$ for CDF, are $M_{W'} > 1\tev$ and $M_{W'} > 788\gev$
respectively. The D0 Collaboration has published results using 0.9
fb$^{-1}$ of data in the $t\bar{b}$ channel, in which the limit for
left-handed $W'$ masses is $M_{W'} > 731 \gev$. Furthermore, for
right-handed $W'$ bosons, the limit is $M_{W'} > 739 \gev$ assuming
the $W'$ boson decays to all fermions, and $M_{W'} > 768\gev$ if it
decays only to quarks. CDF has set the limits, using 1.9 fb$^{-1}$,
$M_{W'} > 800 \gev$ when $M_{W'} > M_{\nu_R}$, the mass of the
right-handed neutrino, and $M_{W'} > 825 \gev$ when $M_{W'} <
M_{\nu_R}$. The limits given are at 95$\%$ confidence level (C.L.). A
recent review on Tevatron searches is contained
in Ref.~\cite{Duperrin:2008in}. More recently, using 36$~\mathrm{pb}^{-1}$ of LHC data at 7~TeV, the CMS
collaboration has published limits for the sequential SM-like $W'$
using the $\mu\nu$ channel of $M_{W'} > 1.40\tev$. This limit was
stated to increase to $1.58\tev$ if the analysis was combined with the
$e\nu$ analysis.  The equivalent limit from the ATLAS collaboration,
extracted using an equivalent amount of data, is $1.47\tev$~\cite{Collaboration:2011fe}. 

Low energy constraints on $W'$'s are strongly model-dependent. If the
$W'$ couples to quarks, then box diagrams involving a SM $W$ and a
$W'$ contribute to meson mixing, for example to $K_L-K_S$ mixing. Then the limit
arising for the left-right symmetric model, based on the symmetry group
$SU(2)_L \times SU(2)_R \times U(1)_{B-L}$, is $M_{W'} > 2.5 \tev$~\cite{Zhang:2007fn}. This can be relaxed if we assume no correlation between the right-handed quark and lepton couplings~\cite{Langacker:1989xa}. Limits also arise from possible contributions of $W'$ bosons in neutrinoless double-beta decay and right-handed neutrino emissions from supernovae~\cite{Nakamura:2010zzi}.
\subsection{Results}\label{sec:results}
\begin{figure}[!htb]
  \vspace{1.6cm}
  \centering 
    \hspace{4.0cm}
    \includegraphics[scale=0.52, angle=90]{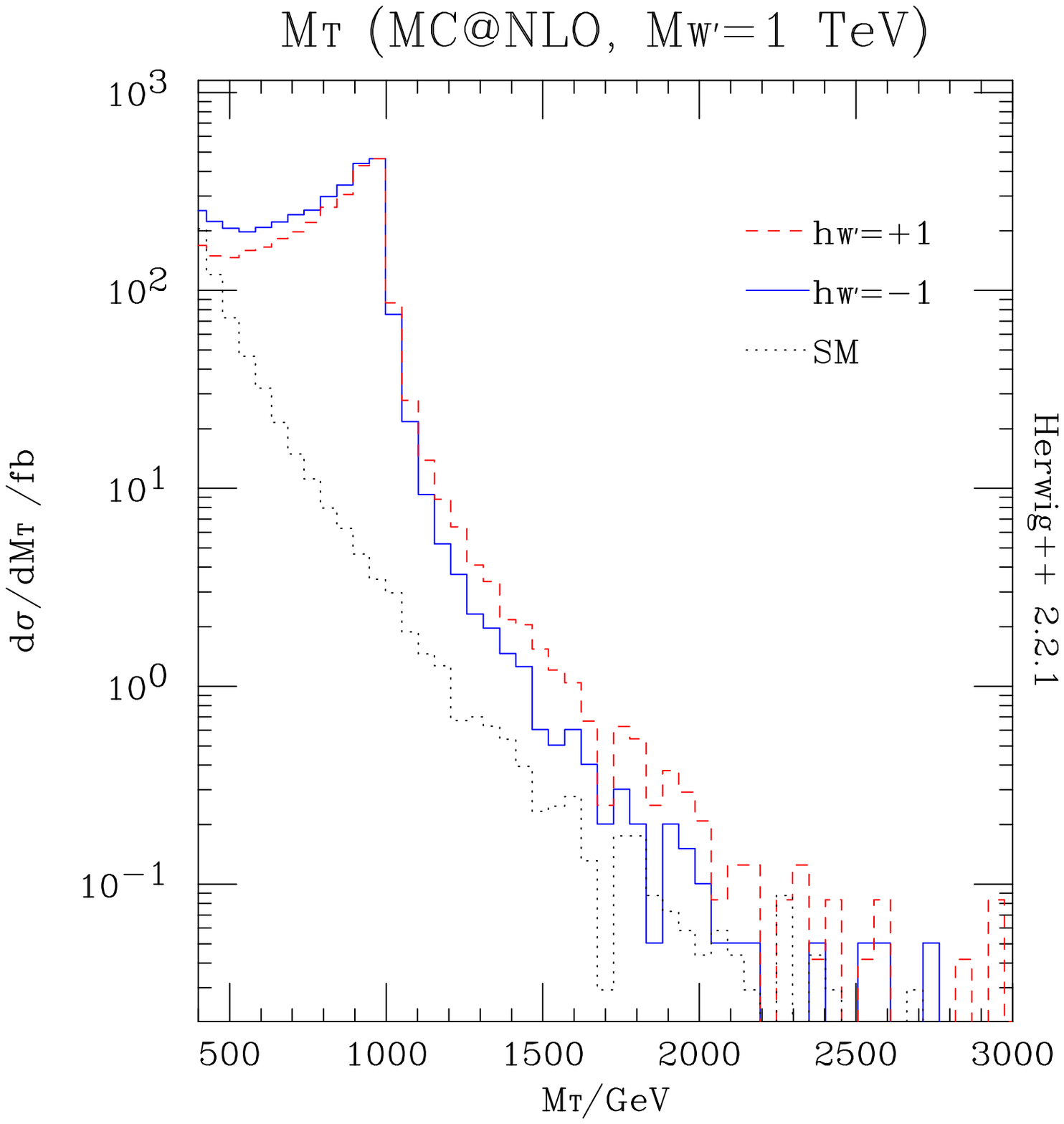}
    \hspace{5.5cm}
   \includegraphics[scale=0.52, angle=90]{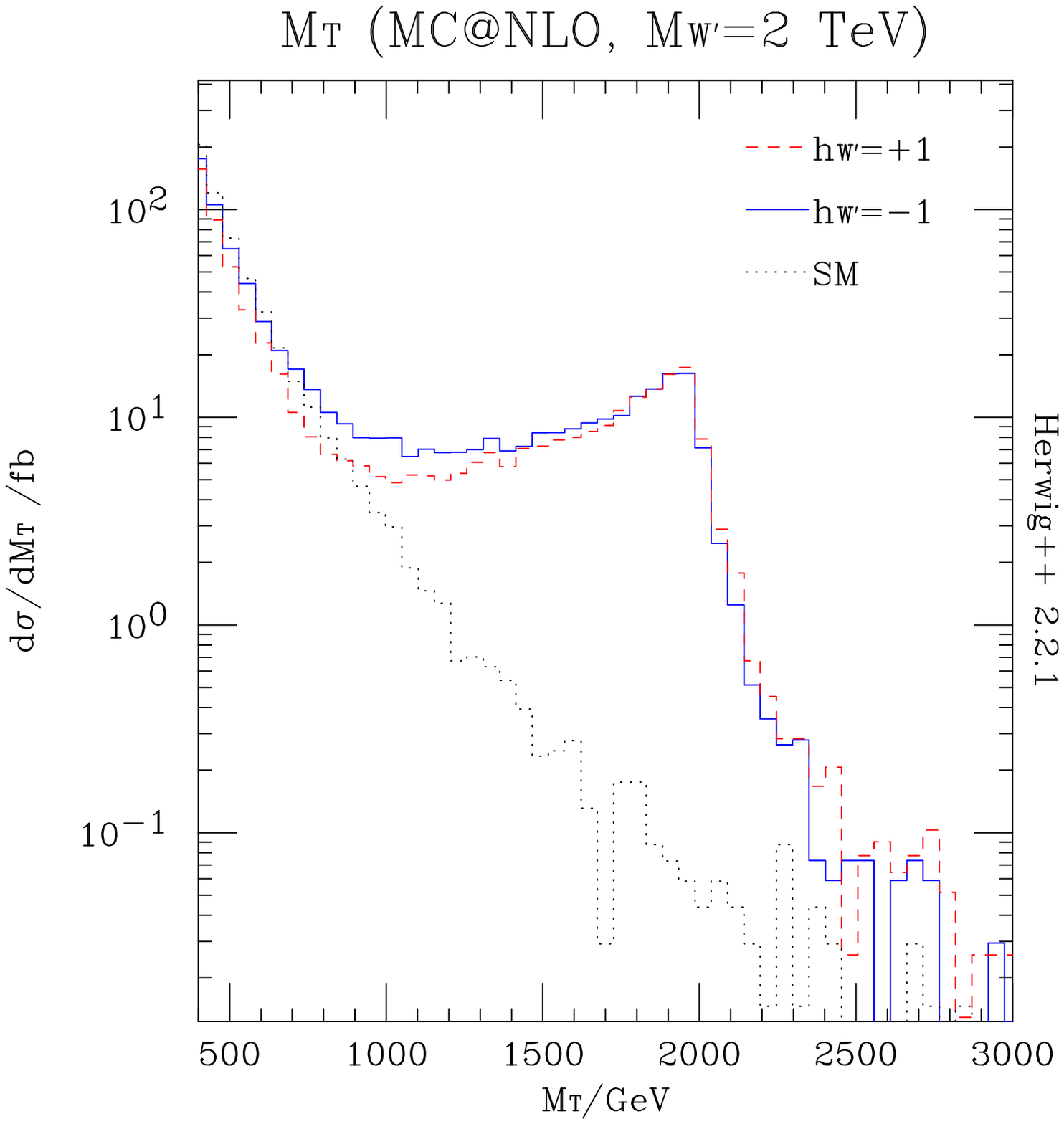}
   \vspace{0.8cm}
\caption{Transverse mass distributions at the LHC obtained using the
  \texttt{MC@NLO} method (cteq5m/\msbar), \texttt{POWHEG} (cteq5m) for
  purely left- and right-handed $W'$'s. The invariant mass range was
  taken to be $(0.4-3.0)\tev$. The plots correspond to masses/widths
  equal to $[1\tev,36\gev]$ and $[2\tev,72\gev]$. The significance of
  the destructive interference can be observed in the left-handed
  case; in the right-handed case the distribution is just the sum of
  the standard model $W$ and right-handed $W'$ contributions. Note
  that the plots are normalised to the NLO cross section for each process.}
\label{fig:mt_mcnlo_LR}
\end{figure}

We present a sample of distributions of variables obtained for
$\sim10^5$ events using the \texttt{Wpnlo} event generator, both at
leading and next-to-leading order, using the \texttt{MC@NLO} and
\texttt{POWHEG} methods. The parton-level \texttt{Wpnlo} output was
interfaced through the Les Houches interface to the general purpose event generator
\Herwigpp, used for showering and hadronization. The $k$-factor (where
$k=\sigma_{NLO} / \sigma_{LO}$), for the considered invariant mass
range and for factorisation/renormalisation scales set to the default
NLO scale $\mu_0 = \sqrt{k_T^2 + Q^2}$ (where $k_T$ and $Q$ are the
dilepton transverse momentum and invariant mass respectively), was
found to be $k\approx1.3$, in all studied cases. The plots have been
normalised to unity (with the exception of Fig.~\ref{fig:mt_mcnlo_LR}) to
emphasise the differences in the shape of the distributions.
\begin{figure}[!htb]
  \vspace{3.0cm}
  \centering 
    \includegraphics[scale=0.55, angle=90]{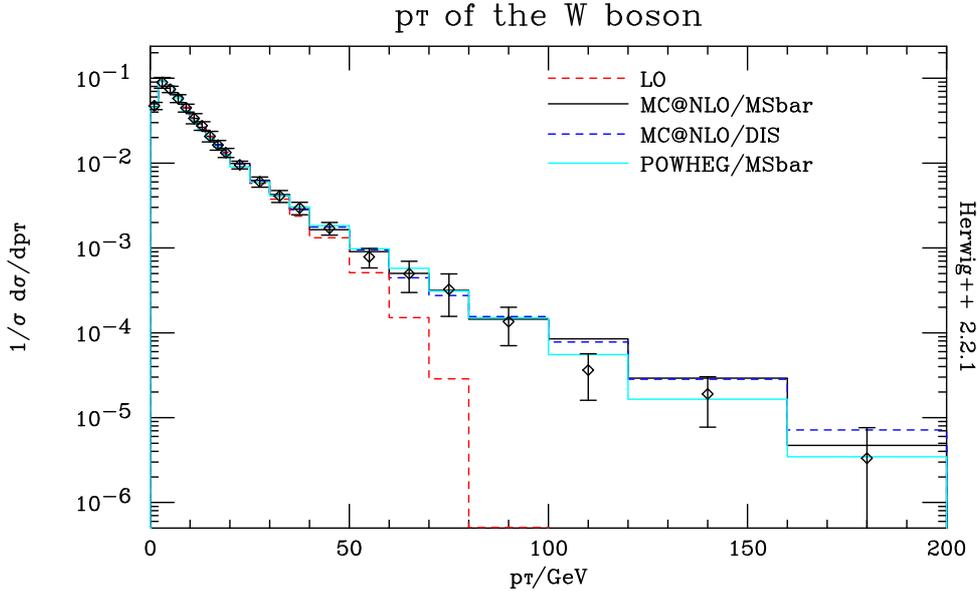}
\caption{Transverse momentum distribution at the Tevatron obtained for
  \texttt{MC@NLO} with \Herwigpp in the DIS and \msbar factorisation
  schemes (PDFs: cteq5d and cteq5m~\cite{Lai:1999wy} respectively), \texttt{POWHEG} \msbar (cteq5m) and LO (PDF: MRST2001LO~\cite{Martin:2001es}), in the mass range $(70-90)\gev$.}
\label{fig:tvt_pt}
\end{figure}
For validation purposes, Fig.~\ref{fig:tvt_pt} presents a comparison
of the $W$ boson transverse momentum distribution, (assuming no $W'$)
between Tevatron data (taken from Ref.~\cite{Affolder:1999jh}) and the
three possible methods: leading order, \texttt{MC@NLO} and
\texttt{POWHEG}. The plots include events in the invariant mass range
$(70-90)\gev$. The \texttt{MC@NLO} and \texttt{POWHEG} distributions
are in agreement with the data within the statistical Monte
Carlo and experimental uncertainties. The leading-order $p_T$
distribution is cut off at the $W$ mass since this provides the only
relevant scale in the parton shower, whereas the \texttt{MC@NLO} and
\texttt{POWHEG} distributions extend to higher transverse
momentum. The different factorisation schemes given, DIS and \msbar,
arise due to the arbitrariness of the prescription in how the finite
contributions to the PDFs are treated when factorising the logarithmic
singularities. The subsequent figures in this section represent simulations made for
the LHC running at 14 TeV proton-proton centre-of-mass energy. 

\begin{figure}[!htb]
  \centering 
  \vspace{1.8cm}
  \hspace{6cm}
    \includegraphics[scale=0.60, angle=90]{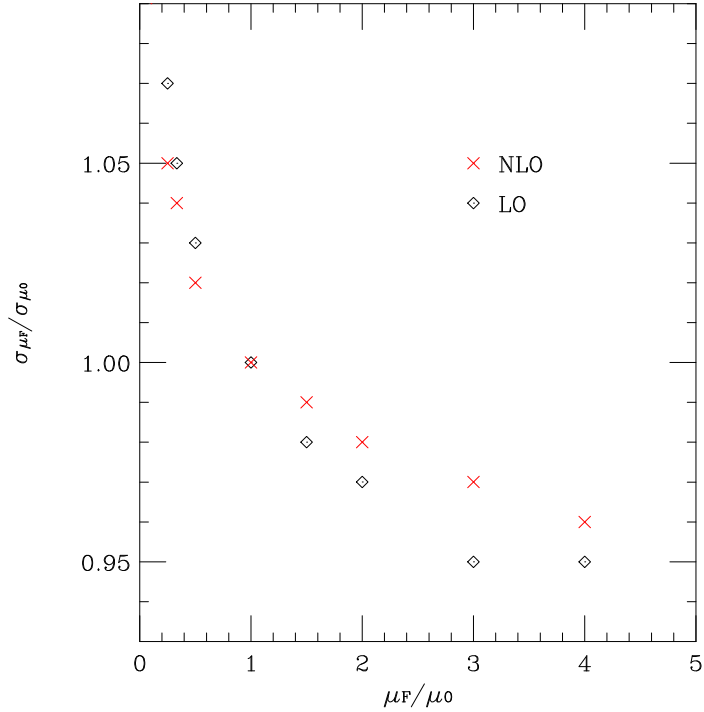}
    \vspace{0.5cm}
\caption{The normalised variation with scale of the cross section calculations at NLO (red crosses) and LO (black diamonds) are shown for a proton-proton collider at 14 TeV, $M_{W'} = 1~\mathrm{TeV}$, $\Gamma_{W'} = 36~\mathrm{GeV}$ and left-handed chirality in the invariant mass range $[400,5000]$ GeV.}
\label{fig:scale}
\end{figure}

\begin{figure}[!htb]
  \centering 
  \vspace{0.8cm}
  \hspace{1.0cm}
    \includegraphics[scale=0.33, angle=90]{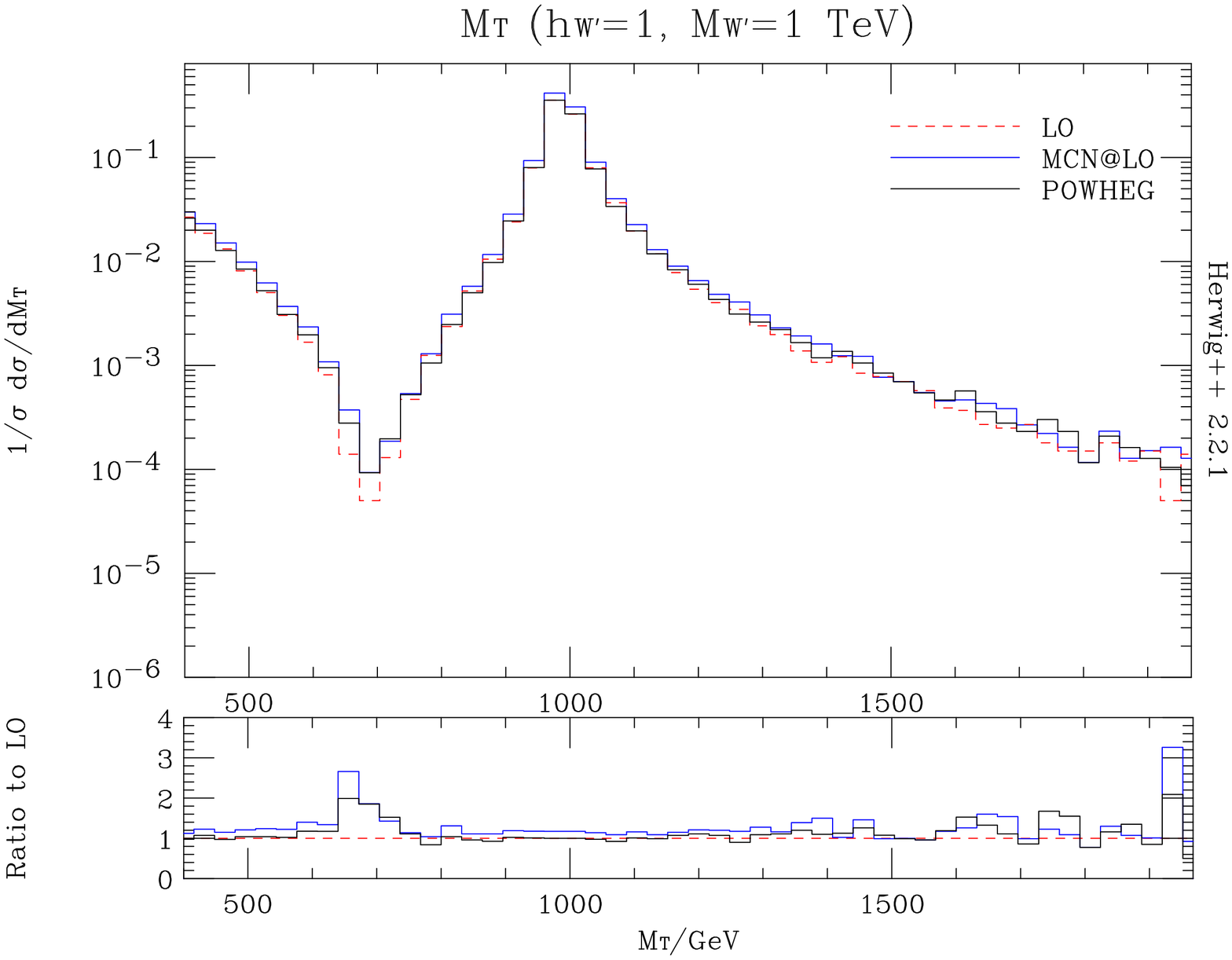}
    \hspace{1.0cm}
   \includegraphics[scale=0.33, angle=90]{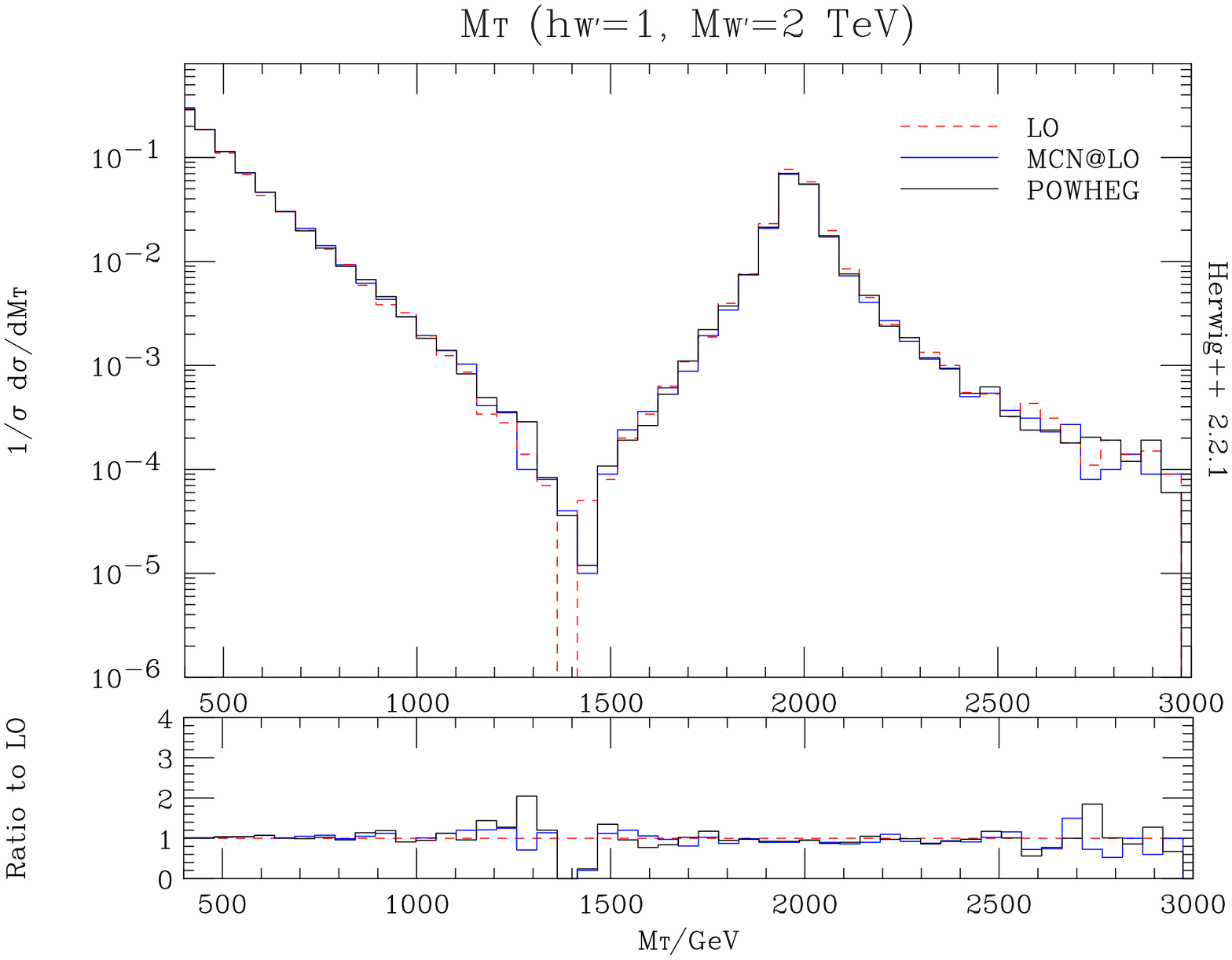}
\caption{Transverse mass distributions at the LHC obtained using the \texttt{MC@NLO} and \texttt{POWHEG} methods (cteq5m/\msbar) and LO (MRST2001LO) for a purely left-handed $W'$. The plots correspond to masses/widths equal to $[1\tev,36\gev]$ (left) and $[2\tev,72\gev]$ (right). The invariant mass range was taken to be $(0.4-3.0)\tev$ for the $1\tev$ case and $(0.4-5.0)\tev$ for the $2\tev$ case. The effect of the destructive interference can be observed in both cases. Note that the plots have been normalised to unity.}
\label{fig:mt_lh}
\end{figure}

\begin{figure}[!htb]
  \centering 
  \vspace{0.8cm}
  \hspace{1.0cm}
  \includegraphics[scale=0.33, angle=90]{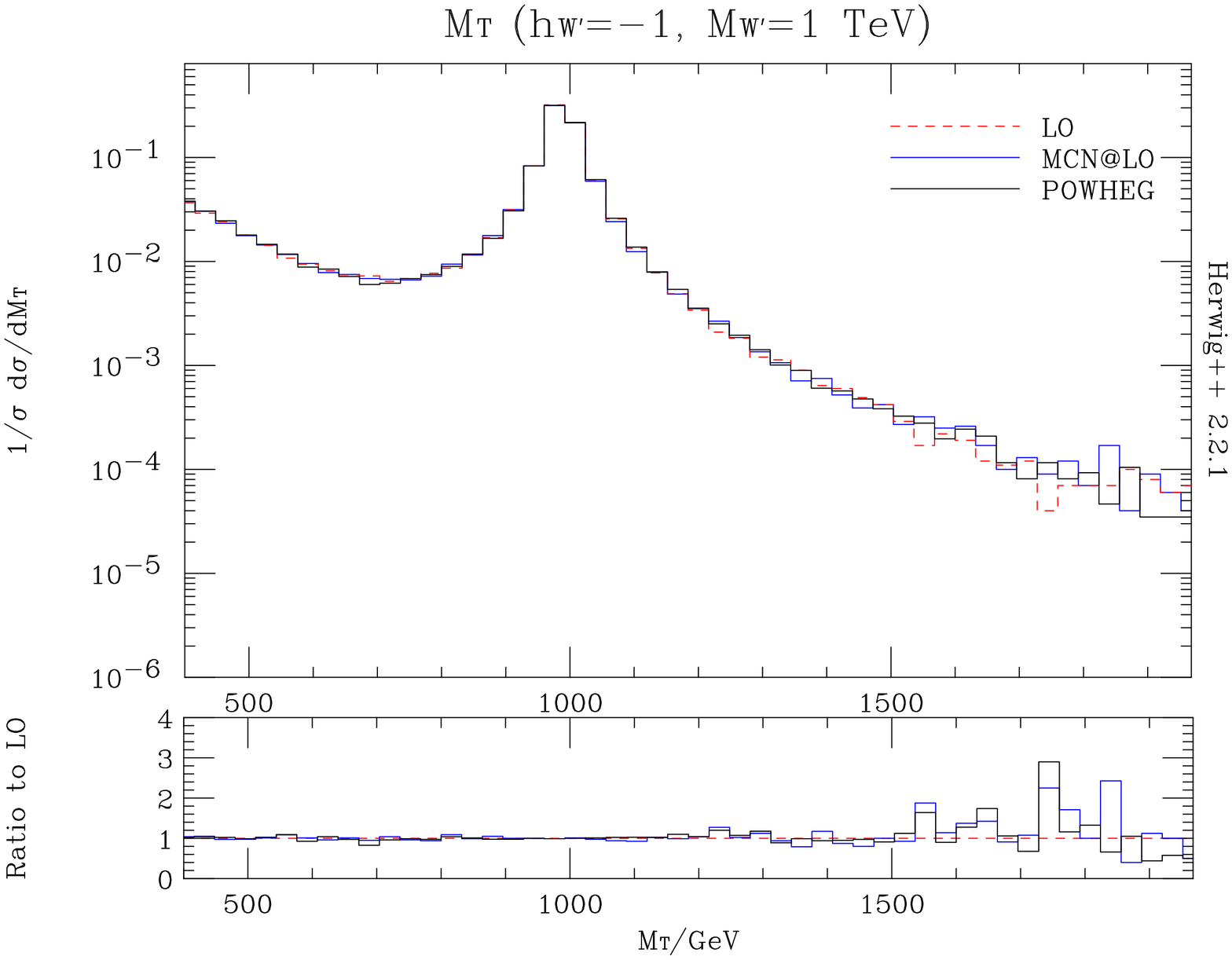}
    \hspace{1.0cm}
   \includegraphics[scale=0.33, angle=90]{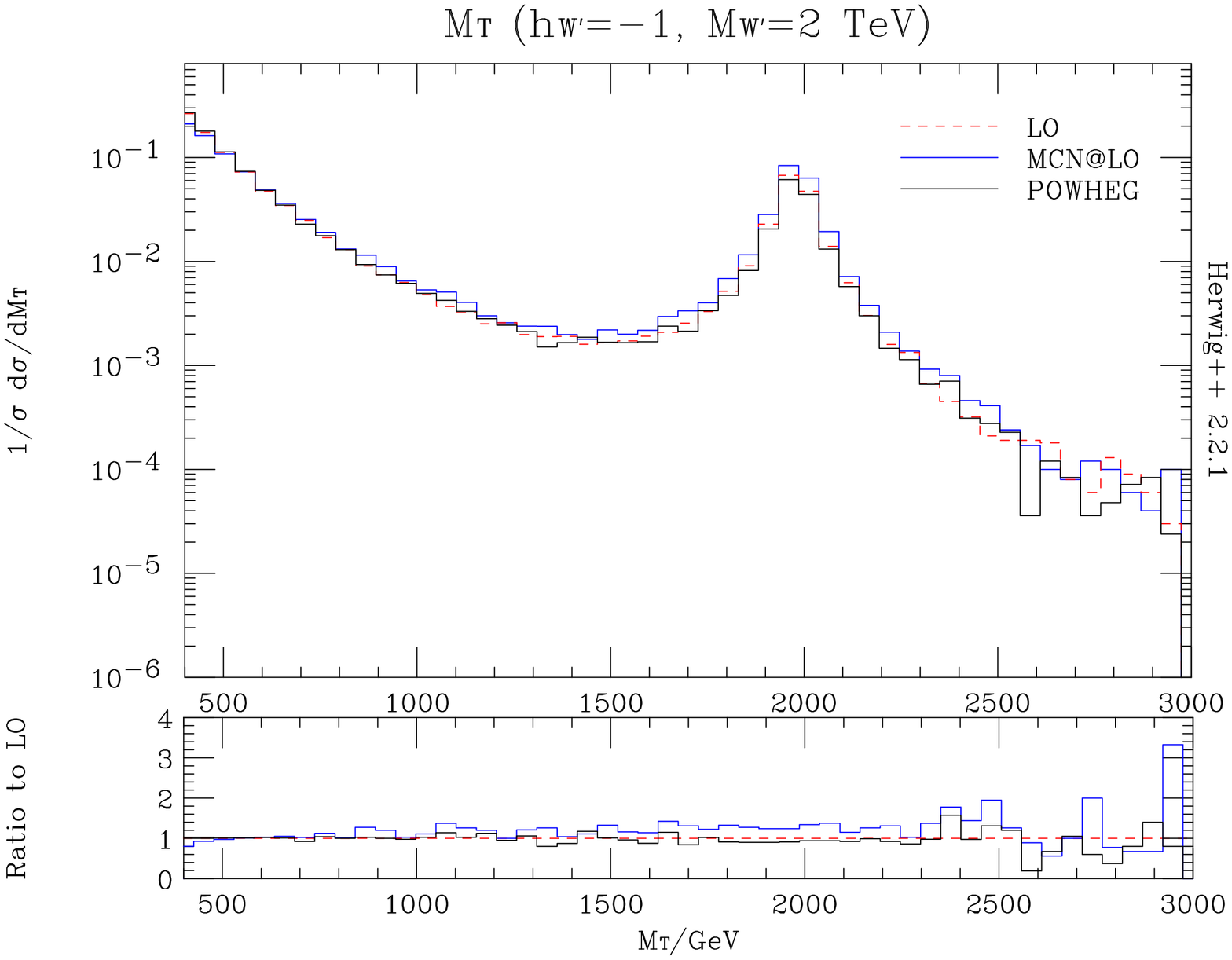}
    \caption{Transverse mass distributions at the LHC obtained using
      the \texttt{MC@NLO} and \texttt{POWHEG} methods (cteq5m/\msbar)
      and LO (MRST2001LO) for a purely right-handed $W'$. The
      invariant mass range, the $W'$ mass and widths are identical to those in the previous figure.}
\label{fig:mt_rh}
\end{figure}

\begin{figure}[!htb]
  \vspace{0.75cm}
  \hspace{1.0cm}
  \centering 
   \includegraphics[scale=0.33, angle=90]{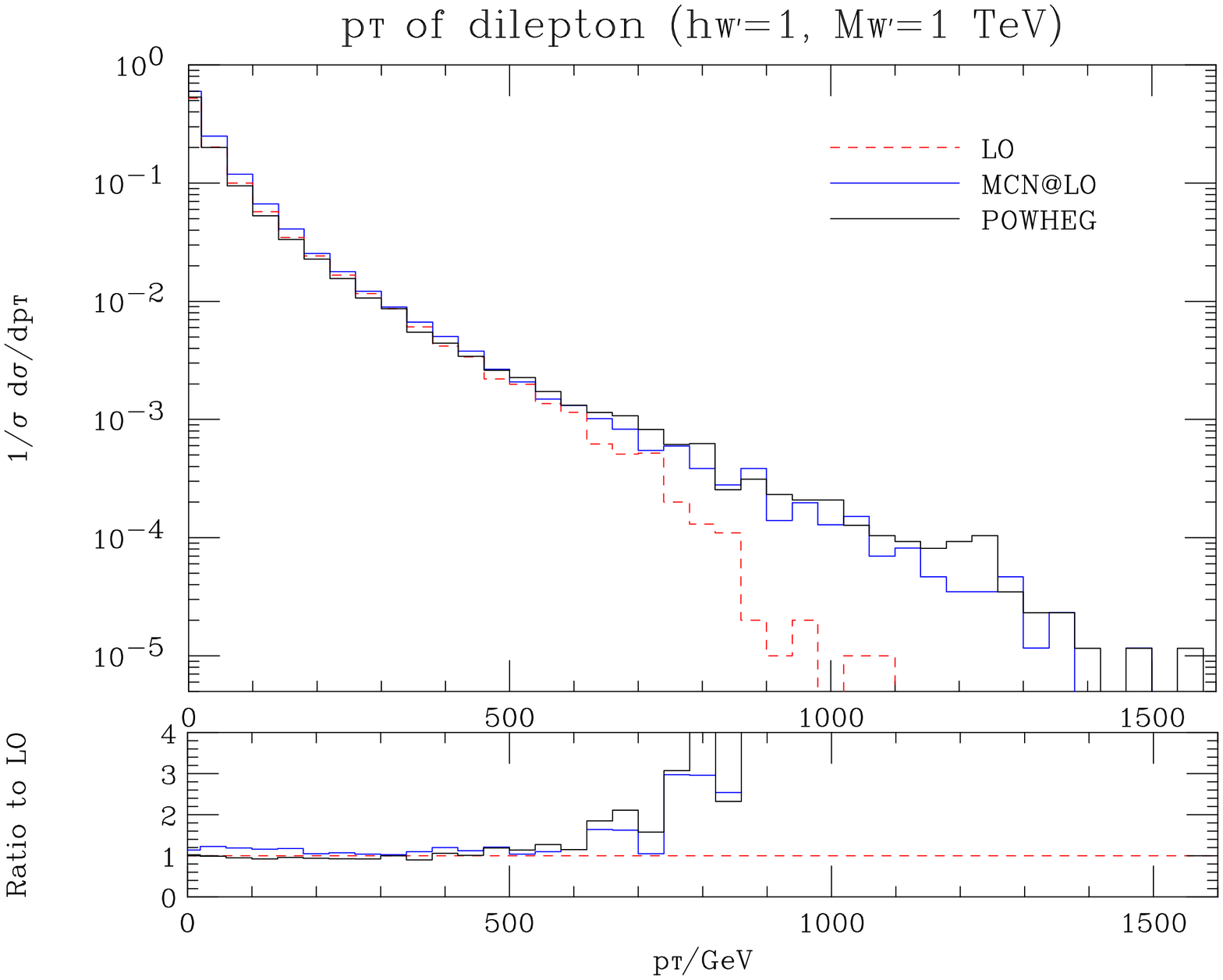}
    \hspace{1.0cm}
   \includegraphics[scale=0.33, angle=90]{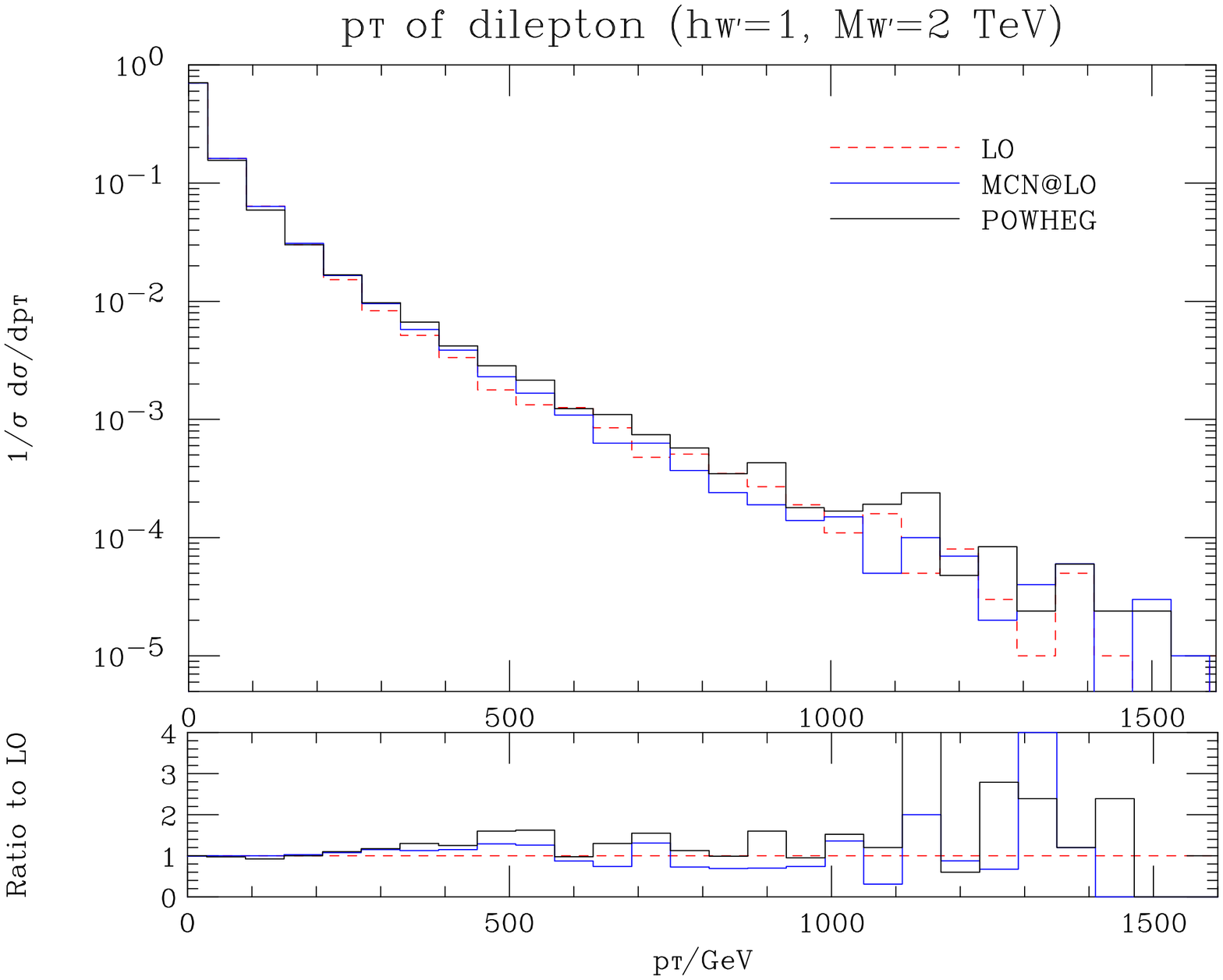}
\caption{Transverse momentum distributions at the LHC obtained using
  the \texttt{MC@NLO} and \texttt{POWHEG} methods (cteq5m/\msbar) and
  LO (MRST2001LO) for a purely left-handed $W'$. The invariant mass
  range, the $W'$ mass and widths are identical to those in the previous figures.}
\label{fig:wpt_lh}
\end{figure}

\begin{figure}[!htb]
  \vspace{0.8cm}
  \hspace{1.0cm}
  \centering 
    \includegraphics[scale=0.33, angle=90]{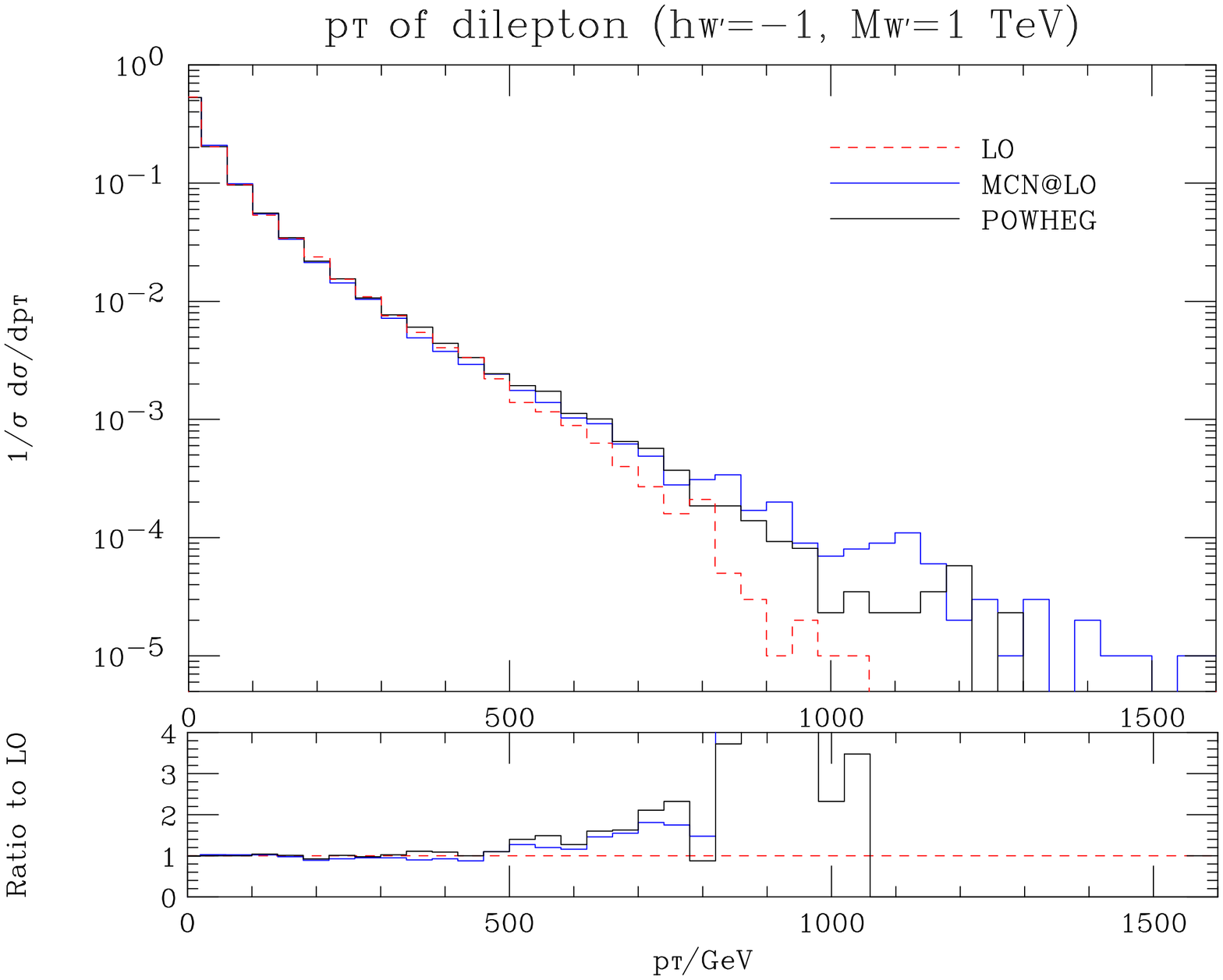}
    \hspace{1.0cm}
    \includegraphics[scale=0.33, angle=90]{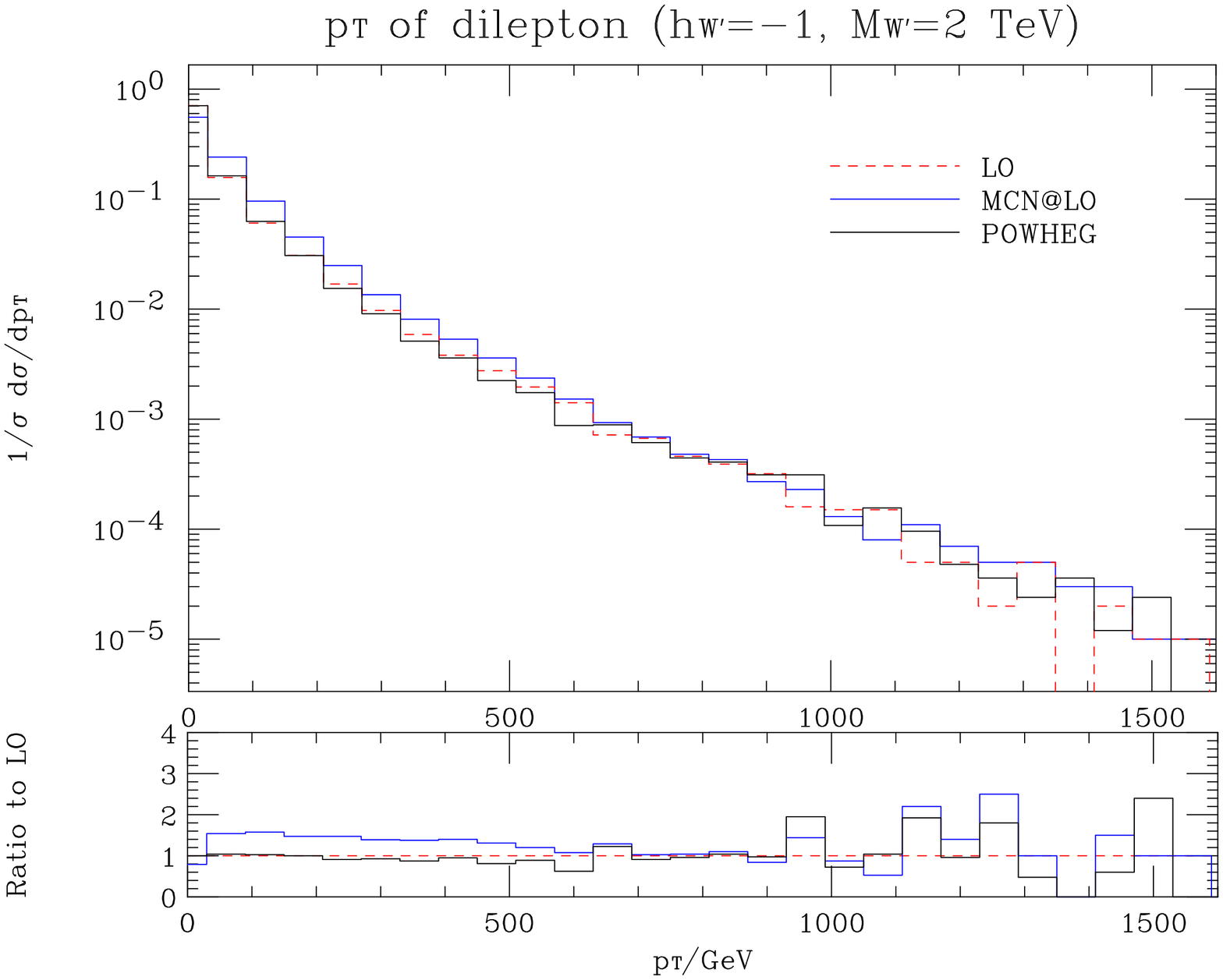}
\caption{Transverse momentum distributions at the LHC obtained using
  the \texttt{MC@NLO} and \texttt{POWHEG} methods (cteq5m/\msbar) and
  LO (MRST2001LO) for a purely right-handed $W'$. The invariant mass
  range, the $W'$ mass and widths are identical to those in the previous figures.}
\label{fig:wpt_rh}
\end{figure}

\begin{figure}[!htb]
  \centering 
  \vspace{0.8cm}
    \includegraphics[scale=0.33, angle=90]{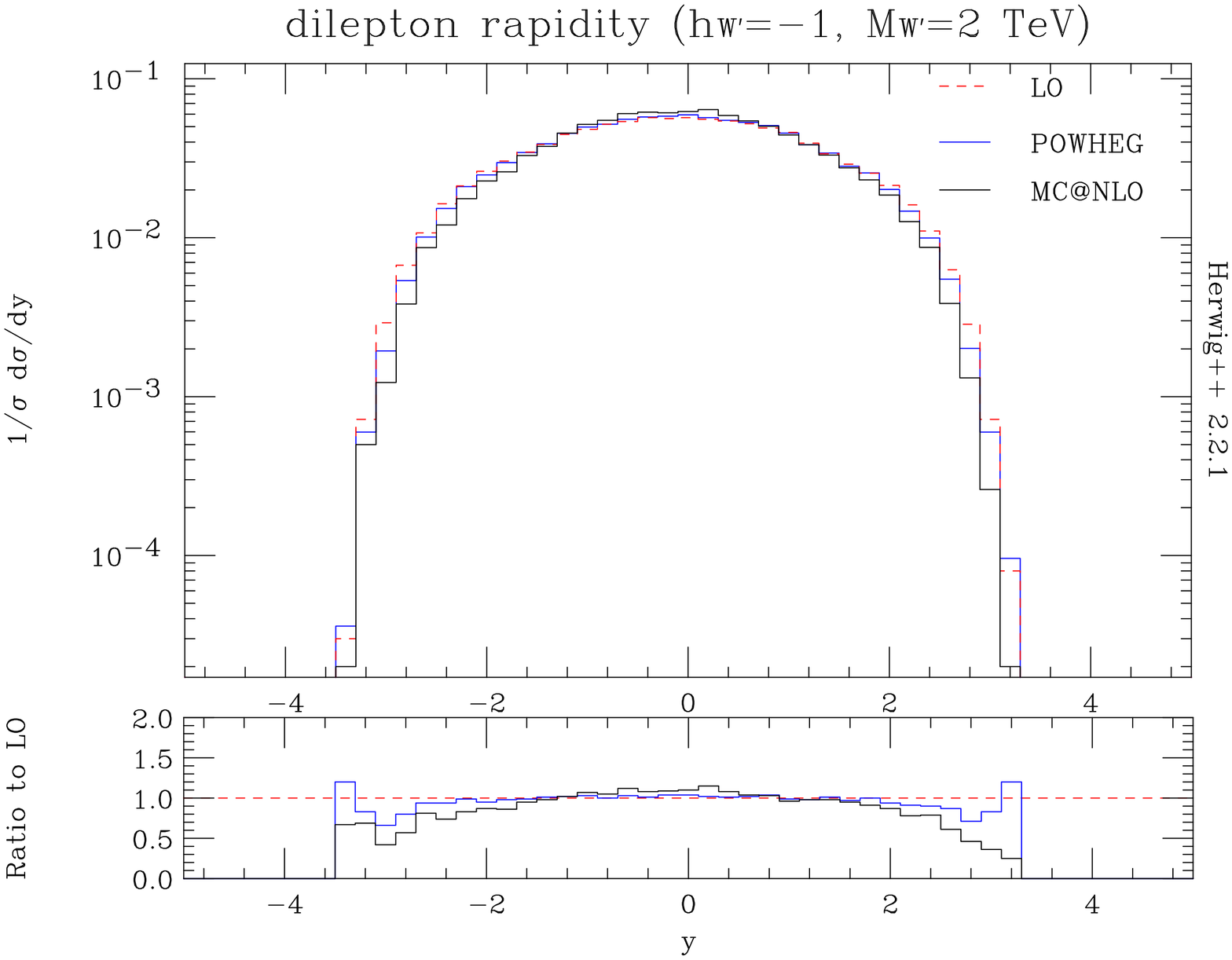}
    \hspace{1.0cm}
    \vspace{1.0cm}
    \includegraphics[scale=0.33, angle=90]{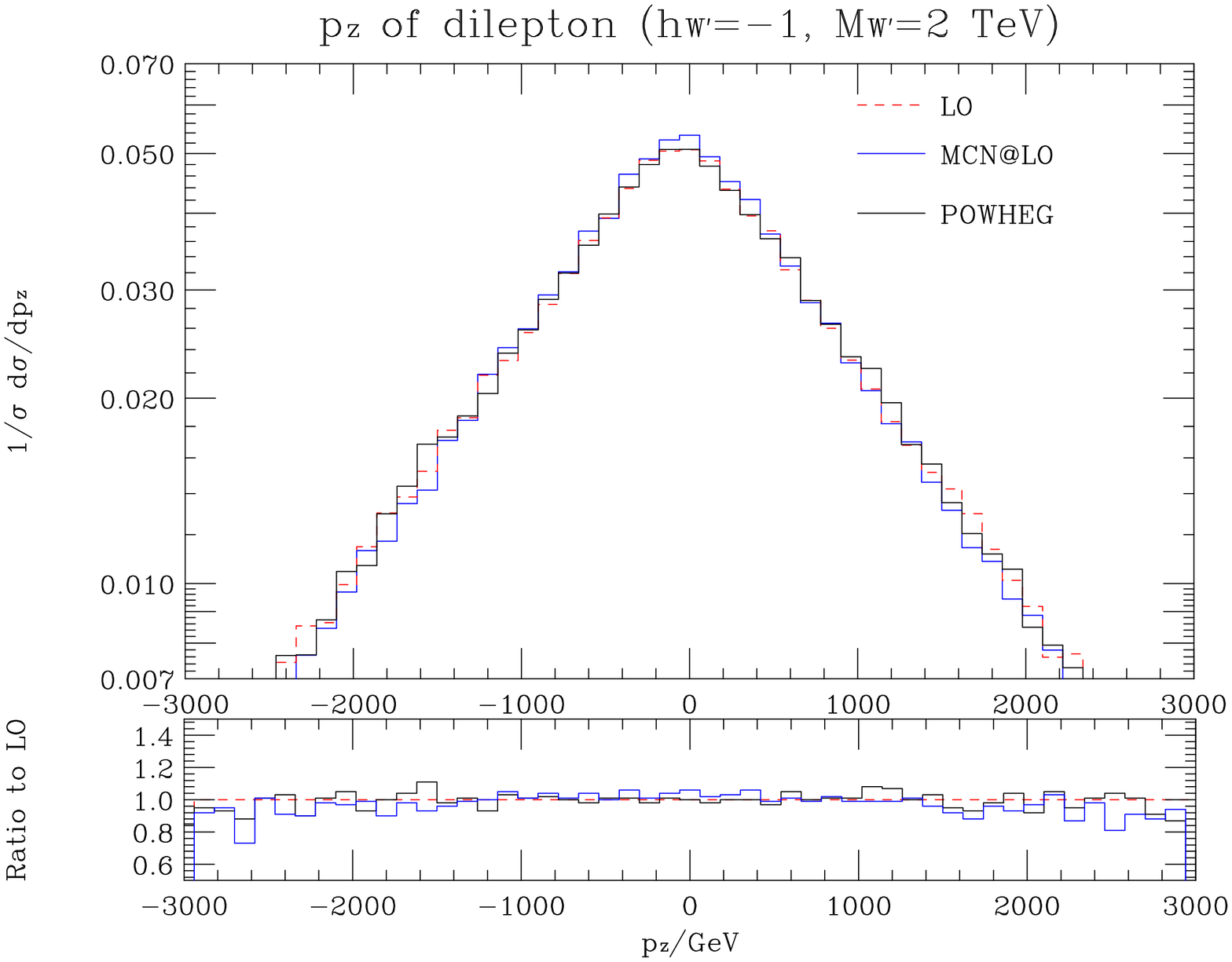}
    \hspace{4.0cm}    
    \includegraphics[scale=0.33, angle=90]{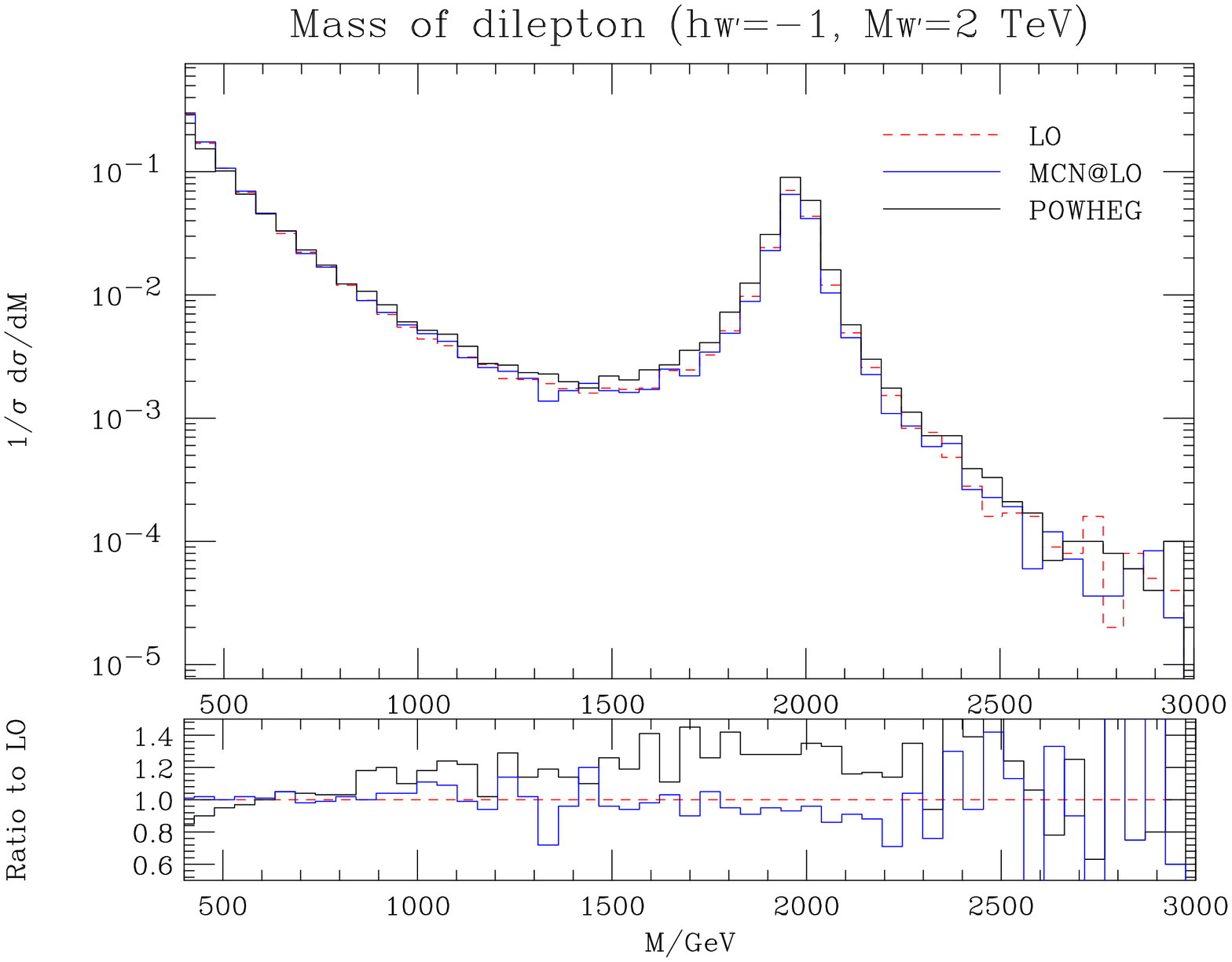}
    \caption{$W/W'$ rapidity (top left), longitudinal momentum (top
      right) and mass (bottom) distributions at the LHC obtained using
      the \texttt{MC@NLO} and \texttt{POWHEG} methods (cteq5m/\msbar)
      and LO (MRST2001LO) for a purely right-handed $W'$ of mass
      $2\tev$ and width $72\gev$. The invariant mass range, the $W'$ mass and widths are identical to those in the previous figures.}
\label{fig:wypzm}
\end{figure}

Figure~\ref{fig:scale} shows the variation of the NLO cross section
for a 1 TeV left-handed $W'$ in the invariant mass range $[400,5000]$
GeV with factorisation scale, $\mu_F$, for a fixed renormalisation scale
using the \msbar scheme. The LO variation with PDF scale is also shown
in an equivalent range.  The values have been normalised to the cross
sections at the default scales $\mu_{0}=\sqrt{k_T^2 + Q^2}$ (default
NLO) and $\mu_{0}=Q$ (default LO). In the NLO case the renormalisation
scale was held fixed at $M_{W'}$. The NLO cross section calculation
appears to be slightly more stable than the LO calculation. The
$k$-factor at $\mu_{0}$ was found to be $k=1.288$ and the LO cross
section at $\mu_{0}=Q$ was found to be $\sigma_{LO}=(2.99\pm0.07)$~pb.
Figures~\ref{fig:mt_lh} and~\ref{fig:mt_rh} show the transverse mass
distributions (in this case we show a `theoretical' transverse mass, defined by $M_T = \sqrt{M^2 + P_T^2}$) at LO and NLO for a $W'$ at masses and widths of
$[1~\mathrm{TeV},36~\mathrm{GeV}]$ and
$[2~\mathrm{TeV},72~\mathrm{GeV}]$, for purely left-handed ($h_{W'} =
1$) and purely right-handed ($h_{W'} = -1$) couplings to fermions
respectively. Figures~\ref{fig:wpt_lh} and~\ref{fig:wpt_rh} show the
corresponding $W/W'$ transverse momentum distributions. In this case
the LO distribution cuts off at the $W'$ mass. The effect is less
visible for higher $W'$ masses. Figure~\ref{fig:wypzm} shows a
comparison between the different methods of the $W/W'$ rapidity, longitudinal momentum and mass distributions for a right-handed $W'$ of mass $2~\mathrm{TeV}$ and width $72~\mathrm{GeV}$ at the LHC. 

Finally, Fig.~\ref{fig:mt_mcnlo_LR} shows a comparison between the
left- and right-handed $W'$ `experimental' transverse mass distributions at NLO (defined by $M_T^2
= 2 \slashed{E}_T E_{T\ell} - 2 \vec{\slashed{p}}_T \cdot
\vec{p}_{T\ell}$, using the missing transverse energy and momentum,
and the lepton transverse energy and momentum), using the \texttt{MC@NLO} method. The importance
of the interference between the SM $W$ and the $W'$ can be clearly
observed: the differential cross section in the region below $M_T =
M_{W'}$ in the purely left-handed ($h_{W'} = 1$) case is reduced in
comparison to the purely right-handed ($h_{W'} = -1$) case. For a transverse mass greater than the on-shell mass of the $W'$, the interference term becomes positive for the left-handed case, although this effect is not significant. The SM contribution, in the absence of a $W'$ boson, is given for comparison in both figures. In the right-handed case the contribution of the $W'$ is simply additive to the SM contribution.

\subsection{Extraction of limits}\label{sec:wprime:limits}
\begin{figure}[!htb]
  \vspace{0.5cm}
  \hspace{1.0cm}
    \includegraphics[scale=0.8]{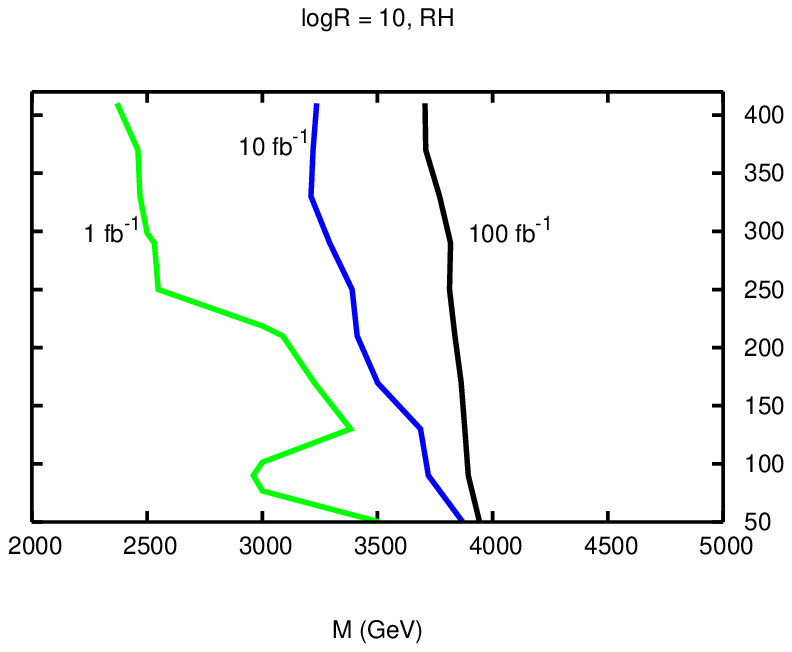}
    \includegraphics[scale=0.8]{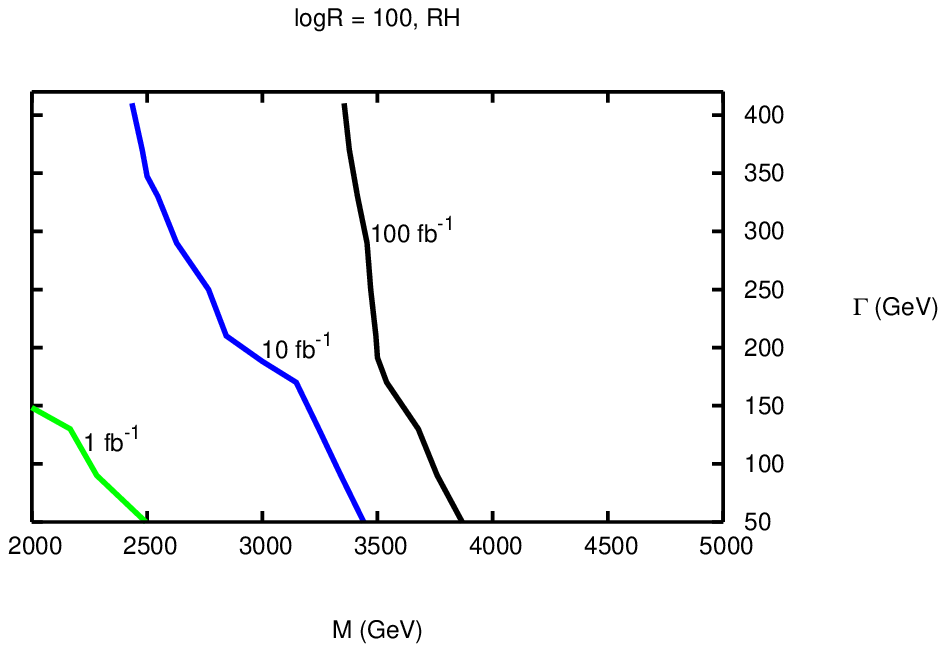}
\caption{The detection reach at the LHC for $\log R = 10$ (left) and $\log R = 100$ (right) at different integrated luminosities for the right-handed case. The colour scheme is: green, blue, black corresponding to the luminosities 1, 10, 100 fb$^{-1}$.}
\label{fig:ex_rh_lhc}
\end{figure}

\begin{figure}[!htb]
  \vspace{0.5cm}
  \hspace{1.0cm}
   \includegraphics[scale=0.8]{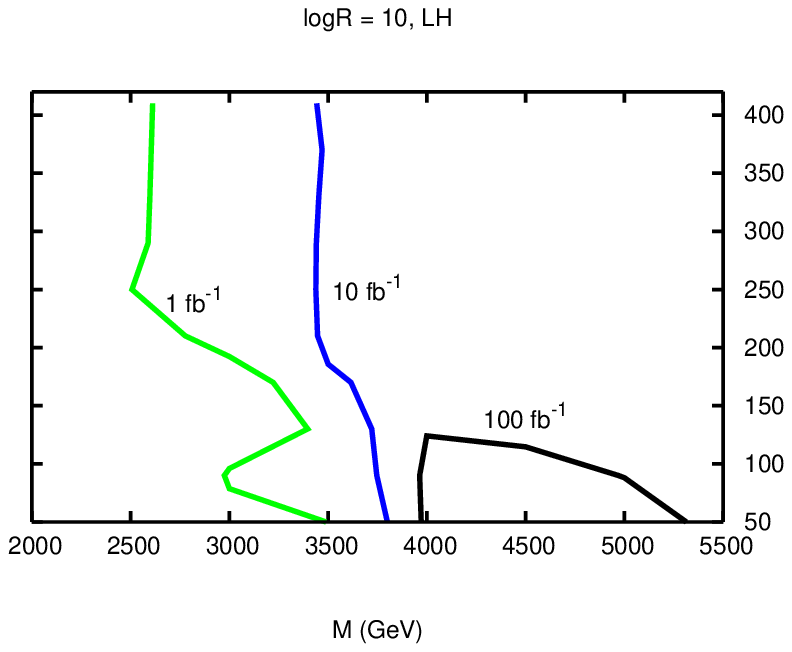}
    \includegraphics[scale=0.8]{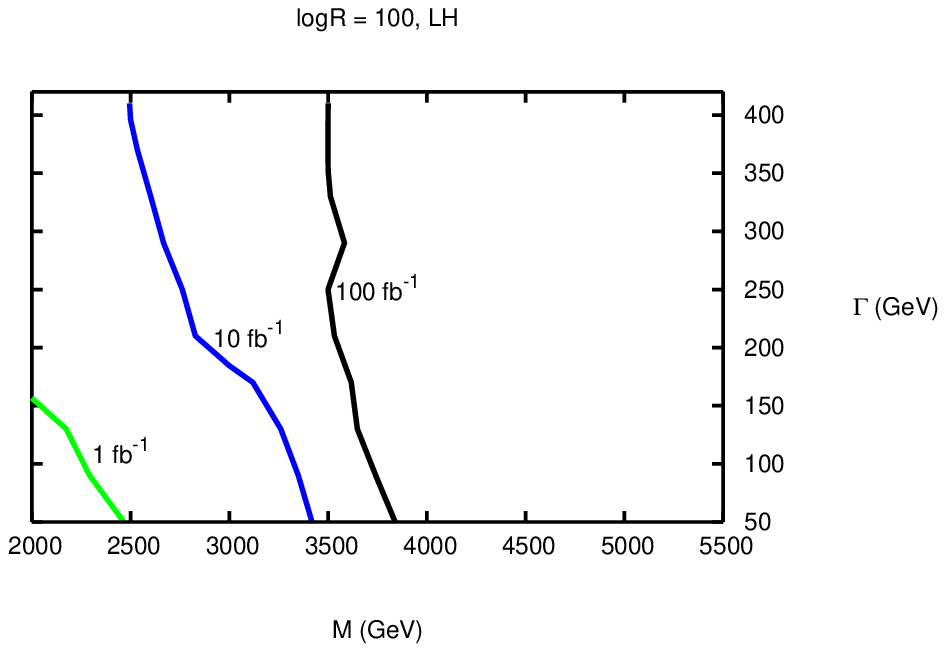}
\caption{The detection reach at the LHC for $\log R = 10$ (left) and $\log R = 100$ (right) at different integrated luminosities for the left-handed case. The colour scheme is identical to the previous figure. In the $\log R = 10$ and 100 fb$^{-1}$ case all points \textit{below} the contour have $\log R < 10$. }
\label{fig:ex_lh_lhc}
\end{figure}

\begin{figure}[!htb]
  \vspace{0.5cm}
  \hspace{1.0cm}
    \includegraphics[scale=0.8]{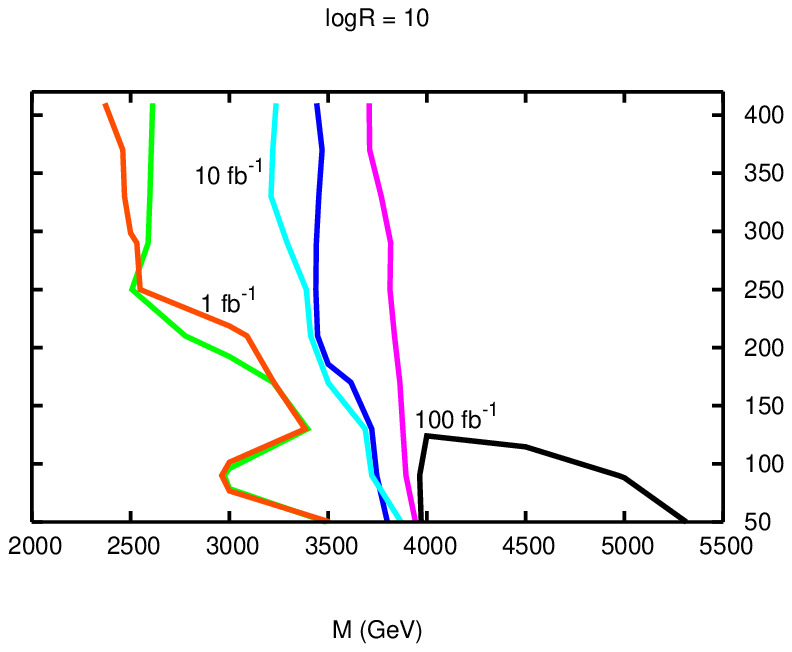}
    \includegraphics[scale=0.8]{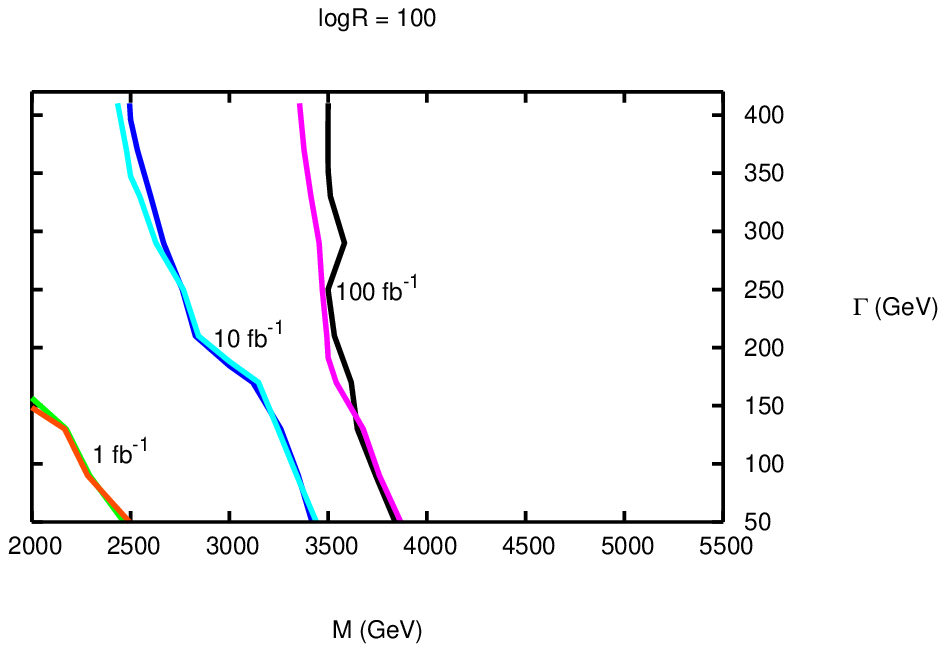}
\caption{The detection reach at the LHC for $\log R = 10$ (left) and $\log R = 100$ (right) at different integrated luminosities for the left- and right-handed cases. The colour scheme is for 1, 10, 100 fb$^{-1}$ is: left-handed: green, blue, black and right-handed: orange, light blue, pink.}
\label{fig:ex_lhc_both}
\end{figure}

\begin{figure}[!htb]
  \vspace{0.5cm}
  \hspace{1.0cm}
    \includegraphics[scale=0.8]{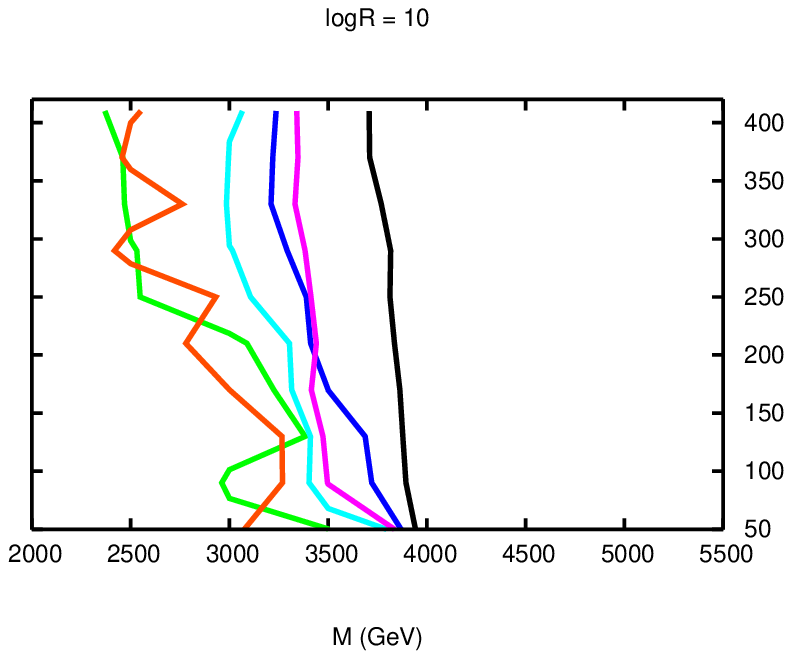}
    \includegraphics[scale=0.8]{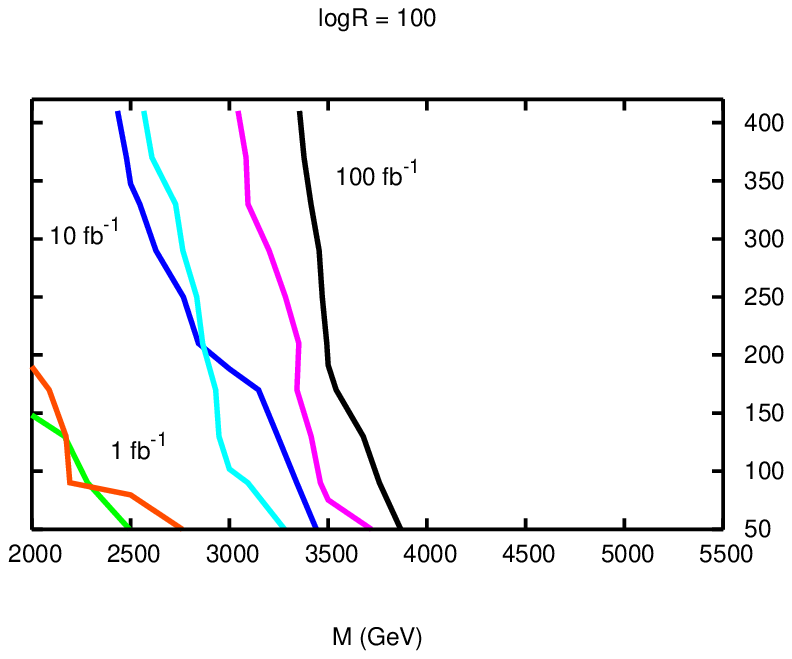}
    \put(10,70){\tiny $\Gamma (\mathrm{GeV})$}
\caption{The detection reach at the LHC for $\log R = 10$ (left) and $\log R = 100$ (right) at different integrated luminosities for the right-handed case compared at LO and NLO. The colour scheme for 1, 10, 100 fb$^{-1}$ is: LO: green, blue, black and NLO: orange, light blue, pink.}
\label{fig:ex_rh_lhclnlo}
\end{figure}

\begin{figure}[!htb]
  \vspace{0.5cm}
  \hspace{1.0cm}
  \includegraphics[scale=0.8]{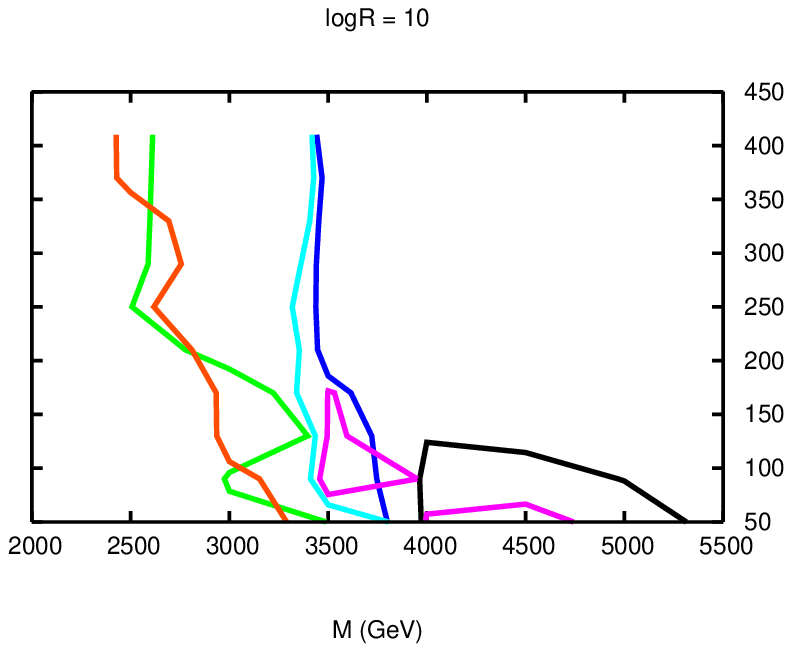}
  \includegraphics[scale=0.8]{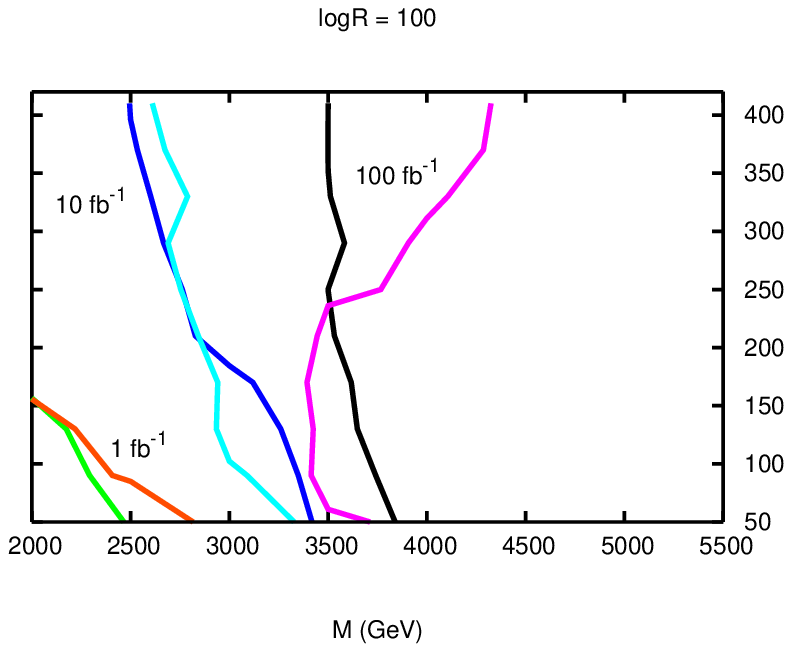}
    \put(10,70){\tiny $\Gamma (\mathrm{GeV})$}
\caption{The detection reach at the LHC for $\log R = 10$ (left) and $\log R = 100$ (right) at different integrated luminosities for the left-handed case compared at LO and NLO. The colour scheme is identical to the previous figure.}
\label{fig:ex_lh_lhc_lnlo}
\end{figure}

\begin{figure}[!htb]
  \vspace{0.5cm}
  \includegraphics[scale=0.8]{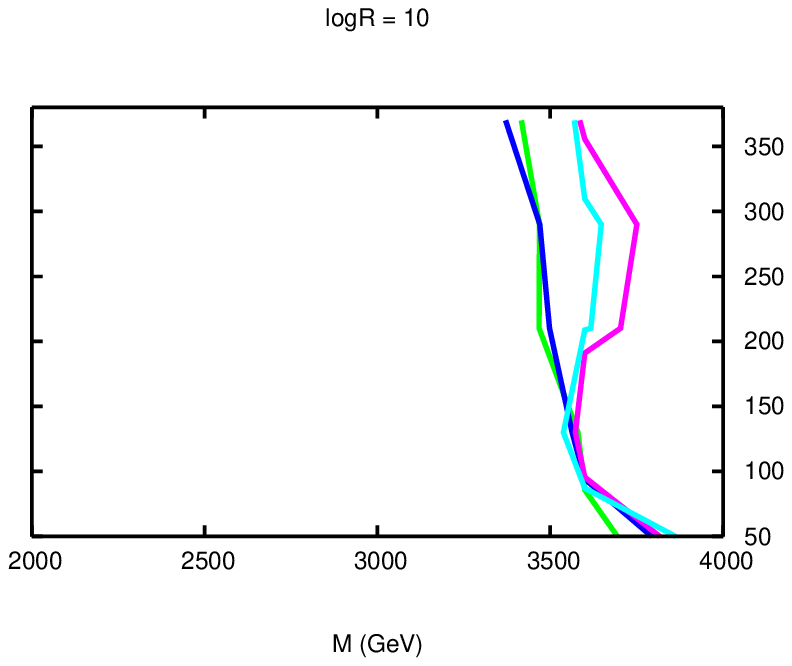}
  \includegraphics[scale=0.8]{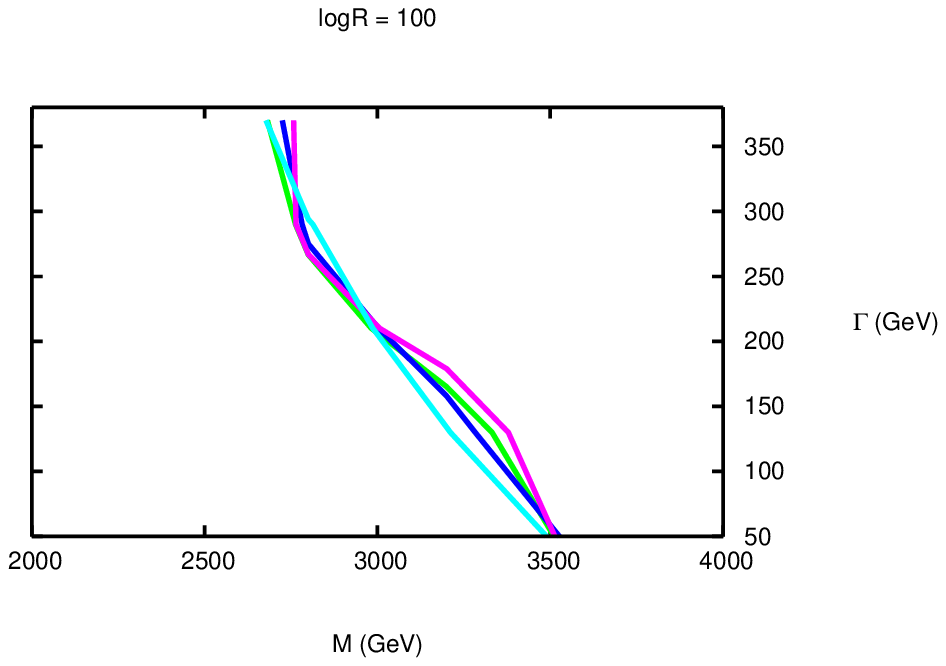}
\caption{The NLO detection reach at the LHC for $\log R = 10$ (left) and $\log R = 100$ (right) for an integrated luminosity of 10 fb$^{-1}$ at different factorisation scale $\mu_F$. The colour scheme for $\mu_F = 0.5\mu_0$, $\mu_0$, $2 \mu_0$ and $4 \mu_0$ is: green, blue, pink and light blue.}
\label{fig:ex_facscale}
\end{figure}

In appendix~\ref{app:modeldiscrimination} we provide a general method
for discriminating between two models given a set of events. Here we
apply this method to extract observation limits on the $W'$ mass and
width at LO. A stand-alone program was written to calculate the
quantity $R$ at matrix element level, given by Eq.~(\ref{eq:Rpoissonexplog}):
\begin{equation}\label{eq:Rpoissonexplog2}
R \frac{p(S)}{p(T)} = \exp{\left(\sum_{i=0}^N \log\left( \frac{ p(M_{T,i}|T) } { p(M_{T,i} | S) } \right) \right)} \times \left( \frac{\bar{N_T}}{\bar{N_S}}\right)^N e^{-(\bar{N_T} - \bar{N_S} ) }\;.\nonumber
\end{equation}
The `true' underlying theory, labelled T, was assumed to contain a $W'$ at a predefined mass and theory S was taken to be the SM. Some comments are appropriate:
\begin{itemize}
\item Although the total $W'$ width was being varied, the decay width to fermions was always assumed to be $\Gamma_{W'\rightarrow ff'} = (4\Gamma_{W}M_{W'}/ 3M_{W}) $.
\item In a real experiment the $W'$ mass would be unknown and maximum likelihood methods should be used to fit the parameters if significant deviation from the SM is found.
\item The $R$ parameter can become very large if a small number of
  unlikely events occur, which favour one theory over the
  other. Experimentally this is unrealistic since `unlikely' events could be the
  effect of background or detector effects. To take these into account, one has to introduce nuisance parameters whose behaviour, at this level of analysis, have to be chosen arbitrarily. Here we avoid the introduction of such arbitrary parameters.
\item The detection curves were drawn for specific data distributions
  and fluctuations are expected. In other words, the plots correspond to a specific `experimental' data set. 
\item The ratio of the prior probability distributions, $p(S)/p(T)$, was taken to be equal to unity throughout this analysis: i.e. we assume both models are equally likely prior to the `experiment'. 
\item A rapidity cut on the leptons corresponding to $y_{cut} = 2.5$ for the LHC and $y_{cut} = 1.3$ for the Tevatron was applied to take into account the acceptance regions of the detectors.
\item The distributions $p(M_T|S)$ and $p(M_T|T)$ were calculated using the Monte Carlo event generator itself at higher statistics ($\sim 10^5$) than the required number of events to reduce the required computer time. The sum over $i$ in Eq.~(\ref{eq:Rpoissonexplog}) was taken over the \textit{bins} of these distributions and not individual events. 
\end{itemize}
The limits were drawn on a width-mass plane as $\log R = C$ detection curves, where $C$ is a
constant. The variable $R$ can be interpreted as a
probability ratio and a discovery curve $\log R = C$ can be
interpreted as the limit where the existence of a $W'$ is discovered
with certainty $1 - e^{-C}$. For example if $C = 10$, then the
detection curve represents the $\sim99.9996\%$ confidence level. The
LO detection curves can be seen, for different integrated luminosities
at the LHC (14 TeV), in Fig~\ref{fig:ex_rh_lhc} for a right-handed
$W'$ and Fig.~\ref{fig:ex_lh_lhc} for a left-handed $W'$. The curves
correspond to a single data sample at each $[M_{W'}, \Gamma_{W'}]$
point, and therefore there are large statistical fluctuations,
particularly in the low-luminosity curves. A comparison between the
curves for a left- and right-handed $W'$ is shown in
Fig.~\ref{fig:ex_lhc_both}. It can be observed that a left-handed $W'$
has a slightly higher detection reach, especially at larger widths. By
examining Figs.~\ref{fig:ex_rh_lhc} and~\ref{fig:ex_lh_lhc}, we deduce
that the maximum detection reach at the LHC, for example using an
integrated luminosity of 100~fb$^{-1}$, for a $W'$ decaying primarily
to fermions ($\Gamma_{W'} \approx \Gamma_{W'\rightarrow ff'}$), is
$\sim 4 \tev$. 
We also show the expected limit at the Tevatron (1.96 TeV) in
Fig.~\ref{fig:ex_tvt_both} with an integrated luminosity of 2
fb$^{-1}$, both at leading and next-to-leading (see below)
orders.  When the $W'$ is only allowed to decay to
fermions, i.e. has width $\Gamma_{W'} \approx 36 \gev$, the predicted
detection limit for $\log R \sim 10$ is $M_{W'} \approx 1.1
\tev$. This is expected to be reduced by experimental effects. Since
the available centre-of-mass energy at the Tevatron is $1.96\tev$, we
expect the saturation of the detection reach to come at about $ M_{W'}
\sim 1 \tev$ without interference and slightly higher in the
left-handed case when interference effects are included. The Tevatron
NLO case does not exhibit any substantial difference from the LO case.
\begin{figure}[!htb]
  \vspace{0.5cm}
  \hspace{1.0cm}
    \includegraphics[scale=0.8]{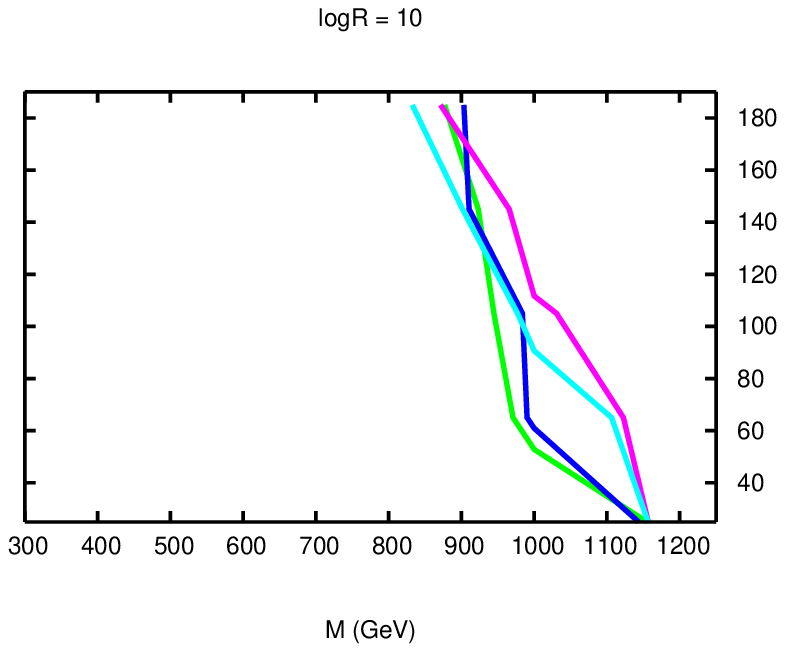}
    \includegraphics[scale=0.8]{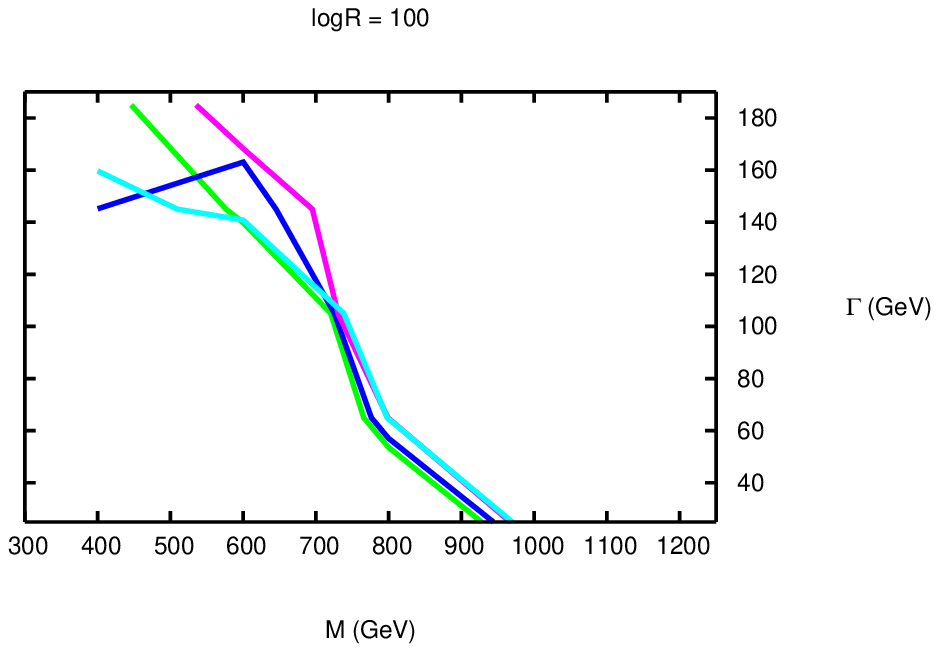}
\caption{The detection reach at the Tevatron for $\log R = 10$ (left) and $\log R = 100$ (right) at 2 fb$^{-1}$, for the left- and right-handed cases, at LO and NLO. The colour scheme is for right-handed and left-handed correspondingly, at LO: green, blue and NLO: light blue, pink.}
\label{fig:ex_tvt_both}
\end{figure}
We have performed an equivalent analysis using the NLO method
\texttt{POWHEG} at matrix element level to improve computational
time. Working at matrix element level with the \texttt{POWHEG} method
is justified since the transverse mass distribution is not
significantly altered after shower and hadronization and no
difficulties arise due to negatively-weighted events, as would be the
case in the \texttt{MC@NLO} case. The comments given at the beginning
of the section for the LO analysis also apply to the NLO analysis. The
results are shown in Figs.~\ref{fig:ex_rh_lhclnlo}
and~\ref{fig:ex_lh_lhc_lnlo} in comparison to the LO results. In the
right-handed chirality scenario, NLO implies a lower detection reach
than indicated at LO. The situation is more complicated in the
left-handed case where the NLO case implies a slightly higher reach
for larger widths.  
To investigate the dependence of the NLO results on the factorisation scale $\mu_F$ we have reproduced the $\log R$ contours for the right-handed $W'$ LHC case with an integrated luminosity of 10 fb$^{-1}$ at different values of $\mu_F$ while keeping the renormalisation scale fixed, using the \msbar scheme. The results are shown in Fig.~\ref{fig:ex_facscale}. The curves show that the factorisation scale does not affect the detection reach substantially, for example only shifting the $\log R = 10$ contour at a width of $\Gamma_{W'} \sim 200\gev$ from $M_{W'} \sim 3500 \gev$ to $M_{W'} \sim 3750\gev$ going from $\mu_F = 0.5 \mu_0$ to $\mu_F = 4 \mu_0$. 
\subsection{Conclusions}\label{sec:wprime:conclusion} 
We have presented a Monte Carlo implementation of the Drell-Yan
production of new charged heavy vector bosons. We have considered the
interference effects with the Standard Model $W$ boson, allowing
arbitrary chiral couplings to the leptons and quarks. Moreover, the
implementation is correct up to next-to-leading order in QCD, via the
\texttt{MC@NLO} and \texttt{POWHEG} methods using the \Herwigpp event
generator. We have presented a sample of results at both leading and
next-to-leading orders. As expected, the LO and NLO boson transverse
momentum distributions were found to differ significantly, the NLO
extending to higher $p_T$. The dilepton transverse mass, invariant
mass, rapidity and $z$-momentum distributions were found not to be
significantly altered by the NLO treatment. The total cross section was found to increase in the NLO case by a factor of $\sim 1.3$ in the region of interest. 

Subsequently we applied a theoretical discrimination method to the $W'$ reference model to obtain mass-width observation curves for left- and right-handed chiralities of the $W'$ both at LO and NLO (\texttt{POWHEG}). The NLO curves were shown not to vary significantly with factorisation scale. The event generator used throughout this analysis, \texttt{Wpnlo}, is fully customisable and publicly available~\cite{wpnlo}.

\section{Searching for third-generation composite leptoquarks}
\label{sec:newphys:leptoquarks}
\subsection{Introduction}\label{sec:newphys:leptointro}
The riddle of electroweak symmetry breaking (EWSB), and specifically the
hierarchy problem we have described in section~\ref{sec:bsmhierarchy},
can be addressed in natural way via the introduction of new physics in
the form of a new strong force. An example of new strong dynamics is
`technicolour' - a theory in which EWSB can be explained without adding any fundamental
scalars and the Higgs boson arises as a composite particle of some new
fermion states. This solution is natural in a literal sense, meaning that we
have already observed an example present in Nature: the hierarchy
between the Planck scale ($10^{19}\gev$) and the proton mass ($\sim 1
\gev$) is a
result of the logarithmic running of the QCD coupling constant and the
onset of the strong coupling in the infrared. 

The Higgs mechanism must explain not only how the masses of the gauge
bosons arise, but also the origin of the fermion masses. We have
already described the conventional mechanism of fermion mass
generation in the SM in section~\ref{sec:sm:fermions}. This is done
conventionally via the so-called `Yukawa' interaction. Before EWSB we have, e.g. for the fermions
of the third generation, couplings of the form:
\begin{equation}\label{eq:lq:yukawasm}
\mathcal{L}_{\mathrm{Yuk.SM}} = \lambda^T \bar{T} \mathcal{O}_H t + \mathrm{h.c.}\;,
\end{equation}
where $\lambda^T$ is the top quark Yukawa coupling, $T$ is the third generation fermion
$SU(2)_L$ doublet, $\mathcal{O}_H$ is the fundamental scalar Higgs operator and
$t$ is the third generation fermion $SU(2)_L$ singlet. In theories of
strong dynamics EWSB it is usual to assume that masses arise via a
similar Yukawa-type interaction:
\begin{equation}\label{eq:lq:yukawastrong}
\mathcal{L}_{\mathrm{Yuk.strong}} = \lambda_s^T \frac{\bar{T} \mathcal{O}_H t}{\Lambda_f^{d-1}} + \mathrm{h.c.}\;,
\end{equation}
where we have included a suppression scale $\Lambda_f$, the scale at which
the flavour physics arises, $d = \left[ \mathcal{O}_h \right]$ is the
mass dimension of the scalar Higgs operator, now taken to be $d \geq
1$ and $\lambda^T_s$ is a `strong' Yukawa coupling (which is
dimensionless like $\lambda^T$). However, in the framework of 
an effective theory, other terms will necessarily need to be added
to the full Lagrangian. One of these terms has the form: 
\begin{equation}\label{eq:lq:extraterms}
\mathcal{L} \supset \frac{ \bar{q}_i q_j \bar{q}_k q_l } {
  \Lambda_f^2}\;.
\end{equation} 
This four-fermion operator is potentially
catastrophic since it can induce flavour-changing neutral currents
(FCNCs) that can contribute to meson mixing. For example, an operator
involving two strange quarks and two down quarks, of the form $\sim
\bar{s}d\bar{d}s$, would contribute to $K^0-\bar{K}^0$ mixing. This
type of FCNCs
have been well-measured experimentally and to satisfy the existing constraints, one
needs to choose the suppression scale to be $\Lambda_f \gtrsim
10^3\gev$. In the SM we can set $\Lambda_f$ as large as we wish and
the FCNCs decouple from the theory since the SM Yukawa interaction term of
Eq.~(\ref{eq:lq:yukawasm}) is not suppressed by any power of $\Lambda_f$. In theories of
strong dynamics, $d > 1$ in general and hence we cannot set the scale
$\Lambda_f$ to be arbitrarily high, otherwise there is a risk of
decoupling the Yukawa term of Eq.~(\ref{eq:lq:yukawastrong}) as well,
rendering the theory incapable of producing fermion masses naturally. This is
particularly true for the large top quark mass.

\begin{figure}[!htb]
\centering
  \includegraphics[scale=0.8]{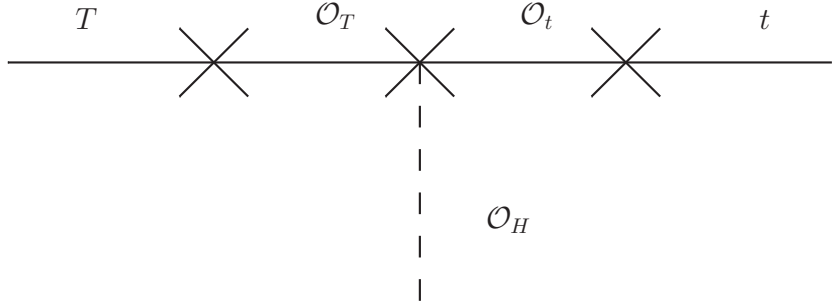}
\caption{A schematic diagram demonstrating how the effective $\bar{T}
  \mathcal{O}_H t$ vertex is formed using the mixing terms and the
  Higgs boson coupling to the composite fermions in the Lagrangian of Eq.~(\ref{eq:lq:mixing}).}
\label{fig:lq:fermionmix}
\end{figure}

A solution to this issue, that was proposed long
ago~\cite{Kaplan:1991dc}, introduces composite fermions which arise due to the
strong dynamics. The elementary
fermions do not couple directly to the scalar Higgs
operator; instead they mix with the composite fermions via
bilinear interactions. Schematically, the Lagrangian terms relevant to the
generation of mass for the third generation is given by
\begin{equation}\label{eq:lq:mixing}
\mathcal{L}_{\mathrm{mix}} \propto m_\rho \left[ \frac{y^T}{g_\rho}
  \bar{T} \mathcal{O}_T + \frac{y^t} { g_\rho } \bar{t} \mathcal{O}_t
  + \bar{\mathcal{O}}_T \mathcal{O}_T + \bar{\mathcal{O}}_t \mathcal{O}_t
\right] + g_\rho \bar{\mathcal{O}}_T \mathcal{O}_H \mathcal{O}_t +
\mathrm{h.c.}\;,
\end{equation}
where $m_\rho$ and $g_\rho$ are the strong coupling mass scale and
coupling respectively, $y^T$ and $y^t$ are the mixing parameters of
the theory, corresponding to the left- and right-handed top quark
multiplets, $\mathcal{O}_T$ and $\mathcal{O}_t$ are composite
left- and right-handed fermions respectively, and $\mathcal{O}_H$ is
the composite scalar Higgs operator.  Note that there are two mixing
parameters for each fermion, one for each chirality. As an example, the fermions
$\mathcal{O}_{T,t}$ could be technibaryons, composites of some new `technifermions' in a technicolour
theory, just as the protons are composites of quarks in QCD. The
Lagrangian $\mathcal{L}_\mathrm{mix}$ produces an effective
Yukawa term for the elementary fermions. By examining Fig.~\ref{fig:lq:fermionmix}:
\begin{equation}
\mathcal{L}_{\mathrm{Yuk.eff}} \propto \frac{y^T y^t} { g_\rho } \bar{T}
\mathcal{O}_H t \equiv \lambda^T \bar{T} \mathcal{O}_H t + \mathrm{h.c.}\;,
\end{equation}
where we have used the mixing parameters to define the top Yukawa coupling:
\begin{equation}\label{eq:lq:topyukawa}
\lambda^T \equiv \frac{y^T y^t} { g_\rho}\;.
\end{equation}
The mechanism is not only capable of producing fermion masses, but
also offers the hope that the observed hierarchies of masses and
mixings of the SM fermions may be related to the electroweak hierarchy
via strong-coupling effects. This is because the third generation can
be considered to be the `most composite' and hence should have masses closest
to the strong dynamics scale~\cite{Kaplan:1991dc}. 

It is obvious now that the new strong sector must know about $SU(3)$
colour and must contain, at the very least, colour-triplet fermionic
resonances that mix with the elementary colour-triplets and make the
observed fermions. It is reasonable to also expect the strong sector
to contain other coloured resonances. It may contain bosonic
resonances that, depending on their gauge charges, may be able to
couple to a lepton and a quark, playing the role of the leptoquark
states. For example, in Ref.~\cite{Kaplan:1991dc}, in which the fermionic
resonances were the technibaryons of $SU(3)$, the model would
also contain technimesons that would be able to act as
leptoquarks. 

\begin{figure}[!htb]
\centering
  \includegraphics[scale=0.8]{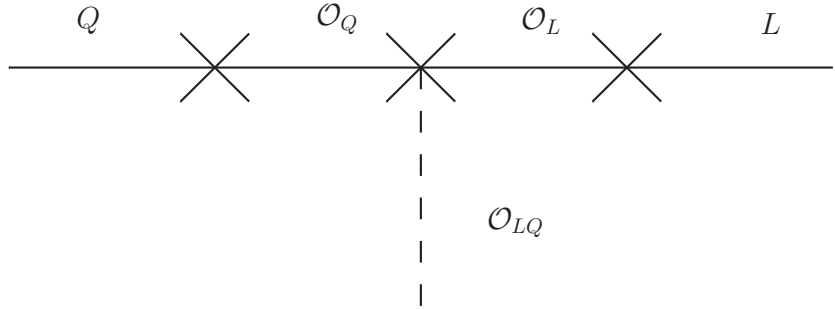}
\caption{A schematic diagram demonstrating how the effective $\bar{Q}
  \mathcal{O}_{LQ} L$ vertex is formed using the mixing terms and the
  leptoquark coupling to the composite quarks and leptons in the Lagrangian of Eq.~(\ref{eq:lq:lqcouplings}).}
\label{fig:lq:lqcouplings}
\end{figure}

It is easy to make an estimate of the magnitude of the leptoquark
couplings to fermions in such models. These arise in much the same way
as the effective Higgs boson couplings, through the following schematic
Lagrangian:
\begin{equation}\label{eq:lq:lqcouplings}
\mathcal{L}_{\mathrm{LQ}} \propto m_\rho \left[ \frac{y^Q}{g_\rho}
  \bar{Q} \mathcal{O}_Q + \frac{y^L} { g_\rho } \bar{L} \mathcal{O}_L
  + \bar{\mathcal{O}}_Q \mathcal{O}_Q + \bar{\mathcal{O}}_L \mathcal{O}_L
\right] + g_\rho \bar{\mathcal{O}}_Q \mathcal{O}_{LQ} \mathcal{O}_L +
\mathrm{h.c.}\;,
\end{equation}
where $y^Q$ is one of the quark mixing parameters, $y^L$ is one of
the lepton mixing parameters and $\mathcal{O}_{LQ}$ is a leptoquark
operator with the correct gauge charges to couple to a composite quark
and a composite lepton. It is important to note that the mixing
parameters that appear in Eq.~(\ref{eq:lq:lqcouplings}) are the same as
those that appear in $\mathcal{L}_{\mathrm{mix}}$ (Eq.~(\ref{eq:lq:mixing})). The leptoquark
effective coupling can then be calculated by considering the schematic
diagram in Fig.~\ref{fig:lq:lqcouplings}. We can deduce that the form
of the coupling is
\begin{equation}\label{eq:lq:lqcoupldef}
\mathcal{L}_{\mathrm{LQff}} \propto \frac{y^Q y^L} { g_\rho } \bar{Q}
\mathcal{O}_{LQ} L \equiv \lambda^{LQ} \bar{Q} \mathcal{O}_{LQ} L +
\mathrm{h.c.}\;,
\end{equation}
where we have defined the leptoquark coupling to quarks and leptons,
$\lambda^{LQ} \equiv  y^Q y^L / g_\rho $. We can make estimates of the
magnitude of the coupling using the measured Yukawa couplings if we
restrict the mixing parameters to be equal for the quarks and leptons
of each generation:
\begin{eqnarray}
y^{Q} &\sim& y^{q} \sim y^{r} \;,\nonumber \\
y^{L} &\sim& y^{\nu} \sim y^{\ell} \;,
\end{eqnarray}
where $Q \in \{U,C,T\}$ and $L \in \{L_1, L_2, L_3\}$
(left-handed doublets), $q \in \{u,c,t\}$
(right-handed up-type singlets), $r \in \{d, s, b\}$ (right-handed
down-type singlets), $\nu \in \{\nu_{eR}, \nu_{\mu R}, \nu_{\tau R}\}$
  (hypothetical right-handed neutrinos) and $\ell \in \{e_R, \mu_R,
  \tau_R\}$. We can then use the measured Yukawa couplings and
  estimate each of the mixing parameters that appear in
  Eq.~(\ref{eq:lq:topyukawa}) for the quarks and leptons:
\begin{eqnarray}
y^Q \sim \sqrt{\lambda^Q g_\rho}\;,\;\; y^L \sim \sqrt{\lambda^L
  g_\rho}\;.
\end{eqnarray}
We can substitute these estimates into the leptoquark coupling to fermions defined in Eq.~(\ref{eq:lq:lqcoupldef}), to obtain an
estimate in terms of the measured Yukawa
couplings:
\begin{equation}
\lambda^{LQ} \sim \sqrt{\lambda^L \lambda^Q}\;.
\end{equation}
From the above equation it is easy to see that couplings to quarks of
the third generation will dominate in this type of models. The SM
fermion Yukawa couplings (taken from Ref.~\cite{Gripaios:2009dq}) are given in Table~\ref{tb:lq:yukawas} and the
resulting estimates of the leptoquark couplings are given in
Table~\ref{tb:lq:couplestimates}. These estimates evade the
constraints coming from flavour experiments, for leptoquark masses
even down to $\sim 200\gev$, which may arise if the leptoquarks appear as
pseudo-Nambu-Goldstone bosons~\cite{Gripaios:2009dq}.
\begin{table}[!t]
\begin{center}
\begin{tabular}{|c|c|} \hline
Fermion & Yukawa, $\lambda^F$ \\\hline 
$e$ & $2.87\times 10^{-6}$\\
$\mu$  & $6.09\times 10^{-4}$ \\
$\tau$ &   $1.02 \times 10^{-2}$ \\\hline
$d$ & $2.30\times 10^{-5}$\\
$s$ &  $3.39\times 10^{-4}$ \\
$b$ & $1.40 \times 10^{-2}$\\\hline
$u$ &  $2.87\times 10^{-6}$\\
$c$ & $6.09\times 10^{-4}$ \\
$t$ & $1.02 \times 10^{-2}$ \\\hline
\end{tabular}
\end{center}
\caption{The Standard Model fermion Yukawa couplings~\cite{Gripaios:2009dq}.}
\label{tb:lq:yukawas}
\end{table}
\begin{table}[!t]
\begin{center}
\begin{tabular}{|c|c|c|c|} \hline
Lepton \ Quark & 1 & 2 & 3 \\\hline 
1 & $8.2 \times 10^{-6}$ & $1.0 \times 10^{-4}$ & $1.5\times 10^{-3}$ \\
2 & $1.2 \times 10^{-4}$ & $1.5 \times 10^{-3}$ & $2.2\times 10^{-2}$\\
3 &  $4.9 \times 10^{-4}$ & $6.0\times 10^{-3}$ & $9.0\times 10^{-2}$  \\\hline
\end{tabular}
\end{center}
\caption{Estimates of the leptoquark couplings to fermions~\cite{Gripaios:2009dq}, using the measured
  Yukawa couplings that appear in Table~\ref{tb:lq:yukawas} and the assumption
  that there is only one mixing parameter for the quarks and leptons
  of each generation.}
\label{tb:lq:couplestimates}
\end{table}

Since the leptoquark couplings to light fermions are highly suppressed, the only relevant couplings for direct collider production and detection are those to third-generation fermions.\footnote{For an alternative scenario with leptoquarks of this type, see \cite{Davidson:2010uu}.} As a result, the leptoquark states will  decay exclusively to third-generation fermions, that is to $t \tau$ or $t \nu_\tau$ or $b \tau$ or $b \nu_\tau$. Na\"{\i}vely, since the leptoquark couplings scale roughly with the Yukawa couplings, and since the bounds preclude a leptoquark mass below $m_t$,\footnote{Searches at D0 for third-generation scalar leptoquarks decaying exclusively to $b \tau$ or $b\nu_{\tau}$ yield bounds of 210 GeV \cite{Abazov:2008jp} and 229 GeV  \cite{Abazov:2007bsa} respectively.} one might conclude that decays involving the top must dominate. However, we shall see later that the gauge quantum numbers sometimes preclude couplings to top quarks and, of course, unknown global symmetries may also preclude one or more couplings. Thus we consider all four possible couplings. 

Since leptoquarks couple dominantly to third-generation quarks and leptons, pair-production through colour gauge interactions will overwhelmingly dominate single-production at the LHC. The channels of interest therefore involve pair-wise combinations of $t \tau$ or $t \nu_\tau$ or $b \tau$ or $b \nu_\tau$.\footnote{Note that third-generation lepton-quark couplings are also possible in $R$-parity-violating supersymmetric theories.}
The $2b 2\tau$ and $2b + \slashed{E_T}$ channels already have been the subject of searches at the Tevatron \cite{Abazov:2007bsa,Abazov:2008jp}, and can be adapted easily for the LHC. The use of novel kinematic variables such as $M_{T2}$ in this $2b+\slashed{E_T}$ channel may well improve the prospects for discovery and mass measurement. The two channels involving the top require more ingenuity, but merit investigation.

In the present section we perform the first detailed phenomenological
study of the possible production of such states at the LHC. In section~\ref{sec:phen} we briefly review their quantum numbers,
couplings and decay modes, which we have implemented in the \Herwigpp
event
generator.\footnote{\Pythia~\cite{Sjostrand:2006za,Sjostrand:2007gs,Sjostrand:2008vc}
  contains an implementation of a single scalar leptoquark of
  arbitrary flavour.}  This allows us to propose and investigate strategies for reconstructing third-generation leptoquark masses from
their decay products, including those that involve top quarks, in the following
sections.  Our conclusions are presented in section~\ref{sec:conc}.

Although we focus here on direct searches at the LHC, there are also promising channels for indirect searches, namely in 
$B_d \rightarrow K \bar{\mu} \mu$ and $B_s\rightarrow \mu \mu$ at LHCb, in $\mu \rightarrow e \gamma$ and $\tau \rightarrow \mu \gamma$, in $\mu-e$ conversion in nuclei, and in  $\tau \rightarrow \eta \mu$ at future B factories \cite{Gripaios:2009dq}.

\subsection{Phenomenology}\label{sec:phen}
\subsubsection{Scalar leptoquark pair-production}
We focus on scalar leptoquarks in the present study since their
bosonic couplings are determined completely by QCD and hence their
production cross sections only depend on their masses. Moreover, the
lightest (and most easily accessible) leptoquarks in these scenarios
arise as scalar pseudo-Nambu-Goldstone bosons. The type of leptoquarks
we are considering are predominantly pair-produced via gluon-gluon
fusion or quark-anti-quark annihilation, due to the fact that they
couple to the third-generation quarks and leptons. Only
charge-conjugate leptoquarks can be produced in this way: associated
production of different leptoquarks is forbidden since it would not
conserve the Standard Model gauge quantum numbers. Single-production
in association with a lepton is allowed but at a $14\tev$ LHC it
becomes dominant at leptoquark masses of about $2.2~\mathrm{TeV}$, at
which point the total cross section, $\sigma \sim
10^{-2}~\mathrm{fb}$, is already too low for discovery.
\subsubsection{Effective Lagrangian for interactions with gluons}
The effective Lagrangian describing the interaction of the scalar leptoquarks with gluons is~\cite{Blumlein:1996qp}
\begin{equation}
\mathcal{L}^g_S = \left(D^\mu_{ij} \Phi^j\right)^\dagger (D^{ik}_\mu \Phi_k)- M_{LQ}^2 \Phi^{i\dagger} \Phi_i\;,
\end{equation}
where $\Phi$ is a scalar leptoquark, $i,j,k$ are colour indices, the field strength tensor of the gluon field is given by
\begin{equation}
\mathcal{G}^a_{\mu \nu} = \partial_\mu \mathcal{A}_\nu^a - \partial_\nu \mathcal{A}^a_\mu + g_s f^{abc} \mathcal{A}_{\mu b} \mathcal{A}_{\nu c}\;,
\end{equation}
and the covariant derivative is
\begin{equation}
D^{ij}_\mu = \partial_\mu \delta^{ij} - i g_s t^{ij}_a \mathcal{A}_\mu^a\;.
\end{equation}
The Feynman rules that result from this Lagrangian and the
diagrams that contribute to pair-production of scalar leptoquarks are
given in appendix~\ref{app:feynman}. Expressions for the cross
sections are given in appendix~\ref{app:cross-sections}.
\subsubsection{Non-derivative fermion couplings}\label{sec:naming}
The effective Lagrangian that describes the possible non-derivative couplings of the scalar leptoquarks to third-generation quarks and leptons is given by~\cite{Belyaev:2005ew}
\begin{eqnarray}
\mathcal{L}_{nd} &=& ( g_{0L} \bar{q}_L^c i \tau _2 \ell_L + g_{0R} \bar{t}^c_R \tau_R ) S_0 \nonumber\\
&+& \tilde{g}_{0R} \bar{b}_R^c \tau_R \tilde{S}_0 + g_{1L} \bar{q}^c_L i \tau _2 \tau_a \ell _L S_1^a \nonumber \\
&+& (h_{1L} \bar{t}_R \ell_L + h_{1R} \bar{q}_L i\tau_2 \tau_R ) S_{1/2} + h_{2L} \bar{b}_R \ell_L \tilde{S}_{1/2} + \mathrm{h.c.}\;,
\label{eq:ells}
\end{eqnarray}
where the $\tau_a$ are the Pauli matrices, $q_L$ and $\ell_L$ are
$SU(2)_L$ quark and lepton doublets respectively and $t_R$, $b_R$ and
$\tau_R$ are the corresponding singlet fields. We denote
charge-conjugate fields by $f^c_{R,L} = (P_{R,L} f ) ^c$, where the
superscript $^c$ implies charge conjugation. In Table~\ref{tb:lquarks}
we give the quantum numbers for the five types of non-derivatively
coupled scalar leptoquarks: the $SU(2)_L$-singlet complex scalars
$S_0$ and $\tilde{S}_0$, the $SU(2)_L$-triplet complex scalar $S_1$ and the $SU(2)_L$-doublets $S_{1/2}$ and $\tilde{S}_{1/2}$. 
\begin{table}[ptb]
\begin{center}
\begin{tabular}
{|c|c|c|c|c|c|c|} \hline
Name & $SU(3)_c$ & $T^3$ & $Y$  & $Q_{\mathrm{em}}$ & Decay mode & \Herwigpp id \\ \hline\hline
$S_0$ & $\bar{3}$ & 0 &  1/3 &   1/3 & $\bar{\tau}_R \bar{t}_R$, $\bar{\tau}_L \bar{t}_L$, $\bar{\nu} _{\tau ,L} \bar{b}_L$  & -9911561  \\ \hline\hline
$\tilde{S}_0$ & $\bar{3}$ & 0 & 4/3 &  4/3 &$ \bar{\tau}_R \bar{b}_R $ & -9921551  \\ \hline \hline
$S_1^{(+)}$ & $\bar{3}$ & +1 &  1/3 &  4/3 &$\bar{\tau}_L \bar{b}_L$  & -9931551  \\ \hline
$S_1^{(0)}$ & $\bar{3}$ & 0 & 1/3 &  1/3 & $\bar{\tau}_L\bar{t_L}$, $\bar{\nu}_{\tau,L} \bar{b}_L$ & -9931561  \\ \hline
$S_1^{(-)}$ & $\bar{3}$ & -1 &  1/3 & -2/3 &$\bar{\nu}_{\tau,L} \bar{t}_L$  & -9931661  \\ \hline\hline
$S_{1/2}^{(+)}$ & $3$ & +1/2 & 7/6 &  5/3 & $t_R \bar{\tau}_L, t_L \bar{\tau}_R$& 9941561  \\ \hline
$S_{1/2}^{(-)}$ & $3$ & -1/2 & 7/6 &  2/3 & $b_L \bar{\tau}_R$, $t_R \bar{\nu}_{\tau,L}$ & 9941551  \\ \hline\hline
$\tilde{S}_{1/2}^{(+)}$ & $3$ & +1/2 & 1/6 &  2/3 & $b_R \bar{\tau}_L$  & 9951551  \\ \hline
$\tilde{S}_{1/2}^{(-)}$ & $3$ & -1/2 & 1/6 &  -1/3 & $b_R \bar{\nu}_{\tau,L}$  & 9951651  \\ \hline
\end{tabular}
\end{center}
\caption{Numbering scheme, charges and possible decay modes for the non-derivatively coupled scalar leptoquarks. $Y$ represents the $U(1)_Y$ charge and $T^3$ is the third component of the $SU(2)_L$ charge. Since $S_1$ is an $SU(2)_L$ triplet, it contains three complex scalars. The $S_{1/2}$ and $\tilde{S}_{1/2}$ are $SU(2)_L$ doublets. The naming convention is explained in the text. The minus sign in the ids of some of the leptoquarks indicates the fact that they are anti-triplets of $SU(3)_c$.}
\label{tb:lquarks}
\end{table}

The numbering scheme used in our implementation of scalar leptoquarks in \Herwigpp is also given in Table~\ref{tb:lquarks}. The particles are numbered as $99NDDDJ$, where $N$ distinguishes the representation of the standard model gauge group, $DDD$ is the lowest possible number chosen to relate the leptoquark to the Particle Data Group (PDG) codes of decaying fermions, and $J = 2 S + 1$, where $S$ is the particle spin. The sign of the PDG code is negative for colour anti-triplets and positive for colour triplets. Hence, $-9911561$ is the `first' type of leptoquark, $S_0$, and can decay to particles with codes $15$ ($\tau$) and $6$ ($t$). 

Notice that the first three kinds of leptoquarks, the $S_0$,
$\tilde{S}_0$ and the $S_1$ triplet are colour anti-triplets and the
particles (as opposed to the anti-particles) decay into an anti-lepton
and an anti-quark. This is contrast to the $S_{1/2}$ and
$\tilde{S}_{1/2}$ doublets, which are colour-triplets and decay into
quarks and anti-leptons. 

\subsubsection{Derivative fermion couplings}\label{sec:derivcoup}
We also consider leptoquarks that couple derivatively to the quarks and leptons. The couplings of the leptoquarks to fermions involve three fields, and hence two independent positions for the derivative to act, modulo integration by parts. Here, we choose to place the derivative on either the quark or the lepton, such that the Lagrangian is given by
\begin{eqnarray}
\mathcal{L}_d &=&  \frac{-i }{\sqrt{2} f} ( g'_{0L,i}\bar{q}_Lp^{\mu,i}  \gamma_\mu  \ell_L + g'_{0R,i} \bar{b}_Rp^{\mu,i}  \gamma_\mu \tau_R )S'_0 \nonumber\\ 
&+& \frac{-i }{\sqrt{2} f} \tilde{g}'_{0R,i} \bar{t}_R p^{\mu,i}  \gamma_\mu \tau_R \tilde{S}'_0 + \frac{-i }{\sqrt{2} f} g'_{1L,i} \bar{q}_Lp^{\mu,i}  \gamma_\mu  \tau _a \ell _L  S_1'^a \nonumber \\
&+& \frac{-i }{\sqrt{2} f} (h'_{1L,i} \bar{b}^c_R p^{\mu,i} \gamma_\mu \ell_L + h'_{1R,i} \bar{q}^c_L p^{\mu,i} \gamma_\mu \tau_R ) S'_{1/2} +\frac{-i }{\sqrt{2} f} h'_{2L,i} \bar{t}^c_R p^{\mu,i} \gamma_\mu \ell_L \tilde{S}'_{1/2} +\mathrm{h.c.}\;,\nonumber\\
\label{eq:lagd}
\end{eqnarray}
where the index $a \in \{1, 2, 3\}$ and $p^{\mu, i}$, $i \in \{l,q\}$, denotes the momentum of the lepton or quark.
\begin{table}[ptb]
\begin{center}
\begin{tabular}
{|c|c|c|c|c|c|c|} \hline 
Name & $SU(3)_c$ & $T^3$ & $Y$  & $Q_{\mathrm{em}}$ & Decay mode & \Herwigpp id \\ \hline\hline
$S_0'$ & $3$ & 0 &  2/3 & 2/3 & $t_R\bar{\nu}_{\tau,L}, b_R \bar{\tau}_L, b_L \bar{\tau}_R$ & 9961551  \\ \hline\hline
$\tilde{S}_0'$ & $3$ & 0 & 5/3 &  5/3 & $t_R \bar{\tau}_L, t_L \bar{\tau}_R$ & 9971561  \\ \hline \hline
$S_1^{'(+)}$ & $3$ & +1 &  2/3 & 5/3  & $t_R \bar{\tau}_L, t_L \bar{\tau}_R$  & 9981561  \\ \hline
$S_1^{'(0)}$ & $3$ & 0 & 2/3 & 2/3  & $t_R\bar{\nu}_{\tau,L}, b_L \bar{\tau}_R, b_R \bar{\tau}_L$ & 9981551  \\ \hline
$S_1^{'(-)}$ & $3$ & -1 &  2/3 & -1/3 & $b_R \bar{\nu}_L$ & 9981651  \\ \hline\hline
$S_{1/2}^{'(+)}$ & $\bar{3}$ & +1/2 & 5/6 & 4/3  & $\bar{b}_L \bar{\tau}_L, \bar{b}_R \bar{\tau}_R$  &-9991551  \\ \hline
$S_{1/2}^{'(-)}$ & $\bar{3}$ & -1/2 & 5/6 & 1/3  & $\bar{b}_L \bar{\nu}_{\tau,L}, \bar{t}_R \bar{\tau}_R, \bar{t}_L \bar{\tau}_L$  &-9991561    \\ \hline\hline
$\tilde{S}_{1/2}^{'(+)}$ & $\bar{3}$ & +1/2 & -1/6 & 1/3  &$\bar{t}_L \bar{\tau}_L, \bar{t}_R \bar{\tau}_R$  & -9901561   \\ \hline
$\tilde{S}_{1/2}^{'(-)}$ & $\bar{3}$ & -1/2 & -1/6 & -2/3  & $\bar{t}_L \bar{\nu}_{\tau,L}$  & -9901661  \\ \hline
\end{tabular}
\end{center}
\caption[]{Numbering scheme, charges and possible decay modes for the derivatively-coupled scalar leptoquarks. The details are as in Table~\ref{tb:lquarks}.}
\label{tb:lquarksprime}
\end{table}

The charges of the primed scalar states appear in Table~\ref{tb:lquarksprime}; they correspond, of course, to those of vector leptoquarks. Notice that whereas the $S_0$ is a colour anti-triplet, $S'_0$ is a colour triplet and so on.

Consider a leptoquark $S_0'$ that couples derivatively to fermions in the following way:
\begin{equation}
\mathcal{L} \sim \frac{1 }{\sqrt{2} f} \left( g'_{0L,i} \bar{t}_L \slashed{p}^i S_0' \nu_L + g'_{0L,i} \bar{b}_L \slashed{p}^i S_0' \tau_L +g'_{0R,i}\bar{b}_R \slashed{p}^iS_0' \tau_R \right)+\mathrm{h.c.}\;, 
\end{equation}
where the $f$ is the sigma model scale for the strong dynamics. Consider the decay of the $S_0'$ to on-shell fermions via the coupling $g'_{0L,i}\bar{b}_L p^{\mu,i}  \gamma_\mu  t_L$. We then have
\begin{eqnarray}
g'_{0L,i} \bar{b}_L \slashed{p}^i \tau_L S_0 '  &=& g'_{0L,q} \bar{b}_L \slashed{p}^q S_0' \ell_L + g'_{0L,\ell}\bar{b}_L \slashed{p}^\ell \tau _L S_0' \nonumber\\
&=& g'_{0L,q} m_b \bar{b}_R S_0 ' \ell_L + g'_{0L,\ell} m_\tau \bar{b}_L S_0' \tau_R\;.
 \label{eq:noconjmanip}
\end{eqnarray}
Note that the chirality of one decay product is reversed in each term
by the mass insertion, which breaks the gauge symmetry. An equivalent
manipulation is given in appendix~\ref{app:conj} for terms that
contain conjugate fields. For simplicity, we choose to set the quark
and lepton primed couplings for each term equal, $g'_\ell = g'_q =
g'$, where $g'$ represents $g'_0$, $g'_1$ or $h'_1$. As a result of the above manipulation, an effective Lagrangian for the on-shell decay of a scalar leptoquark $S'_0$ may be written as
\begin{eqnarray}
\mathcal{L}_{\mathrm{eff.}} &\sim& \frac{1 }{\sqrt{2} f} \left( g'_{0L} m_t \bar{t}_R S_0' \nu_{\tau,L} + g'_{0L} m_b \bar{b}_R S_0' \tau_L + g'_{0L} m_\tau \bar{b}_L S_0' \tau_R \right. \nonumber\\ 
&+&\left.g'_{0R}\bar{b}_L m_b S_0' \tau_R+ g'_{0R}\bar{b}_R m_\tau S_0' \tau_L \right) +\mathrm{h.c.} \\
\Rightarrow \mathcal{L}_{\mathrm{eff.}} &\sim& \left[ \frac{1 }{\sqrt{2} f}  (g'_{0L} m_b+ g'_{0R} m_\tau)\right] \bar{b}_R S_0' \tau_L\nonumber\\
&+& \left[ \frac{1 }{\sqrt{2} f} ( g'_{0L} m_\tau + g'_{0R} m_b )  \right] \bar{b}_L S_0' \tau_R \nonumber \\
&+& \left[  \frac{1 }{\sqrt{2} f} (g'_{0L} m_t)\right]  \bar{t}_R S_0' \nu_{\tau,L}+\mathrm{h.c.}\;,
\end{eqnarray}
converting all the derivative couplings to ones that look
like those for the unprimed leptoquarks, with the lepton or fermion
masses appearing in the coupling. See appendix~\ref{app:conj} for the
full effective Lagrangian. Since the scale $f$ is typically a few
hundred GeV, couplings proportional to the top quark mass are expected
to dominate when the corresponding decays are kinematically
allowed. The on-shell fermion assumption is realistic since the widths
of the fermions are small in comparison to their masses and hence
off-shell effects are negligible.
\subsubsection{Decay widths}
The decay width of non-derivatively coupled scalar leptoquarks in the
limit of \textit{massless} quarks and leptons is given by~\cite{Belyaev:2005ew}
\begin{equation}
\Gamma = \frac{M_{LQ}}{16 \pi} \left( \lambda^2_L (\ell q) + \lambda^2_L (\nu q) + \lambda_R^2 (\nu q) \right) \;,
\label{eq:width}
\end{equation}
where the couplings $\lambda_{L,R}(\ell q)$ for the types of leptoquarks we are considering are given in Table~\ref{tb:lambdas}  in terms of the couplings that appear in the Lagrangian. The couplings are taken to be real. The expression gives, for quark-lepton couplings $g \sim 0.1$ and leptoquark mass of $\sim 400 \gev$, a width of $\sim 0.1 \gev$.
\begin{table}[ptb]
\begin{center}
\begin{tabular}
{|c|c|c|c|} \hline
Name & $\lambda_{L} (\ell q)$ & $\lambda_R(\ell q)$ & $\lambda_L(\nu q) $    \\ \hline\hline
$S_0$ & $g_{0L}$ & $g_{0R}$ & $-g_{0L}$ \\ \hline\hline
$\tilde{S}_0$ & 0 & $\tilde{g}_{0R}$ & 0 \\ \hline \hline
$S_1^{(+)}$ & $\sqrt{2} g_{1L}$ & 0  &  0 \\ \hline 
$S_1^{(0)}$ & $-g_{1L} $ & 0 &  $-g_{1L}$ \\ \hline
$S_1^{(-)}$ & 0 & 0  & $\sqrt{2} g_{1L}$ \\ \hline \hline
$S_{1/2}^{(+)}$ & $h_{1L}$ & $h_{1R}$ & 0 \\ \hline
$S_{1/2}^{(-)}$ & 0 & $-h_{1R}$ & $h_{1L}$ \\ \hline \hline
$\tilde{S}_{1/2}^{(+)}$ & $h_{2L}$ & 0 & 0 \\ \hline 
$\tilde{S}_{1/2}^{(-)}$ & 0& 0 & $h_{2L}$ \\ \hline
\end{tabular}
\end{center}
\caption{The $\lambda_i$ couplings of the non-derivatively-coupled scalar leptoquarks to the different quark-lepton combinations, as they appear in the Lagrangian.}
\label{tb:lambdas}
\end{table}
The decay width to massive $q \ell$ is further suppressed by a phase-space factor compared to the massless quark and lepton width~\cite{Abazov:2007bsa}:
\begin{eqnarray}
F \sim ( 1 - r_q - r_\ell )\sqrt{ 1+(r_q-r_\ell)^2 - 2 r_q - 2 r_l } \;,
\end{eqnarray} 
where $r_{q,\ell}$ are the squared ratios $ m_{q,\ell}^2/M_{LQ}^2$ respectively.

\begin{table}[htb]
\begin{center}
\begin{tabular}
{|c|c|c|c|} \hline
Name & $\lambda_{L} (\ell q) \times \sqrt{2} f$ & $\lambda_R(\ell q)  \times \sqrt{2} f$ & $\lambda_L(\nu q)  \times \sqrt{2} f $    \\ \hline\hline
$S_0'$ & $ g'_{0L,q} m_b + g'_{0R,\ell} m_{\tau} $ & $g'_{0R,q} m_b + g'_{0L,\ell} m_{\tau} $ & $ g'_{0L,q} m_t $ \\ \hline\hline
$\tilde{S}_0'$ & $\tilde{g}'_{0R,\ell} m_{\tau}$ & $\tilde{g}'_{0R,q} m_t$ & 0 \\ \hline \hline
$S_1^{'(+)}$ & $\sqrt{2} g'_{1L,q} m_t $ & $\sqrt{2} g'_{1L,\ell} m_{\tau}$ &  0 \\ \hline 
$S_1^{'(0)}$ & $- g'_{1L,q} m_b $ & $-g'_{1L,\ell} m_\tau$ &  $g'_{1L,q} m_t $ \\ \hline
$S_1^{'(-)}$ & 0 & 0  & $\sqrt{2} g'_{1L,q} m_b$ \\ \hline \hline
$S_{1/2}^{'(+)}$ & $h'_{1L,q} m_b + h'_{1R,\ell} m_{\tau} $ & $h'_{1R,q}  m_b + h'_{1L,\ell} m_{\tau} $  & 0  \\ \hline
$S_{1/2}^{'(-)}$ & $h'_{1R,\ell} m_{\tau} $ & $h'_{1R,q} m_t$ & $h'_{1L,q} m_b$ \\ \hline \hline
$\tilde{S}_{1/2}^{'(+)}$  & $h'_{2L,\ell} m_{\tau}$ & $h'_{2L,q} m_t$  & 0 \\ \hline
$\tilde{S}_{1/2}^{'(-)}$ & 0 & 0 & $h'_{2L,\ell} m_t$ \\ \hline
\end{tabular}
\end{center}
\caption{The $\lambda_i$ couplings of the derivatively-coupled
  (primed) scalar leptoquarks to the different quark-lepton
  combinations, as they appear in the Lagrangian. In our analysis, we have set the
  quark and lepton couplings equal for simplicity.}
\label{tb:lambdasprime}
\end{table}
\begin{table}[htb]
\begin{center}
\begin{tabular}
{|c|c|c|} \hline
Decay mode & Decay width (GeV) & BR   \\ \hline\hline
$\bar{S}_0\rightarrow  \tau^- t$ & 0.1040 & 0.5666 \\ \hline
$\bar{S}_0 \rightarrow \nu_\tau b$ & 0.07956 & 0.4334  \\ \hline \hline
$\bar{\tilde{S}}_0 \rightarrow \tau^- b$ & 0.07956 & 1  \\ \hline \hline
$\bar{S}_1^{(+)} \rightarrow \tau^- b$ & 0.1591 & 1  \\ \hline 
$\bar{S}_1^{(0)} \rightarrow \tau^- t$ & 0.05225 & 0.3964  \\ \hline 
$\bar{S}_1^{(0)} \rightarrow \nu_{\tau} b$ & 0.07956 & 0.6036  \\ \hline 
$\bar{S}_1^{(-)} \rightarrow  \nu_{\tau} t$ & 0.1045 & 1  \\ \hline \hline
$S_{1/2}^{(+)} \rightarrow  \tau^+ t$ & 0.1040 & 1  \\ \hline 
$S_{1/2}^{(-)} \rightarrow  \tau^+ b$ &  0.07956 & 0.6036  \\ \hline 
$S_{1/2}^{(-)} \rightarrow  \bar{\nu}_\tau t$ & 0.05225 & 0.3964  \\ \hline \hline
$\tilde{S}_{1/2}^{(+)} \rightarrow  \tau^+ b$ & 0.07956 & 1  \\ \hline 
$\tilde{S}_{1/2}^{(-)} \rightarrow  \bar{\nu}_\tau b$ & 0.07956 & 1  \\ \hline 
\end{tabular}
\end{center}
\caption{Decay widths for non-derivatively-coupled scalar leptoquarks of mass $M_{LQ} = 400\gev$ and couplings $g = 0.1$.}
\label{tb:widths}
\end{table}
Table~\ref{tb:lambdasprime} shows the couplings for the primed, derivatively-coupled, scalar leptoquarks. The expression for the width given in Eq.~(\ref{eq:width}) remains unchanged in the case of the primed leptoquarks, with the couplings $\lambda_i$ taking the appropriate values.
Tables~\ref{tb:widths} and~\ref{tb:dwidths} show example decay widths and branching ratios for scalar leptoquarks of mass $M_{LQ} = 400 \gev$ and couplings $g = 0.1$. In the case of derivatively coupled leptoquarks we choose a suppression scale $f = 800 \gev$.

\begin{table}[ptb]
\begin{center}
\begin{tabular}
{|c|c|c|} \hline
Decay mode & Decay width (GeV) & BR   \\ \hline\hline
$S'_0 \rightarrow  \tau^- b$ & $4.440\times 10^{-6}$ & 0.0036 \\ \hline 
$S'_0 \rightarrow \nu_\tau t$ & 0.001239 & 0.9964  \\ \hline \hline
$\tilde{S}'_0 \rightarrow \tau^- t$ & 0.001239 & 1  \\ \hline \hline
$S_1'^{(+)} \rightarrow \tau^- t$ & 0.002478 & 1  \\ \hline 
$S_1'^{(0)} \rightarrow \tau^- b$ & $1.292\times 10^{-6}$ & 0.0010  \\ \hline 
$S_1'^{(0)} \rightarrow \nu_{\tau} t$ & 0.001239 & 0.9990  \\ \hline 
$S_1'^{(-)} \rightarrow  \nu_{\tau} b$ & $2.193\times 10^{-6}$ & 1  \\ \hline \hline
$\bar{S}_{1/2}'^{(+)} \rightarrow  \tau^- b$ & $4.440\times10^{-6}$ & 1  \\ \hline 
$\bar{S}_{1/2}'^{(-)} \rightarrow  \tau^- t$ & 0.001239 & 0.9991  \\ \hline 
$\bar{S}_{1/2}'^{(-)} \rightarrow  \nu_\tau b$ & $1.098\times 10^{-6}$ & 0.0009 \\ \hline \hline
$\bar{\tilde{S}}_{1/2}'^{(+)} \rightarrow  \tau^- t$ & 0.001234 & 1  \\ \hline 
$\bar{\tilde{S}}_{1/2}'^{(-)} \rightarrow  \nu_\tau t$ & 0.001239 & 1  \\ \hline 
\end{tabular}
\end{center}
\caption{Decay widths for derivatively-coupled (primed) scalar leptoquarks of mass $M_{LQ} = 400\gev$, couplings $g' = 0.1$ and suppression scale $f = 800 \gev$.}
\label{tb:dwidths}
\end{table}

\subsection{Reconstruction strategies}\label{sec:outlinestrategy}
Table~\ref{tb:strategy} provides an overview of our suggested
reconstruction strategies for the different types of leptoquarks. The
`stransverse' mass variable, $M_{T2}$, which appears in the table has
been defined previously in Ref.~\cite{Lester:1999tx}, for the case of identical semi-invisible pair decays as
\begin{equation}
M_{T2} \equiv
     \min_{ \slashed{\bf c}_T + \slashed{\bf c}'_T = \slashed{\bf p_T} }
     \left\{
     \max { \left( M_T, M_T' \right) }
       \right\} \;,
\end{equation}\label{eq:mt2}where the minimisation is taken over $\slashed{\bf c}_T$ and $\slashed{\bf c}'_T$, the transverse momenta of the invisible particles, with the constraint that their sum equals $\slashed{\bf p_T}$, the total missing transverse momentum, and $M_T$ and $M_T'$ are the transverse masses calculated for the two decay chains. We assume that the invisible particles are massless and use the jet masses in our definitions of $M_{T2}$. The new variables $M_{\mathrm{min}}^{\mathrm{bal}}$ and $M_{\mathrm{min}}$ will be defined in section~\ref{sec:ttaubnurecon}.

We present our analysis of the mass reconstruction techniques for each
pair-production decay mode separately, initially at parton level and then at detector level, including discussion of the relevant backgrounds. We focus on the $S_0$ singlet, $S_1$ triplet and $S_{1/2}$ doublet and outline how to generalise the strategy to all the leptoquark multiplets.

It is evident from Tables~\ref{tb:widths} and~\ref{tb:dwidths} that
the leptoquark decay widths are generally much smaller than the
resolution of the detector components, and hence our analysis is not
sensitive to the decay widths. Throughout the following we have set the
leptoquark couplings to fermions to the value  $g=0.1$. This value is
close to the estimate of the leptoquark coupling to third-generation
quarks and leptons ($\approx 0.09$) derived using the measured fermion
Yukawa couplings and the
assumptions given towards the end of Section~\ref{sec:newphys:leptointro} (see Tables~\ref{tb:lq:yukawas} and~\ref{tb:lq:couplestimates}). The resulting width-to-mass ratio for the leptoquarks corresponding to this coupling, according to Eq.~(\ref{eq:width}), is $\mathcal{O}(10^{-4})$.

We use the \Herwigpp event generator to generate a number of events corresponding to an integrated luminosity of $10$ fb$^{-1}$ of the relevant signal and $t\bar{t}$ background samples. Subsequently we use the \texttt{Delphes} framework~\cite{Ovyn:2009tx} to simulate the detector effects and assess the
feasibility of reconstruction in an experimental
situation.\footnote{\texttt{Delphes} is a framework for fast
  simulation of a general-purpose collider experiment.}
\texttt{Delphes} includes the most crucial experimental features: the
geometry of the central detector, the effect of the magnetic field on
the tracks, reconstruction of photons, leptons, $b$-jets, $\tau$-jets and
missing transverse energy. It contains simplifications such as
idealised geometry, no cracks and no dead material. We use the default
parameter settings in the \texttt{Delphes} package that correspond to
the ATLAS detector. Crucial features of our analysis are
both $b$- and $\tau$-tagging of jets and we caution the reader to take into
consideration that the relevant efficiencies will contain a degree of
uncertainty at the early stages of the LHC experiment. The $b$-tagging present in the \texttt{Delphes} framework assumes an
efficiency of 40\% if the jet has a parent $b$-quark, 10\% if the jet
has a parent $c$-quark and 1\% if the jet is light (i.e. originating
from $u$, $d$, $s$ or $g$). The identification of hadronic $\tau$-jets is consistent
with the one applied in a full detector simulation. The resulting efficiencies for
hadronic $\tau$-jets are in satisfactory agreement with those
assumed by ATLAS and CMS. See~\cite{Ovyn:2009tx} for further details.

Throughout the analysis we
apply transverse momentum cuts of at least $30 \gev$. Since we are always working with high-transverse momentum objects,
we can assume that pile-up arising due to secondary proton-proton
collisions is under experimental control. See, for example, the ATLAS
$t\bar{t} H (\rightarrow b\bar{b})$ study in Ref.~\cite{Aad:2009wy}.
\begin{table}[!htb]
\begin{center}
\begin{tabular}
{|c|c|c|c|c|} \hline
modes & types & technique \\ \hline\hline 
$(t\tau)(b \nu)$ & $S_0$, $S_1^{(0)}$  & $j_{\tau} \parallel \nu_{\tau}$, mass constraints \\
      &       &  $\Rightarrow$ edge reconstruction ($M_{\mathrm{min}}^{\mathrm{bal}}$, $M_{\mathrm{min}}$, $M_{T2}$)       \\ \hline
$(t\tau) (t\tau)$& $S_0$, $S_1^{(0)}$,   & two $j_{\tau} \parallel \nu_{\tau}$, mass constraints  \\   
      &    $ S^{(+)}_{1/2}$, $\tilde{S}'_0$     &  $\Rightarrow$ full reconstruction      \\ \hline
 $(b \nu) (b \nu)$ & $S_0$, $S_1^{(0)}$,    & $M_{T2}$ \\
      &   $\tilde{S}^{(-)}_{1/2}$, $S_1'^{(-)}$    &            \\ \hline
$(b\tau) (b\tau)$ & $S^{(+)}_1$, $\tilde{S}_{1/2}^{(-)}$ & two $j_{\tau} \parallel \nu_{\tau}$, mass constraints   \\
              & $\tilde{S}_0$, $S_{1/2}'^{(+)}$, & $\Rightarrow$ full reconstruction    \\
              & $S_1'^{(0)}$ &  \\ \hline
$(t\nu) (t\nu)$ & $S^{(-)}_1$, $S_{1/2}^{(-)}$ &  \\
              & $S'_0$, $S_{1}'^{(0)}$, &$M_{T2}$    \\
              & $\tilde{S}_{1/2}'^{(-)}$ &  \\ \hline
$(t\nu) (b\tau)$ & $S^{(-)}_{1/2}$, $S_{0}'$ & $j_{\tau} \parallel \nu_{\tau}$, mass constraints   \\
              & $S_1'^{(0)}$ & $\Rightarrow$ edge reconstruction ($M_{\mathrm{min}}^{\mathrm{bal}}$, $M_{\mathrm{min}}$, $M_{T2}$)         \\ \hline
\end{tabular}
\end{center}
\caption{The table outlines the general reconstruction strategy for leptoquark
  pair-production for the different types of leptoquarks. For variable definitions and further details see the respective sections.}
\label{tb:strategy}
\end{table}
\subsection[$(t\tau)(t\tau)$ decay mode]{\boldmath $(t\tau)(t\tau)$ decay mode}\label{sec:tttt}
\begin{figure}[!t]
  \centering 
    \includegraphics[scale=0.60]{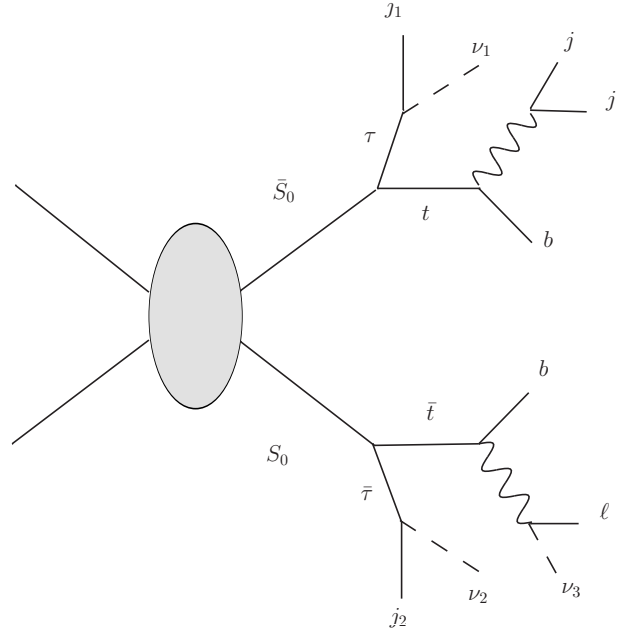}
\caption{Pair-production of $S_0$ leptoquarks with decay to $(t\tau)(t\tau)$, followed by one hadronic and one semi-leptonic top decay.} 
\label{fig:s0topology}
\end{figure}

\begin{figure}[!t]
  \centering 
  \vspace{1.8cm}  
  \hspace{4.0cm}
    \includegraphics[scale=0.60, angle=90]{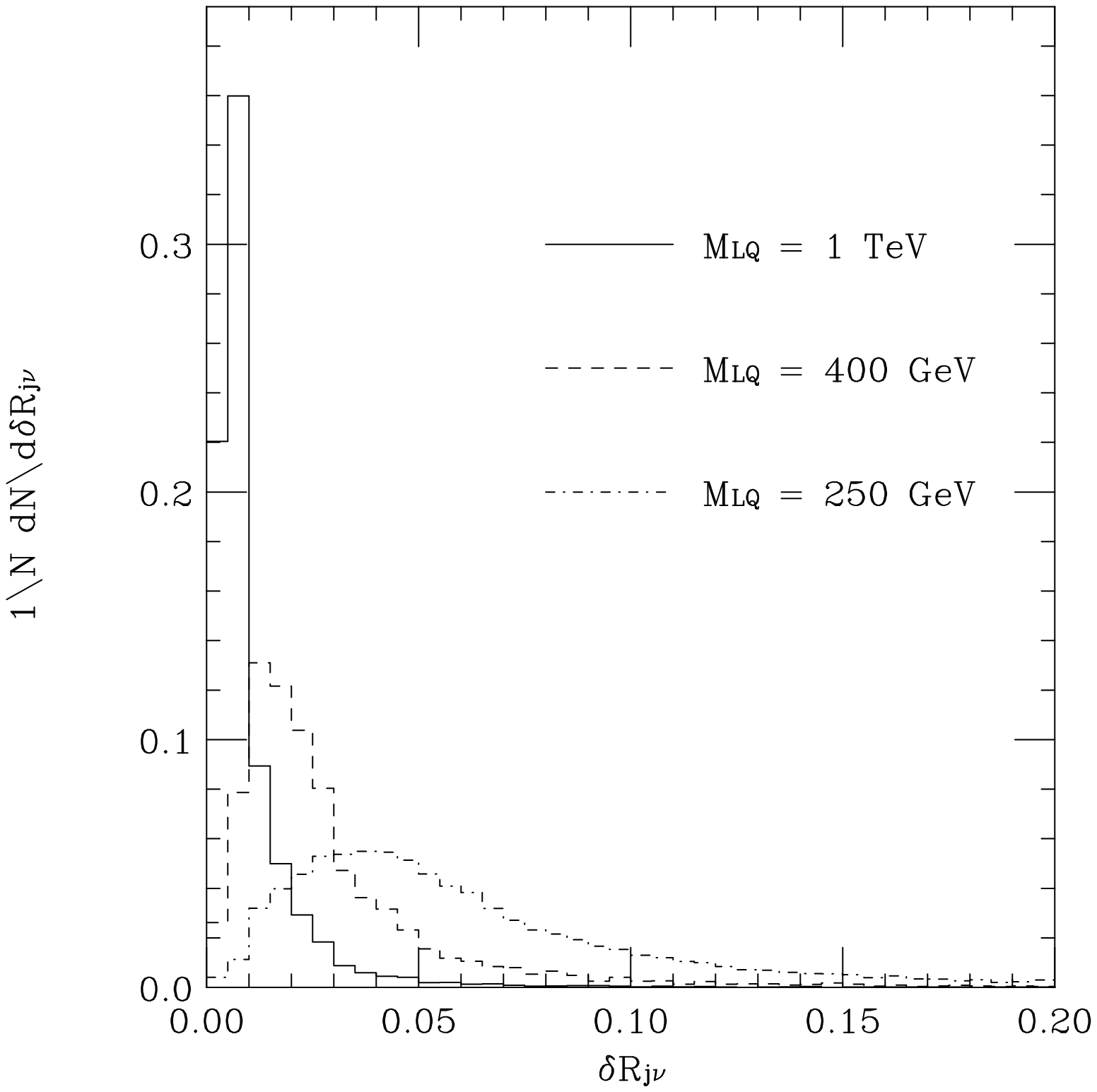}
  \vspace{0.5cm}
\caption{The distribution of the distance in $R$-space ($\delta R =
  \sqrt { \delta \eta ^2 + \delta \phi ^2 }$) between the momenta of
  the $\tau$ jet and the $\tau$ invisibles in $S_0$ pair-production
  for $M_{LQ} = 1, 0.4, 0.25$ TeV.}
\label{fig:tauangle}
\end{figure}

We examine the possibility of full reconstruction of the
topology shown in Fig.~\ref{fig:s0topology}, where we have, for example, $S_0 (\bar{S}_0)
\rightarrow bjj j_1 \nu_1 $ and $S_0 (\bar{S}_0) \rightarrow b \ell \nu_3 j_2
\nu_2$, where $\nu_1$ and $\nu_2$ represent one or more neutrinos coming
from the $\tau$ decays and $\ell$ can be either a muon or an
electron. We can assume that the neutrinos $\nu_{1,2}$ associated with the decays
of the $\tau$s are collinear with the direction of the jets $j_{1,2}$ associated
with them. The validity of this assumption has been confirmed using \Herwigpp, for leptoquarks of masses 1, 0.4 and 0.25 TeV, as may be seen in Fig.~\ref{fig:tauangle}, which shows the distribution of $\delta R =
\sqrt { \delta \eta ^2 + \delta \phi ^2 }$ between the momenta of the
$\tau$ jet partons and the $\tau$ invisibles. The assumption is employed in our reconstruction of any leptoquark decay mode containing a $\tau$-jet.
The top quark branching ratios are $\sim 0.216$ for the
semi-leptonic $e, \mu$ modes and $\sim 0.676$ for the hadronic top
modes.  These appear twice since we have either the $t$ or
$\bar{t}$, resulting in an overall $\sim 0.292$ factor for the top decay modes. The branching ratios and cross sections for
$S_0 \bar{S}_0$ production depend on the leptoquark mass and coupling and are shown in
Table~\ref{tb:s0br}, where the last column is the resulting cross
section for the topology under study. We focus on $400\gev$
leptoquarks since these are clearly not excluded by direct searches at
present and still provide a sufficient number of events to be potentially discovered at a reasonable luminosity (10 fb$^{-1}$) at $14$ TeV.
\begin{table}[!t]
\begin{center}
\begin{tabular}
{|c|c|c|c|} \hline
$M_{S_0}$ (GeV) & $\sigma(pp \rightarrow S_0\bar{S}_0)$ (pb) & BR($t\tau$) & $\sigma(t\tau \bar{t}\bar{\tau} \rightarrow b \bar{b} jj \ell(=e,\mu) \nu \tau \bar{\tau})$ (pb)  \\ \hline\hline 
174.2 (= $m_{top}$) & 141(1)   &   0.  & 0. \\ \hline
250 & 24.3(3) & 0.34 & 0.729 \\ \hline
400 & 2.000(7) & 0.567 & 0.188  \\ \hline
500 & 0.561(6) &  0.606 & 0.06  \\ \hline 
1000 &  5.94(7) $\times 10^{-3}$& 0.65 & $7.3\times 10^{-4}$  \\ \hline  \hline
$M_{top}$ (GeV) & $\sigma(pp \rightarrow t\bar{t})$ (pb) & - & $\sigma(t\bar{t} \rightarrow b \bar{b} jj \ell(=e,\mu) \nu)$ (pb) \\ \hline 
174.2 & 834(1) & - & 242 \\\hline 
\end{tabular}
\end{center}
\caption{$S_0\bar{S}_0$ total cross section at the LHC at 14 TeV $pp$ centre-of-mass energy, branching ratio to
  $t\tau$ and remaining cross section taking into account the top branching ratios. The corresponding $t\bar{t}$ values are given for comparison.}
\label{tb:s0br}
\end{table}

\subsubsection{Kinematic reconstruction}\label{sec:ttauttaurecon}

The final states of $S_0 \bar S_0 \to \bar t \tau^+ t \tau^-$ processes contain 
many decay products including neutrinos.
If the system has a large enough number of kinematical constraints, 
such as mass-shell conditions and the balance of 
the total transverse momentum, we can completely reconstruct the kinematics of the system.
The numbers of unknown variables and constraints are summarised in Table~\ref{tab:ttauttau} 
for each decay pattern of the tops:
(1) both tops decay hadronically, (2) one top decays semi-leptonically and another hadronically and (3)
both tops decay semi-leptonically.
\begin{table}[!t]
\begin{center}
\begin{tabular}{|l|c|c|}
\hline
Decay type&\# of unknowns&\# of constraints\\
\hline
(1) had,had&$1+(0+2)N$&$(2+2)N$\\
\hline
(2) had,lep &$1+ (4+2)N$&$(5+2)N$\\
\hline
(3) lep,lep&$1+(8+2)N$&$(8+2)N$\\
\hline
\end{tabular}
\caption{The numbers of unknown variables 
([$m_{LQ}$], [$\nu$ from top], [energy fraction of tau])
and constraints ([mass-shell conditions], [balance of missing momentum]) 
in $N$ events of each decay type.
The mass-shell conditions that constrain the unknown variables
are counted here, i.e.~the mass-shell conditions on $S_0$, leptonic top, $W$ and 
$\nu$ from leptonic top decay. 
}
\label{tab:ttauttau}
\end{center}
\end{table}%
As mentioned above, we assume 
$\tau$-neutrinos are collinear to the $\tau$-jets,
leaving two unknown parameters associated with the taus, namely the
energy ratios $z_i$ ($i=1,2$, $z_i \ge 1$) which are defined (neglecting masses) by
\begin{eqnarray}\label{eq:zis}
p_{\tau_i} &=& z_i p_{j_i}\;,
\nonumber \\
p_{\nu_i} &=& p_{\tau_i} - p_{j_i} = (z_i - 1) p_{j_i} \;,
\end{eqnarray}
where $p_{\tau_i}$, $p_{j_i}$ and $p_{\nu_i}$ are the four-momenta of the $\tau$ leptons, $\tau$-jets and $\tau$-neutrinos,
respectively. 
Under this assumption, the unknown variables 
in Table~\ref{tab:ttauttau} are the mass of the leptoquark, the 4-momenta of neutrinos
from leptonic top decays and the energy fractions associated with the neutrinos from the tau decays.
The mass-shell conditions that could constrain the unknown variables
are counted in Table~\ref{tab:ttauttau},
i.e.~the mass-shell conditions on $S_0$, leptonic top, $W$ and 
$\nu$ from leptonic top decay.

It is only possible to wholly reconstruct the kinematics of a single event in decay types (1) and (2). 
In decays of type (1), it would be difficult to reconstruct both hadronic tops 
because of the large combinatorial background. 
Thus, we focus on decay type (2) and attempt to determine the leptoquark mass.
As we show in appendix~\ref{app:ttauttau}, in this case one obtains a quartic equation for the energy ratio $z_2$, and hence in general up to four solutions for the leptoquark mass, at least one of which should be close to the true value if the visible momenta and missing transverse momenta are well-measured.

\subsubsection{Parton-level reconstruction}
\begin{figure}[!htb]
  \centering 
  \vspace{1.5cm}
  \hspace{3.0cm}
 \includegraphics[scale=0.35, angle=90]{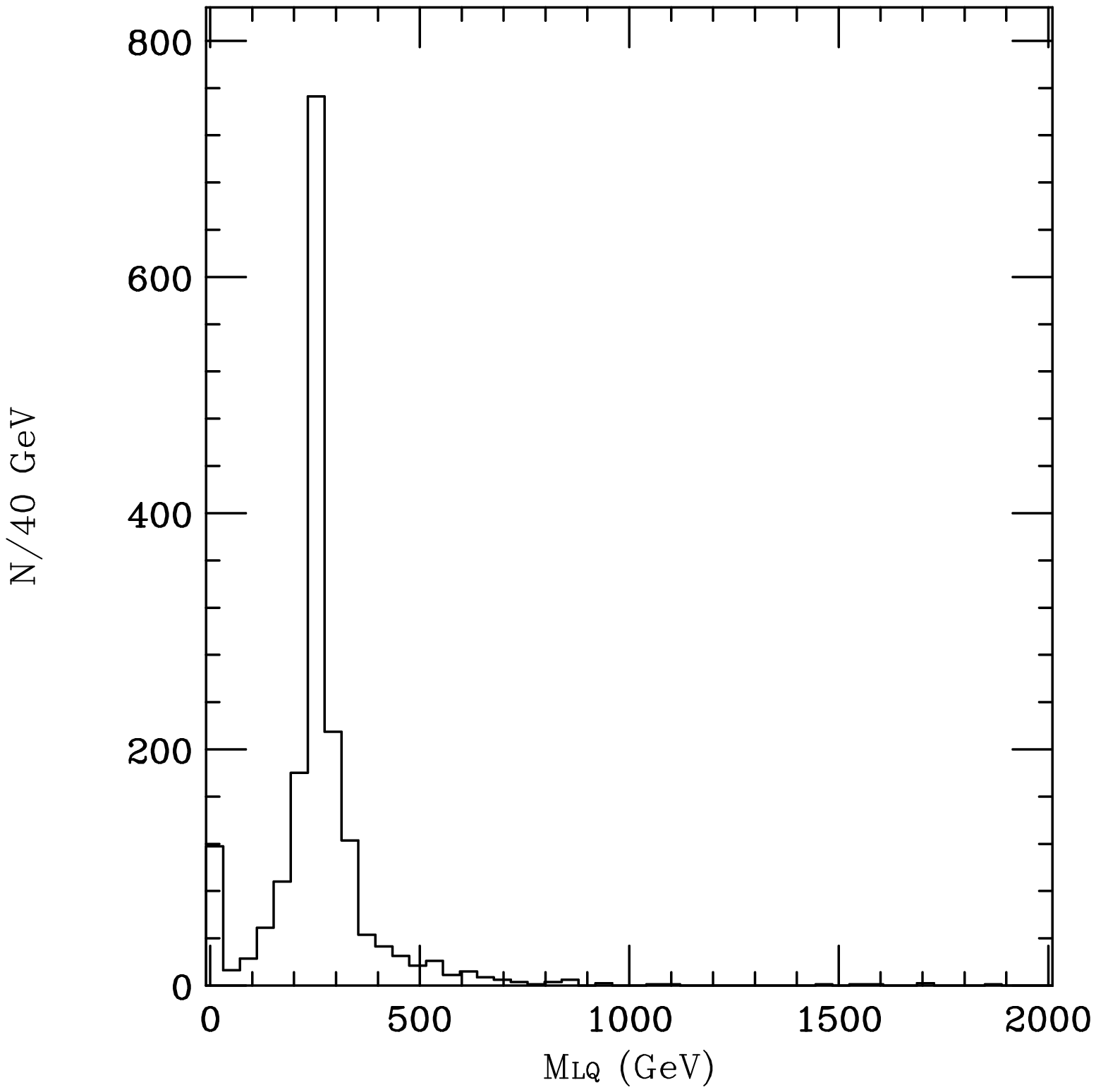}
  \hspace{3.5cm}
  \includegraphics[scale=0.35, angle=90]{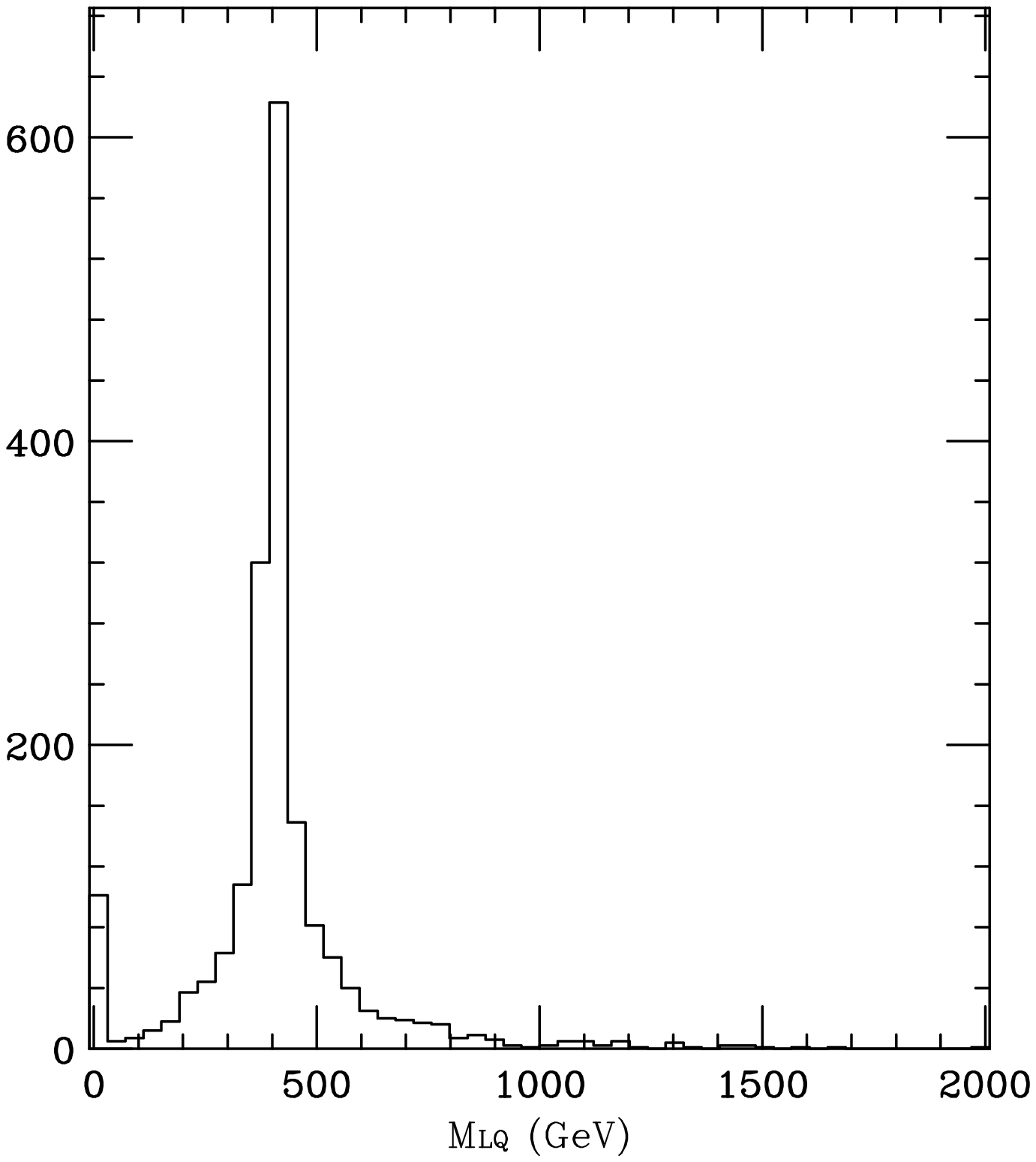}
  \hspace{3.5cm}
 \includegraphics[scale=0.35, angle=90]{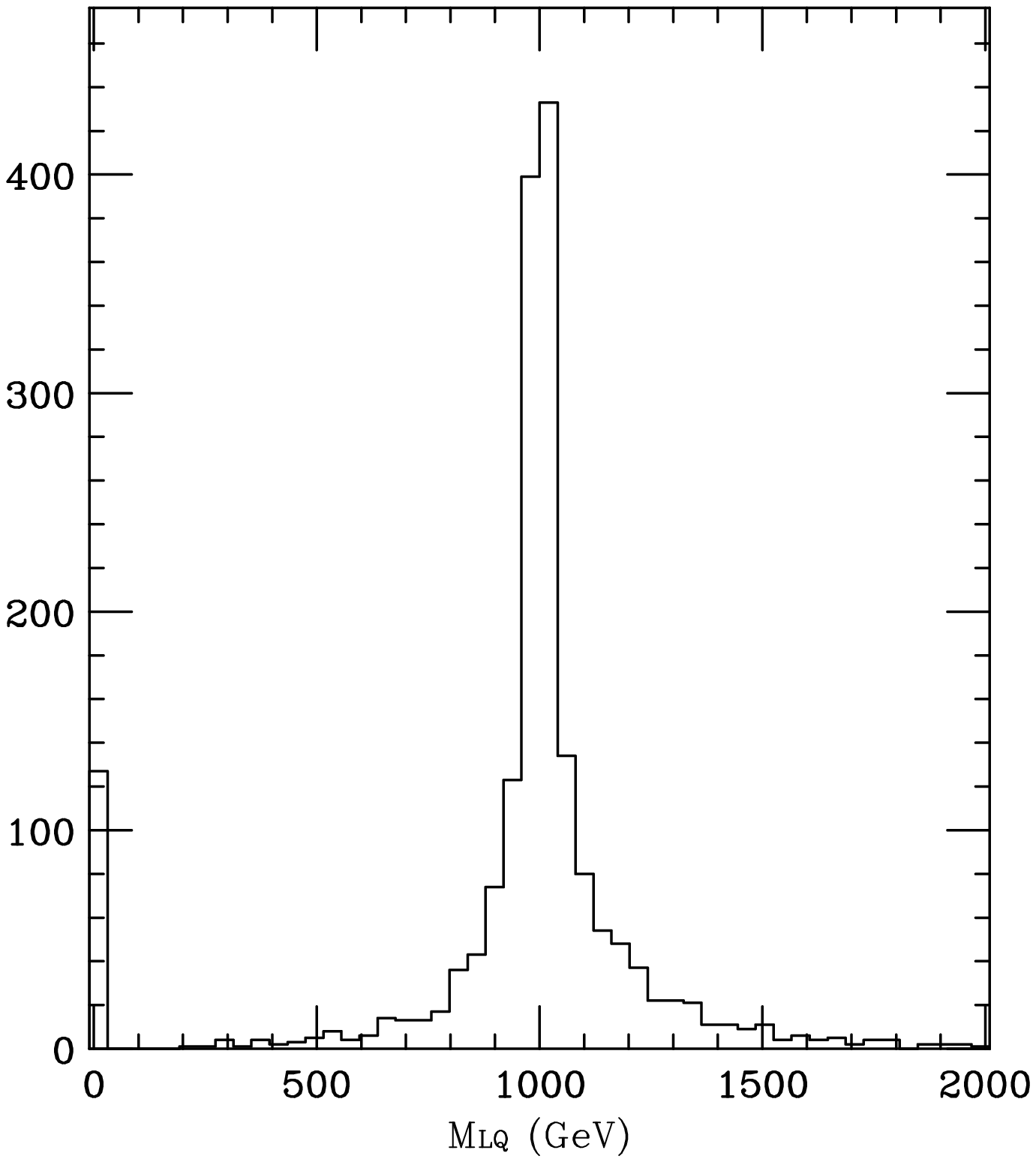}
  \vspace{0.5cm}
\caption{Histograms of the solutions obtained at parton level for
  $S_0\bar{S}_0$ and decay to a semi-leptonic top, a hadronic top and
two tau leptons, for $M_{S_0} =   (0.25, 0.4, 1.0)$ TeV (from left to
right respectively). The first bin (0 GeV) contains the events for which no real
solution has been found.}
\label{fig:mS0parton}
\end{figure}

We first perform the $(t\tau)(t\tau)$ analysis of the hard process (no
initial- or final-state radiation, no underlying event) at parton level without considering experimental or combinatoric effects, to examine its feasibility. For the
 majority of cases there are only two physical, approximately degenerate,
 solutions, which are close to the true leptoquark mass. The numerical
 solution of the quartic equation sometimes fails to yield real roots. The
 results for true leptoquark masses $M_{S_0} = (0.25, 0.4, 1.0)$ TeV
 are shown in Fig.~\ref{fig:mS0parton}, which includes histograms of the solutions
 obtained for $10^3$ events. The histogram includes a bin at
 0 where the events without real solution are placed. These amount to about 10\% of the total events. At this level the reconstruction technique provides a good estimate of the leptoquark mass for all the trial true masses, lying within a few GeV of the true mass. 
\subsubsection{Experimental reconstruction}
We consider an $S_0$ leptoquark with mass $M_{S_0} = 400\gev$, for
which the cross section for production and decay into the topology of
Fig.~\ref{fig:s0topology}, $S_0\bar{S}_0 \rightarrow t\tau
\bar{t}\bar{\tau} \rightarrow b \bar{b} jj \ell\nu \tau
\bar{\tau}$, is $\sigma = 0.187$ pb. The most significant background
in this scenario is $t \bar{t}$ production, with two extra jets misidentified as $\tau$s and subsequent decay of
the tops into $b \bar{b} jj \ell\nu$. The cross section for
this process is $242.4$ pb, overwhelming to begin with. There is also
potentially an irreducible $Ht\bar{t} \rightarrow \tau \bar{\tau} t
\bar{t}$ background which, for a Higgs boson of mass $M_H = 115 \gev$, has a
cross section of approximately 65 fb.  Since one of the main rejection
mechanisms is the reconstruction through the
solution of the kinematic equations, we do not expect this background to
contribute significantly in the signal region.

\begin{figure}[!htb]
  \centering
  \vspace{0.4cm}
  \includegraphics[scale=0.40, angle=270]{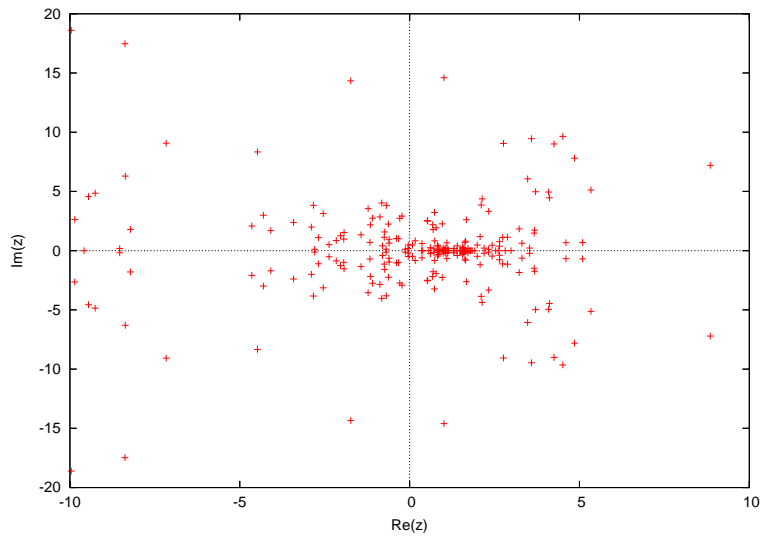}
  \vspace{0.75cm}
  \caption{The plot shows the complex values of the solutions for $z_2$ after solving the quartic equation for the events that have passed the experimental cuts. There exists a higher concentration of events about the positive real axis. The number of entries is 280 (4 solutions included for each of the 70 events).}
  \label{fig:zcomplex}
\end{figure}

\begin{figure}[!htb]
  \centering
  \vspace{1.4cm}
  \hspace{0.5cm}
  \includegraphics[scale=0.45, angle=90]{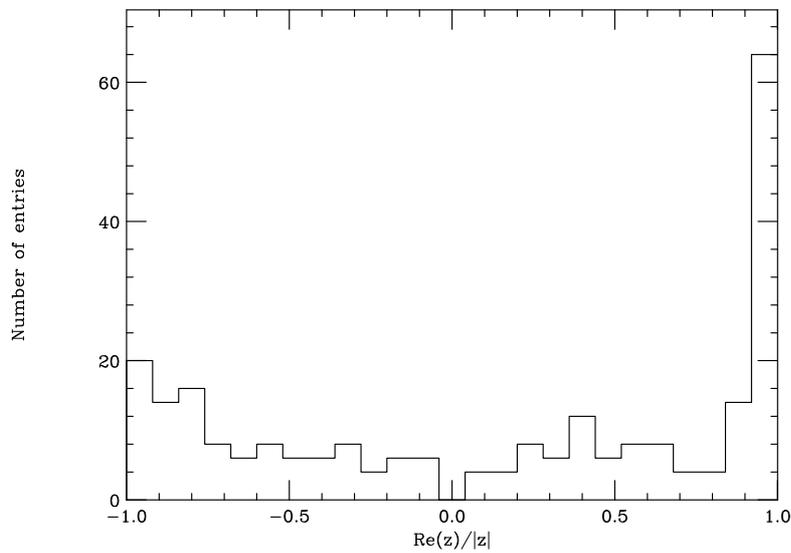}
  \vspace{0.5cm}
  \caption{The plot shows the ratio of the real part of $z_2$ and its modulus. The peak close to 1 demonstrates the clustering of the real positive solutions about the real axis and justifies the use of the real part as an input to the rest of the calculation. The number of entries is 280 (4 solutions included for each of the 70 events).}
  \label{fig:zcomplex2}
\end{figure}

\begin{figure}[!htb]
  \centering 
  \vspace{1.5cm}
  \hspace{6.0cm}
  \includegraphics[scale=0.55, angle=90]{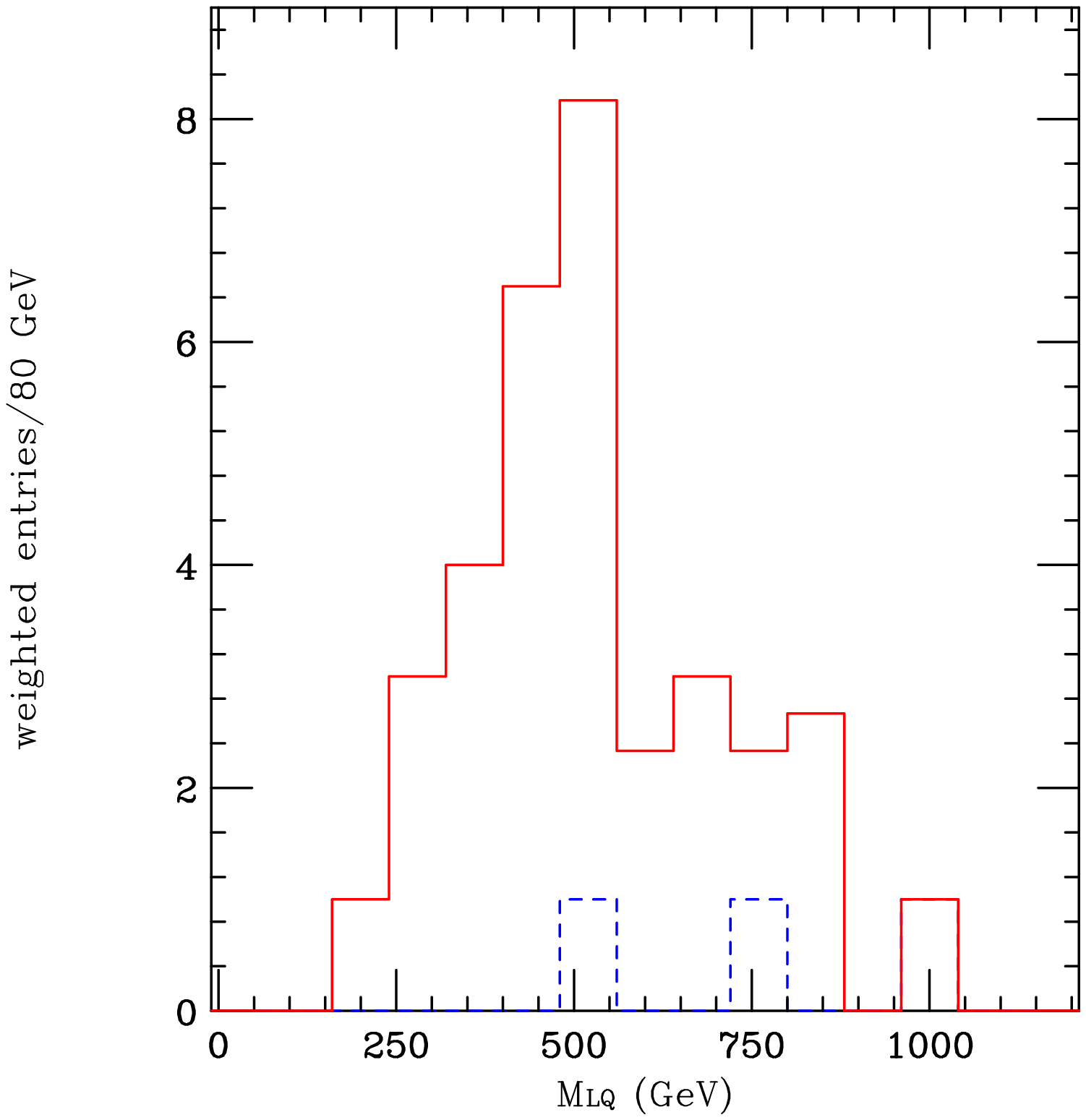}
  \vspace{0.5cm}
  \caption{Experimental reconstruction of the $S_0\bar{S}_0
    \rightarrow t\tau \bar{t}\bar{\tau} \rightarrow b \bar{b} jj
    \ell(=e,\mu) \nu \tau \bar{\tau}$ mode using the method described
    in the text.
  Note that each event has weight 1, distributed
    evenly amongst the solutions it yields. The signal is shown in
    red (35 entries) and the $t\bar{t}$ background in blue dashes (3 entries).}
  \label{fig:mS0ttauttau_btau22}
\end{figure}

We simulate the events with QCD initial-state radiation
(ISR), final-state radiation (FSR) and underlying event (UE). We use the default jet algorithm provided by the \texttt{Delphes}
package for the ATLAS configuration, the anti-$k_t$ with the parameter
set to $R=0.7$. We then
demand a set of relatively loose cuts on the full $t\bar{t}$ and
$S_0\bar{S_0}$ samples, since in a real experiment we would not be able
to separate the different decay modes of the top quark or $S_0$ leptoquark. The cuts applied are the following:
\begin{itemize}
\item{The existence of a lepton in the event, being either a muon or
    electron, with $p_{T,\ell} > 30\gev$.}
\item{A minimum of 6 jets.}
\item{The missing transverse momentum in the event, $\slashed{E_T} > 20
    \gev$.}
\item{Two $\tau$-tagged jets, with the extra requirement that they
    both have $p_{T,\tau} > 30\gev$.}
\item{No jets tagged as both $b$- and $\tau$-jets simultaneously.}
\end{itemize}
We also require that the highest-$p_T$ lepton is at a distance $\delta
R > 0.1$ from the $\tau$-tagged jets, since electrons may create a
candidate in the jet collection as well as the lepton collection. The
analysis then breaks up into different branches according to the number of $b$-tagged jets in an event: 
\begin{itemize}
\item{\textit{Two $b$-tagged jets:} we look for one or two further jets (with $p_T > 30\gev$) that form an invariant mass close to the top mass, within $20 \gev$. One $b$-jet is then associated with the semi-leptonic top decay and the other with the hadronic top decay.}
\item{\textit{One or no $b$-tagged jets:} when there is one $b$-tagged jet, we check whether it will satisfy the top mass conditions with any other (one or two) remaining jets, otherwise we associate it with the semi-leptonic top. If so, we look for any two or three jets that satisfy the top mass conditions, and form the hadronic top within a $20\gev$ mass window. For the remaining $b$-jets (or if there are no $b$-jets) we look for the remaining highest-$p_T$ jets. Any jets that are found in this way and called $b$-jets are required to have a $p_{T,b} > 30\gev$.} 
\end{itemize}
No solutions are found in the sample of 70 signal events passing the cuts, if we require the ratios $z_2$ to be purely real. Hence, the solutions to the quartic equation for the momentum ratios $z_2$,
described in appendix~\ref{app:ttauttau}, are now allowed to be
complex in order to provide some signal. This is valid since even true
leptoquark events are smeared and distorted by detector and
QCD effects. We use the real part of $z_2$ as an input to the
calculation of the rest of the kinematic variables. This is reasonable since the experimental effects are expected to `smear' the position of the true value of $z_2$ in all directions in the complex plane. The effect is shown in Fig.~\ref{fig:zcomplex}, where we plot the real and imaginary parts of $z_2$ for the events that have passed the kinematic cuts. Evidently, there is a concentration of solutions around the positive real axis, an effect exemplified by Fig.~\ref{fig:zcomplex2}, where we show the ratio of the real part of $z_2$ and its modulus. We have further demanded that the resulting momentum fractions are physical: $\mathcal{R}(z_{1,2}) > 1$, resulting in only real solutions for $M_{S_0}$.
Figure~\ref{fig:mS0ttauttau_btau22} shows a reconstruction plot for
leptoquarks of mass $M_{S_0} = 400\gev$. Note that each event was given weight 1, distributed evenly amongst the solutions it yields. In the case of complex $z_2$, we assume there is one solution corresponding to the complex conjugate pair.

Although the cuts applied are relatively weak, most of the background rejection
comes from the requirement of two $\tau$-tagged jets. The background does not produce solutions in the
physical region often enough to be significant.
\subsection[$(q\nu)(q\nu)$ decay modes]{\boldmath $(q\nu)(q\nu)$ decay modes}
We can obtain the mass of the leptoquarks when both of them decay into
$b\nu$ or $t\nu$ using the $M_{T2}$ variable (Eq.~(\ref{eq:mt2})). Examples of these
decay mode are  $S_0\bar{S}_0 \rightarrow \bar{b}\bar{\nu} b\nu$ and
$\bar{S}^{(-)}_{1/2}S^{(-)}_{1/2} \rightarrow \bar{t}\nu t\bar{\nu}$.  

\subsubsection{Parton-level reconstruction}
\begin{figure}[!htb]
  \centering 
  \vspace{1.5cm}  
  \hspace{4.5cm}
    \includegraphics[scale=0.60, angle=90]{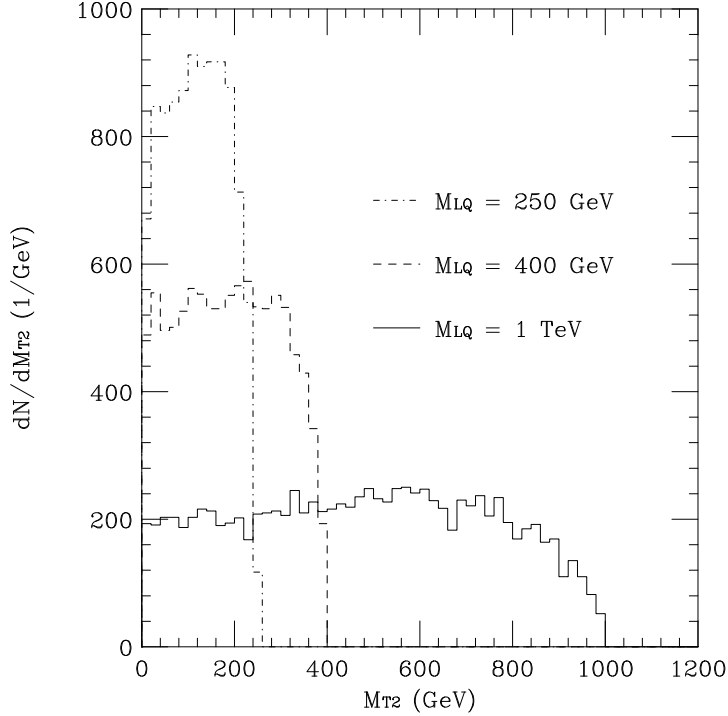}
  \vspace{0.5cm}  
\caption{The parton-level $M_{T2}$ distribution constructed for the
  $S_0\bar{S}_0 \rightarrow \bar{b}\bar{\nu} b\nu$ using the true
  $b$-quark momenta, for $M_{LQ} = 1, 0.4, 0.25$ TeV.}
  \vspace{0.5cm}
\label{fig:bnubnumt2_parton}
\end{figure}

At parton level, the $(t\nu)(t\nu)$ and $(b\nu)(b\nu)$ decay modes are
similar and hence we consider only the latter here. We first construct
the $M_{T2}$ variable using the parton-level $b$-quark 4-momenta, in
the absence of any experimental effects, ISR, FSR or UE. The result is shown in Fig.~\ref{fig:bnubnumt2_parton} for $M_{LQ} = (0.25, 0.4, 1)$ TeV, confirming the expected sharp edge in these idealised conditions.
\subsubsection{Experimental reconstruction}
\begin{figure}[!htb]
  \centering 
  \vspace{1.3cm}
  \hspace{3.0cm}
  \includegraphics[scale=0.40, angle=90]{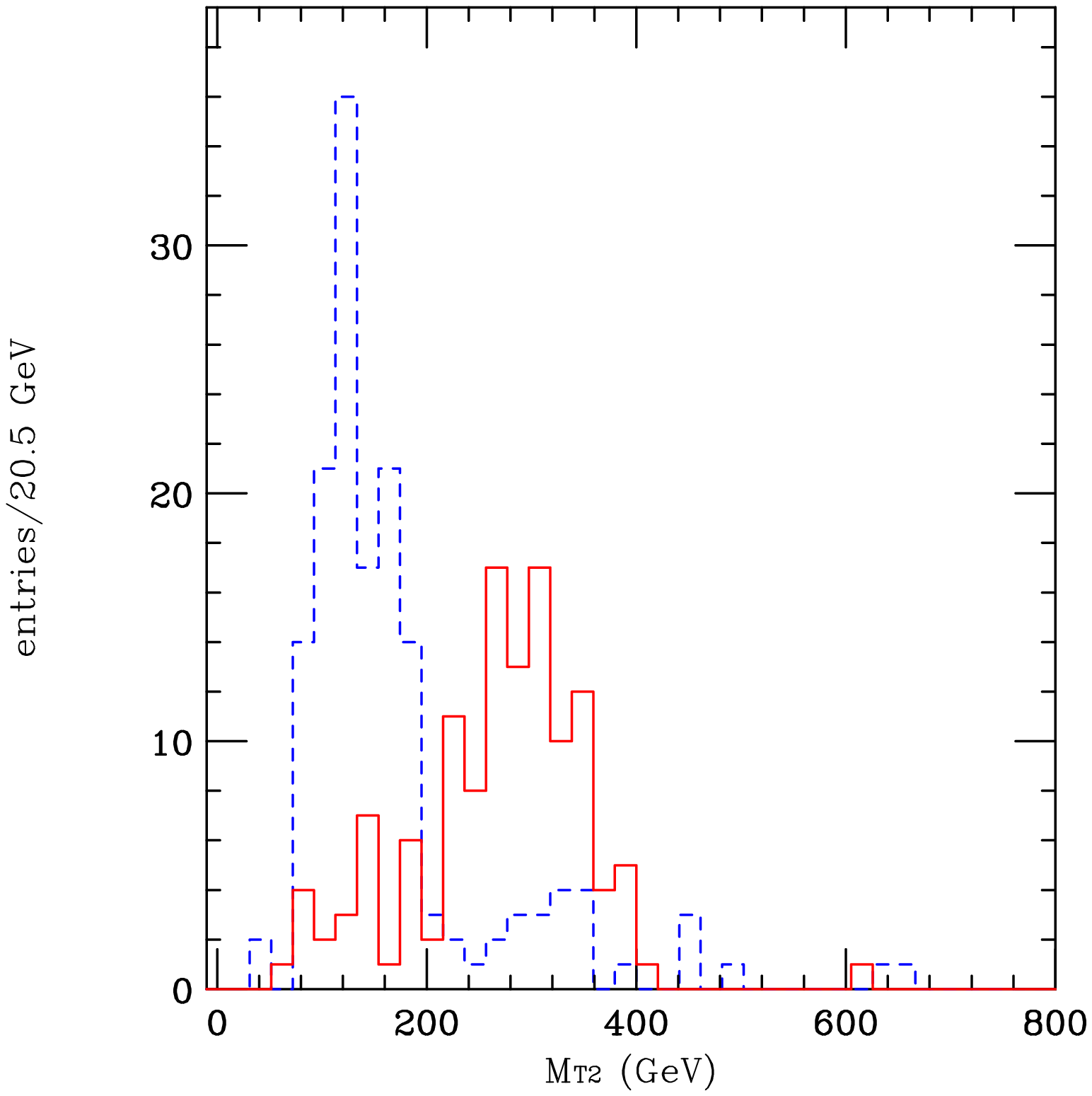}
  \hspace{5.0cm}
  \includegraphics[scale=0.40, angle=90]{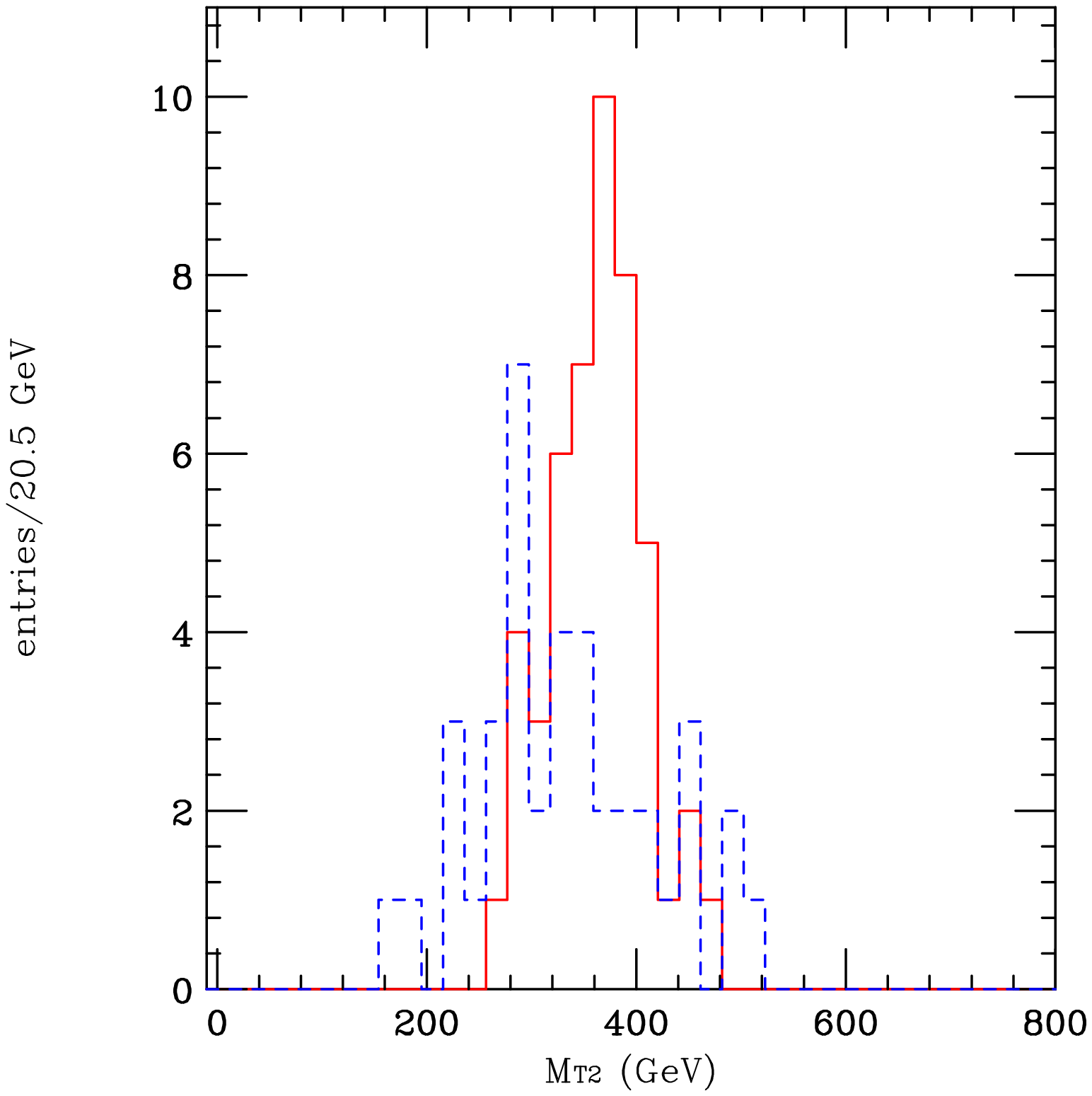}
  \vspace{0.5cm}
  \caption{Experimental reconstruction of the $S_0\bar{S}_0
    \rightarrow b\nu_\tau \bar{b} \bar{\nu_\tau}$ mode (left, 121 background
    events, 125 signal events)
    and $\bar{S}^-_{1/2}S^-_{1/2} \rightarrow \bar{t}\nu t\bar{\nu} $
    mode (right, 39
    background events, 48 signal events) using $M_{T2}$. ISR, FSR and the underlying event have been
    included in the simulation. The signal is given in
    red and the $t\bar{t}$ background in blue dashes.}
  \label{fig:mt2bnu}
\end{figure}

As before, we use the \texttt{Delphes} framework to simulate the
detector effects, with the settings stated in section~\ref{sec:outlinestrategy}. We demand two $b$-tagged jets in
both the $q = b$ and $q = t$ cases. In the latter we search for
combinations of 1 or 2 jets with the $b$-tagged jets which form the
top mass within a window of $10 \gev$.  We require the following
cuts for the $(b\nu)(b\nu)$ case, on the full $S_0 \bar{S}_0$ sample:
\begin{itemize}
\item{Two $b$-tagged jets with $p_{T,b} > 120 \gev$ each.}
\item{No electrons or muons in the event.}
\item{Missing transverse energy $\slashed{E_T} > 250 \gev$.}
\end{itemize}
For the $(t\nu)(t\nu)$ case we require the following cuts on the $\bar{S}^{(-)}_{1/2}S^{(-)}_{1/2}$ sample:
\begin{itemize}
\item{Two $b$-tagged jets with $p_{T,b} > 80 \gev$ each.}
\item{No electrons or muons in the event.}
\item{Missing transverse energy $\slashed{E_T} > 260 \gev$.}
\end{itemize}
The resulting $M_{T2}$ distributions for the signal (blue) and $t\bar{t}$
background (red) can be seen in Fig.~\ref{fig:mt2bnu}.
The $(t\nu)(t\nu)$ mode appears to be more challenging to reconstruct than the
$(b\nu)(b\nu)$ mode. This is due to the fact that the $t\bar{t}$ background is very similar to the
signal and the difficulties that are present in reconstructing hadronic tops. Nevertheless, as the results show, it may
be possible to observe an excess over the $M_{T2}$ distribution of the
background and provide an estimate of the mass.
\subsection[$(q'\tau)(q \nu)$ decay modes]{\boldmath $(q'\tau)(q \nu)$ decay modes}
\begin{figure}[!t]
  \centering 
    \includegraphics[scale=0.60]{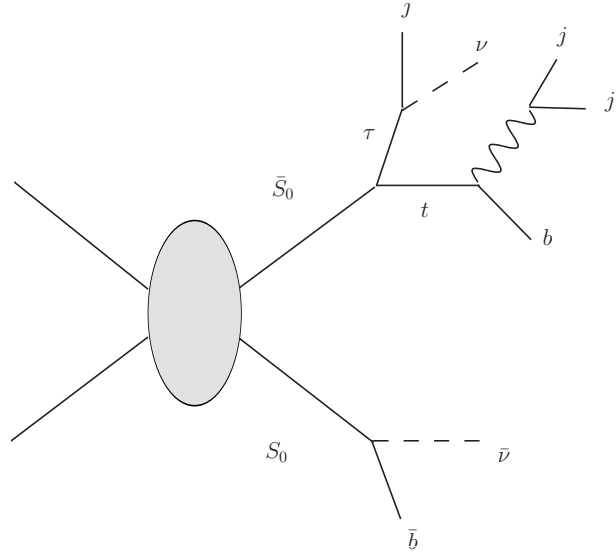}
 \caption{Pair production of $S_0$ leptoquarks with decay to $(t\tau)(b\nu)$, followed by hadronic top decay.}
\label{fig:topo-bntt}
\end{figure}
One possible event topology for the $S_0 \overline S_0 \to \bar b \bar \nu t \tau^-$ processes
is shown in Fig.~\ref{fig:topo-bntt}.
When the top decays hadronically the system has two neutrinos, 
one from an $S_0$ decay and another from a $\tau$ decay. 
If we can reconstruct the hadronic top correctly,
we can simply use $M_{T2}$ to obtain the mass of the leptoquarks. Similar topology is present in the $\bar{S}^{(-)}_{1/2} S^{(-)}_{1/2} \rightarrow (b\bar{\tau}) (\bar{t}\nu)$ decay mode.

It is known that the information from $M_{T2}$ is the same as that from
the `minimal kinetic constraints', in events where 
two identical particles decay to missing particles with the same mass~\cite{Serna:2008zk,Cheng:2008hk,Barr:2009jv}.
As discussed in section~\ref{sec:tttt}, in this type of event,
we can take advantage of the fact that, to a good approximation,
 the neutrino from a $\tau$ decay is travelling almost collinearly to the $\tau$-jet in the lab frame.
By including this constraint, we can define kinematical variables, 
$M_{\rm min}$ and $M_{\rm min}^{\rm bal}$, which perform better than
$M_{T2}$ at parton level,
as we will show in the following subsections.

\subsubsection{Kinematic reconstruction}\label{sec:ttaubnurecon}

In the $\tau$ collinearity approximation, neglecting masses, we can write 
\beq\label{eq:wpj}
p_{\nu_\tau} = w p_j\;,\;\;w>0\,.
\eeq
The second neutrino comes directly from the $S_0$ decay associated with a $b$-jet.
The transverse components of the momentum of this neutrino are constrained by
\beq
{\bf p}_{\nu} = {\bf p}_{\rm miss} - w {\bf p}_j\,.
\eeq  
There are two unknown parameters left, $w$ and $p_{\nu}^z$.  In terms of these,
we define two invariant mass variables:
\beq
m_{t \tau}^2(w) = (p_t + (1+w)p_j)^2 = m_t^2 + 2 (1+w) p_t\cdot p_j
\eeq
and
\beqn\label{eq:mbnu}
m_{b \nu}^2(w,p_{\nu}^z) &=& (p_b + p_{\nu})^2 \nonumber \\
&=&\, 2 E_b \sqrt{(\pmiss - w {\bf p}_j )^2 + (p_{\nu}^z)^2} - 2 {\bf p}_b \cdot (\pmiss -w {\bf p}_j)
- 2 p_b^z p_{\nu}^z\;. \nonumber\\
\eeqn
Note that $m_{t \tau}$ does not depend on $p_{\nu}^z$ and is
 a monotonically increasing function of $w$ because $p_t\cdot p_j > 0$. 
We can now define two $M_{T2}$-like variables: 
\beq
M_{\rm min} = \min [ \max \{ m_{t\tau}, m_{b\nu}  \} ] \geq M_{T2}\;,
\eeq
and
\beq
M_{\rm min}^{\rm bal} = \min_{m_{t\tau} = m_{b\nu} } [m_{b\nu}]\;,
\eeq
where minimisation is taken for all possible ($w$, $p_{\nu}^z$).
By construction, both these quantities have an upper bound equal to the leptoquark mass:
\beq
M_{S_0} \ge M_{\rm min}, ~~~~M_{S_0} \ge M_{\rm min}^{\rm bal}\;.  
\eeq
Furthermore, we show in appendix~\ref{app:mminbal} that
\beq
M_{\rm min}^{\rm bal} \ge M_{\rm min}\;.
\label{eq:mb.gt.m}
\eeq
\subsubsection{Parton-level reconstruction}\label{sec:partonqtauqnu}
\begin{figure}[!t]
 \begin{center}
   \vspace{1.0cm}
  \hspace{3.0cm}
  \includegraphics[width=4.5cm, angle = 90]{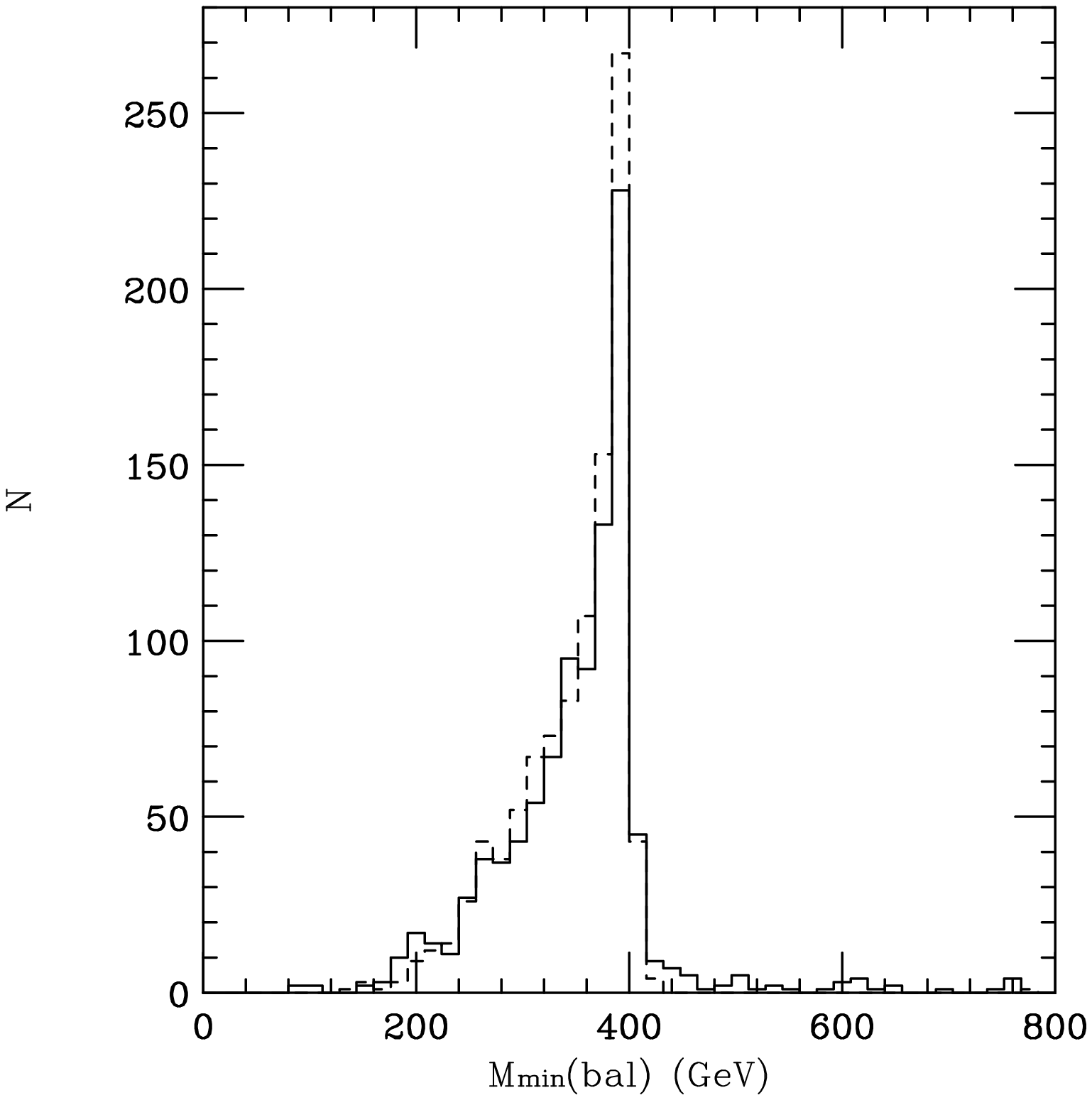}
  \hspace{3.5cm}
  \includegraphics[width=4.5cm, angle = 90]{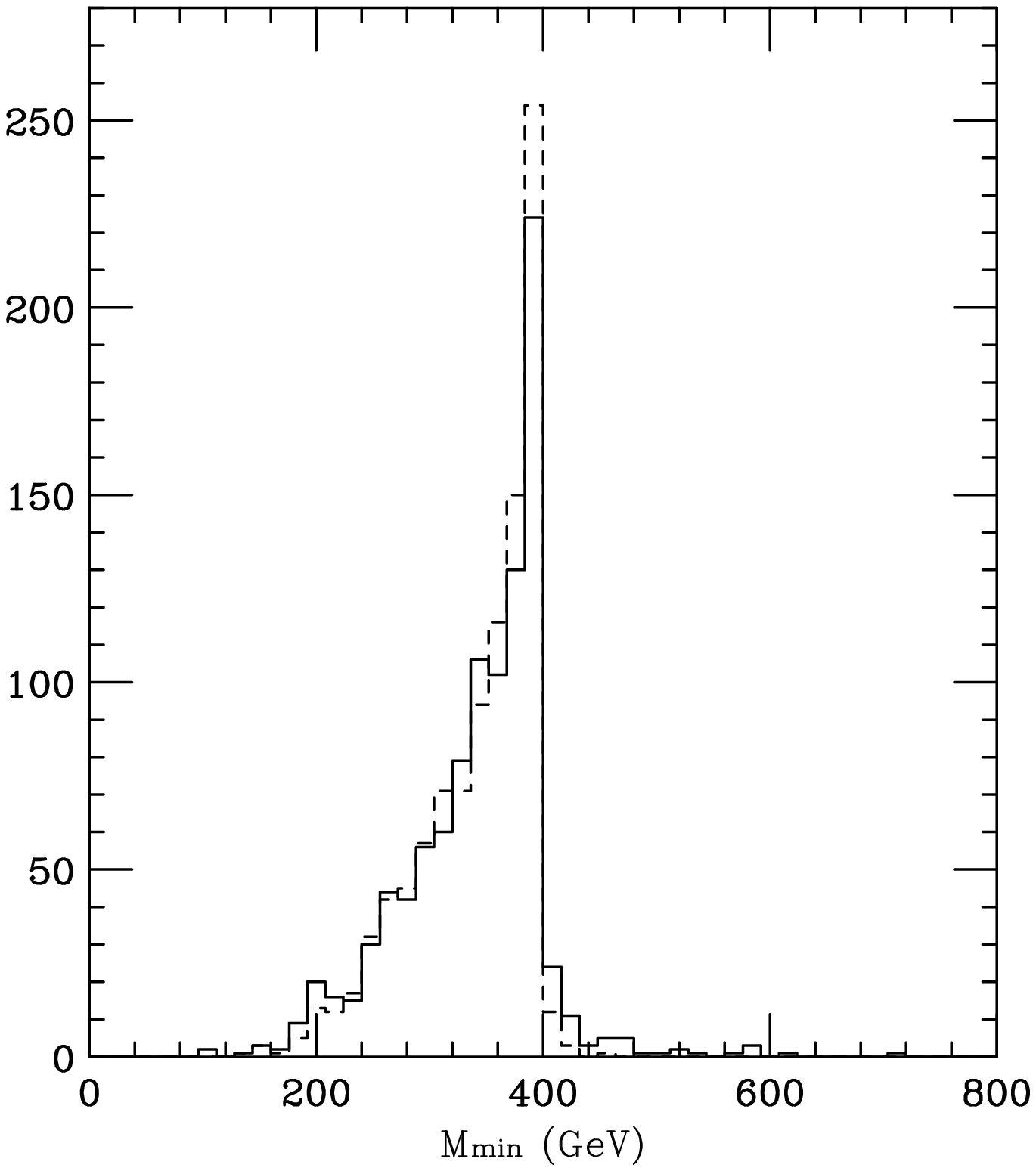}
  \hspace{3.5cm}
  \includegraphics[width=4.5cm, angle = 90]{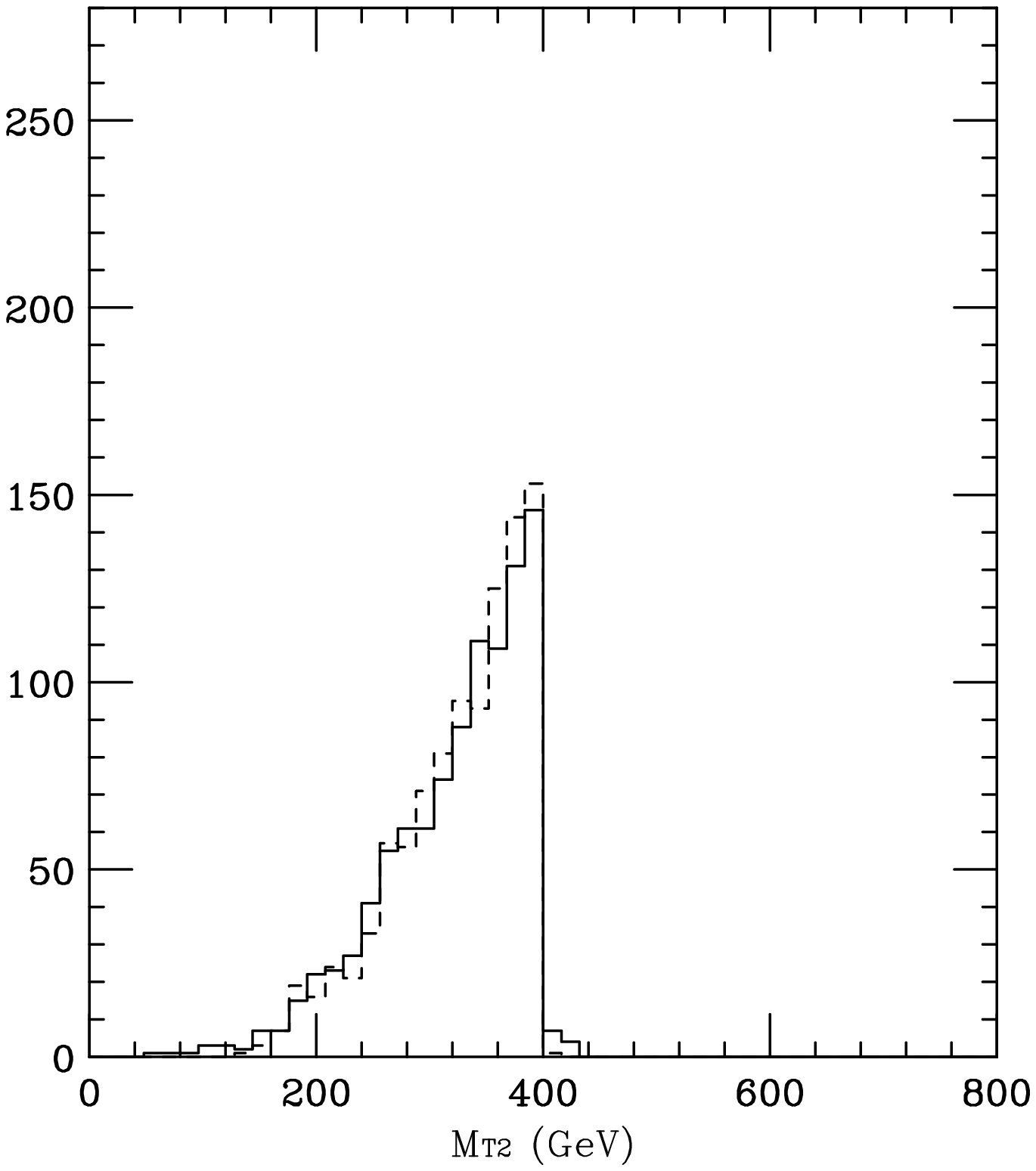}
  \vspace{0.5cm}
 \caption{ Parton-level distributions of $M_{\rm min}^{\rm bal}$ (left), $M_{\rm min}$ (centre) 
 and $M_{T2}$ (right) for $(b\nu)(t\tau)$ (solid curve) and $(t\nu)(b\tau)$ (dashed curve).
 }
\label{fig:mmin}
\end{center}
\end{figure}
Figure~\ref{fig:mmin} shows the parton-level distributions of $M_{\rm
  min}^{\rm bal}$, $M_{\rm min}$ and $M_{T2}$ for 1000 events. We took only the true combination of the jet assignment.
As can be seen, all the distributions have clear edge structures at the input leptoquark mass of 400\,GeV.

\begin{figure}[!t]
 \begin{center}
      \vspace{1.0cm}
   \hspace{4.0cm}
   \includegraphics[width=4.5cm, angle = 90]{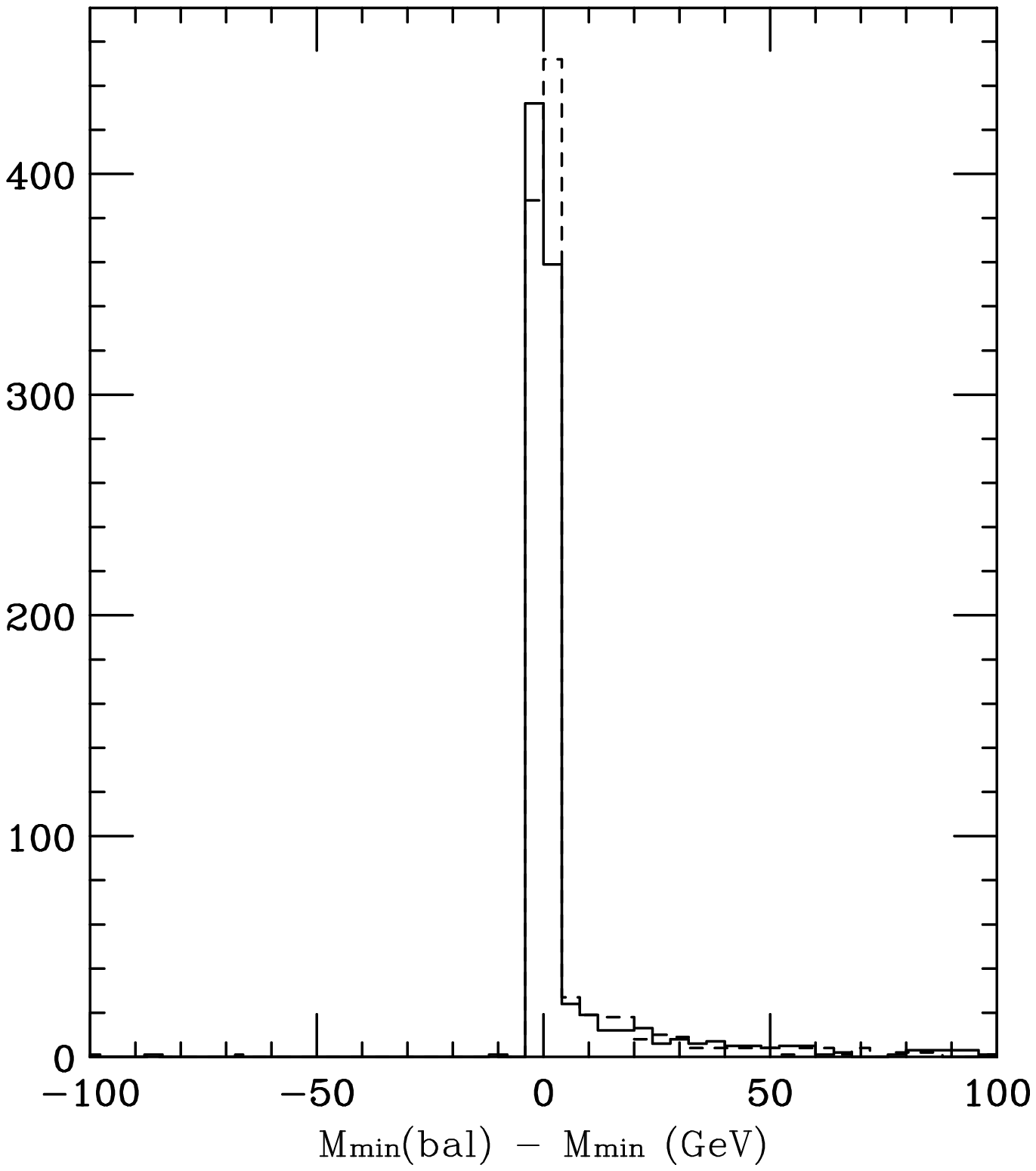}
  	\hspace{4.0cm}
  \includegraphics[width=4.5cm, angle = 90]{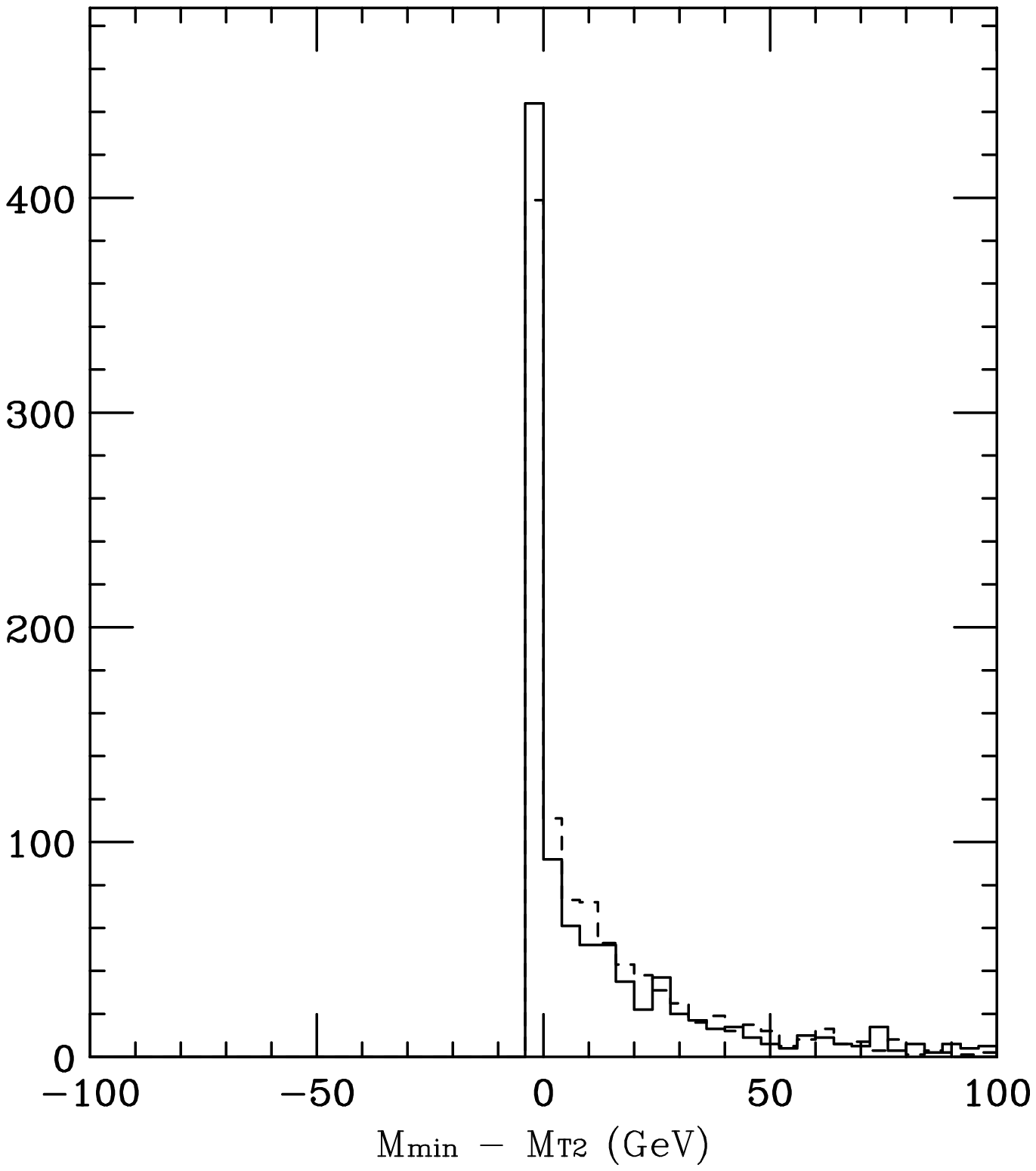}
  \vspace{0.5cm}
 \caption{
 Parton-level
 distributions of $M_{\rm min}^{\rm bal}-M_{\rm min}$ (left) and 
 $M_{\rm min}-M_{T2}$ (right) for $(b\nu)(t\tau)$ (solid curve) and $(t\nu)(b\tau)$ (dashed curve)}
\label{fig:diff}
\end{center}
\end{figure}

In order to compare these variables we took their differences, shown in Fig.~\ref{fig:diff}. 
The relation $M_{\rm min}^{\rm bal} \ge M_{\rm min} \ge M_{T2}$ is seen to
hold on an event-by-event basis.  
This implies that $M_{\rm min}^{\rm bal}$ and $M_{\rm min}$ are more powerful than $M_{T2}$
for determining the mass of the leptoquark, at least at parton level.

\subsubsection{Experimental reconstruction}
The settings for experimental reconstruction used for the
\texttt{Delphes} fast simulation remain unaltered in this analysis (see section~\ref{sec:outlinestrategy}). We apply the following event
selection cuts to the full $S_0 \bar{S_0}$ signal and the $t\bar{t}$
background:
\begin{itemize}
\item{At least four jets found in each event.}
\item{Exactly one $\tau$-tagged jet with $p_T > 120 \gev$.}
\item{No, one or two $b$-tagged jets with $p_T > 60 \gev$.}
\item{Missing transverse energy, $\slashed{E_T} > 200 \gev$.}
\end{itemize}
For the $b$-jet originating from the leptoquark decay,
we choose the highest-$p_T$ $b$-tagged jet when there are
two $b$-tagged jets and the highest-$p_T$ jet (excluding the
$\tau$-tagged jet) when there are no $b$-tagged jets. We use all the
remaining jets with $p_T > 30\gev$, (not identified as the $b$-jet from the leptoquark) to
search for one, two or three jets that form an invariant mass close to the top
mass, within a $20 \gev$ window. We apply the
additional constraint that the difference between the $p_T$ of the $\tau$-tagged jet and the $p_T$ of the $b$-tagged jet, $p_{T,\tau} - p_{T,b} > -
10 \gev$. This eliminates a high fraction of the $t\bar{t}$ background
since the $\tau$s in that sample originate from the $W$ decay and are
expected to have lower $p_T$ on average than the $b$s that originate
directly from the top. On the contrary, in the leptoquark signal the
$\tau$ and $b$ transverse momenta are expected to be of the same magnitude on average. 

\begin{figure}[!t]
 \begin{center}
   \vspace{0.8cm}
  \hspace{4.0cm}
  \includegraphics[width=4.5cm, angle = 90]{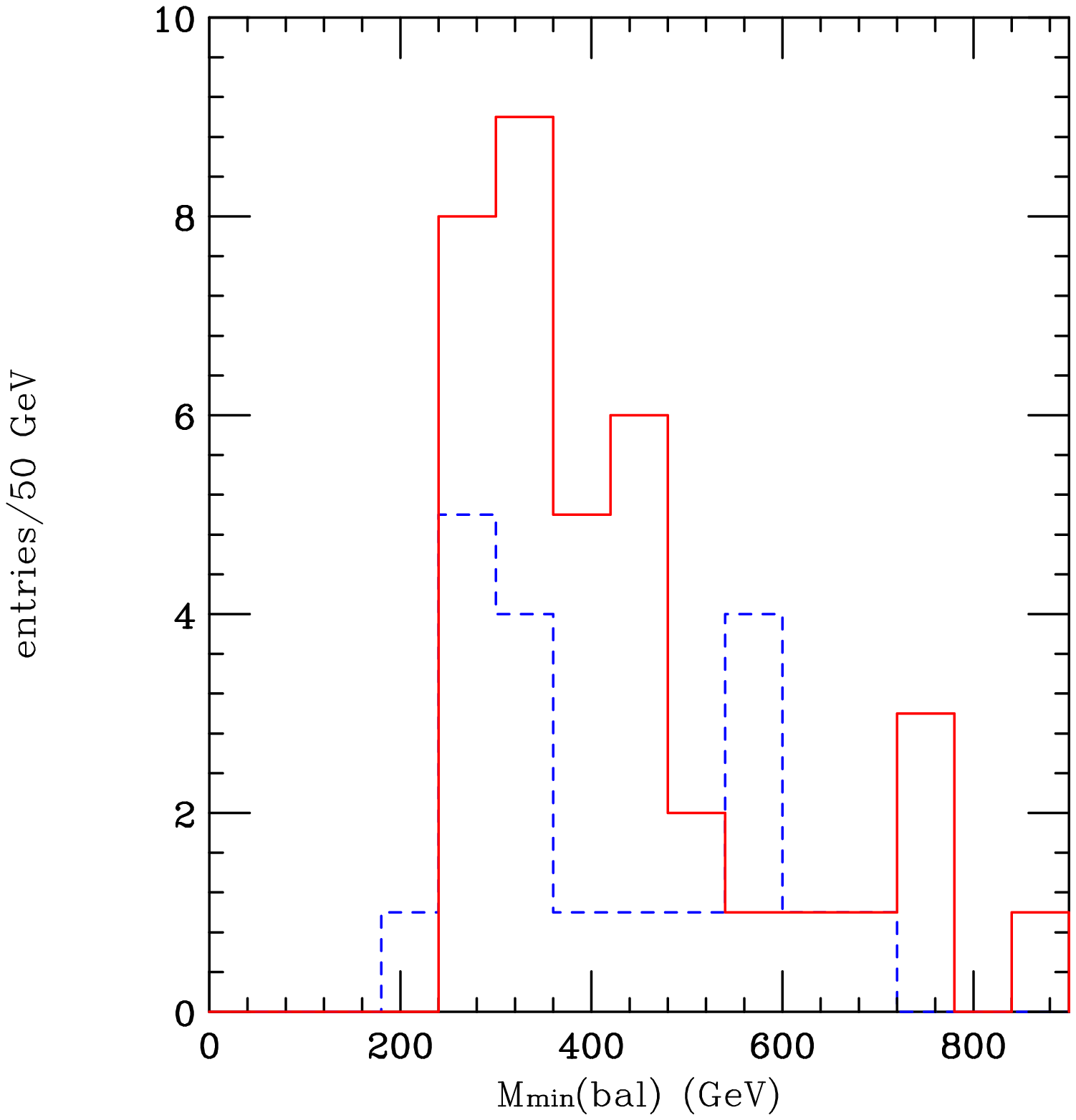}
  \hspace{4.0cm}
  \includegraphics[width=4.5cm, angle = 90]{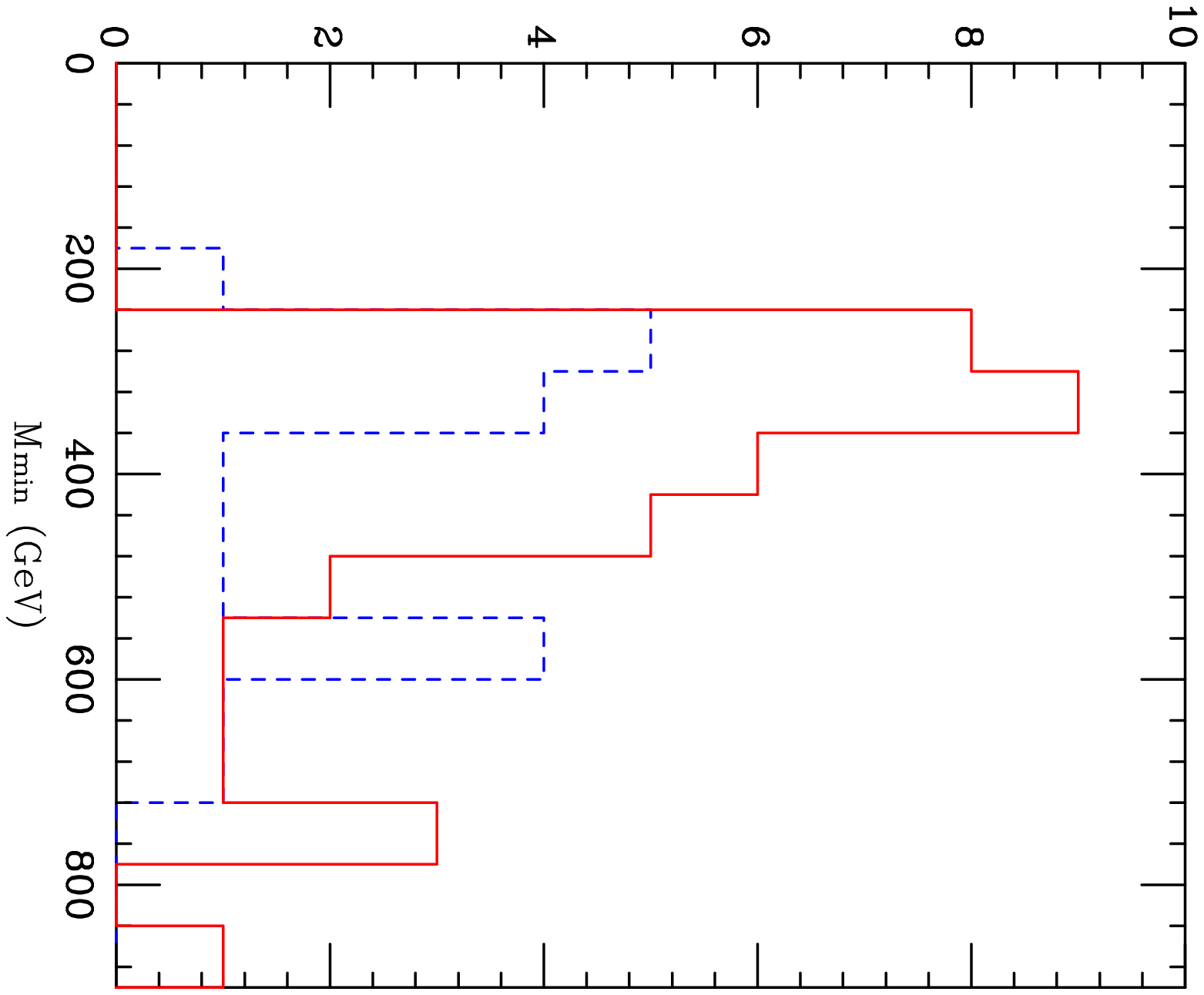}
  \hspace{4.0cm}
  \includegraphics[width=4.5cm, angle =
  90]{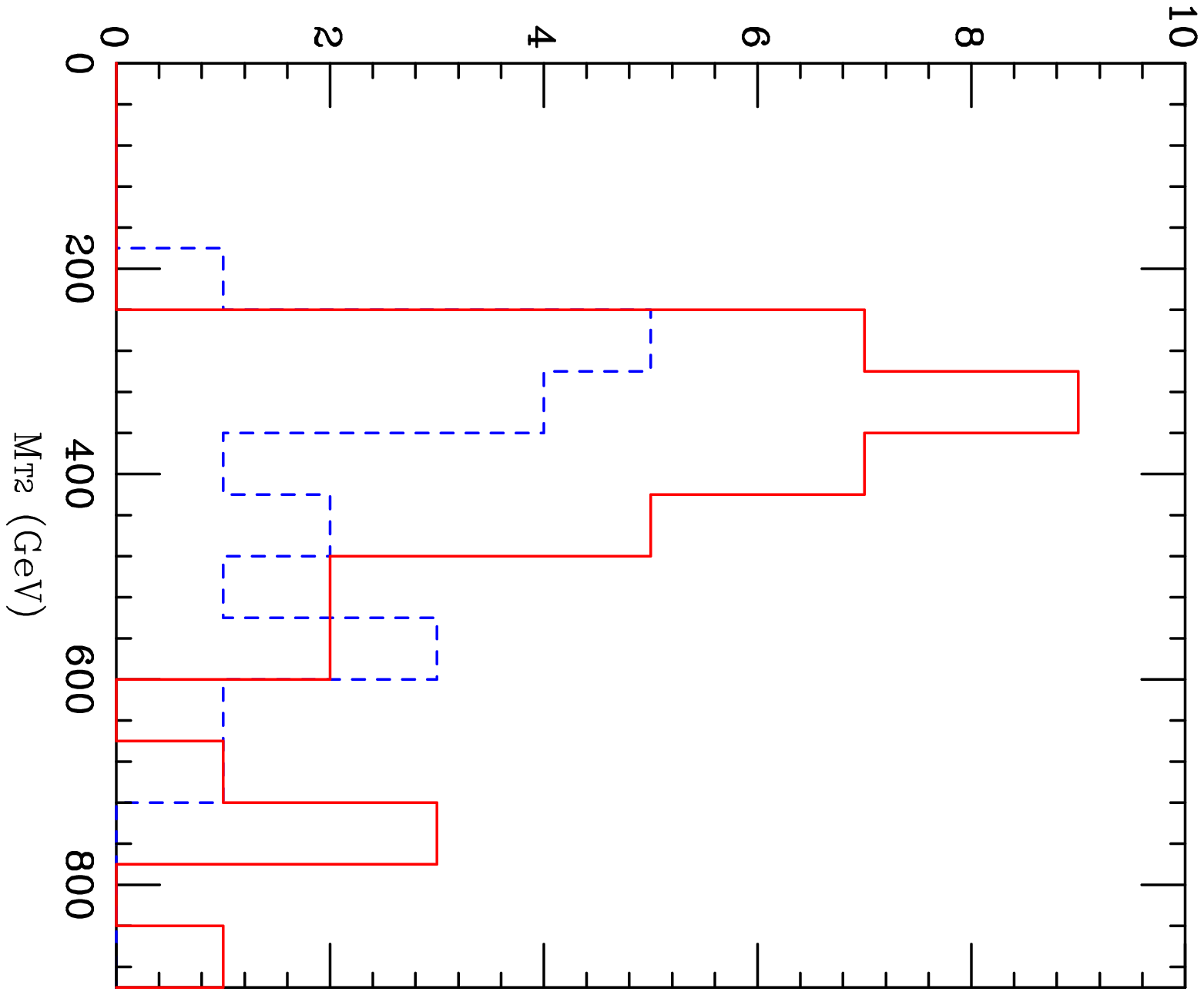}
  \vspace{0.5cm}
  \caption{Reconstructed distributions $M_{\rm min}^{\rm bal}$ (left),
   $M_{\rm min}$ (centre) 
 and $M_{T2}$ (right) for $(b\nu)(t\tau)$ signal (red) and  $t\bar{t}$ background
 events (blue dashes) including ISR, FSR and the underlying event.  There are
37 signal events and 19 background events in all plots.}
\label{fig:bnuttaures}
\end{center}
\end{figure}

The resulting distributions are shown in Fig.~\ref{fig:bnuttaures}. Due to the low
number of events passing the selection cuts, it is not obvious whether the
$M_{\mathrm{min}}^{\mathrm{bal}}$ observable performs better than $M_{\rm min}$
and $M_{T2}$. However, we checked that the three distributions
satisfy the same inequalities presented in Fig.~\ref{fig:diff} for
the parton-level reconstruction.

\begin{figure}[!t]
 \begin{center}
   \vspace{1.0cm}
  \hspace{4.0cm}
  \includegraphics[width=4.5cm, angle = 90]{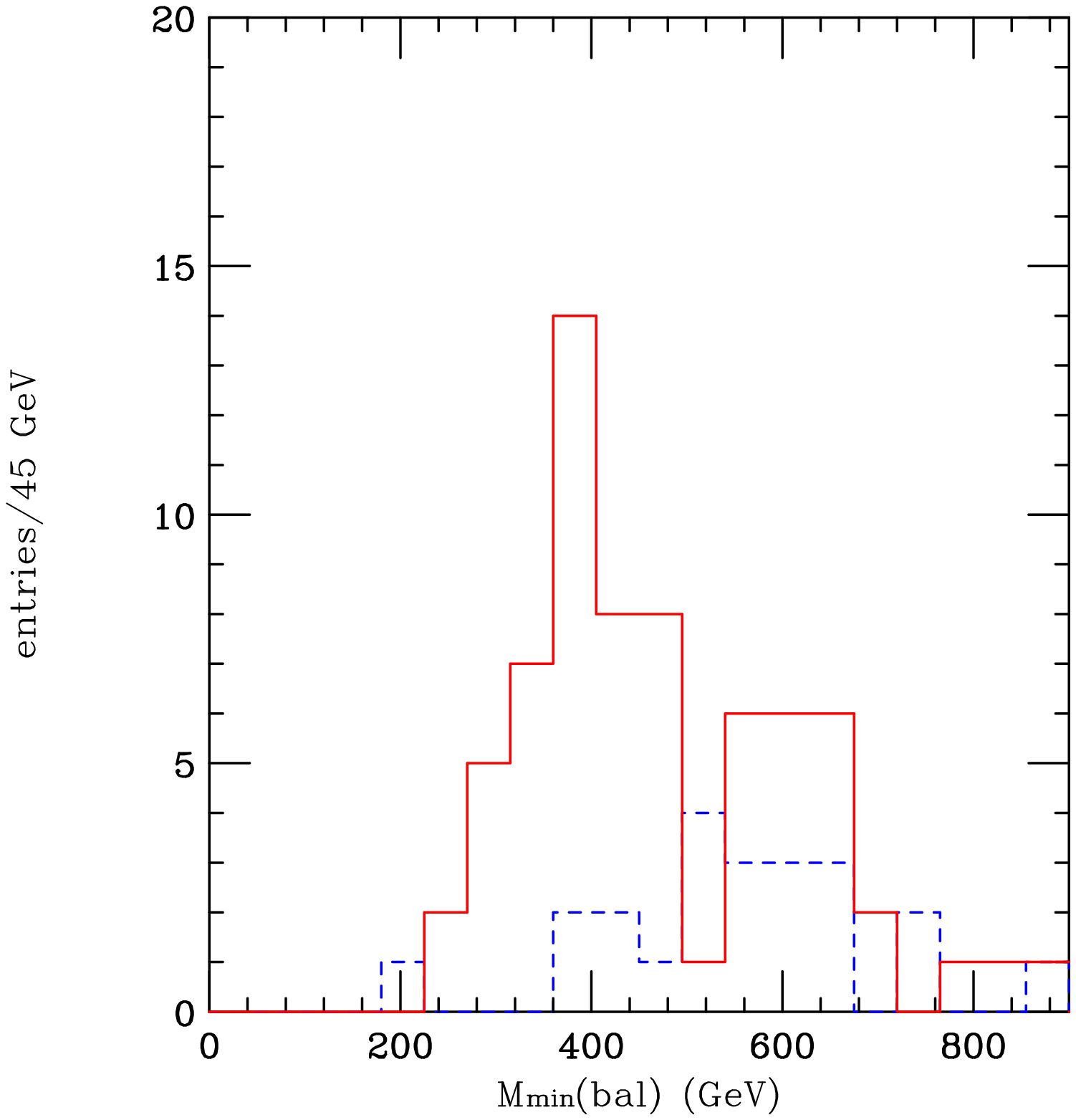}
  \hspace{4.0cm}
  \includegraphics[width=4.5cm, angle = 90]{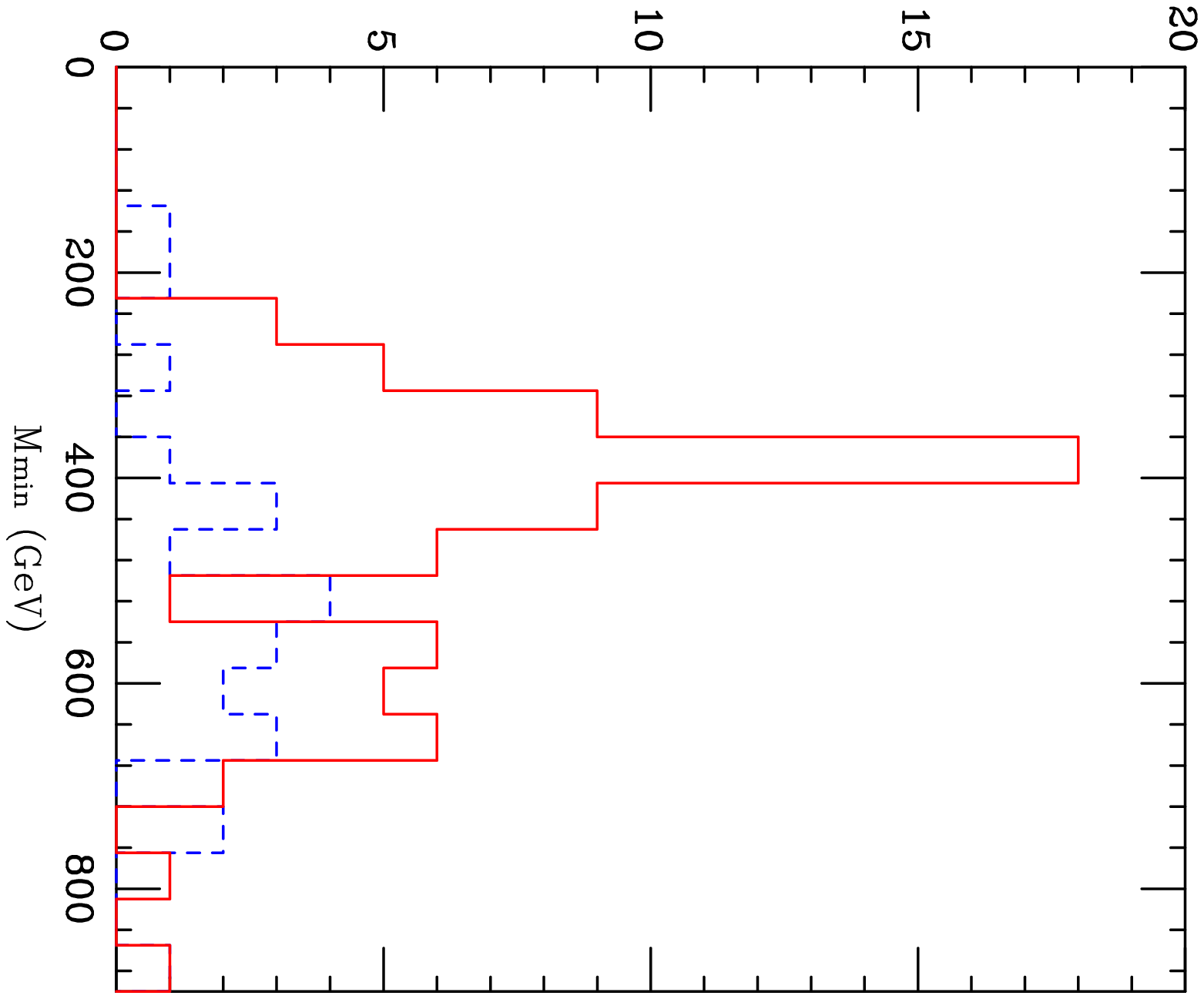}
  \hspace{4.0cm}
  \includegraphics[width=4.5cm, angle =
  90]{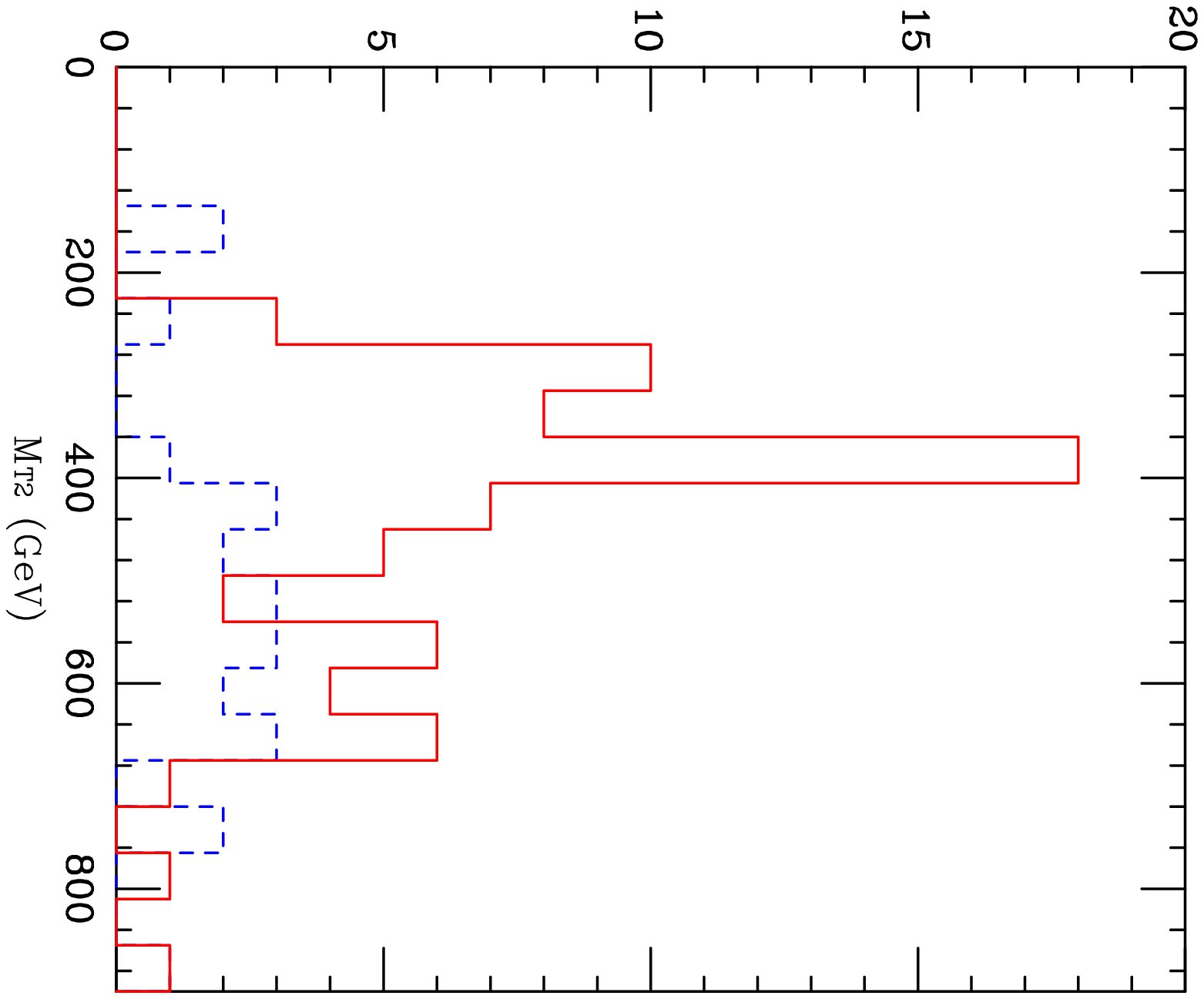}
  \vspace{0.5cm}
  \caption{Reconstructed distributions $M_{\rm min}^{\rm bal}$ (left),
   $M_{\rm min}$ (centre) 
 and $M_{T2}$ (right) for the $(t\nu)(b\tau)$ signal (red) and $t\bar{t}$ background 
 events (blue dashes) including ISR, FSR and the underlying event.  There are
 68, 72 and 72 signal events (left to right) and 22, 23, 23 background events (left to right).}
\label{fig:tnubtaures}
\end{center}
\end{figure}

The reconstruction strategy for the $(t\nu)(b \tau)$ mode follows the technique described in section~\ref{sec:ttaubnurecon} for the $(b\nu)(t \tau)$ case, with the simple replacement $b \leftrightarrow t$. The assignment of $b$-jets and top-jets is performed in the same way as in the $(b\nu)(t \tau)$ analysis, with the following cuts applied to the full $\bar{S}^{(-)}_{1/2} S^{(-)}_{1/2}$ sample:
\begin{itemize}
\item{At least four jets found in each event.}
\item{Exactly one $\tau$-tagged jet with $p_T > 190 \gev$.}
\item{No, one or two $b$-tagged jets with $p_T > 40 \gev$.}
\item{Missing transverse energy, $\slashed{E_T} > 120 \gev$.}
\end{itemize}
There is also a cut on the reconstructed hadronic top jet, of $p_T > 120\gev$ and that its invariant mass lies within $20\gev$ of the top mass. The results are shown in Fig.~\ref{fig:tnubtaures}.  Note that the background that would be present due to the $S^{(+)}_{1/2}$ leptoquark has not been included.

Although at parton level, the variable
$M_{\mathrm{min}}^{\mathrm{bal}}$ performs better than
$M_{\mathrm{min}}$ and $M_{T2}$, it seems to become unstable after
including experimental effects, with some events failing to produce a
value within the range of the plots shown in
Fig.~\ref{fig:tnubtaures}. The origin of the instability is the
additional assumption of the leptoquark masses being equal, which is
satisfied at parton level (up to small width effects) but does not hold exactly after detector simulation. For the events for which no solution is found, we assign  $M_{\mathrm{min}}^{\mathrm{bal}} = M_{\mathrm{min}}$. Even after this readjustment, there are a few events for which a solution for $M_{\mathrm{min}}^{\mathrm{bal}}$ is found and lies outside the region shown. Therefore, $M_{T2}$ and $M_{\mathrm{min}}$ appear to be preferable as experimental observables.
\subsection[$(b\tau)(b\tau)$ decay mode]{\boldmath $(b\tau)(b\tau)$ decay mode}
\subsubsection{Kinematic reconstruction}\label{sec:btaubtaukin}
The $(b\tau)(b\tau)$ mode can be fully reconstructed if one again assumes collinearity of the $\tau$-jets and $\tau$-neutrinos: $p_{\tau,i} = z_i p_{j,i}$ ($i = 1,2$, $z_i>1$). This implies that the missing momentum from each $\tau$ can be written as $\slashed{p_i} = (z_i - 1) p_{j,i}$. Hence, we may write the following equalities for the components of the measured missing transverse momentum:
\begin{eqnarray}
p_{\rm miss}^x &=& p^x_{j1} ( z_1 - 1 ) + p^x_{j2} (z_2 - 1)\;,\nonumber \\
p_{\rm miss}^y &=& p^y_{j1} ( z_1 - 1 ) + p^y_{j2} (z_2 - 1)\;.
\end{eqnarray}
The above equations may be written in matrix form and inverted to give
\begin{eqnarray}\label{eq:z1z2btaubtau}
z_1 &=& 1+ \frac{ p^y_{j2} p^x_{\rm miss} - p^x_{j2} p^y_{\rm miss} } { p^x_{j1} p^y_{j2} - p^y_{j2} p^x_{j2} }\;, \nonumber \\
z_2 &=& 1 -\frac{ p^y_{j1} p^x_{\rm miss} - p^x_{j1} p^y_{\rm miss} } { p^x_{j1} p^y_{j2} - p^y_{j2} p^x_{j2} }\;.
\end{eqnarray}
Now the invariant mass of each of the two leptoquarks may be written as $m_S^2 = (p_b + p_\tau)^2$, resulting in the following expression:
\begin{equation}
m_S^2 = 2 z_i p_{bi} \cdot p_{ji}\;,
\end{equation}
where we have neglected the $\tau$ and $b$-quark mass terms. Using
Eqs.~(\ref{eq:z1z2btaubtau}), we obtain two values of $m_S$ per
event. At parton level, with the correct jet assignments, these solutions approximate the leptoquark mass very closely, up to the collinearity approximation.

\subsubsection{Experimental reconstruction}
\begin{figure}[!htb]
  \centering 
  \vspace{1.4cm}
  \hspace{5.5cm}
  \includegraphics[scale=0.55, angle=90]{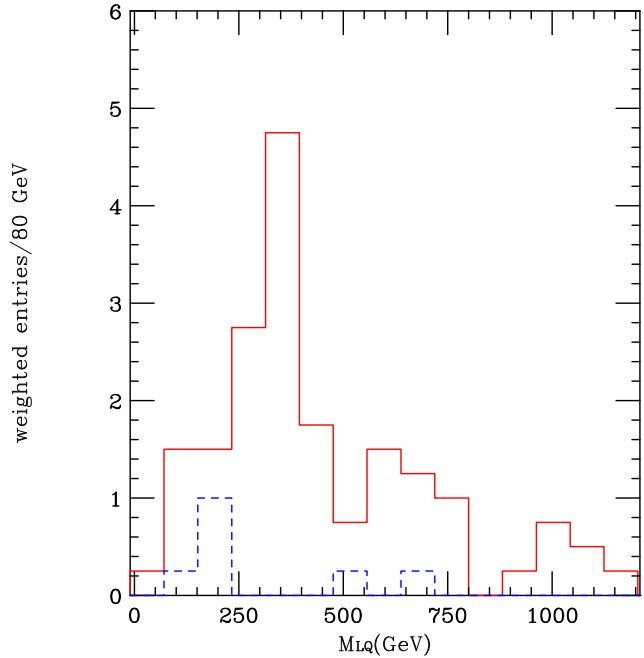}
  \vspace{0.5cm}
  \caption{Experimental reconstruction of the $(b\tau)(b\tau)$ mode using the method described
    in the main text. ISR, FSR and the underlying event have been
    included in the simulation. Note that each solution has weight 0.25. The signal is shown in
    red (18.75 entries) and the $t\bar{t}$ background in blue dashes (1.75 entries).}
  \label{fig:mbtaubtau}
\end{figure}

The \texttt{Delphes} framework has been used with identical settings
as in the previous sections. The following cuts have been applied to the $S^{(+)}_{1} \bar{S}^{(+)}_{1} \rightarrow (\bar{b}\bar{\tau}) (b\tau)$ mode: 
\begin{itemize}
\item{At least 4 jets present in the event.}
\item{Two $\tau$-tagged jets with $p_T > 140 \gev$.}
\item{Missing transverse energy $\slashed{E_T} > 140 \gev$.}
\end{itemize}
We accept events with no, one or two $b$-tagged jets. If there are less than two $b$-jets,
we search for the highest-$p_T$ non-tagged jet(s) to obtain two $b$-jets. We apply a cut of $p_T > 50\gev$ on these. There are two possible assignments of the $b\tau$ combination, resulting in a total of four solutions. The resulting distribution for the mass solutions, as described in section~\ref{sec:btaubtaukin}, is show in Fig.~\ref{fig:mbtaubtau}.
The $t\bar{t}$ background appears to be under control, with 1.75 entries in the mass histogram. 

We also considered the $b\bar{b}jj$ background, for which we generated events using \texttt{Alpgen} v2.13~\cite{Mangano:2002ea}, applying the $p_T$ cuts on the four parton-level objects. We concluded that we can safely ignore this background since the expected number of events with two $\tau$-tagged jets was $\mathcal{O}(1)$, before applying any restrictions on the missing transverse energy, $\slashed{E_T}$. Note that the backgrounds to this decay channel originating from the other members of the leptoquark multiplet have not been included.

\subsection{Determination of quantum numbers}
In the ideal scenario where all of the decay modes of a leptoquark
multiplet are seen, the quantum numbers can be deduced without
ambiguity. For example if we only observe combinations of $(t\tau)$
and $(b\nu)$ decay modes, then the only likely candidate is an $S_0$
singlet. However, if in conjunction with these decay modes we observe $(b\tau)$ and $(t\nu)$ decay modes, with corresponding total rates, then we might guess that we have observed the $S_1$ multiplet. 

The issue is more complicated if some decay modes are missed.  For
example if only the $(t\tau)(t\tau)$ decay mode has been seen, we
might assume that we have observed the pair production of an
$\tilde{S}'_0$ leptoquark. However, we might have observed the
$(t\tau)(t\tau)$ decay of an $\tilde{S}_{1/2}'^{(+)}$ leptoquark pair
and missed the more challenging $(t\nu)(t\nu)$ mode of the
$\tilde{S}_{1/2}'^{(-)}$ leptoquarks. In this case we would need to
examine the helicities and charges of the decay products: the
$\tilde{S}_{1/2}'^{(+)}$ decays to $\bar{t}_L \bar{\tau}_L$ and
$\bar{t}_R \bar{\tau}_R$ whereas the $\tilde{S}'_0$ decays to  $t_R
\bar{\tau}_L$ and $t_L \bar{\tau}_R$.  Since we can reconstruct all
decay products of the top and $\tau$ without combinatorial ambiguity,
using the measured leptoquark mass as an input, there is hope that we could measure top~\cite{Shelton:2008nq, Godbole:2006tq, Dalitz:1991wa} and $\tau$~\cite{Guchait:2008ar, Godbole:2008it} polarisations simultaneously. This would allow us to distinguish these two cases. We leave investigation of the feasibility of this to future work.

\subsection{Conclusions}\label{sec:conc}
If strongly-coupled dynamics solves the hierarchy problem of electroweak symmetry breaking, the question arises of how best to discover it at the LHC. Existing constraints coming from electroweak precision tests tell us that, at least at low energies, any model of strong dynamics must be a lot like the Standard Model (with perhaps a faint hope of observable deviations in the Higgs sector \cite{Low:2009di,Gripaios:2009pe}). 
Existing constraints coming from flavour physics are somewhat different in that, while the data are certainly consistent with the Standard Model, naturalness arguments suggest that strongly-coupled theories should differ from the Standard Model in the flavour sector. Indeed, fermion masses should arise via mixing between elementary and composite fermions of the strongly-coupled sector.

If that is so, then composite leptoquarks (or diquarks) may also
appear, coupled predominantly to third-generation fermions. These
would provide a spectacular signature at the LHC.  Their Standard
Model quantum numbers imply that they would be produced strongly as
conjugate particle-anti-particle pairs, decaying into third-generation
quarks and leptons in the combinations summarised in
Table~\ref{tb:strategy}.  We have proposed a number of new
experimental search strategies adapted to these characteristic final
states, also summarised in Table~\ref{tb:strategy}, and implemented
the relevant processes in the \Herwigpp event generator version 2.5.0~\cite{Gieseke:2011na}
in order to study their effectiveness in the presence of QCD radiation, backgrounds and the underlying event.  We used the \texttt{Delphes} detector simulation to assess the effects of $b$- and $\tau$-tagging efficiencies and  detector resolution.
For definiteness we assumed a leptoquark mass of 400 GeV and an integrated $pp$ luminosity of 10 fb$^{-1}$ at 14 TeV.

In the case of decays of leptoquark pairs to $(q\tau)(q\tau)$ where $q=t$ or $b$, the approximate collinearity of the missing neutrinos and jets from the tau decays allows full reconstruction of the leptoquark mass, even when one top decay is semi-leptonic.
In the former case there is a quartic ambiguity in the resulting mass, although not all of the solutions are real.  After detector resolution smearing, the correct solutions for the momentum fraction $z_2$ may be complex, but we found that using the real parts provides a fair estimate of the mass, with resolution of the order of $\pm 150$ GeV.  For $(b\tau)(b\tau)$ the only ambiguity is combinatoric but the mass resolution is similar.  In both cases the expected background from $q\bar q jj$ is small after cuts and reconstruction.

For decays to $(t\tau)(b\nu)$ or $(t\nu)(b\tau)$, we have proposed an edge reconstruction strategy similar to those developed for supersymmetric models, but using mass variables $M_{\rm min}^{\rm bal}$ and $M_{\rm min}$ that are in principle superior to the classic `stransverse mass' $M_{T2}$.  However, given the limited statistics expected, the difference in performance between these variables was not obvious.  We found cuts to reduce the background from $t\bar t$ to manageable levels, but the edge reconstruction remains challenging without higher statistics.  For $(q\nu)(q\nu)$ the story is similar for edge reconstruction in  $M_{T2}$, the case of $q=t$ being the more difficult owing to the similarity of the distributions of the signal and $t\bar t$ background.  But even in that case a clear excess over background should be visible and would give a rough estimate of the leptoquark mass.

In the event that a discovery is made, one might ask to what extent this provides proof that electroweak symmetry breaking is driven by strongly-coupled, composite dynamics. After all, one can easily imagine weakly-coupled theories with such states, for example, third-generation squarks in $R$-parity-violating supersymmetric models. Ultimately, 
TeV-scale compositeness can only be revealed by experiments probing significantly higher scales; for that, we shall have to wait some time.
In the meantime, the discovery of leptoquarks coupled to third-generation fermions and their {\em de facto} consistency with the multitude of existing flavour experiments would imply very strong bounds on the couplings to other fermions. The scenario in which the observed fermions are partially elementary and partially composite provides, as far as we know, the only mechanism in which the required suppression can be automatically achieved. Moreover, it gives a prediction for the size of the other couplings, some of which are not far from current bounds, which may then be targeted in ongoing flavour experiments. Though circumstantial, this would seem to be the best possible evidence for compositeness that one might hope for in the LHC era.


\chapter{Conclusions and Outlook}
\label{cha:conclusions}
At the time of writing of this thesis, the framework of the Standard
Model (SM) is a well-established set of gauge theories, described by
the product group $SU(3)_c \times SU(2)_L \times U(1)_Y$ and complemented by the Higgs
mechanism, responsible for the breaking of the $SU(2)_L \times U(1)_Y$ down to
$U(1)_\mathrm{em}$. It is well-established in the sense that it
currently provides excellent agreement with experimental
data. However, as we have discussed in
chapter~\ref{cha:smbsm}, there remain serious open questions that
cast doubt on the status of the SM as a `final' theory of Nature. These include the
stability of the scalar Higgs mass against radiative corrections,
known as the hierarchy problem, the `near-miss' of the unification of
the gauge couplings in the SM, and questions originating from
astrophysical and cosmological measurements such as the existence of
non-luminous gravitating matter (dark matter) and a mysterious form of
energy that causes the Universe's expansion rate to accelerate (dark energy). There is
also a strong feeling that the absence of a quantum description of
gravitational interactions in the same framework as the other forces
is an indication that we are far from a complete theoretical description of
Nature. 

Several solutions, of varying degrees of ambition, have been proposed
to address the problems that plague SM. These range from extensions of
space-time symmetries, by adding extra dimensions or supersymmetries to
the Poincar\'e symmetry, models with larger gauge groups with intricate
symmetry breaking patterns resulting in
interesting effective theories (e.g. Little Higgs models) or the
addition of new strong forces (e.g. technicolour). High energy
particle colliders allow us to explore the high
energy realm, and determine which of the proposed theories, if any, is
related to Nature. In these experiments, collisions of particles are performed in a
controlled environment and deductions are made by examining the
products of the scattering reactions. The latest and greatest
experiment is the Large Hadron Collider (LHC), at CERN, near Geneva,
Switzerland. The LHC is a machine that collides protons head-on, with a
design nominal energy of $14\tev$. 

Theoretical predictions are necessary if we wish to squeeze out every drop of physics from the LHC
experiment. The Monte Carlo method that we described in
chapter~\ref{cha:mc}, provides powerful tools that enable us
to make phenomenological predictions, incorporating perturbative
quantum chromodynamics (QCD) and other models inspired by it. Monte
Carlo methods are conceptually easy to associate with
experiments. They provide simulations of particle collisions, starting from the parton-level
theoretical predictions, e.g. derived from a set of Feynman rules, to a
full simulation of the effects of interactions of the resulting
particles with the components of the detector. Such simulations also
include perturbative treatment of initial- and final-state QCD showers, as
well as models of secondary partonic interactions, which form the underlying
event, a phenomenon under intense theoretical and experimental
investigation. The matrix elements and the showers can nowadays be
provided at next-to-leading order (NLO) accuracy, providing more reliable
predictions and better agreement with experiment via the \texttt{POWHEG} and
\texttt{MC@NLO} methods we discussed in
section~\ref{sec:mc:nlomatching}.  

The exploration of physics at high energy hadron colliders is a
non-trivial task. We considered, in the
introduction of chapter~\ref{cha:qcdrad}, the complications that we
have to face, both due
to the complexity of the possible new physics signals and the difficulties that
arise due to the fact that the colliding particles are hadrons. These
are illustrated schematically in Figs.~\ref{fig:qcdrad:challenges1}
and~\ref{fig:qcdrad:challenges2}. In chapter~\ref{cha:qcdrad} we constructed analytical
predictions of the effects of QCD phenomena on certain hadron collider
variables: on a class of variables called `global inclusive
variables', which make use of all observed momenta,
and on the transverse energy of initial-state radiation in Higgs and Drell-Yan
gauge boson production. We compared the analytical predictions to
results obtained from Monte Carlo event generators. In the future, the calculations
of QCD effects on global inclusive variables could be
extended to make use NLO cross sections and splitting
functions. Furthermore, the validity of the calculation of the
distribution of transverse energy
associated with Higgs production is uncertain, even after matching to
NLO. This could be investigated further with matching to the full NNLO
result. 

To be able to cope with the complexity of new physics signals and set
the best possible bounds on model parameters, or improve the potential
of discovering new physics, we have to be adequately prepared. This
involves constructing robust general search strategies and predictions. In chapter~\ref{cha:NEWPHYS} we first improved
the treatment of Drell-Yan production of heavy charged vector bosons,
called $W'$ ($W$ prime) by using the \texttt{POWHEG} and
\texttt{MC@NLO} methods to generate fully exclusive events at NLO. We
also considered the effects of interference of a potential $W'$ with
the SM $W$. Interference may provide extra information to the nature of the new
particle or improve the detection reach. We also investigated an
interesting model of scalar leptoquarks which couple to third-generation
quarks and leptons, inspired by a theory of strong dynamics
electroweak symmetry breaking, although not limited to it. The signals
are challenging at the LHC due to their complexity and potentially
huge backgrounds, but we provided a complete strategy for
reconstruction of all the possible decays in pair-production of
charge-conjugate leptoquark states. Future extensions to this work may
involve using NLO matrix elements for the production of the
leptoquarks or investigating the existence of new diquark states that
might exist in the same models. Strategies may be
developed for determining the quantum numbers of the leptoquarks,
using methods that measure the helicity of the fermions that originate
from their decays, as we pointed out at the end of
section~\ref{sec:newphys:leptoquarks}. 

We are at the beginning of an exciting time for particle physics, and
science in general. The quest for understanding the underlying structure of Nature that started thousands of years ago will surely
enter a new chapter during the Large Hadron Collider era. Our duty is
to be prepared, guided by our intuition and the analytical and
computational tools that we have created, so that we will be able to
comprehend the new results that will be faced with. The methods and
ideas studied in this thesis could contribute in that direction.


\appendix

\chapter{Illustration of a Monte Carlo event}
\begin{figure}[!htb]
  \centering 
  \includegraphics[scale=0.55, angle=0]{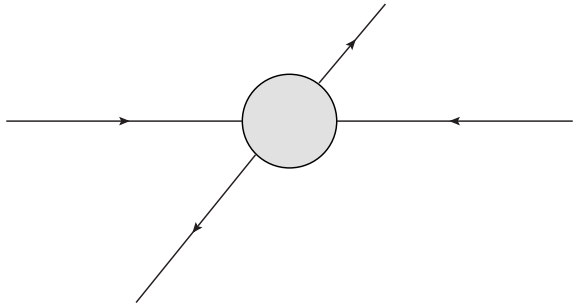}
  \caption[]{\textbf{STEP 1}: Generation of the hard process.}
\label{fig:app:mcillustration:step1}
\end{figure}

\begin{figure}[!htb]
  \centering 
  \includegraphics[scale=0.55, angle=0]{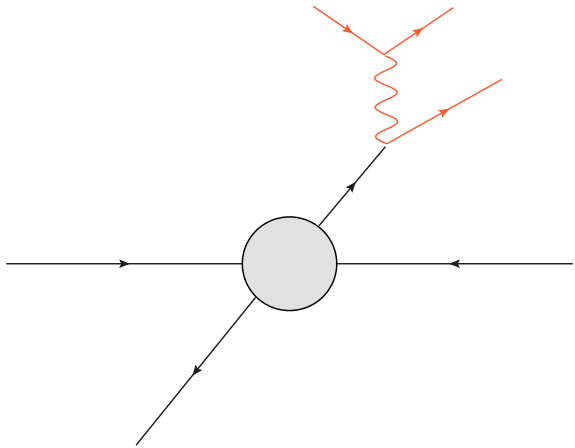}
  \caption[]{\textbf{STEP 2}: Decay of heavy resonances.}
\label{fig:app:mcillustration:step2}
\end{figure}

\begin{figure}[!htb]
  \centering 
  \includegraphics[scale=0.55, angle=0]{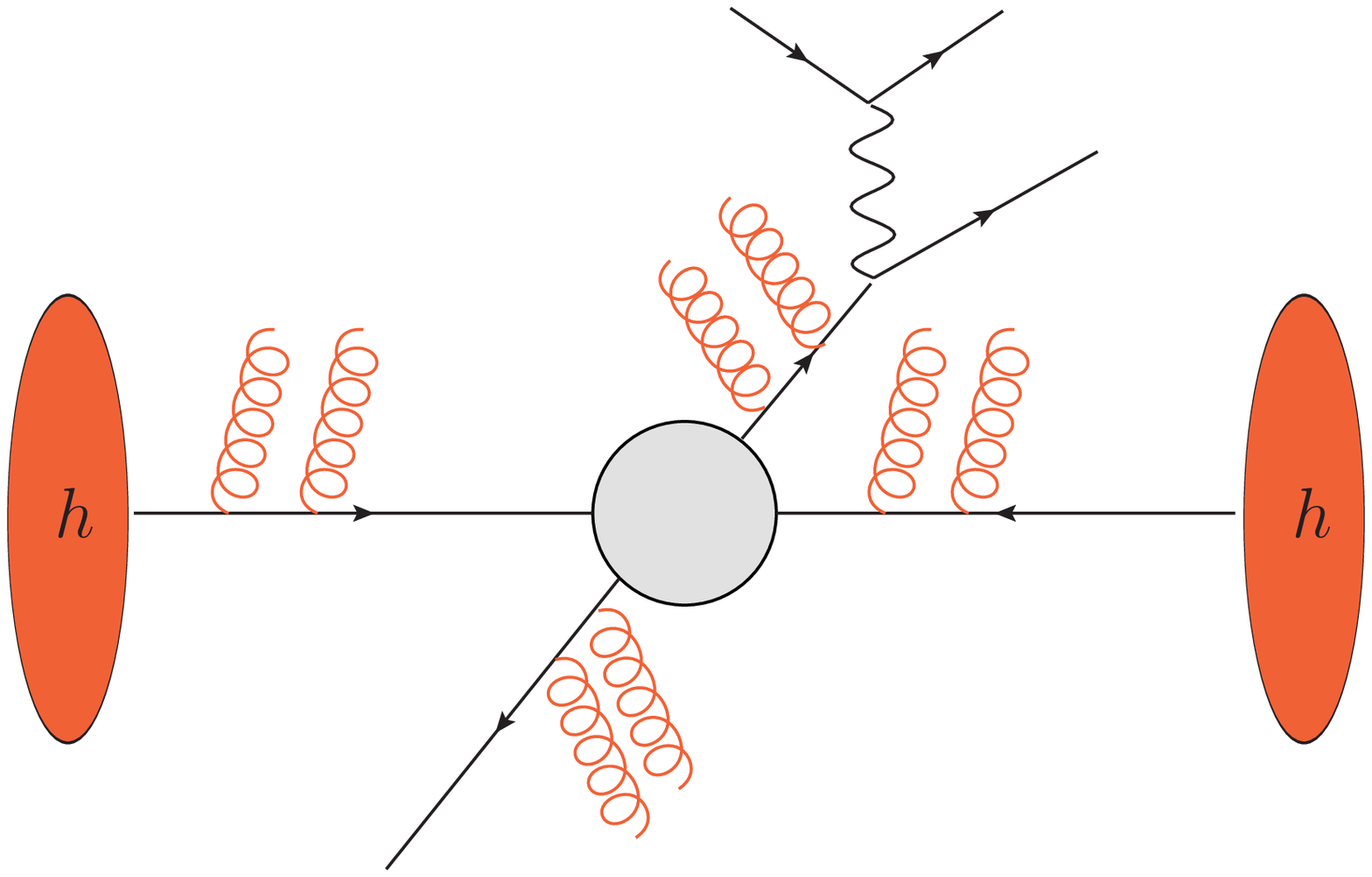}
  \caption[]{\textbf{STEP 3}: Parton showers.}
\label{fig:app:mcillustration:step3}
\end{figure}

\begin{figure}[!htb]
  \centering 
  \includegraphics[scale=0.55, angle=0]{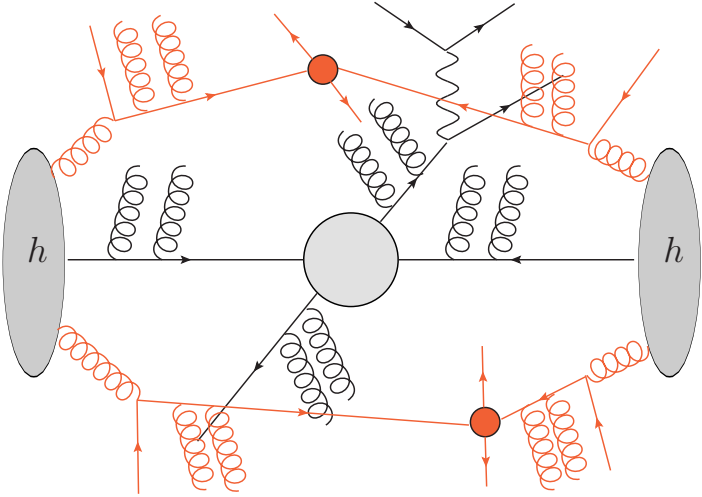}
  \caption[]{\textbf{STEP 4}: Multiple parton interactions.}
\label{fig:app:mcillustration:step4}
\end{figure}

\begin{figure}[!htb]
  \centering 
  \includegraphics[scale=0.55, angle=0]{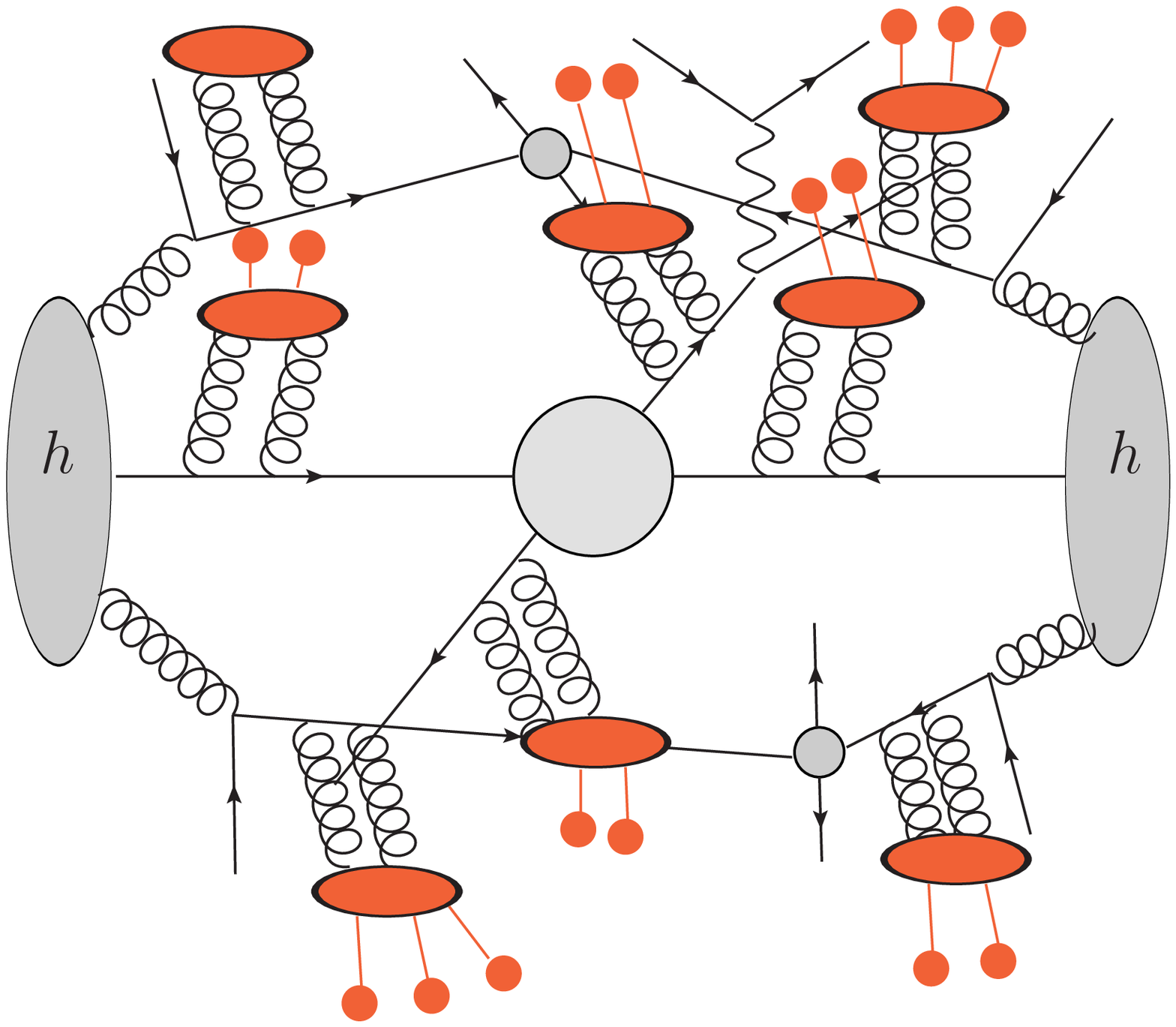}
  \caption[]{\textbf{STEP 4}: Hadronization and hadron decays.}
\label{fig:app:mcillustration:step5}
\end{figure}

In this appendix we illustrate, with the help of schematic diagrams, the set of steps performed by a generic Monte
Carlo event generator when producing a full event
simulation. Figures~\ref{fig:app:mcillustration:step1} to~\ref{fig:app:mcillustration:step5}
demonstrate the various steps. In each step, the newly added features
are shown in red colour. 
\begin{enumerate}
\item{\textbf{Hard process generation,
      Figure~\ref{fig:app:mcillustration:step1}}: The hard process is
    generated by choosing a point on the phase space according to the
    `hit-or-miss' method.}
\item{\textbf{Heavy resonance decay,
      Figure~\ref{fig:app:mcillustration:step2}}: Heavy resonances
    with narrow widths are
    decayed before the parton shower. In this example the heavy
    resonance could be a top quark, decaying to a $\ell \nu_{\ell}$
    and a $b$-quark.} 
\item{\textbf{Parton showers,
      Figure~\ref{fig:app:mcillustration:step3}}: The incoming partons
    are showered by evolving backwards to the incoming hadrons,
    producing initial-state radiation. Any final-state particles
    that are colour-charged also radiate, producing final-state
    radiation.}
\item{\textbf{Multiple parton interactions,
      Figure~\ref{fig:app:mcillustration:step4}}: Secondary
    interactions between partons within the colliding hadrons,
    modelled as QCD $2\rightarrow2$ interactions, are generated. The
    secondary partons are showered and are always evolved backwards to
    gluons in the underlying event model present in \Herwigpp.}
\item{\textbf{Hadronization and hadron decays,
      Figure~\ref{fig:app:mcillustration:step5}}: In the cluster
    model, clusters are formed and hadrons are produced. Unstable
    hadrons are subsequently decayed.}
\end{enumerate}

\label{app:mcillustration}

\chapter{Pair-production cross sections}
\label{app:cross-sections}
The leading order parton-level cross section for QCD pair-production of particles of mass $m_p$ may be written in terms of scaling functions $f_{ij}$ as
\begin{equation}
\hat{\sigma}_{ij} (Q^2) = \frac{\alpha_S^2(Q^2)}{m_p^2} f_{ij}\;.
\end{equation}

For heavy quark pair-production, the functions for gluon-gluon and quark-anti-quark initial states are given by~\cite{Ellis:1996qj}
\begin{eqnarray}
f_{gg} &=& \frac{ \pi \beta \rho } {192 } \left\{ \frac{1}{\beta} ( \rho ^2 + 16 \rho + 16 ) \log\left|\frac{1+\beta}{1 - \beta}\right| - 28 - 31 \rho \right\}\;,\\
f_{q\bar{q}} &=& \frac{ \pi \beta \rho } { 27 } (2 + \rho)\;.
\end{eqnarray}
where $\rho= 4 m_q^2 / Q^2$ and $\beta = \sqrt{1 - \rho}$.

 For the case of gluino pair-production, the equivalent functions $f_{ij}$ are given by~\cite{Beenakker:1995fp}
\begin{eqnarray}
f_{gg} &=& \frac{ \pi m_{\tilde{g}}^2 }{ Q^2 } \left\{ \left[ \frac{9}{4} + \frac{ 9 m_{\tilde{g}}^2 } { Q^2 } - \frac{9 m_{\tilde{g}}^4} { Q^4 } \right] \log \left| \frac{1 + \beta}{1 - \beta} \right| - 3 \beta - \frac{ 51 \beta m_{\tilde{g}}^2 } { 4 Q^2 } \right\}\;,\\ 
f_{q\bar{q}} &=& \frac{ \pi m_{\tilde{g}}^2 }{ Q^2 } \left\{ \beta \left[ \frac{20}{27} +\frac{16 m_{\tilde{g}}^2} { 9 Q^2 } - \frac{ 8 m_{-}^2 } { 3 Q^2 } + \frac{ 32 m_{-}^4 } { 27 ( m_{-}^4 + m_{\tilde{q}}^2 Q^2 ) } \right] \right.\\\nonumber
 &+&\left. \left[ \frac{ 64 m_{\tilde{q}}^2 } { 27 Q^2 } + \frac{ 8 m_{-}^4 } { 3 Q^4 } - \frac{ 16 m_{\tilde{g}}^2 m_{-}^2 }{ 27 Q^2 ( Q^2 - 2 m_{-}^2 ) }  \right]\log \left( \frac{ 1 - \beta - 2 m^2_{-} / Q^2 } { 1 + \beta - 2 m^2_- / Q^2 } \right) \right\}\;,
\end{eqnarray}
where now $\beta=\sqrt{1 - 4 m_{\tilde{g}}^2 / Q^2 }$ and $m_-^2$
represents the mass-squared difference between the gluino and the
$t$-channel squark, $m_{-}^2 = m ^2 _{\tilde{g}} - m ^2 _{\tilde{q}}$.

For the case of scalar leptoquark pair-production, the scaling
functions are given by
\begin{eqnarray}
f_{gg} &=& \frac{ \pi M_{LQ}^2 }{96 \hat{s} } \left\{ \beta ( 41 - 31 \beta ^2 ) - (17 - 18 \beta ^2 + \beta^4) \log \left| \frac{1 + \beta} { 1 - \beta } \right| \right\}\;, \nonumber \\
f_{q\bar{q}} &=& \frac{ 2 \pi M_{LQ}^2 } { 27 \hat{s} } \beta^3 \;,
\end{eqnarray}
where $\beta=\sqrt{1 - 4 M_{LQ}^2 / Q^2 }$. The differential cross sections with respect to the leptoquark
scattering angle in the partonic centre-of-mass frame, $\theta$, are
given by
\begin{eqnarray}
\frac{\mathrm{d} \hat{\sigma}^{gg}_{S\bar{S}}} { \mathrm{d} \cos \theta } &=& \frac{ \pi \alpha _s ^2 } { 6 \hat{s} } \beta \left\{ \frac{1}{32} [ 25 + 9 \beta ^2 \cos ^2 \theta - 18 \beta^2 ] \right.\nonumber \\
 &-& \left. \frac{1}{16} \frac{ (25 - 34 \beta ^2 + 9 \beta^4 ) } { 1 - \beta^2 \cos ^2 \theta } + \frac{ ( 1 - \beta ^2 ) ^ 2} { (1 - \beta^2 \cos ^2 \theta )^2} \right\}\;, \nonumber \\
\frac{\mathrm{d} \hat{\sigma}^{q\bar{q}}_{S\bar{S}}} { \mathrm{d} \cos \theta } &=& \frac{ \pi \alpha _s ^ 2} { 18 \hat{s} } \beta^3 \sin ^2 \theta \;.
\end{eqnarray}

\chapter{The Cabibbo-Kobayashi-Maskawa matrix}
\label{sec:ckm}

The Cabibbo-Kobayashi-Maskawa (CKM) matrix (Eq.~(\ref{eq:sm:ckm})) is a $3\times3$ unitary
matrix that can be parametrized by three mixing angles $\theta_{ij}$
($i,j \in \{1,2,3\}$, $i \neq j$) and a charge-parity (CP) violating phase, $\delta$. A common choice is
\begin{eqnarray}
V &=& \left(\begin{array}{ccc} 
    c_{12} c_{13} & s_{12} c_{13} & s_{13} e^{-i \delta} \\
    -s_{12} c_{23} - c_{12} s_{23} s_{13} e^{i\delta} & c_{12} c_{23}
    - s_{12} s_{23} s_{13} e^{i\delta} & s_{23} c_{13} \\
    s_{12} s_{23} - c_{12} c_{23} s_{13} e^{i\delta} & -c_{12} s_{23}
    - s_{12} c_{23} s_{13} e^i\delta & c_{23} c_{13} \\ 
\end{array}\right)  \nonumber \\
\end{eqnarray}
where we have used the shorthand notation $c_{ij}
= \cos \theta_{ij}$ and $s_{ij} = \sin \theta_{ij}$. The angles
$\theta_{ij}$ have been chosen to lie in the first quadrant.
Experimentally it has been observed that $s_{13} \ll s_{23} \ll s_{12}
\ll 1$, so it is
convenient to demonstrate the hierarchy using the Wolfenstein
parametrisation:
\begin{eqnarray}
s_{12} &=& \lambda = \frac{ |V_{us} | } { \sqrt{ |V_{ud}|^2 + |V_{us}|^2
  } }\;,\;\;\;\; s_{23} = A \lambda^2 = \lambda \left| \frac{V_{cb}}{
    V_{us} } \right| \;,\nonumber \\
s_{13} e^{i \delta} &=& V_{ub}^* = A \lambda^3 ( \rho + i \eta ) =
\frac{ A \lambda^3  ( \bar{\rho} + i \bar{\eta} ) \sqrt{ 1 - A^2
    \lambda^4 } } { \sqrt{ 1 - \lambda^2 }} [1 - A^2 \lambda (\bar{\rho}
    + i \bar{\eta})]\;.
\end{eqnarray}
We then have
\begin{eqnarray}
V &=& \left(\begin{array}{ccc} 
    1 - \lambda^2 / 2& \lambda & A \lambda^3 ( \rho - i \eta) \\
    - \lambda & 1 - \lambda^2 / 2& A \lambda^2\\
    A \lambda^3 ( 1 - \rho - i \eta ) & - A \lambda^2  & 1 \\ 
\end{array}\right) + \mathcal{O} ( \lambda^4 ) \;. 
\end{eqnarray}
The fit for the Wolfenstein parameters defined above gives
\begin{eqnarray}
\lambda = 0.2253 \pm 0.0007\;,\;\;\;\; A =
0.808^{+0.022}_{-0.015}\;,\nonumber \\
\bar{\rho} = 0.132 ^{+0.022}_{-0.014}\;,\;\;\;\; \bar{\eta} = 0.341\pm 0.013\;.
\end{eqnarray}
We can use this to estimate the CP violating phase, $\delta$ to be
$\approx 68.8^\circ$.

The Particle Data Group world average values,
including errors, for the absolute values of the matrix
elements of the CKM matrix are~\cite{Nakamura:2010zzi}:
\begin{eqnarray}
V &=& \left(\begin{array}{ccc} 
    0.97425 \pm 0.00022 & 0.2252 \pm 0.0009 & (3.89 \pm 0.44) \times 10^{-3}\\
    0.230 \pm 0.011 & 1.023 \pm 0.036  & (40.6 \pm 1.3) \times 10^{-3} \\
    (8.4 \pm 0.6) \times 10^{-3} &   (38.7 \pm 2.1) \times 10^{-3} & 0.88 \pm 0.07   \\ 
\end{array}\right)\;.
\end{eqnarray}

\chapter{Supplementary material for $E_T$ resummation}
\section{Relation of $E_T$ resummation to $q_T$ resummation}\label{app:etres1}
Here we demonstrate the equivalence of transverse energy and transverse momentum resummation at $\mathcal{O}(\as)$.  Expanding Eq.~(\ref{formfact}) to this order, using (\ref{eq:nllint}) and substituting into (\ref{resgen}) and (\ref{eq:Wab}), we find terms involving the integrals
\beq\label{eq:Ip}
{\cal I}_p(Q,E_T) = \frac 1{2\pi}\int_{-\infty}^{+\infty} \mrd\tau \; {\rm e}^{-i\tau E_T} 
\ln^p\left(\frac{Q\tau}{i\tau_0}\right)\;,
\eeq
with $p=1,2$.  At this order, evaluating the PDFs at the scale $i\tau_0/\tau$ leads to single-logarithmic terms of the same form when we use (\ref{eq:fabi}) to write
\beq\label{eq:fQ}
f_{a/h}(x,i\tau_0/\tau)) = f_{a/h}(x,Q) -\frac{\as}{\pi}
\ln\left(\frac{Q\tau}{i\tau_0}\right)\sum_b \int_x^1 \frac{\mrd z}z P_{ab}(z) f_{b/h}(x/z,Q)\;.
\eeq

The integral (\ref{eq:Ip}) may be evaluated from
\beq\label{eq:IpIu}
{\cal I}_p(Q,E_T) = \frac{\mrd^p}{\mrd u^p}{\cal I}(Q,E_T;u)|_{u=0}\;,
\eeq
where
\beq\label{eq:Iu}
{\cal I}(Q,E_T;u) = \frac 1{2\pi}\int_{-\infty}^{+\infty} \mrd\tau \; {\rm e}^{-i\tau E_T}
\left(\frac{Q\tau}{i\tau_0}\right)^u \;.
\eeq
Writing $\tau=iz/E_T$, we have
\beq
{\cal I}(Q,E_T;u) = -\frac i{2\pi E_T}\left(\frac Q{E_T\tau_0}\right)^u
\int_{-i\infty}^{+i\infty} \mrd z\; z^u\,{\rm e}^z\;.
\eeq
We can safely deform the integration contour around the branch cut along the negative real axis to obtain
\beq\label{eq:Iufin}
{\cal I}(Q,E_T;u) = -\frac 1{\pi E_T}\left(\frac Q{E_T\tau_0}\right)^u
\sin(\pi u)\,\Gamma(1+u)\;,
\eeq
which, recalling that $\ln\tau_0 =-\gamma_E = \Gamma'(1)$, gives
\beq\label{eq:I12}
{\cal I}_1(Q,E_T) = -\frac 1{E_T}\;,\quad
{\cal I}_2(Q,E_T) = -\frac 2{E_T}\ln\left(\frac Q{E_T}\right)\;.
\eeq

The resummed component of the transverse momentum ($q_T$) distribution takes the form:
\beeq
\label{eq:qtres}
\left[ \frac{\mrd \sigma_{F}}{\mrd Q^2\;\mrd q_T} \right]_{\res} &=& q_T\sum_{a,b}
\int_0^1 \mrd x_1 \int_0^1 \mrd x_2 \int_0^{\infty} \mrd b\,b \,J_0(bq_T)
\;f_{a/h_1}(x_1,b_0/b) \; f_{b/h_2}(x_2,b_0/b) \nn \\
&\times& \overline W_{ab}^{F}(x_1 x_2 s; Q,b)\;,
\eeeq
where $b_0=2\exp(-\gamma_E)$,
\beeq 
\label{eq:qtWab}
\overline W_{ab}^{F}(s; Q,b) &=& \sum_c \int_0^1 \mrd z_1 \int_0^1
\mrd z_2 
\; C_{ca}(\as(b_0/b), z_1) \; C_{{\bar c}b}(\as(b_0/b), z_2)
\; \delta(Q^2 - z_1 z_2 s) \nn\\
&\times& \sigma_{c{\bar c}}^F(Q,\as(Q)) \;\overline S_c(Q,b)\;,
\eeeq
and
\beq
\label{eq:qtff}
\overline S_c(Q,b) = \exp \left\{-2\int_{b_0/b}^Q \frac{\mrd q}q 
\left[ 2A_c(\as(q)) \;\ln \frac{Q}{q} + B_c(\as(q)) \right]\right\} \;.
\eeq
Expanding to $\mathcal{O}(\as)$, we find the same terms as in the $E_T$ resummation except that (\ref{eq:Ip}) is replaced by
\beq\label{eq:qTIp}
\overline {\cal I}_p(Q,q_T) =  q_T\int_0^{\infty} \mrd b\,b \,J_0(bq_T)\ln^p(Qb/b_0)\;.
\eeq
It therefore suffices to show that
\beq\label{eq:qTIpIp}
\overline {\cal I}_p(Q,q_T) =  {\cal I}_p(Q,E_T=q_T)\;,\;\;p=1,2\;.
\eeq
Now corresponding to (\ref{eq:Iu}) we have
\beq\label{eq:qTIu}
\overline {\cal I}(Q,q_T;u) =  q_T\int_0^{\infty} \mrd b\,b \,J_0(bq_T)\left(\frac{Qb}{b_0}\right)^u\;.
\eeq
Using the result:
\beq
\int_0^{\infty} dt\,t^{\mu-1} \,J_0(t) = \frac{2^\mu}{2\pi}\sin\left(\frac{\pi\mu}2\right)
\Gamma^2\left(\frac{\mu}2\right)\;,
\eeq
gives
\beq
\overline {\cal I}(Q,q_T;u) =-\frac{2}{\pi q_T}\left(\frac{2Q}{q_T b_0}\right)^u\sin\left(\frac{\pi u}2\right)
\Gamma^2\left(1+\frac u2\right)\;,
\eeq
and hence
\beq
\overline {\cal I}_1(Q,q_T) = -\frac 1{q_T}\;,\quad
\overline {\cal I}_2(Q,q_T) = -\frac 2{q_T}\ln\left(\frac Q{q_T}\right)\;,
\eeq
in agreement with  (\ref{eq:I12}) and (\ref{eq:qTIpIp}).  Notice, however, that the higher ($p>2$)
derivatives of ${\cal I}$ and $\overline {\cal I}$ differ, corresponding to the difference between $E_T$ and
$q_T$ resummation beyond ${\cal O}(\as)$.
\section{Results for LHC at 7 TeV}\label{app:etres2}

\begin{figure}[!htb]
\begin{center}
  \vspace{0.8cm}
  \includegraphics[scale=0.35, angle=90]{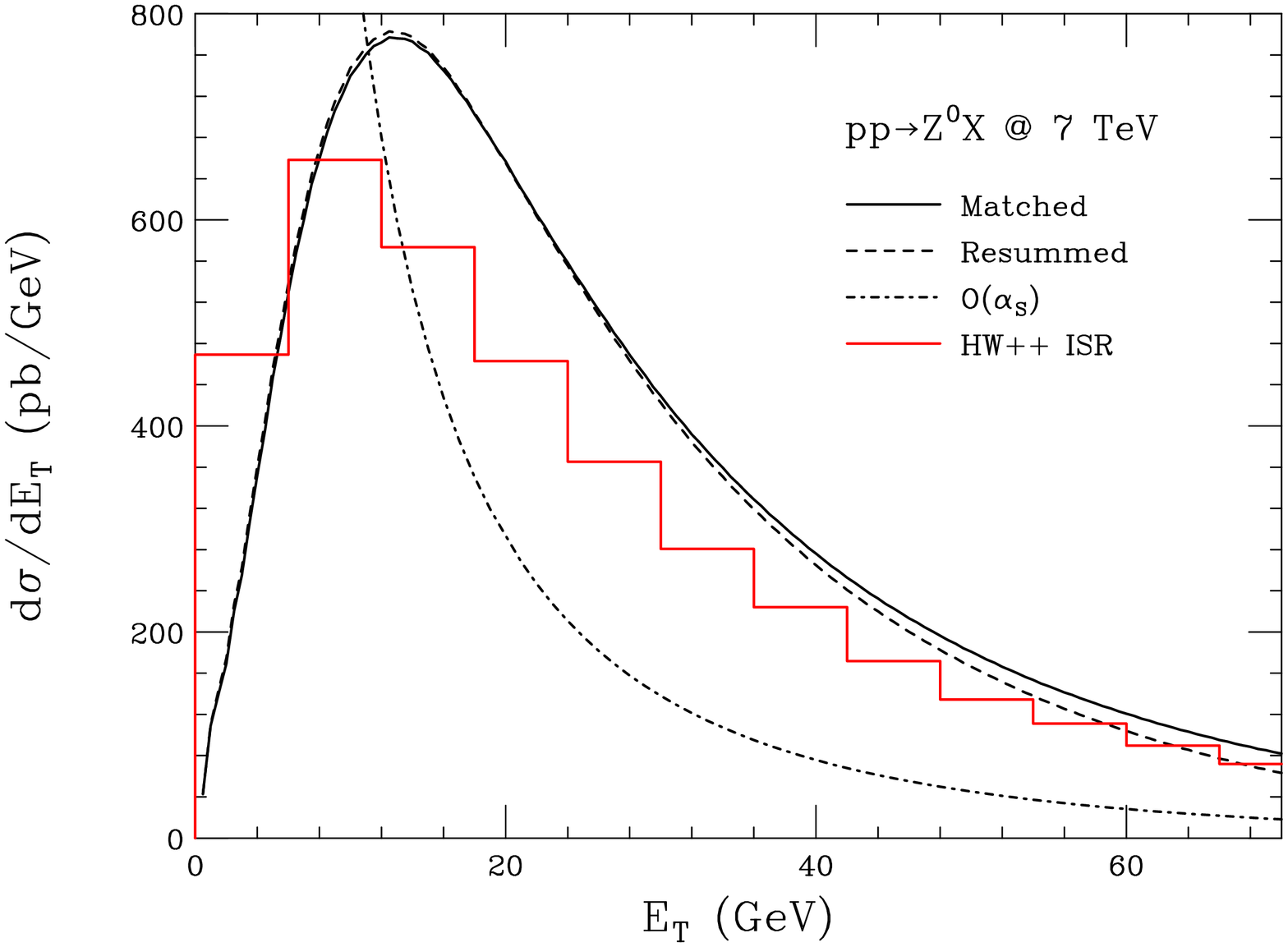}
  \hspace{1.2cm}
  \includegraphics[scale=0.35, angle=90]{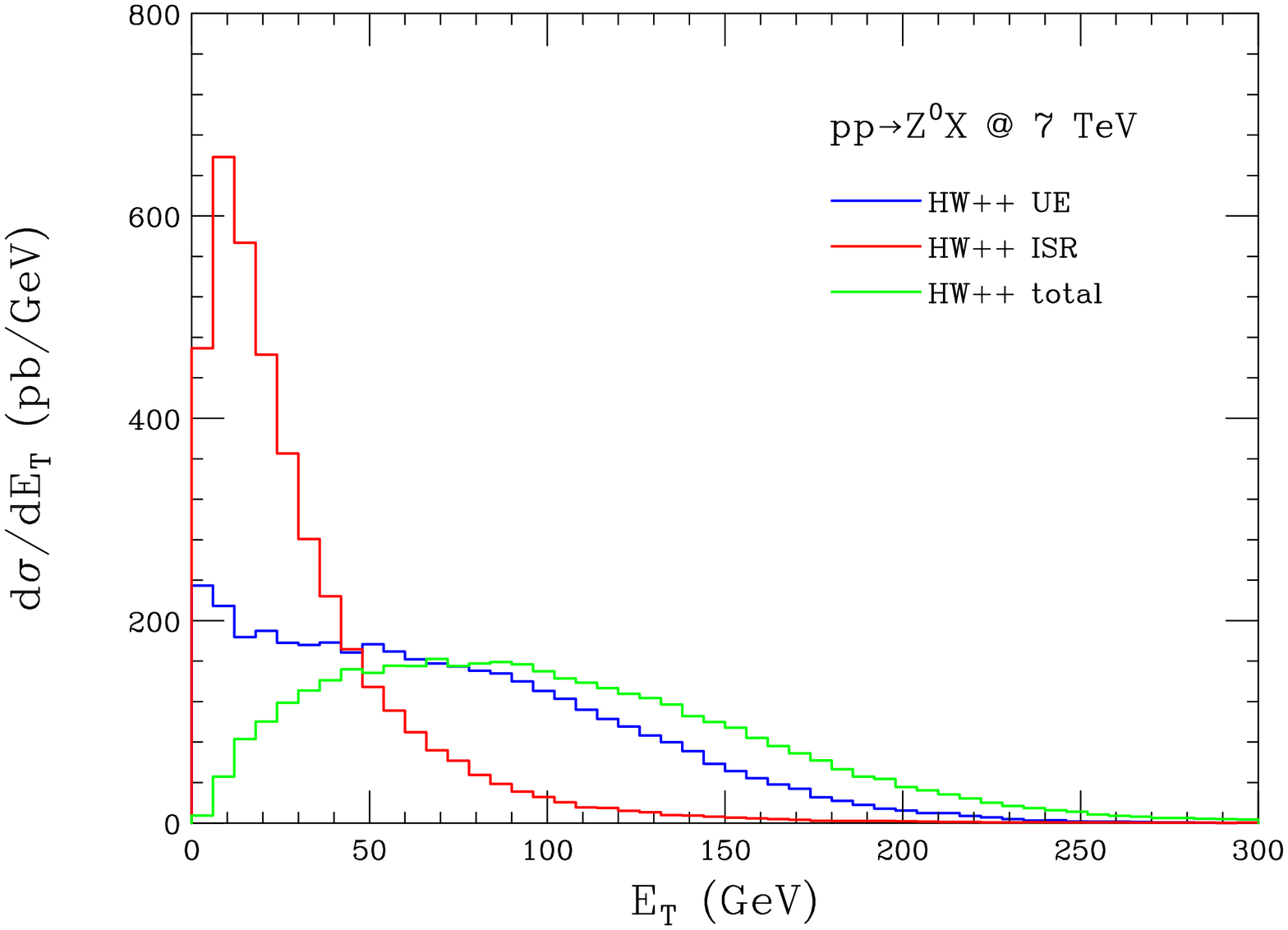}
\end{center}
\caption{Predicted $E_T$ distributions in $Z^0$ production in $pp$ collisions at $\rs =7$ TeV .
\label{fig:LHC7Z} }
\end{figure}
\begin{figure}[!htb]
\vspace{0.4cm}
\begin{center}
  \vspace{0.5cm}
  \includegraphics[scale=0.35, angle=90]{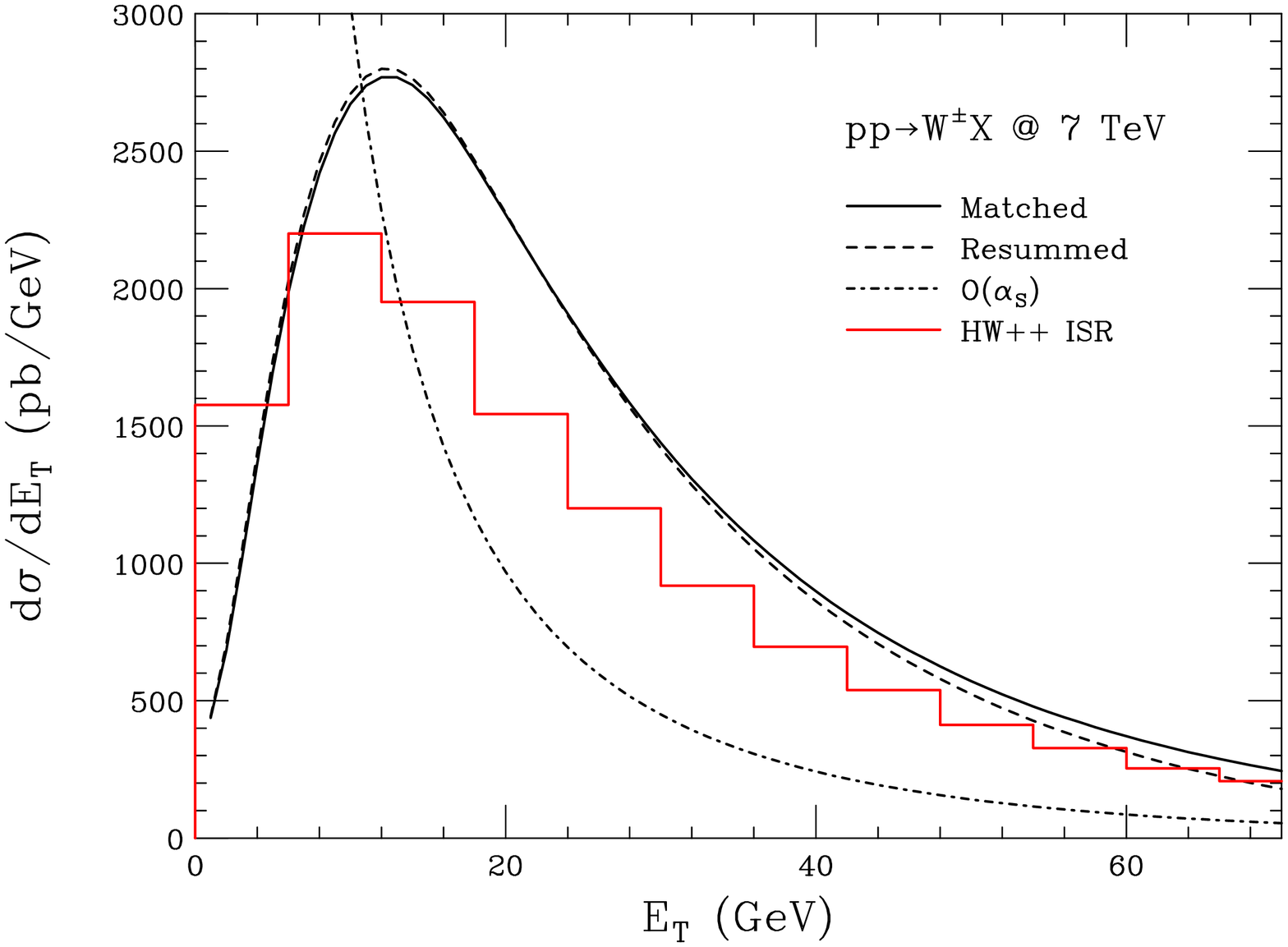}
  \hspace{1.2cm}
  \includegraphics[scale=0.35, angle=90]{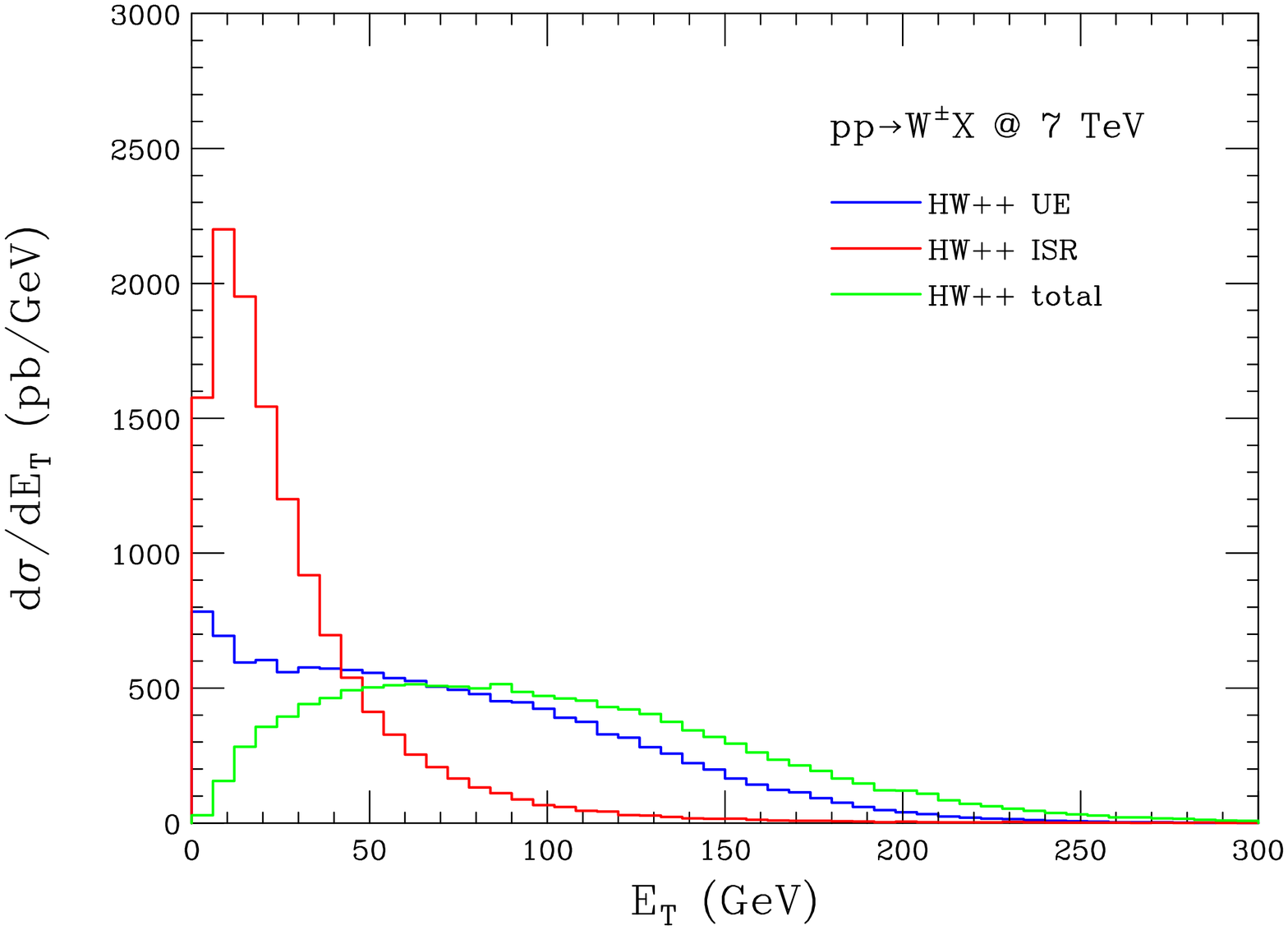}
\end{center}
\caption{Predicted $E_T$ distributions in $W^\pm$ production in $pp$ collisions at $\rs =7$ TeV .
\label{fig:LHC7W} }
\end{figure}
\begin{figure}[!htb]
\vspace{1.2cm}
\begin{center}
  \includegraphics[scale=0.35, angle=90]{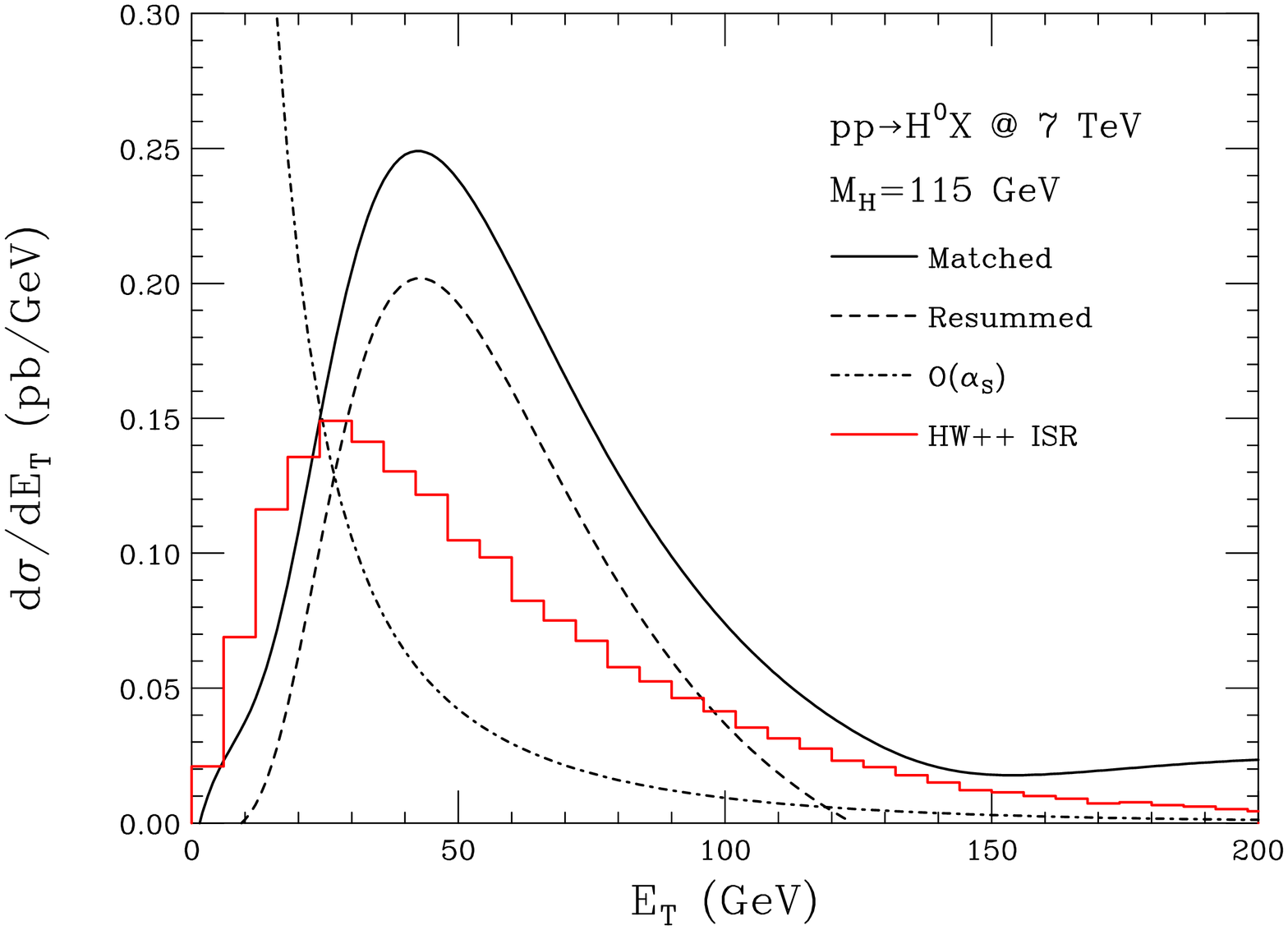}
  \hspace{1.2cm}
  \includegraphics[scale=0.35, angle=90]{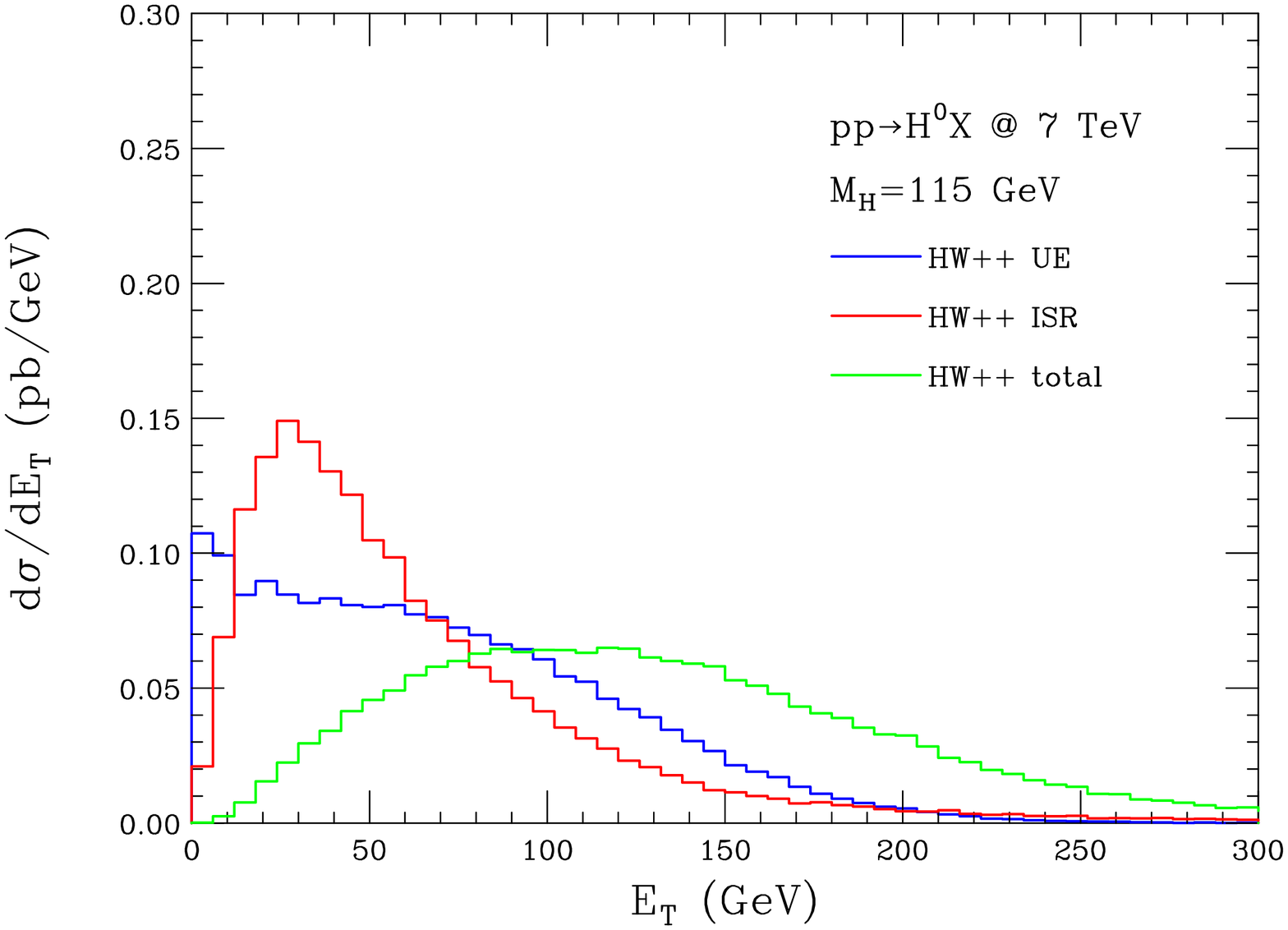}
\end{center}
\caption{Predicted $E_T$ distributions in Higgs boson production in $pp$ collisions at $\rs =7$ TeV .
\label{fig:LHC7H} }
\end{figure}
We show here results for the LHC operating at a centre-of-mass energy
of 7 TeV, corresponding to those shown earlier for 14 TeV.  Apart from
the normalisation, the predictions for the two energies are very
similar, with only a slight downward shift in the position of the peak in the
$E_T$ distribution at the lower energy.
\chapter[Supplementary material for $W'$ study]{\boldmath Supplementary material for $W'$ study}
\section{Model discrimination}\label{app:modeldiscrimination}
The search for new physics often involves the task of discriminating
between two models: one with new physics, the other without. The
actual task of discovering new physics though is laborious: one has to
understand the detector well enough and has to be able to obtain
sufficient statistics to say with certainty that something new has been observed. Here we adopt a rather theoretical approach: we describe a purely statistical method for discriminating between models~\cite{Athanasiou:2006ef, jennie}. This will  essentially yield an upper bound on the detection reach of a heavy particle: detector effects and backgrounds will result in a reduced detection limit. It is useful, however, to be aware of the theoretical possibilities for discovery.
\subsection{Likelihood ratios of probability density functions}\label{sec:likelihood}
Consider $N$ data points, of a mass variable measurement, $\{m_i\}$. Based on these data points, a theoretical model T is $R$ times more likely than another theoretical model S, if
\begin{equation}\label{eq:Rdef}
R = \frac{ p(T|\{m_i\}) }{ p(S|\{m_i\}) }\;,
\end{equation}
where $p(X|\{m_i\})$ is the probability of model X being true given the data set $\{m_i\}$. We may use Bayes' Theorem to rewrite $R$ as
\begin{equation}\label{eq:RBayes}
R = \frac{ p(\{m_i\}|T) p(T) }{ p(\{m_i\}|S) p(S) }\;,
\end{equation}
where $p(T)$ and $p(S)$ are the probabilities that S and T are true
respectively, called \textit{prior} probabilities since they represent any
previous knowledge we may possess on the theories. In the study
performed here, we assume that these quantities are equal: there is no strong evidence for either model. We may simplify Eq.~(\ref{eq:RBayes}) further:
\begin{eqnarray}\label{eq:Rexplog}
R \frac{p(S)}{p(T)} = \frac { \Pi_{i=0}^{N} p(m_i|T) } { \Pi_{j=0}^{N} p(m_j|S) } = \Pi_{i=0}^N \frac{ p(m_i|T) } { p(m_i | S) } \nonumber\\
\Rightarrow R \frac{p(S)}{p(T)} = \exp \sum_{i=0}^N \log \left( \frac{ p(m_i|T) } { p(m_i | S) } \right) \;,
\end{eqnarray}
where we have assumed that events in the data set $\{m_i\}$ are
independent. Eq.~(\ref{eq:Rexplog}) is a discrete version of the
Kullback-Leibler distance~\cite{kldistance}, a useful quantity for
comparing the relative likelihood of two theories according to a given
data sample. It is important to note that the distributions $p(m_i|T)$
and $p(m_i|S)$ are normalised to unity. This means that any difference
in the number of events predicted by the two theories will not be
taken into account. This will obviously underestimate the significance
of a difference in the number of events as predicted by the two
models, for example a substantial excess of events in an invariant
mass peak that may be present. We describe a method which takes this factor into account in the next section.
\subsection{Poisson likelihood ratios}\label{sec:poisson}
In this modification to the method described in the previous section, we simply multiply the variable $R$ defined in Eq.~(\ref{eq:Rdef}) by a ratio of Poisson distributions for the total number of events:
\begin{equation}\label{eq:Rpoissondef}
R = \frac{ p(T|\{m_i\}) }{ p(S|\{m_i\}) } \left( \frac{\bar{N_T}}{\bar{N_S}}\right)^N e^{-(\bar{N_T} - \bar{N_S} ) }\;,
\end{equation}
where $\bar{N_X} = \sigma_X . L$ is the expectation value of the
number of events according to theory X, given by the product of the
cross section, $\sigma_X$, and the integrated luminosity, $L$. This expression can be manipulated in a similar manner to Eq.~(\ref{eq:Rexplog}) to obtain
\begin{equation}\label{eq:Rpoissonexplog}
R \frac{p(S)}{p(T)} = \exp{\left(\sum_{i=0}^N \log\left( \frac{ p(m_i|T) } { p(m_i | S) } \right) \right)} \times \left( \frac{\bar{N_T}}{\bar{N_S}}\right)^N e^{-(\bar{N_T} - \bar{N_S} ) }\;,
\end{equation}
For convenience we may define the `shape' and `Poisson' factors respectively:
\begin{eqnarray}
R_S = \exp \sum_{i=0}^N \log\left( \frac{ p(m_i|T) } { p(m_i | S) } \right)\;,\nonumber \\
R_P = \left( \frac{\bar{N_T}}{\bar{N_S}}\right)^N e^{-(\bar{N_T} - \bar{N_S} ) }\;,
\end{eqnarray}
This method takes into account the difference in the total number of events expected according to each theory at the given integrated luminosity. This is accomplished by reweighing the `shape' factor $R_S$ by a factor $R_P$ which gives the ratio of probabilities to obtain the observed number of events.
\subsection{Application to a toy model}
\begin{figure}[!htb]
  \begin{center}
  \input{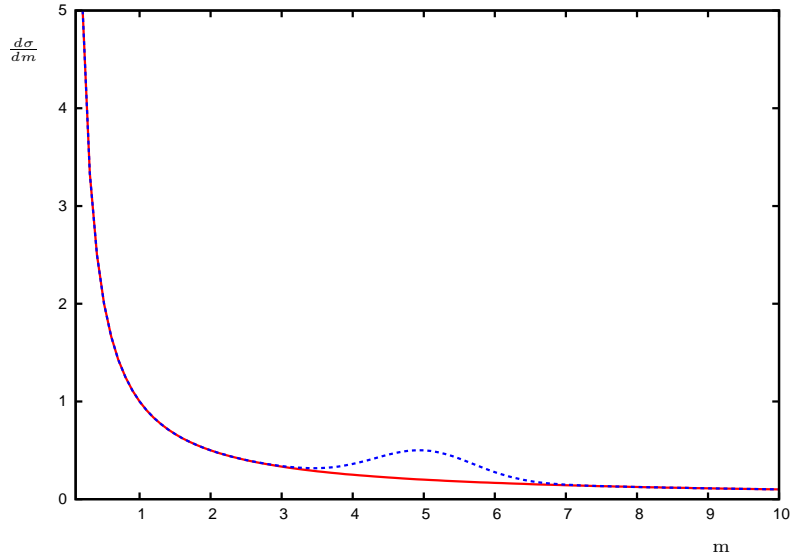}
  \end{center}
\caption{The differential cross sections $\frac{ \mathrm{d}\sigma } { \mathrm{d} m }$ according to two `toy' theories T and S are shown. Theory T possesses a Gaussian `bump', at $m = 5$ whereas S is just a falling distribution, $1/m$. $m$ is in arbitrary mass units and $\sigma$ in equivalent inverse area squared units.}
\label{fig:TS}
\end{figure}
Before applying the method to the full $W'$ model, it is instructive
to present its application to a simple model involving two analytical
`toy' distributions. Events for the two distributions have been
generated by the general Monte Carlo event generation method. The
`differential cross sections' for the two `theories' T and S with
respect to a variable $m$ in arbitrary units, defined in the range $m
\in[0.1,10]$, are given by 
\begin{eqnarray}\label{eq:TSxsection}
\frac{ \mathrm{d}\sigma_T } { \mathrm{d} m } &=& \frac{1}{m} + 0.3 e^{-(m-5)^2}\;,\nonumber\\ 
\frac{ \mathrm{d}\sigma_S } { \mathrm{d} m } &=& \frac{1}{m} \;.
\end{eqnarray}
Theory T has a Gaussian peak at $m=5$ on top of a background falling
as $\sim 1/m$ and theory S falls as $\sim 1/m$. The situation is shown
in Fig.~\ref{fig:TS}. This is qualitatively similar to the SM tail
(theory S) and the SM plus a heavy particle (theory T). The
`cross sections' in the range $m = [0.1,10]$ are $\sigma_T = 5.14$ and
$\sigma_S = 4.60$, in arbitrary area units. Assuming an integrated
`luminosity' of $L=30$ (in equivalent arbitrary inverse area units),
we have an expected number of events $\bar{N_T} = 154$ and $\bar{N_S}
= 138$. Initially, we assume that theory T is the correct underlying
theory: we produce events that are actually distributed according to
it. The result for the variable $R$ if theory T was `true' was then
found to be $R = 62$. This implies that theory T is 62 times more
likely than theory S given this specific data set. If, however, the
underlying theory is chosen to be S, then we get $R = 0.23$. This
implies that in the case that theory T is `true', it is easier to
exclude theory S than to exclude theory T in the case that theory S is
`true'. In other words, it is easier to make a discovery of a new resonance if it is
there than to exclude it if it's not.

\section[The $W'$ Drell-Yan cross section]{\boldmath The $W'$
  Drell-Yan cross section}\label{app:wprime:crosssection}
\begin{figure}[!htb]
  \begin{center}
  \input{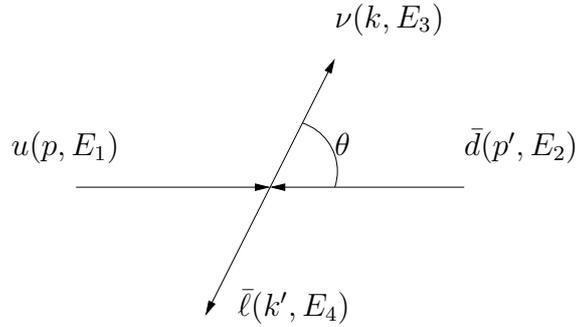}
  \end{center}
\caption{Diagram showing the kinematics for the specific case of $u \bar{d} \rightarrow W_i^+ \rightarrow \bar{\ell} \nu$ in the centre-of-mass frame. The angle $\theta$ is defined to be the scattering angle in the centre-of-mass between the $\nu$ and the $u$, both being fermions.}
\label{fig:kinematics}
\end{figure}

\begin{figure}[!htb]
  \begin{center}
  \input{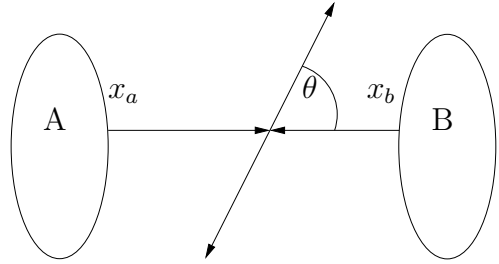}
  \end{center}
\caption{Schematic diagram showing the quark momenta fractions, $x_a$ corresponding to the quark coming from the `left' proton, A, and $x_b$ corresponding to the quark coming from the `right' proton, B.}
\label{fig:protonLR}
\end{figure}

In the present section we give details of the derivation of the leading order Drell-Yan cross
section for $pp (\bar{p}) \rightarrow W/W' \rightarrow \ell \nu X$,
given in Eq.~(\ref{eq:xsection}). 
\begin{figure}[!htb]
  \begin{center}
  \input{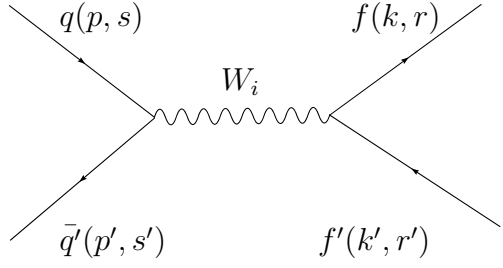}
  \end{center}
\caption{Feynman diagram for $q\bar{q}' \rightarrow W_i \rightarrow f\bar{f'} $. The quantities in the parentheses represent the 4-momentum and spin of the particle respectively.}
\label{fig:qqWff}
\end{figure}
 We reproduce the $W'$ and $W$ couplings to fermions given in
 Eq.~(\ref{eq:coupling}) (with $k_W=k_{W'} = 1$):
\begin{equation}\label{eq:couplingA}
\mathcal{L}_{W_iff'}  = \left(\frac{G_FM_W^2}{\sqrt{2}}\right)^{1/2 }V_{ff'}C_i^{\ell,q}\bar{f}\gamma_{\mu}(1 - h_i \gamma_5)f'W^{\mu}_i + \mathrm{h.c.}\;,
\end{equation}
The propagator for a massive vector boson $i$ is given by
\begin{equation}\label{eq:propagator} 
W_i^{\mu \nu}(q) = \frac{-i}{q^2 - M_i^2 + i M_i \Gamma_i} \left[ g^{\mu \nu} - \frac{ q^{\mu} q^{\nu} } { q^2 - \zeta M_i^2 } (1 - \zeta) \right]\;,
\end{equation}
where $\zeta$ is the gauge fixing parameter. Two possible gauges are
$\zeta = 1$, the Feynman gauge, and $\zeta = 0$, the Landau gauge. Any
observable quantity calculated should be independent of the gauge
fixing parameter. Here we derive the differential cross section using
arbitrary $\zeta$ to show this fact explicitly. The invariant matrix
element for the parton-level process $q\bar{q}' \rightarrow W_i \rightarrow f\bar{f'} $  (Fig.~\ref{fig:qqWff}) is given by
\begin{equation}
\mathcal{M}_i = \frac{G_F M_W^2}{\sqrt{2}} V_{ff '} V_{qq'}
C_i^\ell C_i^q \left[ \bar{f} \gmu (1 - h_i \gamma _5) f'\right] W_i^{\mu \nu} \left[\bar{q'} \gnu (1-h_i \gamma_5) q \right] \;,
\end{equation}
As stated in section~\ref{sec:wprime:refmodel}, $V_{ff '}$ is the unit
matrix when $ff'$ are leptons, $\ell \ell'$, so we set $V_{\ell \ell
  '}=1$ for the allowed lepton combinations. $V_{ff'}$ is a CKM matrix
element when $ff' = qq'$ (see appendix~\ref{sec:ckm} for the CKM matrix element values). We now form: 
\begin{eqnarray}\label{eq:mimj}
\mathcal{M}_i\mathcal{M}_j^* &=& \Omega_{qq'} \left[ \bar{f} \gmu (1 - h_j \gamma_5) f'\right] W_i^{\mu \nu} \left[\bar{q'} \gnu (1-h_j \gamma_5) q \right]\nonumber\\
&\times&\left[ \bar{f'} \gamma_{\lambda} (1 - h_i \gamma_5) f\right] W_j^{\lambda \kappa*} \left[\bar{q} \gamma_{\kappa} (1-h_i \gamma_5) q' \right] \;,
\end{eqnarray}
where we have defined the constant $\Omega_{qq'} = \frac{G_F^2 M_W^4}{2} \left|V_{qq'}\right|^2  (C_iC_j)^\ell (C_iC_j)^q$. We take the sum over the fermion spins $(s,s',r,r')$ and use the identity $\sum_{s}f^s(k) \bar{f}^s(k) = \slashed{k} \pm m$, where $f^s(k)$ are spinors representing particles (or antiparticles) of mass $m$, spin $s$ and 4-momentum $k$. We have
\begin{eqnarray}
&&\sum_{\mathrm{spins}} \left[ \bar{f'} \gamma_{\lambda} (1 - h_i \gamma_5) f\right] \left[ \bar{f} \gmu (1 - h_j \gamma_5) f'\right] \nonumber\\
&=&\sum_{\mathrm{spins}} \bar{f}'_a \gamma_{\lambda}^{ab} (1 - h_i \gamma_5)_{bc} f_c \bar{f}_d \gamma_\mu^{de} (1 - h_j \gamma_5)_{ef} f'_f \nonumber\\
&=&\slashed{k'}_{fa} \gamma_{\lambda}^{ab} (1-h_i \gamma_5)_{bc} \slashed{k}_{cd} \gamma^{de}_\mu (1-h_j \gamma_5)_{ef} \nonumber\\
&=&\mathrm{Tr}\left[\slashed{k'} \gamma_\lambda (1-h_i \gamma_5) \slashed{k} \gamma_\mu (1-h_j \gamma_5) \right]\;,
\end{eqnarray}
and similarly
\begin{eqnarray}
&&\sum_{\mathrm{spins}} \left[\bar{q} \gamma_{\kappa} (1-h_j \gamma_5) q' \right] \left[\bar{q'} \gnu (1-h_j \gamma_5) q \right] \nonumber\\
&=& \mathrm{Tr} \left[\slashed{p} \gamma_\kappa (1-h_i \gamma_5) \slashed{p'} \gamma_{\nu} (1-h_j \gamma_5)\right]\;.
\end{eqnarray}
In the above, we have neglected all fermion masses. Putting everything together we obtain
\begin{eqnarray}\label{eq:beforeform}
\sum_{\mathrm{spins}}\mathcal{M}_i\mathcal{M}_j^* &=& \Omega_{qq'} \mathrm{Tr}\left[\slashed{k'} \gamma_\lambda (1-h_i \gamma_5) \slashed{k} \gamma_\mu (1-h_j \gamma_5) \right] W_j^{\lambda \kappa*} \nonumber\\
&\times&   \mathrm{Tr} \left[\slashed{p} \gamma_\kappa (1-h_i \gamma_5) \slashed{p'} \gamma_{\nu} (1-h_j \gamma_5)\right] W_i^{\mu \nu}\;.
\end{eqnarray}
We can simplify Eq.~(\ref{eq:beforeform}) by performing the traces and contracting, using the mathematical package \texttt{FORM}~\cite{Vermaseren:2000nd}. We obtain
\begin{eqnarray}\label{eq:afterform}
\sum_{\mathrm{spins}}\mathcal{M}_i\mathcal{M}_j^* &=& \Omega_{qq'} \frac{ ( \hat{s} - M_i^2 ) ( \hat{s}- M_j^2 ) + M_i M_j \Gamma_i \Gamma_j } { \left[(\hat{s} - M_i^2)^2 + M_i^2 \Gamma_i^2 \right] \left[i \rightarrow j \right] } \nonumber\\
&\times&  \left[ 8 (1 + h_i h_j)^2 [ (\hat{t} + \hat{s})^2 + \hat{t}^2 ] + 8 \hat{s} ( h_i + h_j )^2 (2 \hat{t} + \hat{s}) \right]\;,
\end{eqnarray}
where $\hat{s} = (p + p')^2$ is the square of the quark centre-of-mass
energy and $\hat{t} = (p - k)^2$. It is reassuring that the
gauge-fixing parameter $\zeta$ does not appear in
Eq.~(\ref{eq:afterform}), as it should not have any physical
significance.
We now consider the kinematics, a schematic diagram of which is shown in Fig.~\ref{fig:kinematics}, where $p^{\mu} = (\left|\vec{p}\right|, \left|\vec{p}\right|, 0, 0)$ for the $u$-type quark, $k^\mu = (|\vec{k}|,|\vec{k}| \cos \theta, |\vec{k}| \sin \theta, 0)$ for the outgoing neutrino, $\nu$. The angle $\theta$ is defined between the $u$-type quark and the neutrino. We have $\hat{s} = (p+p')^2 = 2 p \cdot p'$ and $\hat{t} = ( p - k )^2 = - 2 p \cdot k$. Since we have neglected fermion masses $\left|\vec{p}\right|= |\vec{k}| = \sqrt{\hat{s}}/2$ and hence $\hat{t} = - 2 |\vec{k}| \left|\vec{p}\right| (1 - \cos \theta) = 
- \frac{\hat{s}}{2} (1- \cos \theta)$. Finally, we obtain
\begin{eqnarray}
&&\hat{t}^2 + (\hat{t} + \hat{s})^2 = \frac{\hat{s}^2}{2} (1 + \cos ^2 \theta)\;,\nonumber\\
&&\hat{s} ( 2 \hat{t} + \hat{s} ) = \hat{s}^2 \cos \theta\;.
\end{eqnarray}
With these relations at hand and by using $z = \cos \theta$, we can rewrite $\sum_{\mathrm{spins}}\mathcal{M}_i\mathcal{M}_j^*$ as
\begin{eqnarray}\label{eq:mimjmanip}
\sum_{\mathrm{spins}}\mathcal{M}_i\mathcal{M}_j^* &=& \sum_{qq'} \frac{\Omega_{qq'}}{3} \frac{ ( \hat{s} - M_i^2 ) ( \hat{s}- M_j^2 ) + M_i M_j \Gamma_i \Gamma_j }{ \left[(\hat{s} - M_i^2)^2 + M_i^2 \Gamma_i^2 \right] \left[i \rightarrow j \right] } \nonumber\\
&\times&  \hat{s}^2 \left[ (1+ h_i h_j)^2 (1 + z^2) + 2 ( h_i + h_j )^2 z\right]\;.
\end{eqnarray}
We have averaged over \textit{initial} spins by multiplying by $1/4$ and divided by $3$ to account for the fact that the fusing quark and anti-quark must form a colour singlet. We have also summed over all possible quark flavour combinations, $qq'$. Using Eq.~(\ref{eq:mimjmanip}) we may now write the total matrix element squared as $\left|\mathcal{M}\right|^2 = \left|\mathcal{M}_W\right|^2 + \mathcal{M}_W^* \mathcal{M}_{W'} + \mathcal{M}_{W'}^* \mathcal{M}_{W} +  \left|\mathcal{M}_{W'}\right|^2$. It is not difficult to see that we may group the $z$-even and $z$-odd terms into the functions $S(\hat{s})$ and $A(\hat{s})$, defined in Eqs.~(\ref{eq:Sterm}) and~(\ref{eq:Aterm}) respectively.
If the collision had involved only quarks of constant centre-of-mass
energy then we would simply plug the matrix element squared at parton level into the expression for the $2\rightarrow 2$ scattering differential cross section:
\begin{equation}
\frac{\mathrm{d} \sigma}{\mathrm{d} z \mathrm{d} \phi} = \frac{1}{64 \pi ^2 \hat{s}} \left|\mathcal{M}\right|^2\;.
\end{equation}
Integrating over $\phi$ would give
\begin{equation}
\frac{\mathrm{d} \sigma}{\mathrm{d} z} = \frac{1}{32 \pi \hat{s}} \left|\mathcal{M}\right|^2\;.
\end{equation}
In a collision which involves a quark $q$ and an anti-quark $\bar{q}'$, each can come from either of the two hadrons. Let us identify the two hadrons to `left' (hadron A) when moving in the positive $z$-direction and `right' (hadron B) when moving in the negative $z$-direction. Since we do not have any information about which quark came from which hadron, we have to include both possibilities in the calculation. If the quark $q$ comes from hadron A, and the anti-quark $\bar{q'}$ comes from hadron B then the definition of the angle $\theta$, and hence $z$, remains unchanged. Contrariwise, if $q$ comes from A and $\bar{q}'$ comes from B, we have to take $z \rightarrow -z$ in our expressions to take into account the fact that the $z$-axis definition would change. A schematic diagram can be seen in Fig.~\ref{fig:protonLR}. 
Consider the prior case first. Note that if we consider the lab frame
collision of the quarks, where $p^\mu_{q,\mathrm{lab}} = \frac{\sqrt{s}}{2}
(x_a, 0, 0, x_a)$ and $p'^\mu_{\bar{q}',\mathrm{lab}} = \frac{\sqrt{s}}{2}
(x_b, 0, 0, - x_b)$, we may write $\hat{s}$ as $\hat{s} = (p+p')^2 =
\frac{s}{4} \left[ (x_a+x_b)^2 - (x_a-x_b)^2\right] = x_a x_b s$,
where $s$ is the hadron centre-of-mass collision energy. The quarks
are of course confined in the hadrons and possess a distribution of
momenta, distinct for each quark flavour. We denote the differential
cross section $\sum_{qq'} q\bar{q'} \rightarrow W/W' \rightarrow
f\bar{f'} $ by $\mathrm{d} \hat{\sigma}$ and the full hadronic
differential cross section, including the PDFs, by $\mathrm{d}
\sigma$. To obtain the full hadronic differential cross section we
multiply by the parton density functions, and
integrate over the momentum fractions $x_a$ and $x_b$, using a delta
function which ensures that the quarks have the correct centre-of-mass energy, $\delta (\hat{s} - x_a x_b s)$. For the case of $q$ coming A and $\bar{q'}$ from B:
\begin{eqnarray}\label{eq:fullhadronLR}
\frac{\mathrm{d} \sigma_{LR}}{\mathrm{d} z \mathrm{d} \hat{s}} &=&
\int_0^1 \mathrm{d} x_a \mathrm{d} x_b \delta (\hat{s} - x_a x_b s) f_{q/A}(x_a, \hat{s}) f_{q'/B}(x_b, \hat{s})
\frac{\mathrm{d} \hat{\sigma}}{\mathrm{d} z \mathrm{d} \hat{s}} \nonumber\\
&=& \sum_{qq'} \frac{\Omega_{qq'}}{96} \int_0^1 \mathrm{d} x_a \mathrm{d} x_b \delta (\hat{s} - x_a x_b s) f_{q/A}(x_a, \hat{s}) f_{q'/B}(x_b, \hat{s})\left[ S(\hat{s}) (1+z^2) + 2A(\hat{s}) z \right]  \;.\nonumber\\
\end{eqnarray}
If we now consider the case when the $q$ comes from B and $\bar{q}'$ from A, we have to take $z \rightarrow -z$. This will not change the $z$-even factor, but \textit{will} change the $z$-odd factor:
\begin{eqnarray}\label{eq:fullhadronRL}
\frac{\mathrm{d} \sigma_{RL}}{\mathrm{d} z \mathrm{d} \hat{s}} = \sum_{qq'} \frac{\Omega_{qq'}}{96} \int_0^1 \mathrm{d} x_a \mathrm{d} x_b \delta (\hat{s} - x_a x_b s)f_{q'/A}(x_a, \hat{s}) f_{q/B}(x_b, \hat{s}) \left[ S(\hat{s}) (1+z^2) - 2A(\hat{s}) z \right] \;.\nonumber\\
\end{eqnarray}
To combine the two results into the full differential cross section,
we combine the PDFs into an even and an odd function respectively:
\beq
G^{\pm}_{qq'} = \left[f_{q/A}(x_a, \hat{s})f_{q'/B}(x_b, \hat{s}) \pm
  f_{q/B}(x_b, \hat{s})f_{q'/A}(x_a, \hat{s})\right]\;.
\eeq
This results in the following expression:
\begin{eqnarray}\label{eq:fullhadronnonint}
\frac{\mathrm{d} \sigma}{\mathrm{d} z \mathrm{d} \hat{s}} = \sum_{qq'} \frac{\Omega_{qq'}}{96} \int_0^1 \mathrm{d} x_a \mathrm{d} x_b \delta (\hat{s} - x_a x_b s) \left[ S(\hat{s}) (1+z^2) G^{+}_{qq'} + 2A(\hat{s}) z G^{-}_{qq'} \right ]\;.\nonumber\\
\end{eqnarray}
Since $\hat{s} = x_a x_b s$, we have $\tau = x_a x_b$ and by defining
the boson (or dilepton) rapidity $y \equiv \frac{1}{2} \log(\frac{E+p_z}{E-p_z}) = \frac{1}{2} \log(x_a/x_b)$, we finally arrive at the full hadronic differential cross section:
\begin{equation}\label{eq:fullhadronic}
\frac{\mathrm{d} \sigma}{ \mathrm{d} \tau \mathrm{d} y \mathrm{d} z} =
\frac{G_F^2M_W^4}{192 \pi} \sum_{qq'}\left|V_{qq'}\right|^2 \left[S G^{+}_{qq'}(1+z^2) + 2AG^{-}_{qq'}z\right]\;,
\end{equation}
where we have set $C^{\ell,q}_i = 1$. This is exactly what was given in section~\ref{sec:wprime:refmodel}, Eq.~(\ref{eq:xsection}).
\chapter{Supplementary material for leptoquark study}
\label{sec:lqappendix}
\section{Feynman rules and diagrams}\label{app:feynman}

\begin{figure}[!htb]
\begin{tabular}{cc}
\begin{minipage}{0.5\hsize}
\begin{center}
  \leavevmode
    \includegraphics[scale=1.00]{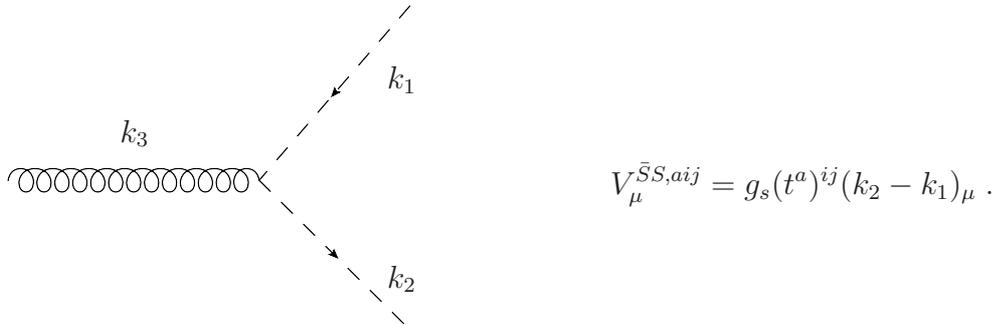}
\put(-110,70){$k_3$}
\put(-10,90){$k_1$}
\put(-10,15){$k_2$}

\end{center}
\end{minipage}
\begin{minipage}{0.5\hsize}
\begin{center}
\begin{eqnarray}
V^{\bar{S}{S},aij}_{\mu} = g_s (t^a)^{ij} ( k_2 - k_1 )_\mu \;.\nonumber
\end{eqnarray}
\end{center}
\end{minipage}
\end{tabular}
\caption{Feynman rule for the vertex scalar leptoquark-scalar anti-leptoquark-gluon. All momenta are incoming and arrows indicate colour flow.}
\label{fig:scalarfeyngss}
\end{figure}

\begin{figure}[!htb]
\begin{tabular}{cc}
\begin{minipage}{0.5\hsize}
\begin{center}
  \leavevmode
    \includegraphics[scale=0.85]{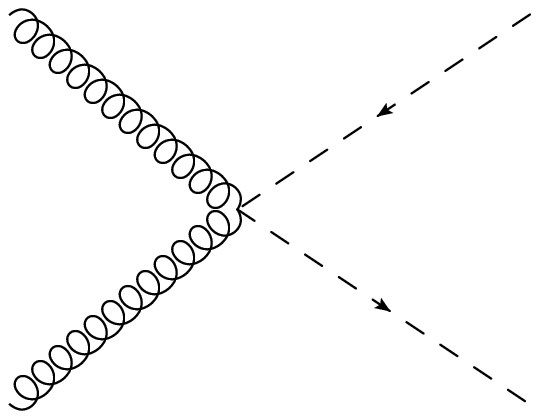}
\put(-135,85){$p_3$}
\put(-135,20){$p_4$}
\put(0,85){$p_1$}
\put(0,20){$p_2$}
\end{center}
\end{minipage}
\begin{minipage}{0.5\hsize}
\begin{center}
\begin{eqnarray}
W^{\bar{S}Sgg,ija_1a_2}(p_1,p_2,p_3,p_4) &=& g_s^2 ( t^{a_1}t^{a_2} +
t^{a_2} t^{a_1})^{ij} \nonumber\\
&\times& g_{\mu _1 \mu_2}\;.\nonumber
\end{eqnarray}
\end{center}
\end{minipage}
\end{tabular}
\caption{Feynman rule for the vertex scalar leptoquark-scalar anti-leptoquark-gluon-gluon. All momenta are incoming.}
\label{fig:scalarfeynggss}
\end{figure}

\begin{figure}[!htb]
 \centering 
    \includegraphics[scale=1.00]{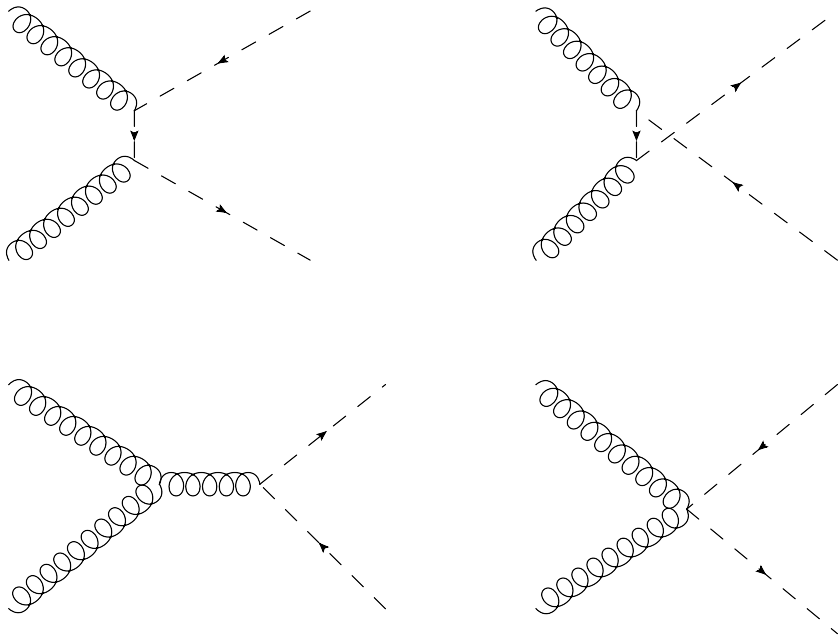}
\caption{Feynman diagrams relevant to scalar leptoquark pair-production with gluon-gluon initial states.}
\label{fig:ggfeynpair}
\end{figure}

\begin{figure}[!htb]
 \centering 
    \includegraphics[scale=1.00]{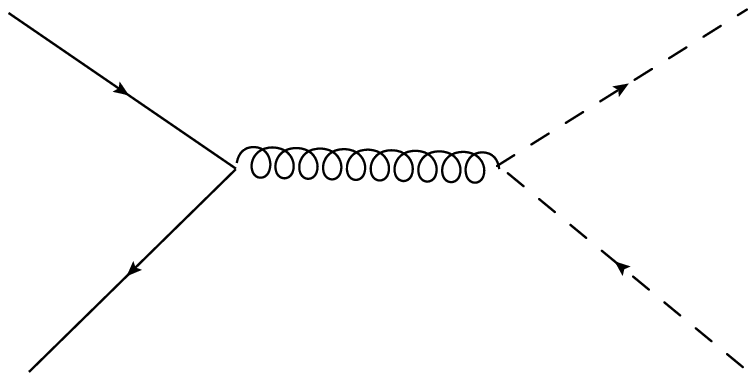}
\caption{Feynman diagram relevant to scalar leptoquark pair-production with guark-anti-quark initial states.}
\label{fig:qqfeynpair}
\end{figure}

The Feynman rules~\cite{Blumlein:1996qp} relevant to the leptoquark
pair-production diagrams are given in Figs.~\ref{fig:scalarfeyngss}
and~\ref{fig:scalarfeynggss}. The relevant parton-level Feynman diagrams are shown in Figs.~\ref{fig:ggfeynpair} and~\ref{fig:qqfeynpair} for gluon-gluon and quark-anti-quark initial states respectively. 


\begin{figure}[!htb]
 \centering 
    \includegraphics[scale=1.00]{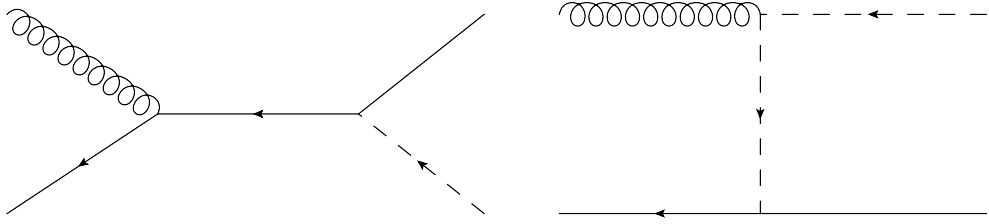}
\caption{Feynman diagrams relevant to scalar leptoquark single production. Solid lines with an arrow indicate quarks, lines without an arrow indicate leptons.}
\label{fig:singlelq}
\end{figure}

Since the couplings to light generations are suppressed in the kind of models we are considering, leptoquark single-production in hadron colliders can proceed only via $b$-quark gluon fusion, as shown in Fig.~\ref{fig:singlelq}.  However this is also heavily suppressed due to the low $b$-quark PDF and the small couplings to fermions, and can be neglected.

\section{The effective Lagrangian for derivatively-coupled leptoquarks}\label{app:conj}
The Lagrangian for derivatively-coupled conjugate fields, which appears in Eq.~(\ref{eq:lagd}), also contains terms involving the conjugate fields, such as
\begin{equation}
\mathcal{L}_{\tilde{S}'_{1/2}} \sim \bar{t}^c_R  \gamma_\mu \tau_Lp^{\mu,q} \tilde{S}'^{(+)}_{1/2}\;.
\end{equation}
To manipulate the above expression for the case of on-shell $\tilde{S}'_{1/2}$ decays as we did in Eq.~(\ref{eq:noconjmanip}), we need to show that
\begin{eqnarray}
\bar{\Psi}^C_{R,L} \slashed{p} = m \bar{\Psi}^C_{L,R}\;,
\label{eq:conjmanip1}
\end{eqnarray}
where $\Psi$ is a 4-component spinor and $\Psi_{L,R}^C = (P_{L,R} \Psi)^C$. This can be demonstrated by using the following identities~\cite{Dreiner:2008tw}:
\begin{eqnarray}
\bar{\Psi}^C &=& - \Psi ^T C^{-1}\;, \nonumber\\
C^{-1} \gamma_\mu &=& - \gamma_\mu^T C^{-1}\;,
\end{eqnarray}
and hence
\begin{eqnarray}
\bar{\Psi}^C_{R,L} = - \left[P_{R,L} \Psi\right]^T C^{-1} \;.
\end{eqnarray}
So the necessary effective Lagrangian for the decay is given by
\begin{eqnarray}
\mathcal{L}_{eff} \sim m_t \bar{t}^c_L  \tau_L \tilde{S}'^{(+)}_{1/2} \;.
\end{eqnarray}
The full list of effective Lagrangians for the primed leptoquarks, from which the decay modes and couplings in Tables~\ref{tb:lquarksprime} and~\ref{tb:lambdasprime} can be derived, is given by
\begin{eqnarray}
\mathcal{L}_{S_0'} &=& \left[ \frac{-i }{\sqrt{2} f}  (g'_{0L} m_b+ g'_{0R} m_\tau)\right] \bar{b}_R S_0' \tau_L\nonumber\\
&+& \left[ \frac{-i }{\sqrt{2} f} ( g'_{0L} m_\tau + g'_{0R} m_b )  \right] \bar{b}_L S_0' \tau_R \nonumber \\
&+& \left[  \frac{-i }{\sqrt{2} f} (g'_{0L} m_t)\right]  \bar{t}_R S_0' \nu_{\tau,L}\;,\\
\mathcal{L}_{\tilde{S}_0'} &=&  \left[ \frac{-i }{\sqrt{2} f} (\tilde{g}'_{0R} m_t \bar{t}_L \tau_R + \tilde{g}'_{0R} m_\tau \bar{t}_R \tau_L) \tilde{S}'_0 \right]\;,\\
\mathcal{L}_{S_1'} &=&  \left[ \frac{-i }{\sqrt{2} f} \sqrt{2} g'_{1L} ( m_t \bar{t}_R \tau_L + m_\tau \bar{t}_L \tau_R ) S'^{(+)}_1 \right. \nonumber \\
&+& \frac{-i }{\sqrt{2} f} \sqrt{2} g'_{1L} m_b \bar{b}_R \nu_L S'^{(-)}_1 \nonumber \\
&+& \left. \frac{-i }{\sqrt{2} f} ( g'_{1L} m_t \bar{t}_R \nu _L - g'_{1L} m_b \bar{b}_R \tau_L - g'_{1L} m_\tau \bar{b}_L \tau_R) S'^{(0)}_1 \right] \;,\\
\mathcal{L}_{S_{1/2}'} &=& \left[ \frac{-i }{\sqrt{2} f} (h'_{1L} m_b \bar{b}^c_L \nu_L + h'_{1R} m_t \bar{t}_R^c \tau_R + h'_{1R} m_\tau \bar{t}^c_L \tau_L) S'^{(-)}_{1/2} \right. \nonumber\\
&+& \left. \frac{-i }{\sqrt{2} f} ( h'_{1L} m_b + h'_{1R} m_\tau )
  \bar{b}^c_L \tau_L S'^{(+)}_{1/2}  \right. \nonumber\\
&+&\left. \frac{-i }{\sqrt{2} f}( h'_{1L} m_\tau + h'_{1R} m_b ) \bar{b}^c_R \tau_R  S'^{(+)}_{1/2} \right] \;,\\
\mathcal{L}_{\tilde{S}_{1/2}'} &=& \left[ \frac{-i }{\sqrt{2} f} h'_{2L} m_t \bar{t}_L^c \nu_L \tilde{S}'^{(-)}_{1/2} \right. \nonumber \\
&+& \left. \frac{-i }{\sqrt{2} f}(h'_{2L} m_t \bar{t}^c_L \tau_L + h'_{2L} m_\tau \bar{t}^c_R \tau_R ) \tilde{S}'^{(+)}_{1/2} \right]\;,
\end{eqnarray}
where we have defined: $S'^{(\pm)}_1 \equiv (S'^{(1)}_1 \mp
iS'^{(2)}_1)/\sqrt{2}$ (and equivalent definitions for $\tilde{S}'^{(\pm)}_{1/2}$) and $S'^{(0)}_1 \equiv S'^{(3)}_1$. We have also used the
fact that the doublet leptoquarks may be written as a vector $S'_{1/2}
=  ( S'^{(-)}_{1/2},  S'^{(+)}_{1/2} )$. We have set the quark and
lepton couplings to equal, $g^q = g^\ell$,\footnote{The implementation in \Herwigpp version 2.5.0 also includes this simplification.} however these can be
reinstated trivially by replacing $g \rightarrow g^q$ where a quark
mass term appears and $g \rightarrow g^\ell$ where a lepton mass term appears. 

Note that terms appearing in this Lagrangian are no longer $SU(2)_L \times U(1)_Y$ gauge-invariant. This is consistent since these terms would appear in the Lagrangian after electroweak symmetry breaking and vanish as the fermion masses tend to zero. The Lagrangian is, of course, $U(1)_{em}$ gauge-invariant.

\section[$(t\tau)(t\tau)$ reconstruction method]{\boldmath $(t\tau)(t\tau)$ reconstruction method}\label{app:ttauttau} 
In terms of the momentum ratios $z_i$ defined in Eq.~(\ref{eq:zis}), the conditions for balancing the total missing transverse momentum can be written as
\beqn
z_1 &=& (\ptmx - ( z_2 -1) p_{j_2}^x - p_{\nu_l}^x) / p_{j_1}^x + 1, \label{eq:z1}\\
p_{j_1}^y p_{\nu_l}^x - p_{j_1}^x p_{\nu_l}^y  &=&
\ptmx p_{j_1}^y - \ptmy p_{j_1}^x + (z_2 -1) (p_{j_1}^x p_{j_2}^y - p_{j_1}^y p_{j_2}^x) .
\label{eq:lin1}
\eeqn
The mass-shell conditions, except for $p_{\nu_l}^2=0$, can be written as
\beqn
m_W^2 &=& (p_l + p_{\nu_l})^2 = 2 p_l \cdot p_{\nu_l}\;,  \label{eq:lin2} \\
m_t^2 &=& (p_b + p_l + p_{\nu_l})^2 = m_W^2 + m_b^2 + 2 p_b \cdot p_{l} + 2 p_b \cdot p_{\nu_l} \label{eq:lin3}\;, \\
m^2_{S_0} &=& (p_t + p_{\tau_1})^2 = \tilde m_t^2 + 2 z_1 p_t \cdot p_{j_1} \label{eq:ms1} \;,\\
m^2_{S_0} &=& (p_b + p_l + p_{\nu_l} + p_{\tau_2})^2 = m_t^2 + 2 z_2 (p_b + p_l) \cdot p_{j_2}
+ 2 z_2 p_{j_2} \cdot p_{\nu_l} \label{eq:ms2} \;,
\eeqn
where $\tilde{m}_t$ is the reconstructed mass of the hadronic top and $m_t$ is the assumed mass of the semi-leptonic top.
By eliminating $z_1$ and $m_{S_0}$ from Eqs.~(\ref{eq:z1}),~(\ref{eq:ms1}) and~(\ref{eq:ms2}), 
one obtains
\beq
z_2 p_{j_2}\cdot p_{\nu_l}+
\frac{p_t \cdot p_{j_1}}{p_{j_1}^x} p_{\nu_l}^x  
= t_3 + u_3 z_2,
\label{eq:lin4}
\eeq
where
\beqn
t_3 &=& \frac{\tilde m_t^2 - m_t^2}{2} + 
\frac{\ptmx + p_{j_1}^x + p_{j_2}^x}{p^x_{j_1}}
p_t\cdot p_{j_1}\;,  \\
u_3 &=& -(p_b+p_l)\cdot p_{j_2} - \frac{p^x_{j_2}}{p_{j_1}^x} p_t \cdot p_{j_1}\;.
\eeqn
Using a vector ${\bf p}_{\nu_l} = (E_{\nu_l},~p_{\nu_l}^x,~p_{\nu_l}^y,~p_{\nu_l}^z)$,
Eqs.~(\ref{eq:lin1}),~(\ref{eq:lin2}),~(\ref{eq:lin3}) and~(\ref{eq:lin4}) can be recasted as
\beq
{\bf A} {\bf P}_{\nu_l} = {\bf S}
\label{eq:aps}
\eeq
where
\begin{eqnarray}
{\bf A} = \left(
\begin{array}{cccc}
E_l & -p_l^x & -p_l^y & - p_l^z  \\
E_b & -p_b^x & -p_b^y & -p_b^z \\
z_2 E_{j_2} & -z_2 p_{j_2}^x + (p_t \cdot p_{j_1})/p_{j_1}^x & -z_2 p_{j_1}^y & -z_2 p_{j_2}^z \\
0 & p_{j_1}^y & -p_{j_1}^x & 0
\end{array}\right),
\end{eqnarray}
and
\begin{eqnarray}
{\bf S} =
\left(\begin{array}{cccc}
\frac{m_W^2}{2}, & \frac{m_t^2 - m_b^2-m_W^2}{2} - p_b \cdot p_l, & t_3+u_3 z_2, & t_4 + u_4 z_2
\end{array}\right).
\end{eqnarray}
$t_4$ and $u_4$ are defined as
\beqn
t_4 &=& (\ptmx + p_{j_2}^x) p_{j_1}^y - (\ptmy + p_{j_2}^y) p_{j_1}^x, \\
u_4 &=& p_{j_1}^x p_{j_2}^y - p_{j_1}^y p_{j_2}^x .
\eeqn
From Eq.~(\ref{eq:aps}), we can determine ${\bf p}_{\nu_l}$
as a function of $z_2$.
Finally, $z_2$ can be determined from the mass-shell condition:
\beq
{\bf p}_{\nu_l}^2 = ( {\bf A}^{-1} {\bf S})^2 = 0 .
\eeq
This provides a quartic equation for $z_2$, and
we can find up to four real solutions in the physical range $z_2 \ge 1$. 
We can then obtain $m_{S_0}$ by substituting $z_2$ into
Eq.~(\ref{eq:ms2}). 

\section[$(q'\tau)(q \nu)$ reconstruction method]{\boldmath $(q'\tau)(q \nu)$ reconstruction method}\label{app:mminbal}
Given $w$ in Eq.~(\ref{eq:mbnu}), $m_{b\nu}(w,p_{\nu}^z)$ can be
minimised in terms of $p_{\nu}^z$. 
The result is
\beqn
[m_{b\nu}^{\rm min}(w)]^2 &=& m^2_{b\nu}(w, \tilde p_{\nu}^z)
\nonumber \\ &=&
2|{\bf p}_b| |{\bf p}_{\rm miss}-w {\bf p}_j| - 2 {\bf p}_b \cdot ({\bf p}_{\rm miss} - w{\bf p}_j)
\nonumber \\ &=&
[m^{b\nu}_{T}(w)]^2,
\label{eq:mbnut}
\eeqn
where
\beq
\tilde p_{\nu}^z \equiv \frac{|{\bf p}_{\rm miss} - w {\bf p}_j|}{|{\bf p}_b|} p_b^z\,
\eeq 
and $m^{b\nu}_{T}(w)$ is the transverse mass of the $b \nu$ system.
This allows us to calculate $M_{\rm min}$ by one-parameter minimisation:
\beq
M_{\rm min} = \min_w [ \max\{ m_{t\tau}(w), m^{b\nu}_T(w) \}]. 
\eeq
Since $m_{t\tau}(w)$ is a monotonically increasing function of $w$,
if $m_{t\tau}(0) \ge  m^{b\nu}_T(0)$:
\beq
M_{\rm min} = m_{t\tau}(0)\,.
\label{eq:mmin1}
\eeq
Furthermore, since there exists a value $\hat p_{\nu}^z$ which fulfils $m_{t\tau}(0) = m_{b\nu}(0,\hat p_{\nu}^z)$,
we find
\beq
M_{\rm min}^{\rm bal} = m_{t\tau}(0)\;.
\label{eq:mminb1}
\eeq
If $m^{b\nu}_T(0) > m_{t\tau}(0)$, we have to search for other values of $w$.
For the true $w$ and $p_{\nu}^z$, say $w^*$ and $p_{\nu}^{z*}$, we have
\beq
m^{b\nu}_T(w^*) < m_{b\nu}(w^*,p_{\nu}^{z*}) = m_{t\tau}(w^*)\,.
\eeq
This assures existence of $\hat w$ which satisfies the relation $m^{b\nu}_T(\hat w)=m_{t\tau}(\hat w)$.
By scanning $w$ from 0 to $\hat w$, one finds:
\beq
M_{\rm min}^{\rm bal} = m^{b\nu}_T(\hat w)\,,
\label{eq:mmin2}
\eeq
and
\beq
M_{\rm min} = \min_{w\in[0-\hat w]} [m^{b\nu}_T(w)]\,.
\label{eq:mminb2}
\eeq
Hence we have
\beq
M_{\rm min}^{\rm bal} \ge M_{\rm min}.
\eeq

\bibliography{references}
\bibliographystyle{utphys}

\end{document}